\documentclass[aps,prb,showpacs,onecolumn,superscriptaddress]{revtex4-2}
\usepackage{bm,color,amsmath,amssymb,mathrsfs,latexsym,graphicx,psfrag,mathtools}
\usepackage{color}
\usepackage[dvipsnames]{xcolor}
\usepackage[hidelinks,colorlinks,linkcolor=blue,
citecolor=blue,urlcolor=blue]{hyperref}
\usepackage{dsfont}

\usepackage{empheq}
\newcommand{\eqb}[1]{\begin{empheq}[box=\fbox]{align}\begin{split} #1 \end{split}\end{empheq}}

\newcommand{\bsl}[1]{\boldsymbol{#1}}

\newcommand{\shpa}{{\mkern3mu\vphantom{\perp}\vrule depth 0pt\mkern2mu\vrule depth 0pt\mkern3mu}}

\newcommand{\bra}[1]{\langle #1|}
\newcommand{\ket}[1]{|#1 \rangle}

\newcommand{\ii}{\mathrm{i}}

\newcommand{\dsZ}{\mathbb{Z}}

\newcommand{\dsR}{\mathbb{R}}

\newcommand{\Tr}{\mathop{\mathrm{Tr}}}
\renewcommand{\Re}{\mathop{\mathrm{Re}}}

\renewcommand{\O}{\mathop{\mathrm{O}}}
\newcommand{\SO}{\mathrm{SO}}
\newcommand{\SU}{\mathrm{SU}}
\newcommand{\U}{\mathrm{U}}

\newcommand{\eqnref}[1]{Eq.\,\eqref{#1}}
\newcommand{\figref}[1]{Fig.\,\ref{#1}}
\newcommand{\tabref}[1]{Tab.\,\ref{#1}}

\newcommand{\appref}[1]{supplementary information\,\ref{#1}}
\newcommand{\refcite}[1]{Ref.\,\cite{#1}}

\newcommand{\mat}[1]{\left(\begin{matrix}#1\end{matrix}\right)}

\newcommand{\eq}[1]{\begin{equation} #1 \end{equation}}

\newcommand{\eqa}[1]{\begin{align}\begin{split} #1 \end{split}\end{align}}

\let\oldAA\AA
\renewcommand{\AA}{\text{\normalfont\oldAA}}

\newcommand{\ie}{{\emph{i.e.}}}
\newcommand{\eg}{{\emph{e.g.}}}
\newcommand{\TR}{\mathcal{T}}

\newcommand{\eV}{\mathrm{eV}}

\newcommand{\Or}[1]{#1}

\newcommand{\W}{\mathcal{W}}
\newcommand{\G}{\mathcal{G}}

\renewcommand{\P}{\mathcal{P}}
\newcommand{\PT}{\mathcal{PT}}

\newcommand{\N}{\mathcal{N}}

\newcommand{\A}{\mathcal{A}}
\newcommand{\V}{\mathcal{V}}
\renewcommand{\L}{\mathcal{L}}
\newcommand{\diag}{\text{diag}}

\newcommand{\K}{\text{K}}
\newcommand{\abi}{\text{\emph{ab initio}}}

\newcommand{\mgb}{\text{MgB$_2$}}
\newcommand{\mcomega}{\left\langle \omega^2 \right\rangle}
\newcommand{\BZ}{\text{1BZ}}

\newtheorem{definition}{Definition}

\newtheorem{assumption}{Assumption}
\newcommand{\asmref}[1]{Asm.\,\ref{#1}}

\usepackage{comment}

\begin{document}
\title{Nontrivial Quantum Geometry and the Strength of Electron-Phonon Coupling}

\author{Jiabin Yu}
\affiliation{Department of Physics, Princeton University, Princeton, NJ 08544, USA}
\affiliation{Condensed Matter Theory Center and Joint Quantum Institute, Department of Physics, University of Maryland, College Park, MD 20742, USA}

\author{Christopher J. Ciccarino}
\affiliation{Department of Materials Science and Engineering, Stanford University, CA 94305, USA}
\affiliation{College of Letters and Science, University of California, Los Angeles, CA 90095, USA}

\author{Raffaello Bianco}
\affiliation{Centro de F\'isica de Materiales (CSIC-UPV/EHU), Manuel de Lardizabal pasealekua 5, 20018 Donostia/San Sebasti\'an, Spain}
\affiliation{ Ruder Boskovi\'c Institute, 10000 Zagreb, Croatia}
\affiliation{ Dipartimento di Scienze Fisiche, Informatiche e Matematiche, Universit\`a di Modena e Reggio Emilia, Via Campi 213/a I-41125 Modena, Italy}
\affiliation{ Centro S3, Istituto Nanoscienze-CNR, Via Campi 213/a, I-41125 Modena, Italy}

\author{Ion Errea}
\affiliation{Centro de F\'isica de Materiales (CSIC-UPV/EHU), Manuel de Lardizabal pasealekua 5, 20018 Donostia/San Sebasti\'an, Spain}
\affiliation{Fisika Aplikatua Saila, Gipuzkoako Ingeniaritza Eskola, University of the Basque Country (UPV/EHU), Europa Plaza 1, 20018 Donostia/San Sebasti\'an, Spain}
\affiliation{Donostia International Physics Center (DIPC), Manuel Lardizabal pasealekua 4, 20018 Donostia/San Sebasti\'an, Spain}

\author{Prineha Narang}
\affiliation{College of Letters and Science, University of California, Los Angeles, CA 90095, USA}

\author{B. Andrei Bernevig}
\email{bernevig@princeton.edu}
\affiliation{Department of Physics, Princeton University, Princeton, NJ 08544, USA}
\affiliation{Donostia International Physics Center (DIPC), Manuel Lardizabal pasealekua 4, 20018 Donostia/San Sebasti\'an, Spain}
\affiliation{IKERBASQUE, Basque Foundation for Science, Bilbao, Spain}

\begin{abstract}
The coupling of electrons to phonons (electron-phonon coupling) is crucial for the existence of various phases of matter, in particular superconductivity and density waves. Here, we devise a theory that incorporates the quantum geometry of the electron bands into the electron-phonon coupling, demonstrating the crucial contributions of the Fubini-Study metric or its orbital selective version to the dimensionless electron-phonon coupling constant. We apply the theory to two materials, graphene and MgB$_2$ where the geometric contributions account for approximately 50\% and 90\% of the total electron-phonon coupling constant, respectively. The quantum geometric contributions in the two systems are further bounded from below by topological contributions. Our results suggest that the nontrivial electron band geometry/topology might favor superconductivity with relatively high critical temperature. 
\end{abstract}
\maketitle

\section{main}

\subsection{Introduction}

\begin{figure}[h]
    \centering
    \includegraphics[width=\columnwidth]{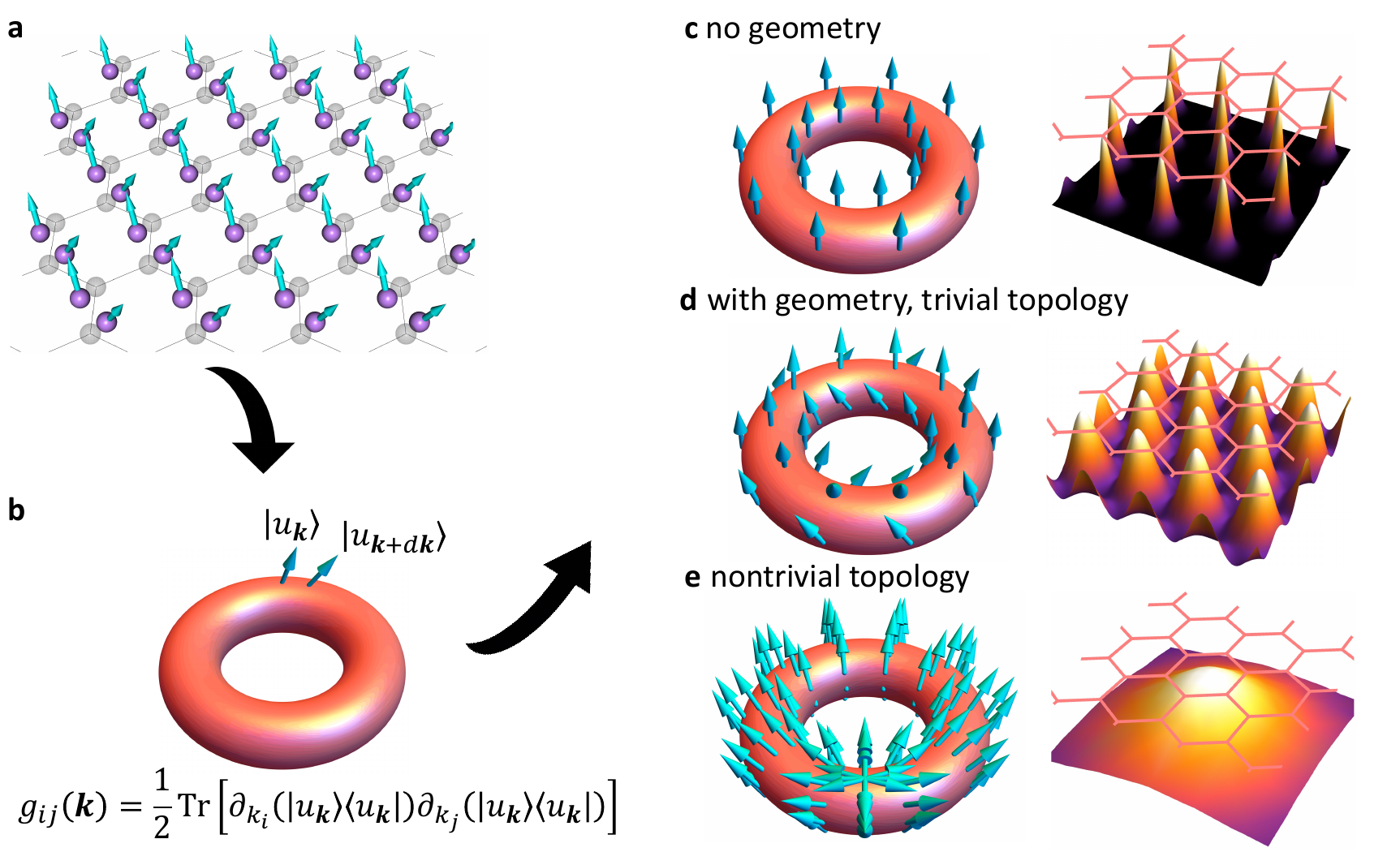}
    \caption{Quantum geometry and EPC. (a) When the ions (pink) move away from the equilibrium positions (gray) due to phonons, electrons (blue arrows) would follow the motions of ions in the tight-binding approximation owing to EPC. 
    (b) The FSM $g_{ij}(\bsl{k})$ provides a measure of quantum geometry, \ie, how the periodic part of Bloch state, $\ket{u_{\bsl{k}}}$, vary in the first Brillouin zone (1BZ, represented by the torus).
    (c) Quantum geometry can vanish (left) for trivial atomic limit (right). 
    (d) Quantum geometry must be strong (left) for obstructed atomic limit (right), even if the band topology is trivial.
    (e) Nontrivial band topology forces the quantum geometry to be strong (left), and leads to power-law decayed Wannier functions (right).
    }
    \label{main_fig:highlevel}
\end{figure}

Topology has been at the forefront of condensed matter physics for the past two decades, influencing our understanding of quantum materials and phenomena. More recently, it has however become clear and appreciated that a more general concept, that of quantum geometry, manifests itself in a series of quantum phenomena involving flat electronic bands. Nontrivial quantum geometry --- expressing change in wavefunctions under infinitesimal change in the Hamiltonian parameters such as momentum (\figref{main_fig:highlevel}(b)) --- appears naturally in multi-band systems~\cite{Provost1980FSMetric,Resta2011QuantumGeometry}. If a band is topologically nontrivial, the quantum metric is bounded from below by the topological invariant of the band (\figref{main_fig:highlevel}(e)). However, even if the band is topologically trivial, but has Wannier states that are not fully localized on the atoms (such as in the obstructed atomic limits~\cite{Bradlyn2017TQC}), the quantum geometry --- usually described up to now by the Fubini Study metric (FSM)--- can be bounded from below (\figref{main_fig:highlevel}(c-d)). For flat electronic bands --- whose flatness comes from quantum interference effects~\cite{Lieb1989LiebLattice,Mielke1991LineGraph, Dumitru2022GeneralConstructionFlatBand} --- it has been shown that the quantum geometry is directly related to superfluid weight~\cite{Torma2015SWBoundChern,Torma2016SuperfluidWeightLieb,Torma2018SelectiveQuantumMetric,Xie2020TopologyBoundSCTBG,Herzogarbeitman2021SWBound,Verma2021FlatBandSC,Rossi2019SFWTBG,Torma2020SFWTBG,Park2020SCHofBut,Herzog2022FlatBandQuantumGeometry,Huhtinen2022FlatBandSCQuantumMetric, Yu2022EOCPTBG,Herzog2022ManyBodySCFlatBand,Huang2022arXivQuantumGeometryPDW,Law2023QuantumMetricLandau,Torma2022ReviewQuantumGeometry,Chowdhury2022SFFlatBandQuantumGeometry,Chowdhury2023FlatbandSFStiffness,Torma2021SFReview} and other phenomena (such as the fractional Chern insulators~\cite{BAB2011FCI,Sondhi2013FCI,Regnault2013FCI,Roy2014FCI,Vishwanath2020FCITBG,Wang2022FCITwistedGraphene,Vishwanath2020FCITBG,Wang2022FCITwistedGraphene}, etc~\cite{Yang2020QuantumDistanceFlatBands,Mera2021FlatBandsKahler,Torma2021FlatBandBEC,Wang2021GeometryFlatBand,Rossi2022QuantumMetricExciton,Holder2022FlatBandQuantumMetricResistivity,Refael2022ShiftCurrentTBG,Oh04252022RhimQuantumDistance,Murakami2003BerryPhaseMSC,Zaletel2020AHTBG,Yu2022EOCPTBG}), mostly within contrived special models. Hence flat bands, previously thought to be detrimental to superconductivity~\cite{Basov2011HighTc}, actually have superfluid weight bound from below if topological.  Experimental investigations of these predictions are ongoing in systems such as magic-angle twisted bilayer graphene~\cite{Cao2018TBGSC,Tian2023QuantumGeoSC}.
Besides flat-band systems, the effect of quantum geometry in dispersive-band systems has also been studied, \eg, \refcite{Hosur2011PhotoBerry,Neupert2013QuantumGeometryCurrentNoise,Niu2014GeometryPositionalShift,Nagaosa2016NonlinearOpticalGeometric,Moore2017PhotoWSM,Paivi2017QuantumMetricAttractiveHubbard,Tomoki2018QuantumMetricPeriodicDrive,Li2018WSMObstructedPairing,Xiao2019QMDipole,Gianfrate01102019QuantumGeometry,Rhim08122019QuantumDistance,Moore2021IntrinsicAnomalousHall,Chen2021QuantumGeometryOpticalAnomaly,Ahn2022OpticalGeometry}.

All the previous works on quantum geometry either do not include the realistic interaction or treat the interaction strength as a tuning parameter.
Up to now, it is unknown how quantum geometry (characterized by, \eg, the FSM) affects the strength of realistic interactions. 
One main and important interaction in solids is the electron-phonon coupling (EPC), which is crucial for superconductivity~\cite{BCS1957SC,Migdal1958EPC,Eliashberg1960EPCSC} and other quantum phases.
For phonon-mediated superconductors, a large $\lambda$ typically leads to a high superconducting transition temperature $T_c$~\cite{McMillan1968SCTc,AllenDynes1975SCTc}.
Therefore, it is natural to ask how $\lambda$ is directly related to the electron band geometry---most importantly to the Fermi surface quantum geometry (characterized by, \eg, the FSM)---which is bounded by topology.
Such relation, if revealed, may help look for new superconductors, given the large number of topological materials~\cite{Bradlyn2017TQC,Po2017SymIndi,Bernevig2019TopoMat,Fang2019TopoMat,Wan2019TopoMat,Bernevig2020MTQCMat,Narang2021TopologyBands}.

In this work, we compute the contribution of electron band geometry and topology to the bulk EPC constant $\lambda$.
First, we introduce a simple (but in many cases remarkably accurate) model --- dubbed the Gaussian approximation (GA) ---  for the EPC to show its deep link to the \emph{electronic} band Hamiltonian. In this approximation, the quantum geometric contribution to $\lambda$ emerges naturally and can be differentiated from the energy dispersion contribution. In particular, we find that the either the FSM or the orbital-selective Fubini-Study metric (OFSM) directly enter the expression of EPC. We show that when the electron states on or near the Fermi surfaces exhibit topology --- such as winding numbers of the wavefunctions --- the geometric contribution (arising from O/FSM) is bounded from below by the topological contribution (arising from topological invariants).
The topological contribution serving as a lower bound of the geometric contribution is in the same spirit as the band topology serving as a lower bound of the band geometry.

To test our theory, we apply it to the EPC of two famous materials: graphene and MgB$_2$, where we find that our approximation becomes (almost) exact; we then identify the quantum geometric contributions to the bulk EPC constant $\lambda$, as well as the topological contributions that bound the geometric ones from below, in the two systems.
We further perform the {\abi} calculation, with two different methods~\cite{baroni2001,Narang2021EPCWeyl,Burch2021EPCWeyl} for {\mgb}, from which we find that the quantum geometric (topological) contribution to $\lambda$ accounts for roughly 50\% (50\%) and 90\% (43\%) of the total value of EPC constant in graphene and {\mgb}, respectively.
Beyond the GA, we introduce an alternative but similar way of identifying the quantum geometric contributions to $\lambda$ based on the symmetry representations (reps) and the short-ranged nature of the hopping, and reproduce our results. 
Since {\mgb} is a phonon-mediated superconductor with $T_c=39\K$~\cite{Jun03012001MgB2SC,Budko02262001MgB2Isotope,Jorgensen05012001MgB2Isotope}, our work on {\mgb} suggests that strong geometric properties or nontrivial topology of the electron Bloch states may favor strong EPC constant $\lambda$ and thus may favor the high superconducting $T_c$, which would serve as a guidance for future search of superconductors.

\subsection{Gaussian Approximation: Geometric Contribution to $\lambda$}
The bulk EPC constant~\cite{McMillan1968SCTc} $\lambda  = 2 \int_0^{\infty} d\omega \frac{\alpha^2F(\omega)}{\omega}$ is obtained from the Eliashberg function \cite{Eliashberg1960EPCSC} $\alpha^2F$.  It can be written as $\lambda = 2 \frac{ D(\mu)}{N} \frac{ \hbar\left\langle \Gamma \right\rangle }{\hbar^2 \mcomega}$
where $D(\mu)$ is the single-particle electron density of states at the chemical potential $\mu$, $N$ is the number of lattice sites, and $\mcomega$ is the McMillan's mean-squared phonon frequency. For a multi-band electron system, we show that the average phonon line width $\left\langle \Gamma \right\rangle$ (up to a factor of $D^2(\mu)$) is the average of 
\begin{align}
\label{main_eq:Gamma}
   &\Gamma_{nm}(\bsl{k}_1,\bsl{k}_2)  \\
&= \frac{\hbar}{2} \sum_{\bsl{\tau},i}  \frac{1}{m_{\bsl{\tau}}} \Tr\left[ P_{n}(\bsl{k}_1)  F_{\bsl{\tau}i}(\bsl{k}_1,\bsl{k}_2) P_{m}(\bsl{k}_2)   F_{\bsl{\tau}i}^\dagger(\bsl{k}_1,\bsl{k}_2) \right] \nonumber ,
 \end{align} over the Fermi surfaces. $\bsl{k}_1$ and $\bsl{k}_2$ are the Bloch momenta of electrons, $\bsl{\tau}$ is the sub-lattice vector, $m_{\bsl{\tau}}$ is the mass of the ion at $\bsl{\tau}$, $i$ labels the spatial directions of the possible ion motions, and crucially $P_{n}(\bsl{k})=U_n(\bsl{k}) U_n^\dagger(\bsl{k})$ is the projection matrix to the $n$th electron band with $U_n(\bsl{k})$ the eigenvector.
 $F_{\bsl{\tau}i}(\bsl{k}_1,\bsl{k}_2)$ in \eqnref{main_eq:Gamma} is the EPC matrix in the electron atomic basis and the ion motion basis, whose general expression can be found in Eq.\,{\color{Blue}(B41)} in supplementary information\,{\color{blue}B}.
 As embedded in 3D space, the ion can move in 3D (\ie, $i=x,y,z$) regardless of the sample dimensionality.

 For time-reversal (TR) invariant systems with negligible Coulomb interaction, we show in supplementary information\,{\color{blue}E} that the mean-field superconducting $k_B T_c \geq 1.13 \epsilon_c e^{-\frac{1}{\lambda}}$ is bounded from below by $\lambda$ regardless of the pairing function, as long as (i) the cutoff $\epsilon_c$ is much larger than the temperature and (ii) the bands cut by the Fermi energy are dispersive with a large Fermi velocity. (We note that the bound relies on the Migdal-Elishberg theory which usually holds in the weak-coupling regime. The Migdal-Elishberg theory is not necessarily reliable in the strong-coupling regime~\cite{Kivelson2018BreakDownofME,Sous2018BipolaronsPeierls,Zhang2023Bipolarons}.)
If the Coulomb interaction is considerable, $T_c$ of phonon-mediated superconductors still typically increases with increasing $\lambda$~\cite{McMillan1968SCTc,AllenDynes1975SCTc}.
 In the expression of $\lambda$, $\mcomega$ can be well approximated by certain phonon frequencies in many cases (\eg, in graphene and {\mgb}), and $D(\mu)$ only involves electrons.
 Thus the main information of the EPC is often in the average phonon line width $\left\langle \Gamma \right\rangle$. 
 To study $\left\langle \Gamma \right\rangle$, we adopt the two-center approximation~\cite{Mitra1969EPC} for the EPC: only the relative motions of two ions matter for the EPC between the electronic orbitals on those two ions.
 As a result, the EPC matrix $F_{\bsl{\tau}i}(\bsl{k}_1,\bsl{k}_2)$ has the following form (supplementary information\,{\color{blue}C}):
 \eqa{
\label{main_eq:f_k_2center}
F_{\bsl{\tau}i}(\bsl{k}_1,\bsl{k}_2) = \chi_{\bsl{\tau}} f_{i}(\bsl{k}_2) - f_{i}(\bsl{k}_1)\chi_{\bsl{\tau}} \ ,
}
where $\chi_{\bsl{\tau}}$ is a diagonal projection matrix with elements being 1 only for the electron degrees of freedom (like orbitals) at $\bsl{\tau}$.
$f_{i}(\bsl{k})$ is a matrix for the case with more than one bands, and is the quantity we want to determine (supplementary information\,{\color{blue}C}), whose deep physical origin is missing in the literature.

We now show that $f_i(\bf{k})$ is intimately related to the \emph{electronic} Hamiltonian. To show this general relation, we introduce the GA.
As a concrete simple illustration, we consider a $3$D system with only one kind of atom and one spinless $s$ orbital per atom. (See generalization in supplementary information\,{\color{blue}F} and supplementary information\,{\color{blue}H}.)
We allow multiple atoms per unit cell so that more than one electron bands can exist.
Under the two-center approximation, the non-interacting electron Hamiltonian and EPC Hamiltonian are directly given by the smooth hopping function $t(\bsl{r})$, which specifies the hopping between two $s$ orbitals separated by $\bsl{r}$.
Explicitly, the electron matrix Hamiltonian reads $\left[ h(\bsl{k}) \right]_{\bsl{\tau}\bsl{\tau}'} = \sum_{\bsl{R}}t(\bsl{R}+\bsl{\tau}-\bsl{\tau}')  e^{-\ii \bsl{k}\cdot (\bsl{R} + \bsl{\tau} - \bsl{\tau}')}$ with $\bsl{R}$ labelling the lattice vectors, and the EPC $f_{i}(\bsl{k})$ in \eqnref{main_eq:f_k_2center} reads $\left[f_{i}(\bsl{k})\right]_{\bsl{\tau}_1 \bsl{\tau}_2 } = \sum_{ \bsl{R} } e^{- \ii \bsl{k} \cdot ( \bsl{R}+ \bsl{\tau}_1 - \bsl{\tau}_2)}   \left. \partial_{r_i} t(\bsl{r})\right|_{\bsl{r} = \bsl{R}+\bsl{\tau}_1-\bsl{\tau}_2}$.
The GA assumes the hopping function to have a Gaussian form:
$t(\bsl{r}) = t_0 \exp[\gamma \frac{|\bsl{r}|^2}{2}]$,
where $\gamma<0$ is determined by the standard deviation. Usual overlaps between orbital in lattices do have exponentially decaying form; hence we expect the GA to be a qualitatively and quantitatively good description of the physics. Other powers of $|\bsl{r}|$ in the exponential are possible, and lead to generalized quantum geometric quantities, but we focus on the GA due to its simplicity. We later show it is exact in the short-range-hopping or $\bsl{k}\cdot\bsl{p}$ models of graphene and {\mgb}. 

Crucially, the GA enables us to uncover a relation between the EPC $f_{i}(\bsl{k})$ and the electron  Hamiltonian $h(\bf{k})$. As $\partial_{r_i} t({\bf{r}}) = \gamma r_{i} t(\bf{r}) $, we Fourier transform to find a simple relation between EPC and the electron Hamiltonian 
\eq{
\label{main_eq:g_Gaussian}
f_{i}(\bsl{k}) = \ii \gamma \partial_{k_i}  h(\bsl{k})\ .
}
With the spectral decomposition $h({\bf{k}}) = \sum_{n} E_n({\bf{k}}) P_{n}({\bf{k}})$ where $E_n(\bsl{k})$ is the $n$th  electron band with projection operator $P_n(\bsl{k})$, we can split the EPC $f_{i}(\bsl{k})$ into the energetic and geometric parts $f_{i}(\bsl{k}) = f_{i}^E(\bsl{k}) + f_{i}^{geo}(\bsl{k})$, where
\eqa{
\label{main_eq:g_E_g_geo_Gaussian}
&  f_{i}^E(\bsl{k}) = \ii \gamma   \sum_{n} \partial_{k_i} E_n(\bsl{k}) P_n(\bsl{k}) \\
& f_{i}^{geo}(\bsl{k}) =  \ii \gamma  \sum_{n}E_n(\bsl{k}) \partial_{k_i} P_n(\bsl{k})\ ,
}
$f_{i}^E(\bsl{k})$ is the energetic part of the EPC since it vanishes if electron bands are exactly flat.
$f_{i}^{geo}(\bsl{k})$ is the geometric part of the EPC since $f_{i}^{geo}(\bsl{k})$ relies on the momentum dependence of $P_n(\bsl{k})$; it vanishes for trivial bands with no $\bsl{k}$ dependence in their eigenstates, or for one-band systems. 
The separation \eqnref{main_eq:g_E_g_geo_Gaussian} allows us the split the bulk EPC $\lambda$ into $3$ parts $\lambda = \lambda_E + \lambda_{geo} + \lambda_{E-geo} $, where $\lambda_E$ is linked to $f_{i}^E(\bsl{k})$, $\lambda_{geo}$ to $f_{i}^{geo}(\bsl{k})$, and $\lambda_{E-geo}$ to both $f_{i}^E(\bsl{k})$ and  $f_{i}^{geo}(\bsl{k})$.
Similar to the names of $f_{i}^{E}(\bsl{k})$ and $f_{i}^{geo}(\bsl{k})$, we call $\lambda_{E}$ and $\lambda_{geo}$ the energetic and geometric contributions to the bulk EPC constant $\lambda$, respectively.
$\lambda_{E-geo}$ is not our focus in this work since it vanishes in graphene and {\mgb} under the approximation that we adopt, though $\lambda_{E-geo}$ also has geometric dependence in it.
(supplementary information\,{\color{blue}A}.)

In particular, $f_{i}^{geo}(\bsl{k})$ is responsible for leading to the FSM/OFSM in $\lambda_{geo}=\lambda_{geo,1}+\lambda_{geo,2}$, where both parts depends on geometric quantities, as discussed in supplementary information\,{\color{blue}A}.
In this work, we will mainly focus on $\lambda_{geo,1}$, since $\lambda_{geo,2}$ is restricted to zero by symmetries for graphene, and is either restricted to zero or can be converted to the same geometric expressions as $\lambda_{geo,1}$ for {\mgb}, as discussed in the next section.
Explicitly, in the two-band case, $\lambda_{geo,1}$ reads
\eqa{
\label{main_eq:lambda_geo,1}
\lambda_{geo,1}& = \frac{2\Omega \gamma^2}{ (2\pi)^3 m \mcomega} \sum_{n,i,\bsl{\tau}} \int_{FS_n} d\sigma_{\bsl{k}}\frac{\Delta E^2(\bsl{k})}{|\nabla_{\bsl{k}} E_n(\bsl{k})|}  a_{\bsl{\tau}} \left[g_{n,\bsl{\tau}}(\bsl{k}) \right]_{ii}
}
where $m$ is the mass of the ion, $\Omega$ is the volume of the unit cell, $d\sigma_{\bsl{k}}$ is the measure on the Fermi surface, $\Delta E(\bsl{k})$ is the difference between two energy bands, $FS_n$ is the Fermi surface given by $E_n(\bsl{k})=\mu$, and $a_{\bsl{\tau}} = \frac{1}{D(\mu) } \sum_m \sum_{\bsl{k}_2}^{\BZ}\delta\left(\mu - E_m(\bsl{k}_2) \right) \left[ P_{m}(\bsl{k}_2) \right]_{\bsl{\tau}\bsl{\tau}}$. 
(supplementary information\,{\color{blue}A}.)
\eq{
\left[ g_{n,\bsl{\tau}}(\bsl{k}) \right]_{ij}=  \frac{1}{2}\Tr\left[ \partial_{k_i} P_{n}(\bsl{k})  P_{n}(\bsl{k})  \partial_{k_j} P_{n}(\bsl{k}) \chi_{\bsl{\tau}} \right] +(i\leftrightarrow j)
}
is the \emph{orbital-selective} Fubini-Study metric (OFSM).
More general definitions of OFSM can be found in supplementary information\,{\color{blue}G}, and similar OFSM generalizations were proposed in \refcite{Torma2018SelectiveQuantumMetric,Herzog2022ManyBodySCFlatBand}.
When symmetries requires $a_{\bsl{\tau}}$ to be the same for all $\bsl{\tau}$ (like graphene), the OFSM would be summed over all $\bsl{\tau}$ and reduce to the conventional FSM.

Although we only discuss the GA for a 3D system with only one kind of atom and one spinless $s$ orbital per atom, the GA can be defined for more complicated cases.
We also introduce an alternative way of identifying the geometric contribution to $\lambda$ based on the symmetry reps for systems with short-range hoppings  (supplementary information\,{\color{blue}D}).
Both methods can be applied to graphene and {\mgb} and give identical results.
Moreover, we also use the most-general symmetry-allowed short-range hopping form to reproduce the results from GA in graphene and {\mgb}.

\Or{
We have not developed a completely general version of GA that is applicable to all systems.
In general, it is unlikely to cover the full \emph{ab initio} results just by allowing other powers of the distance between orbitals in the exponential or in the perfector of the exponential. 
Allowing other powers of the distance can cover the radial part of the EPC, \ie, the EPC matrix elements that correspond to the atomic motions in parallel with the hopping direction; however it cannot always cover the angular part of the EPC, \ie, the EPC matrix elements from the atomic motions in perpendicular to the hopping direction, which might be considerable when the orbitals have strong angular dependence such as $p, d, f$ orbitals. 
As discussed in the next section, graphene is special since $p_z$ orbitals are effectively $s$ orbitals in 2D, and we only need to consider the in-plane motions to the leading order, which therefore involve no angular dependence; MgB$_2$ is also special since the angular part of the EPC has the same expression as the radial part of the EPC to the leading order, which would allow us to use the GA with additional powers in the perfector to cover the whole EPC to the leading order. 
Nevertheless, this is not always true in general. 
Therefore, when studying the geometric contribution to EPC in other systems, one might need certain modification of \eqnref{main_eq:g_Gaussian} beyond what we will do for graphene and {\mgb} in the rest of this paper, and might also need to verify the results with different methods.
Nevertheless, it is, in many case,
}
possible to use certain polynomials of $\bsl{r}$ to re-express the spatial gradient of the hopping functions, which, when the hopping is short-ranged enough, would give momentum derivatives of the electron Hamiltonian after the Fourier transformation and give geometric contribution.

\subsection{Geometric Contribution to $\lambda$ in Graphene and {\mgb}}

We now apply the GA to the specific cases of graphene and {\mgb}.
With the nearest-neighboring-hopping model of graphene~\cite{Neto2009GrapheneRMP}, we find that the EPC form (\eqnref{main_eq:g_Gaussian}) derived from the GA is exact in graphene for the in-plane atom motions.
Due to the mirror symmetry that flips $z$ direction, the out-of-plane atomic motions do not couple to the electrons, and thus we find that the energetic and geometric parts of the EPC for graphene in \eqnref{main_eq:g_E_g_geo_Gaussian} are nonzero only for in-plane $i=x,y$.
Then, we obtain (supplementary information\,{\color{blue}F})
\begin{align}
\label{main_eq:lambda_E_lambda_geo_graphene}
& \lambda_E =  \frac{ \Omega \gamma^2 }{(2\pi)^2 m_{\text{C}} \mcomega}  \int_{FS} d\sigma_{\bsl{k}} |\nabla_{\bsl{k}} E_{n_F}(\bsl{k})| \\
&  \lambda_{geo}  =  \frac{ \Omega \gamma^2 }{(2\pi)^2 m_{\text{C}}  \mcomega} \int_{FS}d\sigma_{\bsl{k}} \frac{\Delta E^2(\bsl{k})}{|\nabla_{\bsl{k}} E_{n_F}(\bsl{k})|} \Tr\left[g_{n_F}(\bsl{k})\right]\nonumber \ ,
\end{align}
where $m_{\text{C}}$ is the mass of carbon atom, $E_{n_F}(\bsl{k})$ is the band that gives the Fermi surface, and $\Delta E(\bsl{k})$ is the absolute difference of two energy bands. %
Remarkably, we find that, as advertised, the FSM of the electron Bloch states--- $\left[g_{n}(\bsl{k})\right]_{ij} =  \Tr[\partial_{k_i} P_{n}(\bsl{k}) \partial_{k_j} P_{n}(\bsl{k}) ]/2$ (equal to the expression in \figref{main_fig:highlevel}(b) under the tight-binding approximation)---directly appears in the $\lambda_{geo}$.
The appearance of the FSM in \eqnref{main_eq:lambda_E_lambda_geo_graphene} comes from $a_{\bsl{\tau}} = 1/2$ in \eqnref{main_eq:lambda_geo,1} and $\lambda_{geo,2}=0$ for graphene, owing to the $C_2\TR$ and $C_3$ symmetry, respectively, where $C_n$ is the n-fold rotational symmetry along $z$-axis and $\TR$ is the time-reversal symmetry.
The symmetries of graphene also requires $\lambda_{E-geo} = 0$.
Therefore, the bulk EPC constant $\lambda$ of graphene only has the energetic and geometric contributions in \eqnref{main_eq:lambda_E_lambda_geo_graphene}, \ie, $\lambda = \lambda_E + \lambda_{geo}$ (supplementary information\,{\color{blue}F}). Analytically, we find (supplementary information\,{\color{blue}F}), $\lambda_{geo}/\lambda$ limits to exactly 50\%  as $\mu$ approaches to the energy of the Dirac points (which is zero). Remarkably, \emph{half} of the EPC strength is supported by the geometric (and as we will show, topological) properties of the graphene Bloch states.

\begin{figure}[h]
    \centering
    \includegraphics[width=\columnwidth]{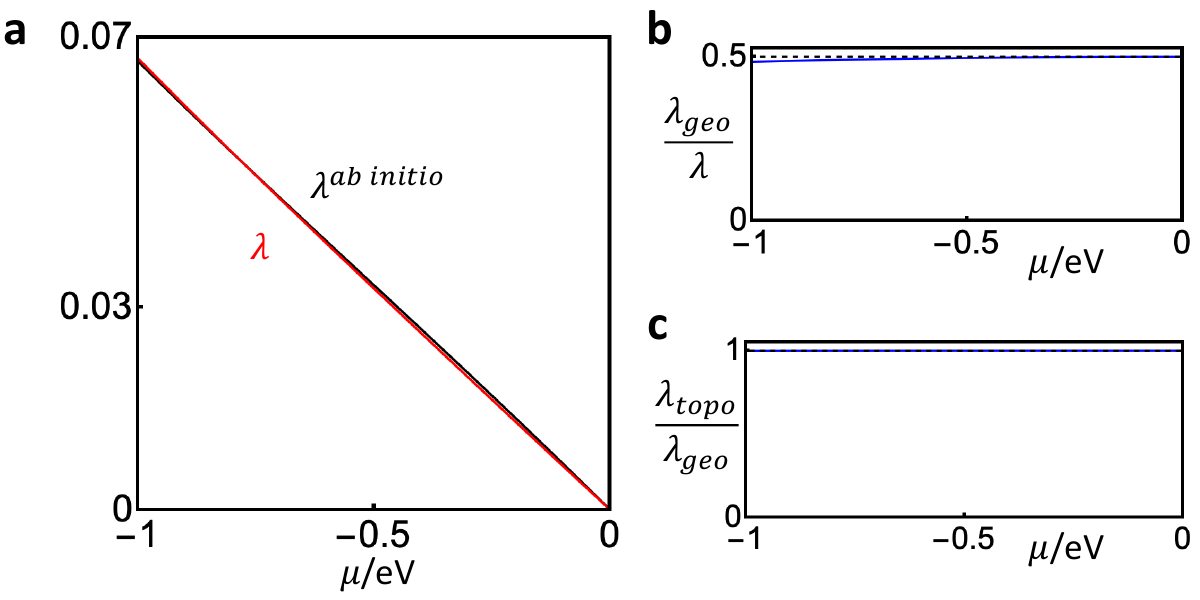}
    \caption{Plots for graphene. The chemical potential $\mu$ ranges from $-1$eV to $0$eV, while setting the Dirac-point energy to be zero.
    (a) is the plot of EPC constants from the {\abi} calculation ($\lambda^{\abi}$, black) and from \eqnref{main_eq:lambda_E_lambda_geo_graphene} ($\lambda$, red).
    (b) and (c) are the plots of $\lambda_{geo}/\lambda$ and $\lambda_{topo}/\lambda$, respectively.
    }
    \label{main_fig:graphene}
\end{figure}

We determine the numerical values of the model parameter $\gamma$ (in addition to the electron nearest-neighboring (NN) hopping and $\langle \omega^2 \rangle$) by matching our model to our ${\abi}$ calculation. (See {\abi} calculation in supplementary information\,{\color{blue}I} and see also \refcite{Narang2018HydroWSM,Narang2019EPCWSM}.)
With the values of the model parameters (supplementary information\,{\color{blue}F}), we find that $\lambda$ from our model almost perfectly matches with that from the {\abi} calculation for a large range of $\mu$ up to $-1$eV, as shown in \figref{main_fig:graphene}(a).
We note that we do not tune the EPC parameter $\gamma$ to fit our analytical $\lambda$  to our $\lambda^{\abi}$; instead we determine the value of $\gamma$ by matching the EPC analytic/${\abi}$ matrix elements at the corners of \BZ.
The match in \figref{main_fig:graphene}(a) is hence not a result of tuning the EPC parameter and shows the great validity of the our GA.
Moreover, our numerical calculation also finds that the geometric contribution is roughly 50\% of the total $\lambda$ (\figref{main_fig:graphene}(b)), consistent with our analytical results.
In \figref{main_fig:graphene}(a), we directly use the value of $\mcomega$ from the {\abi} calculation. 
We find that $\mcomega$ can be approximated by an analytical expression $\mcomega = \frac{2 \omega_{E_{2g}}^2(\Gamma) \omega_{A_1'}^2(\K) }{\omega_{E_{2g}}^2(\Gamma) + \omega_{A_1'}^2(\K)}$ (derived for $\mu\rightarrow 0$) with only $9\%$ error, where $\omega_{E_{2g}}(\Gamma)$ and $\omega_{A_1'}(\K)$ are the frequencies of the $E_{2g}$ phonons at $\Gamma$ and the $A_1'$ phonons at $\K$, respectively (supplementary information\,{\color{blue}F}).
This underscores the excellent agreement of our analytic calculation with realistic ${\abi}$. 

\Or{
Although the direct application of GA is not straightforward for moiré system (which we leave for future work), we indeed find that the mean-field critical temperature of twisted bilayer graphene derived from the EPC can be estimated by a geometric expression similar to \eqnref{main_eq:lambda_E_lambda_geo_graphene} in the first chiral limit~\cite{Tarnopolsky2019MagicAngleChiralLimit,BAB2021TBGIII,Wang2021ChiralMATBG,Bistritzer2011BMModel} based on the topological heavy fermion framework~\cite{Song20211110MATBGHF,CXL2023ElKPhCouplingTBG}. (See supplementary information\,{\color{blue}F} for details.) Our approximated expression relies on the FSM of the flat bands and gives $T_c\approx 0.6$K around the magic angles, which is close to the experimental values~\cite{Cao2018TBGSC}.
}

\begin{figure}[h]
    \centering
    \includegraphics[width=\columnwidth]{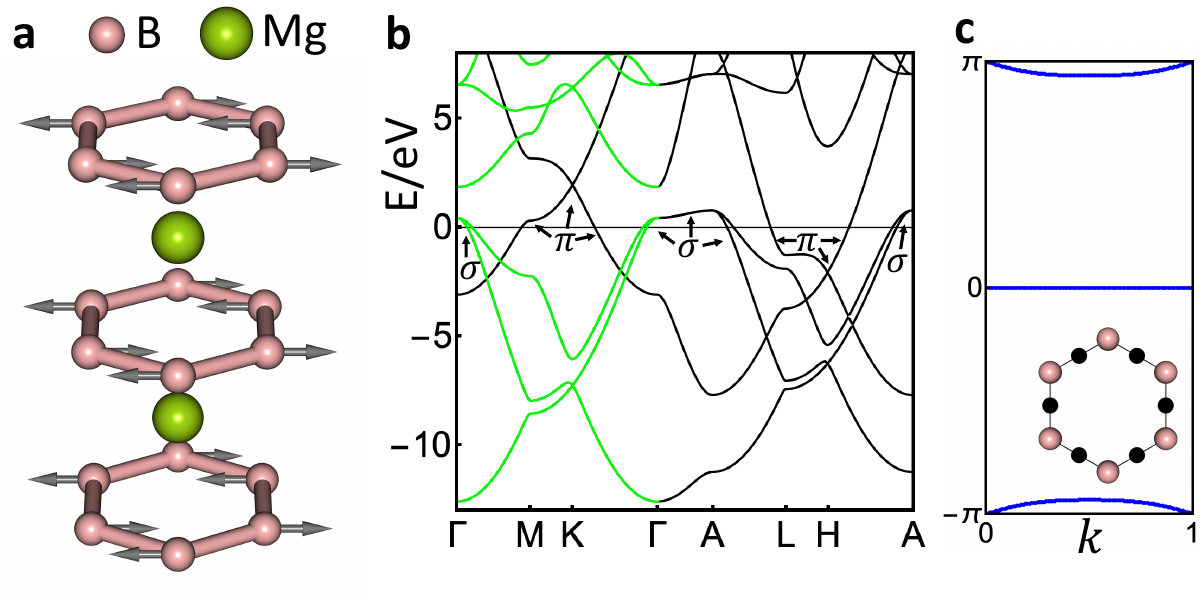}
    \caption{Plots for {\mgb}.
    (a)  structure of {\mgb}. The gray arrows show one type of the ion motion of the $E_{2g}$ phonon at $\Gamma$.
    (b)  {\abi} band structure.
    $\sigma$ and $\pi$ indicate states from $\sigma$ ($p_x/p_y$ orbitals) and  $\pi$-bonding ($p_z$ orbital) among B atoms, respectively~\cite{Boyer01302001MgB2SCBand}. 
    Green lines represent bands in the $m_z$-even subspace on the $k_z = 0$ plane.
    The Fermi energy is at $0$.
    (c) Wilson loop spectrum of the lowest $3$ bands in the $m_z$-even subspace on the $k_z = 0$ plane in (b).
    The black dots of the inset shows the Wannier center of the $3$ bands in one plane of B atoms (pink). 
    }
    \label{main_fig:MgB2}
\end{figure}

While graphene is a relatively ``simple" compound, and one could discount our excellent agreement and the findings that follow as ``accidental," {\mgb} (\figref{main_fig:MgB2}(a)) is a far more complicated system~\cite{Jun03012001MgB2SC} with multiple Fermi surfaces.
The EPC constant $\lambda$ only involves electron states at Fermi energy, which originate from  B atoms~\cite{Boyer01302001MgB2SCBand}.  (\figref{main_fig:MgB2}(b))
In addition, the main phonon contribution to $\lambda$ is from the $E_{2}$ modes along $\Gamma-$A (enhanced to $E_{2g}$ exactly at $\Gamma$ and A), which also  only involve B atoms~\cite{Pickett2011CovalentBondsDriven,Kong02272001MgB2EPC} (\figref{main_fig:MgB2}(a)).
The irrelevance of Mg for $\lambda$ is supported by \refcite{Jorgensen05012001MgB2Isotope} which finds an isotope effect of Mg atoms much smaller than that of the B atoms.
Therefore, we neglect Mg atoms when constructing the models for electrons and EPC.

The bands near the Fermi level originate from the $\sigma$ bonding among B $p_x/p_y$  orbitals and the $\pi$ bonding among B $p_z$ orbitals~\cite{Boyer01302001MgB2SCBand}(\figref{main_fig:MgB2}(b)).
The Fermi surfaces of the two bonding types are separated away from each other by a large in-plane momentum difference (supplementary information\,{\color{blue}H}), while the dominant phonon modes for $\lambda$ (mainly the $E_2$ phonons along $\Gamma$-A which are enhanced to $E_{2g}$ at $\Gamma$ and A) have small in-plane phonon momenta~\cite{Kong02272001MgB2EPC}.
Therefore, for evaluating $\lambda$, we reasonably assume that the $\sigma$-bonding states are decoupled from the $\pi$-bonding states in the electron and EPC Hamiltonian, which is also supported by the small linewidths of the phonons with large in-plane momenta observed in \refcite{Parlinski2003MgB2PhononLineWidth}.
As a result, we have $\lambda = \lambda_{\pi} + \lambda_{\sigma}$
where $\lambda_{\pi}$ ($\lambda_{\sigma}$) is the EPC constant of the $\pi$-bonding ($\sigma$-bonding) states.

The derivation for $\lambda_{\pi}$ is similar to graphene, since the $\pi$-bonding states originate from the $p_z$ orbitals of B atoms arranged as AA-stacking graphite (\figref{main_fig:MgB2}(a)).
The main difference is that the $\pi$-bonding states in {\mgb} have an extra NN hopping along $z$ direction in our model, which mainly affects the energetic contribution $\lambda_{\pi, E}$.
Nevertheless, we can still use GA in $x/y$ directions to derive the energetic and geometric parts of the EPC, which turns out to be the same as \eqnref{main_eq:g_E_g_geo_Gaussian} except that the hopping decay $\gamma_{\pi,z}$ along $z$ which is different from $\gamma_{\pi,\shpa}$ along $x/y$.
We adopt the GA only in the $x-y$ plane because the dominant $E_{2}$ phonons arise from the in-plane ($x-y$) motions of the B atoms~\cite{Kong02272001MgB2EPC}; the EPC Hamiltonian derived from GA exactly matches the actual EPC Hamiltonian with NN terms for the in-plane atomic motions.
We then find $\lambda_{\pi} = \lambda_{\pi,E} + \lambda_{\pi,geo}$, where $\lambda_{\pi, E-geo}$ is zero again due to symmetries.
The geometric $\lambda_{\pi,geo}$ has the same form as $\lambda_{geo}$ in \eqnref{main_eq:lambda_E_lambda_geo_graphene} for graphene (relying on FSM), and $\lambda_{\pi,E}$ just acquires an extra derivative of dispersion with respective to $k_z$ compared to $\lambda_{E}$ in \eqnref{main_eq:lambda_E_lambda_geo_graphene} for graphene, in addition to an extra factor $D_{\pi}(\mu)/D(\mu)$ in $\lambda_{\pi,E}$ and $\lambda_{\pi,geo}$ with $D_{\pi}(\mu)$ the density of the $\pi$-bonding states.  (supplementary information\,{\color{blue}H}.)

We now discuss $\lambda_{\sigma}$ for the $\sigma$-bonding states.
By adopting the GA in the $x/y$ directions and the NN-hopping approximation along $z$, we obtain the energetic and geometric parts of EPC, which are equal to  \eqnref{main_eq:g_E_g_geo_Gaussian} after replacing $\gamma$ by $\gamma_{\sigma,z}$ for the $z$ direction and by $\gamma_{\sigma,\shpa}$ for the $x/y$ directions (supplementary information\,{\color{blue}H}).
The form of the EPC derived from the GA is exact if (i) we only include the NN hopping terms among $p_x/p_y$ orbitals in the $x/y$ plane and along $z$, and (ii) we only keep first order in $\bsl{k}_\shpa$ in the electron Hamiltonian (allowed by small $\bsl{k}_{\shpa}$ on the Fermi surface of the $\sigma$-bonding states shown in \figref{main_fig:MgB2}(b)).
Because of the $\bsl{k}_{\shpa}$-first-order approximation, the effective Hamiltonian has two doubly-degenerate energy bands $E_{eff,n}(\bsl{k})$ with $n=1,2$---the lower $E_{eff,1}(\bsl{k})$ is cut by the Fermi energy.
While the effective model does not capture the splitting between the two bands near the Fermi level away from $\Gamma$-A shown in \figref{main_fig:MgB2}(b), it is a good approximation for the evaluation of the EPC as discussed at the end of this section and in supplementary information\,{\color{blue}H}.

Owing to the $\bsl{k}_{\shpa}$-first-order approximation of the electron Hamiltonian, we find $\lambda_{\sigma,E-geo}=0$, and thus obtain $\lambda_{\sigma}  = \lambda_{\sigma,E}  + \lambda_{\sigma,geo}$, which read
\begin{align}
\label{main_eq:lambda_sigma_E_geo}
\lambda_{\sigma,E} 
& = \frac{D_{\sigma}(\mu)}{D(\mu)} \frac{ \gamma_{\sigma,z}^2 \Omega }{ (2\pi)^3  m_{B} \mcomega}  \int_{FS_{eff,1}} d\sigma_{\bsl{k}} \frac{  [\partial_{k_{z}}   E_{eff,1}(\bsl{k})]^2}{\left|\nabla_{\bsl{k}}E_{eff,1}(\bsl{k})\right|} \nonumber\\
\lambda_{\sigma,geo}  & = \frac{D_{\sigma}(\mu)}{D(\mu)} \frac{ \gamma_{\sigma,\shpa}^2 \Omega }{ (2\pi)^3 m_{B} \mcomega}  \int_{FS_{eff,1}} d\sigma_{\bsl{k}}  \nonumber\\
& \qquad \qquad \times \sum_{i=x,y} \sum_{\alpha}^{-} \frac{  \Delta E_{eff}^2(0)  \left[ g_{eff,1,\alpha}(0) \right]_{ii}}{\left|\nabla_{\bsl{k}}E_{eff,1}(\bsl{k})\right|}
\end{align}
where $m_{\text{B}}$ is the mass of the B atom, $\Delta E_{eff}(\bsl{k}_\shpa) $ is the absolute difference between two doubly degenerate bands of the effective model, $D_{\sigma}(\mu)$ is the density of the $\sigma$-bonding states, and $FS_{eff,1}$ is the Fermi surface given by $E_{eff,1}(\bsl{k}) = \mu$. 
$g_{eff,1,\alpha}( \bsl{k}_\shpa ) $ is an OFSM
\begin{align}
\label{main_eq:orbital_selective_FS}
& \left[ g_{eff,1,\alpha}( \bsl{k}_\shpa ) \right]_{ij} \nonumber\\
& = \frac{1}{2}\Tr\left[ \xi_\alpha \xi^\dagger_\alpha \partial_{k_i} P_{eff,1}(\bsl{k}_\shpa)  P_{eff,1}(\bsl{k}_\shpa) \partial_{k_j} P_{eff,1}(\bsl{k}_\shpa)       \right] \nonumber\\
& \qquad + (i\leftrightarrow j) \ ,
\end{align}
where $\xi_\alpha$ is a normalized vector that represents the electronic orbitals linear combination picked by the relevant phonons for EPC $\lambda$ (\figref{main_fig:MgB2}(a)), and $P_{eff,1}(\bsl{k}_\shpa)$ is the projection matrix for the band $E_{eff,1}(\bsl{k})$.
In $\lambda_{\sigma,geo}$, we only sum $\alpha$ over the parity-odd combinations of $p_x/p_y$ orbitals (labelled by ``$-$" on top of the summation), because only the $E_{2}$ phonons matter under the $\bsl{k}_{\shpa}$-first-order approximation of the electron Hamiltonian and they flip the parity of the parity-even $P_{eff,1}(0)$.
We only have OFSM in $\lambda_{\sigma,geo}$ because $\lambda_{geo,2}$ mentioned above \eqnref{main_eq:lambda_geo,1} (which in general might leads to geometric quantity different from OFSM) turns out to have the same final expression as the OFSM under the approximation of the linear-momentum electron Hamiltonian, which allows us to use OFSM to describe the geometric dependence in $\lambda_{geo,2}$. (supplementary information\,{\color{blue}H})
We only consider the OFSM and $\Delta E_{eff}(\bsl{k}_\shpa) $ with $\bsl{k}_\shpa = 0$ in \eqnref{main_eq:lambda_sigma_E_geo} because the EPC matrix is given by the momentum derivative of the $\bsl{k}_{\shpa}$-first-order electron matrix Hamiltonian and thus is only reliable to zeroth order in $\bsl{k}_{\shpa}$.
We expect $\lambda_{\sigma,E}$ to be small, as it does not involve in-plane motions of B atoms manifested by the absence of momentum derivative along $x$ and $y$ in the numerator (confirmed by our ${\abi}$ calculation).

We determine the hopping decay parameters $\gamma_{\pi,\shpa}$, $\gamma_{\pi,z}$,  $\gamma_{\sigma,\shpa}$ and $\gamma_{\sigma,z}$ by matching the EPC $\Gamma_{nm}(\bsl{k},\bsl{k}+\bsl{q})$ (with $\bsl{k}=\Gamma,\K$ and $\bsl{q}$ along $\Gamma$-A) to our two {\abi} calculations for {\mgb}. (supplementary information\,{\color{blue}H}.)
Then, we obtain the values of various contributions to $\lambda$ as shown in \tabref{main_tab:lambda_MgB2}.
Note that we do not tune $\gamma_{\pi,\shpa}$, $\gamma_{\pi,z}$,  $\gamma_{\sigma,\shpa}$ and $\gamma_{\sigma,z}$ to fit our $\lambda$ (a single value) to the single value $\lambda^{\abi}$ given by the {\abi} calculation.
Therefore,  our value of $\lambda = 0.78$, which is remarkably close to the {\abi} value $\lambda^{\abi} = 0.67$ (17\% error), verifies the validity of our approximations.
Moreover, $\lambda_{\sigma}$ is much larger than $\lambda_{\pi}$, which is consistent with the previous result~\cite{Kong02272001MgB2EPC}.

We find that the quantum geometric contribution is about 92\% of the total $\lambda$, with most originating from the $\sigma$ bonding.
On the other hand, we find the energetic contribution from the $\sigma$ bonding ($\lambda_{\sigma,E}$) to be negligible, consistent with our analytical argument.
Therefore, the quantum geometry of the $\sigma$ bonding states supports the large EPC constant in MgB$_2$. 
The values in \tabref{main_tab:lambda_MgB2} are calculated with the {\abi} value of $\mcomega$ ($\hbar \sqrt{\mcomega} = 68 \text{meV}$), which can be approximated by the frequency of the $E_{2g}$ phonons at $\Gamma$ ($\hbar\omega_{E_{2g}}(\Gamma)= 75.3$meV) with about $10\%$ error.

\subsection{Topological Contributions to $\lambda$ in Graphene and {\mgb}}

The quantum geometric contributions in graphene and {\mgb} can be bounded from below by the topological invariants of the states on or near the Fermi surfaces in these materials, showing a deep connection between EPC and topology.
The graphene $\lambda_{geo}$ in \eqnref{main_eq:lambda_E_lambda_geo_graphene} is bounded from below by the topological contribution $\lambda_{topo}$, \ie, $\lambda_{geo}\geq\lambda_{topo}$, where $\lambda_{topo}$ reads
\eqa{
\label{main_eq:lambda_topo_graphene}
\lambda_{topo}  & =    \frac{ \Omega \gamma^2 }{ 4 m_{\text{C}}  \mcomega}    \frac{\left( |W_{\K}| + |W_{\K'}| \right)^2 }{ \int_{FS} d\sigma_{\bsl{k}}  \frac{  |\nabla_{\bsl{k}}E_n(\bsl{k})|}{\Delta E^2(\bsl{k}) } }\ ,
} 
where the chemical potential is moderate (\eg, within 1eV from $0$).
\Or{We derive \eqnref{main_eq:lambda_topo_graphene} from the $\int_{FS}d\sigma_{\bsl{k}}  \sqrt{\Tr\left[g_{n_F}(\bsl{k})\right]}$ $\geq$  $\pi  (|W_{\K}| + |W_{\K'}|)$ for moderate chemical potential.}
$\lambda_{topo}$ is topological because $ W_{\K}=1$ and $W_{\K'}=-1$ are the integer winding numbers~\cite{Neto2009GrapheneRMP} (or chiralities) of the Dirac cones at $\K$ and $\K'$, respectively.
Other parameters in \eqnref{main_eq:lambda_topo_graphene} are defined below \eqnref{main_eq:lambda_E_lambda_geo_graphene}.
We analytically show that $\lambda_{topo}/\lambda_{geo}$ limits to exactly 1 as $\mu\rightarrow 0$, which is consistent with the numerical calculation in \figref{main_fig:graphene}(c). (supplementary information\,{\color{blue}F}) 

For the $\pi$-bonding states in {\mgb}, the band structure has two $\PT$-protected nodal lines ($\mathcal{P}$ and $\TR$ are inversion and TR symmetries) along $k_z$-directional hinges of the {\BZ}, which carry winding numbers just like Dirac cones of graphene~\cite{Liu10192019MgB2Dirac}.
The winding numbers accounts for the topological contribution $\lambda_{\pi,topo}$ to $\lambda_\pi$, which bounds the geometric $\lambda_{\pi,geo}$ from below in a similar way to \eqnref{main_eq:lambda_topo_graphene}. (supplementary information\,{\color{blue}H})

Besides the nodal lines, we find an obstructed atomic set of bands on the $k_z=0$ plane of {\mgb}, which contains the $\sigma$-bonding states around the Fermi level.
The Bloch Hamiltonian has the mirror symmetry $m_z$ (that flips the $z$ direction) on the $k_z=0$ plane.
In the $m_z$-even subspace, we find that the isolated set of three bands cut by the Fermi energy is the elementary band representation (EBR) $A_{1g}$@3f, which is obstructed atomic since the atoms are not at 3f and which have nonzero $\PT$-protected second Stiefel-Whitney class $w_2 =1$. (\figref{main_fig:MgB2}(c)) 
(Here we follow the conventions in \emph{Bilbao Crystallographic Server}~\cite{Bradlyn2017TQC,Aroyo2006Bilbao}, and general discussion on $w_2$ can be found in \refcite{Ahn2019TBGFragile}.)
$ w_2 = 1 $ can be understood as having a band inversion at $\Gamma$, resulting in the effective Euler number $\Delta \N = 1$ of the $\sigma$-bonding states around $\Gamma$ near the Fermi level. (See details in supplementary information\,{\color{blue}H}.)
Remarkably, the effective Euler number $\Delta \N = 1$ gives a topological $\lambda_{\sigma, topo}$ which bounds the geometric contribution from below, where $\lambda_{\sigma, topo}$ reads
\eqa{
\label{main_eq:lambda_topo_MgB2}
\lambda_{\sigma,topo} & = \frac{D_{\sigma}(\mu)}{D(\mu)} \frac{ 4 \pi \gamma_{\sigma,\shpa}^2 \Omega }{ m_{B} \mcomega c^2} \left[\Delta\N\right]^2\\ 
& \qquad \times  \left[ \int_{FS_{eff,1}} d\sigma_{\bsl{k}} \frac{\left|\nabla_{\bsl{k}}E_{eff,1}(\bsl{k})\right|} {  |\bsl{d}(\bsl{k}_\shpa)|^2 } \right]^{-1} \ ,
}
where $\bsl{d}(\bsl{k}_\shpa) = v \bsl{k}_\shpa a$ couples the states with different parities in the $\sigma$-bonding effective model, and $a$ and $c$ are the lattice constant along $x/y$ and $z$, respectively.
Other parameters in \eqnref{main_eq:lambda_topo_MgB2} are defined below \eqnref{main_eq:lambda_sigma_E_geo}.
We mention that $\sum_{i=x,y} \sum_{\alpha}^{-} g_{eff,1,\alpha}(0)$ itself is not bounded from below since the $\sigma$-bonding states at $\Gamma$ is gapped.
\Or{Instead, we look at the product of the gap squared and the orbital-selective Fubini-Study metric, which is in dependent of the gap. In particular, by using the H\"older inequality, we find that the integration of $\sum_{i=x,y} \sum_{\alpha}^{-} \Delta E_{eff}^2(0)  \left[ g_{eff,1,\alpha}(0) \right]_{ii}/|\bsl{d}(\bsl{k}_\shpa)|^2$ on the Fermi surface is bounded from below by the winding number of $\bsl{d}(\bsl{k}_\shpa)$.
Since the winding number of $\bsl{d}(\bsl{k}_\shpa)$ determines the change of the topological invariant caused by the band inversion at $\Gamma$, it is further bounded from below by the effective Euler number.} (See details in supplementary information\,{\color{blue}H}.)
As shown in \tabref{main_tab:lambda_MgB2}, the total topological contribution $\lambda_{topo}=\lambda_{\pi,topo}+\lambda_{\sigma,topo}$ is about 44\% of the quantum geometric contribution $\lambda_{geo}$.

We note that the topological contribution just tells us that the geometric contribution may be stronger in the topologically nontrivial system.
In principle, there can be trivial bands in real material that have strong geometric properties and have a large geometric contribution.

\section{Discussion}

Our work shows that quantum geometric properties, now at the forefront of flat band research, are also fundamental --- and can in fact be dominant --- in a deep understanding of the different contributions to the EPC in systems with dispersive bands. One future direction is the development of a general framework that specifies the geometric and topological contributions to the bulk EPC constant $\lambda$ for all 2D and 3D systems with any types of topological invariants of states on or near Fermi surface.
Our current results imply that given two systems with similar band dispersion, the system with stronger geometric properties would tend to have stronger EPC, which serves as a guidance for future material search (\eg, one can look for Weyl semimetals that have Fermi surfaces enclosing Weyl points with large net chiralities.)  
The study of the geometric and topological contributions to the bulk EPC constant $\lambda$ in more phonon-mediated superconducting materials is essential for checking the relation between electron band topology/geometry and the superconducting $T_c$. Further work will focus on a {\abi} high-throughput of the quantum geometry effects in the EPC of many other multi-band superconductors. 

We find that the the energetic contribution $\lambda_E$ in graphene can be directly measured from the zero-temperature phonon linewidth of the $E_{2g}$ phonons at $\Gamma$, together with the frequencies of the $E_{2g}$ phonons at $\Gamma$ and the $A_1'$ phonon at K. (supplementary information\,{\color{blue}J}.)
Experimentally, the frequency and linewidth of the $E_{2g}$ phonons at $\Gamma$ can be measured in the Raman spectroscopy~\cite{Yan2007EPCGrapheneEXPRamen}, while the frequency of the $A_1'$ phonon at K can be approximated by the inelastic x-ray scattering measurement in graphite~\cite{Maultzsch11092003Graphite}.
Existing experimental data suggest the experimental value of $\lambda_E$ for $\mu\approx-0.1$eV is $0.0018\sim 0.0034$, whereas the value from our model is $0.0032$, which is within in the current experimental range. More precise measurement can be done in the future. 
Combined with the fact that the total $\lambda$ of graphene may be measured from the Helium scattering~\cite{Zhu01302012EPClambdaHelium,Benedek09082021EPCConstantGrapheneHe}, the geometric contribution $\lambda_{geo}$ may be measured from $\lambda -\lambda_E$.
Furthermore, FSM in graphene may be measured from the current noise spectrum~\cite{Neupert06102013CurrentNoiseFSM} or more generally the first-order optical response~\cite{Sipe06291999OpticalResponseFSM}, owing to the two-band nature of graphene.
Therefore, $\frac{\lambda_{geo}}{\lambda_E} = \frac{ |\mu|}{\pi } \int_{FS} d\sigma_{\bsl{k}} \frac{ \Tr[g_{n_F}(\bsl{k})] }{|\bsl{\nabla}_{\bsl{k}} E_{n_F}(\bsl{k})|} = \frac{h c }{2 \pi^2 e^2 } A(\omega = 2|\mu|/\hbar)$ may be experimentally testable, where $A(\omega)$ is the optical absorption coefficient for photons with frequency $\omega$ in the unit system where $1/(4\pi\epsilon_0)= 1$~\cite{Sipe06291999OpticalResponseFSM,Geim2008OpticalGraphene,Mak2008OpticalGraphene,Mak2012optical}.
If tested, this expression would relate the $\frac{\lambda_{geo}}{\lambda_E}$ in scattering experiments to the response coefficient in the optical response.
Besides graphene, on the surface of the topological insulator Bi$_2$Se$_3$ with the hexagonal distortion~\cite{Chen07102009Bi2Te3}, we can track the momentum-dependence of the geometric quantities (like FSM and OFSM) and the EPC coupling measured in time- and angle-resolved photoemission spectroscopy measurements~\cite{Hofmann04262011EPClambdaARPES,Pan09162011EPClambdaARPES,Sobota2014Bi2Se3EPCArpes,Baldini2023Ta2NiSe5EPCArpes}, as a test of the relation between quantum geometry and the EPC strength.
For 3D meterials like {\mgb}, the EPC constant $\lambda$ can be measured in various ways, \eg, by tracking the temperature behavior of specific heat\cite{Wang05082001EPClambdaCv} and inelastic x-ray scattering experiments~\cite{Parlinski2003MgB2PhononLineWidth}.
It is possible to test our theory in a system with tunable band geometry/topology by measuring the EPC constant while changing the band geometry/topology - for example through gating in 2D or strain in 3D. 

Note Added: During the review process of this manuscript, \refcite{Korshunov04182023ScV6Sn6Kagome} (authored by one of the authors of the current work) was posted online, which applied the GA proposed in this work to Kagome ScV$_6$Sn$_6$ and explained the phonon softenning in the system.

\section{Acknowledgements} 

We would like to acknowledge Edoardo Baldini, Dumitru Calugaru, Meng Cheng, Sankar Das Sarma, Leonid Glazman, F. Duncan M. Haldane, Jonah Herzog-Arbeitman, Haoyu Hu, Lun-Hui Hu, Yves H Kwan, Biao Lian, Cheol-Hwan Park, Zhi-Da Song, Ivo Souza, John Sous, Xiao-Qi Sun,  P\"aivi T\"orm\"a, Yuanfeng Xu, Rui-Xing Zhang and especially Chao-Xing Liu for helpful discussion. B.A.B. research was supported by the European Research Council (ERC) under the European Union’s Horizon 2020 research and innovation program (grant agreement no. 101020833) and partially by the Simons Investigator Grant No. 404513, the Gordon and Betty Moore Foundation through Grant No.GBMF8685 towards the Princeton theory program, the Gordon and Betty Moore Foundation’s EPiQS Initiative (Grant No. GBMF11070), Office of Naval Research (ONR Grant No. N00014-20-1-2303), Global Collaborative Network Grant at Princeton University, BSF Israel US foundation No. 2018226, NSF-MERSEC (Grant No. MERSEC DMR 2011750). J. Y. is supported by the Gordon and Betty Moore Foundation through Grant No. GBMF8685 towards the Princeton theory program. I.E. and B.A.B. are part of the SuperC collaboration.
I.E. has received funding from the European Research Council (ERC) under the European Union’s Horizon 2020 research and innovation programme (grant agreement No 802533) and the Department of Education, Universities and Research of the Eusko Jaurlaritza and the University of the Basque Country UPV/EHU (Grant No. IT1527-22).
Calculations by by C.J.C. and P.N. were supported by the Office of Naval Research grant \#13672292 and by the Gordon and Betty Moore Foundation grant number \#8048.
P.N. gratefully acknowledges support from the Alexander von Humboldt Foundation (Bessel Research Award) and from the John Simon Guggenheim Memorial Foundation (Guggenheim Fellowship).

\section{Author Contributions}

J.Y. and B.A.B conceived and supervised the project.
J.Y. and B.A.B. performed the theoretical analysis.
C.J.C., R.B., I.E., and P.N. performed the ab initio calculations. 
J.Y. and B.A.B. wrote the manuscript with the input from C.J.C., R.B., I.E., and P.N.

\section{Competing Interests}

The authors declare no competing interests.

\clearpage

\begin{table}
    \centering
    \begin{tabular}{|c|c||c|c||c|c|}
    \hline
     $\lambda$ ($\lambda^{\abi}$)   &       0.78 (0.67) & $\lambda_{\pi}$  &  0.16 & $\lambda_{\sigma}$ &  0.62 \\
    \hline
     $\lambda_{E}$ & 0.07 & $\lambda_{\pi,E}$ & 0.07 & $\lambda_{\sigma,E}$  & 0.00\\
    \hline
     $\lambda_{geo}$ & 0.71 & $\lambda_{\pi,geo}$& 0.09   & $\lambda_{\sigma,geo}$ & 0.62 \\ 
    \hline
     $\lambda_{topo}$ & 0.32  & $\lambda_{\pi,topo}$ & 0.01   & $\lambda_{\sigma,topo}$  & 0.31\\
    \hline
    \end{tabular}
    \caption{Numerical values of the $\lambda$ and its various contributions for {\mgb}.
   $\lambda^{\abi} = 0.67$ in the bracket is the {\abi} value  for $\lambda$.
    All other values are calculated from our model with parameter values determined by matching the EPC $\Gamma_{nm}(\bsl{k},\bsl{k}+\bsl{q})$ (with $\bsl{k}=\Gamma,\K$ and $\bsl{q}$ along $\Gamma$-A) and fitting the electron band structure to the {\abi} results.
    We do not fit the single value $\lambda$ to $\lambda^{\abi}$.
    }
    \label{main_tab:lambda_MgB2}
\end{table}

\clearpage

\clearpage

\bibliography{bibfile_references.bib}

\begin{thebibliography}{147}%
\makeatletter
\providecommand \@ifxundefined [1]{%
 \@ifx{#1\undefined}
}%
\providecommand \@ifnum [1]{%
 \ifnum #1\expandafter \@firstoftwo
 \else \expandafter \@secondoftwo
 \fi
}%
\providecommand \@ifx [1]{%
 \ifx #1\expandafter \@firstoftwo
 \else \expandafter \@secondoftwo
 \fi
}%
\providecommand \natexlab [1]{#1}%
\providecommand \enquote  [1]{``#1''}%
\providecommand \bibnamefont  [1]{#1}%
\providecommand \bibfnamefont [1]{#1}%
\providecommand \citenamefont [1]{#1}%
\providecommand \href@noop [0]{\@secondoftwo}%
\providecommand \href [0]{\begingroup \@sanitize@url \@href}%
\providecommand \@href[1]{\@@startlink{#1}\@@href}%
\providecommand \@@href[1]{\endgroup#1\@@endlink}%
\providecommand \@sanitize@url [0]{\catcode `\\12\catcode `\$12\catcode
  `\&12\catcode `\#12\catcode `\^12\catcode `\_12\catcode `\%12\relax}%
\providecommand \@@startlink[1]{}%
\providecommand \@@endlink[0]{}%
\providecommand \url  [0]{\begingroup\@sanitize@url \@url }%
\providecommand \@url [1]{\endgroup\@href {#1}{\urlprefix }}%
\providecommand \urlprefix  [0]{URL }%
\providecommand \Eprint [0]{\href }%
\providecommand \doibase [0]{https://doi.org/}%
\providecommand \selectlanguage [0]{\@gobble}%
\providecommand \bibinfo  [0]{\@secondoftwo}%
\providecommand \bibfield  [0]{\@secondoftwo}%
\providecommand \translation [1]{[#1]}%
\providecommand \BibitemOpen [0]{}%
\providecommand \bibitemStop [0]{}%
\providecommand \bibitemNoStop [0]{.\EOS\space}%
\providecommand \EOS [0]{\spacefactor3000\relax}%
\providecommand \BibitemShut  [1]{\csname bibitem#1\endcsname}%
\let\auto@bib@innerbib\@empty
\bibitem [{\citenamefont {Provost}\ and\ \citenamefont
  {Vallee}(1980)}]{Provost1980FSMetric}%
  \BibitemOpen
  \bibfield  {author} {\bibinfo {author} {\bibfnamefont {J.}~\bibnamefont
  {Provost}}\ and\ \bibinfo {author} {\bibfnamefont {G.}~\bibnamefont
  {Vallee}},\ }\bibfield  {title} {\bibinfo {title} {Riemannian structure on
  manifolds of quantum states},\ }\href@noop {} {\bibfield  {journal} {\bibinfo
   {journal} {Communications in Mathematical Physics}\ }\textbf {\bibinfo
  {volume} {76}},\ \bibinfo {pages} {289} (\bibinfo {year} {1980})}\BibitemShut
  {NoStop}%
\bibitem [{\citenamefont {Resta}(2011)}]{Resta2011QuantumGeometry}%
  \BibitemOpen
  \bibfield  {author} {\bibinfo {author} {\bibfnamefont {R.}~\bibnamefont
  {Resta}},\ }\bibfield  {title} {\bibinfo {title} {The insulating state of
  matter: a geometrical theory},\ }\href
  {https://doi.org/10.1140/epjb/e2010-10874-4} {\bibfield  {journal} {\bibinfo
  {journal} {The European Physical Journal B}\ }\textbf {\bibinfo {volume}
  {79}},\ \bibinfo {pages} {121} (\bibinfo {year} {2011})}\BibitemShut
  {NoStop}%
\bibitem [{\citenamefont {Bradlyn}\ \emph {et~al.}(2017)\citenamefont
  {Bradlyn}, \citenamefont {Elcoro}, \citenamefont {Cano}, \citenamefont
  {Vergniory}, \citenamefont {Wang}, \citenamefont {Felser}, \citenamefont
  {Aroyo},\ and\ \citenamefont {Bernevig}}]{Bradlyn2017TQC}%
  \BibitemOpen
  \bibfield  {author} {\bibinfo {author} {\bibfnamefont {B.}~\bibnamefont
  {Bradlyn}}, \bibinfo {author} {\bibfnamefont {L.}~\bibnamefont {Elcoro}},
  \bibinfo {author} {\bibfnamefont {J.}~\bibnamefont {Cano}}, \bibinfo {author}
  {\bibfnamefont {M.}~\bibnamefont {Vergniory}}, \bibinfo {author}
  {\bibfnamefont {Z.}~\bibnamefont {Wang}}, \bibinfo {author} {\bibfnamefont
  {C.}~\bibnamefont {Felser}}, \bibinfo {author} {\bibfnamefont
  {M.}~\bibnamefont {Aroyo}},\ and\ \bibinfo {author} {\bibfnamefont {B.~A.}\
  \bibnamefont {Bernevig}},\ }\bibfield  {title} {\bibinfo {title} {Topological
  quantum chemistry},\ }\href {https://doi.org/10.1038/nature23268} {\bibfield
  {journal} {\bibinfo  {journal} {Nature}\ }\textbf {\bibinfo {volume} {547}},\
  \bibinfo {pages} {298} (\bibinfo {year} {2017})}\BibitemShut {NoStop}%
\bibitem [{\citenamefont {Lieb}(1989)}]{Lieb1989LiebLattice}%
  \BibitemOpen
  \bibfield  {author} {\bibinfo {author} {\bibfnamefont {E.~H.}\ \bibnamefont
  {Lieb}},\ }\bibfield  {title} {\bibinfo {title} {Two theorems on the hubbard
  model},\ }\href {https://doi.org/10.1103/PhysRevLett.62.1201} {\bibfield
  {journal} {\bibinfo  {journal} {Phys. Rev. Lett.}\ }\textbf {\bibinfo
  {volume} {62}},\ \bibinfo {pages} {1201} (\bibinfo {year}
  {1989})}\BibitemShut {NoStop}%
\bibitem [{\citenamefont {Mielke}(1991)}]{Mielke1991LineGraph}%
  \BibitemOpen
  \bibfield  {author} {\bibinfo {author} {\bibfnamefont {A.}~\bibnamefont
  {Mielke}},\ }\bibfield  {title} {\bibinfo {title} {Ferromagnetism in the
  hubbard model on line graphs and further considerations},\ }\href
  {https://doi.org/10.1088/0305-4470/24/14/018} {\bibfield  {journal} {\bibinfo
   {journal} {Journal of Physics A: Mathematical and General}\ }\textbf
  {\bibinfo {volume} {24}},\ \bibinfo {pages} {3311} (\bibinfo {year}
  {1991})}\BibitemShut {NoStop}%
\bibitem [{\citenamefont {C{\u{a}}lug{\u{a}}ru}\ \emph
  {et~al.}(2022)\citenamefont {C{\u{a}}lug{\u{a}}ru}, \citenamefont {Chew},
  \citenamefont {Elcoro}, \citenamefont {Xu}, \citenamefont {Regnault},
  \citenamefont {Song},\ and\ \citenamefont
  {Bernevig}}]{Dumitru2022GeneralConstructionFlatBand}%
  \BibitemOpen
  \bibfield  {author} {\bibinfo {author} {\bibfnamefont {D.}~\bibnamefont
  {C{\u{a}}lug{\u{a}}ru}}, \bibinfo {author} {\bibfnamefont {A.}~\bibnamefont
  {Chew}}, \bibinfo {author} {\bibfnamefont {L.}~\bibnamefont {Elcoro}},
  \bibinfo {author} {\bibfnamefont {Y.}~\bibnamefont {Xu}}, \bibinfo {author}
  {\bibfnamefont {N.}~\bibnamefont {Regnault}}, \bibinfo {author}
  {\bibfnamefont {Z.-D.}\ \bibnamefont {Song}},\ and\ \bibinfo {author}
  {\bibfnamefont {B.~A.}\ \bibnamefont {Bernevig}},\ }\bibfield  {title}
  {\bibinfo {title} {General construction and topological classification of
  crystalline flat bands},\ }\href {https://doi.org/10.1038/s41567-021-01445-3}
  {\bibfield  {journal} {\bibinfo  {journal} {Nature Physics}\ }\textbf
  {\bibinfo {volume} {18}},\ \bibinfo {pages} {185} (\bibinfo {year}
  {2022})}\BibitemShut {NoStop}%
\bibitem [{\citenamefont {Peotta}\ and\ \citenamefont
  {T{\"o}rm{\"a}}(2015)}]{Torma2015SWBoundChern}%
  \BibitemOpen
  \bibfield  {author} {\bibinfo {author} {\bibfnamefont {S.}~\bibnamefont
  {Peotta}}\ and\ \bibinfo {author} {\bibfnamefont {P.}~\bibnamefont
  {T{\"o}rm{\"a}}},\ }\bibfield  {title} {\bibinfo {title} {Superfluidity in
  topologically nontrivial flat bands},\ }\href
  {https://doi.org/10.1038/ncomms9944} {\bibfield  {journal} {\bibinfo
  {journal} {Nature Communications}\ }\textbf {\bibinfo {volume} {6}},\
  \bibinfo {pages} {8944} (\bibinfo {year} {2015})}\BibitemShut {NoStop}%
\bibitem [{\citenamefont {Julku}\ \emph {et~al.}(2016)\citenamefont {Julku},
  \citenamefont {Peotta}, \citenamefont {Vanhala}, \citenamefont {Kim},\ and\
  \citenamefont {T\"orm\"a}}]{Torma2016SuperfluidWeightLieb}%
  \BibitemOpen
  \bibfield  {author} {\bibinfo {author} {\bibfnamefont {A.}~\bibnamefont
  {Julku}}, \bibinfo {author} {\bibfnamefont {S.}~\bibnamefont {Peotta}},
  \bibinfo {author} {\bibfnamefont {T.~I.}\ \bibnamefont {Vanhala}}, \bibinfo
  {author} {\bibfnamefont {D.-H.}\ \bibnamefont {Kim}},\ and\ \bibinfo {author}
  {\bibfnamefont {P.}~\bibnamefont {T\"orm\"a}},\ }\bibfield  {title} {\bibinfo
  {title} {Geometric origin of superfluidity in the lieb-lattice flat band},\
  }\href {https://doi.org/10.1103/PhysRevLett.117.045303} {\bibfield  {journal}
  {\bibinfo  {journal} {Phys. Rev. Lett.}\ }\textbf {\bibinfo {volume} {117}},\
  \bibinfo {pages} {045303} (\bibinfo {year} {2016})}\BibitemShut {NoStop}%
\bibitem [{\citenamefont {T\"orm\"a}\ \emph {et~al.}(2018)\citenamefont
  {T\"orm\"a}, \citenamefont {Liang},\ and\ \citenamefont
  {Peotta}}]{Torma2018SelectiveQuantumMetric}%
  \BibitemOpen
  \bibfield  {author} {\bibinfo {author} {\bibfnamefont {P.}~\bibnamefont
  {T\"orm\"a}}, \bibinfo {author} {\bibfnamefont {L.}~\bibnamefont {Liang}},\
  and\ \bibinfo {author} {\bibfnamefont {S.}~\bibnamefont {Peotta}},\
  }\bibfield  {title} {\bibinfo {title} {Quantum metric and effective mass of a
  two-body bound state in a flat band},\ }\href
  {https://doi.org/10.1103/PhysRevB.98.220511} {\bibfield  {journal} {\bibinfo
  {journal} {Phys. Rev. B}\ }\textbf {\bibinfo {volume} {98}},\ \bibinfo
  {pages} {220511} (\bibinfo {year} {2018})}\BibitemShut {NoStop}%
\bibitem [{\citenamefont {Xie}\ \emph {et~al.}(2020)\citenamefont {Xie},
  \citenamefont {Song}, \citenamefont {Lian},\ and\ \citenamefont
  {Bernevig}}]{Xie2020TopologyBoundSCTBG}%
  \BibitemOpen
  \bibfield  {author} {\bibinfo {author} {\bibfnamefont {F.}~\bibnamefont
  {Xie}}, \bibinfo {author} {\bibfnamefont {Z.}~\bibnamefont {Song}}, \bibinfo
  {author} {\bibfnamefont {B.}~\bibnamefont {Lian}},\ and\ \bibinfo {author}
  {\bibfnamefont {B.~A.}\ \bibnamefont {Bernevig}},\ }\bibfield  {title}
  {\bibinfo {title} {Topology-bounded superfluid weight in twisted bilayer
  graphene},\ }\href {https://doi.org/10.1103/PhysRevLett.124.167002}
  {\bibfield  {journal} {\bibinfo  {journal} {Phys. Rev. Lett.}\ }\textbf
  {\bibinfo {volume} {124}},\ \bibinfo {pages} {167002} (\bibinfo {year}
  {2020})}\BibitemShut {NoStop}%
\bibitem [{\citenamefont {Herzog-Arbeitman}\ \emph {et~al.}(2021)\citenamefont
  {Herzog-Arbeitman}, \citenamefont {Peri}, \citenamefont {Schindler},
  \citenamefont {Huber},\ and\ \citenamefont
  {Bernevig}}]{Herzogarbeitman2021SWBound}%
  \BibitemOpen
  \bibfield  {author} {\bibinfo {author} {\bibfnamefont {J.}~\bibnamefont
  {Herzog-Arbeitman}}, \bibinfo {author} {\bibfnamefont {V.}~\bibnamefont
  {Peri}}, \bibinfo {author} {\bibfnamefont {F.}~\bibnamefont {Schindler}},
  \bibinfo {author} {\bibfnamefont {S.~D.}\ \bibnamefont {Huber}},\ and\
  \bibinfo {author} {\bibfnamefont {B.~A.}\ \bibnamefont {Bernevig}},\
  }\href@noop {} {\bibinfo {title} {Superfluid weight bounds from symmetry and
  quantum geometry in flat bands}} (\bibinfo {year} {2021}),\ \Eprint
  {https://arxiv.org/abs/2110.14663} {arXiv:2110.14663 [cond-mat.mes-hall]}
  \BibitemShut {NoStop}%
\bibitem [{\citenamefont {Verma}\ \emph {et~al.}(2021)\citenamefont {Verma},
  \citenamefont {Hazra},\ and\ \citenamefont {Randeria}}]{Verma2021FlatBandSC}%
  \BibitemOpen
  \bibfield  {author} {\bibinfo {author} {\bibfnamefont {N.}~\bibnamefont
  {Verma}}, \bibinfo {author} {\bibfnamefont {T.}~\bibnamefont {Hazra}},\ and\
  \bibinfo {author} {\bibfnamefont {M.}~\bibnamefont {Randeria}},\ }\bibfield
  {title} {\bibinfo {title} {Optical spectral weight, phase stiffness, and
  <i>t</i><sub><i>c</i></sub> bounds for trivial and topological flat band
  superconductors},\ }\href {https://doi.org/10.1073/pnas.2106744118}
  {\bibfield  {journal} {\bibinfo  {journal} {Proceedings of the National
  Academy of Sciences}\ }\textbf {\bibinfo {volume} {118}},\ \bibinfo {pages}
  {e2106744118} (\bibinfo {year} {2021})},\ \Eprint
  {https://arxiv.org/abs/https://www.pnas.org/doi/pdf/10.1073/pnas.2106744118}
  {https://www.pnas.org/doi/pdf/10.1073/pnas.2106744118} \BibitemShut {NoStop}%
\bibitem [{\citenamefont {Hu}\ \emph {et~al.}(2019)\citenamefont {Hu},
  \citenamefont {Hyart}, \citenamefont {Pikulin},\ and\ \citenamefont
  {Rossi}}]{Rossi2019SFWTBG}%
  \BibitemOpen
  \bibfield  {author} {\bibinfo {author} {\bibfnamefont {X.}~\bibnamefont
  {Hu}}, \bibinfo {author} {\bibfnamefont {T.}~\bibnamefont {Hyart}}, \bibinfo
  {author} {\bibfnamefont {D.~I.}\ \bibnamefont {Pikulin}},\ and\ \bibinfo
  {author} {\bibfnamefont {E.}~\bibnamefont {Rossi}},\ }\bibfield  {title}
  {\bibinfo {title} {Geometric and conventional contribution to the superfluid
  weight in twisted bilayer graphene},\ }\href
  {https://doi.org/10.1103/PhysRevLett.123.237002} {\bibfield  {journal}
  {\bibinfo  {journal} {Phys. Rev. Lett.}\ }\textbf {\bibinfo {volume} {123}},\
  \bibinfo {pages} {237002} (\bibinfo {year} {2019})}\BibitemShut {NoStop}%
\bibitem [{\citenamefont {Julku}\ \emph {et~al.}(2020)\citenamefont {Julku},
  \citenamefont {Peltonen}, \citenamefont {Liang}, \citenamefont {Heikkil\"a},\
  and\ \citenamefont {T\"orm\"a}}]{Torma2020SFWTBG}%
  \BibitemOpen
  \bibfield  {author} {\bibinfo {author} {\bibfnamefont {A.}~\bibnamefont
  {Julku}}, \bibinfo {author} {\bibfnamefont {T.~J.}\ \bibnamefont {Peltonen}},
  \bibinfo {author} {\bibfnamefont {L.}~\bibnamefont {Liang}}, \bibinfo
  {author} {\bibfnamefont {T.~T.}\ \bibnamefont {Heikkil\"a}},\ and\ \bibinfo
  {author} {\bibfnamefont {P.}~\bibnamefont {T\"orm\"a}},\ }\bibfield  {title}
  {\bibinfo {title} {Superfluid weight and berezinskii-kosterlitz-thouless
  transition temperature of twisted bilayer graphene},\ }\href
  {https://doi.org/10.1103/PhysRevB.101.060505} {\bibfield  {journal} {\bibinfo
   {journal} {Phys. Rev. B}\ }\textbf {\bibinfo {volume} {101}},\ \bibinfo
  {pages} {060505} (\bibinfo {year} {2020})}\BibitemShut {NoStop}%
\bibitem [{\citenamefont {Park}\ \emph {et~al.}(2020)\citenamefont {Park},
  \citenamefont {Kim},\ and\ \citenamefont {Lee}}]{Park2020SCHofBut}%
  \BibitemOpen
  \bibfield  {author} {\bibinfo {author} {\bibfnamefont {M.~J.}\ \bibnamefont
  {Park}}, \bibinfo {author} {\bibfnamefont {Y.~B.}\ \bibnamefont {Kim}},\ and\
  \bibinfo {author} {\bibfnamefont {S.}~\bibnamefont {Lee}},\ }\bibfield
  {title} {\bibinfo {title} {Geometric superconductivity in 3d hofstadter
  butterfly},\ }\href {https://arxiv.org/abs/2007.16205} {\bibfield  {journal}
  {\bibinfo  {journal} {arXiv:2007.16205}\ } (\bibinfo {year}
  {2020})}\BibitemShut {NoStop}%
\bibitem [{\citenamefont {Herzog-Arbeitman}\ \emph
  {et~al.}(2022{\natexlab{a}})\citenamefont {Herzog-Arbeitman}, \citenamefont
  {Peri}, \citenamefont {Schindler}, \citenamefont {Huber},\ and\ \citenamefont
  {Bernevig}}]{Herzog2022FlatBandQuantumGeometry}%
  \BibitemOpen
  \bibfield  {author} {\bibinfo {author} {\bibfnamefont {J.}~\bibnamefont
  {Herzog-Arbeitman}}, \bibinfo {author} {\bibfnamefont {V.}~\bibnamefont
  {Peri}}, \bibinfo {author} {\bibfnamefont {F.}~\bibnamefont {Schindler}},
  \bibinfo {author} {\bibfnamefont {S.~D.}\ \bibnamefont {Huber}},\ and\
  \bibinfo {author} {\bibfnamefont {B.~A.}\ \bibnamefont {Bernevig}},\
  }\bibfield  {title} {\bibinfo {title} {Superfluid weight bounds from symmetry
  and quantum geometry in flat bands},\ }\href
  {https://doi.org/10.1103/PhysRevLett.128.087002} {\bibfield  {journal}
  {\bibinfo  {journal} {Phys. Rev. Lett.}\ }\textbf {\bibinfo {volume} {128}},\
  \bibinfo {pages} {087002} (\bibinfo {year} {2022}{\natexlab{a}})}\BibitemShut
  {NoStop}%
\bibitem [{\citenamefont {Huhtinen}\ \emph {et~al.}(2022)\citenamefont
  {Huhtinen}, \citenamefont {Herzog-Arbeitman}, \citenamefont {Chew},
  \citenamefont {Bernevig},\ and\ \citenamefont
  {T\"orm\"a}}]{Huhtinen2022FlatBandSCQuantumMetric}%
  \BibitemOpen
  \bibfield  {author} {\bibinfo {author} {\bibfnamefont {K.-E.}\ \bibnamefont
  {Huhtinen}}, \bibinfo {author} {\bibfnamefont {J.}~\bibnamefont
  {Herzog-Arbeitman}}, \bibinfo {author} {\bibfnamefont {A.}~\bibnamefont
  {Chew}}, \bibinfo {author} {\bibfnamefont {B.~A.}\ \bibnamefont {Bernevig}},\
  and\ \bibinfo {author} {\bibfnamefont {P.}~\bibnamefont {T\"orm\"a}},\
  }\bibfield  {title} {\bibinfo {title} {Revisiting flat band
  superconductivity: Dependence on minimal quantum metric and band touchings},\
  }\href {https://doi.org/10.1103/PhysRevB.106.014518} {\bibfield  {journal}
  {\bibinfo  {journal} {Phys. Rev. B}\ }\textbf {\bibinfo {volume} {106}},\
  \bibinfo {pages} {014518} (\bibinfo {year} {2022})}\BibitemShut {NoStop}%
\bibitem [{\citenamefont {Yu}\ \emph {et~al.}(2023{\natexlab{a}})\citenamefont
  {Yu}, \citenamefont {Xie}, \citenamefont {Wu},\ and\ \citenamefont
  {Das~Sarma}}]{Yu2022EOCPTBG}%
  \BibitemOpen
  \bibfield  {author} {\bibinfo {author} {\bibfnamefont {J.}~\bibnamefont
  {Yu}}, \bibinfo {author} {\bibfnamefont {M.}~\bibnamefont {Xie}}, \bibinfo
  {author} {\bibfnamefont {F.}~\bibnamefont {Wu}},\ and\ \bibinfo {author}
  {\bibfnamefont {S.}~\bibnamefont {Das~Sarma}},\ }\bibfield  {title} {\bibinfo
  {title} {Euler-obstructed nematic nodal superconductivity in twisted bilayer
  graphene},\ }\href {https://doi.org/10.1103/PhysRevB.107.L201106} {\bibfield
  {journal} {\bibinfo  {journal} {Phys. Rev. B}\ }\textbf {\bibinfo {volume}
  {107}},\ \bibinfo {pages} {L201106} (\bibinfo {year}
  {2023}{\natexlab{a}})}\BibitemShut {NoStop}%
\bibitem [{\citenamefont {Herzog-Arbeitman}\ \emph
  {et~al.}(2022{\natexlab{b}})\citenamefont {Herzog-Arbeitman}, \citenamefont
  {Chew}, \citenamefont {Huhtinen}, \citenamefont {T{\"o}rm{\"a}},\ and\
  \citenamefont {Bernevig}}]{Herzog2022ManyBodySCFlatBand}%
  \BibitemOpen
  \bibfield  {author} {\bibinfo {author} {\bibfnamefont {J.}~\bibnamefont
  {Herzog-Arbeitman}}, \bibinfo {author} {\bibfnamefont {A.}~\bibnamefont
  {Chew}}, \bibinfo {author} {\bibfnamefont {K.-E.}\ \bibnamefont {Huhtinen}},
  \bibinfo {author} {\bibfnamefont {P.}~\bibnamefont {T{\"o}rm{\"a}}},\ and\
  \bibinfo {author} {\bibfnamefont {B.~A.}\ \bibnamefont {Bernevig}},\
  }\bibfield  {title} {\bibinfo {title} {Many-body superconductivity in
  topological flat bands},\ }\href@noop {} {\bibfield  {journal} {\bibinfo
  {journal} {arXiv preprint arXiv:2209.00007}\ } (\bibinfo {year}
  {2022}{\natexlab{b}})}\BibitemShut {NoStop}%
\bibitem [{\citenamefont {{Chen}}\ and\ \citenamefont
  {{Huang}}(2022)}]{Huang2022arXivQuantumGeometryPDW}%
  \BibitemOpen
  \bibfield  {author} {\bibinfo {author} {\bibfnamefont {W.}~\bibnamefont
  {{Chen}}}\ and\ \bibinfo {author} {\bibfnamefont {W.}~\bibnamefont
  {{Huang}}},\ }\bibfield  {title} {\bibinfo {title} {{Pair density wave
  facilitated by Bloch quantum geometry in nearly flat band multiorbital
  superconductors}},\ }\href {https://doi.org/10.48550/arXiv.2208.02285}
  {\bibfield  {journal} {\bibinfo  {journal} {arXiv e-prints}\ ,\ \bibinfo
  {eid} {arXiv:2208.02285}} (\bibinfo {year} {2022})},\ \Eprint
  {https://arxiv.org/abs/2208.02285} {arXiv:2208.02285 [cond-mat.supr-con]}
  \BibitemShut {NoStop}%
\bibitem [{\citenamefont {Chen}\ and\ \citenamefont
  {Law}(2023)}]{Law2023QuantumMetricLandau}%
  \BibitemOpen
  \bibfield  {author} {\bibinfo {author} {\bibfnamefont {S.~A.}\ \bibnamefont
  {Chen}}\ and\ \bibinfo {author} {\bibfnamefont {K.}~\bibnamefont {Law}},\
  }\bibfield  {title} {\bibinfo {title} {Towards a ginzburg-landau theory of
  the quantum geometric effect in superconductors},\ }\href@noop {} {\bibfield
  {journal} {\bibinfo  {journal} {arXiv preprint arXiv:2303.15504}\ } (\bibinfo
  {year} {2023})}\BibitemShut {NoStop}%
\bibitem [{\citenamefont {T{\"o}rm{\"a}}\ \emph {et~al.}(2022)\citenamefont
  {T{\"o}rm{\"a}}, \citenamefont {Peotta},\ and\ \citenamefont
  {Bernevig}}]{Torma2022ReviewQuantumGeometry}%
  \BibitemOpen
  \bibfield  {author} {\bibinfo {author} {\bibfnamefont {P.}~\bibnamefont
  {T{\"o}rm{\"a}}}, \bibinfo {author} {\bibfnamefont {S.}~\bibnamefont
  {Peotta}},\ and\ \bibinfo {author} {\bibfnamefont {B.~A.}\ \bibnamefont
  {Bernevig}},\ }\bibfield  {title} {\bibinfo {title} {Superconductivity,
  superfluidity and quantum geometry in twisted multilayer systems},\ }\href
  {https://doi.org/10.1038/s42254-022-00466-y} {\bibfield  {journal} {\bibinfo
  {journal} {Nature Reviews Physics}\ }\textbf {\bibinfo {volume} {4}},\
  \bibinfo {pages} {528} (\bibinfo {year} {2022})}\BibitemShut {NoStop}%
\bibitem [{\citenamefont {{Hofmann}}\ \emph {et~al.}(2022)\citenamefont
  {{Hofmann}}, \citenamefont {{Berg}},\ and\ \citenamefont
  {{Chowdhury}}}]{Chowdhury2022SFFlatBandQuantumGeometry}%
  \BibitemOpen
  \bibfield  {author} {\bibinfo {author} {\bibfnamefont {J.~S.}\ \bibnamefont
  {{Hofmann}}}, \bibinfo {author} {\bibfnamefont {E.}~\bibnamefont {{Berg}}},\
  and\ \bibinfo {author} {\bibfnamefont {D.}~\bibnamefont {{Chowdhury}}},\
  }\bibfield  {title} {\bibinfo {title} {{Superconductivity, charge density
  wave, and supersolidity in flat bands with tunable quantum metric}},\ }\href
  {https://doi.org/10.48550/arXiv.2204.02994} {\bibfield  {journal} {\bibinfo
  {journal} {arXiv e-prints}\ ,\ \bibinfo {eid} {arXiv:2204.02994}} (\bibinfo
  {year} {2022})},\ \Eprint {https://arxiv.org/abs/2204.02994}
  {arXiv:2204.02994 [cond-mat.str-el]} \BibitemShut {NoStop}%
\bibitem [{\citenamefont {Mao}\ and\ \citenamefont
  {Chowdhury}(2023)}]{Chowdhury2023FlatbandSFStiffness}%
  \BibitemOpen
  \bibfield  {author} {\bibinfo {author} {\bibfnamefont {D.}~\bibnamefont
  {Mao}}\ and\ \bibinfo {author} {\bibfnamefont {D.}~\bibnamefont
  {Chowdhury}},\ }\bibfield  {title} {\bibinfo {title} {Diamagnetic response
  and phase stiffness for interacting isolated narrow bands},\ }\href
  {https://doi.org/10.1073/pnas.2217816120} {\bibfield  {journal} {\bibinfo
  {journal} {Proceedings of the National Academy of Sciences}\ }\textbf
  {\bibinfo {volume} {120}},\ \bibinfo {pages} {e2217816120} (\bibinfo {year}
  {2023})},\ \Eprint
  {https://arxiv.org/abs/https://www.pnas.org/doi/pdf/10.1073/pnas.2217816120}
  {https://www.pnas.org/doi/pdf/10.1073/pnas.2217816120} \BibitemShut {NoStop}%
\bibitem [{\citenamefont {Törmä}\ \emph {et~al.}(2021)\citenamefont
  {Törmä}, \citenamefont {Peotta},\ and\ \citenamefont
  {Bernevig}}]{Torma2021SFReview}%
  \BibitemOpen
  \bibfield  {author} {\bibinfo {author} {\bibfnamefont {P.}~\bibnamefont
  {Törmä}}, \bibinfo {author} {\bibfnamefont {S.}~\bibnamefont {Peotta}},\
  and\ \bibinfo {author} {\bibfnamefont {B.~A.}\ \bibnamefont {Bernevig}},\
  }\href@noop {} {\bibinfo {title} {Superfluidity and quantum geometry in
  twisted multilayer systems}} (\bibinfo {year} {2021}),\ \Eprint
  {https://arxiv.org/abs/2111.00807} {arXiv:2111.00807 [cond-mat.supr-con]}
  \BibitemShut {NoStop}%
\bibitem [{\citenamefont {Regnault}\ and\ \citenamefont
  {Bernevig}(2011)}]{BAB2011FCI}%
  \BibitemOpen
  \bibfield  {author} {\bibinfo {author} {\bibfnamefont {N.}~\bibnamefont
  {Regnault}}\ and\ \bibinfo {author} {\bibfnamefont {B.~A.}\ \bibnamefont
  {Bernevig}},\ }\bibfield  {title} {\bibinfo {title} {Fractional chern
  insulator},\ }\href {https://doi.org/10.1103/PhysRevX.1.021014} {\bibfield
  {journal} {\bibinfo  {journal} {Phys. Rev. X}\ }\textbf {\bibinfo {volume}
  {1}},\ \bibinfo {pages} {021014} (\bibinfo {year} {2011})}\BibitemShut
  {NoStop}%
\bibitem [{\citenamefont {Parameswaran}\ \emph {et~al.}(2013)\citenamefont
  {Parameswaran}, \citenamefont {Roy},\ and\ \citenamefont
  {Sondhi}}]{Sondhi2013FCI}%
  \BibitemOpen
  \bibfield  {author} {\bibinfo {author} {\bibfnamefont {S.~A.}\ \bibnamefont
  {Parameswaran}}, \bibinfo {author} {\bibfnamefont {R.}~\bibnamefont {Roy}},\
  and\ \bibinfo {author} {\bibfnamefont {S.~L.}\ \bibnamefont {Sondhi}},\
  }\bibfield  {title} {\bibinfo {title} {Fractional quantum hall physics in
  topological flat bands},\ }\href
  {https://doi.org/https://doi.org/10.1016/j.crhy.2013.04.003} {\bibfield
  {journal} {\bibinfo  {journal} {Comptes Rendus Physique}\ }\textbf {\bibinfo
  {volume} {14}},\ \bibinfo {pages} {816} (\bibinfo {year} {2013})},\ \bibinfo
  {note} {topological insulators / Isolants topologiques}\BibitemShut {NoStop}%
\bibitem [{\citenamefont {Dobard\ifmmode \check{z}\else
  \v{z}\fi{}i\ifmmode~\acute{c}\else \'{c}\fi{}}\ \emph
  {et~al.}(2013)\citenamefont {Dobard\ifmmode \check{z}\else
  \v{z}\fi{}i\ifmmode~\acute{c}\else \'{c}\fi{}}, \citenamefont
  {Milovanovi\ifmmode~\acute{c}\else \'{c}\fi{}},\ and\ \citenamefont
  {Regnault}}]{Regnault2013FCI}%
  \BibitemOpen
  \bibfield  {author} {\bibinfo {author} {\bibfnamefont {E.}~\bibnamefont
  {Dobard\ifmmode \check{z}\else \v{z}\fi{}i\ifmmode~\acute{c}\else
  \'{c}\fi{}}}, \bibinfo {author} {\bibfnamefont {M.~V.}\ \bibnamefont
  {Milovanovi\ifmmode~\acute{c}\else \'{c}\fi{}}},\ and\ \bibinfo {author}
  {\bibfnamefont {N.}~\bibnamefont {Regnault}},\ }\bibfield  {title} {\bibinfo
  {title} {Geometrical description of fractional chern insulators based on
  static structure factor calculations},\ }\href
  {https://doi.org/10.1103/PhysRevB.88.115117} {\bibfield  {journal} {\bibinfo
  {journal} {Phys. Rev. B}\ }\textbf {\bibinfo {volume} {88}},\ \bibinfo
  {pages} {115117} (\bibinfo {year} {2013})}\BibitemShut {NoStop}%
\bibitem [{\citenamefont {Roy}(2014)}]{Roy2014FCI}%
  \BibitemOpen
  \bibfield  {author} {\bibinfo {author} {\bibfnamefont {R.}~\bibnamefont
  {Roy}},\ }\bibfield  {title} {\bibinfo {title} {Band geometry of fractional
  topological insulators},\ }\href {https://doi.org/10.1103/PhysRevB.90.165139}
  {\bibfield  {journal} {\bibinfo  {journal} {Phys. Rev. B}\ }\textbf {\bibinfo
  {volume} {90}},\ \bibinfo {pages} {165139} (\bibinfo {year}
  {2014})}\BibitemShut {NoStop}%
\bibitem [{\citenamefont {Ledwith}\ \emph {et~al.}(2020)\citenamefont
  {Ledwith}, \citenamefont {Tarnopolsky}, \citenamefont {Khalaf},\ and\
  \citenamefont {Vishwanath}}]{Vishwanath2020FCITBG}%
  \BibitemOpen
  \bibfield  {author} {\bibinfo {author} {\bibfnamefont {P.~J.}\ \bibnamefont
  {Ledwith}}, \bibinfo {author} {\bibfnamefont {G.}~\bibnamefont
  {Tarnopolsky}}, \bibinfo {author} {\bibfnamefont {E.}~\bibnamefont
  {Khalaf}},\ and\ \bibinfo {author} {\bibfnamefont {A.}~\bibnamefont
  {Vishwanath}},\ }\bibfield  {title} {\bibinfo {title} {Fractional chern
  insulator states in twisted bilayer graphene: An analytical approach},\
  }\href {https://doi.org/10.1103/PhysRevResearch.2.023237} {\bibfield
  {journal} {\bibinfo  {journal} {Phys. Rev. Res.}\ }\textbf {\bibinfo {volume}
  {2}},\ \bibinfo {pages} {023237} (\bibinfo {year} {2020})}\BibitemShut
  {NoStop}%
\bibitem [{\citenamefont {Wang}\ and\ \citenamefont
  {Liu}(2022)}]{Wang2022FCITwistedGraphene}%
  \BibitemOpen
  \bibfield  {author} {\bibinfo {author} {\bibfnamefont {J.}~\bibnamefont
  {Wang}}\ and\ \bibinfo {author} {\bibfnamefont {Z.}~\bibnamefont {Liu}},\
  }\bibfield  {title} {\bibinfo {title} {Hierarchy of ideal flatbands in chiral
  twisted multilayer graphene models},\ }\href
  {https://doi.org/10.1103/PhysRevLett.128.176403} {\bibfield  {journal}
  {\bibinfo  {journal} {Phys. Rev. Lett.}\ }\textbf {\bibinfo {volume} {128}},\
  \bibinfo {pages} {176403} (\bibinfo {year} {2022})}\BibitemShut {NoStop}%
\bibitem [{\citenamefont {Rhim}\ \emph
  {et~al.}(2020{\natexlab{a}})\citenamefont {Rhim}, \citenamefont {Kim},\ and\
  \citenamefont {Yang}}]{Yang2020QuantumDistanceFlatBands}%
  \BibitemOpen
  \bibfield  {author} {\bibinfo {author} {\bibfnamefont {J.-W.}\ \bibnamefont
  {Rhim}}, \bibinfo {author} {\bibfnamefont {K.}~\bibnamefont {Kim}},\ and\
  \bibinfo {author} {\bibfnamefont {B.-J.}\ \bibnamefont {Yang}},\ }\bibfield
  {title} {\bibinfo {title} {Quantum distance and anomalous landau levels of
  flat bands},\ }\href {https://doi.org/10.1038/s41586-020-2540-1} {\bibfield
  {journal} {\bibinfo  {journal} {Nature}\ }\textbf {\bibinfo {volume} {584}},\
  \bibinfo {pages} {59} (\bibinfo {year} {2020}{\natexlab{a}})}\BibitemShut
  {NoStop}%
\bibitem [{\citenamefont {Mera}\ and\ \citenamefont
  {Ozawa}(2021)}]{Mera2021FlatBandsKahler}%
  \BibitemOpen
  \bibfield  {author} {\bibinfo {author} {\bibfnamefont {B.}~\bibnamefont
  {Mera}}\ and\ \bibinfo {author} {\bibfnamefont {T.}~\bibnamefont {Ozawa}},\
  }\bibfield  {title} {\bibinfo {title} {Engineering geometrically flat chern
  bands with fubini-study k\"ahler structure},\ }\href
  {https://doi.org/10.1103/PhysRevB.104.115160} {\bibfield  {journal} {\bibinfo
   {journal} {Phys. Rev. B}\ }\textbf {\bibinfo {volume} {104}},\ \bibinfo
  {pages} {115160} (\bibinfo {year} {2021})}\BibitemShut {NoStop}%
\bibitem [{\citenamefont {Julku}\ \emph {et~al.}(2021)\citenamefont {Julku},
  \citenamefont {Bruun},\ and\ \citenamefont
  {T\"orm\"a}}]{Torma2021FlatBandBEC}%
  \BibitemOpen
  \bibfield  {author} {\bibinfo {author} {\bibfnamefont {A.}~\bibnamefont
  {Julku}}, \bibinfo {author} {\bibfnamefont {G.~M.}\ \bibnamefont {Bruun}},\
  and\ \bibinfo {author} {\bibfnamefont {P.}~\bibnamefont {T\"orm\"a}},\
  }\bibfield  {title} {\bibinfo {title} {Quantum geometry and flat band
  bose-einstein condensation},\ }\href
  {https://doi.org/10.1103/PhysRevLett.127.170404} {\bibfield  {journal}
  {\bibinfo  {journal} {Phys. Rev. Lett.}\ }\textbf {\bibinfo {volume} {127}},\
  \bibinfo {pages} {170404} (\bibinfo {year} {2021})}\BibitemShut {NoStop}%
\bibitem [{\citenamefont {Wang}\ \emph
  {et~al.}(2021{\natexlab{a}})\citenamefont {Wang}, \citenamefont {Cano},
  \citenamefont {Millis}, \citenamefont {Liu},\ and\ \citenamefont
  {Yang}}]{Wang2021GeometryFlatBand}%
  \BibitemOpen
  \bibfield  {author} {\bibinfo {author} {\bibfnamefont {J.}~\bibnamefont
  {Wang}}, \bibinfo {author} {\bibfnamefont {J.}~\bibnamefont {Cano}}, \bibinfo
  {author} {\bibfnamefont {A.~J.}\ \bibnamefont {Millis}}, \bibinfo {author}
  {\bibfnamefont {Z.}~\bibnamefont {Liu}},\ and\ \bibinfo {author}
  {\bibfnamefont {B.}~\bibnamefont {Yang}},\ }\bibfield  {title} {\bibinfo
  {title} {Exact landau level description of geometry and interaction in a
  flatband},\ }\href {https://doi.org/10.1103/PhysRevLett.127.246403}
  {\bibfield  {journal} {\bibinfo  {journal} {Phys. Rev. Lett.}\ }\textbf
  {\bibinfo {volume} {127}},\ \bibinfo {pages} {246403} (\bibinfo {year}
  {2021}{\natexlab{a}})}\BibitemShut {NoStop}%
\bibitem [{\citenamefont {Hu}\ \emph {et~al.}(2022)\citenamefont {Hu},
  \citenamefont {Hyart}, \citenamefont {Pikulin},\ and\ \citenamefont
  {Rossi}}]{Rossi2022QuantumMetricExciton}%
  \BibitemOpen
  \bibfield  {author} {\bibinfo {author} {\bibfnamefont {X.}~\bibnamefont
  {Hu}}, \bibinfo {author} {\bibfnamefont {T.}~\bibnamefont {Hyart}}, \bibinfo
  {author} {\bibfnamefont {D.~I.}\ \bibnamefont {Pikulin}},\ and\ \bibinfo
  {author} {\bibfnamefont {E.}~\bibnamefont {Rossi}},\ }\bibfield  {title}
  {\bibinfo {title} {Quantum-metric-enabled exciton condensate in double
  twisted bilayer graphene},\ }\href
  {https://doi.org/10.1103/PhysRevB.105.L140506} {\bibfield  {journal}
  {\bibinfo  {journal} {Phys. Rev. B}\ }\textbf {\bibinfo {volume} {105}},\
  \bibinfo {pages} {L140506} (\bibinfo {year} {2022})}\BibitemShut {NoStop}%
\bibitem [{\citenamefont {Mitscherling}\ and\ \citenamefont
  {Holder}(2022)}]{Holder2022FlatBandQuantumMetricResistivity}%
  \BibitemOpen
  \bibfield  {author} {\bibinfo {author} {\bibfnamefont {J.}~\bibnamefont
  {Mitscherling}}\ and\ \bibinfo {author} {\bibfnamefont {T.}~\bibnamefont
  {Holder}},\ }\bibfield  {title} {\bibinfo {title} {Bound on resistivity in
  flat-band materials due to the quantum metric},\ }\href
  {https://doi.org/10.1103/PhysRevB.105.085154} {\bibfield  {journal} {\bibinfo
   {journal} {Phys. Rev. B}\ }\textbf {\bibinfo {volume} {105}},\ \bibinfo
  {pages} {085154} (\bibinfo {year} {2022})}\BibitemShut {NoStop}%
\bibitem [{\citenamefont {Chaudhary}\ \emph {et~al.}(2022)\citenamefont
  {Chaudhary}, \citenamefont {Lewandowski},\ and\ \citenamefont
  {Refael}}]{Refael2022ShiftCurrentTBG}%
  \BibitemOpen
  \bibfield  {author} {\bibinfo {author} {\bibfnamefont {S.}~\bibnamefont
  {Chaudhary}}, \bibinfo {author} {\bibfnamefont {C.}~\bibnamefont
  {Lewandowski}},\ and\ \bibinfo {author} {\bibfnamefont {G.}~\bibnamefont
  {Refael}},\ }\bibfield  {title} {\bibinfo {title} {Shift-current response as
  a probe of quantum geometry and electron-electron interactions in twisted
  bilayer graphene},\ }\href {https://doi.org/10.1103/PhysRevResearch.4.013164}
  {\bibfield  {journal} {\bibinfo  {journal} {Phys. Rev. Res.}\ }\textbf
  {\bibinfo {volume} {4}},\ \bibinfo {pages} {013164} (\bibinfo {year}
  {2022})}\BibitemShut {NoStop}%
\bibitem [{\citenamefont {Oh}\ \emph {et~al.}(2022)\citenamefont {Oh},
  \citenamefont {Cho}, \citenamefont {Park},\ and\ \citenamefont
  {Rhim}}]{Oh04252022RhimQuantumDistance}%
  \BibitemOpen
  \bibfield  {author} {\bibinfo {author} {\bibfnamefont {C.-g.}\ \bibnamefont
  {Oh}}, \bibinfo {author} {\bibfnamefont {D.}~\bibnamefont {Cho}}, \bibinfo
  {author} {\bibfnamefont {S.~Y.}\ \bibnamefont {Park}},\ and\ \bibinfo
  {author} {\bibfnamefont {J.-W.}\ \bibnamefont {Rhim}},\ }\bibfield  {title}
  {\bibinfo {title} {Bulk-interface correspondence from quantum distance in
  flat band systems},\ }\href {https://doi.org/10.1038/s42005-022-01102-y}
  {\bibfield  {journal} {\bibinfo  {journal} {Communications Physics}\ }\textbf
  {\bibinfo {volume} {5}},\ \bibinfo {pages} {320} (\bibinfo {year}
  {2022})}\BibitemShut {NoStop}%
\bibitem [{\citenamefont {Murakami}\ and\ \citenamefont
  {Nagaosa}(2003)}]{Murakami2003BerryPhaseMSC}%
  \BibitemOpen
  \bibfield  {author} {\bibinfo {author} {\bibfnamefont {S.}~\bibnamefont
  {Murakami}}\ and\ \bibinfo {author} {\bibfnamefont {N.}~\bibnamefont
  {Nagaosa}},\ }\bibfield  {title} {\bibinfo {title} {Berry phase in magnetic
  superconductors},\ }\href {https://doi.org/10.1103/PhysRevLett.90.057002}
  {\bibfield  {journal} {\bibinfo  {journal} {Phys. Rev. Lett.}\ }\textbf
  {\bibinfo {volume} {90}},\ \bibinfo {pages} {057002} (\bibinfo {year}
  {2003})}\BibitemShut {NoStop}%
\bibitem [{\citenamefont {Bultinck}\ \emph {et~al.}(2020)\citenamefont
  {Bultinck}, \citenamefont {Chatterjee},\ and\ \citenamefont
  {Zaletel}}]{Zaletel2020AHTBG}%
  \BibitemOpen
  \bibfield  {author} {\bibinfo {author} {\bibfnamefont {N.}~\bibnamefont
  {Bultinck}}, \bibinfo {author} {\bibfnamefont {S.}~\bibnamefont
  {Chatterjee}},\ and\ \bibinfo {author} {\bibfnamefont {M.~P.}\ \bibnamefont
  {Zaletel}},\ }\bibfield  {title} {\bibinfo {title} {Mechanism for anomalous
  hall ferromagnetism in twisted bilayer graphene},\ }\href
  {https://doi.org/10.1103/PhysRevLett.124.166601} {\bibfield  {journal}
  {\bibinfo  {journal} {Phys. Rev. Lett.}\ }\textbf {\bibinfo {volume} {124}},\
  \bibinfo {pages} {166601} (\bibinfo {year} {2020})}\BibitemShut {NoStop}%
\bibitem [{\citenamefont {Basov}\ and\ \citenamefont
  {Chubukov}(2011)}]{Basov2011HighTc}%
  \BibitemOpen
  \bibfield  {author} {\bibinfo {author} {\bibfnamefont {D.~N.}\ \bibnamefont
  {Basov}}\ and\ \bibinfo {author} {\bibfnamefont {A.~V.}\ \bibnamefont
  {Chubukov}},\ }\bibfield  {title} {\bibinfo {title} {Manifesto for a higher
  tc},\ }\href {https://doi.org/10.1038/nphys1975} {\bibfield  {journal}
  {\bibinfo  {journal} {Nature Physics}\ }\textbf {\bibinfo {volume} {7}},\
  \bibinfo {pages} {272} (\bibinfo {year} {2011})}\BibitemShut {NoStop}%
\bibitem [{\citenamefont {Cao}\ \emph {et~al.}(2018)\citenamefont {Cao},
  \citenamefont {Fatemi}, \citenamefont {Fang}, \citenamefont {Watanabe},
  \citenamefont {Taniguchi}, \citenamefont {Kaxiras},\ and\ \citenamefont
  {Jarillo-Herrero}}]{Cao2018TBGSC}%
  \BibitemOpen
  \bibfield  {author} {\bibinfo {author} {\bibfnamefont {Y.}~\bibnamefont
  {Cao}}, \bibinfo {author} {\bibfnamefont {V.}~\bibnamefont {Fatemi}},
  \bibinfo {author} {\bibfnamefont {S.}~\bibnamefont {Fang}}, \bibinfo {author}
  {\bibfnamefont {K.}~\bibnamefont {Watanabe}}, \bibinfo {author}
  {\bibfnamefont {T.}~\bibnamefont {Taniguchi}}, \bibinfo {author}
  {\bibfnamefont {E.}~\bibnamefont {Kaxiras}},\ and\ \bibinfo {author}
  {\bibfnamefont {P.}~\bibnamefont {Jarillo-Herrero}},\ }\bibfield  {title}
  {\bibinfo {title} {Unconventional superconductivity in magic-angle graphene
  superlattices},\ }\href {https://doi.org/10.1038/nature26160} {\bibfield
  {journal} {\bibinfo  {journal} {Nature}\ }\textbf {\bibinfo {volume} {556}},\
  \bibinfo {pages} {43} (\bibinfo {year} {2018})}\BibitemShut {NoStop}%
\bibitem [{\citenamefont {Tian}\ \emph {et~al.}(2023)\citenamefont {Tian},
  \citenamefont {Gao}, \citenamefont {Zhang}, \citenamefont {Che},
  \citenamefont {Xu}, \citenamefont {Cheung}, \citenamefont {Watanabe},
  \citenamefont {Taniguchi}, \citenamefont {Randeria}, \citenamefont {Zhang},
  \citenamefont {Lau},\ and\ \citenamefont {Bockrath}}]{Tian2023QuantumGeoSC}%
  \BibitemOpen
  \bibfield  {author} {\bibinfo {author} {\bibfnamefont {H.}~\bibnamefont
  {Tian}}, \bibinfo {author} {\bibfnamefont {X.}~\bibnamefont {Gao}}, \bibinfo
  {author} {\bibfnamefont {Y.}~\bibnamefont {Zhang}}, \bibinfo {author}
  {\bibfnamefont {S.}~\bibnamefont {Che}}, \bibinfo {author} {\bibfnamefont
  {T.}~\bibnamefont {Xu}}, \bibinfo {author} {\bibfnamefont {P.}~\bibnamefont
  {Cheung}}, \bibinfo {author} {\bibfnamefont {K.}~\bibnamefont {Watanabe}},
  \bibinfo {author} {\bibfnamefont {T.}~\bibnamefont {Taniguchi}}, \bibinfo
  {author} {\bibfnamefont {M.}~\bibnamefont {Randeria}}, \bibinfo {author}
  {\bibfnamefont {F.}~\bibnamefont {Zhang}}, \bibinfo {author} {\bibfnamefont
  {C.~N.}\ \bibnamefont {Lau}},\ and\ \bibinfo {author} {\bibfnamefont {M.~W.}\
  \bibnamefont {Bockrath}},\ }\bibfield  {title} {\bibinfo {title} {Evidence
  for dirac flat band superconductivity enabled by quantum geometry},\ }\href
  {https://doi.org/10.1038/s41586-022-05576-2} {\bibfield  {journal} {\bibinfo
  {journal} {Nature}\ }\textbf {\bibinfo {volume} {614}},\ \bibinfo {pages}
  {440} (\bibinfo {year} {2023})}\BibitemShut {NoStop}%
\bibitem [{\citenamefont {Hosur}(2011)}]{Hosur2011PhotoBerry}%
  \BibitemOpen
  \bibfield  {author} {\bibinfo {author} {\bibfnamefont {P.}~\bibnamefont
  {Hosur}},\ }\bibfield  {title} {\bibinfo {title} {Circular photogalvanic
  effect on topological insulator surfaces: Berry-curvature-dependent
  response},\ }\href {https://doi.org/10.1103/PhysRevB.83.035309} {\bibfield
  {journal} {\bibinfo  {journal} {Phys. Rev. B}\ }\textbf {\bibinfo {volume}
  {83}},\ \bibinfo {pages} {035309} (\bibinfo {year} {2011})}\BibitemShut
  {NoStop}%
\bibitem [{\citenamefont {Neupert}\ \emph
  {et~al.}(2013{\natexlab{a}})\citenamefont {Neupert}, \citenamefont {Chamon},\
  and\ \citenamefont {Mudry}}]{Neupert2013QuantumGeometryCurrentNoise}%
  \BibitemOpen
  \bibfield  {author} {\bibinfo {author} {\bibfnamefont {T.}~\bibnamefont
  {Neupert}}, \bibinfo {author} {\bibfnamefont {C.}~\bibnamefont {Chamon}},\
  and\ \bibinfo {author} {\bibfnamefont {C.}~\bibnamefont {Mudry}},\ }\bibfield
   {title} {\bibinfo {title} {Measuring the quantum geometry of bloch bands
  with current noise},\ }\href {https://doi.org/10.1103/PhysRevB.87.245103}
  {\bibfield  {journal} {\bibinfo  {journal} {Phys. Rev. B}\ }\textbf {\bibinfo
  {volume} {87}},\ \bibinfo {pages} {245103} (\bibinfo {year}
  {2013}{\natexlab{a}})}\BibitemShut {NoStop}%
\bibitem [{\citenamefont {Gao}\ \emph {et~al.}(2014)\citenamefont {Gao},
  \citenamefont {Yang},\ and\ \citenamefont
  {Niu}}]{Niu2014GeometryPositionalShift}%
  \BibitemOpen
  \bibfield  {author} {\bibinfo {author} {\bibfnamefont {Y.}~\bibnamefont
  {Gao}}, \bibinfo {author} {\bibfnamefont {S.~A.}\ \bibnamefont {Yang}},\ and\
  \bibinfo {author} {\bibfnamefont {Q.}~\bibnamefont {Niu}},\ }\bibfield
  {title} {\bibinfo {title} {Field induced positional shift of bloch electrons
  and its dynamical implications},\ }\href
  {https://doi.org/10.1103/PhysRevLett.112.166601} {\bibfield  {journal}
  {\bibinfo  {journal} {Phys. Rev. Lett.}\ }\textbf {\bibinfo {volume} {112}},\
  \bibinfo {pages} {166601} (\bibinfo {year} {2014})}\BibitemShut {NoStop}%
\bibitem [{\citenamefont {Morimoto}\ and\ \citenamefont
  {Nagaosa}(2016)}]{Nagaosa2016NonlinearOpticalGeometric}%
  \BibitemOpen
  \bibfield  {author} {\bibinfo {author} {\bibfnamefont {T.}~\bibnamefont
  {Morimoto}}\ and\ \bibinfo {author} {\bibfnamefont {N.}~\bibnamefont
  {Nagaosa}},\ }\bibfield  {title} {\bibinfo {title} {Topological nature of
  nonlinear optical effects in solids},\ }\href
  {https://doi.org/10.1126/sciadv.1501524} {\bibfield  {journal} {\bibinfo
  {journal} {Science Advances}\ }\textbf {\bibinfo {volume} {2}},\ \bibinfo
  {pages} {e1501524} (\bibinfo {year} {2016})},\ \Eprint
  {https://arxiv.org/abs/https://www.science.org/doi/pdf/10.1126/sciadv.1501524}
  {https://www.science.org/doi/pdf/10.1126/sciadv.1501524} \BibitemShut
  {NoStop}%
\bibitem [{\citenamefont {de~Juan}\ \emph {et~al.}(2017)\citenamefont
  {de~Juan}, \citenamefont {Grushin}, \citenamefont {Morimoto},\ and\
  \citenamefont {Moore}}]{Moore2017PhotoWSM}%
  \BibitemOpen
  \bibfield  {author} {\bibinfo {author} {\bibfnamefont {F.}~\bibnamefont
  {de~Juan}}, \bibinfo {author} {\bibfnamefont {A.~G.}\ \bibnamefont
  {Grushin}}, \bibinfo {author} {\bibfnamefont {T.}~\bibnamefont {Morimoto}},\
  and\ \bibinfo {author} {\bibfnamefont {J.~E.}\ \bibnamefont {Moore}},\
  }\bibfield  {title} {\bibinfo {title} {Quantized circular photogalvanic
  effect in weyl semimetals},\ }\href {https://doi.org/10.1038/ncomms15995}
  {\bibfield  {journal} {\bibinfo  {journal} {Nature Communications}\ }\textbf
  {\bibinfo {volume} {8}},\ \bibinfo {pages} {15995} (\bibinfo {year}
  {2017})}\BibitemShut {NoStop}%
\bibitem [{\citenamefont {Liang}\ \emph {et~al.}(2017)\citenamefont {Liang},
  \citenamefont {Vanhala}, \citenamefont {Peotta}, \citenamefont {Siro},
  \citenamefont {Harju},\ and\ \citenamefont
  {T\"orm\"a}}]{Paivi2017QuantumMetricAttractiveHubbard}%
  \BibitemOpen
  \bibfield  {author} {\bibinfo {author} {\bibfnamefont {L.}~\bibnamefont
  {Liang}}, \bibinfo {author} {\bibfnamefont {T.~I.}\ \bibnamefont {Vanhala}},
  \bibinfo {author} {\bibfnamefont {S.}~\bibnamefont {Peotta}}, \bibinfo
  {author} {\bibfnamefont {T.}~\bibnamefont {Siro}}, \bibinfo {author}
  {\bibfnamefont {A.}~\bibnamefont {Harju}},\ and\ \bibinfo {author}
  {\bibfnamefont {P.}~\bibnamefont {T\"orm\"a}},\ }\bibfield  {title} {\bibinfo
  {title} {Band geometry, berry curvature, and superfluid weight},\ }\href
  {https://doi.org/10.1103/PhysRevB.95.024515} {\bibfield  {journal} {\bibinfo
  {journal} {Phys. Rev. B}\ }\textbf {\bibinfo {volume} {95}},\ \bibinfo
  {pages} {024515} (\bibinfo {year} {2017})}\BibitemShut {NoStop}%
\bibitem [{\citenamefont {Ozawa}\ and\ \citenamefont
  {Goldman}(2018)}]{Tomoki2018QuantumMetricPeriodicDrive}%
  \BibitemOpen
  \bibfield  {author} {\bibinfo {author} {\bibfnamefont {T.}~\bibnamefont
  {Ozawa}}\ and\ \bibinfo {author} {\bibfnamefont {N.}~\bibnamefont
  {Goldman}},\ }\bibfield  {title} {\bibinfo {title} {Extracting the quantum
  metric tensor through periodic driving},\ }\href
  {https://doi.org/10.1103/PhysRevB.97.201117} {\bibfield  {journal} {\bibinfo
  {journal} {Phys. Rev. B}\ }\textbf {\bibinfo {volume} {97}},\ \bibinfo
  {pages} {201117} (\bibinfo {year} {2018})}\BibitemShut {NoStop}%
\bibitem [{\citenamefont {Li}\ and\ \citenamefont
  {Haldane}(2018)}]{Li2018WSMObstructedPairing}%
  \BibitemOpen
  \bibfield  {author} {\bibinfo {author} {\bibfnamefont {Y.}~\bibnamefont
  {Li}}\ and\ \bibinfo {author} {\bibfnamefont {F.~D.~M.}\ \bibnamefont
  {Haldane}},\ }\bibfield  {title} {\bibinfo {title} {Topological nodal cooper
  pairing in doped weyl metals},\ }\href
  {https://doi.org/10.1103/PhysRevLett.120.067003} {\bibfield  {journal}
  {\bibinfo  {journal} {Phys. Rev. Lett.}\ }\textbf {\bibinfo {volume} {120}},\
  \bibinfo {pages} {067003} (\bibinfo {year} {2018})}\BibitemShut {NoStop}%
\bibitem [{\citenamefont {Gao}\ and\ \citenamefont
  {Xiao}(2019)}]{Xiao2019QMDipole}%
  \BibitemOpen
  \bibfield  {author} {\bibinfo {author} {\bibfnamefont {Y.}~\bibnamefont
  {Gao}}\ and\ \bibinfo {author} {\bibfnamefont {D.}~\bibnamefont {Xiao}},\
  }\bibfield  {title} {\bibinfo {title} {Nonreciprocal directional dichroism
  induced by the quantum metric dipole},\ }\href
  {https://doi.org/10.1103/PhysRevLett.122.227402} {\bibfield  {journal}
  {\bibinfo  {journal} {Phys. Rev. Lett.}\ }\textbf {\bibinfo {volume} {122}},\
  \bibinfo {pages} {227402} (\bibinfo {year} {2019})}\BibitemShut {NoStop}%
\bibitem [{\citenamefont {Gianfrate}\ \emph {et~al.}(2020)\citenamefont
  {Gianfrate}, \citenamefont {Bleu}, \citenamefont {Dominici}, \citenamefont
  {Ardizzone}, \citenamefont {De~Giorgi}, \citenamefont {Ballarini},
  \citenamefont {Lerario}, \citenamefont {West}, \citenamefont {Pfeiffer},
  \citenamefont {Solnyshkov}, \citenamefont {Sanvitto},\ and\ \citenamefont
  {Malpuech}}]{Gianfrate01102019QuantumGeometry}%
  \BibitemOpen
  \bibfield  {author} {\bibinfo {author} {\bibfnamefont {A.}~\bibnamefont
  {Gianfrate}}, \bibinfo {author} {\bibfnamefont {O.}~\bibnamefont {Bleu}},
  \bibinfo {author} {\bibfnamefont {L.}~\bibnamefont {Dominici}}, \bibinfo
  {author} {\bibfnamefont {V.}~\bibnamefont {Ardizzone}}, \bibinfo {author}
  {\bibfnamefont {M.}~\bibnamefont {De~Giorgi}}, \bibinfo {author}
  {\bibfnamefont {D.}~\bibnamefont {Ballarini}}, \bibinfo {author}
  {\bibfnamefont {G.}~\bibnamefont {Lerario}}, \bibinfo {author} {\bibfnamefont
  {K.~W.}\ \bibnamefont {West}}, \bibinfo {author} {\bibfnamefont {L.~N.}\
  \bibnamefont {Pfeiffer}}, \bibinfo {author} {\bibfnamefont {D.~D.}\
  \bibnamefont {Solnyshkov}}, \bibinfo {author} {\bibfnamefont
  {D.}~\bibnamefont {Sanvitto}},\ and\ \bibinfo {author} {\bibfnamefont
  {G.}~\bibnamefont {Malpuech}},\ }\bibfield  {title} {\bibinfo {title}
  {Measurement of the quantum geometric tensor and of the anomalous hall
  drift},\ }\href {https://doi.org/10.1038/s41586-020-1989-2} {\bibfield
  {journal} {\bibinfo  {journal} {Nature}\ }\textbf {\bibinfo {volume} {578}},\
  \bibinfo {pages} {381} (\bibinfo {year} {2020})}\BibitemShut {NoStop}%
\bibitem [{\citenamefont {Rhim}\ \emph
  {et~al.}(2020{\natexlab{b}})\citenamefont {Rhim}, \citenamefont {Kim},\ and\
  \citenamefont {Yang}}]{Rhim08122019QuantumDistance}%
  \BibitemOpen
  \bibfield  {author} {\bibinfo {author} {\bibfnamefont {J.-W.}\ \bibnamefont
  {Rhim}}, \bibinfo {author} {\bibfnamefont {K.}~\bibnamefont {Kim}},\ and\
  \bibinfo {author} {\bibfnamefont {B.-J.}\ \bibnamefont {Yang}},\ }\bibfield
  {title} {\bibinfo {title} {Quantum distance and anomalous landau levels of
  flat bands},\ }\href {https://doi.org/10.1038/s41586-020-2540-1} {\bibfield
  {journal} {\bibinfo  {journal} {Nature}\ }\textbf {\bibinfo {volume} {584}},\
  \bibinfo {pages} {59} (\bibinfo {year} {2020}{\natexlab{b}})}\BibitemShut
  {NoStop}%
\bibitem [{\citenamefont {Kozii}\ \emph {et~al.}(2021)\citenamefont {Kozii},
  \citenamefont {Avdoshkin}, \citenamefont {Zhong},\ and\ \citenamefont
  {Moore}}]{Moore2021IntrinsicAnomalousHall}%
  \BibitemOpen
  \bibfield  {author} {\bibinfo {author} {\bibfnamefont {V.}~\bibnamefont
  {Kozii}}, \bibinfo {author} {\bibfnamefont {A.}~\bibnamefont {Avdoshkin}},
  \bibinfo {author} {\bibfnamefont {S.}~\bibnamefont {Zhong}},\ and\ \bibinfo
  {author} {\bibfnamefont {J.~E.}\ \bibnamefont {Moore}},\ }\bibfield  {title}
  {\bibinfo {title} {Intrinsic anomalous hall conductivity in a nonuniform
  electric field},\ }\href {https://doi.org/10.1103/PhysRevLett.126.156602}
  {\bibfield  {journal} {\bibinfo  {journal} {Phys. Rev. Lett.}\ }\textbf
  {\bibinfo {volume} {126}},\ \bibinfo {pages} {156602} (\bibinfo {year}
  {2021})}\BibitemShut {NoStop}%
\bibitem [{\citenamefont {Chen}\ and\ \citenamefont
  {Huang}(2021)}]{Chen2021QuantumGeometryOpticalAnomaly}%
  \BibitemOpen
  \bibfield  {author} {\bibinfo {author} {\bibfnamefont {W.}~\bibnamefont
  {Chen}}\ and\ \bibinfo {author} {\bibfnamefont {W.}~\bibnamefont {Huang}},\
  }\bibfield  {title} {\bibinfo {title} {Quantum-geometry-induced intrinsic
  optical anomaly in multiorbital superconductors},\ }\href
  {https://doi.org/10.1103/PhysRevResearch.3.L042018} {\bibfield  {journal}
  {\bibinfo  {journal} {Phys. Rev. Res.}\ }\textbf {\bibinfo {volume} {3}},\
  \bibinfo {pages} {L042018} (\bibinfo {year} {2021})}\BibitemShut {NoStop}%
\bibitem [{\citenamefont {Ahn}\ \emph {et~al.}(2022)\citenamefont {Ahn},
  \citenamefont {Guo}, \citenamefont {Nagaosa},\ and\ \citenamefont
  {Vishwanath}}]{Ahn2022OpticalGeometry}%
  \BibitemOpen
  \bibfield  {author} {\bibinfo {author} {\bibfnamefont {J.}~\bibnamefont
  {Ahn}}, \bibinfo {author} {\bibfnamefont {G.-Y.}\ \bibnamefont {Guo}},
  \bibinfo {author} {\bibfnamefont {N.}~\bibnamefont {Nagaosa}},\ and\ \bibinfo
  {author} {\bibfnamefont {A.}~\bibnamefont {Vishwanath}},\ }\bibfield  {title}
  {\bibinfo {title} {Riemannian geometry of resonant optical responses},\
  }\href {https://doi.org/10.1038/s41567-021-01465-z} {\bibfield  {journal}
  {\bibinfo  {journal} {Nature Physics}\ }\textbf {\bibinfo {volume} {18}},\
  \bibinfo {pages} {290} (\bibinfo {year} {2022})}\BibitemShut {NoStop}%
\bibitem [{\citenamefont {Bardeen}\ \emph {et~al.}(1957)\citenamefont
  {Bardeen}, \citenamefont {Cooper},\ and\ \citenamefont
  {Schrieffer}}]{BCS1957SC}%
  \BibitemOpen
  \bibfield  {author} {\bibinfo {author} {\bibfnamefont {J.}~\bibnamefont
  {Bardeen}}, \bibinfo {author} {\bibfnamefont {L.~N.}\ \bibnamefont
  {Cooper}},\ and\ \bibinfo {author} {\bibfnamefont {J.~R.}\ \bibnamefont
  {Schrieffer}},\ }\bibfield  {title} {\bibinfo {title} {Theory of
  superconductivity},\ }\href {https://doi.org/10.1103/PhysRev.108.1175}
  {\bibfield  {journal} {\bibinfo  {journal} {Phys. Rev.}\ }\textbf {\bibinfo
  {volume} {108}},\ \bibinfo {pages} {1175} (\bibinfo {year}
  {1957})}\BibitemShut {NoStop}%
\bibitem [{\citenamefont {Migdal}(1958)}]{Migdal1958EPC}%
  \BibitemOpen
  \bibfield  {author} {\bibinfo {author} {\bibfnamefont {A.}~\bibnamefont
  {Migdal}},\ }\bibfield  {title} {\bibinfo {title} {Interaction between
  electrons and lattice vibrations in a normal metal},\ }\href@noop {}
  {\bibfield  {journal} {\bibinfo  {journal} {Sov. Phys. JETP}\ }\textbf
  {\bibinfo {volume} {7}},\ \bibinfo {pages} {996} (\bibinfo {year}
  {1958})}\BibitemShut {NoStop}%
\bibitem [{\citenamefont {Eliashberg}(1960)}]{Eliashberg1960EPCSC}%
  \BibitemOpen
  \bibfield  {author} {\bibinfo {author} {\bibfnamefont {G.}~\bibnamefont
  {Eliashberg}},\ }\bibfield  {title} {\bibinfo {title} {Interactions between
  electrons and lattice vibrations in a superconductor},\ }\href@noop {}
  {\bibfield  {journal} {\bibinfo  {journal} {Sov. Phys. JETP}\ }\textbf
  {\bibinfo {volume} {11}},\ \bibinfo {pages} {696} (\bibinfo {year}
  {1960})}\BibitemShut {NoStop}%
\bibitem [{\citenamefont {McMillan}(1968)}]{McMillan1968SCTc}%
  \BibitemOpen
  \bibfield  {author} {\bibinfo {author} {\bibfnamefont {W.~L.}\ \bibnamefont
  {McMillan}},\ }\bibfield  {title} {\bibinfo {title} {Transition temperature
  of strong-coupled superconductors},\ }\href
  {https://doi.org/10.1103/PhysRev.167.331} {\bibfield  {journal} {\bibinfo
  {journal} {Phys. Rev.}\ }\textbf {\bibinfo {volume} {167}},\ \bibinfo {pages}
  {331} (\bibinfo {year} {1968})}\BibitemShut {NoStop}%
\bibitem [{\citenamefont {Allen}\ and\ \citenamefont
  {Dynes}(1975)}]{AllenDynes1975SCTc}%
  \BibitemOpen
  \bibfield  {author} {\bibinfo {author} {\bibfnamefont {P.~B.}\ \bibnamefont
  {Allen}}\ and\ \bibinfo {author} {\bibfnamefont {R.~C.}\ \bibnamefont
  {Dynes}},\ }\bibfield  {title} {\bibinfo {title} {Transition temperature of
  strong-coupled superconductors reanalyzed},\ }\href
  {https://doi.org/10.1103/PhysRevB.12.905} {\bibfield  {journal} {\bibinfo
  {journal} {Phys. Rev. B}\ }\textbf {\bibinfo {volume} {12}},\ \bibinfo
  {pages} {905} (\bibinfo {year} {1975})}\BibitemShut {NoStop}%
\bibitem [{\citenamefont {Po}\ \emph {et~al.}(2017)\citenamefont {Po},
  \citenamefont {Vishwanath},\ and\ \citenamefont {Watanabe}}]{Po2017SymIndi}%
  \BibitemOpen
  \bibfield  {author} {\bibinfo {author} {\bibfnamefont {H.~C.}\ \bibnamefont
  {Po}}, \bibinfo {author} {\bibfnamefont {A.}~\bibnamefont {Vishwanath}},\
  and\ \bibinfo {author} {\bibfnamefont {H.}~\bibnamefont {Watanabe}},\
  }\bibfield  {title} {\bibinfo {title} {Symmetry-based indicators of band
  topology in the 230 space groups},\ }\href
  {https://doi.org/10.1038/s41467-017-00133-2} {\bibfield  {journal} {\bibinfo
  {journal} {Nature communications}\ }\textbf {\bibinfo {volume} {8}},\
  \bibinfo {pages} {50} (\bibinfo {year} {2017})}\BibitemShut {NoStop}%
\bibitem [{\citenamefont {Vergniory}\ \emph {et~al.}(2019)\citenamefont
  {Vergniory}, \citenamefont {Elcoro}, \citenamefont {Felser}, \citenamefont
  {Regnault}, \citenamefont {Bernevig},\ and\ \citenamefont
  {Wang}}]{Bernevig2019TopoMat}%
  \BibitemOpen
  \bibfield  {author} {\bibinfo {author} {\bibfnamefont {M.~G.}\ \bibnamefont
  {Vergniory}}, \bibinfo {author} {\bibfnamefont {L.}~\bibnamefont {Elcoro}},
  \bibinfo {author} {\bibfnamefont {C.}~\bibnamefont {Felser}}, \bibinfo
  {author} {\bibfnamefont {N.}~\bibnamefont {Regnault}}, \bibinfo {author}
  {\bibfnamefont {B.~A.}\ \bibnamefont {Bernevig}},\ and\ \bibinfo {author}
  {\bibfnamefont {Z.}~\bibnamefont {Wang}},\ }\bibfield  {title} {\bibinfo
  {title} {A complete catalogue of high-quality topological materials},\ }\href
  {https://doi.org/10.1038/s41586-019-0954-4} {\bibfield  {journal} {\bibinfo
  {journal} {Nature}\ }\textbf {\bibinfo {volume} {566}},\ \bibinfo {pages}
  {480} (\bibinfo {year} {2019})}\BibitemShut {NoStop}%
\bibitem [{\citenamefont {Zhang}\ \emph {et~al.}(2019)\citenamefont {Zhang},
  \citenamefont {Jiang}, \citenamefont {Song}, \citenamefont {Huang},
  \citenamefont {He}, \citenamefont {Fang}, \citenamefont {Weng},\ and\
  \citenamefont {Fang}}]{Fang2019TopoMat}%
  \BibitemOpen
  \bibfield  {author} {\bibinfo {author} {\bibfnamefont {T.}~\bibnamefont
  {Zhang}}, \bibinfo {author} {\bibfnamefont {Y.}~\bibnamefont {Jiang}},
  \bibinfo {author} {\bibfnamefont {Z.}~\bibnamefont {Song}}, \bibinfo {author}
  {\bibfnamefont {H.}~\bibnamefont {Huang}}, \bibinfo {author} {\bibfnamefont
  {Y.}~\bibnamefont {He}}, \bibinfo {author} {\bibfnamefont {Z.}~\bibnamefont
  {Fang}}, \bibinfo {author} {\bibfnamefont {H.}~\bibnamefont {Weng}},\ and\
  \bibinfo {author} {\bibfnamefont {C.}~\bibnamefont {Fang}},\ }\bibfield
  {title} {\bibinfo {title} {Catalogue of topological electronic materials},\
  }\href {https://doi.org/10.1038/s41586-019-0944-6} {\bibfield  {journal}
  {\bibinfo  {journal} {Nature}\ }\textbf {\bibinfo {volume} {566}},\ \bibinfo
  {pages} {475} (\bibinfo {year} {2019})}\BibitemShut {NoStop}%
\bibitem [{\citenamefont {Tang}\ \emph {et~al.}(2019)\citenamefont {Tang},
  \citenamefont {Po}, \citenamefont {Vishwanath},\ and\ \citenamefont
  {Wan}}]{Wan2019TopoMat}%
  \BibitemOpen
  \bibfield  {author} {\bibinfo {author} {\bibfnamefont {F.}~\bibnamefont
  {Tang}}, \bibinfo {author} {\bibfnamefont {H.~C.}\ \bibnamefont {Po}},
  \bibinfo {author} {\bibfnamefont {A.}~\bibnamefont {Vishwanath}},\ and\
  \bibinfo {author} {\bibfnamefont {X.}~\bibnamefont {Wan}},\ }\bibfield
  {title} {\bibinfo {title} {Comprehensive search for topological materials
  using symmetry indicators},\ }\href
  {https://doi.org/10.1038/s41586-019-0937-5} {\bibfield  {journal} {\bibinfo
  {journal} {Nature}\ }\textbf {\bibinfo {volume} {566}},\ \bibinfo {pages}
  {486} (\bibinfo {year} {2019})}\BibitemShut {NoStop}%
\bibitem [{\citenamefont {Xu}\ \emph {et~al.}(2020)\citenamefont {Xu},
  \citenamefont {Elcoro}, \citenamefont {Song}, \citenamefont {Wieder},
  \citenamefont {Vergniory}, \citenamefont {Regnault}, \citenamefont {Chen},
  \citenamefont {Felser},\ and\ \citenamefont
  {Bernevig}}]{Bernevig2020MTQCMat}%
  \BibitemOpen
  \bibfield  {author} {\bibinfo {author} {\bibfnamefont {Y.}~\bibnamefont
  {Xu}}, \bibinfo {author} {\bibfnamefont {L.}~\bibnamefont {Elcoro}}, \bibinfo
  {author} {\bibfnamefont {Z.}~\bibnamefont {Song}}, \bibinfo {author}
  {\bibfnamefont {B.~J.}\ \bibnamefont {Wieder}}, \bibinfo {author}
  {\bibfnamefont {M.}~\bibnamefont {Vergniory}}, \bibinfo {author}
  {\bibfnamefont {N.}~\bibnamefont {Regnault}}, \bibinfo {author}
  {\bibfnamefont {Y.}~\bibnamefont {Chen}}, \bibinfo {author} {\bibfnamefont
  {C.}~\bibnamefont {Felser}},\ and\ \bibinfo {author} {\bibfnamefont {B.~A.}\
  \bibnamefont {Bernevig}},\ }\bibfield  {title} {\bibinfo {title}
  {High-throughput calculations of antiferromagnetic topological materials from
  magnetic topological quantum chemistry},\ }\href
  {https://arxiv.org/abs/2003.00012} {\bibfield  {journal} {\bibinfo  {journal}
  {arXiv:2003.00012}\ } (\bibinfo {year} {2020})}\BibitemShut {NoStop}%
\bibitem [{\citenamefont {Narang}\ \emph {et~al.}(2021)\citenamefont {Narang},
  \citenamefont {Garcia},\ and\ \citenamefont
  {Felser}}]{Narang2021TopologyBands}%
  \BibitemOpen
  \bibfield  {author} {\bibinfo {author} {\bibfnamefont {P.}~\bibnamefont
  {Narang}}, \bibinfo {author} {\bibfnamefont {C.~A.~C.}\ \bibnamefont
  {Garcia}},\ and\ \bibinfo {author} {\bibfnamefont {C.}~\bibnamefont
  {Felser}},\ }\bibfield  {title} {\bibinfo {title} {The topology of electronic
  band structures},\ }\href {https://doi.org/10.1038/s41563-020-00820-4}
  {\bibfield  {journal} {\bibinfo  {journal} {Nature Materials}\ }\textbf
  {\bibinfo {volume} {20}},\ \bibinfo {pages} {293} (\bibinfo {year}
  {2021})}\BibitemShut {NoStop}%
\bibitem [{\citenamefont {Baroni}\ \emph {et~al.}(2001)\citenamefont {Baroni},
  \citenamefont {de~Gironcoli}, \citenamefont {Dal~Corso},\ and\ \citenamefont
  {Giannozzi}}]{baroni2001}%
  \BibitemOpen
  \bibfield  {author} {\bibinfo {author} {\bibfnamefont {S.}~\bibnamefont
  {Baroni}}, \bibinfo {author} {\bibfnamefont {S.}~\bibnamefont
  {de~Gironcoli}}, \bibinfo {author} {\bibfnamefont {A.}~\bibnamefont
  {Dal~Corso}},\ and\ \bibinfo {author} {\bibfnamefont {P.}~\bibnamefont
  {Giannozzi}},\ }\bibfield  {title} {\bibinfo {title} {Phonons and related
  crystal properties from density-functional perturbation theory},\ }\href
  {https://doi.org/10.1103/RevModPhys.73.515} {\bibfield  {journal} {\bibinfo
  {journal} {Rev. Mod. Phys.}\ }\textbf {\bibinfo {volume} {73}},\ \bibinfo
  {pages} {515} (\bibinfo {year} {2001})}\BibitemShut {NoStop}%
\bibitem [{\citenamefont {Garcia}\ \emph {et~al.}(2021)\citenamefont {Garcia},
  \citenamefont {Nenno}, \citenamefont {Varnavides},\ and\ \citenamefont
  {Narang}}]{Narang2021EPCWeyl}%
  \BibitemOpen
  \bibfield  {author} {\bibinfo {author} {\bibfnamefont {C.~A.~C.}\
  \bibnamefont {Garcia}}, \bibinfo {author} {\bibfnamefont {D.~M.}\
  \bibnamefont {Nenno}}, \bibinfo {author} {\bibfnamefont {G.}~\bibnamefont
  {Varnavides}},\ and\ \bibinfo {author} {\bibfnamefont {P.}~\bibnamefont
  {Narang}},\ }\bibfield  {title} {\bibinfo {title} {Anisotropic
  phonon-mediated electronic transport in chiral weyl semimetals},\ }\href
  {https://doi.org/10.1103/PhysRevMaterials.5.L091202} {\bibfield  {journal}
  {\bibinfo  {journal} {Phys. Rev. Mater.}\ }\textbf {\bibinfo {volume} {5}},\
  \bibinfo {pages} {L091202} (\bibinfo {year} {2021})}\BibitemShut {NoStop}%
\bibitem [{\citenamefont {Osterhoudt}\ \emph {et~al.}(2021)\citenamefont
  {Osterhoudt}, \citenamefont {Wang}, \citenamefont {Garcia}, \citenamefont
  {Plisson}, \citenamefont {Gooth}, \citenamefont {Felser}, \citenamefont
  {Narang},\ and\ \citenamefont {Burch}}]{Burch2021EPCWeyl}%
  \BibitemOpen
  \bibfield  {author} {\bibinfo {author} {\bibfnamefont {G.~B.}\ \bibnamefont
  {Osterhoudt}}, \bibinfo {author} {\bibfnamefont {Y.}~\bibnamefont {Wang}},
  \bibinfo {author} {\bibfnamefont {C.~A.~C.}\ \bibnamefont {Garcia}}, \bibinfo
  {author} {\bibfnamefont {V.~M.}\ \bibnamefont {Plisson}}, \bibinfo {author}
  {\bibfnamefont {J.}~\bibnamefont {Gooth}}, \bibinfo {author} {\bibfnamefont
  {C.}~\bibnamefont {Felser}}, \bibinfo {author} {\bibfnamefont
  {P.}~\bibnamefont {Narang}},\ and\ \bibinfo {author} {\bibfnamefont {K.~S.}\
  \bibnamefont {Burch}},\ }\bibfield  {title} {\bibinfo {title} {Evidence for
  dominant phonon-electron scattering in weyl semimetal ${\mathrm{wp}}_{2}$},\
  }\href {https://doi.org/10.1103/PhysRevX.11.011017} {\bibfield  {journal}
  {\bibinfo  {journal} {Phys. Rev. X}\ }\textbf {\bibinfo {volume} {11}},\
  \bibinfo {pages} {011017} (\bibinfo {year} {2021})}\BibitemShut {NoStop}%
\bibitem [{\citenamefont {Nagamatsu}\ \emph {et~al.}(2001)\citenamefont
  {Nagamatsu}, \citenamefont {Nakagawa}, \citenamefont {Muranaka},
  \citenamefont {Zenitani},\ and\ \citenamefont
  {Akimitsu}}]{Jun03012001MgB2SC}%
  \BibitemOpen
  \bibfield  {author} {\bibinfo {author} {\bibfnamefont {J.}~\bibnamefont
  {Nagamatsu}}, \bibinfo {author} {\bibfnamefont {N.}~\bibnamefont {Nakagawa}},
  \bibinfo {author} {\bibfnamefont {T.}~\bibnamefont {Muranaka}}, \bibinfo
  {author} {\bibfnamefont {Y.}~\bibnamefont {Zenitani}},\ and\ \bibinfo
  {author} {\bibfnamefont {J.}~\bibnamefont {Akimitsu}},\ }\bibfield  {title}
  {\bibinfo {title} {Superconductivity at 39{\thinspace}k in magnesium
  diboride},\ }\href {https://doi.org/10.1038/35065039} {\bibfield  {journal}
  {\bibinfo  {journal} {Nature}\ }\textbf {\bibinfo {volume} {410}},\ \bibinfo
  {pages} {63} (\bibinfo {year} {2001})}\BibitemShut {NoStop}%
\bibitem [{\citenamefont {Bud'ko}\ \emph {et~al.}(2001)\citenamefont {Bud'ko},
  \citenamefont {Lapertot}, \citenamefont {Petrovic}, \citenamefont
  {Cunningham}, \citenamefont {Anderson},\ and\ \citenamefont
  {Canfield}}]{Budko02262001MgB2Isotope}%
  \BibitemOpen
  \bibfield  {author} {\bibinfo {author} {\bibfnamefont {S.~L.}\ \bibnamefont
  {Bud'ko}}, \bibinfo {author} {\bibfnamefont {G.}~\bibnamefont {Lapertot}},
  \bibinfo {author} {\bibfnamefont {C.}~\bibnamefont {Petrovic}}, \bibinfo
  {author} {\bibfnamefont {C.~E.}\ \bibnamefont {Cunningham}}, \bibinfo
  {author} {\bibfnamefont {N.}~\bibnamefont {Anderson}},\ and\ \bibinfo
  {author} {\bibfnamefont {P.~C.}\ \bibnamefont {Canfield}},\ }\bibfield
  {title} {\bibinfo {title} {Boron isotope effect in superconducting
  ${\mathrm{mgb}}_{2}$},\ }\href {https://doi.org/10.1103/PhysRevLett.86.1877}
  {\bibfield  {journal} {\bibinfo  {journal} {Phys. Rev. Lett.}\ }\textbf
  {\bibinfo {volume} {86}},\ \bibinfo {pages} {1877} (\bibinfo {year}
  {2001})}\BibitemShut {NoStop}%
\bibitem [{\citenamefont {Hinks}\ \emph {et~al.}(2001)\citenamefont {Hinks},
  \citenamefont {Claus},\ and\ \citenamefont
  {Jorgensen}}]{Jorgensen05012001MgB2Isotope}%
  \BibitemOpen
  \bibfield  {author} {\bibinfo {author} {\bibfnamefont {D.~G.}\ \bibnamefont
  {Hinks}}, \bibinfo {author} {\bibfnamefont {H.}~\bibnamefont {Claus}},\ and\
  \bibinfo {author} {\bibfnamefont {J.~D.}\ \bibnamefont {Jorgensen}},\
  }\bibfield  {title} {\bibinfo {title} {The complex nature of
  superconductivity in mgb2 as revealed by the reduced total isotope effect},\
  }\href {https://doi.org/10.1038/35078037} {\bibfield  {journal} {\bibinfo
  {journal} {Nature}\ }\textbf {\bibinfo {volume} {411}},\ \bibinfo {pages}
  {457} (\bibinfo {year} {2001})}\BibitemShut {NoStop}%
\bibitem [{\citenamefont {Esterlis}\ \emph {et~al.}(2018)\citenamefont
  {Esterlis}, \citenamefont {Nosarzewski}, \citenamefont {Huang}, \citenamefont
  {Moritz}, \citenamefont {Devereaux}, \citenamefont {Scalapino},\ and\
  \citenamefont {Kivelson}}]{Kivelson2018BreakDownofME}%
  \BibitemOpen
  \bibfield  {author} {\bibinfo {author} {\bibfnamefont {I.}~\bibnamefont
  {Esterlis}}, \bibinfo {author} {\bibfnamefont {B.}~\bibnamefont
  {Nosarzewski}}, \bibinfo {author} {\bibfnamefont {E.~W.}\ \bibnamefont
  {Huang}}, \bibinfo {author} {\bibfnamefont {B.}~\bibnamefont {Moritz}},
  \bibinfo {author} {\bibfnamefont {T.~P.}\ \bibnamefont {Devereaux}}, \bibinfo
  {author} {\bibfnamefont {D.~J.}\ \bibnamefont {Scalapino}},\ and\ \bibinfo
  {author} {\bibfnamefont {S.~A.}\ \bibnamefont {Kivelson}},\ }\bibfield
  {title} {\bibinfo {title} {Breakdown of the migdal-eliashberg theory: A
  determinant quantum monte carlo study},\ }\href
  {https://doi.org/10.1103/PhysRevB.97.140501} {\bibfield  {journal} {\bibinfo
  {journal} {Phys. Rev. B}\ }\textbf {\bibinfo {volume} {97}},\ \bibinfo
  {pages} {140501} (\bibinfo {year} {2018})}\BibitemShut {NoStop}%
\bibitem [{\citenamefont {Sous}\ \emph {et~al.}(2018)\citenamefont {Sous},
  \citenamefont {Chakraborty}, \citenamefont {Krems},\ and\ \citenamefont
  {Berciu}}]{Sous2018BipolaronsPeierls}%
  \BibitemOpen
  \bibfield  {author} {\bibinfo {author} {\bibfnamefont {J.}~\bibnamefont
  {Sous}}, \bibinfo {author} {\bibfnamefont {M.}~\bibnamefont {Chakraborty}},
  \bibinfo {author} {\bibfnamefont {R.~V.}\ \bibnamefont {Krems}},\ and\
  \bibinfo {author} {\bibfnamefont {M.}~\bibnamefont {Berciu}},\ }\bibfield
  {title} {\bibinfo {title} {Light bipolarons stabilized by peierls
  electron-phonon coupling},\ }\href
  {https://doi.org/10.1103/PhysRevLett.121.247001} {\bibfield  {journal}
  {\bibinfo  {journal} {Phys. Rev. Lett.}\ }\textbf {\bibinfo {volume} {121}},\
  \bibinfo {pages} {247001} (\bibinfo {year} {2018})}\BibitemShut {NoStop}%
\bibitem [{\citenamefont {Zhang}\ \emph {et~al.}(2023)\citenamefont {Zhang},
  \citenamefont {Sous}, \citenamefont {Reichman}, \citenamefont {Berciu},
  \citenamefont {Millis}, \citenamefont {Prokof'ev},\ and\ \citenamefont
  {Svistunov}}]{Zhang2023Bipolarons}%
  \BibitemOpen
  \bibfield  {author} {\bibinfo {author} {\bibfnamefont {C.}~\bibnamefont
  {Zhang}}, \bibinfo {author} {\bibfnamefont {J.}~\bibnamefont {Sous}},
  \bibinfo {author} {\bibfnamefont {D.~R.}\ \bibnamefont {Reichman}}, \bibinfo
  {author} {\bibfnamefont {M.}~\bibnamefont {Berciu}}, \bibinfo {author}
  {\bibfnamefont {A.~J.}\ \bibnamefont {Millis}}, \bibinfo {author}
  {\bibfnamefont {N.~V.}\ \bibnamefont {Prokof'ev}},\ and\ \bibinfo {author}
  {\bibfnamefont {B.~V.}\ \bibnamefont {Svistunov}},\ }\bibfield  {title}
  {\bibinfo {title} {Bipolaronic high-temperature superconductivity},\ }\href
  {https://doi.org/10.1103/PhysRevX.13.011010} {\bibfield  {journal} {\bibinfo
  {journal} {Phys. Rev. X}\ }\textbf {\bibinfo {volume} {13}},\ \bibinfo
  {pages} {011010} (\bibinfo {year} {2023})}\BibitemShut {NoStop}%
\bibitem [{\citenamefont {Mitra}(1969)}]{Mitra1969EPC}%
  \BibitemOpen
  \bibfield  {author} {\bibinfo {author} {\bibfnamefont {T.}~\bibnamefont
  {Mitra}},\ }\bibfield  {title} {\bibinfo {title} {Electron-phonon interaction
  in the modified tight-binding approximation},\ }\href@noop {} {\bibfield
  {journal} {\bibinfo  {journal} {Journal of Physics C: Solid State Physics}\
  }\textbf {\bibinfo {volume} {2}},\ \bibinfo {pages} {52} (\bibinfo {year}
  {1969})}\BibitemShut {NoStop}%
\bibitem [{\citenamefont {Castro~Neto}\ \emph {et~al.}(2009)\citenamefont
  {Castro~Neto}, \citenamefont {Guinea}, \citenamefont {Peres}, \citenamefont
  {Novoselov},\ and\ \citenamefont {Geim}}]{Neto2009GrapheneRMP}%
  \BibitemOpen
  \bibfield  {author} {\bibinfo {author} {\bibfnamefont {A.~H.}\ \bibnamefont
  {Castro~Neto}}, \bibinfo {author} {\bibfnamefont {F.}~\bibnamefont {Guinea}},
  \bibinfo {author} {\bibfnamefont {N.~M.~R.}\ \bibnamefont {Peres}}, \bibinfo
  {author} {\bibfnamefont {K.~S.}\ \bibnamefont {Novoselov}},\ and\ \bibinfo
  {author} {\bibfnamefont {A.~K.}\ \bibnamefont {Geim}},\ }\bibfield  {title}
  {\bibinfo {title} {The electronic properties of graphene},\ }\href
  {https://doi.org/10.1103/RevModPhys.81.109} {\bibfield  {journal} {\bibinfo
  {journal} {Rev. Mod. Phys.}\ }\textbf {\bibinfo {volume} {81}},\ \bibinfo
  {pages} {109} (\bibinfo {year} {2009})}\BibitemShut {NoStop}%
\bibitem [{\citenamefont {Coulter}\ \emph {et~al.}(2018)\citenamefont
  {Coulter}, \citenamefont {Sundararaman},\ and\ \citenamefont
  {Narang}}]{Narang2018HydroWSM}%
  \BibitemOpen
  \bibfield  {author} {\bibinfo {author} {\bibfnamefont {J.}~\bibnamefont
  {Coulter}}, \bibinfo {author} {\bibfnamefont {R.}~\bibnamefont
  {Sundararaman}},\ and\ \bibinfo {author} {\bibfnamefont {P.}~\bibnamefont
  {Narang}},\ }\bibfield  {title} {\bibinfo {title} {Microscopic origins of
  hydrodynamic transport in the type-ii weyl semimetal ${\mathrm{wp}}_{2}$},\
  }\href {https://doi.org/10.1103/PhysRevB.98.115130} {\bibfield  {journal}
  {\bibinfo  {journal} {Phys. Rev. B}\ }\textbf {\bibinfo {volume} {98}},\
  \bibinfo {pages} {115130} (\bibinfo {year} {2018})}\BibitemShut {NoStop}%
\bibitem [{\citenamefont {Coulter}\ \emph {et~al.}(2019)\citenamefont
  {Coulter}, \citenamefont {Osterhoudt}, \citenamefont {Garcia}, \citenamefont
  {Wang}, \citenamefont {Plisson}, \citenamefont {Shen}, \citenamefont {Ni},
  \citenamefont {Burch},\ and\ \citenamefont {Narang}}]{Narang2019EPCWSM}%
  \BibitemOpen
  \bibfield  {author} {\bibinfo {author} {\bibfnamefont {J.}~\bibnamefont
  {Coulter}}, \bibinfo {author} {\bibfnamefont {G.~B.}\ \bibnamefont
  {Osterhoudt}}, \bibinfo {author} {\bibfnamefont {C.~A.~C.}\ \bibnamefont
  {Garcia}}, \bibinfo {author} {\bibfnamefont {Y.}~\bibnamefont {Wang}},
  \bibinfo {author} {\bibfnamefont {V.~M.}\ \bibnamefont {Plisson}}, \bibinfo
  {author} {\bibfnamefont {B.}~\bibnamefont {Shen}}, \bibinfo {author}
  {\bibfnamefont {N.}~\bibnamefont {Ni}}, \bibinfo {author} {\bibfnamefont
  {K.~S.}\ \bibnamefont {Burch}},\ and\ \bibinfo {author} {\bibfnamefont
  {P.}~\bibnamefont {Narang}},\ }\bibfield  {title} {\bibinfo {title}
  {Uncovering electron-phonon scattering and phonon dynamics in type-i weyl
  semimetals},\ }\href {https://doi.org/10.1103/PhysRevB.100.220301} {\bibfield
   {journal} {\bibinfo  {journal} {Phys. Rev. B}\ }\textbf {\bibinfo {volume}
  {100}},\ \bibinfo {pages} {220301} (\bibinfo {year} {2019})}\BibitemShut
  {NoStop}%
\bibitem [{\citenamefont {Tarnopolsky}\ \emph {et~al.}(2019)\citenamefont
  {Tarnopolsky}, \citenamefont {Kruchkov},\ and\ \citenamefont
  {Vishwanath}}]{Tarnopolsky2019MagicAngleChiralLimit}%
  \BibitemOpen
  \bibfield  {author} {\bibinfo {author} {\bibfnamefont {G.}~\bibnamefont
  {Tarnopolsky}}, \bibinfo {author} {\bibfnamefont {A.~J.}\ \bibnamefont
  {Kruchkov}},\ and\ \bibinfo {author} {\bibfnamefont {A.}~\bibnamefont
  {Vishwanath}},\ }\bibfield  {title} {\bibinfo {title} {Origin of magic angles
  in twisted bilayer graphene},\ }\href
  {https://doi.org/10.1103/PhysRevLett.122.106405} {\bibfield  {journal}
  {\bibinfo  {journal} {Phys. Rev. Lett.}\ }\textbf {\bibinfo {volume} {122}},\
  \bibinfo {pages} {106405} (\bibinfo {year} {2019})}\BibitemShut {NoStop}%
\bibitem [{\citenamefont {Bernevig}\ \emph {et~al.}(2021)\citenamefont
  {Bernevig}, \citenamefont {Song}, \citenamefont {Regnault},\ and\
  \citenamefont {Lian}}]{BAB2021TBGIII}%
  \BibitemOpen
  \bibfield  {author} {\bibinfo {author} {\bibfnamefont {B.~A.}\ \bibnamefont
  {Bernevig}}, \bibinfo {author} {\bibfnamefont {Z.-D.}\ \bibnamefont {Song}},
  \bibinfo {author} {\bibfnamefont {N.}~\bibnamefont {Regnault}},\ and\
  \bibinfo {author} {\bibfnamefont {B.}~\bibnamefont {Lian}},\ }\bibfield
  {title} {\bibinfo {title} {Twisted bilayer graphene. iii. interacting
  hamiltonian and exact symmetries},\ }\href
  {https://doi.org/10.1103/PhysRevB.103.205413} {\bibfield  {journal} {\bibinfo
   {journal} {Phys. Rev. B}\ }\textbf {\bibinfo {volume} {103}},\ \bibinfo
  {pages} {205413} (\bibinfo {year} {2021})}\BibitemShut {NoStop}%
\bibitem [{\citenamefont {Wang}\ \emph
  {et~al.}(2021{\natexlab{b}})\citenamefont {Wang}, \citenamefont {Zheng},
  \citenamefont {Millis},\ and\ \citenamefont {Cano}}]{Wang2021ChiralMATBG}%
  \BibitemOpen
  \bibfield  {author} {\bibinfo {author} {\bibfnamefont {J.}~\bibnamefont
  {Wang}}, \bibinfo {author} {\bibfnamefont {Y.}~\bibnamefont {Zheng}},
  \bibinfo {author} {\bibfnamefont {A.~J.}\ \bibnamefont {Millis}},\ and\
  \bibinfo {author} {\bibfnamefont {J.}~\bibnamefont {Cano}},\ }\bibfield
  {title} {\bibinfo {title} {Chiral approximation to twisted bilayer graphene:
  Exact intravalley inversion symmetry, nodal structure, and implications for
  higher magic angles},\ }\href
  {https://doi.org/10.1103/PhysRevResearch.3.023155} {\bibfield  {journal}
  {\bibinfo  {journal} {Phys. Rev. Res.}\ }\textbf {\bibinfo {volume} {3}},\
  \bibinfo {pages} {023155} (\bibinfo {year} {2021}{\natexlab{b}})}\BibitemShut
  {NoStop}%
\bibitem [{\citenamefont {Bistritzer}\ and\ \citenamefont
  {MacDonald}(2011)}]{Bistritzer2011BMModel}%
  \BibitemOpen
  \bibfield  {author} {\bibinfo {author} {\bibfnamefont {R.}~\bibnamefont
  {Bistritzer}}\ and\ \bibinfo {author} {\bibfnamefont {A.~H.}\ \bibnamefont
  {MacDonald}},\ }\bibfield  {title} {\bibinfo {title} {Moir{\'e} bands in
  twisted double-layer graphene},\ }\href@noop {} {\bibfield  {journal}
  {\bibinfo  {journal} {Proceedings of the National Academy of Sciences}\
  }\textbf {\bibinfo {volume} {108}},\ \bibinfo {pages} {12233} (\bibinfo
  {year} {2011})}\BibitemShut {NoStop}%
\bibitem [{\citenamefont {Song}\ and\ \citenamefont
  {Bernevig}(2022)}]{Song20211110MATBGHF}%
  \BibitemOpen
  \bibfield  {author} {\bibinfo {author} {\bibfnamefont {Z.-D.}\ \bibnamefont
  {Song}}\ and\ \bibinfo {author} {\bibfnamefont {B.~A.}\ \bibnamefont
  {Bernevig}},\ }\bibfield  {title} {\bibinfo {title} {Magic-angle twisted
  bilayer graphene as a topological heavy fermion problem},\ }\href
  {https://doi.org/10.1103/PhysRevLett.129.047601} {\bibfield  {journal}
  {\bibinfo  {journal} {Phys. Rev. Lett.}\ }\textbf {\bibinfo {volume} {129}},\
  \bibinfo {pages} {047601} (\bibinfo {year} {2022})}\BibitemShut {NoStop}%
\bibitem [{\citenamefont {Liu}\ \emph {et~al.}(2023)\citenamefont {Liu},
  \citenamefont {Chen}, \citenamefont {Yazdani},\ and\ \citenamefont
  {Bernevig}}]{CXL2023ElKPhCouplingTBG}%
  \BibitemOpen
  \bibfield  {author} {\bibinfo {author} {\bibfnamefont {C.-X.}\ \bibnamefont
  {Liu}}, \bibinfo {author} {\bibfnamefont {Y.}~\bibnamefont {Chen}}, \bibinfo
  {author} {\bibfnamefont {A.}~\bibnamefont {Yazdani}},\ and\ \bibinfo {author}
  {\bibfnamefont {B.~A.}\ \bibnamefont {Bernevig}},\ }\bibfield  {title}
  {\bibinfo {title} {Electron-k-phonon interaction in twisted bilayer
  graphene},\ }\href {https://arxiv.org/abs/2303.15551} {\bibfield  {journal}
  {\bibinfo  {journal} {arXiv:2303.15551}\ } (\bibinfo {year}
  {2023})}\BibitemShut {NoStop}%
\bibitem [{\citenamefont {Kortus}\ \emph {et~al.}(2001)\citenamefont {Kortus},
  \citenamefont {Mazin}, \citenamefont {Belashchenko}, \citenamefont
  {Antropov},\ and\ \citenamefont {Boyer}}]{Boyer01302001MgB2SCBand}%
  \BibitemOpen
  \bibfield  {author} {\bibinfo {author} {\bibfnamefont {J.}~\bibnamefont
  {Kortus}}, \bibinfo {author} {\bibfnamefont {I.~I.}\ \bibnamefont {Mazin}},
  \bibinfo {author} {\bibfnamefont {K.~D.}\ \bibnamefont {Belashchenko}},
  \bibinfo {author} {\bibfnamefont {V.~P.}\ \bibnamefont {Antropov}},\ and\
  \bibinfo {author} {\bibfnamefont {L.~L.}\ \bibnamefont {Boyer}},\ }\bibfield
  {title} {\bibinfo {title} {Superconductivity of metallic boron in
  ${\mathrm{mgb}}_{2}$},\ }\href {https://doi.org/10.1103/PhysRevLett.86.4656}
  {\bibfield  {journal} {\bibinfo  {journal} {Phys. Rev. Lett.}\ }\textbf
  {\bibinfo {volume} {86}},\ \bibinfo {pages} {4656} (\bibinfo {year}
  {2001})}\BibitemShut {NoStop}%
\bibitem [{\citenamefont {An}\ and\ \citenamefont
  {Pickett}(2001)}]{Pickett2011CovalentBondsDriven}%
  \BibitemOpen
  \bibfield  {author} {\bibinfo {author} {\bibfnamefont {J.~M.}\ \bibnamefont
  {An}}\ and\ \bibinfo {author} {\bibfnamefont {W.~E.}\ \bibnamefont
  {Pickett}},\ }\bibfield  {title} {\bibinfo {title} {Superconductivity of
  ${\mathrm{mgb}}_{2}$: Covalent bonds driven metallic},\ }\href
  {https://doi.org/10.1103/PhysRevLett.86.4366} {\bibfield  {journal} {\bibinfo
   {journal} {Phys. Rev. Lett.}\ }\textbf {\bibinfo {volume} {86}},\ \bibinfo
  {pages} {4366} (\bibinfo {year} {2001})}\BibitemShut {NoStop}%
\bibitem [{\citenamefont {Kong}\ \emph {et~al.}(2001)\citenamefont {Kong},
  \citenamefont {Dolgov}, \citenamefont {Jepsen},\ and\ \citenamefont
  {Andersen}}]{Kong02272001MgB2EPC}%
  \BibitemOpen
  \bibfield  {author} {\bibinfo {author} {\bibfnamefont {Y.}~\bibnamefont
  {Kong}}, \bibinfo {author} {\bibfnamefont {O.~V.}\ \bibnamefont {Dolgov}},
  \bibinfo {author} {\bibfnamefont {O.}~\bibnamefont {Jepsen}},\ and\ \bibinfo
  {author} {\bibfnamefont {O.~K.}\ \bibnamefont {Andersen}},\ }\bibfield
  {title} {\bibinfo {title} {Electron-phonon interaction in the normal and
  superconducting states of ${\mathrm{mgb}}_{2}$},\ }\href
  {https://doi.org/10.1103/PhysRevB.64.020501} {\bibfield  {journal} {\bibinfo
  {journal} {Phys. Rev. B}\ }\textbf {\bibinfo {volume} {64}},\ \bibinfo
  {pages} {020501} (\bibinfo {year} {2001})}\BibitemShut {NoStop}%
\bibitem [{\citenamefont {Shukla}\ \emph {et~al.}(2003)\citenamefont {Shukla},
  \citenamefont {Calandra}, \citenamefont {d'Astuto}, \citenamefont {Lazzeri},
  \citenamefont {Mauri}, \citenamefont {Bellin}, \citenamefont {Krisch},
  \citenamefont {Karpinski}, \citenamefont {Kazakov}, \citenamefont {Jun},
  \citenamefont {Daghero},\ and\ \citenamefont
  {Parlinski}}]{Parlinski2003MgB2PhononLineWidth}%
  \BibitemOpen
  \bibfield  {author} {\bibinfo {author} {\bibfnamefont {A.}~\bibnamefont
  {Shukla}}, \bibinfo {author} {\bibfnamefont {M.}~\bibnamefont {Calandra}},
  \bibinfo {author} {\bibfnamefont {M.}~\bibnamefont {d'Astuto}}, \bibinfo
  {author} {\bibfnamefont {M.}~\bibnamefont {Lazzeri}}, \bibinfo {author}
  {\bibfnamefont {F.}~\bibnamefont {Mauri}}, \bibinfo {author} {\bibfnamefont
  {C.}~\bibnamefont {Bellin}}, \bibinfo {author} {\bibfnamefont
  {M.}~\bibnamefont {Krisch}}, \bibinfo {author} {\bibfnamefont
  {J.}~\bibnamefont {Karpinski}}, \bibinfo {author} {\bibfnamefont {S.~M.}\
  \bibnamefont {Kazakov}}, \bibinfo {author} {\bibfnamefont {J.}~\bibnamefont
  {Jun}}, \bibinfo {author} {\bibfnamefont {D.}~\bibnamefont {Daghero}},\ and\
  \bibinfo {author} {\bibfnamefont {K.}~\bibnamefont {Parlinski}},\ }\bibfield
  {title} {\bibinfo {title} {Phonon dispersion and lifetimes in
  ${\mathrm{m}\mathrm{g}\mathrm{b}}_{2}$},\ }\href
  {https://doi.org/10.1103/PhysRevLett.90.095506} {\bibfield  {journal}
  {\bibinfo  {journal} {Phys. Rev. Lett.}\ }\textbf {\bibinfo {volume} {90}},\
  \bibinfo {pages} {095506} (\bibinfo {year} {2003})}\BibitemShut {NoStop}%
\bibitem [{\citenamefont {Jin}\ \emph {et~al.}(2019)\citenamefont {Jin},
  \citenamefont {Huang}, \citenamefont {Mei}, \citenamefont {Liu},
  \citenamefont {Lim},\ and\ \citenamefont {Liu}}]{Liu10192019MgB2Dirac}%
  \BibitemOpen
  \bibfield  {author} {\bibinfo {author} {\bibfnamefont {K.-H.}\ \bibnamefont
  {Jin}}, \bibinfo {author} {\bibfnamefont {H.}~\bibnamefont {Huang}}, \bibinfo
  {author} {\bibfnamefont {J.-W.}\ \bibnamefont {Mei}}, \bibinfo {author}
  {\bibfnamefont {Z.}~\bibnamefont {Liu}}, \bibinfo {author} {\bibfnamefont
  {L.-K.}\ \bibnamefont {Lim}},\ and\ \bibinfo {author} {\bibfnamefont
  {F.}~\bibnamefont {Liu}},\ }\bibfield  {title} {\bibinfo {title} {Topological
  superconducting phase in high-tc superconductor mgb2 with dirac--nodal-line
  fermions},\ }\href {https://doi.org/10.1038/s41524-019-0191-2} {\bibfield
  {journal} {\bibinfo  {journal} {npj Computational Materials}\ }\textbf
  {\bibinfo {volume} {5}},\ \bibinfo {pages} {57} (\bibinfo {year}
  {2019})}\BibitemShut {NoStop}%
\bibitem [{\citenamefont {Aroyo}\ \emph {et~al.}(2006)\citenamefont {Aroyo},
  \citenamefont {Perez-Mato}, \citenamefont {Capillas}, \citenamefont
  {Kroumova}, \citenamefont {Ivantchev}, \citenamefont {Madariaga},
  \citenamefont {Kirov},\ and\ \citenamefont {Wondratschek}}]{Aroyo2006Bilbao}%
  \BibitemOpen
  \bibfield  {author} {\bibinfo {author} {\bibfnamefont {M.~I.}\ \bibnamefont
  {Aroyo}}, \bibinfo {author} {\bibfnamefont {J.~M.}\ \bibnamefont
  {Perez-Mato}}, \bibinfo {author} {\bibfnamefont {C.}~\bibnamefont
  {Capillas}}, \bibinfo {author} {\bibfnamefont {E.}~\bibnamefont {Kroumova}},
  \bibinfo {author} {\bibfnamefont {S.}~\bibnamefont {Ivantchev}}, \bibinfo
  {author} {\bibfnamefont {G.}~\bibnamefont {Madariaga}}, \bibinfo {author}
  {\bibfnamefont {A.}~\bibnamefont {Kirov}},\ and\ \bibinfo {author}
  {\bibfnamefont {H.}~\bibnamefont {Wondratschek}},\ }\bibfield  {title}
  {\bibinfo {title} {Bilbao crystallographic server: I. databases and
  crystallographic computing programs},\ }\href@noop {} {\bibfield  {journal}
  {\bibinfo  {journal} {Zeitschrift f{\"u}r Kristallographie-Crystalline
  Materials}\ }\textbf {\bibinfo {volume} {221}},\ \bibinfo {pages} {15}
  (\bibinfo {year} {2006})}\BibitemShut {NoStop}%
\bibitem [{\citenamefont {Ahn}\ \emph {et~al.}(2019)\citenamefont {Ahn},
  \citenamefont {Park},\ and\ \citenamefont {Yang}}]{Ahn2019TBGFragile}%
  \BibitemOpen
  \bibfield  {author} {\bibinfo {author} {\bibfnamefont {J.}~\bibnamefont
  {Ahn}}, \bibinfo {author} {\bibfnamefont {S.}~\bibnamefont {Park}},\ and\
  \bibinfo {author} {\bibfnamefont {B.-J.}\ \bibnamefont {Yang}},\ }\bibfield
  {title} {\bibinfo {title} {Failure of nielsen-ninomiya theorem and fragile
  topology in two-dimensional systems with space-time inversion symmetry:
  Application to twisted bilayer graphene at magic angle},\ }\href
  {https://doi.org/10.1103/PhysRevX.9.021013} {\bibfield  {journal} {\bibinfo
  {journal} {Phys. Rev. X}\ }\textbf {\bibinfo {volume} {9}},\ \bibinfo {pages}
  {021013} (\bibinfo {year} {2019})}\BibitemShut {NoStop}%
\bibitem [{\citenamefont {Yan}\ \emph {et~al.}(2007)\citenamefont {Yan},
  \citenamefont {Zhang}, \citenamefont {Kim},\ and\ \citenamefont
  {Pinczuk}}]{Yan2007EPCGrapheneEXPRamen}%
  \BibitemOpen
  \bibfield  {author} {\bibinfo {author} {\bibfnamefont {J.}~\bibnamefont
  {Yan}}, \bibinfo {author} {\bibfnamefont {Y.}~\bibnamefont {Zhang}}, \bibinfo
  {author} {\bibfnamefont {P.}~\bibnamefont {Kim}},\ and\ \bibinfo {author}
  {\bibfnamefont {A.}~\bibnamefont {Pinczuk}},\ }\bibfield  {title} {\bibinfo
  {title} {Electric field effect tuning of electron-phonon coupling in
  graphene},\ }\href {https://doi.org/10.1103/PhysRevLett.98.166802} {\bibfield
   {journal} {\bibinfo  {journal} {Phys. Rev. Lett.}\ }\textbf {\bibinfo
  {volume} {98}},\ \bibinfo {pages} {166802} (\bibinfo {year}
  {2007})}\BibitemShut {NoStop}%
\bibitem [{\citenamefont {Maultzsch}\ \emph {et~al.}(2004)\citenamefont
  {Maultzsch}, \citenamefont {Reich}, \citenamefont {Thomsen}, \citenamefont
  {Requardt},\ and\ \citenamefont {Ordej\'on}}]{Maultzsch11092003Graphite}%
  \BibitemOpen
  \bibfield  {author} {\bibinfo {author} {\bibfnamefont {J.}~\bibnamefont
  {Maultzsch}}, \bibinfo {author} {\bibfnamefont {S.}~\bibnamefont {Reich}},
  \bibinfo {author} {\bibfnamefont {C.}~\bibnamefont {Thomsen}}, \bibinfo
  {author} {\bibfnamefont {H.}~\bibnamefont {Requardt}},\ and\ \bibinfo
  {author} {\bibfnamefont {P.}~\bibnamefont {Ordej\'on}},\ }\bibfield  {title}
  {\bibinfo {title} {Phonon dispersion in graphite},\ }\href
  {https://doi.org/10.1103/PhysRevLett.92.075501} {\bibfield  {journal}
  {\bibinfo  {journal} {Phys. Rev. Lett.}\ }\textbf {\bibinfo {volume} {92}},\
  \bibinfo {pages} {075501} (\bibinfo {year} {2004})}\BibitemShut {NoStop}%
\bibitem [{\citenamefont {Zhu}\ \emph {et~al.}(2012)\citenamefont {Zhu},
  \citenamefont {Santos}, \citenamefont {Howard}, \citenamefont {Sankar},
  \citenamefont {Chou}, \citenamefont {Chamon},\ and\ \citenamefont
  {El-Batanouny}}]{Zhu01302012EPClambdaHelium}%
  \BibitemOpen
  \bibfield  {author} {\bibinfo {author} {\bibfnamefont {X.}~\bibnamefont
  {Zhu}}, \bibinfo {author} {\bibfnamefont {L.}~\bibnamefont {Santos}},
  \bibinfo {author} {\bibfnamefont {C.}~\bibnamefont {Howard}}, \bibinfo
  {author} {\bibfnamefont {R.}~\bibnamefont {Sankar}}, \bibinfo {author}
  {\bibfnamefont {F.~C.}\ \bibnamefont {Chou}}, \bibinfo {author}
  {\bibfnamefont {C.}~\bibnamefont {Chamon}},\ and\ \bibinfo {author}
  {\bibfnamefont {M.}~\bibnamefont {El-Batanouny}},\ }\bibfield  {title}
  {\bibinfo {title} {Electron-phonon coupling on the surface of the topological
  insulator ${\mathrm{bi}}_{2}{\mathrm{se}}_{3}$ determined from surface-phonon
  dispersion measurements},\ }\href
  {https://doi.org/10.1103/PhysRevLett.108.185501} {\bibfield  {journal}
  {\bibinfo  {journal} {Phys. Rev. Lett.}\ }\textbf {\bibinfo {volume} {108}},\
  \bibinfo {pages} {185501} (\bibinfo {year} {2012})}\BibitemShut {NoStop}%
\bibitem [{\citenamefont {Benedek}\ \emph {et~al.}(2021)\citenamefont
  {Benedek}, \citenamefont {Manson},\ and\ \citenamefont
  {Miret-Art{\'e}s}}]{Benedek09082021EPCConstantGrapheneHe}%
  \BibitemOpen
  \bibfield  {author} {\bibinfo {author} {\bibfnamefont {G.}~\bibnamefont
  {Benedek}}, \bibinfo {author} {\bibfnamefont {J.~R.}\ \bibnamefont
  {Manson}},\ and\ \bibinfo {author} {\bibfnamefont {S.}~\bibnamefont
  {Miret-Art{\'e}s}},\ }\bibfield  {title} {\bibinfo {title} {The
  electron--phonon coupling constant for single-layer graphene on metal
  substrates determined from he atom scattering},\ }\href@noop {} {\bibfield
  {journal} {\bibinfo  {journal} {Physical Chemistry Chemical Physics}\
  }\textbf {\bibinfo {volume} {23}},\ \bibinfo {pages} {7575} (\bibinfo {year}
  {2021})}\BibitemShut {NoStop}%
\bibitem [{\citenamefont {Neupert}\ \emph
  {et~al.}(2013{\natexlab{b}})\citenamefont {Neupert}, \citenamefont {Chamon},\
  and\ \citenamefont {Mudry}}]{Neupert06102013CurrentNoiseFSM}%
  \BibitemOpen
  \bibfield  {author} {\bibinfo {author} {\bibfnamefont {T.}~\bibnamefont
  {Neupert}}, \bibinfo {author} {\bibfnamefont {C.}~\bibnamefont {Chamon}},\
  and\ \bibinfo {author} {\bibfnamefont {C.}~\bibnamefont {Mudry}},\ }\bibfield
   {title} {\bibinfo {title} {Measuring the quantum geometry of bloch bands
  with current noise},\ }\href {https://doi.org/10.1103/PhysRevB.87.245103}
  {\bibfield  {journal} {\bibinfo  {journal} {Phys. Rev. B}\ }\textbf {\bibinfo
  {volume} {87}},\ \bibinfo {pages} {245103} (\bibinfo {year}
  {2013}{\natexlab{b}})}\BibitemShut {NoStop}%
\bibitem [{\citenamefont {Sipe}\ and\ \citenamefont
  {Shkrebtii}(2000)}]{Sipe06291999OpticalResponseFSM}%
  \BibitemOpen
  \bibfield  {author} {\bibinfo {author} {\bibfnamefont {J.~E.}\ \bibnamefont
  {Sipe}}\ and\ \bibinfo {author} {\bibfnamefont {A.~I.}\ \bibnamefont
  {Shkrebtii}},\ }\bibfield  {title} {\bibinfo {title} {Second-order optical
  response in semiconductors},\ }\href
  {https://doi.org/10.1103/PhysRevB.61.5337} {\bibfield  {journal} {\bibinfo
  {journal} {Phys. Rev. B}\ }\textbf {\bibinfo {volume} {61}},\ \bibinfo
  {pages} {5337} (\bibinfo {year} {2000})}\BibitemShut {NoStop}%
\bibitem [{\citenamefont {Nair}\ \emph {et~al.}(2008)\citenamefont {Nair},
  \citenamefont {Blake}, \citenamefont {Grigorenko}, \citenamefont {Novoselov},
  \citenamefont {Booth}, \citenamefont {Stauber}, \citenamefont {Peres},\ and\
  \citenamefont {Geim}}]{Geim2008OpticalGraphene}%
  \BibitemOpen
  \bibfield  {author} {\bibinfo {author} {\bibfnamefont {R.~R.}\ \bibnamefont
  {Nair}}, \bibinfo {author} {\bibfnamefont {P.}~\bibnamefont {Blake}},
  \bibinfo {author} {\bibfnamefont {A.~N.}\ \bibnamefont {Grigorenko}},
  \bibinfo {author} {\bibfnamefont {K.~S.}\ \bibnamefont {Novoselov}}, \bibinfo
  {author} {\bibfnamefont {T.~J.}\ \bibnamefont {Booth}}, \bibinfo {author}
  {\bibfnamefont {T.}~\bibnamefont {Stauber}}, \bibinfo {author} {\bibfnamefont
  {N.~M.~R.}\ \bibnamefont {Peres}},\ and\ \bibinfo {author} {\bibfnamefont
  {A.~K.}\ \bibnamefont {Geim}},\ }\bibfield  {title} {\bibinfo {title} {Fine
  structure constant defines visual transparency of graphene},\ }\href
  {https://doi.org/10.1126/science.1156965} {\bibfield  {journal} {\bibinfo
  {journal} {Science}\ }\textbf {\bibinfo {volume} {320}},\ \bibinfo {pages}
  {1308} (\bibinfo {year} {2008})},\ \Eprint
  {https://arxiv.org/abs/https://www.science.org/doi/pdf/10.1126/science.1156965}
  {https://www.science.org/doi/pdf/10.1126/science.1156965} \BibitemShut
  {NoStop}%
\bibitem [{\citenamefont {Mak}\ \emph {et~al.}(2008)\citenamefont {Mak},
  \citenamefont {Sfeir}, \citenamefont {Wu}, \citenamefont {Lui}, \citenamefont
  {Misewich},\ and\ \citenamefont {Heinz}}]{Mak2008OpticalGraphene}%
  \BibitemOpen
  \bibfield  {author} {\bibinfo {author} {\bibfnamefont {K.~F.}\ \bibnamefont
  {Mak}}, \bibinfo {author} {\bibfnamefont {M.~Y.}\ \bibnamefont {Sfeir}},
  \bibinfo {author} {\bibfnamefont {Y.}~\bibnamefont {Wu}}, \bibinfo {author}
  {\bibfnamefont {C.~H.}\ \bibnamefont {Lui}}, \bibinfo {author} {\bibfnamefont
  {J.~A.}\ \bibnamefont {Misewich}},\ and\ \bibinfo {author} {\bibfnamefont
  {T.~F.}\ \bibnamefont {Heinz}},\ }\bibfield  {title} {\bibinfo {title}
  {Measurement of the optical conductivity of graphene},\ }\href
  {https://doi.org/10.1103/PhysRevLett.101.196405} {\bibfield  {journal}
  {\bibinfo  {journal} {Phys. Rev. Lett.}\ }\textbf {\bibinfo {volume} {101}},\
  \bibinfo {pages} {196405} (\bibinfo {year} {2008})}\BibitemShut {NoStop}%
\bibitem [{\citenamefont {Mak}\ \emph {et~al.}(2012)\citenamefont {Mak},
  \citenamefont {Ju}, \citenamefont {Wang},\ and\ \citenamefont
  {Heinz}}]{Mak2012optical}%
  \BibitemOpen
  \bibfield  {author} {\bibinfo {author} {\bibfnamefont {K.~F.}\ \bibnamefont
  {Mak}}, \bibinfo {author} {\bibfnamefont {L.}~\bibnamefont {Ju}}, \bibinfo
  {author} {\bibfnamefont {F.}~\bibnamefont {Wang}},\ and\ \bibinfo {author}
  {\bibfnamefont {T.~F.}\ \bibnamefont {Heinz}},\ }\bibfield  {title} {\bibinfo
  {title} {Optical spectroscopy of graphene: From the far infrared to the
  ultraviolet},\ }\href@noop {} {\bibfield  {journal} {\bibinfo  {journal}
  {Solid State Communications}\ }\textbf {\bibinfo {volume} {152}},\ \bibinfo
  {pages} {1341} (\bibinfo {year} {2012})}\BibitemShut {NoStop}%
\bibitem [{\citenamefont {Chen}\ \emph {et~al.}(2009)\citenamefont {Chen},
  \citenamefont {Analytis}, \citenamefont {Chu}, \citenamefont {Liu},
  \citenamefont {Mo}, \citenamefont {Qi}, \citenamefont {Zhang}, \citenamefont
  {Lu}, \citenamefont {Dai}, \citenamefont {Fang} \emph
  {et~al.}}]{Chen07102009Bi2Te3}%
  \BibitemOpen
  \bibfield  {author} {\bibinfo {author} {\bibfnamefont {Y.}~\bibnamefont
  {Chen}}, \bibinfo {author} {\bibfnamefont {J.~G.}\ \bibnamefont {Analytis}},
  \bibinfo {author} {\bibfnamefont {J.-H.}\ \bibnamefont {Chu}}, \bibinfo
  {author} {\bibfnamefont {Z.}~\bibnamefont {Liu}}, \bibinfo {author}
  {\bibfnamefont {S.-K.}\ \bibnamefont {Mo}}, \bibinfo {author} {\bibfnamefont
  {X.-L.}\ \bibnamefont {Qi}}, \bibinfo {author} {\bibfnamefont
  {H.}~\bibnamefont {Zhang}}, \bibinfo {author} {\bibfnamefont
  {D.}~\bibnamefont {Lu}}, \bibinfo {author} {\bibfnamefont {X.}~\bibnamefont
  {Dai}}, \bibinfo {author} {\bibfnamefont {Z.}~\bibnamefont {Fang}}, \emph
  {et~al.},\ }\bibfield  {title} {\bibinfo {title} {Experimental realization of
  a three-dimensional topological insulator, bi2te3},\ }\href@noop {}
  {\bibfield  {journal} {\bibinfo  {journal} {science}\ }\textbf {\bibinfo
  {volume} {325}},\ \bibinfo {pages} {178} (\bibinfo {year}
  {2009})}\BibitemShut {NoStop}%
\bibitem [{\citenamefont {Hatch}\ \emph {et~al.}(2011)\citenamefont {Hatch},
  \citenamefont {Bianchi}, \citenamefont {Guan}, \citenamefont {Bao},
  \citenamefont {Mi}, \citenamefont {Iversen}, \citenamefont {Nilsson},
  \citenamefont {Hornek\ae{}r},\ and\ \citenamefont
  {Hofmann}}]{Hofmann04262011EPClambdaARPES}%
  \BibitemOpen
  \bibfield  {author} {\bibinfo {author} {\bibfnamefont {R.~C.}\ \bibnamefont
  {Hatch}}, \bibinfo {author} {\bibfnamefont {M.}~\bibnamefont {Bianchi}},
  \bibinfo {author} {\bibfnamefont {D.}~\bibnamefont {Guan}}, \bibinfo {author}
  {\bibfnamefont {S.}~\bibnamefont {Bao}}, \bibinfo {author} {\bibfnamefont
  {J.}~\bibnamefont {Mi}}, \bibinfo {author} {\bibfnamefont {B.~B.}\
  \bibnamefont {Iversen}}, \bibinfo {author} {\bibfnamefont {L.}~\bibnamefont
  {Nilsson}}, \bibinfo {author} {\bibfnamefont {L.}~\bibnamefont
  {Hornek\ae{}r}},\ and\ \bibinfo {author} {\bibfnamefont {P.}~\bibnamefont
  {Hofmann}},\ }\bibfield  {title} {\bibinfo {title} {Stability of the
  ${\mathbf{bi}}_{2}{\mathbf{se}}_{3}$(111) topological state: Electron-phonon
  and electron-defect scattering},\ }\href
  {https://doi.org/10.1103/PhysRevB.83.241303} {\bibfield  {journal} {\bibinfo
  {journal} {Phys. Rev. B}\ }\textbf {\bibinfo {volume} {83}},\ \bibinfo
  {pages} {241303} (\bibinfo {year} {2011})}\BibitemShut {NoStop}%
\bibitem [{\citenamefont {Pan}\ \emph {et~al.}(2012)\citenamefont {Pan},
  \citenamefont {Fedorov}, \citenamefont {Gardner}, \citenamefont {Lee},
  \citenamefont {Chu},\ and\ \citenamefont
  {Valla}}]{Pan09162011EPClambdaARPES}%
  \BibitemOpen
  \bibfield  {author} {\bibinfo {author} {\bibfnamefont {Z.-H.}\ \bibnamefont
  {Pan}}, \bibinfo {author} {\bibfnamefont {A.~V.}\ \bibnamefont {Fedorov}},
  \bibinfo {author} {\bibfnamefont {D.}~\bibnamefont {Gardner}}, \bibinfo
  {author} {\bibfnamefont {Y.~S.}\ \bibnamefont {Lee}}, \bibinfo {author}
  {\bibfnamefont {S.}~\bibnamefont {Chu}},\ and\ \bibinfo {author}
  {\bibfnamefont {T.}~\bibnamefont {Valla}},\ }\bibfield  {title} {\bibinfo
  {title} {Measurement of an exceptionally weak electron-phonon coupling on the
  surface of the topological insulator ${\mathrm{bi}}_{2}{\mathrm{se}}_{3}$
  using angle-resolved photoemission spectroscopy},\ }\href
  {https://doi.org/10.1103/PhysRevLett.108.187001} {\bibfield  {journal}
  {\bibinfo  {journal} {Phys. Rev. Lett.}\ }\textbf {\bibinfo {volume} {108}},\
  \bibinfo {pages} {187001} (\bibinfo {year} {2012})}\BibitemShut {NoStop}%
\bibitem [{\citenamefont {Sobota}\ \emph {et~al.}(2014)\citenamefont {Sobota},
  \citenamefont {Yang}, \citenamefont {Leuenberger}, \citenamefont {Kemper},
  \citenamefont {Analytis}, \citenamefont {Fisher}, \citenamefont {Kirchmann},
  \citenamefont {Devereaux},\ and\ \citenamefont
  {Shen}}]{Sobota2014Bi2Se3EPCArpes}%
  \BibitemOpen
  \bibfield  {author} {\bibinfo {author} {\bibfnamefont {J.~A.}\ \bibnamefont
  {Sobota}}, \bibinfo {author} {\bibfnamefont {S.-L.}\ \bibnamefont {Yang}},
  \bibinfo {author} {\bibfnamefont {D.}~\bibnamefont {Leuenberger}}, \bibinfo
  {author} {\bibfnamefont {A.~F.}\ \bibnamefont {Kemper}}, \bibinfo {author}
  {\bibfnamefont {J.~G.}\ \bibnamefont {Analytis}}, \bibinfo {author}
  {\bibfnamefont {I.~R.}\ \bibnamefont {Fisher}}, \bibinfo {author}
  {\bibfnamefont {P.~S.}\ \bibnamefont {Kirchmann}}, \bibinfo {author}
  {\bibfnamefont {T.~P.}\ \bibnamefont {Devereaux}},\ and\ \bibinfo {author}
  {\bibfnamefont {Z.-X.}\ \bibnamefont {Shen}},\ }\bibfield  {title} {\bibinfo
  {title} {Distinguishing bulk and surface electron-phonon coupling in the
  topological insulator ${\mathrm{bi}}_{2}{\mathrm{se}}_{3}$ using
  time-resolved photoemission spectroscopy},\ }\href
  {https://doi.org/10.1103/PhysRevLett.113.157401} {\bibfield  {journal}
  {\bibinfo  {journal} {Phys. Rev. Lett.}\ }\textbf {\bibinfo {volume} {113}},\
  \bibinfo {pages} {157401} (\bibinfo {year} {2014})}\BibitemShut {NoStop}%
\bibitem [{\citenamefont {Baldini}\ \emph {et~al.}(2023)\citenamefont
  {Baldini}, \citenamefont {Zong}, \citenamefont {Choi}, \citenamefont {Lee},
  \citenamefont {Michael}, \citenamefont {Windgaetter}, \citenamefont {Mazin},
  \citenamefont {Latini}, \citenamefont {Azoury}, \citenamefont {Lv} \emph
  {et~al.}}]{Baldini2023Ta2NiSe5EPCArpes}%
  \BibitemOpen
  \bibfield  {author} {\bibinfo {author} {\bibfnamefont {E.}~\bibnamefont
  {Baldini}}, \bibinfo {author} {\bibfnamefont {A.}~\bibnamefont {Zong}},
  \bibinfo {author} {\bibfnamefont {D.}~\bibnamefont {Choi}}, \bibinfo {author}
  {\bibfnamefont {C.}~\bibnamefont {Lee}}, \bibinfo {author} {\bibfnamefont
  {M.~H.}\ \bibnamefont {Michael}}, \bibinfo {author} {\bibfnamefont
  {L.}~\bibnamefont {Windgaetter}}, \bibinfo {author} {\bibfnamefont {I.~I.}\
  \bibnamefont {Mazin}}, \bibinfo {author} {\bibfnamefont {S.}~\bibnamefont
  {Latini}}, \bibinfo {author} {\bibfnamefont {D.}~\bibnamefont {Azoury}},
  \bibinfo {author} {\bibfnamefont {B.}~\bibnamefont {Lv}}, \emph {et~al.},\
  }\bibfield  {title} {\bibinfo {title} {The spontaneous symmetry breaking in
  ta2nise5 is structural in nature},\ }\href
  {https://www.pnas.org/doi/abs/10.1073/pnas.2221688120} {\bibfield  {journal}
  {\bibinfo  {journal} {Proceedings of the National Academy of Sciences}\
  }\textbf {\bibinfo {volume} {120}},\ \bibinfo {pages} {e2221688120} (\bibinfo
  {year} {2023})}\BibitemShut {NoStop}%
\bibitem [{\citenamefont {Wang}\ \emph {et~al.}(2001)\citenamefont {Wang},
  \citenamefont {Plackowski},\ and\ \citenamefont
  {Junod}}]{Wang05082001EPClambdaCv}%
  \BibitemOpen
  \bibfield  {author} {\bibinfo {author} {\bibfnamefont {Y.}~\bibnamefont
  {Wang}}, \bibinfo {author} {\bibfnamefont {T.}~\bibnamefont {Plackowski}},\
  and\ \bibinfo {author} {\bibfnamefont {A.}~\bibnamefont {Junod}},\ }\bibfield
   {title} {\bibinfo {title} {Specific heat in the superconducting and normal
  state (2–300 k, 0–16 t), and magnetic susceptibility of the 38 k
  superconductor mgb2: evidence for a multicomponent gap},\ }\href
  {https://doi.org/https://doi.org/10.1016/S0921-4534(01)00617-7} {\bibfield
  {journal} {\bibinfo  {journal} {Physica C: Superconductivity}\ }\textbf
  {\bibinfo {volume} {355}},\ \bibinfo {pages} {179} (\bibinfo {year}
  {2001})}\BibitemShut {NoStop}%
\bibitem [{\citenamefont {{Korshunov}}\ \emph {et~al.}(2023)\citenamefont
  {{Korshunov}}, \citenamefont {{Hu}}, \citenamefont {{Subires}}, \citenamefont
  {{Jiang}}, \citenamefont {{C{\u{a}}lug{\u{a}}ru}}, \citenamefont {{Feng}},
  \citenamefont {{Rajapitamahuni}}, \citenamefont {{Yi}}, \citenamefont
  {{Roychowdhury}}, \citenamefont {{Vergniory}}, \citenamefont {{Strempfer}},
  \citenamefont {{Shekhar}}, \citenamefont {{Vescovo}}, \citenamefont
  {{Chernyshov}}, \citenamefont {{Said}}, \citenamefont {{Bosak}},
  \citenamefont {{Felser}}, \citenamefont {{Bernevig}},\ and\ \citenamefont
  {{Blanco-Canosa}}}]{Korshunov04182023ScV6Sn6Kagome}%
  \BibitemOpen
  \bibfield  {author} {\bibinfo {author} {\bibfnamefont {A.}~\bibnamefont
  {{Korshunov}}}, \bibinfo {author} {\bibfnamefont {H.}~\bibnamefont {{Hu}}},
  \bibinfo {author} {\bibfnamefont {D.}~\bibnamefont {{Subires}}}, \bibinfo
  {author} {\bibfnamefont {Y.}~\bibnamefont {{Jiang}}}, \bibinfo {author}
  {\bibfnamefont {D.}~\bibnamefont {{C{\u{a}}lug{\u{a}}ru}}}, \bibinfo {author}
  {\bibfnamefont {X.}~\bibnamefont {{Feng}}}, \bibinfo {author} {\bibfnamefont
  {A.}~\bibnamefont {{Rajapitamahuni}}}, \bibinfo {author} {\bibfnamefont
  {C.}~\bibnamefont {{Yi}}}, \bibinfo {author} {\bibfnamefont {S.}~\bibnamefont
  {{Roychowdhury}}}, \bibinfo {author} {\bibfnamefont {M.~G.}\ \bibnamefont
  {{Vergniory}}}, \bibinfo {author} {\bibfnamefont {J.}~\bibnamefont
  {{Strempfer}}}, \bibinfo {author} {\bibfnamefont {C.}~\bibnamefont
  {{Shekhar}}}, \bibinfo {author} {\bibfnamefont {E.}~\bibnamefont
  {{Vescovo}}}, \bibinfo {author} {\bibfnamefont {D.}~\bibnamefont
  {{Chernyshov}}}, \bibinfo {author} {\bibfnamefont {A.~H.}\ \bibnamefont
  {{Said}}}, \bibinfo {author} {\bibfnamefont {A.}~\bibnamefont {{Bosak}}},
  \bibinfo {author} {\bibfnamefont {C.}~\bibnamefont {{Felser}}}, \bibinfo
  {author} {\bibfnamefont {B.~A.}\ \bibnamefont {{Bernevig}}},\ and\ \bibinfo
  {author} {\bibfnamefont {S.}~\bibnamefont {{Blanco-Canosa}}},\ }\bibfield
  {title} {\bibinfo {title} {{Softening of a flat phonon mode in the kagome
  ScV$_6$Sn$_6$}},\ }\href {https://doi.org/10.48550/arXiv.2304.09173}
  {\bibfield  {journal} {\bibinfo  {journal} {arXiv e-prints}\ ,\ \bibinfo
  {eid} {arXiv:2304.09173}} (\bibinfo {year} {2023})},\ \Eprint
  {https://arxiv.org/abs/2304.09173} {arXiv:2304.09173 [cond-mat.str-el]}
  \BibitemShut {NoStop}%
\bibitem [{\citenamefont {Schrieffer}\ and\ \citenamefont
  {Wolff}(1966)}]{SchriefferWolffTransformation1966}%
  \BibitemOpen
  \bibfield  {author} {\bibinfo {author} {\bibfnamefont {J.~R.}\ \bibnamefont
  {Schrieffer}}\ and\ \bibinfo {author} {\bibfnamefont {P.~A.}\ \bibnamefont
  {Wolff}},\ }\bibfield  {title} {\bibinfo {title} {Relation between the
  anderson and kondo hamiltonians},\ }\href
  {https://doi.org/10.1103/PhysRev.149.491} {\bibfield  {journal} {\bibinfo
  {journal} {Phys. Rev.}\ }\textbf {\bibinfo {volume} {149}},\ \bibinfo {pages}
  {491} (\bibinfo {year} {1966})}\BibitemShut {NoStop}%
\bibitem [{\citenamefont {Sigrist}\ and\ \citenamefont
  {Ueda}(1991)}]{Sigrist1991SC}%
  \BibitemOpen
  \bibfield  {author} {\bibinfo {author} {\bibfnamefont {M.}~\bibnamefont
  {Sigrist}}\ and\ \bibinfo {author} {\bibfnamefont {K.}~\bibnamefont {Ueda}},\
  }\bibfield  {title} {\bibinfo {title} {Phenomenological theory of
  unconventional superconductivity},\ }\href
  {https://doi.org/10.1103/RevModPhys.63.239} {\bibfield  {journal} {\bibinfo
  {journal} {Rev. Mod. Phys.}\ }\textbf {\bibinfo {volume} {63}},\ \bibinfo
  {pages} {239} (\bibinfo {year} {1991})}\BibitemShut {NoStop}%
\bibitem [{\citenamefont {Altland}\ and\ \citenamefont
  {Simons}(2010)}]{altland2010condensed}%
  \BibitemOpen
  \bibfield  {author} {\bibinfo {author} {\bibfnamefont {A.}~\bibnamefont
  {Altland}}\ and\ \bibinfo {author} {\bibfnamefont {B.~D.}\ \bibnamefont
  {Simons}},\ }\href@noop {} {\emph {\bibinfo {title} {Condensed matter field
  theory}}}\ (\bibinfo  {publisher} {Cambridge University Press},\ \bibinfo
  {year} {2010})\BibitemShut {NoStop}%
\bibitem [{\citenamefont {Srednicki}(2007)}]{Srednicki2007QFT}%
  \BibitemOpen
  \bibfield  {author} {\bibinfo {author} {\bibfnamefont {M.}~\bibnamefont
  {Srednicki}},\ }\href@noop {} {\emph {\bibinfo {title} {Quantum field
  theory}}}\ (\bibinfo  {publisher} {Cambridge University Press},\ \bibinfo
  {year} {2007})\BibitemShut {NoStop}%
\bibitem [{\citenamefont {Alexandradinata}\ \emph {et~al.}(2014)\citenamefont
  {Alexandradinata}, \citenamefont {Dai},\ and\ \citenamefont
  {Bernevig}}]{Aris2014Wilsonloop}%
  \BibitemOpen
  \bibfield  {author} {\bibinfo {author} {\bibfnamefont {A.}~\bibnamefont
  {Alexandradinata}}, \bibinfo {author} {\bibfnamefont {X.}~\bibnamefont
  {Dai}},\ and\ \bibinfo {author} {\bibfnamefont {B.~A.}\ \bibnamefont
  {Bernevig}},\ }\bibfield  {title} {\bibinfo {title} {Wilson-loop
  characterization of inversion-symmetric topological insulators},\ }\href
  {https://doi.org/10.1103/PhysRevB.89.155114} {\bibfield  {journal} {\bibinfo
  {journal} {Phys. Rev. B}\ }\textbf {\bibinfo {volume} {89}},\ \bibinfo
  {pages} {155114} (\bibinfo {year} {2014})}\BibitemShut {NoStop}%
\bibitem [{\citenamefont {Yu}\ \emph {et~al.}(2021)\citenamefont {Yu},
  \citenamefont {Chen},\ and\ \citenamefont {Sarma}}]{Yu2021EOCP}%
  \BibitemOpen
  \bibfield  {author} {\bibinfo {author} {\bibfnamefont {J.}~\bibnamefont
  {Yu}}, \bibinfo {author} {\bibfnamefont {Y.-A.}\ \bibnamefont {Chen}},\ and\
  \bibinfo {author} {\bibfnamefont {S.~D.}\ \bibnamefont {Sarma}},\ }\href
  {https://arxiv.org/abs/2109.02685} {\bibinfo {title} {Euler obstructed cooper
  pairing: Nodal superconductivity and hinge majorana zero modes}} (\bibinfo
  {year} {2021}),\ \Eprint {https://arxiv.org/abs/2109.02685} {arXiv:2109.02685
  [cond-mat.supr-con]} \BibitemShut {NoStop}%
\bibitem [{\citenamefont {Thingstad}\ \emph {et~al.}(2020)\citenamefont
  {Thingstad}, \citenamefont {Kamra}, \citenamefont {Wells},\ and\
  \citenamefont {Sudb\o{}}}]{Thingstad20191211EPCSCGraphenehBN}%
  \BibitemOpen
  \bibfield  {author} {\bibinfo {author} {\bibfnamefont {E.}~\bibnamefont
  {Thingstad}}, \bibinfo {author} {\bibfnamefont {A.}~\bibnamefont {Kamra}},
  \bibinfo {author} {\bibfnamefont {J.~W.}\ \bibnamefont {Wells}},\ and\
  \bibinfo {author} {\bibfnamefont {A.}~\bibnamefont {Sudb\o{}}},\ }\bibfield
  {title} {\bibinfo {title} {Phonon-mediated superconductivity in doped
  monolayer materials},\ }\href {https://doi.org/10.1103/PhysRevB.101.214513}
  {\bibfield  {journal} {\bibinfo  {journal} {Phys. Rev. B}\ }\textbf {\bibinfo
  {volume} {101}},\ \bibinfo {pages} {214513} (\bibinfo {year}
  {2020})}\BibitemShut {NoStop}%
\bibitem [{\citenamefont {Slater}\ and\ \citenamefont
  {Koster}(1954)}]{Slater1954LCAO}%
  \BibitemOpen
  \bibfield  {author} {\bibinfo {author} {\bibfnamefont {J.~C.}\ \bibnamefont
  {Slater}}\ and\ \bibinfo {author} {\bibfnamefont {G.~F.}\ \bibnamefont
  {Koster}},\ }\bibfield  {title} {\bibinfo {title} {Simplified lcao method for
  the periodic potential problem},\ }\href
  {https://doi.org/10.1103/PhysRev.94.1498} {\bibfield  {journal} {\bibinfo
  {journal} {Phys. Rev.}\ }\textbf {\bibinfo {volume} {94}},\ \bibinfo {pages}
  {1498} (\bibinfo {year} {1954})}\BibitemShut {NoStop}%
\bibitem [{\citenamefont {Chen}\ \emph {et~al.}(2023)\citenamefont {Chen},
  \citenamefont {Nuckolls}, \citenamefont {Ding}, \citenamefont {Miao},
  \citenamefont {Wong}, \citenamefont {Oh}, \citenamefont {Lee}, \citenamefont
  {He}, \citenamefont {Peng}, \citenamefont {Pei} \emph
  {et~al.}}]{Chen2023ElKPhCouplingTBG}%
  \BibitemOpen
  \bibfield  {author} {\bibinfo {author} {\bibfnamefont {C.}~\bibnamefont
  {Chen}}, \bibinfo {author} {\bibfnamefont {K.~P.}\ \bibnamefont {Nuckolls}},
  \bibinfo {author} {\bibfnamefont {S.}~\bibnamefont {Ding}}, \bibinfo {author}
  {\bibfnamefont {W.}~\bibnamefont {Miao}}, \bibinfo {author} {\bibfnamefont
  {D.}~\bibnamefont {Wong}}, \bibinfo {author} {\bibfnamefont {M.}~\bibnamefont
  {Oh}}, \bibinfo {author} {\bibfnamefont {R.~L.}\ \bibnamefont {Lee}},
  \bibinfo {author} {\bibfnamefont {S.}~\bibnamefont {He}}, \bibinfo {author}
  {\bibfnamefont {C.}~\bibnamefont {Peng}}, \bibinfo {author} {\bibfnamefont
  {D.}~\bibnamefont {Pei}}, \emph {et~al.},\ }\bibfield  {title} {\bibinfo
  {title} {Strong inter-valley electron-phonon coupling in magic-angle twisted
  bilayer graphene},\ }\href@noop {} {\bibfield  {journal} {\bibinfo  {journal}
  {arXiv preprint arXiv:2303.14903}\ } (\bibinfo {year} {2023})}\BibitemShut
  {NoStop}%
\bibitem [{\citenamefont {Kim}\ \emph {et~al.}(2018)\citenamefont {Kim},
  \citenamefont {Ray}, \citenamefont {Bahr},\ and\ \citenamefont
  {Lordi}}]{Lordi05102018MgB2ElectronBands}%
  \BibitemOpen
  \bibfield  {author} {\bibinfo {author} {\bibfnamefont {C.-E.}\ \bibnamefont
  {Kim}}, \bibinfo {author} {\bibfnamefont {K.~G.}\ \bibnamefont {Ray}},
  \bibinfo {author} {\bibfnamefont {D.~F.}\ \bibnamefont {Bahr}},\ and\
  \bibinfo {author} {\bibfnamefont {V.}~\bibnamefont {Lordi}},\ }\bibfield
  {title} {\bibinfo {title} {Electronic structure and surface properties of
  ${\mathrm{mgb}}_{2}$(0001) upon oxygen adsorption},\ }\href
  {https://doi.org/10.1103/PhysRevB.97.195416} {\bibfield  {journal} {\bibinfo
  {journal} {Phys. Rev. B}\ }\textbf {\bibinfo {volume} {97}},\ \bibinfo
  {pages} {195416} (\bibinfo {year} {2018})}\BibitemShut {NoStop}%
\bibitem [{\citenamefont {Pizzi}\ \emph {et~al.}(2020)\citenamefont {Pizzi},
  \citenamefont {Vitale}, \citenamefont {Arita}, \citenamefont {Blügel},
  \citenamefont {Freimuth}, \citenamefont {Géranton}, \citenamefont
  {Gibertini}, \citenamefont {Gresch}, \citenamefont {Johnson}, \citenamefont
  {Koretsune}, \citenamefont {Ibañez-Azpiroz}, \citenamefont {Lee},
  \citenamefont {Lihm}, \citenamefont {Marchand}, \citenamefont {Marrazzo},
  \citenamefont {Mokrousov}, \citenamefont {Mustafa}, \citenamefont {Nohara},
  \citenamefont {Nomura}, \citenamefont {Paulatto}, \citenamefont {Poncé},
  \citenamefont {Ponweiser}, \citenamefont {Qiao}, \citenamefont {Thöle},
  \citenamefont {Tsirkin}, \citenamefont {Wierzbowska}, \citenamefont
  {Marzari}, \citenamefont {Vanderbilt}, \citenamefont {Souza}, \citenamefont
  {Mostofi},\ and\ \citenamefont {Yates}}]{wannier90}%
  \BibitemOpen
  \bibfield  {author} {\bibinfo {author} {\bibfnamefont {G.}~\bibnamefont
  {Pizzi}}, \bibinfo {author} {\bibfnamefont {V.}~\bibnamefont {Vitale}},
  \bibinfo {author} {\bibfnamefont {R.}~\bibnamefont {Arita}}, \bibinfo
  {author} {\bibfnamefont {S.}~\bibnamefont {Blügel}}, \bibinfo {author}
  {\bibfnamefont {F.}~\bibnamefont {Freimuth}}, \bibinfo {author}
  {\bibfnamefont {G.}~\bibnamefont {Géranton}}, \bibinfo {author}
  {\bibfnamefont {M.}~\bibnamefont {Gibertini}}, \bibinfo {author}
  {\bibfnamefont {D.}~\bibnamefont {Gresch}}, \bibinfo {author} {\bibfnamefont
  {C.}~\bibnamefont {Johnson}}, \bibinfo {author} {\bibfnamefont
  {T.}~\bibnamefont {Koretsune}}, \bibinfo {author} {\bibfnamefont
  {J.}~\bibnamefont {Ibañez-Azpiroz}}, \bibinfo {author} {\bibfnamefont
  {H.}~\bibnamefont {Lee}}, \bibinfo {author} {\bibfnamefont {J.-M.}\
  \bibnamefont {Lihm}}, \bibinfo {author} {\bibfnamefont {D.}~\bibnamefont
  {Marchand}}, \bibinfo {author} {\bibfnamefont {A.}~\bibnamefont {Marrazzo}},
  \bibinfo {author} {\bibfnamefont {Y.}~\bibnamefont {Mokrousov}}, \bibinfo
  {author} {\bibfnamefont {J.~I.}\ \bibnamefont {Mustafa}}, \bibinfo {author}
  {\bibfnamefont {Y.}~\bibnamefont {Nohara}}, \bibinfo {author} {\bibfnamefont
  {Y.}~\bibnamefont {Nomura}}, \bibinfo {author} {\bibfnamefont
  {L.}~\bibnamefont {Paulatto}}, \bibinfo {author} {\bibfnamefont
  {S.}~\bibnamefont {Poncé}}, \bibinfo {author} {\bibfnamefont
  {T.}~\bibnamefont {Ponweiser}}, \bibinfo {author} {\bibfnamefont
  {J.}~\bibnamefont {Qiao}}, \bibinfo {author} {\bibfnamefont {F.}~\bibnamefont
  {Thöle}}, \bibinfo {author} {\bibfnamefont {S.~S.}\ \bibnamefont {Tsirkin}},
  \bibinfo {author} {\bibfnamefont {M.}~\bibnamefont {Wierzbowska}}, \bibinfo
  {author} {\bibfnamefont {N.}~\bibnamefont {Marzari}}, \bibinfo {author}
  {\bibfnamefont {D.}~\bibnamefont {Vanderbilt}}, \bibinfo {author}
  {\bibfnamefont {I.}~\bibnamefont {Souza}}, \bibinfo {author} {\bibfnamefont
  {A.~A.}\ \bibnamefont {Mostofi}},\ and\ \bibinfo {author} {\bibfnamefont
  {J.~R.}\ \bibnamefont {Yates}},\ }\bibfield  {title} {\bibinfo {title}
  {Wannier90 as a community code: new features and applications},\ }\href
  {https://doi.org/10.1088/1361-648X/ab51ff} {\bibfield  {journal} {\bibinfo
  {journal} {Journal of Physics: Condensed Matter}\ }\textbf {\bibinfo {volume}
  {32}},\ \bibinfo {pages} {165902} (\bibinfo {year} {2020})}\BibitemShut
  {NoStop}%
\bibitem [{\citenamefont {Ahn}\ \emph {et~al.}(2018)\citenamefont {Ahn},
  \citenamefont {Kim}, \citenamefont {Kim},\ and\ \citenamefont
  {Yang}}]{Ahn2018MonopoleNLSM}%
  \BibitemOpen
  \bibfield  {author} {\bibinfo {author} {\bibfnamefont {J.}~\bibnamefont
  {Ahn}}, \bibinfo {author} {\bibfnamefont {D.}~\bibnamefont {Kim}}, \bibinfo
  {author} {\bibfnamefont {Y.}~\bibnamefont {Kim}},\ and\ \bibinfo {author}
  {\bibfnamefont {B.-J.}\ \bibnamefont {Yang}},\ }\bibfield  {title} {\bibinfo
  {title} {Band topology and linking structure of nodal line semimetals with
  ${Z}_{2}$ monopole charges},\ }\href
  {https://doi.org/10.1103/PhysRevLett.121.106403} {\bibfield  {journal}
  {\bibinfo  {journal} {Phys. Rev. Lett.}\ }\textbf {\bibinfo {volume} {121}},\
  \bibinfo {pages} {106403} (\bibinfo {year} {2018})}\BibitemShut {NoStop}%
\bibitem [{\citenamefont {Po}\ \emph {et~al.}(2018)\citenamefont {Po},
  \citenamefont {Watanabe},\ and\ \citenamefont
  {Vishwanath}}]{Po2018FragileTopo}%
  \BibitemOpen
  \bibfield  {author} {\bibinfo {author} {\bibfnamefont {H.~C.}\ \bibnamefont
  {Po}}, \bibinfo {author} {\bibfnamefont {H.}~\bibnamefont {Watanabe}},\ and\
  \bibinfo {author} {\bibfnamefont {A.}~\bibnamefont {Vishwanath}},\ }\bibfield
   {title} {\bibinfo {title} {Fragile topology and wannier obstructions},\
  }\href {https://doi.org/10.1103/PhysRevLett.121.126402} {\bibfield  {journal}
  {\bibinfo  {journal} {Phys. Rev. Lett.}\ }\textbf {\bibinfo {volume} {121}},\
  \bibinfo {pages} {126402} (\bibinfo {year} {2018})}\BibitemShut {NoStop}%
\bibitem [{\citenamefont {Cano}\ \emph {et~al.}(2018)\citenamefont {Cano},
  \citenamefont {Bradlyn}, \citenamefont {Wang}, \citenamefont {Elcoro},
  \citenamefont {Vergniory}, \citenamefont {Felser}, \citenamefont {Aroyo},\
  and\ \citenamefont {Bernevig}}]{Cano2018DisEBR}%
  \BibitemOpen
  \bibfield  {author} {\bibinfo {author} {\bibfnamefont {J.}~\bibnamefont
  {Cano}}, \bibinfo {author} {\bibfnamefont {B.}~\bibnamefont {Bradlyn}},
  \bibinfo {author} {\bibfnamefont {Z.}~\bibnamefont {Wang}}, \bibinfo {author}
  {\bibfnamefont {L.}~\bibnamefont {Elcoro}}, \bibinfo {author} {\bibfnamefont
  {M.~G.}\ \bibnamefont {Vergniory}}, \bibinfo {author} {\bibfnamefont
  {C.}~\bibnamefont {Felser}}, \bibinfo {author} {\bibfnamefont {M.~I.}\
  \bibnamefont {Aroyo}},\ and\ \bibinfo {author} {\bibfnamefont {B.~A.}\
  \bibnamefont {Bernevig}},\ }\bibfield  {title} {\bibinfo {title} {Topology of
  disconnected elementary band representations},\ }\href
  {https://doi.org/10.1103/PhysRevLett.120.266401} {\bibfield  {journal}
  {\bibinfo  {journal} {Phys. Rev. Lett.}\ }\textbf {\bibinfo {volume} {120}},\
  \bibinfo {pages} {266401} (\bibinfo {year} {2018})}\BibitemShut {NoStop}%
\bibitem [{\citenamefont {Fan}\ \emph {et~al.}(2002)\citenamefont {Fan},
  \citenamefont {Ru-Shan}, \citenamefont {Ning-Hua},\ and\ \citenamefont
  {Wei}}]{Fan03052002MgB2}%
  \BibitemOpen
  \bibfield  {author} {\bibinfo {author} {\bibfnamefont {Y.}~\bibnamefont
  {Fan}}, \bibinfo {author} {\bibfnamefont {H.}~\bibnamefont {Ru-Shan}},
  \bibinfo {author} {\bibfnamefont {T.}~\bibnamefont {Ning-Hua}},\ and\
  \bibinfo {author} {\bibfnamefont {G.}~\bibnamefont {Wei}},\ }\bibfield
  {title} {\bibinfo {title} {Electronic structural properties and
  superconductivity of diborides in the mgb2 structure},\ }\href@noop {}
  {\bibfield  {journal} {\bibinfo  {journal} {Chinese physics letters}\
  }\textbf {\bibinfo {volume} {19}},\ \bibinfo {pages} {1336} (\bibinfo {year}
  {2002})}\BibitemShut {NoStop}%
\bibitem [{\citenamefont {Yu}\ \emph {et~al.}(2023{\natexlab{b}})\citenamefont
  {Yu}, \citenamefont {Ciccarino}, \citenamefont {Bianco}, \citenamefont
  {Errea}, \citenamefont {Narang},\ and\ \citenamefont
  {Bernevig}}]{yu2023reply}%
  \BibitemOpen
  \bibfield  {author} {\bibinfo {author} {\bibfnamefont {J.}~\bibnamefont
  {Yu}}, \bibinfo {author} {\bibfnamefont {C.~J.}\ \bibnamefont {Ciccarino}},
  \bibinfo {author} {\bibfnamefont {R.}~\bibnamefont {Bianco}}, \bibinfo
  {author} {\bibfnamefont {I.}~\bibnamefont {Errea}}, \bibinfo {author}
  {\bibfnamefont {P.}~\bibnamefont {Narang}},\ and\ \bibinfo {author}
  {\bibfnamefont {B.~A.}\ \bibnamefont {Bernevig}},\ }\href@noop {} {\bibinfo
  {title} {Reply to "comment on 'nontrivial quantum geometry and the strength
  of electron-phonon coupling', arxiv:2305.02340, j. yu, c. j. ciccarino, r.
  bianco, i. errea, p. narang, b. a. bernevig"}} (\bibinfo {year}
  {2023}{\natexlab{b}}),\ \Eprint {https://arxiv.org/abs/2306.03814}
  {arXiv:2306.03814 [cond-mat.supr-con]} \BibitemShut {NoStop}%
\bibitem [{\citenamefont {Hamann}\ \emph {et~al.}(1979)\citenamefont {Hamann},
  \citenamefont {Schl\"uter},\ and\ \citenamefont {Chiang}}]{ONCV}%
  \BibitemOpen
  \bibfield  {author} {\bibinfo {author} {\bibfnamefont {D.~R.}\ \bibnamefont
  {Hamann}}, \bibinfo {author} {\bibfnamefont {M.}~\bibnamefont {Schl\"uter}},\
  and\ \bibinfo {author} {\bibfnamefont {C.}~\bibnamefont {Chiang}},\
  }\bibfield  {title} {\bibinfo {title} {Norm-conserving pseudopotentials},\
  }\href {https://doi.org/10.1103/PhysRevLett.43.1494} {\bibfield  {journal}
  {\bibinfo  {journal} {Phys. Rev. Lett.}\ }\textbf {\bibinfo {volume} {43}},\
  \bibinfo {pages} {1494} (\bibinfo {year} {1979})}\BibitemShut {NoStop}%
\bibitem [{\citenamefont {Garrity}\ \emph {et~al.}(2014)\citenamefont
  {Garrity}, \citenamefont {Bennett}, \citenamefont {Rabe},\ and\ \citenamefont
  {Vanderbilt}}]{USPP}%
  \BibitemOpen
  \bibfield  {author} {\bibinfo {author} {\bibfnamefont {K.~F.}\ \bibnamefont
  {Garrity}}, \bibinfo {author} {\bibfnamefont {J.~W.}\ \bibnamefont
  {Bennett}}, \bibinfo {author} {\bibfnamefont {K.~M.}\ \bibnamefont {Rabe}},\
  and\ \bibinfo {author} {\bibfnamefont {D.}~\bibnamefont {Vanderbilt}},\
  }\bibfield  {title} {\bibinfo {title} {Pseudopotentials for high-throughput
  dft calculations},\ }\href {https://doi.org/10.1016/j.commatsci.2013.08.053}
  {\bibfield  {journal} {\bibinfo  {journal} {Comp. Mat. Sci.}\ }\textbf
  {\bibinfo {volume} {81}},\ \bibinfo {pages} {446} (\bibinfo {year}
  {2014})}\BibitemShut {NoStop}%
\bibitem [{\citenamefont {Giustino}\ \emph {et~al.}(2007)\citenamefont
  {Giustino}, \citenamefont {Cohen},\ and\ \citenamefont
  {Louie}}]{Giustino:2007}%
  \BibitemOpen
  \bibfield  {author} {\bibinfo {author} {\bibfnamefont {F.}~\bibnamefont
  {Giustino}}, \bibinfo {author} {\bibfnamefont {M.~L.}\ \bibnamefont
  {Cohen}},\ and\ \bibinfo {author} {\bibfnamefont {S.~G.}\ \bibnamefont
  {Louie}},\ }\bibfield  {title} {\bibinfo {title} {Electron-phonon interaction
  using wannier functions},\ }\href
  {https://doi.org/10.1103/PhysRevB.76.165108} {\bibfield  {journal} {\bibinfo
  {journal} {Phys. Rev. B}\ }\textbf {\bibinfo {volume} {76}},\ \bibinfo
  {pages} {165108} (\bibinfo {year} {2007})}\BibitemShut {NoStop}%
\bibitem [{\citenamefont {Giustino}(2017)}]{Giustino2017}%
  \BibitemOpen
  \bibfield  {author} {\bibinfo {author} {\bibfnamefont {F.}~\bibnamefont
  {Giustino}},\ }\bibfield  {title} {\bibinfo {title} {Electron-phonon
  interactions from first principles},\ }\href
  {https://doi.org/10.1103/RevModPhys.89.015003} {\bibfield  {journal}
  {\bibinfo  {journal} {Rev. Mod. Phys.}\ }\textbf {\bibinfo {volume} {89}},\
  \bibinfo {pages} {015003} (\bibinfo {year} {2017})}\BibitemShut {NoStop}%
\bibitem [{\citenamefont {Sundararaman}\ \emph {et~al.}(2017)\citenamefont
  {Sundararaman}, \citenamefont {Letchworth-Weaver}, \citenamefont {Schwarz},
  \citenamefont {Gunceler}, \citenamefont {Yalcin},\ and\ \citenamefont
  {Arias}}]{JDFTx}%
  \BibitemOpen
  \bibfield  {author} {\bibinfo {author} {\bibfnamefont {R.}~\bibnamefont
  {Sundararaman}}, \bibinfo {author} {\bibfnamefont {K.}~\bibnamefont
  {Letchworth-Weaver}}, \bibinfo {author} {\bibfnamefont {K.~A.}\ \bibnamefont
  {Schwarz}}, \bibinfo {author} {\bibfnamefont {D.}~\bibnamefont {Gunceler}},
  \bibinfo {author} {\bibfnamefont {O.}~\bibnamefont {Yalcin}},\ and\ \bibinfo
  {author} {\bibfnamefont {T.}~\bibnamefont {Arias}},\ }\bibfield  {title}
  {\bibinfo {title} {Jdftx: software for joint density-functional theory},\
  }\href {https://doi.org/10.1016/j.softx.2017.10.006} {\bibfield  {journal}
  {\bibinfo  {journal} {SoftwareX}\ }\textbf {\bibinfo {volume} {6}},\ \bibinfo
  {pages} {278} (\bibinfo {year} {2017})}\BibitemShut {NoStop}%
\bibitem [{\citenamefont {Giannozzi}\ \emph {et~al.}(2009)\citenamefont
  {Giannozzi}, \citenamefont {Baroni}, \citenamefont {Bonini}, \citenamefont
  {Calandra}, \citenamefont {Car}, \citenamefont {Cavazzoni}, \citenamefont
  {{Davide Ceresoli}}, \citenamefont {Chiarotti}, \citenamefont {Cococcioni},
  \citenamefont {Dabo}, \citenamefont {Corso}, \citenamefont {de~Gironcoli},
  \citenamefont {Fabris}, \citenamefont {Fratesi}, \citenamefont {Gebauer},
  \citenamefont {Gerstmann}, \citenamefont {Gougoussis}, \citenamefont
  {Kokalj}, \citenamefont {Lazzeri}, \citenamefont {Martin-Samos},
  \citenamefont {Marzari}, \citenamefont {Mauri}, \citenamefont {Mazzarello},
  \citenamefont {{Stefano Paolini}}, \citenamefont {Pasquarello}, \citenamefont
  {Paulatto}, \citenamefont {Sbraccia}, \citenamefont {Scandolo}, \citenamefont
  {Sclauzero}, \citenamefont {Seitsonen}, \citenamefont {Smogunov},
  \citenamefont {Umari},\ and\ \citenamefont {Wentzcovitch}}]{qe1}%
  \BibitemOpen
  \bibfield  {author} {\bibinfo {author} {\bibfnamefont {P.}~\bibnamefont
  {Giannozzi}}, \bibinfo {author} {\bibfnamefont {S.}~\bibnamefont {Baroni}},
  \bibinfo {author} {\bibfnamefont {N.}~\bibnamefont {Bonini}}, \bibinfo
  {author} {\bibfnamefont {M.}~\bibnamefont {Calandra}}, \bibinfo {author}
  {\bibfnamefont {R.}~\bibnamefont {Car}}, \bibinfo {author} {\bibfnamefont
  {C.}~\bibnamefont {Cavazzoni}}, \bibinfo {author} {\bibnamefont {{Davide
  Ceresoli}}}, \bibinfo {author} {\bibfnamefont {G.~L.}\ \bibnamefont
  {Chiarotti}}, \bibinfo {author} {\bibfnamefont {M.}~\bibnamefont
  {Cococcioni}}, \bibinfo {author} {\bibfnamefont {I.}~\bibnamefont {Dabo}},
  \bibinfo {author} {\bibfnamefont {A.~D.}\ \bibnamefont {Corso}}, \bibinfo
  {author} {\bibfnamefont {S.}~\bibnamefont {de~Gironcoli}}, \bibinfo {author}
  {\bibfnamefont {S.}~\bibnamefont {Fabris}}, \bibinfo {author} {\bibfnamefont
  {G.}~\bibnamefont {Fratesi}}, \bibinfo {author} {\bibfnamefont
  {R.}~\bibnamefont {Gebauer}}, \bibinfo {author} {\bibfnamefont
  {U.}~\bibnamefont {Gerstmann}}, \bibinfo {author} {\bibfnamefont
  {C.}~\bibnamefont {Gougoussis}}, \bibinfo {author} {\bibfnamefont
  {A.}~\bibnamefont {Kokalj}}, \bibinfo {author} {\bibfnamefont
  {M.}~\bibnamefont {Lazzeri}}, \bibinfo {author} {\bibfnamefont
  {L.}~\bibnamefont {Martin-Samos}}, \bibinfo {author} {\bibfnamefont
  {N.}~\bibnamefont {Marzari}}, \bibinfo {author} {\bibfnamefont
  {F.}~\bibnamefont {Mauri}}, \bibinfo {author} {\bibfnamefont
  {R.}~\bibnamefont {Mazzarello}}, \bibinfo {author} {\bibnamefont {{Stefano
  Paolini}}}, \bibinfo {author} {\bibfnamefont {A.}~\bibnamefont
  {Pasquarello}}, \bibinfo {author} {\bibfnamefont {L.}~\bibnamefont
  {Paulatto}}, \bibinfo {author} {\bibfnamefont {C.}~\bibnamefont {Sbraccia}},
  \bibinfo {author} {\bibfnamefont {S.}~\bibnamefont {Scandolo}}, \bibinfo
  {author} {\bibfnamefont {G.}~\bibnamefont {Sclauzero}}, \bibinfo {author}
  {\bibfnamefont {A.~P.}\ \bibnamefont {Seitsonen}}, \bibinfo {author}
  {\bibfnamefont {A.}~\bibnamefont {Smogunov}}, \bibinfo {author}
  {\bibfnamefont {P.}~\bibnamefont {Umari}},\ and\ \bibinfo {author}
  {\bibfnamefont {R.~M.}\ \bibnamefont {Wentzcovitch}},\ }\bibfield  {title}
  {\bibinfo {title} {{QUANTUM} {ESPRESSO}: a modular and open-source software
  project for quantum simulations of materials},\ }\href
  {https://doi.org/10.1088/0953-8984/21/39/395502} {\bibfield  {journal}
  {\bibinfo  {journal} {Journal of Physics: Condensed Matter}\ }\textbf
  {\bibinfo {volume} {21}},\ \bibinfo {pages} {395502} (\bibinfo {year}
  {2009})}\BibitemShut {NoStop}%
\bibitem [{\citenamefont {Giannozzi}\ \emph {et~al.}(2017)\citenamefont
  {Giannozzi}, \citenamefont {Andreussi}, \citenamefont {Brumme}, \citenamefont
  {Bunau}, \citenamefont {Nardelli}, \citenamefont {Calandra}, \citenamefont
  {Car}, \citenamefont {Cavazzoni}, \citenamefont {{D Ceresoli}}, \citenamefont
  {Cococcioni}, \citenamefont {Colonna}, \citenamefont {Carnimeo},
  \citenamefont {Corso}, \citenamefont {de~Gironcoli}, \citenamefont {Delugas},
  \citenamefont {Jr}, \citenamefont {{A Ferretti}}, \citenamefont {Floris},
  \citenamefont {Fratesi}, \citenamefont {Fugallo}, \citenamefont {Gebauer},
  \citenamefont {Gerstmann}, \citenamefont {Giustino}, \citenamefont {Gorni},
  \citenamefont {Jia}, \citenamefont {Kawamura}, \citenamefont {{H-Y Ko}},
  \citenamefont {Kokalj}, \citenamefont {K{\"{u}}{\c{c}}{\"{u}}kbenli},
  \citenamefont {Lazzeri}, \citenamefont {Marsili}, \citenamefont {Marzari},
  \citenamefont {Mauri}, \citenamefont {Nguyen}, \citenamefont {Nguyen},
  \citenamefont {{A Otero-de-la-Roza}}, \citenamefont {Paulatto}, \citenamefont
  {Ponc{\'{e}}}, \citenamefont {Rocca}, \citenamefont {Sabatini}, \citenamefont
  {Santra}, \citenamefont {Schlipf}, \citenamefont {Seitsonen}, \citenamefont
  {Smogunov}, \citenamefont {{I Timrov}}, \citenamefont {Thonhauser},
  \citenamefont {Umari}, \citenamefont {Vast}, \citenamefont {Wu},\ and\
  \citenamefont {Baroni}}]{qe2}%
  \BibitemOpen
  \bibfield  {author} {\bibinfo {author} {\bibfnamefont {P.}~\bibnamefont
  {Giannozzi}}, \bibinfo {author} {\bibfnamefont {O.}~\bibnamefont
  {Andreussi}}, \bibinfo {author} {\bibfnamefont {T.}~\bibnamefont {Brumme}},
  \bibinfo {author} {\bibfnamefont {O.}~\bibnamefont {Bunau}}, \bibinfo
  {author} {\bibfnamefont {M.~B.}\ \bibnamefont {Nardelli}}, \bibinfo {author}
  {\bibfnamefont {M.}~\bibnamefont {Calandra}}, \bibinfo {author}
  {\bibfnamefont {R.}~\bibnamefont {Car}}, \bibinfo {author} {\bibfnamefont
  {C.}~\bibnamefont {Cavazzoni}}, \bibinfo {author} {\bibnamefont {{D
  Ceresoli}}}, \bibinfo {author} {\bibfnamefont {M.}~\bibnamefont
  {Cococcioni}}, \bibinfo {author} {\bibfnamefont {N.}~\bibnamefont {Colonna}},
  \bibinfo {author} {\bibfnamefont {I.}~\bibnamefont {Carnimeo}}, \bibinfo
  {author} {\bibfnamefont {A.~D.}\ \bibnamefont {Corso}}, \bibinfo {author}
  {\bibfnamefont {S.}~\bibnamefont {de~Gironcoli}}, \bibinfo {author}
  {\bibfnamefont {P.}~\bibnamefont {Delugas}}, \bibinfo {author} {\bibfnamefont
  {R.~A.~D.}\ \bibnamefont {Jr}}, \bibinfo {author} {\bibnamefont {{A
  Ferretti}}}, \bibinfo {author} {\bibfnamefont {A.}~\bibnamefont {Floris}},
  \bibinfo {author} {\bibfnamefont {G.}~\bibnamefont {Fratesi}}, \bibinfo
  {author} {\bibfnamefont {G.}~\bibnamefont {Fugallo}}, \bibinfo {author}
  {\bibfnamefont {R.}~\bibnamefont {Gebauer}}, \bibinfo {author} {\bibfnamefont
  {U.}~\bibnamefont {Gerstmann}}, \bibinfo {author} {\bibfnamefont
  {F.}~\bibnamefont {Giustino}}, \bibinfo {author} {\bibfnamefont
  {T.}~\bibnamefont {Gorni}}, \bibinfo {author} {\bibfnamefont
  {J.}~\bibnamefont {Jia}}, \bibinfo {author} {\bibfnamefont {M.}~\bibnamefont
  {Kawamura}}, \bibinfo {author} {\bibnamefont {{H-Y Ko}}}, \bibinfo {author}
  {\bibfnamefont {A.}~\bibnamefont {Kokalj}}, \bibinfo {author} {\bibfnamefont
  {E.}~\bibnamefont {K{\"{u}}{\c{c}}{\"{u}}kbenli}}, \bibinfo {author}
  {\bibfnamefont {M.}~\bibnamefont {Lazzeri}}, \bibinfo {author} {\bibfnamefont
  {M.}~\bibnamefont {Marsili}}, \bibinfo {author} {\bibfnamefont
  {N.}~\bibnamefont {Marzari}}, \bibinfo {author} {\bibfnamefont
  {F.}~\bibnamefont {Mauri}}, \bibinfo {author} {\bibfnamefont {N.~L.}\
  \bibnamefont {Nguyen}}, \bibinfo {author} {\bibfnamefont {H.-V.}\
  \bibnamefont {Nguyen}}, \bibinfo {author} {\bibnamefont {{A
  Otero-de-la-Roza}}}, \bibinfo {author} {\bibfnamefont {L.}~\bibnamefont
  {Paulatto}}, \bibinfo {author} {\bibfnamefont {S.}~\bibnamefont
  {Ponc{\'{e}}}}, \bibinfo {author} {\bibfnamefont {D.}~\bibnamefont {Rocca}},
  \bibinfo {author} {\bibfnamefont {R.}~\bibnamefont {Sabatini}}, \bibinfo
  {author} {\bibfnamefont {B.}~\bibnamefont {Santra}}, \bibinfo {author}
  {\bibfnamefont {M.}~\bibnamefont {Schlipf}}, \bibinfo {author} {\bibfnamefont
  {A.~P.}\ \bibnamefont {Seitsonen}}, \bibinfo {author} {\bibfnamefont
  {A.}~\bibnamefont {Smogunov}}, \bibinfo {author} {\bibnamefont {{I Timrov}}},
  \bibinfo {author} {\bibfnamefont {T.}~\bibnamefont {Thonhauser}}, \bibinfo
  {author} {\bibfnamefont {P.}~\bibnamefont {Umari}}, \bibinfo {author}
  {\bibfnamefont {N.}~\bibnamefont {Vast}}, \bibinfo {author} {\bibfnamefont
  {X.}~\bibnamefont {Wu}},\ and\ \bibinfo {author} {\bibfnamefont
  {S.}~\bibnamefont {Baroni}},\ }\bibfield  {title} {\bibinfo {title}
  {{Advanced capabilities for materials modelling with {Q}uantum {ESPRESSO}}},\
  }\href {https://doi.org/10.1088/1361-648X/aa8f79} {\bibfield  {journal}
  {\bibinfo  {journal} {Journal of Physics: Condensed Matter}\ }\textbf
  {\bibinfo {volume} {29}},\ \bibinfo {pages} {465901} (\bibinfo {year}
  {2017})}\BibitemShut {NoStop}%
\bibitem [{\citenamefont {Marzari}\ and\ \citenamefont
  {Vanderbilt}(1997)}]{Marzari1997}%
  \BibitemOpen
  \bibfield  {author} {\bibinfo {author} {\bibfnamefont {N.}~\bibnamefont
  {Marzari}}\ and\ \bibinfo {author} {\bibfnamefont {D.}~\bibnamefont
  {Vanderbilt}},\ }\bibfield  {title} {\bibinfo {title} {Maximally localized
  generalized wannier functions for composite energy bands},\ }\href
  {https://doi.org/10.1103/PhysRevB.56.12847} {\bibfield  {journal} {\bibinfo
  {journal} {Phys. Rev. B}\ }\textbf {\bibinfo {volume} {56}},\ \bibinfo
  {pages} {12847} (\bibinfo {year} {1997})}\BibitemShut {NoStop}%
\bibitem [{\citenamefont {Souza}\ \emph {et~al.}(2001)\citenamefont {Souza},
  \citenamefont {Marzari},\ and\ \citenamefont {Vanderbilt}}]{SouzaEntangled}%
  \BibitemOpen
  \bibfield  {author} {\bibinfo {author} {\bibfnamefont {I.}~\bibnamefont
  {Souza}}, \bibinfo {author} {\bibfnamefont {N.}~\bibnamefont {Marzari}},\
  and\ \bibinfo {author} {\bibfnamefont {D.}~\bibnamefont {Vanderbilt}},\
  }\bibfield  {title} {\bibinfo {title} {Maximally localized wannier functions
  for entangled energy bands},\ }\href
  {https://doi.org/10.1103/PhysRevB.65.035109} {\bibfield  {journal} {\bibinfo
  {journal} {Phys. Rev. B}\ }\textbf {\bibinfo {volume} {65}},\ \bibinfo
  {pages} {035109} (\bibinfo {year} {2001})}\BibitemShut {NoStop}%
\bibitem [{\citenamefont {Marzari}\ \emph {et~al.}(2012)\citenamefont
  {Marzari}, \citenamefont {Mostofi}, \citenamefont {Yates}, \citenamefont
  {Souza},\ and\ \citenamefont {Vanderbilt}}]{marzari2011}%
  \BibitemOpen
  \bibfield  {author} {\bibinfo {author} {\bibfnamefont {N.}~\bibnamefont
  {Marzari}}, \bibinfo {author} {\bibfnamefont {A.~A.}\ \bibnamefont
  {Mostofi}}, \bibinfo {author} {\bibfnamefont {J.~R.}\ \bibnamefont {Yates}},
  \bibinfo {author} {\bibfnamefont {I.}~\bibnamefont {Souza}},\ and\ \bibinfo
  {author} {\bibfnamefont {D.}~\bibnamefont {Vanderbilt}},\ }\bibfield  {title}
  {\bibinfo {title} {Maximally localized wannier functions: Theory and
  applications},\ }\href {https://doi.org/10.1103/RevModPhys.84.1419}
  {\bibfield  {journal} {\bibinfo  {journal} {Rev. Mod. Phys.}\ }\textbf
  {\bibinfo {volume} {84}},\ \bibinfo {pages} {1419} (\bibinfo {year}
  {2012})}\BibitemShut {NoStop}%
\bibitem [{\citenamefont {He}\ and\ \citenamefont {Vanderbilt}(2001)}]{He2001}%
  \BibitemOpen
  \bibfield  {author} {\bibinfo {author} {\bibfnamefont {L.}~\bibnamefont
  {He}}\ and\ \bibinfo {author} {\bibfnamefont {D.}~\bibnamefont
  {Vanderbilt}},\ }\bibfield  {title} {\bibinfo {title} {Exponential decay
  properties of wannier functions and related quantities},\ }\href
  {https://doi.org/10.1103/PhysRevLett.86.5341} {\bibfield  {journal} {\bibinfo
   {journal} {Phys. Rev. Lett.}\ }\textbf {\bibinfo {volume} {86}},\ \bibinfo
  {pages} {5341} (\bibinfo {year} {2001})}\BibitemShut {NoStop}%
\bibitem [{\citenamefont {Sundararaman}\ and\ \citenamefont
  {Arias}(2013)}]{TruncatedEXX}%
  \BibitemOpen
  \bibfield  {author} {\bibinfo {author} {\bibfnamefont {R.}~\bibnamefont
  {Sundararaman}}\ and\ \bibinfo {author} {\bibfnamefont {T.~A.}\ \bibnamefont
  {Arias}},\ }\bibfield  {title} {\bibinfo {title} {Regularization of the
  coulomb singularity in exact exchange by wigner-seitz truncated interactions:
  Towards chemical accuracy in nontrivial systems},\ }\href
  {https://doi.org/10.1103/PhysRevB.87.165122} {\bibfield  {journal} {\bibinfo
  {journal} {Phys. Rev. B}\ }\textbf {\bibinfo {volume} {87}},\ \bibinfo
  {pages} {165122} (\bibinfo {year} {2013})}\BibitemShut {NoStop}%
\bibitem [{\citenamefont {Perdew}\ \emph {et~al.}(2008)\citenamefont {Perdew},
  \citenamefont {Ruzsinszky}, \citenamefont {Csonka}, \citenamefont {Vydrov},
  \citenamefont {Scuseria}, \citenamefont {Constantin}, \citenamefont {Zhou},\
  and\ \citenamefont {Burke}}]{PBEsol}%
  \BibitemOpen
  \bibfield  {author} {\bibinfo {author} {\bibfnamefont {J.~P.}\ \bibnamefont
  {Perdew}}, \bibinfo {author} {\bibfnamefont {A.}~\bibnamefont {Ruzsinszky}},
  \bibinfo {author} {\bibfnamefont {G.~I.}\ \bibnamefont {Csonka}}, \bibinfo
  {author} {\bibfnamefont {O.~A.}\ \bibnamefont {Vydrov}}, \bibinfo {author}
  {\bibfnamefont {G.~E.}\ \bibnamefont {Scuseria}}, \bibinfo {author}
  {\bibfnamefont {L.~A.}\ \bibnamefont {Constantin}}, \bibinfo {author}
  {\bibfnamefont {X.}~\bibnamefont {Zhou}},\ and\ \bibinfo {author}
  {\bibfnamefont {K.}~\bibnamefont {Burke}},\ }\bibfield  {title} {\bibinfo
  {title} {Restoring the density-gradient expansion for exchange in solids and
  surfaces},\ }\href {https://doi.org/10.1103/PhysRevLett.100.136406}
  {\bibfield  {journal} {\bibinfo  {journal} {Phys. Rev. Lett.}\ }\textbf
  {\bibinfo {volume} {100}},\ \bibinfo {pages} {136406} (\bibinfo {year}
  {2008})}\BibitemShut {NoStop}%
\bibitem [{\citenamefont {Methfessel}\ and\ \citenamefont
  {Paxton}(1989)}]{mpSmearing}%
  \BibitemOpen
  \bibfield  {author} {\bibinfo {author} {\bibfnamefont {M.}~\bibnamefont
  {Methfessel}}\ and\ \bibinfo {author} {\bibfnamefont {A.~T.}\ \bibnamefont
  {Paxton}},\ }\bibfield  {title} {\bibinfo {title} {High-precision sampling
  for brillouin-zone integration in metals},\ }\href
  {https://doi.org/10.1103/PhysRevB.40.3616} {\bibfield  {journal} {\bibinfo
  {journal} {Phys. Rev. B}\ }\textbf {\bibinfo {volume} {40}},\ \bibinfo
  {pages} {3616} (\bibinfo {year} {1989})}\BibitemShut {NoStop}%
\bibitem [{\citenamefont {Poncé}\ \emph {et~al.}(2016)\citenamefont {Poncé},
  \citenamefont {Margine}, \citenamefont {Verdi},\ and\ \citenamefont
  {Giustino}}]{EPW}%
  \BibitemOpen
  \bibfield  {author} {\bibinfo {author} {\bibfnamefont {S.}~\bibnamefont
  {Poncé}}, \bibinfo {author} {\bibfnamefont {E.~R.}\ \bibnamefont {Margine}},
  \bibinfo {author} {\bibfnamefont {C.}~\bibnamefont {Verdi}},\ and\ \bibinfo
  {author} {\bibfnamefont {F.}~\bibnamefont {Giustino}},\ }\bibfield  {title}
  {\bibinfo {title} {{EPW}: {Electron}-phonon coupling, transport and
  superconducting properties using maximally localized {Wannier} functions},\
  }\href {https://doi.org/10.1016/j.cpc.2016.07.028} {\bibfield  {journal}
  {\bibinfo  {journal} {Computer Physics Communications}\ }\textbf {\bibinfo
  {volume} {209}},\ \bibinfo {pages} {116} (\bibinfo {year}
  {2016})}\BibitemShut {NoStop}%
\bibitem [{\citenamefont {{van Setten}}\ \emph {et~al.}(2018)\citenamefont
  {{van Setten}}, \citenamefont {Giantomassi}, \citenamefont {Bousquet},
  \citenamefont {Verstraete}, \citenamefont {Hamann}, \citenamefont {Gonze},\
  and\ \citenamefont {Rignanese}}]{VANSETTEN201839}%
  \BibitemOpen
  \bibfield  {author} {\bibinfo {author} {\bibfnamefont {M.}~\bibnamefont {{van
  Setten}}}, \bibinfo {author} {\bibfnamefont {M.}~\bibnamefont {Giantomassi}},
  \bibinfo {author} {\bibfnamefont {E.}~\bibnamefont {Bousquet}}, \bibinfo
  {author} {\bibfnamefont {M.}~\bibnamefont {Verstraete}}, \bibinfo {author}
  {\bibfnamefont {D.}~\bibnamefont {Hamann}}, \bibinfo {author} {\bibfnamefont
  {X.}~\bibnamefont {Gonze}},\ and\ \bibinfo {author} {\bibfnamefont {G.-M.}\
  \bibnamefont {Rignanese}},\ }\bibfield  {title} {\bibinfo {title} {The
  pseudodojo: Training and grading a 85 element optimized norm-conserving
  pseudopotential table},\ }\href
  {https://doi.org/https://doi.org/10.1016/j.cpc.2018.01.012} {\bibfield
  {journal} {\bibinfo  {journal} {Computer Physics Communications}\ }\textbf
  {\bibinfo {volume} {226}},\ \bibinfo {pages} {39} (\bibinfo {year}
  {2018})}\BibitemShut {NoStop}%
\bibitem [{\citenamefont {Perdew}\ \emph {et~al.}(1996)\citenamefont {Perdew},
  \citenamefont {Burke},\ and\ \citenamefont {Ernzerhof}}]{PBE}%
  \BibitemOpen
  \bibfield  {author} {\bibinfo {author} {\bibfnamefont {J.~P.}\ \bibnamefont
  {Perdew}}, \bibinfo {author} {\bibfnamefont {K.}~\bibnamefont {Burke}},\ and\
  \bibinfo {author} {\bibfnamefont {M.}~\bibnamefont {Ernzerhof}},\ }\bibfield
  {title} {\bibinfo {title} {Generalized gradient approximation made simple},\
  }\href {https://doi.org/10.1103/PhysRevLett.77.3865} {\bibfield  {journal}
  {\bibinfo  {journal} {Phys. Rev. Lett.}\ }\textbf {\bibinfo {volume} {77}},\
  \bibinfo {pages} {3865} (\bibinfo {year} {1996})}\BibitemShut {NoStop}%
\bibitem [{\citenamefont {Lazzeri}\ \emph {et~al.}(2006)\citenamefont
  {Lazzeri}, \citenamefont {Piscanec}, \citenamefont {Mauri}, \citenamefont
  {Ferrari},\ and\ \citenamefont
  {Robertson}}]{Lazzeri08292005PhononLinewidthGraphene}%
  \BibitemOpen
  \bibfield  {author} {\bibinfo {author} {\bibfnamefont {M.}~\bibnamefont
  {Lazzeri}}, \bibinfo {author} {\bibfnamefont {S.}~\bibnamefont {Piscanec}},
  \bibinfo {author} {\bibfnamefont {F.}~\bibnamefont {Mauri}}, \bibinfo
  {author} {\bibfnamefont {A.~C.}\ \bibnamefont {Ferrari}},\ and\ \bibinfo
  {author} {\bibfnamefont {J.}~\bibnamefont {Robertson}},\ }\bibfield  {title}
  {\bibinfo {title} {Phonon linewidths and electron-phonon coupling in graphite
  and nanotubes},\ }\href {https://doi.org/10.1103/PhysRevB.73.155426}
  {\bibfield  {journal} {\bibinfo  {journal} {Phys. Rev. B}\ }\textbf {\bibinfo
  {volume} {73}},\ \bibinfo {pages} {155426} (\bibinfo {year}
  {2006})}\BibitemShut {NoStop}%
\bibitem [{\citenamefont {Bonini}\ \emph {et~al.}(2007)\citenamefont {Bonini},
  \citenamefont {Lazzeri}, \citenamefont {Marzari},\ and\ \citenamefont
  {Mauri}}]{Bonini04072007PhononLinewidthGraphene}%
  \BibitemOpen
  \bibfield  {author} {\bibinfo {author} {\bibfnamefont {N.}~\bibnamefont
  {Bonini}}, \bibinfo {author} {\bibfnamefont {M.}~\bibnamefont {Lazzeri}},
  \bibinfo {author} {\bibfnamefont {N.}~\bibnamefont {Marzari}},\ and\ \bibinfo
  {author} {\bibfnamefont {F.}~\bibnamefont {Mauri}},\ }\bibfield  {title}
  {\bibinfo {title} {Phonon anharmonicities in graphite and graphene},\ }\href
  {https://doi.org/10.1103/PhysRevLett.99.176802} {\bibfield  {journal}
  {\bibinfo  {journal} {Phys. Rev. Lett.}\ }\textbf {\bibinfo {volume} {99}},\
  \bibinfo {pages} {176802} (\bibinfo {year} {2007})}\BibitemShut {NoStop}%
\bibitem [{\citenamefont {Han}\ \emph {et~al.}(2022)\citenamefont {Han},
  \citenamefont {Yang}, \citenamefont {Sullivan}, \citenamefont {Feng},
  \citenamefont {Shi}, \citenamefont {Li},\ and\ \citenamefont
  {Ruan}}]{Han07012021GrapheneFWHMDFT}%
  \BibitemOpen
  \bibfield  {author} {\bibinfo {author} {\bibfnamefont {Z.}~\bibnamefont
  {Han}}, \bibinfo {author} {\bibfnamefont {X.}~\bibnamefont {Yang}}, \bibinfo
  {author} {\bibfnamefont {S.~E.}\ \bibnamefont {Sullivan}}, \bibinfo {author}
  {\bibfnamefont {T.}~\bibnamefont {Feng}}, \bibinfo {author} {\bibfnamefont
  {L.}~\bibnamefont {Shi}}, \bibinfo {author} {\bibfnamefont {W.}~\bibnamefont
  {Li}},\ and\ \bibinfo {author} {\bibfnamefont {X.}~\bibnamefont {Ruan}},\
  }\bibfield  {title} {\bibinfo {title} {Raman linewidth contributions from
  four-phonon and electron-phonon interactions in graphene},\ }\href
  {https://doi.org/10.1103/PhysRevLett.128.045901} {\bibfield  {journal}
  {\bibinfo  {journal} {Phys. Rev. Lett.}\ }\textbf {\bibinfo {volume} {128}},\
  \bibinfo {pages} {045901} (\bibinfo {year} {2022})}\BibitemShut {NoStop}%
\end{thebibliography}%

\clearpage

\section{Data availability}
The datasets generated during and/or analyzed during the current study are available from the authors on reasonable request.

\section{Code availability}

The code generated during and/or analyzed for the current study are available from the authors on reasonable request.

\appendix
\onecolumngrid

\tableofcontents

\section{Gaussian Approximation: Geometric Contribution to the EPC Constant $\lambda$}
\label{app:gaussian}

In this section, we introduce the Gaussian approximation (GA), which shows a simple intuitive reason why the electron-phonon coupling (EPC) is related to the electron Hamiltonian. We also show how to identify the geometric contribution to the EPC constant $\lambda$.
The GA is valid in graphene and {\mgb}, as discussed in \appref{app:graphene} and \appref{app:MgB2}.
In this section, we will not discuss the two realistic cases; instead, we will consider a simple $3$D system with only one kind of atom and one spinless $s$ orbital per atom, as a concrete illustration.
Note that we allow multiple atoms per unit cell so that we are allowed to have more than one electron band in momentum space.

Under the tight-binding approximation and Frohlich two-center approximation~\cite{Mitra1969EPC}, the non-interacting electron Hamiltonian and the EPC Hamiltonian are directly given by the smooth hopping function $t(\bsl{r})$, which satisfies $[t(\bsl{r})]^* = t(-\bsl{r})$ to guarantee Hermiticity.
Here $t(\bsl{r})$ does not carry any orbital, sublattice or spin indices since we only care about one kind of atom and one spinless $s$ orbital per atom, though there can be more than one atom per unit cell.

With the hopping function, the electron Hamiltonian (without the Coulomb interaction) that takes into account the atom motions reads 
\eq{
\label{eq:H_el+elph_Frohlich}
H_{el+ion-motions}=\sum_{\bsl{R}\bsl{\tau}, \bsl{R}'\bsl{\tau}'}  
t(\bsl{R}+\bsl{\tau}+\bsl{u}_{\bsl{R}+\bsl{\tau}}-\bsl{R}'-\bsl{\tau}'-\bsl{u}_{\bsl{R}'+\bsl{\tau}'})
c^\dagger_{\bsl{R}+\bsl{\tau}} c_{\bsl{R}'+\bsl{\tau}'}\ ,
}
where $\bsl{R}$ labels the lattice point, $\bsl{\tau}$ labels the positions of the sublattices in the $\bsl{R}=0$ unit cell, $c^\dagger_{\bsl{R}+\bsl{\tau}}$ creates an electron in the spinless $s$ orbital at $\bsl{R}+\bsl{\tau}$, and $\bsl{u}_{\bsl{R}+\bsl{\tau}}$ is the motion of the atom at $\bsl{R}+\bsl{\tau}$.
Since the hopping function normally exponentially decays as $|\bsl{r}|$ becomes large, $t(\bsl{R}+\bsl{\tau}+\bsl{u}_{\bsl{R}+\bsl{\tau}}-\bsl{R}'-\bsl{\tau}'-\bsl{u}_{\bsl{R}'+\bsl{\tau}'})$ can be expanded in series of $(\bsl{u}_{\bsl{R}+\bsl{\tau}}-\bsl{u}_{\bsl{R}'+\bsl{\tau}'})$.
The zeroth-order term gives the non-interacting electron Hamiltonian under the tight-binding approximation:
\eq{
\label{eq:H_el_s}
H_{el}=\sum_{\bsl{R}\bsl{\tau}, \bsl{R}'\bsl{\tau}'}  
t(\bsl{R}+\bsl{\tau}-\bsl{R}'-\bsl{\tau}')
c^\dagger_{\bsl{R}+\bsl{\tau}} 
c_{\bsl{R}'+\bsl{\tau}'}\ ,
}
and the first-order term is the leading order term for EPC that reads
\eq{
\label{eq:H_elph_Frohlich}
H_{el-ph} = \sum_{\bsl{R}\bsl{\tau}, \bsl{R}'\bsl{\tau}'}  \sum_{ \alpha_{\bsl{\tau}} \alpha'_{\bsl{\tau}'}}
(\bsl{u}_{\bsl{R}+\bsl{\tau}}-\bsl{u}_{\bsl{R}'+\bsl{\tau}'})\cdot \left.\nabla_{\bsl{r}} t(\bsl{r})\right|_{\bsl{r} = \bsl{R}+\bsl{\tau}-\bsl{R}'-\bsl{\tau}'}
c^\dagger_{ \bsl{R}+\bsl{\tau},\alpha_{\bsl{\tau}}} 
c_{\bsl{R}'+\bsl{\tau}',\alpha_{\bsl{\tau}'}'}\ .
}
The higher-order terms are usually neglected.

The GA is to assume that the hopping function has the Gaussian form:
\eq{
t(\bsl{r}) = t_0 \exp[\gamma \frac{r^2}{2}]\ ,
}
where $r=|\bsl{r}|$, and $\gamma<0$ is determined by the standard deviation of the Gaussian function.
As a result, we have
\eq{
\label{eq:t_gradient_Gaussian_s}
\nabla_{\bsl{r}} t(\bsl{r}) = \gamma \bsl{r} t(\bsl{r})\ .
}

\eqnref{eq:t_gradient_Gaussian_s} converts the spatial derivative to the position in the EPC Hamiltonian (\eqnref{eq:H_elph_Frohlich}) together with the extra factor $\gamma$.
To better see how this conversion relates the EPC Hamiltonian to electron Hamiltonian, we transform the Hamiltonian to momentum space.
Specifically, the Fourier transformation rule of the basis reads
\eqa{
\label{eq:FT_Gaussian}
& c_{\bsl{k},\bsl{\tau}}^\dagger = \frac{1}{\sqrt{N}} \sum_{\bsl{R}} e^{\ii \bsl{k}\cdot (\bsl{R}+\bsl{\tau})} c_{\bsl{R}+\bsl{\tau}}^\dagger\ ,\ u_{\bsl{q}\bsl{\tau} i} = \frac{1}{\sqrt{N}} \sum_{\bsl{R}} e^{-\ii \bsl{q}\cdot (\bsl{R}+\bsl{\tau})} u_{\bsl{R}+\bsl{\tau}, i}\ ,
}
from which we know 
\eq{
u_{\bsl{q}\bsl{\tau} i}^\dagger = u_{-\bsl{q}\bsl{\tau} i}\ .
}
For the electron Hamiltonian,
\eqa{
\label{eq:H_el_gen_k_gaussian_s}
H_{el} & = \sum_{\bsl{R}\bsl{\tau}, \bsl{R}'\bsl{\tau}'} t(\bsl{R}+\bsl{\tau}-\bsl{R}'-\bsl{\tau}')
\frac{1}{\sqrt{N}} \sum_{\bsl{k}}^{\text{\BZ}} e^{-\ii \bsl{k}\cdot (\bsl{R}+\bsl{\tau})} c_{\bsl{k},\bsl{\tau}}^\dagger  \frac{1}{\sqrt{N}} \sum_{\bsl{k}'}^{\text{\BZ}} e^{\ii \bsl{k}'\cdot (\bsl{R}'+\bsl{\tau}')} c_{\bsl{k}',\bsl{\tau}'}\\
& = \sum_{\bsl{k}}^{\text{\BZ}} c^\dagger_{\bsl{k}} h(\bsl{k}) c_{\bsl{k}} = \sum_{\bsl{k}}^{\text{\BZ}}\sum_{n} E_{n}(\bsl{k}) \gamma^\dagger_{\bsl{k},n} \gamma_{\bsl{k},n}\ ,
}
where 
$c^\dagger_{\bsl{k}}= (..., c^\dagger_{\bsl{k},\bsl{\tau}} , ...)$,
\eq{
\label{eq:h_k_gaussian_s}
\left[ h(\bsl{k}) \right]_{\bsl{\tau}\bsl{\tau}'} = \sum_{\bsl{R}}t(\bsl{R}+\bsl{\tau}-\bsl{\tau}')  e^{-\ii \bsl{k}\cdot (\bsl{R} + \bsl{\tau} - \bsl{\tau}')}\ ,
}
\eq{
 h(\bsl{k}) U_{n}(\bsl{k}) = E_n(\bsl{k}) U_{n}(\bsl{k})\ ,
}
and $\gamma^\dagger_{\bsl{k},n} = c^\dagger_{\bsl{k}} U_{n}(\bsl{k})$.
Under the tight-binding approximation, the so-called quantum geometry (band geometry) generally refers to the momentum dependence of \eq{
\label{eq:P_n_Gaussian}
P_n(\bsl{k}) = U_n(\bsl{k}) U_n^\dagger(\bsl{k})\ .
}
One specific quantity that measures the quantum geometry is the Fubini-Study metric (FSM), which reads
\eqa{
\label{eq:FSM_g_two_expressions_Gaussian}
\left[g_{n}(\bsl{k})\right]_{ij} & = \frac{1}{2}\Tr\left[ \partial_{k_i} P_n(\bsl{k})  \partial_{k_j} P_n(\bsl{k}) \right] \\
& = \frac{1}{2}\Tr\left[ \partial_{k_i} P_n(\bsl{k}) P_n(\bsl{k}) \partial_{k_j} P_n(\bsl{k}) \right] + (i\leftrightarrow j)\ .
}
We note that the general definition of the FSM is given by the periodic part of the Bloch state $\ket{u_{n,k}}$ instead of the eigenvector $U_n(\bsl{k})$, and the general definition will reduce to \eqnref{eq:FSM_g_two_expressions_Gaussian} under the tight-binding approximation for the Fourier transformation in \eqnref{eq:FT_Gaussian}. (See detailed discussion in \appref{app:general_EPC}.)

For EPC in the momentum space, we have
\eqa{
\label{eq:H_el-ph_Frolich_k}
 H_{el-ph}    & = \frac{1}{\sqrt{N}}\sum_{\bsl{k}_1}^{\BZ} \sum_{\bsl{k}_2}^{\BZ}  \sum_{\bsl{\tau}_1,\bsl{\tau}_2 ,  i}     c^\dagger_{\bsl{k}_1,\bsl{\tau}_1} c_{\bsl{k}_2,\bsl{\tau}_2}   \left[  u^\dagger_{\bsl{k}_2 - \bsl{k}_1,\bsl{\tau}_1,i}  [f_{i}(\bsl{k}_2)]_{\bsl{\tau}_1\bsl{\tau}_2} - [f_{i}(\bsl{k}_1)]_{\bsl{\tau}_1\bsl{\tau}_2} u^\dagger_{\bsl{k}_2 - \bsl{k}_1,\bsl{\tau}_2,i} \right] \\
 & = \frac{1}{\sqrt{N}}\sum_{\bsl{k}_1}^{\BZ} \sum_{\bsl{k}_2}^{\BZ}  \sum_{ \bsl{\tau},\bsl{\tau}_1,\bsl{\tau}_2 ,  i}     c^\dagger_{\bsl{k}_1,\bsl{\tau}_1} c_{\bsl{k}_2,\bsl{\tau}_2}   u^\dagger_{\bsl{k}_2 - \bsl{k}_1,\bsl{\tau},i} \left[ \delta_{\bsl{\tau}_1\bsl{\tau}} [f_{i}(\bsl{k}_2)]_{\bsl{\tau}_1\bsl{\tau}_2} - [f_{i}(\bsl{k}_1)]_{\bsl{\tau}_1\bsl{\tau}_2} \delta_{\bsl{\tau}_2\bsl{\tau}} \right] \\
 & = \frac{1}{\sqrt{N}}\sum_{\bsl{k}_1}^{\BZ} \sum_{\bsl{k}_2}^{\BZ}  \sum_{ \bsl{\tau},\bsl{\tau}_1,\bsl{\tau}_2 ,  i}     c^\dagger_{\bsl{k}_1,\bsl{\tau}_1}\left[ \chi_{\bsl{\tau}} f_{i}(\bsl{k}_2) - f_{i}(\bsl{k}_1)\chi_{\bsl{\tau}} \right]_{\bsl{\tau}_1\bsl{\tau}_2} c_{\bsl{k}_2,\bsl{\tau}_2}   u^\dagger_{\bsl{k}_2 - \bsl{k}_1,\bsl{\tau},i}\\
 & = \frac{1}{\sqrt{N}}\sum_{\bsl{k}_1}^{\BZ} \sum_{\bsl{k}_2}^{\BZ}  \sum_{ \bsl{\tau} ,  i}     c^\dagger_{\bsl{k}_1}\left[ \chi_{\bsl{\tau}} f_{i}(\bsl{k}_2) - f_{i}(\bsl{k}_1)\chi_{\bsl{\tau}} \right] c_{\bsl{k}_2}   u^\dagger_{\bsl{k}_2 - \bsl{k}_1,\bsl{\tau},i} \ ,
}
where 
\eq{
\left[ \chi_{\bsl{\tau}} \right]_{\bsl{\tau}_1 \bsl{\tau}_2} = \delta_{ \bsl{\tau}, \bsl{\tau}_{1}} \delta_{\bsl{\tau}_1 \bsl{\tau}_2 }
}
is the projection matrix onto the $\bsl{\tau}$ sublattice,
\eq{
\label{eq:g_k_Frolich}
\left[f_{i}(\bsl{k})\right]_{\bsl{\tau}_1 \bsl{\tau}_2 } = \sum_{ \bsl{R} } e^{- \ii \bsl{k} \cdot ( \bsl{R}+ \bsl{\tau}_1 - \bsl{\tau}_2)}   \left. \partial_{r_i} t(\bsl{r})\right|_{\bsl{r} = \bsl{R}+\bsl{\tau}_1-\bsl{\tau}_2}  \ ,
}
and $i=x,y,z$ labels the spatial direction.
Clearly, we can see that the form of the EPC Hamiltonian in \eqnref{eq:H_el-ph_Frolich_k} is determined by $f_{i}(\bsl{k})$, which we focus on below.

Owing to \eqnref{eq:t_gradient_Gaussian_s}, the EPC $f_{i}(\bsl{k})$ is related to the electron matrix Hamiltonian $h(\bsl{k})$ as
\eq{
\left[f_{i}(\bsl{k})\right]_{\bsl{\tau}_1 \bsl{\tau}_2 } = \gamma \sum_{ \bsl{R} } e^{- \ii \bsl{k} \cdot ( \bsl{R}+ \bsl{\tau}_1 - \bsl{\tau}_2)}   (\bsl{R}+\bsl{\tau}_1-\bsl{\tau}_2)_i  t(\bsl{R}+\bsl{\tau}_1-\bsl{\tau}_2) = \ii \gamma \partial_{k_i} \left[h(\bsl{k})\right]_{\bsl{\tau}_1 \bsl{\tau}_2 }\ ,
}
meaning that 
\eq{
\label{eq:g_k_Frolich_Gaussian_s}
f_{i}(\bsl{k}) = \ii \gamma \partial_{k_i}  h(\bsl{k})\ .
}
We call \eqnref{eq:g_k_Frolich_Gaussian_s} the Gaussian form of the EPC.

The Gaussian form of the EPC allows us to define the energetic and geometric parts of the EPC.
Note that the electron matrix Hamiltonian contains the information of both the bands and the projection matrix, \ie,
\eq{
h(\bsl{k}) = \sum_{n}E_n(\bsl{k}) P_n(\bsl{k})\ ,
}
where the projection matrix $P_n(\bsl{k})$ is defined in \eqnref{eq:P_n_Gaussian}.
Then, 
\eq{
\label{eq:g_g_E_g_geo_Gaussian_s}
f_{i}(\bsl{k}) = \ii \gamma \partial_{k_i}  h(\bsl{k}) = \ii \gamma   \sum_{n} \partial_{k_i} E_n(\bsl{k}) P_n(\bsl{k}) + \ii \gamma  \sum_{n}E_n(\bsl{k}) \partial_{k_i} P_n(\bsl{k})  = f_{i}^E(\bsl{k}) + f^{geo}(\bsl{k})\ ,
}
where 
\eq{
\label{eq:g_E_Gaussian_s}
f_{i}^E(\bsl{k}) = \ii \gamma   \sum_{n} \partial_{k_i} E_n(\bsl{k}) P_n(\bsl{k}) 
}
is the energetic part of the EPC as it vanishes for systems with all electron bands exactly flat, and 
\eq{
\label{eq:g_geo_Gaussian_s}
f_{i}^{geo}(\bsl{k}) =  \ii \gamma  \sum_{n}E_n(\bsl{k}) \partial_{k_i} P_n(\bsl{k})
}
is the geometric part of the EPC as $f_{i}^{geo}(\bsl{k})$ relies on the geometric properties of the Bloch eigenvector $U_n(\bsl{k})$ (\ie, the momentum dependence of $P_n(\bsl{k})$).
If we consider the one-band case (\ie, only one atom in the unit cell), $f_{i}^{geo}(\bsl{k})$ must be vanishing since $P_n(\bsl{k})=1$ is independent of momentum ($n$ can only take one value in the one-band case), while $f_{i}^E(\bsl{k})$ can be nonvanishing since the energy band can still disperse.

The key quantity that we study is the dimensionless EPC constant $\lambda$, which, according to \appref{app:EPC_Constant}, reads
\eqa{
\label{eq:lambda_omegabar_Gaussian_s}
\lambda  = \frac{2}{ N} D(\mu)  \frac{1}{\hbar \mcomega}  \left\langle \Gamma \right\rangle \ ,
}
where $\mu$ is the chemical potential, $D(\mu)$ is the electron density of states at the chemical potential $\mu$, $\mcomega$ is the mean-squared phonon frequency defined in \refcite{McMillan1968SCTc},
\eqa{
\label{eq:Gamma_nm_Gaussian_s}
  \Gamma_{n n'}(\bsl{k},\bsl{k}') & = \frac{\hbar}{2 m} \sum_{\bsl{\tau},i}  \Tr\left\{ P_{n}(\bsl{k})  \left[ \chi_{\bsl{\tau}} f_{i}(\bsl{k}') - f_{i}(\bsl{k})\chi_{\bsl{\tau}} \right] P_{n'}(\bsl{k}')   \left[ \chi_{\bsl{\tau}} f_{i}(\bsl{k}) - f_{i}(\bsl{k}')\chi_{\bsl{\tau}} \right] \right\} \ ,
 }
$m$ is the mass of the atom, and 
 \eq{
\left\langle \Gamma \right\rangle =\frac{ \sum_{\bsl{k},\bsl{k}'}^{\BZ}\sum_{n,n'} \delta\left(\mu - E_n(\bsl{k}) \right) \delta\left(\mu - E_{n'}(\bsl{k}') \right) \Gamma_{n n'}(\bsl{k},\bsl{k}') }{\sum_{\bsl{k},\bsl{k}'}^{\BZ}\sum_{n,n'} \delta\left(\mu - E_n(\bsl{k}) \right) \delta\left(\mu - E_{n'}(\bsl{k}') \right) }\ .
}
As discussed in \appref{app:EPC_Constant}, we will focus on $\left\langle \Gamma \right\rangle$ and will treat $\mcomega$ as a parameter determined by the first-principle calculation, mainly because $\mcomega$ can be well approximated by the frequency of specific phonon modes in graphene and {\mgb} (\appref{app:graphene} and \appref{app:MgB2}).
By combining \eqnref{eq:Gamma_nm_Gaussian_s} with \eqnref{eq:g_g_E_g_geo_Gaussian_s}, we can define 
\eq{
  \Gamma_{n n'}(\bsl{k},\bsl{k}') = \Gamma_{n n'}^{E-E}(\bsl{k},\bsl{k}')  + \Gamma_{n n'}^{geo-geo}(\bsl{k},\bsl{k}') + \Gamma_{n n'}^{E-geo}(\bsl{k},\bsl{k}') \ ,
 }
where
\eqa{
\label{eq:Gamma_X}
\Gamma_{n n'}^{E-E}(\bsl{k},\bsl{k}') & = \frac{\hbar}{2 m} \sum_{\bsl{\tau},i}  \Tr\left\{ P_{n}(\bsl{k})  \left[ \chi_{\bsl{\tau}} f_{i}^E(\bsl{k}') - f_{i}^E(\bsl{k})\chi_{\bsl{\tau}} \right] P_{n'}(\bsl{k}')   \left[ \chi_{\bsl{\tau}} f_{i}^E(\bsl{k}) - f_{i}^E(\bsl{k}')\chi_{\bsl{\tau}} \right] \right\} \\
\Gamma_{n n'}^{geo-geo}(\bsl{k},\bsl{k}') & = \frac{\hbar}{2 m} \sum_{\bsl{\tau},i}  \Tr\left\{ P_{n}(\bsl{k})  \left[ \chi_{\bsl{\tau}} f_{i}^{geo}(\bsl{k}') - f_{i}^{geo}(\bsl{k})\chi_{\bsl{\tau}} \right] P_{n'}(\bsl{k}')   \left[ \chi_{\bsl{\tau}} f_{i}^{geo}(\bsl{k}) - f_{i}^{geo}(\bsl{k}')\chi_{\bsl{\tau}} \right] \right\} \\
\Gamma_{n n'}^{E-geo}(\bsl{k},\bsl{k}') & = \frac{\hbar}{2 m} \sum_{\bsl{\tau},i}  \Tr\left\{ P_{n}(\bsl{k})  \left[ \chi_{\bsl{\tau}} f_{i}^{E}(\bsl{k}') - f_{i}^{E}(\bsl{k})\chi_{\bsl{\tau}} \right] P_{n'}(\bsl{k}')   \left[ \chi_{\bsl{\tau}} f_{i}^{geo}(\bsl{k}) - f_{i}^{geo}(\bsl{k}')\chi_{\bsl{\tau}} \right] \right\} + c.c.\ .
}
By defining
\eq{
\label{eq:Gamma_mu_X}
\left\langle \Gamma \right\rangle^{X} =\frac{ \sum_{\bsl{k},\bsl{k}'}^{\BZ}\sum_{n,n'} \delta\left(\mu - E_n(\bsl{k}) \right) \delta\left(\mu - E_{n'}(\bsl{k}') \right) \Gamma_{n n'}^X(\bsl{k},\bsl{k}') }{\sum_{\bsl{k},\bsl{k}'}^{\BZ}\sum_{n,n'} \delta\left(\mu - E_n(\bsl{k}) \right) \delta\left(\mu - E_{n'}(\bsl{k}') \right) } \text{ for } X= E-E,\ geo-geo,\ E-geo\ ,
}
we arrive at 
\eq{
\lambda = \lambda_E + \lambda_{geo} + \lambda_{E-geo}\ ,
}
where
\eqa{
\label{eq:lambda_E}
\lambda_E = \frac{2}{ N} D(\mu)  \frac{1}{\hbar \mcomega} \left\langle \Gamma \right\rangle^{E-E}
}
is the energetic contribution to $\lambda$ since it relies on $f^E_i$ not on $f^{geo}_i$, 
\eqa{
\label{eq:lambda_geo}
\lambda_{geo} = \frac{2}{ N} D(\mu)  \frac{1}{\hbar \mcomega} \left\langle \Gamma \right\rangle^{geo-geo}
}
is the geometric contribution to $\lambda$ since it relies on $f^{geo}_i$ not on $f^E_i$,
and 
\eqa{
\label{eq:lambda_E_geo}
\lambda_{E-geo} = \frac{2}{ N} D(\mu)  \frac{1}{\hbar \mcomega} \left\langle \Gamma \right\rangle^{E-geo}
}
is the cross contribution to $\lambda$ since it relies on both $f^{geo}_i$ and $f^E_i$.

Now we discuss more about the expressions of $\left\langle \Gamma \right\rangle^{geo-geo}$ and $\lambda_{geo}$.
We can split $\left\langle \Gamma \right\rangle^{geo-geo}$ into two parts:
\eqa{
  \left\langle \Gamma \right\rangle^{geo-geo} &   =\left\{ \frac{1}{D^2(\mu)}  \frac{\hbar}{2} \sum_{\bsl{\tau},i}  \frac{1}{m} \Tr\left[ \left(\sum_{\bsl{k}_1}^{\BZ}\sum_{n}  \delta\left(\mu - E_n(\bsl{k}_1) \right) \chi_{\bsl{\tau}} f_{i}^{geo}(\bsl{k}_1)  P_{n}(\bsl{k}_1) \right)^2    \right] +c.c.\right\}\\
  & \quad  - \frac{\hbar}{D^2(\mu)} \sum_{\bsl{\tau},i}  \frac{1}{m}  \sum_{\bsl{k}_1,\bsl{k}_2}^{\BZ}\sum_{n,m} \delta\left(\mu - E_n(\bsl{k}_1) \right) \delta\left(\mu - E_m(\bsl{k}_2) \right)  \Tr\left[ f_{i}^{geo}(\bsl{k}_1) P_{n}(\bsl{k}_1)  f_{i}^{geo}(\bsl{k}_1)\chi_{\bsl{\tau}} P_{m}(\bsl{k}_2)   \chi_{\bsl{\tau}}  \right] \\
  & = \left\langle \Gamma \right\rangle^{geo-geo,2} +  \left\langle \Gamma \right\rangle^{geo-geo,1}\ ,
}
where 
\eqa{
  \left\langle \Gamma \right\rangle^{geo-geo,1} &  =  - \frac{\hbar}{D^2(\mu)} \sum_{\bsl{\tau},i}  \frac{1}{m}  \sum_{\bsl{k}_1,\bsl{k}_2}^{\BZ}\sum_{n,m} \delta\left(\mu - E_n(\bsl{k}_1) \right) \delta\left(\mu - E_m(\bsl{k}_2) \right)  \Tr\left[ f_{i}^{geo}(\bsl{k}_1) P_{n}(\bsl{k}_1)  f_{i}^{geo}(\bsl{k}_1)\chi_{\bsl{\tau}} P_{m}(\bsl{k}_2)   \chi_{\bsl{\tau}}  \right]\\
  \left\langle \Gamma \right\rangle^{geo-geo,2} &   =\left\{ \frac{1}{D^2(\mu)}  \frac{\hbar}{2} \sum_{\bsl{\tau},i}  \frac{1}{m} \Tr\left[ \left(\sum_{\bsl{k}_1}^{\BZ}\sum_{n}  \delta\left(\mu - E_n(\bsl{k}_1) \right) \chi_{\bsl{\tau}} f_{i}^{geo}(\bsl{k}_1)  P_{n}(\bsl{k}_1) \right)^2    \right] +c.c.\right\}\ .
}
Similarly, $\lambda_{geo}$ can be split into two parts:
\eq{
\lambda_{geo} = \lambda_{geo,1}+ \lambda_{geo,2}\ ,
}
where 
\eq{
\lambda_{geo,1/2} = \frac{2}{ N} D(\mu)  \frac{1}{\hbar \mcomega} \left\langle \Gamma \right\rangle^{geo-geo,1/2}\ .
}

For convenience of illustrating the explicit geometric dependence in \appref{eq:GA_twoband_oneatom_sorbital}, we re-write $\left\langle \Gamma \right\rangle^{geo-geo,1/2}$.
$\left\langle \Gamma \right\rangle^{geo-geo,1}$ can be re-written as
\eqa{
\label{eq:Gamma_ave_geo-geo,1_GA}
  \left\langle \Gamma \right\rangle^{geo-geo,1} &  =   \frac{\hbar \gamma^2}{D(\mu)} \sum_{i}  \frac{1}{m}  \sum_{\bsl{k}}^{\BZ}\sum_{n,n_1,n_2} \delta\left(\mu - E_n(\bsl{k}) \right) E_{n_1}(\bsl{k}) E_{n_2}(\bsl{k}) \Tr\left[ \partial_{k_i} P_{n_1}(\bsl{k})  P_{n}(\bsl{k})  \partial_{k_i} P_{n_2}(\bsl{k}) M    \right]\ ,
}
where
\eq{
\label{eq:M_expression_GA}
M  = \frac{1}{D(\mu) }\sum_{\bsl{\tau}} \sum_m \sum_{\bsl{k}_2}^{\BZ}\delta\left(\mu - E_m(\bsl{k}_2) \right) \chi_{\bsl{\tau}} P_{m}(\bsl{k}_2)   \chi_{\bsl{\tau}} = \sum_{\bsl{\tau}} a_{\bsl{\tau}} \chi_{\bsl{\tau}}\ ,
}
and
\eq{
\label{eq:a_tau_expression_GA}
a_{\bsl{\tau}} = \frac{1}{D(\mu) } \sum_m \sum_{\bsl{k}_2}^{\BZ}\delta\left(\mu - E_m(\bsl{k}_2) \right) \left[ P_{m}(\bsl{k}_2) \right]_{\bsl{\tau}\bsl{\tau}} 
}

$\left\langle \Gamma \right\rangle^{geo-geo,2}$ can be re-written as
\eqa{
\label{eq:Gamma_ave_geo-geo,2_GA}
\left\langle \Gamma \right\rangle^{geo-geo,2} &  = \frac{1}{D^2(\mu)}  \frac{\hbar}{2} \sum_{\bsl{\tau},i}  \frac{1}{m} \Tr\left[ \left(\sum_{\bsl{k}_1}^{\BZ}\sum_{n}  \delta\left(\mu - E_n(\bsl{k}_1) \right) \chi_{\bsl{\tau}} f_{i}^{geo}(\bsl{k}_1)  P_{n}(\bsl{k}_1) \chi_{\bsl{\tau}}\right)^2    \right] + c.c. \\
& = \frac{1}{D^2(\mu)}  \frac{\hbar}{2} \sum_{\bsl{\tau},i}  \frac{1}{m} \left(\sum_{\bsl{k}_1}^{\BZ}\sum_{n}  \delta\left(\mu - E_n(\bsl{k}_1) \right) \left[ f_{i}^{geo}(\bsl{k}_1)  P_{n}(\bsl{k}_1) \right]_{\bsl{\tau}\bsl{\tau}} \right)^2  \Tr\left[\chi_{\bsl{\tau}}^2\right] + c.c. \\
& = \frac{1}{D^2(\mu)}  \frac{\hbar}{2} \sum_{\bsl{\tau},i}  \frac{1}{m} \left(\sum_{\bsl{k}_1}^{\BZ}\sum_{n}  \delta\left(\mu - E_n(\bsl{k}_1) \right) \Tr\left[ \chi_{\bsl{\tau}}f_{i}^{geo}(\bsl{k}_1)  P_{n}(\bsl{k}_1) \right] \right)^2  + c.c. 
}

\subsection{Two-Band Case}
\label{eq:GA_twoband_oneatom_sorbital}

The explicit geometric dependence is transparent in the two-band case (\ie, the system has only two electron bands).

For $\left\langle \Gamma \right\rangle^{geo-geo,1}$, we have 
\eqa{
  \left\langle \Gamma \right\rangle^{geo-geo,1} &  =   \frac{\hbar\gamma^2}{D(\mu)} \sum_{i}  \frac{1}{m}  \sum_{\bsl{k}}^{\BZ}\sum_{n} \delta\left(\mu - E_n(\bsl{k}) \right) \Delta E^2(\bsl{k}) \Tr\left[ \partial_{k_i} P_{n}(\bsl{k})  P_{n}(\bsl{k})  \partial_{k_i} P_{n}(\bsl{k}) M    \right]\\
  &  =   \frac{\hbar\gamma^2}{D(\mu) m } \sum_{i} \sum_{n} \frac{\V}{(2\pi)^3}\int_{FS_n} d\sigma_{\bsl{k}}\frac{\Delta E^2(\bsl{k})}{|\nabla_{\bsl{k}} E_n(\bsl{k})|}  \Tr\left[ \partial_{k_i} P_{n}(\bsl{k})  P_{n}(\bsl{k})  \partial_{k_i} P_{n}(\bsl{k}) M    \right]\ ,
}
resulting in

\eqa{
\label{eq:lambda_geo_1}
\lambda_{geo,1} & = \frac{2\Omega \gamma^2}{ (2\pi)^3 m \mcomega} \sum_{n,i} \int_{FS_n} d\sigma_{\bsl{k}}\frac{\Delta E^2(\bsl{k})}{|\nabla_{\bsl{k}} E_n(\bsl{k})|}  \Tr\left[ \partial_{k_i} P_{n}(\bsl{k})  P_{n}(\bsl{k})  \partial_{k_i} P_{n}(\bsl{k}) M \right] \\
& = \frac{2\Omega \gamma^2}{ (2\pi)^3 m \mcomega} \sum_{n,i,\bsl{\tau}} \int_{FS_n} d\sigma_{\bsl{k}}\frac{\Delta E^2(\bsl{k})}{|\nabla_{\bsl{k}} E_n(\bsl{k})|}  a_{\bsl{\tau}} \left[g_{n,\bsl{\tau}}(\bsl{k}) \right]_{ii}\ ,
}
where $\Delta E(\bsl{k}) = | E_2(\bsl{k}) - E_1(\bsl{k})|$, 
\eq{
\left[ g_{n,\bsl{\tau}}(\bsl{k}) \right]_{ij}=  \frac{1}{2}\Tr\left[ \partial_{k_i} P_{n}(\bsl{k})  P_{n}(\bsl{k})  \partial_{k_j} P_{n}(\bsl{k}) \chi_{\bsl{\tau}} \right] +(i\leftrightarrow j)\ ,
}
and $a_{\bsl{\tau}}$ is in \eqnref{eq:a_tau_expression_GA}.
As discussed in \appref{app:OFSM}, $g_{n,\bsl{\tau}}(\bsl{k}) $ is an OFSM, since it is defined by inserting the projection matrix $\chi_{\bsl{\tau}}$ into the original definition of FSM.
Therefore, in the two-band case, $\lambda_{geo,1}$ directly relies on the linear combination of the OFSM.

In particular, if we have $P\TR$ symmetry that flips the sublattice index (\ie, the inversion $P$ changes one sublattice to the other one), we would have $P\TR c^\dagger_{\bsl{k}} (P\TR)^{-1} = c^\dagger_{\bsl{k}} \tau_x$ with $\tau_x$ the $x$ Pauli matrix in the sublattice subspace, and then $\left[ P_{n}(\bsl{k}) \right]_{\bsl{\tau} \bsl{\tau}} = \frac{1}{2}$, leading to $a_{\bsl{\tau}}=1/2$ and
\eq{
\lambda_{geo,1} = \frac{\Omega \gamma^2}{ (2\pi)^3 m \mcomega} \sum_{n,i} \int_{FS_n} d\sigma_{\bsl{k}}\frac{\Delta E^2(\bsl{k})}{|\nabla_{\bsl{k}} E_n(\bsl{k})|}   \left[g_{n}(\bsl{k}) \right]_{ii}\ .
}
Therefore, the extra $P\TR$ symmetry that flips the sublattice index would make $\lambda_{geo,1}$ directly depend on the FSM.
This is the case for graphene (\appref{app:graphene}) and for the $\pi$-bonding states of {\mgb} (\appref{app:lambda_pz_topo_geo}).

To simplify $\left\langle \Gamma \right\rangle^{geo-geo,2}$ in \eqnref{eq:Gamma_ave_geo-geo,1_GA}, we first note that $f_{i}^{geo}(\bsl{k}) P_n(\bsl{k})$ is simplified to 
\eq{
\label{eq:g_geo_Gaussian_s_twoband}
f_{i}^{geo}(\bsl{k}) P_n(\bsl{k}) =  \ii \gamma  \sum_{m}E_m(\bsl{k}) \partial_{k_i} P_m(\bsl{k}) P_n(\bsl{k}) = \ii \gamma   (-1)^n \Delta E(\bsl{k}) \partial_{k_i} P_n(\bsl{k}) P_n(\bsl{k})\ ,
}
where we have choose $E_1(\bsl{k})\leq E_2(\bsl{k})$ without loss of generality.
Then, $\left\langle \Gamma \right\rangle^{geo-geo,2}$ becomes
\eqa{
\left\langle \Gamma \right\rangle^{geo-geo,2} &  = \frac{\gamma^2}{D^2(\mu)}  \frac{\hbar}{2} \sum_{\bsl{\tau},i}  \frac{1}{m} \left( \frac{\V}{(2\pi)^3} \sum_{n} \int_{FS_n} d\sigma_{\bsl{k}}\frac{\Delta E(\bsl{k})}{|\nabla_{\bsl{k}} E_n(\bsl{k})|}   \Tr\left[ \chi_{\bsl{\tau}} \partial_{k_{i}} P_n(\bsl{k}) P_n(\bsl{k}) \right] \right)^2  + c.c. \\
&  = \frac{\gamma^2}{D^2(\mu)}  \frac{\hbar}{2} \sum_{\bsl{\tau},i}  \frac{1}{m} \left( \frac{\V}{(2\pi)^3} \sum_{n} \int_{FS_n} d\sigma_{\bsl{k}}\frac{\Delta E(\bsl{k})}{|\nabla_{\bsl{k}} E_n(\bsl{k})|}  \mathcal{A}_{i,n,\bsl{\tau}}(\bsl{k})\right)^2  + c.c. \\
&  = \frac{\gamma^2}{D^2(\mu)}  \frac{\hbar}{2} \sum_{\bsl{\tau},i}  \frac{2}{m} \left( \frac{\V}{(2\pi)^3} \sum_{n} \int_{FS_n} d\sigma_{\bsl{k}}\frac{\Delta E(\bsl{k})}{|\nabla_{\bsl{k}} E_n(\bsl{k})|}   \text{Re}\left(\mathcal{A}_{i,n,\bsl{\tau}}(\bsl{k}) \right) \right)^2  \\
& \quad - \frac{\gamma^2}{D^2(\mu)}  \frac{\hbar}{2} \sum_{\bsl{\tau},i}  \frac{2}{m} \left( \frac{\V}{(2\pi)^3} \sum_{n} \int_{FS_n} d\sigma_{\bsl{k}}\frac{\Delta E(\bsl{k})}{|\nabla_{\bsl{k}} E_n(\bsl{k})|}   \text{Im}\left(\mathcal{A}_{i,n,\bsl{\tau}}(\bsl{k}) \right) \right)^2\ ,
}
where 
\eq{
\label{eq:A_n_tau}
\bsl{\mathcal{A}}_{n,\bsl{\tau}}(\bsl{k}) =  \Tr\left[ \chi_{\bsl{\tau}} \nabla_{\bsl{k}} P_n(\bsl{k}) P_n(\bsl{k}) \right] 
}
is an orbital-selective complex vector field.
Explicitly,
\eq{
\text{Re}\left(\mathcal{A}_{i,n,\bsl{\tau}}(\bsl{k}) \right)=\text{Re}\left( \Tr\left[ \chi_{\bsl{\tau}} \partial_{k_{i}} P_n(\bsl{k}) P_n(\bsl{k}) \right] \right)= \frac{1}{2}\Tr\left[ \chi_{\bsl{\tau}} \partial_{k_{i}} P_n(\bsl{k}) P_n(\bsl{k}) \right] + \frac{1}{2}\Tr\left[ \chi_{\bsl{\tau}} P_n(\bsl{k}) \partial_{k_{i}} P_n(\bsl{k}) \right] = \frac{1}{2} \Tr\left[ \chi_{\bsl{\tau}} \partial_{k_{i}} P_n(\bsl{k}) \right]
}
and 
\eqa{
\text{Im}\left(\mathcal{A}_{i,n,\bsl{\tau}}(\bsl{k}) \right) & =\text{Im}\left( \Tr\left[ \chi_{\bsl{\tau}} \partial_{k_{i}} P_n(\bsl{k}) P_n(\bsl{k}) \right] \right) = \frac{1}{2\ii}\Tr\left[ \chi_{\bsl{\tau}} \partial_{k_{i}} P_n(\bsl{k}) P_n(\bsl{k}) \right] - \frac{1}{2\ii}\Tr\left[ \chi_{\bsl{\tau}}   P_n(\bsl{k})  \partial_{k_{i}} P_n(\bsl{k}) \right] \\
& = \frac{1}{2 \ii} \Tr\left( \chi_{\bsl{\tau}} [\partial_{k_{i}}P_n(\bsl{k}),  P_n(\bsl{k})] \right)\ .
}
Then, 
\eq{
\label{eq:lambda_geo,2_GA_twoband}
\lambda_{geo,2} = \frac{N \gamma^2}{ m D(\mu) \mcomega}  \sum_{\bsl{\tau},i} \left( \frac{\Omega}{(2\pi)^3} \sum_{n}\int_{FS_n} d\sigma_{\bsl{k}}\frac{\Delta E(\bsl{k})}{|\nabla_{\bsl{k}} E_n(\bsl{k})|}  \mathcal{A}_{i,n,\bsl{\tau}}(\bsl{k})\right)^2  + c.c.  \ .
}
Therefore, the explicit geometric dependence is $\lambda_{geo,2}$ in the orbital-selective complex vector field.

Symmetry may restrict $\lambda_{geo,2}$ to zero.
For example, if the system has $C_{3z}$ and $m_z$ symmetries that do not change sublattice index, then we have $\bsl{\mathcal{A}}_{n,\bsl{\tau}}(C_3\bsl{k}) = C_3 \bsl{\mathcal{A}}_{n,\bsl{\tau}}(\bsl{k})$ and $\bsl{\mathcal{A}}_{n,\bsl{\tau}}(m_z\bsl{k}) = m_z \bsl{\mathcal{A}}_{n,\bsl{\tau}}(\bsl{k})$.
As a result, $ \frac{\V}{(2\pi)^3} \sum_{n} \int_{FS_n} d\sigma_{\bsl{k}}\frac{\Delta E(\bsl{k})}{|\nabla_{\bsl{k}} E_n(\bsl{k})|}  \mathcal{A}_{i,n,\bsl{\tau}}(\bsl{k})=0$, which means $\lambda_{geo,2} = 0$.
This is the case for graphene (\appref{app:graphene}) and for the $\pi$-bonding states of {\mgb} (\appref{app:lambda_pz_topo_geo}).

For the $\sigma$-bonding states of MgB2, we do not have two-band model anymore, but $\lambda_{geo}$ is still split into two parts: one part relies on the OFSM and the other part relies on a orbital-selective complex vector field. Nevertheless, the second part (that relies on orbital-selective complex vector field) has the same final expression as the first part (that relies on OFSM) under the linear-momentum approximation that we use for the electron Hamiltonian, which allows us to merge the two parts and have a simple expression with only OFSM. (\appref{app:4-band_Gaussian_lambda_spxpy}) Therefore, in the examples studied in this work, the orbital-selective complex vector field is not explicitly necessary. Nevertheless, in general, it can be important (for example, if we go beyond the linear-momentum approximation that we use for the electron Hamiltonian of $\sigma$-bonding states.)

In general, $\lambda_{E-geo}$ also has dependence on the band geometry.
In the two-band case, we explicitly derive the geometric dependence of $\lambda_{E-geo}$.
Explicitly for $\Gamma_{n n'}^{E-geo}(\bsl{k},\bsl{k}')$, we have
\eqa{
\Gamma_{n n'}^{E-geo}(\bsl{k},\bsl{k}') & = \frac{\hbar}{2 m} \sum_{\bsl{\tau},i}  \Tr\left\{ P_{n}(\bsl{k})  \left[ \chi_{\bsl{\tau}} f_{i}^{E}(\bsl{k}') - f_{i}^{E}(\bsl{k})\chi_{\bsl{\tau}} \right] P_{n'}(\bsl{k}')   \left[ \chi_{\bsl{\tau}} f_{i}^{geo}(\bsl{k}) - f_{i}^{geo}(\bsl{k}')\chi_{\bsl{\tau}} \right] \right\} + c.c.\\
 & = -\gamma^2 \frac{\hbar}{2 m} \sum_{\bsl{\tau},i} \left[\partial_{k'_i}E_{n'}(\bsl{k}')  -  \partial_{k_i}E_{n}(\bsl{k})  \right] \Tr\left\{ P_{n}(\bsl{k}) \chi_{\bsl{\tau}}   P_{n'}(\bsl{k}') \left[ \chi_{\bsl{\tau}} f_{i}^{geo}(\bsl{k}) - f_{i}^{geo}(\bsl{k}')\chi_{\bsl{\tau}} \right] \right\} + c.c.\\
  & = -\gamma^2 \frac{\hbar}{2 m} \sum_{\bsl{\tau},i} \left[\partial_{k'_i}E_{n'}(\bsl{k}')  -  \partial_{k_i}E_{n}(\bsl{k})  \right] \\
  & \qquad \times \Tr\left\{ P_{n}(\bsl{k}) \chi_{\bsl{\tau}}   P_{n'}(\bsl{k}')  \left[  \chi_{\bsl{\tau}} \sum_{n_1} E_{n_1}(\bsl{k}) \partial_{k_i}P_{n_1}(\bsl{k}) - \sum_{n'_1} E_{n'_1}(\bsl{k}') \partial_{k_i'}P_{n'_1}(\bsl{k}') \chi_{\bsl{\tau}} \right] \right\} + c.c.\\
  & = -\gamma^2 \frac{\hbar}{2 m} \sum_{\bsl{\tau},i} \left[\partial_{k'_i}E_{n'}(\bsl{k}')  -  \partial_{k_i}E_{n}(\bsl{k})  \right] \Tr\left\{   P_{n}(\bsl{k}) \chi_{\bsl{\tau}}   P_{n'}(\bsl{k}')   \chi_{\bsl{\tau}} (-1)^n \Delta E(\bsl{k}) \partial_{k_i}P_{n}(\bsl{k})  \right\} \\
  & \quad  +\gamma^2 \frac{\hbar}{2 m} \sum_{\bsl{\tau},i} \left[\partial_{k'_i}E_{n'}(\bsl{k}')  -  \partial_{k_i}E_{n}(\bsl{k})  \right]  \Tr\left\{  P_{n}(\bsl{k}) \chi_{\bsl{\tau}}   P_{n'}(\bsl{k}') (-1)^{n'} \Delta E(\bsl{k}') \partial_{k_i'}P_{n'}(\bsl{k}')  \chi_{\bsl{\tau}} \right\} + c.c. \\
  & = -\gamma^2 \frac{\hbar}{2 m} \sum_{\bsl{\tau},i} \left[\partial_{k'_i}E_{n'}(\bsl{k}')  -  \partial_{k_i}E_{n}(\bsl{k})  \right] (-1)^n \Delta E(\bsl{k})  \mathcal{A}_{\bsl{\tau},n,i} (\bsl{k}) \Tr\left[   P_{n'}(\bsl{k}') \chi_{\bsl{\tau}} \right] \\
  & \quad  +\gamma^2 \frac{\hbar}{2 m} \sum_{\bsl{\tau},i} \left[\partial_{k'_i}E_{n'}(\bsl{k}')  -  \partial_{k_i}E_{n}(\bsl{k})  \right](-1)^{n'} \Delta E(\bsl{k}') \mathcal{A}_{\bsl{\tau},n',i}^* (\bsl{k}') \Tr\left[  P_{n}(\bsl{k})  \chi_{\bsl{\tau}} \right] + c.c. 
}
and thus $\left\langle \Gamma \right\rangle^{E-geo}$ should read
\eqa{
\left\langle \Gamma \right\rangle^{E-geo} & =\frac{1}{D^2(\mu)}  \sum_{\bsl{k},\bsl{k}'}^{\BZ}\sum_{n,n'} \delta\left(\mu - E_n(\bsl{k}) \right) \delta\left(\mu - E_{n'}(\bsl{k}') \right) \Gamma_{n n'}^{E-geo}(\bsl{k},\bsl{k}') \\
& = -\gamma^2 \frac{1}{D^2(\mu)}  \sum_{\bsl{k},\bsl{k}'}^{\BZ}\sum_{n,n'} \delta\left(\mu - E_n(\bsl{k}) \right) \delta\left(\mu - E_{n'}(\bsl{k}') \right)  \frac{\hbar}{2 m} \sum_{\bsl{\tau},i}  \partial_{k'_i}E_{n'}(\bsl{k}')   (-1)^n \Delta E(\bsl{k})  4 \Re[\mathcal{A}_{\bsl{\tau},n} (\bsl{k})] \Tr\left[   P_{n'}(\bsl{k}') \chi_{\bsl{\tau}} \right] \\
& \quad + \gamma^2 \frac{1}{D^2(\mu)}  \sum_{\bsl{k},\bsl{k}'}^{\BZ}\sum_{n,n'} \delta\left(\mu - E_n(\bsl{k}) \right) \delta\left(\mu - E_{n'}(\bsl{k}') \right)  \frac{\hbar}{2 m} \sum_{\bsl{\tau},i}    \partial_{k_i}E_{n}(\bsl{k})  (-1)^n \Delta E(\bsl{k})  4 \Re[\mathcal{A}_{\bsl{\tau},n} (\bsl{k})] \Tr\left[   P_{n'}(\bsl{k}') \chi_{\bsl{\tau}} \right] \\
& = -\gamma^2 \frac{1}{D(\mu)}  \sum_{\bsl{k}}^{\BZ}\sum_{n} \delta\left(\mu - E_n(\bsl{k}) \right)  \frac{\hbar}{2 m} \sum_{\bsl{\tau},i}     (-1)^n \Delta E(\bsl{k})  4 \Re[\mathcal{A}_{\bsl{\tau},n,i} (\bsl{k})] b_{\bsl{\tau},i}   \\
& \quad + \gamma^2 \frac{1}{D(\mu)}  \sum_{\bsl{k}}^{\BZ}\sum_{n} \delta\left(\mu - E_n(\bsl{k}) \right) \frac{\hbar}{2 m} \sum_{\bsl{\tau},i}    \partial_{k_i}E_{n}(\bsl{k})  (-1)^n \Delta E(\bsl{k})  4 \Re[\mathcal{A}_{\bsl{\tau},n,i} (\bsl{k})] a_{\bsl{\tau}}\\
& = -\frac{2 \hbar \gamma^2}{m} \frac{1}{D(\mu)}  \frac{\V}{(2\pi)^3} \sum_{n} \int_{FS_n} d\sigma_{\bsl{k}}\frac{\Delta E(\bsl{k})}{|\nabla_{\bsl{k}} E_n(\bsl{k})|}  (-1)^n \sum_{\bsl{\tau},i}        \Re[\mathcal{A}_{\bsl{\tau},n,i} (\bsl{k})] \left( b_{\bsl{\tau},i} - \partial_{k_i}E_{n}(\bsl{k}) a_{\bsl{\tau}} \right) \ ,
}
where $a_{\bsl{\tau}}$ is defined in \eqnref{eq:a_tau_expression_GA}, $\mathcal{A}_{\bsl{\tau},n,i} (\bsl{k})$ is the orbital-selective complex vector field in \eqnref{eq:A_n_tau}, and
\eq{
b_{\bsl{\tau},i} =\frac{1}{D(\mu)}  \sum_{\bsl{k}'}^{\BZ}\sum_{n'}   \delta\left(\mu - E_{n'}(\bsl{k}') \right)  \Tr\left[  \partial_{k'_i}E_{n'}(\bsl{k}') P_{n'}(\bsl{k}') \chi_{\bsl{\tau}} \right]\ .
}
As a result, we have
\eq{
\lambda_{E-geo} =  \frac{\Omega}{(2\pi)^3} \frac{4  \gamma^2}{m \mcomega}  \sum_{n} \int_{FS_n} d\sigma_{\bsl{k}}\frac{\Delta E(\bsl{k})}{|\nabla_{\bsl{k}} E_n(\bsl{k})|}  (-1)^n \sum_{\bsl{\tau},i}        \Re[\mathcal{A}_{\bsl{\tau},n,i} (\bsl{k})] \left(  \partial_{k_i}E_{n}(\bsl{k}) a_{\bsl{\tau}} - b_{\bsl{\tau},i}  \right)\ .
}
Therefore, we can see the explicit geometric quantity in the $\lambda_{E-geo}$ is again the orbital-selective complex vector field similar to $\lambda_{geo,2}$.
Although $\lambda_{E-geo}$ is irrelevant in the graphene and {\mgb}, it is possible that $\lambda_{E-geo}$ can be relevant in other cases.

Although we only discuss the GA for a 3D system with only one kind of atom and one spinless $s$ orbital per atom, the GA can also be defined and used for more complicated cases as discussed in \appref{app:graphene} and \appref{app:MgB2}.
In particular, GA becomes (almost) exact for graphene and {\mgb} as discussed in \appref{app:graphene} and \appref{app:MgB2}, respectively.
For graphene and {\mgb}, the geometric contribution $\lambda_{geo}$ will be simplified and we can clearly see that specific band-geometric quantities (FSM or its orbital selective version) appear in $\lambda_{geo}$.
Furthermore, $\lambda_{geo}$ will be bounded from below by a topological contribution for graphene and {\mgb}. 
Before talking about graphene and {\mgb}, we will first present some general discussions on EPC in \appref{app:general_EPC}, \appref{app:2center} and \appref{app:geo_EPC_symmetry-rep}, as a preparation.

\section{General Discussions on Electron-Phonon Coupling}

\label{app:general_EPC}

We start with the general discussions on the electron-phonon-coupled model.

\subsection{Real Space}
\label{app:EPC_real_space}

The electron-phonon-coupled model that we consider is defined on a lattice.
We use $\bsl{R}$ to label the lattice point and use $\bsl{\tau}$ to label the positions of the sublattices in the $\bsl{R}=0$ unit cell.
The spatial dimension of the model is defined by the number of primitive basis lattice vectors---if the model has $d$ primitive basis lattice vectors, we say the model is in $d$D.
Throughout this work, we consider $d\leq 3$.
Regardless of the value of $d$, $\bsl{R}$ and $\bsl{\tau}$ are always embedded in 3D---$\bsl{R}$ and $\bsl{\tau}$ always have three components; for example, in 2D, we always write $\bsl{R}=(R_x,R_y,0)^T$.

At each position $\bsl{R}+\bsl{\tau}$, we have the fermionic creation operator for electrons as $c^\dagger_{\bsl{R}+\bsl{\tau}}$, and the bosonic Hermitian operators $P_{\bsl{R}+\bsl{\tau},i}$ and $u_{\bsl{R}+\bsl{\tau},i}$ that eventually give rise to phonons.
Here $\alpha_{\bsl{\tau}}$ labels onsite degrees of freedom for electrons such as orbitals and spins, and $i$ labels the orthogonal spatial directions.
Moreover, $P_{\bsl{R}+\bsl{\tau},i}$ and $u_{\bsl{R}+\bsl{\tau},i}$ satisfy 
\eqa{
\label{eq:u_P_commutation}
& [u_{\bsl{R}+\bsl{\tau},i}, P_{\bsl{R}'+\bsl{\tau}',i'}] = \ii \hbar \delta_{\bsl{R}\bsl{R}'}\delta_{\bsl{\tau}\bsl{\tau}'} \delta_{ii'} \\
& [u_{\bsl{R}+\bsl{\tau},i}, u_{\bsl{R}'+\bsl{\tau}',i'}] =0 \\
& [P_{\bsl{R}+\bsl{\tau},i}, P_{\bsl{R}'+\bsl{\tau}',i'}] = 0\ .
}
The commutation relation in \eqnref{eq:u_P_commutation} suggests that $u_{\bsl{R}+\bsl{\tau},i}$ and $u_{\bsl{R}'+\bsl{\tau}',i'}$ are defined in different Hilbert spaces for $\bsl{R}+\bsl{\tau}\neq \bsl{R}'+\bsl{\tau}'$.
In other words, it means that we treat $P_{\bsl{R}+\bsl{\tau},i}$ and $u_{\bsl{R}+\bsl{\tau},i}$ as ``internal" degrees of freedom at $\bsl{R}+\bsl{\tau}$.
We are allowed to do so for the description of phonons, because (i) $u_{\bsl{R}+\bsl{\tau},i}$ classically is the displacement of the ion at $\bsl{R}+\bsl{\tau}$ due to phonons, and (ii) the ion displacement is typically small compared to the unit cell.

Owing to the lattice structure of system, the basis of the Hamiltonian always furnishes a representation (rep) of a crystalline symmetry group $\G$.
For any $g\in\G$, we have $g=\{ R | \bsl{d} \}$, where $R$ is the point group operation a rotation or rotoinversion and $\bsl{d}$ is a translation of the position (not necessarily a lattice translation).
For example, $g$ acts on a generic position $\bsl{x}$ as $g\bsl{x} = R \bsl{x} + \bsl{d}$.

Under $g$, the atom position $\bsl{R}+\bsl{\tau}$ becomes another atom position $\bsl{R}_{\bsl{\tau},g} + \bsl{\tau}_g = R(\bsl{R}+\bsl{\tau}) + \bsl{d} $.
Here $\bsl{\tau}_g$ is the unique sublattice vector that differs from $R \bsl{\tau} + \bsl{d}$ by a lattice vector.
In other words, $\bsl{\tau}_g$ is the unique sublattice vector that satisfies
\eq{
\label{eq:R_g_tau_g}
R \bsl{\tau} + \bsl{d}  = \bsl{\tau}_g + \Delta \bsl{R}_{\bsl{\tau},g} 
}
with $\Delta \bsl{R}_{\bsl{\tau},g} $ a lattice vector.
Moreover, $\bsl{R}_{\bsl{\tau},g}  = R \bsl{R} +  \Delta \bsl{R}_{\bsl{\tau},g} $ is a lattice vector.
Formally, given $g$, there is a one-to-one correspondence between $\bsl{\tau}_g$ and $\bsl{\tau}$. 
Given $g$ and $\bsl{\tau}$, there is a one-to-one correspondence between $\bsl{R}_{\bsl{\tau},g}  $ and $ \bsl{R}$.
The existence of the one-to-one correspondences is because $g$ is an element of the Euclidean group.

The reps of $g$ furnished by the fermion and boson basis operators are
\eqa{
\label{eq:sym_rep_g_R}
& g c^\dagger_{\bsl{R}+\bsl{\tau}} g^{-1} = c^\dagger_{\bsl{R}_{\bsl{\tau},g} + \bsl{\tau}_g} U_{g}^{\bsl{\tau}_g\bsl{\tau}}\\
& g P_{\bsl{R}+\bsl{\tau},i} g^{-1} = \sum_{i'} P_{ \bsl{R}_{\bsl{\tau},g} + \bsl{\tau}_g ,i'} R_{i'i} \\
& g u_{\bsl{R}+\bsl{\tau},i} g^{-1} = \sum_{i'} u_{ \bsl{R}_{\bsl{\tau},g} + \bsl{\tau}_g ,i'} R_{i'i} \ ,
}
where $c^\dagger_{\bsl{R}+\bsl{\tau}}$ is a row vector of creation operations with components labelled by $\alpha_{\bsl{\tau}}$, \ie, 
\eq{
\label{eq:psi_R+tau}
c^\dagger_{\bsl{R}+\bsl{\tau}} = (...,c^\dagger_{\bsl{R}+\bsl{\tau},\alpha_{\bsl{\tau}}},...) \text{
with ``..." ranging over the values of $\alpha_{\bsl{\tau}}$.
}
}
Moreover, for the time-reversal (TR) symmetry $\TR$, the reps are
\eqa{
\label{eq:sym_rep_TR_R}
& \TR c^\dagger_{\bsl{R}+\bsl{\tau}} \TR^{-1} = c^\dagger_{\bsl{R} + \bsl{\tau} } U_{\TR}^{\bsl{\tau}\bsl{\tau}}\\
& \TR P_{\bsl{R}+\bsl{\tau},i} \TR^{-1} = -P_{\bsl{R}+\bsl{\tau},i} \\
& \TR u_{\bsl{R}+\bsl{\tau},i} \TR^{-1} = u_{\bsl{R}+\bsl{\tau},i}\ .
}
In \eqnref{eq:sym_rep_TR_R}, we have assumed that the fermion basis furnishes a rep of TR symmetry, which can always be guaranteed if we include enough orbitals and spins.
Throughout this work, we assume that the Hamiltonian must at least preserve the lattice translations in $\G$.

Now we introduce the Hamiltonian of a generic electron-phonon-coupled model.
The electron part of the electron-phonon-coupled model is free, and described by the tight-binding model as
\eq{
\label{eq:H_el_gen}
H_{el}=\sum_{\bsl{R}\bsl{\tau}, \bsl{R}'\bsl{\tau}'}  \sum_{ \alpha_{\bsl{\tau}} \alpha'_{\bsl{\tau}'}}
t^{\alpha_{\bsl{\tau}}  \alpha'_{\bsl{\tau}'}}_{\bsl{\tau} \bsl{\tau}'}(\bsl{R}+\bsl{\tau}-\bsl{R}'-\bsl{\tau}')
c^\dagger_{\bsl{R}+\bsl{\tau},\alpha_{\bsl{\tau}}} 
c_{\bsl{R}'+\bsl{\tau}',\alpha_{\bsl{\tau}'}'}\ ,
}
where we have incorporated the lattice translations.
Moreover, Hermiticity is equivalent to
\eq{
[t^{\alpha_{\bsl{\tau}}  \alpha'_{\bsl{\tau}'}}_{\bsl{\tau} \bsl{\tau}'}(\bsl{R}+\bsl{\tau}-\bsl{R}'-\bsl{\tau}')]^* = t^{  \alpha'_{\bsl{\tau}'} \alpha_{\bsl{\tau}}}_{\bsl{\tau}' \bsl{\tau}}(\bsl{R}'+\bsl{\tau}'-\bsl{R}-\bsl{\tau})\ .
}

Under the Harmonic approximation, the phonon part of the electron-phonon-coupled model is described by
\eqa{
\label{eq:H_ph_gen}
H_{ph} & = \sum_{\bsl{R},\bsl{\tau},i} \frac{P_{\bsl{R}+\bsl{\tau},i}^2}{ 2 m_{\bsl{\tau}} }  + \frac{1}{2}\sum_{\bsl{R}\bsl{R}',\bsl{\tau}\bsl{\tau}', i i'} D_{\bsl{\tau} i , \bsl{\tau}' i'}(\bsl{R}+\bsl{\tau}-\bsl{R}'-\bsl{\tau}') u_{\bsl{R}+\bsl{\tau},i} u_{\bsl{R}'+\bsl{\tau}',i'}\ ,
}
where we have incorporated the lattice translations.
$D_{\bsl{\tau} i , \bsl{\tau}' i'}(\bsl{R}+\bsl{\tau}-\bsl{R}'-\bsl{\tau}')$ is called the force-constant matrix.
Based on the commutation relation of $u_{\bsl{R}+\bsl{\tau},i}$, we can always choose
$D_{\bsl{\tau} i , \bsl{\tau}' i'}(\bsl{R}+\bsl{\tau}-\bsl{R}'-\bsl{\tau}') = D_{ \bsl{\tau}' i' , \bsl{\tau} i}(-\bsl{R}-\bsl{\tau}+\bsl{R}'+\bsl{\tau}')$ without loss of generality, since the anti-symmetric part does not contribute to the Hamiltonian.
Then combined with $H_{ph}^\dagger=H_{ph}$, we obtain 
\eq{
\label{eq:D_hc}
D_{\bsl{\tau}\bsl{\tau}',i i'}^*(\bsl{R}+\bsl{\tau}-\bsl{R}'-\bsl{\tau}') = D_{\bsl{\tau}\bsl{\tau}',i i'}(\bsl{R}+\bsl{\tau}-\bsl{R}'-\bsl{\tau}')\ .
}
Owing to \eqnref{eq:D_hc} and \eqnref{eq:sym_rep_TR_R}, the bare phonon Hamiltonian $H_{ph}$ always preserves TR symmetry as shown in the following:
\eqa{
\label{eq:H_ph_TR}
\TR H_{ph} \TR^{-1} & =  \sum_{\bsl{R},\bsl{\tau},i} \frac{P_{\bsl{R}+\bsl{\tau},i}^2}{ 2 m_{\bsl{\tau}} }  + \frac{1}{2}\sum_{\bsl{R}\bsl{R}',\bsl{\tau}\bsl{\tau}', i i'} D_{\bsl{\tau} i , \bsl{\tau}' i'}^*(\bsl{R}+\bsl{\tau}-\bsl{R}'-\bsl{\tau}') u_{\bsl{R}+\bsl{\tau},i} u_{\bsl{R}'+\bsl{\tau}',i'} \\
& =  \sum_{\bsl{R},\bsl{\tau},i} \frac{P_{\bsl{R}+\bsl{\tau},i}^2}{ 2 m_{\bsl{\tau}} }  + \frac{1}{2}\sum_{\bsl{R}\bsl{R}',\bsl{\tau}\bsl{\tau}', i i'} D_{\bsl{\tau} i , \bsl{\tau}' i'}(\bsl{R}+\bsl{\tau}-\bsl{R}'-\bsl{\tau}') u_{\bsl{R}+\bsl{\tau},i} u_{\bsl{R}'+\bsl{\tau}',i'} \\
& =  H_{ph}\ .
}

To the leading order, the EPC in the electron-phonon-coupled model is captured by
\eq{
\label{eq:H_el-ph_gen}
H_{el-ph} = \sum_{\bsl{R}_1\bsl{R}_2 \bsl{R}}\sum_{\bsl{\tau}_1\bsl{\tau}_2\bsl{\tau}}  \sum_{\alpha_{\bsl{\tau}_1} \alpha_{\bsl{\tau}_2}' i}c^\dagger_{\bsl{R}_1+\bsl{\tau}_1,\alpha_{\bsl{\tau}_1}}c_{\bsl{R}_2+\bsl{\tau}_2,\alpha'_{\bsl{\tau}_2}} u_{\bsl{R}+\bsl{\tau},i} \  F^{\alpha_{\bsl{\tau}_1}\alpha_{\bsl{\tau}_2}'i}_{\bsl{R}_1 + \bsl{\tau}_1 , \bsl{R}_2 + \bsl{\tau}_2,  \bsl{R} + \bsl{\tau}} \ .
}
Hermiticity is equivalent to
\eq{
\label{eq:F_hc}
[F^{\alpha_{\bsl{\tau}_1}\alpha_{\bsl{\tau}_2}'i}_{\bsl{R}_1 + \bsl{\tau}_1 , \bsl{R}_2 + \bsl{\tau}_2,  \bsl{R} + \bsl{\tau}}]^* = F^{\alpha_{\bsl{\tau}_2}'\alpha_{\bsl{\tau}_1}i}_{\bsl{R}_2 + \bsl{\tau}_2, \bsl{R}_1 + \bsl{\tau}_1 , \bsl{R} + \bsl{\tau}}\ ,
}
and the invariance under lattice translations is equivalent to
\eq{
\label{eq:F_latt}
F^{\alpha_{\bsl{\tau}_1}\alpha_{\bsl{\tau}_2}'i}_{\bsl{R}_1+\bsl{R}_0 + \bsl{\tau}_1 , \bsl{R}_2+\bsl{R}_0 + \bsl{\tau}_2,  \bsl{R} +\bsl{R}_0 + \bsl{\tau}} = F^{\alpha_{\bsl{\tau}_1}\alpha_{\bsl{\tau}_2}'i}_{\bsl{R}_1 + \bsl{\tau}_1 , \bsl{R}_2 + \bsl{\tau}_2,  \bsl{R} + \bsl{\tau}}\ \forall\ \text{lattice vector }\bsl{R}_0\ .
}
Based on locality, we assume that $F^{\alpha_{\bsl{\tau}_1}\alpha_{\bsl{\tau}_2}'i}_{\bsl{R}_1 + \bsl{\tau}_1 , \bsl{R}_2 + \bsl{\tau}_2, \bsl{R}+\bsl{\tau}}$ decays at least exponentially as $|\bsl{R}_1 + \bsl{\tau}_1-\bsl{R}-\bsl{\tau}|$ or $|\bsl{R}_2 + \bsl{\tau}_2-\bsl{R}-\bsl{\tau}|$ or $|\bsl{R}_1 + \bsl{\tau}_1-\bsl{R}_2 - \bsl{\tau}_2|$ limits to infinity.

With the above discussion, we can provide a specific definition of the electron-phonon model.
\begin{definition}[Electron-Phonon-Coupled Models]
A Hamiltonian $H$ is defined to be an electron-phonon-coupled model if and only if $H=H_{el}+H_{ph}+H_{el-ph}$, where $H_{el}$ is defined in \eqnref{eq:H_el_gen}, $H_{ph}$ is defined in \eqnref{eq:H_ph_gen}, and $H_{el-ph}$ is defined in \eqnref{eq:H_el-ph_gen}.
\end{definition}
Clearly, the charge $\U(1)$ symmetry is automatically preserved in any electron-phonon-coupled model.

\subsection{Momentum Space}

Given a generic electron-phonon-coupled model, we can perform Fourier transformations to the electron and phonon-related operators as
\eqa{
\label{eq:FT_rule}
& c_{\bsl{k},\bsl{\tau},\alpha_{\bsl{\tau}}}^\dagger = \frac{1}{\sqrt{N}} \sum_{\bsl{R}} e^{\ii \bsl{k}\cdot (\bsl{R}+\bsl{\tau})} c_{\bsl{R}+\bsl{\tau},\alpha_{\bsl{\tau}}}^\dagger\ ,\ u_{\bsl{q}\bsl{\tau} i} = \frac{1}{\sqrt{N}} \sum_{\bsl{R}} e^{-\ii \bsl{q}\cdot (\bsl{R}+\bsl{\tau})} u_{\bsl{R}+\bsl{\tau}, i} \ ,\  P_{\bsl{q}\bsl{\tau} i} = \frac{1}{\sqrt{N}} \sum_{\bsl{R}} e^{-\ii \bsl{q}\cdot (\bsl{R}+\bsl{\tau})} P_{\bsl{R}+\bsl{\tau}, i}\ ,
}
from which we find 
\eq{
u_{\bsl{q}\bsl{\tau} i}^\dagger = u_{-\bsl{q}\bsl{\tau} i}\ ,\ P_{\bsl{q}\bsl{\tau} i}^\dagger = P_{-\bsl{q}\bsl{\tau} i}\ ,
}
and 
\eqa{
\left[ u_{\bsl{q}\bsl{\tau} i}  , P_{\bsl{q}'\bsl{\tau}' i'}^\dagger \right] & = \frac{1}{\sqrt{N}} \sum_{\bsl{R}} e^{-\ii \bsl{q}\cdot (\bsl{R}+\bsl{\tau})}  \frac{1}{\sqrt{N}} \sum_{\bsl{R}'} e^{\ii \bsl{q}'\cdot (\bsl{R}'+\bsl{\tau}')} \left[ u_{\bsl{R}\bsl{\tau} i} , P_{\bsl{R}'\bsl{\tau}' i'} \right] \\
& =\ii \hbar  \delta_{\bsl{q},\bsl{q}'} \delta_{\bsl{\tau}\bsl{\tau}'} \delta_{i i'}\ .
}
Then, $\forall\ g=\{ R | \bsl{d} \}\in\G$, the symmetry reps become
\eqa{
\label{eq:sym_rep_g_k}
& g c^\dagger_{\bsl{k}} g^{-1} = c^\dagger_{R \bsl{k}} U_{g} e^{-\ii R\bsl{k}\cdot\bsl{d}}\\
& g P_{\bsl{q},\bsl{\tau},i}^\dagger g^{-1} = \sum_{i'} P_{ R\bsl{q}, \bsl{\tau}_g ,i'} R_{i'i} e^{-\ii R\bsl{q}\cdot\bsl{d}} \\
& g u_{\bsl{q},\bsl{\tau},i}^\dagger g^{-1} = \sum_{i'} u_{ R\bsl{q}, \bsl{\tau}_g ,i'} R_{i'i} e^{-\ii R\bsl{q}\cdot\bsl{d}} \ ,
}
where $c^\dagger_{\bsl{k}}$ is a row vector of creation operations with components labelled by $\bsl{\tau}$ and $\alpha_{\bsl{\tau}}$, \ie, 
\eq{
 c^\dagger_{\bsl{k}} = (..., c^\dagger_{\bsl{k},\bsl{\tau},\alpha_{\bsl{\tau}}} ,...) \text{with ``..." ranging over the values of $\bsl{\tau}$ and $\alpha_{\bsl{\tau}}$}
}
and
\eq{
\label{eq:U_g}
\left[U_{g}\right]_{\bsl{\tau}' \alpha_{\bsl{\tau}'}', \bsl{\tau} \alpha_{\bsl{\tau}}} = \left[U^{\bsl{\tau}'\bsl{\tau}}_g\right]_{\alpha_{\bsl{\tau}'}' \alpha_{\bsl{\tau}}} \text{ with $U^{\bsl{\tau}'\bsl{\tau}}_g = 0 $ for $\bsl{\tau}'\neq \bsl{\tau}_g$}\ .
}
For TR symmetry, the symmetry reps become
\eqa{
\label{eq:sym_rep_TR_k}
& \TR c^\dagger_{\bsl{k}}\TR^{-1} = c^\dagger_{- \bsl{k}} U_{\TR} \\
& \TR P_{\bsl{q},\bsl{\tau},i}^\dagger \TR^{-1} = - P_{ -\bsl{q}, \bsl{\tau} ,i} \\
& \TR u_{\bsl{q},\bsl{\tau},i}^\dagger \TR^{-1} = u_{ -\bsl{q}, \bsl{\tau} ,i} \ ,
}
where 
\eq{
\label{eq:U_TR}
\left[U_{\TR}\right]_{\bsl{\tau}' \alpha_{\bsl{\tau}'}', \bsl{\tau} \alpha_{\bsl{\tau}}} =  \left[U^{\bsl{\tau}' \bsl{\tau}}_{\TR}\right]_{\alpha_{\bsl{\tau}'}' \alpha_{\bsl{\tau}}}
\text{ with $U^{\bsl{\tau}' \bsl{\tau}}_{\TR} = 0 $ for $\bsl{\tau}'\neq \bsl{\tau}$.}
}

For the electron Hamiltonian,
\eqa{
\label{eq:H_el_gen_k}
H_{el} & = \sum_{\bsl{R}\bsl{\tau}, \bsl{R}'\bsl{\tau}'}  \sum_{ \alpha_{\bsl{\tau}} \alpha'_{\bsl{\tau}'}}
t_{\bsl{\tau}\bsl{\tau}'}^{\alpha_{\bsl{\tau}}\alpha_{\bsl{\tau}'}'}(\bsl{R}+\bsl{\tau}-\bsl{R}'-\bsl{\tau}')
\frac{1}{\sqrt{N}} \sum_{\bsl{k}}^{\text{\BZ}} e^{-\ii \bsl{k}\cdot (\bsl{R}+\bsl{\tau})} c_{\bsl{k},\bsl{\tau},\alpha_{\bsl{\tau}}}^\dagger  \frac{1}{\sqrt{N}} \sum_{\bsl{k}'}^{\text{\BZ}} e^{\ii \bsl{k}'\cdot (\bsl{R}'+\bsl{\tau}')} c_{\bsl{k}',\bsl{\tau}',\alpha_{\bsl{\tau}'}'}\\
& = \sum_{\bsl{k}}^{\text{\BZ}} c^\dagger_{\bsl{k}} h(\bsl{k}) c_{\bsl{k}}  = \sum_{\bsl{k}}^{\text{\BZ}}\sum_{n} E_{n}(\bsl{k}) \gamma^\dagger_{\bsl{k},n} \gamma_{\bsl{k},n}\ ,
}
where 
\eq{
\label{eq:h_k}
\left[ h(\bsl{k}) \right]_{\bsl{\tau}\alpha_{\bsl{\tau}},\bsl{\tau}'\alpha_{\bsl{\tau}'}'} = \sum_{\bsl{R}}t_{\bsl{\tau}\bsl{\tau}'}^{\alpha_{\bsl{\tau}}\alpha_{\bsl{\tau}'}'}(\bsl{R}+\bsl{\tau}-\bsl{\tau}')  e^{-\ii \bsl{k}\cdot (\bsl{R} + \bsl{\tau} - \bsl{\tau}')}\ ,
}
\eq{
\label{eq:el_eigen}
 h(\bsl{k}) U_{n}(\bsl{k}) = E_n(\bsl{k}) U_{n}(\bsl{k})\ ,
}
and $\gamma^\dagger_{\bsl{k},n} = c^\dagger_{\bsl{k}} U_{n}(\bsl{k})$.
\eqnref{eq:h_k} shows that for any generic reciprocal lattice vector $\bsl{G}$,
\eq{
\label{eq:h_k+G}
h(\bsl{k}+\bsl{G}) = V_{\bsl{G}}^\dagger h(\bsl{k}) V_{\bsl{G}}\ ,
}
where $V_{\bsl{G}}$ is the embedding matrix
\eq{
\label{eq:V_G}
[V_{\bsl{G}}]_{\bsl{\tau}\alpha_{\bsl{\tau}},\bsl{\tau}'\alpha_{\bsl{\tau}'}'} = e^{\ii \bsl{\tau} \cdot \bsl{G} } \delta_{\bsl{\tau}\alpha_{\bsl{\tau}},\bsl{\tau}'\alpha_{\bsl{\tau}'}'}\ .
}
We always choose $U_{n}(\bsl{k})$ such that the eigen-operators satisfy $\gamma^\dagger_{\bsl{k}+\bsl{G},n}=\gamma^\dagger_{\bsl{k},n}$ for all reciprocal lattice vectors $\bsl{G}$.

For the phonon part,
\eqa{
\label{eq:H_ph_gen_k}
H_{ph}
& =  \sum_{\bsl{q}}^{\text{\BZ}}\sum_{\bsl{\tau},i} \frac{P_{\bsl{q}\bsl{\tau} i}  P_{\bsl{q}\bsl{\tau} i}^\dagger }{ 2 m_{\bsl{\tau}} }  + \frac{1}{2}\sum_{\bsl{q}}^{\text{\BZ}} \sum_{\bsl{\tau}\bsl{\tau}', i i'}  D_{\bsl{\tau}\bsl{\tau}', i i'}(\bsl{q})  u_{\bsl{q}\bsl{\tau} i}^\dagger u_{\bsl{q}\bsl{\tau}' i'} \\
& =  \sum_{\bsl{q}}^{\text{\BZ}}\sum_{\bsl{\tau},i} \frac{\widetilde{P}_{\bsl{q}\bsl{\tau} i}  \widetilde{P}_{\bsl{q}\bsl{\tau} i}^\dagger }{ 2 }  + \frac{1}{2}\sum_{\bsl{q}}^{\text{\BZ}} \sum_{\bsl{\tau}\bsl{\tau}', i i'}  \widetilde{D}_{\bsl{\tau}\bsl{\tau}', i i'}(\bsl{q})  \widetilde{u}_{\bsl{q}\bsl{\tau} i}^\dagger  \widetilde{u}_{\bsl{q}\bsl{\tau}' i'}\\
& =\sum_{\bsl{q}}^{\text{\BZ}}\sum_{l} \left[ \frac{\widetilde{P}_{\bsl{q},l}^\dagger \widetilde{P}_{\bsl{q},l}}{2} +\frac{\omega_l^2(\bsl{q})}{2} \widetilde{u}^\dagger_{\bsl{q},l}  \widetilde{u}_{\bsl{q},l}\right]
\ ,
}
where $i,i'$ range over $x,y,z$ for spatial directions, 
\eq{
\label{eq:Force_constant_FT}
D_{\bsl{\tau}\bsl{\tau}', i i'}(\bsl{q}) = \sum_{\bsl{R}}  D_{\bsl{\tau}i , \bsl{\tau}'i'}(\bsl{R}+\bsl{\tau}-\bsl{\tau}') e^{-\ii \bsl{q}\cdot (\bsl{R} +\bsl{\tau}-\bsl{\tau}')}
}
$\widetilde{P}_{\bsl{q}\bsl{\tau}i}= P_{\bsl{q}\bsl{\tau}i}/\sqrt{m_{\bsl{\tau}}}$, $\widetilde{u}_{\bsl{q}\bsl{\tau}i}= u_{\bsl{q}\bsl{\tau}i} \sqrt{m_{\bsl{\tau}}}$,
\eq{
\label{eq:dynemical_matrix}
\widetilde{D}_{\bsl{\tau}\bsl{\tau}', i i'}(\bsl{q}) = \frac{1}{\sqrt{m_{\bsl{\tau}}}} \frac{1}{\sqrt{m_{\bsl{\tau}'}}} D_{\bsl{\tau}\bsl{\tau}', i i'}(\bsl{q}) \ ,
}
$\widetilde{D}(\bsl{q}) v_l(\bsl{q}) = \omega^2_l(\bsl{q})  v_l(\bsl{q})$, $\widetilde{P}_{\bsl{q},l}^\dagger= \widetilde{P}_{\bsl{q}}^\dagger v_l(\bsl{q})$, and $\widetilde{u}_{\bsl{q},l}^\dagger= \widetilde{u}_{\bsl{q}}^\dagger v_l(\bsl{q})$.
As shown in \eqnref{eq:H_ph_gen_k}, the definitions of $\widetilde{P}_{\bsl{q},l}^\dagger$ and $\widetilde{u}_{\bsl{q},l}^\dagger$ are used to rewrite the phonon Hamiltonian as independent Harmonic oscillators. 
We always choose $v_{l}(\bsl{q})$ such that $\widetilde{P}_{\bsl{q}+\bsl{G},l}^\dagger=\widetilde{P}_{\bsl{q},l}^\dagger$ and $\widetilde{u}_{\bsl{q}+\bsl{G},l}^\dagger=\widetilde{u}_{\bsl{q},l}^\dagger$ for all $\bsl{G}$ reciprocal lattice vectors.
Owing to
\eq{
\left[ \widetilde{u}_{\bsl{q}\bsl{\tau}i}, \widetilde{P}_{\bsl{q}'\bsl{\tau}'i'}^\dagger \right] = \sqrt{ \frac{m_{\bsl{\tau}}}{ m_{\bsl{\tau}'} }}\left[ u_{\bsl{q}\bsl{\tau} i}  , P_{\bsl{q}'\bsl{\tau}' i'}^\dagger \right] =  \ii \hbar  \delta_{\bsl{q},\bsl{q}'} \delta_{\bsl{\tau}\bsl{\tau}'} \delta_{i i'}\ ,
}
we have 
\eq{
\left[ \widetilde{u}_{\bsl{q},l} , \widetilde{P}_{\bsl{q}',l'}^\dagger \right] = \sum_{\bsl{\tau},i,\bsl{\tau}',i'} [v_{\bsl{q},l}]_{\bsl{\tau},i}  [v_{\bsl{q}',l'}]^*_{\bsl{\tau}',i'} \left[ \widetilde{u}_{\bsl{q}\bsl{\tau}i} , \widetilde{P}_{\bsl{q}'\bsl{\tau}'i'}^\dagger \right] =   \ii \hbar  \sum_{\bsl{\tau},i} [v_{\bsl{q},l}]_{\bsl{\tau},i}  [v_{\bsl{q},l'}]^*_{\bsl{\tau},i} \delta_{\bsl{q},\bsl{q}'} = \ii \hbar  \delta_{ll'}\delta_{\bsl{q},\bsl{q}'}\ .
}
Owing to $\widetilde{D}^*(\bsl{q}) = \widetilde{D}(-\bsl{q})$ derived from the fact that the bare phonon Hamiltonian always has TR symmetry (\eqnref{eq:H_ph_TR}), we can always choose the following convention for $v_l$: 
\eq{
\label{eq:v_l_TR}
v_l^*(\bsl{q}) = v_l(-\bsl{q}) 
}
for all $\bsl{q}$ and all $l$.
As a result, we have 
\eq{
 \omega_l(\bsl{q})= \omega_l(-\bsl{q})\ ,\ \widetilde{u}_{\bsl{q},l} =  v_l^\dagger(\bsl{q})\widetilde{u}_{\bsl{q}} = v_l^T(-\bsl{q})[\widetilde{u}_{-\bsl{q}}^\dagger]^T = \widetilde{u}_{-\bsl{q}}^\dagger v_l(-\bsl{q}) = \widetilde{u}_{-\bsl{q},l}^\dagger \ \text{and}\ \widetilde{P}_{\bsl{q},l}^\dagger = \widetilde{P}_{-\bsl{q},l}\ .
}

By defining 
\eq{
b^\dagger_{\bsl{q},l}=\sqrt{ \frac{\omega_l(\bsl{q})}{2\hbar} } \left[ \widetilde{u}_{\bsl{q},l}^\dagger - \frac{\ii}{ \omega_l(\bsl{q})}\widetilde{P}^\dagger_{\bsl{q},l} \right]\text{ for $\omega_l(\bsl{q})\neq 0$}\ ,
}
\eqnref{eq:H_ph_gen_k} can be written as 
\eqa{
H_{ph} 
& = \sum_{\bsl{q}\in \text{\BZ},l}^{\omega_l(\bsl{q})\neq 0} \hbar \omega_l(\bsl{q}) \left[ b^\dagger_{\bsl{q},l} b_{\bsl{q},l} + \frac{1}{2} \right] +\sum_{\bsl{q}\in \text{\BZ},l}^{\omega_l(\bsl{q})= 0} \frac{\widetilde{P}_{\bsl{q},l}^\dagger \widetilde{P}_{\bsl{q},l}}{2}\\
& = \sum_{\bsl{q}\in \text{\BZ},l} \hbar \omega_l(\bsl{q}) \left[ b^\dagger_{\bsl{q},l} b_{\bsl{q},l} + \frac{1}{2} \right]\ ,
}
where the last equality assumes that $\omega_l(\bsl{q})=0$ only happens at measure-zero region of $\bsl{q}$ in which there is no Dirac-delta-type contribution, and 
\eqa{
[b_{\bsl{q},l},b^\dagger_{\bsl{q}',l'}]  = \delta_{\bsl{q}\bsl{q}'}\delta_{ll'}\ \text{and}\ [b_{\bsl{q},l} ,b_{\bsl{q}',l'}]= 0\ .
}
Since the phonon Hamiltonian always has TR symmetry, we have
\eq{
\label{eq:TR_phonon}
\TR \widetilde{u}_{\bsl{q},l}^\dagger \TR^{-1} = \widetilde{u}_{-\bsl{q},l}^\dagger\ ,\ 
\TR \widetilde{P}_{\bsl{q},l}^\dagger \TR^{-1} = -\widetilde{P}_{-\bsl{q},l}^\dagger\ ,\ 
\TR b^\dagger_{\bsl{q},l} \TR^{-1} = b^\dagger_{-\bsl{q},l}\ .
}
A useful expression for the latter discussion is
\eq{
\widetilde{u}_{\bsl{q},l}^\dagger=\sqrt{\frac{\hbar}{2\omega_l(\bsl{q})}}(b^\dagger_{\bsl{q},l}+b_{-\bsl{q},l})\text{ for $\omega_l(\bsl{q})\neq 0$}\ .
}

For the EPC, we have
\eqa{
& H_{el-ph} = \sum_{\bsl{R}_1\bsl{R}_2 \bsl{R}}\sum_{\bsl{\tau}_1\bsl{\tau}_2\bsl{\tau}} \sum_{\alpha_{\bsl{\tau}_1} \alpha_{\bsl{\tau}_2}' i}\sum_{\bsl{k}_1,\bsl{k}_2,\bsl{q}}^{\text{\BZ}}\frac{1}{N^{3/2}}e^{-\ii \bsl{k}_1\cdot(\bsl{R}_1+\bsl{\tau}_1)} e^{\ii \bsl{k}_2\cdot(\bsl{R}_2+\bsl{\tau}_2)} e^{\ii \bsl{q}\cdot(\bsl{R}+\bsl{\tau})}\\
& \qquad \times  c^\dagger_{\bsl{k}_1,\bsl{\tau}_1,\alpha_{\bsl{\tau}_1}}c_{\bsl{k}_2,\bsl{\tau}_2,\alpha'_{\bsl{\tau}_2}} u_{\bsl{q},\bsl{\tau},i} \  F^{\alpha_{\bsl{\tau}_1}\alpha_{\bsl{\tau}_2}'i}_{\bsl{R}_1 + \bsl{\tau}_1 , \bsl{R}_2 + \bsl{\tau}_2,  \bsl{R} + \bsl{\tau}}\\
& = \sum_{\bsl{R}_1\bsl{R}_2 \bsl{R}}\sum_{\bsl{\tau}_1\bsl{\tau}_2\bsl{\tau}} \sum_{\alpha_{\bsl{\tau}_1} \alpha_{\bsl{\tau}_2}' i}\sum_{\bsl{k}_1,\bsl{k}_2,\bsl{q}}^{\text{\BZ}}\frac{1}{N^{3/2}}e^{-\ii \bsl{k}_1\cdot(\bsl{R}_1+\bsl{\tau}_1)} e^{\ii \bsl{k}_2\cdot(\bsl{R}_2+\bsl{\tau}_2)} e^{\ii \bsl{q}\cdot \bsl{\tau}} e^{\ii (\bsl{q}-\bsl{k}_1+\bsl{k}_2)\cdot \bsl{R} }\\
& \qquad \times  c^\dagger_{\bsl{k}_1,\bsl{\tau}_1,\alpha_{\bsl{\tau}_1}}c_{\bsl{k}_2,\bsl{\tau}_2,\alpha'_{\bsl{\tau}_2}} u_{\bsl{q},\bsl{\tau},i} \  F^{\alpha_{\bsl{\tau}_1}\alpha_{\bsl{\tau}_2}'i}_{\bsl{R}_1 + \bsl{\tau}_1 , \bsl{R}_2 + \bsl{\tau}_2,  \bsl{\tau}}\\
& = \sum_{\bsl{R}_1\bsl{R}_2 }\sum_{\bsl{\tau}_1\bsl{\tau}_2\bsl{\tau}} \sum_{\alpha_{\bsl{\tau}_1} \alpha_{\bsl{\tau}_2}' i}\sum_{\bsl{k}_1,\bsl{k}_2,\bsl{q}}^{\text{\BZ}}\frac{1}{\sqrt{N}}e^{-\ii \bsl{k}_1\cdot(\bsl{R}_1+\bsl{\tau}_1)} e^{\ii \bsl{k}_2\cdot(\bsl{R}_2+\bsl{\tau}_2)} e^{\ii \bsl{q}\cdot \bsl{\tau}} \sum_{\bsl{G}}\delta_{\bsl{q}+\bsl{G},\bsl{k}_1-\bsl{k}_2}\\
& \qquad \times  c^\dagger_{\bsl{k}_1,\bsl{\tau}_1,\alpha_{\bsl{\tau}_1}}c_{\bsl{k}_2,\bsl{\tau}_2,\alpha'_{\bsl{\tau}_2}} u_{\bsl{q},\bsl{\tau},i} \  F^{\alpha_{\bsl{\tau}_1}\alpha_{\bsl{\tau}_2}'i}_{\bsl{R}_1 + \bsl{\tau}_1 , \bsl{R}_2 + \bsl{\tau}_2,  \bsl{\tau}}\\
& = \sum_{\bsl{R}_1\bsl{R}_2 }\sum_{\bsl{\tau}_1\bsl{\tau}_2\bsl{\tau}} \sum_{\alpha_{\bsl{\tau}_1} \alpha_{\bsl{\tau}_2}' i}\sum_{\bsl{k}_1,\bsl{k}_2,\bsl{q}}^{\text{\BZ}}\sum_{\bsl{G}}\frac{1}{\sqrt{N}}e^{-\ii \bsl{k}_1\cdot(\bsl{R}_1+\bsl{\tau}_1)} e^{\ii \bsl{k}_2\cdot(\bsl{R}_2+\bsl{\tau}_2)} e^{\ii (\bsl{q}+\bsl{G})\cdot \bsl{\tau}} \delta_{\bsl{q}+\bsl{G},\bsl{k}_1-\bsl{k}_2}\\
& \qquad \times  c^\dagger_{\bsl{k}_1,\bsl{\tau}_1,\alpha_{\bsl{\tau}_1}}c_{\bsl{k}_2,\bsl{\tau}_2,\alpha'_{\bsl{\tau}_2}} u_{\bsl{q}+\bsl{G},\bsl{\tau},i} \  F^{\alpha_{\bsl{\tau}_1}\alpha_{\bsl{\tau}_2}'i}_{\bsl{R}_1 + \bsl{\tau}_1 , \bsl{R}_2 + \bsl{\tau}_2,  \bsl{\tau}}\\
& = \sum_{\bsl{R}_1\bsl{R}_2 }\sum_{\bsl{\tau}_1\bsl{\tau}_2\bsl{\tau}} \sum_{\alpha_{\bsl{\tau}_1} \alpha_{\bsl{\tau}_2}' i}\sum_{\bsl{k}_1,\bsl{k}_2}^{\text{\BZ}}\frac{1}{\sqrt{N}}e^{-\ii \bsl{k}_1\cdot(\bsl{R}_1+\bsl{\tau}_1-\bsl{\tau})} e^{\ii \bsl{k}_2\cdot(\bsl{R}_2+\bsl{\tau}_2-\bsl{\tau})} \\
& \qquad \times  c^\dagger_{\bsl{k}_1,\bsl{\tau}_1,\alpha_{\bsl{\tau}_1}}c_{\bsl{k}_2,\bsl{\tau}_2,\alpha'_{\bsl{\tau}_2}} u_{\bsl{k}_1-\bsl{k}_2,\bsl{\tau},i} \  F^{\alpha_{\bsl{\tau}_1}\alpha_{\bsl{\tau}_2}'i}_{\bsl{R}_1 + \bsl{\tau}_1 , \bsl{R}_2 + \bsl{\tau}_2,  \bsl{\tau}}\ ,
}
which leads to 
\eq{
\label{eq:H_el-ph_k}
H_{el-ph} = \frac{1}{\sqrt{N}} \sum_{\bsl{\tau},i} \sum_{\bsl{k}_1,\bsl{k}_2}^{\text{\BZ}}  c^\dagger_{\bsl{k}_1}F_{\bsl{\tau}i}(\bsl{k}_1,\bsl{k}_2)c_{\bsl{k}_2} u_{\bsl{k}_2-\bsl{k}_1,\bsl{\tau},i}^\dagger \ ,
}
where we used \eqnref{eq:F_latt} and we also used
\eqa{
\label{eq:f_k}
& \left[ F_{\bsl{\tau}i}(\bsl{k}_1,\bsl{k}_2) \right]_{\bsl{\tau}_1\alpha_{\bsl{\tau}_1},\bsl{\tau}_2\alpha_{\bsl{\tau}_2}'}  =  \sum_{\bsl{R}_1\bsl{R}_2 }e^{-\ii \bsl{k}_1\cdot(\bsl{R}_1+\bsl{\tau}_1-\bsl{\tau})} e^{\ii \bsl{k}_2\cdot(\bsl{R}_2+\bsl{\tau}_2-\bsl{\tau})} F^{\alpha_{\bsl{\tau}_1}\alpha_{\bsl{\tau}_2}'i}_{\bsl{R}_1 + \bsl{\tau}_1 , \bsl{R}_2 + \bsl{\tau}_2,  \bsl{\tau}} \ .
}

According to \eqnref{eq:F_hc}, Hermiticity requires and leads to
\eqa{
F_{\bsl{\tau}i}^\dagger (\bsl{k}_1,\bsl{k}_2) = F_{\bsl{\tau}i}(\bsl{k}_2, \bsl{k}_1)\ ,
}
and for any generic reciprocal lattice vectors $\bsl{G}_1$ and $\bsl{G}_2$,
\eq{
\label{eq:f_k+G}
F_{\bsl{\tau}i}(\bsl{k}_1+\bsl{G}_1,\bsl{k}_2+\bsl{G}_2) =  V_{\bsl{G}_1}^\dagger F_{\bsl{\tau}i}(\bsl{k}_1,\bsl{k}_2)  V_{\bsl{G}_2} e^{-\ii \bsl{\tau}\cdot(\bsl{G}_1-\bsl{G}_2)}\ ,
}
where $V_{\bsl{G}}$ is defined in \eqnref{eq:V_G}. 
Owing to the local property of $F^{\alpha_{\bsl{\tau}_1}\alpha_{\bsl{\tau}_2}'i}_{\bsl{R}_1 + \bsl{\tau}_1 , \bsl{R}_2 + \bsl{\tau}_2, \bsl{R}+\bsl{\tau}}$,  $\left[F_{\bsl{\tau}i}(\bsl{k}_1,\bsl{k}_2) \right]_{\bsl{\tau}_1\alpha_{\bsl{\tau}_1},\bsl{\tau}_2\alpha_{\bsl{\tau}_2}'}$ must be a smooth function of $(\bsl{k}_1,\bsl{k}_2)$ for all $\bsl{\tau}_1,\alpha_{\bsl{\tau}_1},\bsl{\tau}_2,\alpha_{\bsl{\tau}_2}'$.
Furthermore, we have
\eqa{
\label{eq:H_el_ph_gen}
& H_{el-ph} = \frac{1}{\sqrt{N}} \sum_{l}\sum_{\bsl{\tau},i} \sum_{\bsl{\tau}',i'}\sum_{\bsl{k}_1,\bsl{k}_2}^{\text{\BZ}}  c^\dagger_{\bsl{k}_1}F_{\bsl{\tau}'i'}(\bsl{k}_1,\bsl{k}_2)c_{\bsl{k}_2} \frac{1}{\sqrt{m_{\bsl{\tau}'}}} [v_l^*(\bsl{k}_2-\bsl{k}_1)]_{\bsl{\tau}'i'}  [v_l(\bsl{k}_2-\bsl{k}_1)]_{\bsl{\tau}i} \sqrt{m_{\bsl{\tau}}} u_{\bsl{k}_2-\bsl{k}_1,\bsl{\tau},i}^\dagger \\
& =\frac{1}{\sqrt{N}} \sum_{l} \sum_{\bsl{k}_1,\bsl{k}_2}^{\text{\BZ}}  c^\dagger_{\bsl{k}_1}\widetilde{F}_{l}(\bsl{k}_1,\bsl{k}_2)c_{\bsl{k}_2} \widetilde{u}_{\bsl{k}_2-\bsl{k}_1,l}^\dagger \\
& =\frac{1}{\sqrt{N}} \sum_{nml} \sum_{\bsl{k}_1,\bsl{k}_2}^{\text{\BZ}}  \widetilde{G}_{nml}(\bsl{k}_1,\bsl{k}_2) \gamma^\dagger_{\bsl{k}_1,n}\gamma_{\bsl{k}_2,m} \widetilde{u}_{\bsl{k}_2-\bsl{k}_1,l}^\dagger \\
& \Rightarrow H_{el-ph} =\frac{1}{\sqrt{N}} \sum_{nml} \sum_{\bsl{k}_1,\bsl{k}_2}^{\text{\BZ}}  G_{nml}(\bsl{k}_1,\bsl{k}_2) \gamma^\dagger_{\bsl{k}_1,n}\gamma_{\bsl{k}_2,m} (b^\dagger_{\bsl{k}_2-\bsl{k}_1,l}+b_{-\bsl{k}_2+\bsl{k}_1,l})\ ,
}
where
\eq{
\label{eq:ft_l}
\widetilde{F}_{l}(\bsl{k}_1,\bsl{k}_2) = \sum_{\bsl{\tau}',i'}  F_{\bsl{\tau}'i'}(\bsl{k}_1,\bsl{k}_2)\frac{1}{\sqrt{m_{\bsl{\tau}'}}} [v_l^*(\bsl{k}_2-\bsl{k}_1)]_{\bsl{\tau}'i'}
}
\eq{
\label{eq:Gt_nml}
\widetilde{G}_{nml}(\bsl{k},\bsl{k}') = U_n^\dagger(\bsl{k})  \widetilde{F}_{l}(\bsl{k},\bsl{k}') U_m(\bsl{k}')\ ,
}
\eq{
\label{eq:G_nml}
G_{nml}(\bsl{k},\bsl{k}') = \sqrt{ \frac{\hbar }{2\omega_l(\bsl{k}'-\bsl{k})} } \widetilde{G}_{nml}(\bsl{k},\bsl{k}')\ ,
} 
and the last equality assumes that $\omega_l(\bsl{q})=0$ only happens at measure-zero region of $\bsl{q}$ in which there is no Dirac-delta-type contribution.
Based on the above expressions, the convention that we choose for the phonon eigenvectors (\eqnref{eq:v_l_TR} derived from TR symmetry of the bare phonon Hamiltonian) and Hermiticity of $H_{el-ph}$ lead to
\eq{
\widetilde{F}_{l}^\dagger(\bsl{k}',\bsl{k}) = \widetilde{F}_{l}(\bsl{k},\bsl{k}')\ ,\ \widetilde{G}_{mnl}^*(\bsl{k}',\bsl{k}) = \widetilde{G}_{nml}(\bsl{k},\bsl{k})\ ,\ G_{mnl}^*(\bsl{k}',\bsl{k}) = G_{nml}(\bsl{k},\bsl{k}')\ .
}
For the convenience of later discussion, we note that $F_{\bsl{\tau}i}(\bsl{k},\bsl{k}')$ can be expressed in terms of $G_{nml}(\bsl{k},\bsl{k}')$ as 
\eq{
\label{eq:from_G_to_f}
F_{\bsl{\tau}i}(\bsl{k},\bsl{k}')=
\sqrt{\frac{m_{\bsl{\tau}}}{\hbar }} \sum_{nml}\sqrt{ 2\hbar  \omega_l(\bsl{k}'-\bsl{k})} U_n(\bsl{k})  G_{nml}(\bsl{k},\bsl{k}')U_m^\dagger(\bsl{k}')  [v_l(\bsl{k}'-\bsl{k})]_{\bsl{\tau}i}\ .
}

In general, $\widetilde{G}_{nml}(\bsl{k},\bsl{k}')$ are gauge dependent, where ``gauge" means that the unitary transformations that mix the electron or phonon eigenvectors, \eg, the random phase factors in front of the eigenvectors.
In general, given an isolated set of electron bands (labelled by $S_{el}$, isolated in a subregion of 1BZ), the gauge transformation for the corresponding basis takes the form of
\eq{
\mat{ ... & U_n(\bsl{k}) & ... } \rightarrow \mat{ ... & U_n(\bsl{k}) & ... }  R^{el}(\bsl{k})\ ,
}
where $R^{el}(\bsl{k})$ is unitary and $R^{el}(\bsl{k}+\forall\bsl{G}) = R^{el}(\bsl{k})$.
Here an isolated set of bands means that the set of bands are gapped from other bands in the subregion of 1BZ of interest.
Similarly, given an isolated set of phonon bands (labelled by $S_{ph}$, isolated in a subregion of 1BZ), the gauge transformation for the corresponding basis takes the form of
\eq{
\mat{ ... & v_l(\bsl{q}) & ... } \rightarrow  \mat{ ... & v_l(\bsl{q}) & ... }  R^{ph}(\bsl{q})\ ,
}
where $R^{ph}(\bsl{q})$ is unitary and $R^{ph}(\bsl{k}+\forall\bsl{G}) = R^{ph}(\bsl{k})$.
Note that the basis may not be the eigen-basis after the gauge transformation, but the transformation keeps the projectors on to the set of bands untouched.

The behaviors of $\widetilde{G}_{nml}(\bsl{k},\bsl{k}')$ under the gauge transformations can be derived from the fact that
\eq{
\sum_{n\in S_{el,1},m\in S_{el,2},l\in S_{ph}} \widetilde{G}_{nml}(\bsl{k},\bsl{k}') \gamma^\dagger_{\bsl{k},n}  \gamma_{\bsl{k}',m} \widetilde{u}_{\bsl{k}'-\bsl{k},l}^\dagger
}
is gauge invariant, where we make sure $\bsl{k}$ is in the subregion of 1BZ where $S_{el,1}$ is isolated, $\bsl{k}'$ is in the subregion of 1BZ where $S_{el,2}$ is isolated, and $\bsl{k}'-\bsl{k}$ is in the subregion of 1BZ where $S_{ph}$ is isolated.
In other words, 
\eq{
\label{eq:GUUV}
\sum_{n\in S_{el,1},m\in S_{el,2},l\in S_{ph}} \widetilde{G}_{nml}(\bsl{k},\bsl{k}') U_{n}(\bsl{k}) \otimes U^*_{m}(\bsl{k}') \otimes v_{l}(\bsl{k}'-\bsl{k})
}
is gauge invariant. 
Then, we know the following expression is gauge invariant:
\eqa{
\label{eq:Gamma_nml}
& \Gamma_{S_{el,1} S_{el,2} S_{ph}}(\bsl{k},\bsl{k}') \\
& =  \frac{\hbar}{2} \sum_{n,n'\in S_{el,1}}\sum_{m,m'\in S_{el,2}}\sum_{l,l'\in S_{ph}} \widetilde{G}_{nml}(\bsl{k},\bsl{k}')  \widetilde{G}_{n'm'l'}^*(\bsl{k},\bsl{k}') (U_{n'}(\bsl{k}) \otimes U^*_{m'}(\bsl{k}') \otimes v_{l'}(\bsl{k}'-\bsl{k}))^\dagger U_{n}(\bsl{k}) \otimes U^*_{m}(\bsl{k}') \otimes v_{l}(\bsl{k}'-\bsl{k}) \\
& = \frac{\hbar}{2}\sum_{n\in S_{el,1}}\sum_{m\in S_{el,2}}\  \sum_{l\in S_{ph}} \left|\widetilde{G}_{nml}(\bsl{k},\bsl{k}')\right|^2  = \sum_{n\in S_{el,1}}\sum_{m\in S_{el,2}}\  \sum_{l\in S_{ph}} \left|G_{nml}(\bsl{k},\bsl{k}')\right|^2 \omega_l(\bsl{k}'-\bsl{k}) \ ,
}
which is useful later.

Furthermore, \eqnref{eq:GUUV} (and thus \eqnref{eq:Gamma_nml}) are smooth at $(\bsl{k},\bsl{k}')$ as long as we make sure $\bsl{k}$ is in the subregion of 1BZ where $S_{el,1}$ is isolated, $\bsl{k}'$ is in the subregion of 1BZ where $S_{el,2}$ is isolated, and $\bsl{k}'-\bsl{k}$ is in the subregion of 1BZ where $S_{ph}$ is isolated.
To see this, we rewrite \eqnref{eq:GUUV} as
\eqa{
& \sum_{n\in S_{el,1},m\in S_{el,2},l\in S_{ph}} \widetilde{G}_{nml}(\bsl{k},\bsl{k}') U_{n}(\bsl{k}) \otimes U^*_{m}(\bsl{k}') \otimes v_{l}(\bsl{k}'-\bsl{k}) \\
& =\sum_{n\in S_{el,1},m\in S_{el,2},l\in S_{ph}}\sum_{\bsl{\tau},i}\frac{1}{\sqrt{m_{\bsl{\tau}}}} \left[ v_l^*(\bsl{k}'-\bsl{k}) \right]_{\bsl{\tau}i} U^\dagger_{n}(\bsl{k})F_{\bsl{\tau}',i'}(\bsl{k},\bsl{k}')U_{m}(\bsl{k}') U_{n}(\bsl{k}) \otimes U^*_{m}(\bsl{k}') \otimes v_{l}(\bsl{k}'-\bsl{k})\ ,
}
where we used \eqnref{eq:ft_l} and \eqnref{eq:Gt_nml}.
We know the projector $\sum_{n\in S_{el}} U_{n}(\bsl{k})U^\dagger_{n}(\bsl{k})$ is smooth in the subregion of 1BZ where $S_{el}$ is isolated, and $\sum_{l\in S_{ph}} v_{l}(\bsl{q})U^\dagger_{l}(\bsl{q})$ is smooth in the subregion of 1BZ where $S_{ph}$ is isolated.
Combined with the fact that $F_{\bsl{\tau},i}(\bsl{k},\bsl{k}')$ is smooth, we know \eqnref{eq:GUUV} is smooth in the subregion of 1BZ in which $S_{el,1}$, $S_{el,2}$ and $S_{ph}$ are all isolated, and thus so is \eqnref{eq:Gamma_nml}.

Besides the gauge freedom, there is another freedom which is the choice of basis of the Hamiltonian in the momentum space, \eg, choices of Fourier transformation rule.
\eqnref{eq:FT_rule} is just one way to fix the basis.
In general, we have the freedom of choosing basis as
\eq{
\label{eq:basis_choice}
c^\dagger_{\bsl{k}} \rightarrow c^\dagger_{\bsl{k}} R_{\bsl{k}}\ ,\ u^\dagger_{\bsl{q}} \rightarrow u^\dagger_{\bsl{q}} R_{\bsl{q}}'\ ,
}
where $R_{\bsl{k}}$ and $R_{\bsl{q}}'$ are unitary and smooth.
Note that although the electron eigenvector $U_n(\bsl{k})$ is not invariant under \eqnref{eq:basis_choice}, the creation operator $\gamma^\dagger_{\bsl{k},n}$ is invariant under \eqnref{eq:basis_choice}; it is because if we perform $c^\dagger_{\bsl{k},\bsl{\tau}} \rightarrow c^\dagger_{\bsl{k}} R_{\bsl{k}}$, we can correspondingly perform $U_{n}(\bsl{k})\rightarrow R_{\bsl{k}}^\dagger U_{n}(\bsl{k})$  to keep $\gamma^\dagger_{\bsl{k},n}$ invariant.
Based on $\gamma^\dagger_{\bsl{k},n}$, the periodic part of the Bloch state reads 
\eq{
\label{eq:u_ket}
\ket{u_{\bsl{k},n}} = e^{-\ii\bsl{k}\cdot\hat{\bsl{r}}}\gamma^\dagger_{\bsl{k},n}\ket{0}\ ,
}
where $\hat{\bsl{r}}$ is the electron position operator in the continuous space, \ie, $\hat{\bsl{r}} = \int d^dr  \bsl{r} \hat{\rho}(\bsl{r})$ with $\hat{\rho}(\bsl{r})$ the electron density operator.
$\ket{u_{\bsl{k},n}}$ is independent of the basis choice, and any final physical/geometric/topological expressions that we eventually use must be invariant under \eqnref{eq:basis_choice}.

Nevertheless, the specific basis choice specified in the Fourier transformation \eqnref{eq:FT_rule} is convenient if we use the tight-binding approximation.
Explicitly, under the tight-binding approximation, $c^\dagger_{\bsl{R}+\bsl{\tau},\alpha_{\bsl{\tau}}}$ is so localized at $\bsl{R}+\bsl{\tau}$ that we are allowed to use $ \hat{\bsl{r}} c^\dagger_{\bsl{R}+\bsl{\tau},\alpha_{\bsl{\tau}}} \ket{0}= (\bsl{R}+\bsl{\tau}) c^\dagger_{\bsl{R}+\bsl{\tau},\alpha_{\bsl{\tau}}} \ket{0}$.
Then, with \eqnref{eq:FT_rule} and the tight-binding approximation, we have
\eqa{
\label{eq:u_ket_U}
\ket{u_{\bsl{k},n}} & = e^{-\ii\bsl{k}\cdot\bsl{r}}c^\dagger_{\bsl{k}} \ket{0} U_n(\bsl{k}) \\
& = e^{-\ii\bsl{k}\cdot\bsl{r}}\sum_{\bsl{\tau},\alpha_{\bsl{\tau}}} c^\dagger_{\bsl{k},\bsl{\tau},\alpha_{\bsl{\tau}}} \ket{0} \left[ U_n(\bsl{k}) \right]_{\bsl{\tau}\alpha_{\bsl{\tau}}} \\
& = \sum_{\bsl{\tau},\alpha_{\bsl{\tau}},\bsl{R}} \frac{1}{\sqrt{N}} e^{\ii \bsl{k}\cdot (\bsl{R}+\bsl{\tau})} e^{-\ii\bsl{k}\cdot\bsl{r}} c^\dagger_{\bsl{R}+\bsl{\tau},\alpha_{\bsl{\tau}}} \ket{0} \left[ U_n(\bsl{k}) \right]_{\bsl{\tau}\alpha_{\bsl{\tau}}} \\
& = \sum_{\bsl{\tau},\alpha_{\bsl{\tau}},\bsl{R}} \frac{1}{\sqrt{N}} e^{\ii \bsl{k}\cdot (\bsl{R}+\bsl{\tau})} e^{-\ii\bsl{k}\cdot(\bsl{R}+\bsl{\tau})} c^\dagger_{\bsl{R}+\bsl{\tau},\alpha_{\bsl{\tau}}} \ket{0} \left[ U_n(\bsl{k}) \right]_{\bsl{\tau}\alpha_{\bsl{\tau}}} \\
& = \sum_{\bsl{\tau},\alpha_{\bsl{\tau}},\bsl{R}} \frac{1}{\sqrt{N}} c^\dagger_{\bsl{R}+\bsl{\tau},\alpha_{\bsl{\tau}}} \ket{0} \left[ U_n(\bsl{k}) \right]_{\bsl{\tau}\alpha_{\bsl{\tau}}} \\
& = \sum_{\bsl{\tau},\alpha_{\bsl{\tau}}} c^\dagger_{\bsl{k}=0,\bsl{\tau},\alpha_{\bsl{\tau}}} \ket{0} \left[ U_n(\bsl{k}) \right]_{\bsl{\tau}\alpha_{\bsl{\tau}}} \\
& =  c^\dagger_{\bsl{k}=0} \ket{0} U_n(\bsl{k}) \ .
}
Therefore, when using \eqnref{eq:FT_rule} and the tight-binding approximation, the momentum dependence of $\ket{u_{\bsl{k},n}}$ is all in $U_n(\bsl{k})$.
This will allows us to evaluate the FSM with the vectors $U_n(\bsl{k})$.
Explicitly, the general definition of the FSM reads
\eq{
\label{eq:FS_metric}
[g_{n}(\bsl{k})]_{j_1 j_2} = \frac{1}{2}\Tr[\partial_{k_{j_1}} \hat{P}_{n}(\bsl{k}) \partial_{k_{j_2}} \hat{P}_{n}(\bsl{k})] \ ,
}
where 
\eq{
\hat{P}_{n}(\bsl{k}) = \ket{u_{n,\bsl{k}}}\bra{u_{n,\bsl{k}}}\ .
}
Owing to the simplification in \eqnref{eq:u_ket_U} brought by the tight-binding approximation and the Fourier transformation (or the choice of basis) in \eqnref{eq:FT_rule}, we have
\eq{
\label{eq:FS_metric_U}
\hat{P}_{n}(\bsl{k}) =  c^\dagger_{\bsl{k}=0} \ket{0} U_n(\bsl{k})U_n^\dagger(\bsl{k}) \bra{0}c_{\bsl{k}=0} =  c^\dagger_{\bsl{k}=0} \ket{0} P_n(\bsl{k}) \bra{0}c_{\bsl{k}=0}
}
with $P_n(\bsl{k})= U_n(\bsl{k})U_n^\dagger(\bsl{k})$, resulting in 
\eq{
[g_{n}(\bsl{k})]_{j_1 j_2}= \frac{1}{2}\Tr[\partial_{k_{j_1}} \hat{P}_{n}(\bsl{k}) \partial_{k_{j_2}} \hat{P}_{n}(\bsl{k})] = \frac{1}{2}\Tr[\partial_{k_{j_1}} P_{n}(\bsl{k}) \partial_{k_{j_2}} P_{n}(\bsl{k})] \ .
}
Since we always use the tight-binding approximation and the Fourier transformation (or the choice of basis) in \eqnref{eq:FT_rule} unless specified otherwise, we will directly use \eqnref{eq:FS_metric_U} in this work unless specified otherwise.

\subsection{Crystalline and TR Symmetries of Electron Hamiltonian and EPC}

In this part, we discuss the Hermiticity, crystalline symmetry, and TR symmetry of the electron Hamiltonian and EPC.

Hermiticity requires
\eqa{
\label{eq:t_h_hc}
& [t_{\bsl{\tau} \bsl{\tau}'}(\bsl{R}+\bsl{\tau}-\bsl{R}'-\bsl{\tau}')]^\dagger = t_{\bsl{\tau}' \bsl{\tau}}(\bsl{R}'+\bsl{\tau}'-\bsl{R}-\bsl{\tau})\\
& h^\dagger(\bsl{k})   =h(\bsl{k})\ ,
}
and
\eqa{
\label{eq:F_f_hc}
& [F^{i}_{\bsl{R}_1 + \bsl{\tau}_1 , \bsl{R}_2 + \bsl{\tau}_2,  \bsl{R} + \bsl{\tau}}]^\dagger = F^{i}_{\bsl{R}_2 + \bsl{\tau}_2, \bsl{R}_1 + \bsl{\tau}_1 , \bsl{R} + \bsl{\tau}}\\
& F_{\bsl{\tau}i}^\dagger(\bsl{k}_1,\bsl{k}_2) = F_{\bsl{\tau}i}(\bsl{k}_2,\bsl{k}_1)\ ,
}
where $t_{\bsl{\tau} \bsl{\tau}'}(\bsl{R}+\bsl{\tau}-\bsl{R}'-\bsl{\tau}')$ is a matrix defined by
\eq{
\label{eq:t_block}
\left[ t_{\bsl{\tau} \bsl{\tau}'}(\bsl{R}+\bsl{\tau}-\bsl{R}'-\bsl{\tau}') \right]_{\alpha_{\bsl{\tau}}\alpha_{\bsl{\tau}'}'} =  t_{\bsl{\tau} \bsl{\tau}'}^{\alpha_{\bsl{\tau}}\alpha_{\bsl{\tau}'}'}(\bsl{R}+\bsl{\tau}-\bsl{R}'-\bsl{\tau}') \ ,
}
$t_{\bsl{\tau} \bsl{\tau}'}^{\alpha_{\bsl{\tau}}\alpha_{\bsl{\tau}'}'}(\bsl{R}+\bsl{\tau}-\bsl{R}'-\bsl{\tau}')$ is defined in \eqnref{eq:H_el_gen}, $h(\bsl{k})$ is defined in \eqnref{eq:h_k}, $F^{i}_{\bsl{R}_1 + \bsl{\tau}_1 , \bsl{R}_2 + \bsl{\tau}_2,  \bsl{R} + \bsl{\tau}}$ is a matrix defined as
\eq{
\label{eq:F_block}
\left[ F^{i}_{\bsl{R}_1 + \bsl{\tau}_1 , \bsl{R}_2 + \bsl{\tau}_2,  \bsl{R} + \bsl{\tau}} \right]_{\alpha_{\bsl{\tau}_1}\alpha_{\bsl{\tau}_2}'} = F^{\alpha_{\bsl{\tau}_1}\alpha_{\bsl{\tau}_2}'i}_{\bsl{R}_1 + \bsl{\tau}_1 , \bsl{R}_2 + \bsl{\tau}_2,  \bsl{R} + \bsl{\tau}}\ ,
}
$F^{\alpha_{\bsl{\tau}_1}\alpha_{\bsl{\tau}_2}'i}_{\bsl{R}_1 + \bsl{\tau}_1 , \bsl{R}_2 + \bsl{\tau}_2,  \bsl{R} + \bsl{\tau}}$ is defined in \eqnref{eq:H_el-ph_gen}, and $F_{\bsl{\tau}i}^\dagger(\bsl{k}_1,\bsl{k}_2)$ is defined in \eqnref{eq:f_k}.

Suppose the Hamiltonian has a crystalline symmetry $g=\{ R | \bsl{d} \}$, where $R$ is the point group part of $g$ and $\bsl{d}$ labels the translational part of $g$.
According to \eqnref{eq:sym_rep_g_R} and \eqnref{eq:sym_rep_g_k}, the matrix Hamiltonians of electrons (\ie, $t_{\bsl{\tau}\bsl{\tau}'}(\bsl{R}+\bsl{\tau}-\bsl{R}'-\bsl{\tau}')$ in \eqnref{eq:t_block} and $h(\bsl{k})$ in \eqnref{eq:H_el_gen_k}) satisfy
\eqa{
\label{eq:t_h_sym_g}
& U^{\bsl{\tau}_{1,g}\bsl{\tau}_1}_g t_{\bsl{\tau}_1 \bsl{\tau}_2 }(\bsl{R}_1+\bsl{\tau}_1-\bsl{R}_2-\bsl{\tau}_2) \left[U^{\bsl{\tau}_{2,g}\bsl{\tau}_2}_g\right]^\dagger   = t_{\bsl{\tau}_{1,g} \bsl{\tau}_{2,g} }(\bsl{R}_{1,\bsl{\tau}_1,g} + \bsl{\tau}_{1,g} - \bsl{R}_{2,\bsl{\tau}_2,g} - \bsl{\tau}_{2,g}) \\
& U_{g}^\dagger h(\bsl{k}) U_{g}  =h(R\bsl{k})\ ,
}
where $U_g$ is defined in \eqnref{eq:U_g}, and $U^{\bsl{\tau}_{g}\bsl{\tau}}_g$ is defined in \eqnref{eq:sym_rep_g_R}.
$F^{\alpha_{\bsl{\tau}_1}\alpha_{\bsl{\tau}_2}'i}_{\bsl{R}_1 + \bsl{\tau}_1 , \bsl{R}_2 + \bsl{\tau}_2,  \bsl{R} + \bsl{\tau}}$ in \eqnref{eq:H_el-ph_gen} and $F_{\bsl{\tau}i}(\bsl{k}_1,\bsl{k}_2)$ in \eqnref{eq:H_el-ph_k} satisfy
\eqa{
\label{eq:F_f_sym_g}
& \sum_{i'}U^{\bsl{\tau}_{1,g}\bsl{\tau}_1}_g F^{i'}_{\bsl{R}_1 + \bsl{\tau}_1 , \bsl{R}_2 + \bsl{\tau}_2,  \bsl{R} + \bsl{\tau}}  \left[U^{\bsl{\tau}_{2,g}\bsl{\tau}_2}_g\right]^\dagger R_{ii'} = F^{i}_{\bsl{R}_{1,\bsl{\tau}_1,g} + \bsl{\tau}_{1,g} , \bsl{R}_{2,\bsl{\tau}_2,g} + \bsl{\tau}_{2,g},  \bsl{R}_g + \bsl{\tau}_g} \\
& \sum_{i'} U_{g}^\dagger F_{\bsl{\tau}i'}(\bsl{k}_1,\bsl{k}_2) U_{g}  R_{ii'} = F_{\bsl{\tau}_g i}(R\bsl{k}_1,R\bsl{k}_2)\ .
}

Suppose the Hamiltonian has TR symmetry $\TR$.
According to \eqnref{eq:sym_rep_TR_R} and \eqnref{eq:sym_rep_TR_k}, the matrix Hamiltonians of electrons (\ie, $t_{\bsl{\tau}\alpha_{\bsl{\tau}},\bsl{\tau}'\alpha_{\bsl{\tau}'}'}(\bsl{R}+\bsl{\tau}-\bsl{R}'-\bsl{\tau}')$ in \eqnref{eq:t_block} and $h(\bsl{k})$ in \eqnref{eq:H_el_gen_k}) satisfy
\eqa{
\label{eq:t_h_sym_TR}
& U^{\bsl{\tau}_1 \bsl{\tau}_1}_{\TR} \left[t_{\bsl{\tau}_1 \bsl{\tau}_2 }(\bsl{R}+\bsl{\tau}-\bsl{R}'-\bsl{\tau}')\right]^* \left[U^{\bsl{\tau}_2 \bsl{\tau}_2}_{\TR}\right]^\dagger  =t_{\bsl{\tau}_1 \bsl{\tau}_2}(\bsl{R}+\bsl{\tau}-\bsl{R}'-\bsl{\tau}')  \\
& U_{\TR}^\dagger h^*(\bsl{k}) U_{\TR}  =h(-\bsl{k})\ .
}
$F^{\alpha_{\bsl{\tau}_1}\alpha_{\bsl{\tau}_2}'i}_{\bsl{R}_1 + \bsl{\tau}_1 , \bsl{R}_2 + \bsl{\tau}_2,  \bsl{R} + \bsl{\tau}}$ in \eqnref{eq:H_el-ph_gen} and $F_{\bsl{\tau}i}(\bsl{k}_1,\bsl{k}_2)$ in \eqnref{eq:H_el-ph_k} satisfy
\eqa{
\label{eq:F_f_sym_TR}
&  U^{\bsl{\tau}_1 \bsl{\tau}_1}_{\TR} \left[F^{i}_{\bsl{R}_1 + \bsl{\tau}_1 , \bsl{R}_2 + \bsl{\tau}_2,  \bsl{R} + \bsl{\tau}}\right]^*  \left[U^{\bsl{\tau}_2 \bsl{\tau}_2}_{\TR}\right]^\dagger  = F^{i}_{\bsl{R}_{1} + \bsl{\tau}_{1} , \bsl{R}_{2} + \bsl{\tau}_{2},  \bsl{R} + \bsl{\tau}} \\
&  U_{\TR}^\dagger F_{\bsl{\tau}i}^*(\bsl{k}_1,\bsl{k}_2) U_{\TR}    = F_{\bsl{\tau} i}(-\bsl{k}_1,-\bsl{k}_2)\ .
}

\subsection{EPC Constant}
\label{app:EPC_Constant}

The EPC strengh is commonly characterized by the dimensionless EPC constant $\lambda$~\cite{McMillan1968SCTc}, which is defined as
\eq{
\label{eq:lambda}
\lambda = \frac{2}{D(\mu) N} \sum_{\bsl{k},\bsl{k}'}^{\BZ} \sum_{nml}\frac{|G_{nml}(\bsl{k},\bsl{k}')|^2}{\hbar \omega_l(\bsl{k}'-\bsl{k})} \delta\left(\mu - E_n(\bsl{k}) \right) \delta\left(\mu - E_m(\bsl{k}') \right)\ ,
}
where $n,m$ sum over all the electron bands, $l$ sums over all the phonon bands,
\eq{
\label{eq:D_mu}
D(\mu) = \sum_{\bsl{k}}^{\BZ} \sum_{n} \delta\left(\mu - E_n(\bsl{k}) \right)\ 
}
is the electronic density of states at $\mu$.
For the convenience of the later discussion, we note that given any function $F(\bsl{k})$ such that $F(\bsl{k}+\forall \bsl{G}) = F(\bsl{k})$, we have in the thermodynamic limit
\eq{
\label{eq:sum_delta_E_n_mu}
\sum_{\bsl{k}\in\BZ} \delta\left(\mu - E_n(\bsl{k}) \right) F(\bsl{k}) = \frac{\V}{(2\pi)^d} \int_{FS_n} d\sigma_{\bsl{k}}\frac{1}{|\nabla_{\bsl{k}} E_{n}(\bsl{k})|}  F(\bsl{k})\ ,
}
where $\V$ is the volume of the system, and $FS_n$ is the Fermi surface given by $E_n(\bsl{k})=\mu$.
This will be useful for the calculation of the energetic/geometric contributions to $\lambda$.

Another commonly used expression of $\lambda$ reads~\cite{McMillan1968SCTc}
\eq{
\lambda = 2 \int_0^{+\infty} d\omega\ \frac{1}{\omega}\ \alpha^2F(\omega)\ ,
}
where 
\eq{
\label{eq:alpha2F}
\alpha^2 F(\omega) = \frac{1}{D(\mu) N} \sum_{\bsl{k},\bsl{k}'}^{\BZ} \sum_{nml}\frac{|G_{nml}(\bsl{k},\bsl{k}')|^2}{\hbar } \delta\left(\mu - E_n(\bsl{k}) \right) \delta\left(\mu - E_m(\bsl{k}') \right) \delta(\omega - \omega_l(\bsl{k}'-\bsl{k}))
}
is the Eliashberg function.
Interestingly, \refcite{McMillan1968SCTc} defines a mean squared phonon frequency as
\eq{
\label{eq:McMillan_omega_square_ave}
\mcomega = \frac{ \int_0^{+\infty} d\omega\ \omega^2 \ \frac{1}{\omega}\ \alpha^2F(\omega) }{ \int_0^{+\infty} d\omega\ \frac{1}{\omega}\ \alpha^2F(\omega) }\ .
}
With \eqnref{eq:McMillan_omega_square_ave}, we can rewrite $\lambda$ as
\eqa{
\label{eq:lambda_omegabar}
\lambda  & = 2 \frac{1}{\mcomega} \int_0^{+\infty} d\omega\ \omega\ \alpha^2F(\omega)  \\
& = \frac{2 }{\mcomega}  \frac{1}{D(\mu) N} \sum_{\bsl{k},\bsl{k}'}^{\BZ} \sum_{nml}\frac{|G_{nml}(\bsl{k},\bsl{k}')|^2 \omega_l(\bsl{k}'-\bsl{k})}{\hbar } \delta\left(\mu - E_n(\bsl{k}) \right) \delta\left(\mu - E_m(\bsl{k}') \right) \\
& = \frac{2 }{\hbar \mcomega} \frac{D(\mu)}{ N} \left\langle \Gamma \right\rangle \ ,
}
where 
\eq{
\label{eq:Gamma_ave_mu}
\left\langle \Gamma \right\rangle =  \frac{ \sum_{\bsl{k},\bsl{k}'}^{\BZ} \sum_{nm } \Gamma_{nm}(\bsl{k},\bsl{k}')  \delta\left(\mu - E_n(\bsl{k}) \right) \delta\left(\mu - E_m(\bsl{k}') \right) }{ \sum_{\bsl{k}_1,\bsl{k}_1'}^{\BZ} \sum_{nm }\delta\left(\mu - E_n(\bsl{k}_1) \right) \delta\left(\mu - E_m(\bsl{k}_1') \right)} , 
}
\eqa{
\label{eq:Gamma_nm}
  \Gamma_{nm}(\bsl{k},\bsl{k}') & = \sum_{l}|G_{nml}(\bsl{k},\bsl{k}')|^2 \omega_l(\bsl{k}'-\bsl{k}) = \frac{\hbar}{2}\sum_{l}|\widetilde{G}_{nml}(\bsl{k},\bsl{k}')|^2 \\
  & = \frac{\hbar}{2} \sum_l \Tr\left[ P_{n}(\bsl{k})  \widetilde{F}_{l}(\bsl{k},\bsl{k}') P_{m}(\bsl{k}') \widetilde{F}_{l}^\dagger(\bsl{k},\bsl{k}') \right]\\
  & = \frac{\hbar}{2} \sum_l \Tr\left[ P_{n}(\bsl{k})  \sum_{\bsl{\tau}',i'}  F_{\bsl{\tau}'i'}(\bsl{k},\bsl{k}')\frac{1}{\sqrt{m_{\bsl{\tau}'}}} [v_l^*(\bsl{k}'-\bsl{k})]_{\bsl{\tau}'i'}P_{m}(\bsl{k}')  \sum_{\bsl{\tau}_1',i_1'}  F_{\bsl{\tau}_1'i_1'}^\dagger(\bsl{k},\bsl{k}')\frac{1}{\sqrt{m_{\bsl{\tau}_1'}}} [v_l(\bsl{k}'-\bsl{k})]_{\bsl{\tau}_1'i_1'} \right] \\
  & = \frac{\hbar}{2} \sum_{\bsl{\tau},i}  \frac{1}{m_{\bsl{\tau}}} \Tr\left[ P_{n}(\bsl{k})  F_{\bsl{\tau}i}(\bsl{k},\bsl{k}') P_{m}(\bsl{k}')   F_{\bsl{\tau}i}^\dagger(\bsl{k},\bsl{k}') \right] \ ,
 }
 and 
 \eq{
 \label{eq:P_U}
 P_{n}(\bsl{k}) = U_{n}(\bsl{k})U^\dagger_{n}(\bsl{k})
 }
 is the electron projection matrix ($U_n$ is defined in \eqnref{eq:el_eigen}).
 \eqnref{eq:Gamma_nm} is invariant under the change of the basis (\eqnref{eq:basis_choice}) since the first expression is in the eigenbasis.

As shown in \eqnref{eq:McMillan_omega_square_ave}, $\mcomega$ formally depends on the EPC, the phonon frequencies/eigenvectors, and the electron bands/eigenvectors.
For graphene and {\mgb}, specific phonon modes dominate EPC, and thus $\mcomega$ can be well approximated by the frequency of those dominant phonon modes (as discussed in \appref{app:graphene} and \appref{app:MgB2}, respectively).
Therefore, in this work, we will treat $\mcomega$ as a parameter and focus on the remaining part of $\lambda$ in \eqnref{eq:lambda_omegabar}.
In the expression of $\lambda$ (\eqnref{eq:lambda_omegabar}), all information on the phonon frequencies or phonon eigenvectors is contained in $\mcomega$.
Since we only want to study the effect of the electron band geometry or topology on the EPC, it is also for our simplicity to treat $\mcomega$ as a parameter determined in the \emph{ab initio} calculation and to focus on the rest.

\section{Two-Center Approximation of EPC}

\label{app:2center}

Throughout this work, we adopt the two-center approximation of EPC.
In this section, we provide more details on the two-center approximation and set the stage for the discussions of geometric part of EPC in \appref{app:geo_EPC_symmetry-rep}.

\subsection{Real Space}
\label{app:2center_real}

The idea of two-center approximation was discussed in \refcite{Mitra1969EPC}. 
Our discussion here will look different from that in \refcite{Mitra1969EPC}, since we will not care about the detailed form of the EPC and only focus on the key assumptions.
Nevertheless the essential idea of the two-center approximation used here is the same as that in \refcite{Mitra1969EPC}.
Specifically, we adopt the following two-center approximation:
\begin{assumption}[Two-center Approximation of EPC]
\label{asm:2center}
Given any two electron (spinful) orbitals, the EPC for them is determined by the relative motions of two ions at which the two electron orbitals are localized.
\end{assumption}
\asmref{asm:2center} is a generalized version of the two-center approximation used in \refcite{Mitra1969EPC}.
Clearly, if two electrons are localized at the same ion, the EPC of them is zero under \asmref{asm:2center}.

In terms of equations, \asmref{asm:2center} is equivalent to assuming
\eqa{
\label{eq:2center}
& F^{i}_{\bsl{R}_1 + \bsl{\tau}_1 , \bsl{R}_2 + \bsl{\tau}_2,  \bsl{R} + \bsl{\tau}} = 0 \text{ for } \bsl{R} + \bsl{\tau} \neq \bsl{R}_1 + \bsl{\tau}_1 \text{ or } \bsl{R} + \bsl{\tau} \neq \bsl{R}_2 + \bsl{\tau}_2 \\ 
& F^{i}_{\bsl{R}_1 + \bsl{\tau}_1 , \bsl{R}_2 + \bsl{\tau}_2,  \bsl{R}_1 + \bsl{\tau}_1} = - F^{i}_{\bsl{R}_1 + \bsl{\tau}_1 , \bsl{R}_2 + \bsl{\tau}_2,  \bsl{R}_2 + \bsl{\tau}_2}\ ,
}
where $F^{i}_{\bsl{R}_1 + \bsl{\tau}_1 , \bsl{R}_2 + \bsl{\tau}_2,  \bsl{R} + \bsl{\tau}}$ is related to $F^{\alpha_{\bsl{\tau}_1}\alpha_{\bsl{\tau}_2}'i}_{\bsl{R}_1 + \bsl{\tau}_1 , \bsl{R}_2 + \bsl{\tau}_2,  \bsl{R} + \bsl{\tau}}$ in \eqnref{eq:H_el-ph_gen} by \eqnref{eq:F_block}.
For convenience, we can define 
\eq{
\label{eq:g_R_ini}
f^{i}_{\bsl{R}_1 + \bsl{\tau}_1 , \bsl{R}_2 + \bsl{\tau}_2} = \frac{1}{2} \left( F^{i}_{\bsl{R}_1 + \bsl{\tau}_1 , \bsl{R}_2 + \bsl{\tau}_2,  \bsl{R}_1 + \bsl{\tau}_1} - F^{i}_{\bsl{R}_1 + \bsl{\tau}_1 , \bsl{R}_2 + \bsl{\tau}_2,  \bsl{R}_2 + \bsl{\tau}_2} \right)\ ,
}
which gives 
\eq{
 F^{i}_{\bsl{R}_1 + \bsl{\tau}_1 , \bsl{R}_2 + \bsl{\tau}_2,  \bsl{R} + \bsl{\tau}} = f^{i}_{\bsl{R}_1 + \bsl{\tau}_1 , \bsl{R}_2 + \bsl{\tau}_2}(\delta_{\bsl{R} + \bsl{\tau}, \bsl{R}_1 + \bsl{\tau}_1} -\delta_{\bsl{R} + \bsl{\tau}, \bsl{R}_2 + \bsl{\tau}_2})\ .
}
Then, under the two-center approximation, we have 
\eq{
\label{eq:H_el-ph_2center_ini}
H_{el-ph} = \sum_{\bsl{R}_1\bsl{R}_2 }\sum_{\bsl{\tau}_1\bsl{\tau}_2}  \sum_{ i}c^\dagger_{\bsl{R}_1+\bsl{\tau}_1} f^{i}_{\bsl{R}_1 + \bsl{\tau}_1 , \bsl{R}_2 + \bsl{\tau}_2} c_{\bsl{R}_2+\bsl{\tau}_2} \  (u_{\bsl{R}_1 + \bsl{\tau}_1,i} -u_{\bsl{R}_2 + \bsl{\tau}_2,i} )\ ,
}
where $c^\dagger_{\bsl{R}+\bsl{\tau}}$ is defined in \eqnref{eq:psi_R+tau}.
\eqnref{eq:H_el-ph_2center_ini} clearly shows that the EPC for $c^\dagger_{\bsl{R}_1+\bsl{\tau}_1}$ and $c_{\bsl{R}_2+\bsl{\tau}_2}$ only involves the relative motion $u_{\bsl{R}_1 + \bsl{\tau}_1,i} -u_{\bsl{R}_2 + \bsl{\tau}_2,i}$, and shows that 
\eq{
\label{eq:f_zero_onsite_ini}
f^i_{\bsl{R} + \bsl{\tau} , \bsl{R} + \bsl{\tau}} = 0\ ,
}
which is consistent with \asmref{asm:2center}.

According to \eqnref{eq:F_hc} and \eqnref{eq:F_latt}, the lattice translation requires
\eq{
\label{eq:g_latt_ini}
f^{i}_{\bsl{R}_1 + \bsl{\tau}_1 , \bsl{R}_2 + \bsl{\tau}_2} = f^{i}_{\bsl{R}_1 + \bsl{\tau}_1 - \bsl{R}_2 , \bsl{\tau}_2}\ ,
}
and the Hermiticity requires 
\eq{
\label{eq:g_hc_ini}
\left( f^{i}_{ \bsl{R}_2 + \bsl{\tau}_2 , \bsl{R}_1 + \bsl{\tau}_1 }\right)^\dagger = - f^{i}_{\bsl{R}_1 + \bsl{\tau}_1 , \bsl{R}_2 + \bsl{\tau}_2}\ .
}
For convenience of the latter discussion, we define a matrix $f_{\bsl{\tau}_1 \bsl{\tau}_2, i}(\bsl{R}_1 + \bsl{\tau}_1 - \bsl{R}_2 - \bsl{\tau}_2 )$ as
\eq{
\label{eq:g_R}
  f_{\bsl{\tau}_1 \bsl{\tau}_2,i}(\bsl{R}_1 + \bsl{\tau}_1 - \bsl{R}_2 - \bsl{\tau}_2 )   = f^{i}_{\bsl{R}_1 + \bsl{\tau}_1 - \bsl{R}_2 , \bsl{\tau}_2}\ ,
}
which satisfies 
\eq{
\label{eq:f_zero_onsite}
f_{\bsl{\tau} \bsl{\tau},i}( 0 ) = 0
}
according to \eqnref{eq:f_zero_onsite_ini}.
Then, \eqnref{eq:H_el-ph_2center_ini} can be rewritten as
\eq{
\label{eq:H_el-ph_2center_R}
H_{el-ph} = \sum_{\bsl{R}_1\bsl{R}_2 }\sum_{\bsl{\tau}_1\bsl{\tau}_2}  \sum_{ i}c^\dagger_{\bsl{R}_1+\bsl{\tau}_1}f_{\bsl{\tau}_1 \bsl{\tau}_2,i}(\bsl{R}_1 + \bsl{\tau}_1 - \bsl{R}_2 - \bsl{\tau}_2 )c_{\bsl{R}_2+\bsl{\tau}_2}  (u_{\bsl{R}_1 + \bsl{\tau}_1,i} -u_{\bsl{R}_2 + \bsl{\tau}_2,i} )\ ,
}
where \eqnref{eq:g_latt_ini} has been used.

When we use smooth hopping functions $t^{\alpha_{\bsl{\tau}_1}  \alpha'_{\bsl{\tau}_2}}_{\bsl{\tau}_1 \bsl{\tau}_2}(\bsl{r})$ to describe the hopping among electrons in \eqnref{eq:H_el_gen}, $f_{\bsl{\tau}_1 \bsl{\tau}_2,i}(\bsl{R}_1 + \bsl{\tau}_1 - \bsl{R}_2 - \bsl{\tau}_2 )$ takes the following form
\eq{
\label{eq:f_i_partial_i_t_real_space}
\left[ f_{\bsl{\tau}_1 \bsl{\tau}_2,i}(\bsl{R}_1 + \bsl{\tau}_1 - \bsl{R}_2 - \bsl{\tau}_2 )\right]_{\alpha_{\bsl{\tau}_1}  \alpha'_{\bsl{\tau}_2}} = \left. \partial_{r_i} t^{\alpha_{\bsl{\tau}_1}  \alpha'_{\bsl{\tau}_2}}_{\bsl{\tau}_1 \bsl{\tau}_2}(\bsl{r}) \right|_{\bsl{r} = \bsl{R}_1 + \bsl{\tau}_1 - \bsl{R}_2 - \bsl{\tau}_2}\ .
}

\subsection{Momentum Space}
\label{app:2center_momentum}

The discussion in \appref{app:2center_real} is in real space; now we convert \eqnref{eq:H_el-ph_2center_R} and \eqnref{eq:g_R} to momentum space.
To do so, we first Fourier transform \eqnref{eq:g_R} into momentum space and obtain 
\eq{
\label{eq:g_k_block}
f_{\bsl{\tau}_1\bsl{\tau}_2,i}(\bsl{k}) = \sum_{\Delta\bsl{R} } e^{- \ii \bsl{k} \cdot (\Delta\bsl{R}+ \bsl{\tau}_1 - \bsl{\tau}_2)}   f_{\bsl{\tau}_1 \bsl{\tau}_2,i}(\Delta\bsl{R} + \bsl{\tau}_1  - \bsl{\tau}_2 )  \ ,
}
and we define 
\eq{
\label{eq:g_k}
\left[ f_{i}(\bsl{k}) \right]_{\bsl{\tau}_1\alpha_{\bsl{\tau}_1}, \bsl{\tau}_2 \alpha_{\bsl{\tau}_2}'} = \left[ f_{\bsl{\tau}_1\bsl{\tau}_2,i}(\bsl{k}) \right]_{\alpha_{\bsl{\tau}_1} \alpha_{\bsl{\tau}_2}'} \ .
}
\eqnref{eq:g_k} defines $f_i(\bsl{k})$ to be a matrix; $f_{\bsl{\tau}_1\bsl{\tau}_2,i}(\bsl{k})$ is nothing but the $\bsl{\tau}_1 \bsl{\tau}_2$ block of $f_i(\bsl{k})$.

Furthermore, we Fourier transform $c^\dagger_{\bsl{R}+\bsl{\tau}}$ in \eqnref{eq:psi_R+tau} to momentum space and obtain 
\eq{
\label{eq:psi_block_FT_rule}
 c_{\bsl{k},\bsl{\tau}}^\dagger =  \frac{1}{\sqrt{N}} \sum_{\bsl{R}} e^{\ii \bsl{k}\cdot (\bsl{R}+\bsl{\tau})} c_{\bsl{R}+\bsl{\tau}}^\dagger\ ,
}
which means that 
\eq{
 c_{\bsl{k},\bsl{\tau}}^\dagger = (..., c_{\bsl{k},\bsl{\tau},\alpha_{\bsl{\tau}}}^\dagger,...)
}
with ``..." ranging over the orbitals or spins (labeled by $\alpha_{\bsl{\tau}}$) at the position $\bsl{\tau}$ in the unit cell, where $c_{\bsl{k},\bsl{\tau},\alpha_{\bsl{\tau}}}^\dagger$ is define in \eqnref{eq:FT_rule}.
With \eqnref{eq:psi_block_FT_rule} and \eqnref{eq:g_k_block}, we can finally transform \eqnref{eq:H_el-ph_2center_R} to momentum space and obtain
\eqa{
H_{el-ph} & = \sum_{\bsl{R}_1\bsl{R}_2  }\sum_{\bsl{\tau}_1\bsl{\tau}_2 }  \sum_{i}c^\dagger_{\bsl{R}_1+\bsl{\tau}_1} f_{\bsl{\tau}_1 \bsl{\tau}_2,i}(\bsl{R}_1 + \bsl{\tau}_1 - \bsl{R}_2 - \bsl{\tau}_2 ) c_{\bsl{R}_2+\bsl{\tau}_2}  (u_{\bsl{R}_1 + \bsl{\tau}_1,i} -u_{\bsl{R}_2 + \bsl{\tau}_2,i} ) \\
& = \sum_{\bsl{k}_1}^{\BZ} \sum_{\bsl{k}_2}^{\BZ} \sum_{\bsl{q}}^{\BZ} \sum_{\bsl{R}_1\bsl{R}_2  }\sum_{\bsl{\tau}_1\bsl{\tau}_2 }  \sum_{i} \frac{1}{\sqrt{N}} e^{-\ii \bsl{k}_1 \cdot (\bsl{R}_1+\bsl{\tau}_1)} \frac{1}{\sqrt{N}} e^{\ii \bsl{k}_2 \cdot (\bsl{R}_2+\bsl{\tau}_2)}  c^\dagger_{\bsl{k}_1,\bsl{\tau}_1} f_{\bsl{\tau}_1 \bsl{\tau}_2,i}(\bsl{R}_1 + \bsl{\tau}_1 - \bsl{R}_2 - \bsl{\tau}_2 ) c_{\bsl{k}_2,\bsl{\tau}_2} \\
& (\frac{1}{\sqrt{N}} e^{\ii \bsl{q} \cdot (\bsl{R}_1+\bsl{\tau}_1)} u_{\bsl{q},\bsl{\tau}_1,i} - \frac{1}{\sqrt{N}} e^{\ii \bsl{q} \cdot (\bsl{R}_2+\bsl{\tau}_2)} u_{\bsl{q},\bsl{\tau}_2,i})\\
& = \frac{1}{\sqrt{N}}\sum_{\bsl{k}_1}^{\BZ} \sum_{\bsl{k}_2}^{\BZ} \sum_{\bsl{q}}^{\BZ} \sum_{\Delta\bsl{R} }\sum_{\bsl{\tau}_1\bsl{\tau}_2 }  \sum_{i} \frac{1}{N}\sum_{ \bsl{R}_2  } e^{-\ii \bsl{k}_1 \cdot (\Delta\bsl{R} + \bsl{R}_2+\bsl{\tau}_1)}  e^{\ii \bsl{k}_2 \cdot (\bsl{R}_2+\bsl{\tau}_2)}  c^\dagger_{\bsl{k}_1,\bsl{\tau}_1} f_{\bsl{\tau}_1 \bsl{\tau}_2,i}(\Delta\bsl{R} + \bsl{\tau}_1 - \bsl{\tau}_2 ) c_{\bsl{k}_2,\bsl{\tau}_2} \\
& \quad ( e^{\ii \bsl{q} \cdot (\Delta\bsl{R}+\bsl{R}_2+\bsl{\tau}_1)} u_{\bsl{q},\bsl{\tau}_1,i} -  e^{\ii \bsl{q} \cdot (\bsl{R}_2+\bsl{\tau}_2)} u_{\bsl{q},\bsl{\tau}_2,i}) \\
& = \frac{1}{\sqrt{N}}\sum_{\bsl{k}_1}^{\BZ} \sum_{\bsl{k}_2}^{\BZ} \sum_{\bsl{q}}^{\BZ} \sum_{\Delta\bsl{R} }\sum_{\bsl{\tau}_1\bsl{\tau}_2 }  \sum_{i} \frac{1}{N}\sum_{ \bsl{R}_2  }  e^{\ii (- \bsl{k}_1 +\bsl{k}_2 + \bsl{q}) \cdot  \bsl{R}_2 }   e^{-\ii \bsl{k}_1 \cdot (\Delta\bsl{R} +\bsl{\tau}_1)}  e^{\ii \bsl{k}_2 \cdot  \bsl{\tau}_2 }  c^\dagger_{\bsl{k}_1,\bsl{\tau}_1}f_{\bsl{\tau}_1 \bsl{\tau}_2,i}(\Delta\bsl{R} + \bsl{\tau}_1 - \bsl{\tau}_2 )c_{\bsl{k}_2,\bsl{\tau}_2} \\
& \quad  ( e^{\ii \bsl{q} \cdot (\Delta\bsl{R}+ \bsl{\tau}_1)} u_{\bsl{q},\bsl{\tau}_1,i} -  e^{\ii \bsl{q} \cdot ( \bsl{\tau}_2)} u_{\bsl{q},\bsl{\tau}_2,i}) \\
& = \frac{1}{\sqrt{N}}\sum_{\bsl{k}_1}^{\BZ} \sum_{\bsl{k}_2}^{\BZ}   \sum_{\Delta\bsl{R} }\sum_{\bsl{\tau}_1\bsl{\tau}_2 }  \sum_{i}    e^{-\ii \bsl{k}_1 \cdot (\Delta\bsl{R} +\bsl{\tau}_1)}  e^{\ii \bsl{k}_2 \cdot  \bsl{\tau}_2 }  c^\dagger_{\bsl{k}_1,\bsl{\tau}_1}f_{\bsl{\tau}_1 \bsl{\tau}_2,i}(\Delta\bsl{R} + \bsl{\tau}_1 - \bsl{\tau}_2 )c_{\bsl{k}_2,\bsl{\tau}_2} \\
& \quad  ( e^{\ii (\bsl{k}_1 - \bsl{k}_2) \cdot (\Delta\bsl{R}+ \bsl{\tau}_1)} u_{\bsl{k}_1 - \bsl{k}_2,\bsl{\tau}_1,i} -  e^{\ii (\bsl{k}_1 - \bsl{k}_2) \cdot  \bsl{\tau}_2 } u_{\bsl{k}_1 - \bsl{k}_2,\bsl{\tau}_2,i}) \\
& = \frac{1}{\sqrt{N}}\sum_{\bsl{k}_1}^{\BZ} \sum_{\bsl{k}_2}^{\BZ}   \sum_{\Delta\bsl{R} }\sum_{\bsl{\tau}_1\bsl{\tau}_2 }  \sum_{i}     c^\dagger_{\bsl{k}_1,\bsl{\tau}_1}f_{\bsl{\tau}_1 \bsl{\tau}_2,i}(\Delta\bsl{R} + \bsl{\tau}_1 - \bsl{\tau}_2 )c_{\bsl{k}_2,\bsl{\tau}_2} \\
& \quad  (    e^{- \ii \bsl{k}_2 \cdot (\Delta\bsl{R}+ \bsl{\tau}_1 - \bsl{\tau}_2)} u_{\bsl{k}_1 - \bsl{k}_2,\bsl{\tau}_1,i} -  e^{-\ii \bsl{k}_1 \cdot (\Delta\bsl{R} +\bsl{\tau}_1-\bsl{\tau}_2)}    u_{\bsl{k}_1 - \bsl{k}_2,\bsl{\tau}_2,i}) \\
& = \frac{1}{\sqrt{N}}\sum_{\bsl{k}_1}^{\BZ} \sum_{\bsl{k}_2}^{\BZ}   \sum_{\bsl{\tau}_1\bsl{\tau}_2 \bsl{\tau}}  \sum_{i}     c^\dagger_{\bsl{k}_1,\bsl{\tau}_1} (     f_{\bsl{\tau}_1 \bsl{\tau}_2, i}(\bsl{k}_2)   \delta_{\bsl{\tau},\bsl{\tau}_1} -    f_{\bsl{\tau}_1\bsl{\tau}_2,i}(\bsl{k}_1)    \delta_{\bsl{\tau},\bsl{\tau}_2}  )  c_{\bsl{k}_2,\bsl{\tau}_2}   u^\dagger_{\bsl{k}_2 - \bsl{k}_1,\bsl{\tau},i} \ ,
}
resulting in 
\eq{
\label{eq:H_el-ph_2center_k}
 H_{el-ph} = \frac{1}{\sqrt{N}}\sum_{\bsl{k}_1}^{\BZ} \sum_{\bsl{k}_2}^{\BZ}  \sum_{ \bsl{\tau} ,  i}     c^\dagger_{\bsl{k}_1}\left[ \chi_{\bsl{\tau}} f_{i}(\bsl{k}_2) - f_{i}(\bsl{k}_1)\chi_{\bsl{\tau}} \right] c_{\bsl{k}_2}   u^\dagger_{\bsl{k}_2 - \bsl{k}_1,\bsl{\tau},i}\ ,
}
where 
\eq{
\label{eq:chi_gen}
\left[ \chi_{\bsl{\tau}} \right]_{\bsl{\tau}_1 \alpha_{\bsl{\tau}_1} ,\bsl{\tau}_2 \alpha_{\bsl{\tau}_{2}}' } = \delta_{ \bsl{\tau}, \bsl{\tau}_{1}} \delta_{\bsl{\tau}_1 \alpha_{\bsl{\tau}_1},  \bsl{\tau}_2 \alpha_{\bsl{\tau}_{2}}' }\ .
}
\eqnref{eq:H_el-ph_2center_k} can be alternatively derived from \eqnref{eq:H_el-ph_k} by substituting 
\eqnref{eq:2center}, \eqnref{eq:g_latt_ini}, \eqnref{eq:g_R} and \eqnref{eq:g_k} into \eqnref{eq:f_k}, which leads to the following expression of $F_{\bsl{\tau}i}(\bsl{k}_1,\bsl{k}_2)$:
\eqa{
\label{eq:f_k_2center}
F_{\bsl{\tau}i}(\bsl{k}_1,\bsl{k}_2) = \chi_{\bsl{\tau}} f_{i}(\bsl{k}_2) - f_{i}(\bsl{k}_1)\chi_{\bsl{\tau}} \ ,
}
where $F_{\bsl{\tau}i}(\bsl{k}_1,\bsl{k}_2)$ is defined in \eqnref{eq:f_k}.

As discussed in \appref{app:EPC_Constant}, the key quantity that we will study is $\left\langle \Gamma \right\rangle $ in \eqnref{eq:Gamma_ave_mu}.
Then, it is useful to see how \eqnref{eq:Gamma_ave_mu} looks under the two-center approximation (or more precisely under \eqnref{eq:f_k_2center}).
To do so, first we substitute \eqnref{eq:f_k_2center} into  \eqnref{eq:Gamma_nm} and obtain the following expression of  $\Gamma_{nm}(\bsl{k}_1,\bsl{k}_2)$:
\eqa{
\label{eq:Gamma_nm_2center_intermedia}
  \Gamma_{nm}(\bsl{k}_1,\bsl{k}_2) & = \frac{\hbar}{2} \sum_{\bsl{\tau},i}  \frac{1}{m_{\bsl{\tau}}} \Tr\left[ P_{n}(\bsl{k}_1) (\chi_{\bsl{\tau}} f_{i}(\bsl{k}_2) - f_{i}(\bsl{k}_1)\chi_{\bsl{\tau}}) P_{m}(\bsl{k}_2)  ( \chi_{\bsl{\tau}} f_{i}(\bsl{k}_1)  - f_{i}(\bsl{k}_2) \chi_{\bsl{\tau}} ) \right] \\
  & = \frac{\hbar}{2} \sum_{\bsl{\tau},i}  \frac{1}{m_{\bsl{\tau}}} \Tr\left[ P_{n}(\bsl{k}_1) \chi_{\bsl{\tau}} f_{i}(\bsl{k}_2)  P_{m}(\bsl{k}_2)   \chi_{\bsl{\tau}} f_{i}(\bsl{k}_1)   \right] 
     -  \frac{\hbar}{2} \sum_{\bsl{\tau},i}  \frac{1}{m_{\bsl{\tau}}} \Tr\left[ P_{n}(\bsl{k}_1) \chi_{\bsl{\tau}} f_{i}(\bsl{k}_2) P_{m}(\bsl{k}_2)  f_{i}(\bsl{k}_2) \chi_{\bsl{\tau}}  \right]\\
  &\quad  -  \frac{\hbar}{2} \sum_{\bsl{\tau},i}  \frac{1}{m_{\bsl{\tau}}} \Tr\left[ P_{n}(\bsl{k}_1)  f_{i}(\bsl{k}_1)\chi_{\bsl{\tau}} P_{m}(\bsl{k}_2)   \chi_{\bsl{\tau}} f_{i}(\bsl{k}_1) \right]
     +  \frac{\hbar}{2} \sum_{\bsl{\tau},i}  \frac{1}{m_{\bsl{\tau}}} \Tr\left[ P_{n}(\bsl{k}_1)  f_{i}(\bsl{k}_1)\chi_{\bsl{\tau}} P_{m}(\bsl{k}_2)   f_{i}(\bsl{k}_2) \chi_{\bsl{\tau}}  \right]\ ,
     }
resulting in
\eqa{
\label{eq:Gamma_nm_2center}
  \Gamma_{nm}(\bsl{k}_1,\bsl{k}_2) 
  & 
  = \frac{\hbar}{2} \sum_{\bsl{\tau},i}  \frac{1}{m_{\bsl{\tau}}} \Tr\left[ \chi_{\bsl{\tau}} f_{i}(\bsl{k}_1)  P_{n}(\bsl{k}_1) \chi_{\bsl{\tau}} f_{i}(\bsl{k}_2)  P_{m}(\bsl{k}_2)     \right]
  +  \frac{\hbar}{2} \sum_{\bsl{\tau},i}  \frac{1}{m_{\bsl{\tau}}} \Tr\left[ P_{n}(\bsl{k}_1)  f_{i}(\bsl{k}_1)\chi_{\bsl{\tau}} P_{m}(\bsl{k}_2)   f_{i}(\bsl{k}_2) \chi_{\bsl{\tau}}  \right] \\
  &\quad  
  -  \frac{\hbar}{2} \sum_{\bsl{\tau},i}  \frac{1}{m_{\bsl{\tau}}} \Tr\left[ f_{i}(\bsl{k}_1) P_{n}(\bsl{k}_1)  f_{i}(\bsl{k}_1)\chi_{\bsl{\tau}} P_{m}(\bsl{k}_2)   \chi_{\bsl{\tau}}  \right]-  \frac{\hbar}{2} \sum_{\bsl{\tau},i}  \frac{1}{m_{\bsl{\tau}}} \Tr\left[  f_{i}(\bsl{k}_2) P_{m}(\bsl{k}_2)  f_{i}(\bsl{k}_2) \chi_{\bsl{\tau}} P_{n}(\bsl{k}_1) \chi_{\bsl{\tau}} \right]     \ ,
 }
 where $P_{n}(\bsl{k})$  is the electron projection matrix defined in \eqnref{eq:P_U}.
Then, $\left\langle \Gamma \right\rangle $ in \eqnref{eq:lambda_omegabar} becomes
\eqa{
\left\langle \Gamma \right\rangle &  =\frac{1}{D^2(\mu)} \sum_{\bsl{k}_1,\bsl{k}_2}^{\BZ}\sum_{n,m} \delta\left(\mu - E_n(\bsl{k}_1) \right) \delta\left(\mu - E_m(\bsl{k}_2) \right) \\
& \quad \times \left\{ \frac{\hbar}{2} \sum_{\bsl{\tau},i}  \frac{1}{m_{\bsl{\tau}}} \Tr\left[ \chi_{\bsl{\tau}} f_{i}(\bsl{k}_1)  P_{n}(\bsl{k}_1) \chi_{\bsl{\tau}} f_{i}(\bsl{k}_2)  P_{m}(\bsl{k}_2)     \right]
  +  \frac{\hbar}{2} \sum_{\bsl{\tau},i}  \frac{1}{m_{\bsl{\tau}}} \Tr\left[ P_{n}(\bsl{k}_1)  f_{i}(\bsl{k}_1)\chi_{\bsl{\tau}} P_{m}(\bsl{k}_2)   f_{i}(\bsl{k}_2) \chi_{\bsl{\tau}}  \right] \right.\\
  & \quad  \left.-  \frac{\hbar}{2} \sum_{\bsl{\tau},i}  \frac{1}{m_{\bsl{\tau}}} \Tr\left[ f_{i}(\bsl{k}_1) P_{n}(\bsl{k}_1)  f_{i}(\bsl{k}_1)\chi_{\bsl{\tau}} P_{m}(\bsl{k}_2)   \chi_{\bsl{\tau}}  \right]-  \frac{\hbar}{2} \sum_{\bsl{\tau},i}  \frac{1}{m_{\bsl{\tau}}} \Tr\left[  f_{i}(\bsl{k}_2) P_{m}(\bsl{k}_2)  f_{i}(\bsl{k}_2) \chi_{\bsl{\tau}} P_{n}(\bsl{k}_1) \chi_{\bsl{\tau}} \right] \right\} \\
  &  =\frac{1}{D^2(\mu)} \sum_{\bsl{k}_1,\bsl{k}_2}^{\BZ}\sum_{n,m} \left\{ \delta\left(\mu - E_n(\bsl{k}_1) \right) \delta\left(\mu - E_m(\bsl{k}_2) \right)  \frac{\hbar}{2} \sum_{\bsl{\tau},i}  \frac{1}{m_{\bsl{\tau}}} \Tr\left[ \chi_{\bsl{\tau}} f_{i}(\bsl{k}_1)  P_{n}(\bsl{k}_1) \chi_{\bsl{\tau}} f_{i}(\bsl{k}_2)  P_{m}(\bsl{k}_2)     \right] +c.c.\right\}\\
  & \quad  - \frac{\hbar}{D^2(\mu)} \sum_{\bsl{k}_1,\bsl{k}_2}^{\BZ}\sum_{n,m} \delta\left(\mu - E_n(\bsl{k}_1) \right) \delta\left(\mu - E_m(\bsl{k}_2) \right)  \Tr\left[ f_{i}(\bsl{k}_1) P_{n}(\bsl{k}_1)  f_{i}(\bsl{k}_1)\chi_{\bsl{\tau}} P_{m}(\bsl{k}_2)   \chi_{\bsl{\tau}}  \right]\ ,
  }
resulting in 
  \eqa{
  \label{eq:Gamma_ave_mu_2center}
  \left\langle \Gamma \right\rangle &   =\left\{ \frac{1}{D^2(\mu)}  \frac{\hbar}{2} \sum_{\bsl{\tau},i}  \frac{1}{m_{\bsl{\tau}}} \Tr\left[ \left(\sum_{\bsl{k}_1}^{\BZ}\sum_{n}  \delta\left(\mu - E_n(\bsl{k}_1) \right) \chi_{\bsl{\tau}} f_{i}(\bsl{k}_1)  P_{n}(\bsl{k}_1) \right)^2    \right] +c.c.\right\}\\
  & \quad  - \frac{\hbar}{D^2(\mu)} \sum_{\bsl{\tau},i}  \frac{1}{m_{\bsl{\tau}}}  \sum_{\bsl{k}_1,\bsl{k}_2}^{\BZ}\sum_{n,m} \delta\left(\mu - E_n(\bsl{k}_1) \right) \delta\left(\mu - E_m(\bsl{k}_2) \right)  \Tr\left[ f_{i}(\bsl{k}_1) P_{n}(\bsl{k}_1)  f_{i}(\bsl{k}_1)\chi_{\bsl{\tau}} P_{m}(\bsl{k}_2)   \chi_{\bsl{\tau}}  \right]\ .
}
For graphene, the second line of the expression gives the geometric and topological lower bounds, as discussed in \appref{app:graphene}.
For {\mgb}, all terms are important, as discussed in \appref{app:MgB2}.

As shown by \eqnref{eq:Gamma_ave_mu_2center}, $\left\langle \Gamma \right\rangle$ is determined by $ f_{i}(\bsl{k})$, in addition to the electron bands $E_n(\bsl{k})$, the electron projection matrix $P_{n}(\bsl{k})$, the chemical potential and the atomic masses.
$E_n(\bsl{k})$ and $P_{n}(\bsl{k})$ can be derived from the electron Hamiltonian \eqnref{eq:H_el_gen_k}.
In the following, we discuss the symmetry properties of $ f_{i}(\bsl{k})$ (and $f_{\bsl{\tau}_1\bsl{\tau}_2,i}(\bsl{R}_1 + \bsl{\tau}_1 - \bsl{R}_2 - \bsl{\tau}_2 )$ in \eqnref{eq:g_R}), which will be crucial for the discussion in \appref{app:geo_EPC_symmetry-rep}.

\subsection{Symmetry Properties of $f_{\bsl{\tau}_1\bsl{\tau}_2,i}(\bsl{R}_1 + \bsl{\tau}_1 - \bsl{R}_2 - \bsl{\tau}_2 )$ and $f_{i}(\bsl{k})$}

In this part, we will discuss the symmetry properties of $f_{\bsl{\tau}_1\bsl{\tau}_2,i}(\bsl{R}_1 + \bsl{\tau}_1 - \bsl{R}_2 - \bsl{\tau}_2 )$ in \eqnref{eq:g_R} and $f_{i}(\bsl{k})$ in \eqnref{eq:g_k}, which will be crucial for the discussion in \appref{app:geo_EPC_symmetry-rep}.

Owing to \eqnref{eq:g_hc_ini} and \eqnref{eq:g_k}, Hermiticity requires and leads to
\eqa{
\label{eq:g_hc}
& f_{\bsl{\tau}_2 \bsl{\tau}_1,i}^\dagger(\bsl{R}_2 + \bsl{\tau}_2 -\bsl{R}_1 - \bsl{\tau}_1) = - f_{\bsl{\tau}_1 \bsl{\tau}_2,i}(\bsl{R}_1 + \bsl{\tau}_1 - \bsl{R}_2 - \bsl{\tau}_2 ) \\
& f_i(\bsl{k})^\dagger = - f_i(\bsl{k})\ ,
}
where $i=x,y,z$ regardless of the spacial dimension $d\leq 3$.
Specifically for $f_i(\bsl{k})$, we have
\eqa{
\left[ f_{i}(\bsl{k}) \right]_{ \bsl{\tau}_2 \alpha_{\bsl{\tau}_2}', \bsl{\tau}_1\alpha_{\bsl{\tau}_1} }^* & = \sum_{\Delta\bsl{R} } e^{ \ii \bsl{k} \cdot (\Delta\bsl{R}+ \bsl{\tau}_2 - \bsl{\tau}_1)}   \left[ f_{\bsl{\tau}_2 \bsl{\tau}_1 ,i}(\Delta\bsl{R} + \bsl{\tau}_2  - \bsl{\tau}_1 ) \right]_{\alpha_{\bsl{\tau}_2}' \alpha_{\bsl{\tau}_1}}^* \\ & = \sum_{\Delta\bsl{R} } e^{- \ii \bsl{k} \cdot (\Delta\bsl{R}+ \bsl{\tau}_1 - \bsl{\tau}_2)}   \left[ f_{\bsl{\tau}_2 \bsl{\tau}_1 ,i}( - \Delta\bsl{R} + \bsl{\tau}_2   - \bsl{\tau}_1 ) \right]_{\alpha_{\bsl{\tau}_2}' \alpha_{\bsl{\tau}_1}}^* \\
& = \sum_{\Delta\bsl{R} } e^{- \ii \bsl{k} \cdot (\Delta\bsl{R}+ \bsl{\tau}_1 - \bsl{\tau}_2)}  
\left[ f^{i}_{- \Delta\bsl{R}+ \bsl{\tau}_2 , \bsl{\tau}_1}\right]_{\alpha_{\bsl{\tau}_2}' \alpha_{\bsl{\tau}_1}}^* \\
& = -\sum_{\Delta\bsl{R} } e^{- \ii \bsl{k} \cdot (\Delta\bsl{R}+ \bsl{\tau}_1 - \bsl{\tau}_2)}  
\left[ f^{i}_{ \bsl{\tau}_1, - \Delta\bsl{R}+ \bsl{\tau}_2}\right]_{ \alpha_{\bsl{\tau}_1} \alpha_{\bsl{\tau}_2}'} \\
& = -\sum_{\Delta\bsl{R} } e^{- \ii \bsl{k} \cdot (\Delta\bsl{R}+ \bsl{\tau}_1 - \bsl{\tau}_2)}  
\left[ f^{i}_{\Delta\bsl{R}+  \bsl{\tau}_1,  \bsl{\tau}_2}\right]_{ \alpha_{\bsl{\tau}_1} \alpha_{\bsl{\tau}_2}'} \\
& = -\sum_{\Delta\bsl{R} } e^{- \ii \bsl{k} \cdot (\Delta\bsl{R}+ \bsl{\tau}_1 - \bsl{\tau}_2)}  
\left[ f_{\bsl{\tau}_1 \bsl{\tau}_2,i}(\Delta\bsl{R}+ \bsl{\tau}_1 - \bsl{\tau}_2 )\right]_{ \alpha_{\bsl{\tau}_1} \alpha_{\bsl{\tau}_2}'} \\
& = - \left[ f_{i}(\bsl{k}) \right]_{\bsl{\tau}_1\alpha_{\bsl{\tau}_1}, \bsl{\tau}_2 \alpha_{\bsl{\tau}_2}'}\ ,
}
where we used \eqnref{eq:g_R} and \eqnref{eq:g_hc_ini}.

Suppose the system has a crystalline symmetry $g=\{ R | \bsl{d} \}$, where $R$ is the point group part of $g$ and $\bsl{d}$ labels the translational part of $g$.
According to \eqnref{eq:F_f_sym_g} and \eqnref{eq:g_R_ini}, we have
\eqa{
&  \sum_{ i'} U^{\bsl{\tau}_{1,g}\bsl{\tau}_1}_g f^{i'}_{\bsl{R}_1 + \bsl{\tau}_1 , \bsl{R}_2 + \bsl{\tau}_2}   \left[U^{\bsl{\tau}_{2,g}\bsl{\tau}_2}_g\right]^\dagger R_{ii'}\\
&= \frac{1}{2} \sum_{\alpha_{\bsl{\tau}_1}\alpha_{\bsl{\tau}_2}' i'} U^{\bsl{\tau}_{1,g}\bsl{\tau}_1}_g F^{ i'}_{\bsl{R}_1 + \bsl{\tau}_1 , \bsl{R}_2 + \bsl{\tau}_2,  \bsl{R}_1 + \bsl{\tau}_1} \left[U^{\bsl{\tau}_{2,g}\bsl{\tau}_2}_g\right]^\dagger R_{ii'}  -\frac{1}{2} \sum_{ i'} U^{\bsl{\tau}_{1,g}\bsl{\tau}_1}_g F^{i'}_{\bsl{R}_1 + \bsl{\tau}_1 , \bsl{R}_2 + \bsl{\tau}_2,  \bsl{R}_2 + \bsl{\tau}_2}  \left[U^{\bsl{\tau}_{2,g}\bsl{\tau}_2}_g\right]^\dagger R_{ii'} \\
& = \frac{1}{2}  F^{i}_{\bsl{R}_{1,\bsl{\tau}_1,g} + \bsl{\tau}_{1,g} , \bsl{R}_{2,\bsl{\tau}_2,g} + \bsl{\tau}_{2,g},  \bsl{R}_{1,\bsl{\tau}_1,g} + \bsl{\tau}_{1,g} } -  \frac{1}{2}  F^{ i}_{\bsl{R}_{1,\bsl{\tau}_1,g} + \bsl{\tau}_{1,g} , \bsl{R}_{2,\bsl{\tau}_2,g} + \bsl{\tau}_{2,g},  \bsl{R}_{2,\bsl{\tau}_2,g} + \bsl{\tau}_{2,g} } \\
& = f^i_{\bsl{R}_{1,\bsl{\tau}_1,g} + \bsl{\tau}_{1,g} , \bsl{R}_{2,\bsl{\tau}_2,g} + \bsl{\tau}_{2,g}}\ ,
}
which, combined with \eqnref{eq:g_R} and \eqnref{eq:g_latt_ini}, leads to
\eqa{
\label{eq:g_R_sym_g}
& \sum_{i'}  U^{\bsl{\tau}_{1,g}\bsl{\tau}_1}_g f_{\bsl{\tau}_1\bsl{\tau}_2,i'}(\bsl{R}_1 + \bsl{\tau}_1 - \bsl{R}_2 - \bsl{\tau}_2 ) \left[U^{\bsl{\tau}_{2,g}\bsl{\tau}_2}_g\right]^\dagger R_{ii'}  =   f_{\bsl{\tau}_{1,g}\bsl{\tau}_{2,g}, i}(\bsl{R}_{1,\bsl{\tau}_1,g} + \bsl{\tau}_{1,g} - \bsl{R}_{2,\bsl{\tau}_2,g} - \bsl{\tau}_{2,g})  \ ,
}
where $\bsl{R}_{\bsl{\tau},g}$ and $\bsl{\tau}_{g}$ are defined in \eqnref{eq:R_g_tau_g}, and 
\eq{
\bsl{R}_{1,\bsl{\tau}_1,g} + \bsl{\tau}_{1,g} - \bsl{R}_{2,\bsl{\tau}_2,g} - \bsl{\tau}_{2,g} = R(\bsl{R}_1 + \bsl{\tau}_1 - \bsl{R}_2 - \bsl{\tau}_2)
}
owing to \eqnref{eq:R_g_tau_g}.
Then, combining \eqnref{eq:g_R_sym_g} with \eqnref{eq:g_k_block} and \eqnref{eq:U_g}, we obtain
\eqa{
& \sum_{\bsl{\tau}_1 \bsl{\tau}_2 i'} U_g^{\bsl{\tau}_1'\bsl{\tau}_1} f_{\bsl{\tau}_1\bsl{\tau}_2,i'}(\bsl{k}) \left[U_g^{\bsl{\tau}_2'\bsl{\tau}_2}\right]^\dagger R_{ii'}\\
& =\sum_{\bsl{\tau}_1\bsl{\tau}_2} \delta_{\bsl{\tau}_1', \bsl{\tau}_{1,g}}  \delta_{\bsl{\tau}_2', \bsl{\tau}_{2,g}} \sum_{ i'}U_g^{\bsl{\tau}_{1,g}\bsl{\tau}_1} f_{\bsl{\tau}_1\bsl{\tau}_2,i'}(\bsl{k}) \left[U_g^{\bsl{\tau}_{2,g}\bsl{\tau}_2}\right]^\dagger R_{ii'}\\
& = \sum_{\bsl{\tau}_1\bsl{\tau}_2}  \delta_{\bsl{\tau}_1', \bsl{\tau}_{1,g}}  \delta_{\bsl{\tau}_2', \bsl{\tau}_{2,g}} \sum_{\Delta\bsl{R} } e^{- \ii \bsl{k} \cdot (\Delta\bsl{R}+ \bsl{\tau}_1 - \bsl{\tau}_2)} \sum_{i'}  U_g^{\bsl{\tau}_{1,g}\bsl{\tau}_1} f_{\bsl{\tau}_1 \bsl{\tau}_2,i'}(\Delta\bsl{R} + \bsl{\tau}_1  - \bsl{\tau}_2 )  \left[U_g^{\bsl{\tau}_{2,g}\bsl{\tau}_2}\right]^\dagger
R_{ii'}\\
& = \sum_{\bsl{\tau}_1\bsl{\tau}_2}  \delta_{\bsl{\tau}_1', \bsl{\tau}_{1,g}}  \delta_{\bsl{\tau}_2', \bsl{\tau}_{2,g}} \sum_{\Delta\bsl{R} } e^{- \ii R \bsl{k} \cdot R(\Delta\bsl{R}+ \bsl{\tau}_1 - \bsl{\tau}_2)} 
  f_{\bsl{\tau}_{1,g} \bsl{\tau}_{2,g},i}(R\Delta \bsl{R}+\Delta\bsl{R}_{\bsl{\tau}_{1},g}-\Delta\bsl{R}_{\bsl{\tau}_{2},g} + \bsl{\tau}_{1,g} - \bsl{\tau}_{2,g})   \\
& = \sum_{\bsl{\tau}_1\bsl{\tau}_2}  \delta_{\bsl{\tau}_1', \bsl{\tau}_{1,g}}  \delta_{\bsl{\tau}_2', \bsl{\tau}_{2,g}} \sum_{ \Delta \bsl{R} } e^{- \ii R \bsl{k} \cdot (R\Delta \bsl{R}+\Delta\bsl{R}_{\bsl{\tau}_{1},g}-\Delta\bsl{R}_{\bsl{\tau}_{2},g} + \bsl{\tau}_{1,g} - \bsl{\tau}_{2,g})} 
 f_{\bsl{\tau}_{1,g} \bsl{\tau}_{2,g} ,i}(R\Delta \bsl{R}+\Delta\bsl{R}_{\bsl{\tau}_{1},g}-\Delta\bsl{R}_{\bsl{\tau}_{2},g} + \bsl{\tau}_{1,g} - \bsl{\tau}_{2,g})  \\
 & = \sum_{\bsl{\tau}_1\bsl{\tau}_2}  \delta_{\bsl{\tau}_1', \bsl{\tau}_{1,g}}  \delta_{\bsl{\tau}_2', \bsl{\tau}_{2,g}} \sum_{ \Delta \bsl{R} } e^{- \ii R \bsl{k} \cdot ( \Delta \bsl{R} + \bsl{\tau}_{1,g} - \bsl{\tau}_{2,g})} 
 f_{\bsl{\tau}_{1,g} \bsl{\tau}_{2,g} ,i}( \Delta \bsl{R} + \bsl{\tau}_{1,g} - \bsl{\tau}_{2,g})  \\
& = \sum_{\bsl{\tau}_1\bsl{\tau}_2}  \delta_{\bsl{\tau}_1', \bsl{\tau}_{1,g}}  \delta_{\bsl{\tau}_2', \bsl{\tau}_{2,g}}  f_{\bsl{\tau}_{1,g} \bsl{\tau}_{2,g}, i}(R\bsl{k})  \\
& = f_{\bsl{\tau}_1' \bsl{\tau}_2', i}(R \bsl{k}) \text{  for any values of $\bsl{\tau}_1'$ and $\bsl{\tau}_2'$}\ ,
}
resulting in 
\eq{
\label{eq:g_k_sym_g}
\sum_{i'}  U_g f_{i'}(\bsl{k}) U_g^\dagger  R_{ii'} = f_{i}(R \bsl{k})\ ,
}
where $\bsl{\tau}_{1,g} - \bsl{\tau}_{2,g}= R(\bsl{\tau}_{1} - \bsl{\tau}_{2}) - (\Delta\bsl{R}_{\bsl{\tau}_{1},g}-\Delta\bsl{R}_{\bsl{\tau}_{2},g})$ according to \eqnref{eq:R_g_tau_g}, and we used $\bsl{\tau}_1'$ and $\bsl{\tau}_2'$ because we include the case where $\bsl{\tau}_1'\neq \bsl{\tau}_{1,g}$ and $\bsl{\tau}_2'\neq \bsl{\tau}_{2,g}$, for which the equation trivially holds as both sides are zero.

Suppose the system has TR symmetry $\TR$.
According to \eqnref{eq:F_f_sym_TR} and \eqnref{eq:g_R_ini}, we have
\eqa{
&  U^{\bsl{\tau}_1 \bsl{\tau}_1}_{\TR} \left[f^{i}_{\bsl{R}_1 + \bsl{\tau}_1 , \bsl{R}_2 + \bsl{\tau}_2} \right]^* \  \left[U^{\bsl{\tau}_2 \bsl{\tau}_2}_{\TR}\right]^\dagger\\
&= \frac{1}{2}   U^{\bsl{\tau}_1 \bsl{\tau}_1}_{\TR} \left[F^{i}_{\bsl{R}_1 + \bsl{\tau}_1 , \bsl{R}_2 + \bsl{\tau}_2,  \bsl{R}_1 + \bsl{\tau}_1}\right]^*  \left[U^{\bsl{\tau}_2 \bsl{\tau}_2}_{\TR}\right]^\dagger -\frac{1}{2}  U^{\bsl{\tau}_1 \bsl{\tau}_1}_{\TR} \left[F^{i}_{\bsl{R}_1 + \bsl{\tau}_1 , \bsl{R}_2 + \bsl{\tau}_2,  \bsl{R}_2 + \bsl{\tau}_2}\right]^*  \left[U^{\bsl{\tau}_2 \bsl{\tau}_2}_{\TR}\right]^\dagger \\
& = \frac{1}{2}   F^{i}_{\bsl{R}_{1} + \bsl{\tau}_{1} , \bsl{R}_{2} + \bsl{\tau}_{2},  \bsl{R}_1 + \bsl{\tau}_1} -  \frac{1}{2}   F^{i}_{\bsl{R}_{1} + \bsl{\tau}_{1} , \bsl{R}_{2} + \bsl{\tau}_{2},  \bsl{R}_2 + \bsl{\tau}_2} \\
& = f^{i}_{\bsl{R}_1 + \bsl{\tau}_1 , \bsl{R}_2 + \bsl{\tau}_2}\ ,
}
which, combined with the lattice translations (\eqnref{eq:g_latt_ini}) and the definition of $f_{\bsl{\tau}_1 \bsl{\tau}_2,i}(\bsl{R}_1 + \bsl{\tau}_1 - \bsl{R}_2 - \bsl{\tau}_2 ) $ in  \eqnref{eq:g_R}, leads to
\eqa{
\label{eq:g_R_sym_TR}
& U^{\bsl{\tau}_1 \bsl{\tau}_1}_{\TR}\left[ f_{\bsl{\tau}_1 \bsl{\tau}_2,i}(\bsl{R}_1 + \bsl{\tau}_1 - \bsl{R}_2 - \bsl{\tau}_2 ) \right]^* \left[U^{\bsl{\tau}_2 \bsl{\tau}_2}_{\TR}\right]^\dagger =  f_{\bsl{\tau}_1 \bsl{\tau}_2,i}(\bsl{R}_1 + \bsl{\tau}_1 - \bsl{R}_2 - \bsl{\tau}_2 ) \ .
}
Then, combining \eqnref{eq:g_R_sym_TR} with \eqnref{eq:g_k_block} and \eqnref{eq:U_TR}, we obtain
\eqa{
& \sum_{\bsl{\tau}_1 \bsl{\tau}_2 } U_{\TR}^{\bsl{\tau}_1'\bsl{\tau}_1} f_{\bsl{\tau}_1\bsl{\tau}_2,i}^*(\bsl{k}) \left[U_{\TR}^{\bsl{\tau}_2'\bsl{\tau}_2}\right]^\dagger \\
& =\sum_{\bsl{\tau}_1\bsl{\tau}_2} \delta_{\bsl{\tau}_1', \bsl{\tau}_{1}}  \delta_{\bsl{\tau}_2', \bsl{\tau}_{2}} U_{\TR}^{\bsl{\tau}_1\bsl{\tau}_1} f_{\bsl{\tau}_1\bsl{\tau}_2,i}^*(\bsl{k}) \left[U_{\TR}^{\bsl{\tau}_2\bsl{\tau}_2}\right]^\dagger  \\
& = \sum_{\bsl{\tau}_1\bsl{\tau}_2}  \delta_{\bsl{\tau}_1', \bsl{\tau}_{1}}  \delta_{\bsl{\tau}_2', \bsl{\tau}_{2}} \sum_{\Delta\bsl{R} } e^{ \ii \bsl{k} \cdot (\Delta\bsl{R}+ \bsl{\tau}_1 - \bsl{\tau}_2)} U_{\TR}^{\bsl{\tau}_1\bsl{\tau}_1} f_{\bsl{\tau}_1\bsl{\tau}_2,i}^*(\Delta\bsl{R} + \bsl{\tau}_1 - \bsl{\tau}_2 ) \left[U_{\TR}^{\bsl{\tau}_2\bsl{\tau}_2}\right]^\dagger \\
& = \sum_{\bsl{\tau}_1\bsl{\tau}_2}  \delta_{\bsl{\tau}_1', \bsl{\tau}_{1}}  \delta_{\bsl{\tau}_2', \bsl{\tau}_{2}} \sum_{\Delta\bsl{R} } e^{ \ii \bsl{k} \cdot (\Delta\bsl{R}+ \bsl{\tau}_1 - \bsl{\tau}_2)} f_{\bsl{\tau}_1\bsl{\tau}_2,i}(\Delta\bsl{R} + \bsl{\tau}_1 - \bsl{\tau}_2 ) \\
& = \sum_{\bsl{\tau}_1\bsl{\tau}_2}  \delta_{\bsl{\tau}_1', \bsl{\tau}_{1}}  \delta_{\bsl{\tau}_2', \bsl{\tau}_{2}}   f_{\bsl{\tau}_1\bsl{\tau}_2,i}(-\bsl{k} )  \\
& =   f_{\bsl{\tau}_1'\bsl{\tau}_2',i}(-\bsl{k} ) \ ,
}
resulting in 
\eq{
\label{eq:g_k_sym_TR}
U_{\TR} f_{i}^*(\bsl{k}) U_{\TR}^\dagger = f_{i}(-\bsl{k} )\ .
}

The symmetry properties of $f_{\bsl{\tau}_1\bsl{\tau}_2,i}(\bsl{R}_1 + \bsl{\tau}_1 - \bsl{R}_2 - \bsl{\tau}_2 )$ are summarized in 
\eqa{
\label{eq:g_R_sym}
& f_{\bsl{\tau}_2 \bsl{\tau}_1,i}^\dagger(\bsl{R}_2 + \bsl{\tau}_2 -\bsl{R}_1 - \bsl{\tau}_1) = - f_{\bsl{\tau}_1 \bsl{\tau}_2,i}(\bsl{R}_1 + \bsl{\tau}_1 - \bsl{R}_2 - \bsl{\tau}_2 ) \text{ for Hermiticity}  \\
& \sum_{i'}  U^{\bsl{\tau}_{1,g}\bsl{\tau}_1}_g f_{\bsl{\tau}_1\bsl{\tau}_2,i'}(\bsl{R}_1 + \bsl{\tau}_1 - \bsl{R}_2 - \bsl{\tau}_2 ) \left[U^{\bsl{\tau}_{2,g}\bsl{\tau}_2}_g\right]^\dagger R_{ii'}  =   f_{\bsl{\tau}_{1,g}\bsl{\tau}_{2,g}, i}(\bsl{R}_{1,\bsl{\tau}_1,g} + \bsl{\tau}_{1,g} - \bsl{R}_{2,\bsl{\tau}_2,g} - \bsl{\tau}_{2,g}) \text{ for $g=\{ R | \bsl{d} \}$}\\
& U^{\bsl{\tau}_1 \bsl{\tau}_1}_{\TR}\left[ f_{\bsl{\tau}_1 \bsl{\tau}_2,i}(\bsl{R}_1 + \bsl{\tau}_1 - \bsl{R}_2 - \bsl{\tau}_2 ) \right]^* \left[U^{\bsl{\tau}_2 \bsl{\tau}_2}_{\TR}\right]^\dagger =  f_{\bsl{\tau}_1 \bsl{\tau}_2,i}(\bsl{R}_1 + \bsl{\tau}_1 - \bsl{R}_2 - \bsl{\tau}_2 ) \text{ for  $\TR$} \ ,
}
where $U^{\bsl{\tau}_{g}\bsl{\tau}}_g$ is defined in \eqnref{eq:sym_rep_g_R}, $U^{\bsl{\tau} \bsl{\tau}}_{\TR}$ is defined in \eqnref{eq:sym_rep_TR_R}, $\bsl{R}_{\bsl{\tau},g}$ and $\bsl{\tau}_{g}$ are defined in \eqnref{eq:R_g_tau_g}, and 
\eq{
\bsl{R}_{1,\bsl{\tau}_1,g} + \bsl{\tau}_{1,g} - \bsl{R}_{2,\bsl{\tau}_2,g} - \bsl{\tau}_{2,g} = R(\bsl{R}_1 + \bsl{\tau}_1 - \bsl{R}_2 - \bsl{\tau}_2)
}
owing to \eqnref{eq:R_g_tau_g}.
The symmetry properties of $f_{i}(\bsl{k})$ are summarized in 
\eqb{
\label{eq:g_k_sym}
& f_i(\bsl{k})^\dagger = - f_i(\bsl{k}) \text{ for Hermiticity}  \\
& \sum_{i'}  U_g f_{i'}(\bsl{k}) U_g^\dagger  R_{ii'} = f_{i}(R \bsl{k}) \text{ for $g=\{ R | \bsl{d} \}$}\\
& U_{\TR} f_{i}^*(\bsl{k}) U_{\TR}^\dagger = f_{i}(-\bsl{k} )  \text{ for  $\TR$} \ ,
}
where $U_g$ is defined in \eqnref{eq:U_g} and $U_\TR$ is defined in \eqnref{eq:U_TR}.

Finally, from \eqnref{eq:g_k}, we can clearly see that for any generic reciprocal lattice vector $\bsl{G}$,
\eq{
\label{eq:g_G}
f_{i}(\bsl{k}+\bsl{G}) = V_{\bsl{G}}^\dagger f_{i}(\bsl{k}) V_{\bsl{G}}\ ,
}
where $V_{\bsl{G}}$ is the embedding matrix defined in \eqnref{eq:V_G}.
This means that $f_{i}(\bsl{k})$ has the same transformation rule as the matrix Hamiltonian $h(\bsl{k})$ of the electrons in \eqnref{eq:h_k} under the shifts of the reciprocal lattice vectors.
We emphasize that the appearance of only two embedding matrices of \eqnref{eq:g_G} relies on the two-center approximation of the EPC Hamiltonian (\eqnref{eq:H_el-ph_2center_R}).

\section{Symmetry-Rep Method: Geometric Contribution to the EPC Constant $\lambda$}

\label{app:geo_EPC_symmetry-rep}

In \appref{app:gaussian}, we have shown that under the GA, we can identify the geometric contribution to $\lambda$. 
In this section, we will discuss an alternative way of defining the geometric contribution to $\lambda$, which we call the symmetry-rep method.
We emphasize that unlike the GA, the symmetry-rep method is just a formal way of defining the geometric contribution.
The symmetry-rep method repeatedly uses the fact that two functions of the same symmetry properties can be expressed in terms of the same combinations of irreps.
As a result, the symmetry-rep method would need the explicit function form of the electron Hamiltonian in terms of not only the momentum but also the hopping parameters as discussed in \appref{app:symmetry-rep_detials}.
In practice, when we use the symmetry-rep method, we always need to combine it with the short-range nature of the hoppings, since the symmetry-rep method does not involve any information of locality; when combined with short-range nature of the hoppings, the number of basis functions would be dramatically reduced, allowing us to obtain explicit results.
Then, in graphene and {\mgb}, we find that the symmetry-rep method combined with the short-range hoppings give exactly the same results as the GA, as discussed in \appref{app:graphene} and \appref{app:MgB2}.

Before going into details, let us discuss more generally what the geometric contribution to $\lambda$ means.
First, the geometric part of EPC means the part of $f_{i}(\bsl{k})$ in \eqnref{eq:H_el-ph_2center_k} that relies on the geometric properties of the electron Bloch states.
Here the geometric properties of the electron Bloch states refer to the momentum dependence of the projection operator $\hat{P}_{n}(\bsl{k}) = \ket{u_{n,\bsl{k}}}\bra{u_{n,\bsl{k}}}$ on the periodic part of the Bloch state $\ket{u_{n,\bsl{k}}}$ (which is defined in \eqnref{eq:u_ket}).
More explicitly, the momentum derivatives of $\hat{P}_{n}(\bsl{k})$ characterize the momentum dependence of $\hat{P}_{n}(\bsl{k})$.
Owing to the tight-binding approximation that we use, \eqnref{eq:u_ket_U} suggests that 
the momentum dependence of $\hat{P}_{n}(\bsl{k})$ is all in the projection matrix $P_{n}(\bsl{k})$ defined in \eqnref{eq:P_U} for the convention \eqnref{eq:FT_rule}.
Therefore, in this work, the geometric part of EPC specifically means the part of $f_{i}(\bsl{k})$ in \eqnref{eq:H_el-ph_2center_k} that relies on the momentum derivatives of the projection matrix $P_{n}(\bsl{k})$ defined in \eqnref{eq:P_U} for the convention \eqnref{eq:FT_rule}.
Furthermore, we can also define the energetic part of $f_{i}(\bsl{k})$ in \eqnref{eq:H_el-ph_2center_k}, which is just the part of $f_{i}(\bsl{k})$ in \eqnref{eq:H_el-ph_2center_k} that relies on the momentum derivatives of the electron bands $E_{n}(\bsl{k})$ \eqnref{eq:el_eigen} but not on the momentum derivatives of $P_{n}(\bsl{k})$.
The energetic (geometric) contribution to $\lambda$ is the part of $\lambda$ that solely depends on the energetic (geometric) part of $f_{i}(\bsl{k})$.

In the following, we will present more details on the symmetry-rep method.
We will only discuss certain 2D systems.
It is nontrivial to define the geometric part of EPC in generic dimensions based on the symmetry-rep method, which we leave to the future.
Yet, the physical relation between EPC and the quantum geometry has been revealed by the GA.

\subsection{Momentum Derivatives}
\label{app:geo_EPC_dk}

As discussed in the above introductory part of this section, the final expressions of the geometric contribution to $\lambda$ should contain momentum derivatives of $P_{n}(\bsl{k})$.
In the GA discussed in \appref{app:gaussian}, the momentum derivatives naturally appear because of the Gaussian form of the hopping function.
In this part, we will show where the momentum derivatives originate from in the symmetry-rep method.
We note that unlike the GA discussed in \appref{app:gaussian}, the discussion in this part is just a re-formulation of $f_{i}(\bsl{k})$ in \eqnref{eq:H_el-ph_2center_k}.

First, we should start from real space by using \eqnref{eq:g_k_block}.
Owing to \eqnref{eq:f_zero_onsite}, \eqnref{eq:g_k_block} becomes
\eqa{
  f_{\bsl{\tau}_1 \bsl{\tau}_2,i}(\bsl{k})  
  & = \sum_{\Delta\bsl{R} } e^{- \ii \bsl{k} \cdot (\Delta\bsl{R}+ \bsl{\tau}_1 - \bsl{\tau}_2)}   f_{\bsl{\tau}_1 \bsl{\tau}_2,i}(\Delta\bsl{R} + \bsl{\tau}_1 - \bsl{\tau}_2 )    \\
& = \sum_{\Delta\bsl{R} }^{\Delta\bsl{R} + \bsl{\tau}_1 - \bsl{\tau}_2  \neq 0} e^{- \ii \bsl{k} \cdot (\Delta\bsl{R}+ \bsl{\tau}_1 - \bsl{\tau}_2)}   f_{\bsl{\tau}_1 \bsl{\tau}_2,i}(\Delta\bsl{R} + \bsl{\tau}_1 - \bsl{\tau}_2 )  \ .
}
\eqnref{eq:f_zero_onsite} shows that the terms for $\Delta\bsl{R} + \bsl{\tau}_1 - \bsl{\tau}_2  = 0$ vanish and thus can be safely omitted.
For simplicity, we only consider the 2D systems that do not have two atoms with the same in-plane coordinates, \ie, $\bsl{\tau}_1=\bsl{\tau}_2 \Leftrightarrow |\bsl{\tau}_1-\bsl{\tau}_2|_{\shpa} = 0$ with $|(x,y,z)|_\shpa=\sqrt{x^2+y^2}$.
Equivalently, we have
\eq{
\label{eq:2D_asm}
\bsl{R}_1 + \bsl{\tau}_1 = \bsl{R}_2 + \bsl{\tau}_2 \Leftrightarrow |\bsl{R}_1 + \bsl{\tau}_1 - \bsl{R}_2 - \bsl{\tau}_2|_{\shpa} = 0\ ,
}
since $\bsl{R}_1$ and $\bsl{R}_2$ only have the in-plane components.

To proceed, we will use the following expression
\eq{
\label{eq:delta_split}
\delta_{ii'} = \left(\frac{\bsl{r}}{|\bsl{r}|_\shpa}\right)_i \left(\frac{\bsl{r}}{|\bsl{r}|_\shpa}\right)_{i'} + \left(\frac{\bsl{e}_z\times \bsl{r}}{|\bsl{r}|_\shpa}\right)_i \left(\frac{\bsl{e}_z\times \bsl{r}}{|\bsl{r}|_\shpa}\right)_{i'} \ \forall\ i,i' \in \{ x,y \} \text{ and } |\bsl{r}|_\shpa\neq 0\ ,
}
where $\bsl{r}=(r_x,r_y,r_z)^T$, and $\bsl{e}_{z}$ is the unit vector in the $z$ direction.
Combined with \eqnref{eq:2D_asm}, we can rewrite $f_{\bsl{\tau}_1 \bsl{\tau}_2,i}(\bsl{k})$ with $i=x,y$ as 
\eqa{
\label{eq:g_2D_partial_k}
&   f_{\bsl{\tau}_1 \bsl{\tau}_2,i=x,y}(\bsl{k}) \\
& = \sum_{\Delta\bsl{R} }^{\Delta\bsl{R} + \bsl{\tau}_1 - \bsl{\tau}_2  \neq 0} e^{- \ii \bsl{k} \cdot (\Delta\bsl{R}+ \bsl{\tau}_1 - \bsl{\tau}_2)}  \sum_{i'=x,y}   f_{\bsl{\tau}_1 \bsl{\tau}_2, i'}(\Delta\bsl{R} + \bsl{\tau}_1 - \bsl{\tau}_2 )   \delta_{ii'}  \\
& = \sum_{\Delta\bsl{R} }^{\Delta\bsl{R} + \bsl{\tau}_1 - \bsl{\tau}_2  \neq 0} e^{- \ii \bsl{k} \cdot (\Delta\bsl{R}+ \bsl{\tau}_1 - \bsl{\tau}_2)}  \sum_{i'=x,y}   f_{\bsl{\tau}_1 \bsl{\tau}_2,i'}(\Delta\bsl{R} + \bsl{\tau}_1 - \bsl{\tau}_2 )  \\
& \quad \times  \left[ \left(\frac{\Delta\bsl{R} + \bsl{\tau}_1  - \bsl{\tau}_2}{|\Delta\bsl{R} + \bsl{\tau}_1  - \bsl{\tau}_2|_{\shpa}}\right)_i \left(\frac{\Delta\bsl{R} + \bsl{\tau}_1  - \bsl{\tau}_2}{|\Delta\bsl{R} + \bsl{\tau}_1  - \bsl{\tau}_2|_{\shpa}}\right)_{i'}  + \left(\frac{\bsl{e}_z\times (\Delta\bsl{R} + \bsl{\tau}_1  - \bsl{\tau}_2)}{|\Delta\bsl{R} + \bsl{\tau}_1  - \bsl{\tau}_2|_\shpa}\right)_i \left(\frac{\bsl{e}_z\times (\Delta\bsl{R} + \bsl{\tau}_1  - \bsl{\tau}_2)}{|\Delta\bsl{R} + \bsl{\tau}_1  - \bsl{\tau}_2|_\shpa}\right)_{i'} \right] \\
& = \sum_{\Delta\bsl{R} }^{\Delta\bsl{R} + \bsl{\tau}_1 - \bsl{\tau}_2  \neq 0} e^{- \ii \bsl{k} \cdot (\Delta\bsl{R}+ \bsl{\tau}_1 - \bsl{\tau}_2)}   \left(  \widetilde{f}_{\bsl{\tau}_1\bsl{\tau}_2,\shpa}(\Delta\bsl{R} + \bsl{\tau}_1 - \bsl{\tau}_2 )   \left[ \Delta\bsl{R} + \bsl{\tau}_1  - \bsl{\tau}_2 \right]_i \right. \\
& \quad +\left.    \widetilde{f}_{\bsl{\tau}_1 \bsl{\tau}_2,\perp}(\Delta\bsl{R} + \bsl{\tau}_1 - \bsl{\tau}_2 )   \left[\bsl{e}_z\times (\Delta\bsl{R} + \bsl{\tau}_1  - \bsl{\tau}_2)\right]_i \right) \\
& = \sum_{\Delta\bsl{R} }  e^{- \ii \bsl{k} \cdot (\Delta\bsl{R}+ \bsl{\tau}_1 - \bsl{\tau}_2)}   \left(   \widetilde{f}_{\bsl{\tau}_1\bsl{\tau}_2,\shpa}(\Delta\bsl{R} + \bsl{\tau}_1 - \bsl{\tau}_2 )   \left[ \Delta\bsl{R} + \bsl{\tau}_1  - \bsl{\tau}_2 \right]_i \right. \\
& \quad +\left.    \widetilde{f}_{\bsl{\tau}_1 \bsl{\tau}_2,\perp}(\Delta\bsl{R} + \bsl{\tau}_1 - \bsl{\tau}_2 )   \left[\bsl{e}_z\times (\Delta\bsl{R} + \bsl{\tau}_1  - \bsl{\tau}_2)\right]_i \right)
}
where we use \eqnref{eq:2D_asm} for the second equality, we define
\eqa{
\label{eq:g_beta_R}
&    \widetilde{f}_{\bsl{\tau}_1\bsl{\tau}_2,\shpa}(\Delta\bsl{R} + \bsl{\tau}_1 - \bsl{\tau}_2\neq 0 )  = 
\sum_{i'=x,y}   f_{\bsl{\tau}_1 \bsl{\tau}_2, i'}(\Delta\bsl{R} + \bsl{\tau}_1 - \bsl{\tau}_2 )  \left(\frac{\Delta\bsl{R} + \bsl{\tau}_1  - \bsl{\tau}_2}{|\Delta\bsl{R} + \bsl{\tau}_1  - \bsl{\tau}_2|_{\shpa}^2}\right)_{i'}  \\
&    \widetilde{f}_{\bsl{\tau}_1 \bsl{\tau}_2,\perp}(\Delta\bsl{R} + \bsl{\tau}_1 - \bsl{\tau}_2 \neq 0)   = 
 \sum_{i'=x,y} f_{\bsl{\tau}_1 \bsl{\tau}_2,i'}(\Delta\bsl{R} + \bsl{\tau}_1 - \bsl{\tau}_2 ) \left(\frac{\bsl{e}_z\times (\Delta\bsl{R} + \bsl{\tau}_1  - \bsl{\tau}_2)}{|\Delta\bsl{R} + \bsl{\tau}_1  - \bsl{\tau}_2|_{\shpa}^2}\right)_{i'} 
}
for the third equality, and we define 
\eq{
\label{eq:g_beta_onsite_zero}
 \widetilde{f}_{\bsl{\tau} \bsl{\tau},\shpa}(0)   =  \widetilde{f}_{\bsl{\tau} \bsl{\tau},\perp}(0)   = 0
}for the last equality.
By further defining 
\eq{
\label{eq:g_beta_k_block}
  \widetilde{f}_{\bsl{\tau}_1 \bsl{\tau}_2,\beta}( \bsl{k} )  =  \sum_{\Delta\bsl{R} }  e^{- \ii \bsl{k} \cdot (\Delta\bsl{R}+ \bsl{\tau}_1 - \bsl{\tau}_2)}  \widetilde{f}_{\bsl{\tau}_1 \bsl{\tau}_2,\beta}(\Delta\bsl{R} + \bsl{\tau}_1 - \bsl{\tau}_2 ) \ \forall\ \beta = \shpa,\perp\ ,
} 
we eventually have
\eq{
\label{eq:g_i_g_beta}
 f_{i}(\bsl{k})  = \left[ \ii \partial_{k_i}  \widetilde{f}_{\shpa}( \bsl{k} )  + \ii \sum_{i'=x,y} \epsilon_{i'i}\partial_{k_{i'}}    \widetilde{f}_{\perp}(\bsl{k} ) \right](\delta_{ix} + \delta_{iy}) +\delta_{iz} f_{z}(\bsl{k})\ ,
}
where
\eq{
\label{eq:g_beta_k}
 \left[  \widetilde{f}_{\beta}( \bsl{k} ) \right]_{\bsl{\tau}_1\alpha_{\bsl{\tau}_1} ,\bsl{\tau}_2\alpha_{\bsl{\tau}_2}'} = \left[   \widetilde{f}_{\bsl{\tau}_1 \bsl{\tau}_2,\beta}( \bsl{k} ) \right]_{\alpha_{\bsl{\tau}_1}\alpha_{\bsl{\tau}_2}'} \ \forall\ \beta = \shpa,\perp\ .
} 
So we have rewritten the form of $f_{i}(\bsl{k})$ (\eqnref{eq:H_el-ph_2center_k}) to make momentum derivatives explicit in it, since the quantities that capture quantum geometry rely on the momentum derivatives as shown, \eg, in \eqnref{eq:FSM_g_two_expressions_Gaussian}.

\subsection{Symmetry properties of $  \widetilde{f}_{\bsl{\tau}_1 \bsl{\tau}_2,\beta}(\Delta\bsl{R} + \bsl{\tau}_1 - \bsl{\tau}_2 )  $  and $ \widetilde{f}_{\beta}(\bsl{k})$}
\label{app:geo_EPC_g_beta_sym}

To reveal the geometric part of the EPC, the symmetry properties of $  \widetilde{f}_{\bsl{\tau}_1 \bsl{\tau}_2,\beta}(\Delta\bsl{R} + \bsl{\tau}_1 - \bsl{\tau}_2 )  $ in \eqnref{eq:g_beta_R} and $ \widetilde{f}_{\beta}(\bsl{k})$ in \eqnref{eq:g_beta_k} are particularly important, since the method described in this section relies on the symmetry reps.
Here $\beta=\shpa,\perp$.
$\widetilde{f}_{\bsl{\tau}_1 \bsl{\tau}_2, \shpa}(\Delta\bsl{R} + \bsl{\tau}_1 - \bsl{\tau}_2 )$ and $\widetilde{f}_{\bsl{\tau}_1 \bsl{\tau}_2, \perp}(\Delta\bsl{R} + \bsl{\tau}_1 - \bsl{\tau}_2 )$ in \eqnref{eq:g_beta_R} can be rewritten into one expression:
\eq{
\label{eq:g_beta_sim}
\widetilde{f}_{\bsl{\tau}_1 \bsl{\tau}_2, \beta}(\Delta\bsl{R} + \bsl{\tau}_1 - \bsl{\tau}_2\neq 0 )   = 
\sum_{i,i'=x,y}   f_{\bsl{\tau}_1 \bsl{\tau}_2,i}(\Delta\bsl{R} + \bsl{\tau}_1 - \bsl{\tau}_2 )  S^\beta_{ii'} \frac{\left(\Delta\bsl{R} + \bsl{\tau}_1  - \bsl{\tau}_2\right)_{i'}}{|\Delta\bsl{R} + \bsl{\tau}_1  - \bsl{\tau}_2|_\shpa^2} \ ,
}
where  
\eq{
S^\shpa = \mat{ 1 & 0 \\ 0 & 1}\ ,\ S^\perp = \mat{ 0 & 1 \\ -1 & 0}\ .
}

First, let us discuss the Hermiticity.
\eqnref{eq:g_R_sym} gives
\eqa{
\widetilde{f}_{ \bsl{\tau}_2\bsl{\tau}_1, \beta}^\dagger (-\Delta\bsl{R} + \bsl{\tau}_2- \bsl{\tau}_1  \neq 0 )  &  = - \sum_{i,i'=x,y}   f_{\bsl{\tau}_2\bsl{\tau}_1 ,i}^\dagger(-\Delta\bsl{R} - \bsl{\tau}_1 + \bsl{\tau}_2 )  S^\beta_{ii'} \frac{\left(\Delta\bsl{R} + \bsl{\tau}_1  - \bsl{\tau}_2\right)_{i'}}{|\Delta\bsl{R} + \bsl{\tau}_1  - \bsl{\tau}_2|_\shpa^2} \\
& =   \sum_{i,i'=x,y}   f_{\bsl{\tau}_1 \bsl{\tau}_2,i}(\Delta\bsl{R} + \bsl{\tau}_1 - \bsl{\tau}_2 )  S^\beta_{ii'} \frac{\left(\Delta\bsl{R} + \bsl{\tau}_1  - \bsl{\tau}_2\right)_{i'}}{|\Delta\bsl{R} + \bsl{\tau}_1  - \bsl{\tau}_2|_\shpa^2} \\
& = \widetilde{f}_{ \bsl{\tau}_1 \bsl{\tau}_2, \beta} (\Delta\bsl{R} + \bsl{\tau}_1 - \bsl{\tau}_2 \neq 0 )\ .
}
In addition, $f_{ \bsl{\tau}\bsl{\tau}, \beta}^\dagger (0) = 0 = f_{ \bsl{\tau}\bsl{\tau}, \beta} (0)$.
Then, we have 
\eq{
\widetilde{f}_{ \bsl{\tau}_2\bsl{\tau}_1, \beta}^\dagger (-\Delta\bsl{R} + \bsl{\tau}_2- \bsl{\tau}_1 ) = \widetilde{f}_{ \bsl{\tau}_1 \bsl{\tau}_2, \beta} (\Delta\bsl{R} + \bsl{\tau}_1 - \bsl{\tau}_2 )\ .
}
Combined with \eqnref{eq:g_beta_k_block}, we further have the Hermiticity of $ \widetilde{f}_{\beta}(\bsl{k})$, which reads
\eqa{
\widetilde{f}_{\bsl{\tau}_2\bsl{\tau}_1,\beta}^\dagger(\bsl{k}) & = \sum_{\Delta\bsl{R} }  e^{ \ii \bsl{k} \cdot (\Delta\bsl{R}+ \bsl{\tau}_2 - \bsl{\tau}_1)} \widetilde{f}_{\bsl{\tau}_2 \bsl{\tau}_1 ,\beta}^\dagger(\Delta\bsl{R} + \bsl{\tau}_2 - \bsl{\tau}_1) \\
& = \sum_{\Delta\bsl{R} }  e^{ -\ii \bsl{k} \cdot (\Delta\bsl{R}+ \bsl{\tau}_1 - \bsl{\tau}_2)} \widetilde{f}_{\bsl{\tau}_1 \bsl{\tau}_2  ,\beta}(\Delta\bsl{R} + \bsl{\tau}_1 - \bsl{\tau}_2) \\
& =  \widetilde{f}_{\bsl{\tau}_1 \bsl{\tau}_2,\beta}(\bsl{k}) \ ,
}
resulting in
\eq{
 \widetilde{f}_{\beta}^\dagger(\bsl{k}) =  \widetilde{f}_{\beta}(\bsl{k})\ .
}
The Hermiticity requirement of $ \widetilde{f}_{\beta}$ can be summarized as
\eqa{
\label{eq:g_beta_hc}
& \widetilde{f}_{ \bsl{\tau}_2\bsl{\tau}_1, \beta}^\dagger (-\Delta\bsl{R} + \bsl{\tau}_2- \bsl{\tau}_1 ) = \widetilde{f}_{ \bsl{\tau}_1 \bsl{\tau}_2, \beta} (\Delta\bsl{R} + \bsl{\tau}_1 - \bsl{\tau}_2 )\\
&  \widetilde{f}_{\beta}^\dagger(\bsl{k}) =  \widetilde{f}_{\beta}(\bsl{k})\ .
}

Based on \eqnref{eq:g_beta_sim}, if the Hamiltonian preserves a crystalline symmetry $g=\{ R|\bsl{d} \}$, we have
\eqa{
\label{eq:g_beta_sym_g_ini}
&  U^{\bsl{\tau}_{1,g}\bsl{\tau}_1}_g \widetilde{f}_{\bsl{\tau}_1 \bsl{\tau}_2, \beta}(\bsl{R}_1 + \bsl{\tau}_1 - \bsl{R}_2 - \bsl{\tau}_2\neq 0 ) \left[U^{\bsl{\tau}_{2,g}\bsl{\tau}_2}_g\right]^\dagger \\
& = \sum_{i,i'=x,y}   U^{\bsl{\tau}_{1,g}\bsl{\tau}_1}_g f_{\bsl{\tau}_1 \bsl{\tau}_2,i}( \bsl{R}_1 + \bsl{\tau}_1 - \bsl{R}_2 - \bsl{\tau}_2 )\left[U^{\bsl{\tau}_{2,g}\bsl{\tau}_2}_g\right]^\dagger   S^\beta_{ii'} \frac{\left(\bsl{R}_1 + \bsl{\tau}_1 - \bsl{R}_2  - \bsl{\tau}_2\right)_{i'}}{|\bsl{R}_1 + \bsl{\tau}_1 - \bsl{R}_2 - \bsl{\tau}_2|_\shpa^2} \\
& = \sum_{i,i'=x,y} \sum_{i_1,i_1'}  f_{\bsl{\tau}_{1,g}\bsl{\tau}_{2,g}, i_1}(R(\bsl{R}_1 + \bsl{\tau}_1 - \bsl{R}_2 - \bsl{\tau}_2))  R_{i_1i}S^\beta_{ii'} R_{i_1'i'}\frac{\left(R(\bsl{R}_1 + \bsl{\tau}_1 - \bsl{R}_2  - \bsl{\tau}_2)\right)_{i_1'}}{|\bsl{R}_1 + \bsl{\tau}_1 - \bsl{R}_2 - \bsl{\tau}_2|_\shpa^2} \ ,
}
where we have used the fact that $R$ is a real orthogonal matrix.
Since we are considering 2D systems, $R_{xz}=R_{yz} = 0$ for any crystalline symmetry $g$ preserved by the system, \ie,
\eq{
\label{eq:R_2D}
R = \mat{ R_{2D} & \\ & R_{zz} } \text{with $R_{2D}$ a $2\times 2$ real orthogonal matrix and $R_{zz}=\pm 1$}\ .
}
Therefore, we have 
\eqa{
\label{eq:V_beta_R}
& |\bsl{R}_1 + \bsl{\tau}_1 - \bsl{R}_2 - \bsl{\tau}_2|_\shpa = |R (\bsl{R}_1 + \bsl{\tau}_1 - \bsl{R}_2 - \bsl{\tau}_2) |_\shpa \\
&   \sum_{i,i'=x,y} R_{i_1i}S^\shpa_{ii'} R_{i_1'i'} = \sum_{i=x,y} R_{i_1i} R_{i_1'i} = \delta_{i_1 i_1'} = z_{R,\shpa} S^\shpa_{i_1i_1'}\ \forall\ i_1,i_1' = x,y\\
&   \sum_{i,i'=x,y} R_{i_1i}S^\perp_{ii'} R_{i_1'i'} = \sum_{i,i'=x,y} \epsilon_{ii'} (R_{2D})_{i_1i} (R_{2D})_{i_1'i'} =  \epsilon_{i_1i_1'} \det(R_{2D}) = z_{R,\perp} S^\perp_{i_1i_1'} \ \forall\ i_1,i_1' = x,y\\
&   \sum_{i,i'=x,y} R_{i_1i}S^\beta_{ii'} R_{i_1'i'} = 0 \text{ if $i_1=z$ or $i_1'=z$\ ,}
}
where
\eq{
\label{eq:z_factor}
z_{R,\shpa} = 1 \ \ ,\ z_{R,\perp} = \det(R_{2D})\ ,
}
$g=\{ R | \bsl{d} \}$, and $R_{2D}$ is in \eqnref{eq:R_2D}.
Then, with \eqnref{eq:R_2D} and \eqnref{eq:V_beta_R}, \eqnref{eq:g_beta_sym_g_ini} becomes 
\eqa{
&  U^{\bsl{\tau}_{1,g}\bsl{\tau}_1}_g \widetilde{f}_{\bsl{\tau}_1 \bsl{\tau}_2, \beta}(\bsl{R}_1 + \bsl{\tau}_1 - \bsl{R}_2 - \bsl{\tau}_2\neq 0 ) \left[U^{\bsl{\tau}_{2,g}\bsl{\tau}_2}_g\right]^\dagger \\
& = \sum_{i_1,i_1'=x,y}  f_{\bsl{\tau}_{1,g}\bsl{\tau}_{2,g}, i_1}(R(\bsl{R}_1 + \bsl{\tau}_1 - \bsl{R}_2 - \bsl{\tau}_2))  z_{R,\beta} S^\beta_{i_1i_1'}\frac{\left(R(\bsl{R}_1 + \bsl{\tau}_1 - \bsl{R}_2  - \bsl{\tau}_2)\right)_{i_1'}}{|R(\bsl{R}_1 + \bsl{\tau}_1 - \bsl{R}_2 - \bsl{\tau}_2)|_\shpa^2} \\
& = \sum_{i,i'=x,y}  f_{\bsl{\tau}_{1,g}\bsl{\tau}_{2,g}, i}( \bsl{R}_{1,\bsl{\tau}_1,g} + \bsl{\tau}_{1,g} - \bsl{R}_{2,\bsl{\tau}_2,g} - \bsl{\tau}_{2,g}))  S^\beta_{ii'}\frac{\left(\bsl{R}_{1,\bsl{\tau}_1,g} + \bsl{\tau}_{1,g} - \bsl{R}_{2,\bsl{\tau}_2,g} - \bsl{\tau}_{2,g}\right)_{i'}}{|\bsl{R}_{1,\bsl{\tau}_1,g} + \bsl{\tau}_{1,g} - \bsl{R}_{2,\bsl{\tau}_2,g} - \bsl{\tau}_{2,g}|_\shpa^2} \\
& =  z_{R,\beta} \widetilde{f}_{\bsl{\tau}_{1,g}\bsl{\tau}_{2,g}, \beta}( \bsl{R}_{1,\bsl{\tau}_1,g} + \bsl{\tau}_{1,g} - \bsl{R}_{2,\bsl{\tau}_2,g} - \bsl{\tau}_{2,g})\ .
}
On the other hand, we have
\eq{
U^{\bsl{\tau}_{g}\bsl{\tau}}_g \widetilde{f}_{\bsl{\tau} \bsl{\tau}, \beta}(0 ) \left[U^{\bsl{\tau}_{g}\bsl{\tau}}_g\right]^\dagger  = 0 =\widetilde{f}_{\bsl{\tau}_g \bsl{\tau}_g, \beta}(0 )\ .
}
Hence, if we require the Hamiltonian to have the symmetry $g$, $\widetilde{f}_{\bsl{\tau}_1 \bsl{\tau}_2, \beta}(\Delta\bsl{R}_1 + \bsl{\tau}_1 - \bsl{R}_2 - \bsl{\tau}_2 )$ must satisfy
\eq{
\label{eq:g_beta_sym_g_R}
U^{\bsl{\tau}_{1,g}\bsl{\tau}_1}_g \widetilde{f}_{\bsl{\tau}_1 \bsl{\tau}_2, \beta}(\bsl{R}_1 + \bsl{\tau}_1 - \bsl{R}_2 - \bsl{\tau}_2 ) \left[U^{\bsl{\tau}_{2,g}\bsl{\tau}_2}_g\right]^\dagger=  z_{R,\beta}\  \widetilde{f}_{\bsl{\tau}_{1,g}\bsl{\tau}_{2,g}, \beta}( \bsl{R}_{1,\bsl{\tau}_1,g} + \bsl{\tau}_{1,g} - \bsl{R}_{2,\bsl{\tau}_2,g} - \bsl{\tau}_{2,g})\ ,
}
where $z_{R,\beta}$ is defined in \eqnref{eq:z_factor}.
For $ \widetilde{f}_{\beta}(\bsl{k})$, we use \eqnref{eq:g_beta_k_block} and \eqnref{eq:U_g} to convert \eqnref{eq:g_beta_sym_g_R} to momentum space and obtain
\eqa{
& \sum_{\bsl{\tau}_1 \bsl{\tau}_2}U^{\bsl{\tau}_{1}'\bsl{\tau}_1}_g  \widetilde{f}_{\bsl{\tau}_1 \bsl{\tau}_2,\beta}( \bsl{k} )  \left[U^{\bsl{\tau}_{2}'\bsl{\tau}_2}_g\right]^\dagger \\
&  = \sum_{\bsl{\tau}_1 \bsl{\tau}_2} \delta_{\bsl{\tau}_{1}'\bsl{\tau}_{1,g}}\delta_{\bsl{\tau}_{2}'\bsl{\tau}_{2,g}}U^{\bsl{\tau}_{1,g}\bsl{\tau}_1}_g  \widetilde{f}_{\bsl{\tau}_1 \bsl{\tau}_2,\beta}( \bsl{k} )  \left[U^{\bsl{\tau}_{2,g}\bsl{\tau}_2}_g\right]^\dagger \\
& = \sum_{\bsl{\tau}_1 \bsl{\tau}_2} \delta_{\bsl{\tau}_{1}'\bsl{\tau}_{1,g}}\delta_{\bsl{\tau}_{2}'\bsl{\tau}_{2,g}} \sum_{\Delta\bsl{R} }  e^{- \ii \bsl{k} \cdot (\Delta\bsl{R}+ \bsl{\tau}_1 - \bsl{\tau}_2)} U^{\bsl{\tau}_{1,g}\bsl{\tau}_1}_g  \widetilde{f}_{\bsl{\tau}_1 \bsl{\tau}_2,\beta}(\Delta\bsl{R} + \bsl{\tau}_1 - \bsl{\tau}_2 ) \left[U^{\bsl{\tau}_{2,g}\bsl{\tau}_2}_g\right]^\dagger \\
& = \sum_{\bsl{\tau}_1 \bsl{\tau}_2} \delta_{\bsl{\tau}_{1}'\bsl{\tau}_{1,g}}\delta_{\bsl{\tau}_{2}'\bsl{\tau}_{2,g}} \sum_{\Delta\bsl{R} }  e^{- \ii R \bsl{k} \cdot (R \Delta\bsl{R}+\Delta\bsl{R}_{\bsl{\tau}_1,g}  - \Delta\bsl{R}_{\bsl{\tau}_2,g} + \bsl{\tau}_{1,g} - \bsl{\tau}_{2,g})} \\
& \qquad \times z_{R,\beta}\  \widetilde{f}_{\bsl{\tau}_{1,g}\bsl{\tau}_{2,g}, \beta}( R \Delta\bsl{R} + \Delta\bsl{R}_{\bsl{\tau}_1,g}  - \Delta\bsl{R}_{\bsl{\tau}_2,g} + \bsl{\tau}_{1,g} - \bsl{\tau}_{2,g}) \\
& = z_{R,\beta} \sum_{\bsl{\tau}_1 \bsl{\tau}_2} \delta_{\bsl{\tau}_{1}'\bsl{\tau}_{1,g}}\delta_{\bsl{\tau}_{2}'\bsl{\tau}_{2,g}} \sum_{\Delta\bsl{R} }  e^{- \ii R \bsl{k} \cdot (\Delta\bsl{R} + \bsl{\tau}_{1,g} - \bsl{\tau}_{2,g})} \  \widetilde{f}_{\bsl{\tau}_{1,g}\bsl{\tau}_{2,g}, \beta}( \Delta\bsl{R} + \bsl{\tau}_{1,g} - \bsl{\tau}_{2,g}) \\
& =  z_{R,\beta} \sum_{\bsl{\tau}_1 \bsl{\tau}_2} \delta_{\bsl{\tau}_{1}'\bsl{\tau}_{1,g}}\delta_{\bsl{\tau}_{2}'\bsl{\tau}_{2,g}} \widetilde{f}_{\bsl{\tau}_{1,g}\bsl{\tau}_{2,g}, \beta}( R \bsl{k} ) \\
& =  z_{R,\beta} \widetilde{f}_{\bsl{\tau}_{1}'\bsl{\tau}_{2}', \beta}( R \bsl{k} ) \ ,
}
which, combined with \eqnref{eq:g_beta_k}, leads to 
\eq{
\label{eq:g_beta_sym_g_k}
U_g\  \widetilde{f}_{\beta}( \bsl{k} )  \left[U_g\right]^\dagger =  z_{R,\beta}\  \widetilde{f}_{\beta}( R \bsl{k} )\ .
}

Based on \eqnref{eq:g_beta_sim}, if the Hamiltonian preserves TR symmetry, we have
\eqa{
\label{eq:g_beta_sym_TR_ini}
&  U^{\bsl{\tau}_{1}\bsl{\tau}_1}_{\TR} \widetilde{f}_{\bsl{\tau}_1 \bsl{\tau}_2, \beta}^*(\bsl{R}_1 + \bsl{\tau}_1 - \bsl{R}_2 - \bsl{\tau}_2\neq 0 ) \left[U^{\bsl{\tau}_{2}\bsl{\tau}_2}_{\TR}\right]^\dagger \\
& = \sum_{i,i'=x,y}U^{\bsl{\tau}_{1}\bsl{\tau}_1}_{\TR}  f_{\bsl{\tau}_1 \bsl{\tau}_2,i}^*( \bsl{R}_1 + \bsl{\tau}_1 - \bsl{R}_2 - \bsl{\tau}_2 )\left[U^{\bsl{\tau}_{2}\bsl{\tau}_2}_{\TR}\right]^\dagger  S^\beta_{ii'} \frac{\left(\bsl{R}_1 + \bsl{\tau}_1 - \bsl{R}_2  - \bsl{\tau}_2\right)_{i'}}{|\bsl{R}_1 + \bsl{\tau}_1 - \bsl{R}_2 - \bsl{\tau}_2|_\shpa^2} \\
& = \sum_{i,i'=x,y} f_{\bsl{\tau}_1 \bsl{\tau}_2,i} ( \bsl{R}_1 + \bsl{\tau}_1 - \bsl{R}_2 - \bsl{\tau}_2 )   S^\beta_{ii'} \frac{\left(\bsl{R}_1 + \bsl{\tau}_1 - \bsl{R}_2  - \bsl{\tau}_2\right)_{i'}}{|\bsl{R}_1 + \bsl{\tau}_1 - \bsl{R}_2 - \bsl{\tau}_2|_\shpa^2}\\
& = \widetilde{f}_{\bsl{\tau}_1 \bsl{\tau}_2, \beta}(\bsl{R}_1 + \bsl{\tau}_1 - \bsl{R}_2 - \bsl{\tau}_2\neq 0)\ .
}
where we have used the fact that $R$ is a real orthogonal matrix.
In addition, we have
\eq{
U^{\bsl{\tau}\bsl{\tau}}_{\TR} \widetilde{f}_{\bsl{\tau} \bsl{\tau}, \beta}^*(0) \left[U^{\bsl{\tau}\bsl{\tau}}_{\TR}\right]^\dagger  = 0 =\widetilde{f}_{\bsl{\tau}\bsl{\tau}, \beta}(0 )\ .
}
Eventually, we have the symmetry requirement of $\widetilde{f}_{\bsl{\tau}_1 \bsl{\tau}_2, \beta}(\Delta\bsl{R}_1 + \bsl{\tau}_1 - \bsl{R}_2 - \bsl{\tau}_2 )$ under $\TR$:
\eq{
\label{eq:g_beta_sym_TR_R}
U^{\bsl{\tau}_{1}\bsl{\tau}_1}_{\TR} \widetilde{f}_{\bsl{\tau}_1 \bsl{\tau}_2, \beta}^*(\bsl{R}_1 + \bsl{\tau}_1 - \bsl{R}_2 - \bsl{\tau}_2\neq 0 ) \left[U^{\bsl{\tau}_{2}\bsl{\tau}_2}_{\TR}\right]^\dagger  =  \widetilde{f}_{\bsl{\tau}_1 \bsl{\tau}_2, \beta}^*(\bsl{R}_1 + \bsl{\tau}_1 - \bsl{R}_2 - \bsl{\tau}_2)\ .
}
For $ \widetilde{f}_{\beta}(\bsl{k})$, we use \eqnref{eq:g_beta_k_block} and \eqnref{eq:U_TR} to convert \eqnref{eq:g_beta_sym_TR_R} to momentum space and obtain
\eqa{
& U^{\bsl{\tau}_{1}\bsl{\tau}_1}_\TR  \widetilde{f}_{\bsl{\tau}_1 \bsl{\tau}_2,\beta}^*( \bsl{k} )  \left[U^{\bsl{\tau}_{2}\bsl{\tau}_2}_\TR\right]^\dagger \\
&  =   \sum_{\Delta\bsl{R} }  e^{ \ii \bsl{k} \cdot (\Delta\bsl{R}+ \bsl{\tau}_1 - \bsl{\tau}_2)}  U^{\bsl{\tau}_{1}\bsl{\tau}_1}_\TR   \widetilde{f}_{\bsl{\tau}_1 \bsl{\tau}_2,\beta}^*(\Delta\bsl{R} + \bsl{\tau}_1 - \bsl{\tau}_2 ) \left[U^{\bsl{\tau}_{2}\bsl{\tau}_2}_\TR\right]^\dagger \\
& =   \sum_{\Delta\bsl{R} }  e^{ \ii \bsl{k} \cdot (\Delta\bsl{R}+ \bsl{\tau}_1 - \bsl{\tau}_2)}     \widetilde{f}_{\bsl{\tau}_1 \bsl{\tau}_2,\beta}(\Delta\bsl{R} + \bsl{\tau}_1 - \bsl{\tau}_2 ) \\
& =   \widetilde{f}_{\bsl{\tau}_{1}\bsl{\tau}_{2}, \beta}( - \bsl{k} )  \ ,
}
which, combined with \eqnref{eq:g_beta_k}, leads to 
\eq{
\label{eq:g_beta_sym_TR_k}
U_\TR\  \widetilde{f}_{\beta}^*( \bsl{k} )  \left[U_\TR\right]^\dagger =  \widetilde{f}_{\beta}( - \bsl{k} )\ .
}

In sum, the symmetry properties of $ \widetilde{f}_{\beta}$ are
\eqb{
\label{eq:g_beta_sym}
& \widetilde{f}_{ \bsl{\tau}_2\bsl{\tau}_1, \beta}^\dagger (-\Delta\bsl{R} + \bsl{\tau}_2- \bsl{\tau}_1 ) = \widetilde{f}_{ \bsl{\tau}_1 \bsl{\tau}_2, \beta} (\Delta\bsl{R} + \bsl{\tau}_1 - \bsl{\tau}_2 )\\
&  \widetilde{f}_{\beta}^\dagger(\bsl{k}) =  \widetilde{f}_{\beta}(\bsl{k})\\
& U^{\bsl{\tau}_{1,g}\bsl{\tau}_1}_g  \widetilde{f}_{\bsl{\tau}_1 \bsl{\tau}_2, \beta}(\bsl{R}_1 + \bsl{\tau}_1 - \bsl{R}_2 - \bsl{\tau}_2 ) \left[U^{\bsl{\tau}_{2,g}\bsl{\tau}_2}_g\right]^\dagger=  z_{R,\beta}\   \widetilde{f}_{\bsl{\tau}_{1,g}\bsl{\tau}_{2,g}, \beta}( \bsl{R}_{1,\bsl{\tau}_1,g} + \bsl{\tau}_{1,g} - \bsl{R}_{2,\bsl{\tau}_2,g} - \bsl{\tau}_{2,g})\\
& U_g\  \widetilde{f}_{\beta}( \bsl{k} )  \left[U_g\right]^\dagger =  z_{R,\beta}\  \widetilde{f}_{\beta}( R \bsl{k} )\\
& U^{\bsl{\tau}_{1}\bsl{\tau}_1}_{\TR} \widetilde{f}_{\bsl{\tau}_1 \bsl{\tau}_2, \beta}^*(\bsl{R}_1 + \bsl{\tau}_1 - \bsl{R}_2 - \bsl{\tau}_2\neq 0 ) \left[U^{\bsl{\tau}_{2}\bsl{\tau}_2}_{\TR}\right]^\dagger  =  \widetilde{f}_{\bsl{\tau}_1 \bsl{\tau}_2, \beta}^*(\bsl{R}_1 + \bsl{\tau}_1 - \bsl{R}_2 - \bsl{\tau}_2)\\
& U_\TR\  \widetilde{f}_{\beta}^*( \bsl{k} )  \left[U_\TR\right]^\dagger =  \widetilde{f}_{\beta}( - \bsl{k} )\ ,
}
where $g=\{ R|\bsl{d}\}$ is a crystalline symmetry, $z_{R,\shpa}$ is defined in \eqnref{eq:z_factor}, $U^{\bsl{\tau}_{g}\bsl{\tau}}_g$ is defined in \eqnref{eq:sym_rep_g_R}, $U^{\bsl{\tau} \bsl{\tau}}_{\TR}$ is defined in \eqnref{eq:sym_rep_TR_R}, $\bsl{R}_{\bsl{\tau},g}$ and $\bsl{\tau}_g$ are defined in \eqnref{eq:R_g_tau_g}, $U_g$ is defined in \eqref{eq:U_g}, $U_{\TR}$ is defined in \eqref{eq:U_TR}, and we have assumed that the system preserves $\TR$ and $g$.

\subsection{Symmetry-Rep Method}
\label{app:symmetry-rep_detials}

With the preparation in \appref{app:geo_EPC_dk} and \appref{app:geo_EPC_g_beta_sym}, we now introduce the symmetry-rep method to define the geometric part of $f_i(\bsl{k})$ (in \eqnref{eq:H_el-ph_2center_k}).
In this part, we will just present the symmetry-rep method in an abstract way.
Applications of the symmetry-rep method to explicit examples (\ie, graphene and {\mgb}) are in \appref{app:graphene} and \appref{app:MgB2}, where, we find that the symmetry-rep method combined with the short-range hoppings give exactly the same results as the GA.

Our strategy is to replace terms in $ \widetilde{f}_{\beta}(\bsl{k})$ (\eqnref{eq:g_beta_k}) by terms that involve the electron Hamiltonian $h(\bsl{k})$, which will become the terms that involve $\partial_{k_x} h(\bsl{k})$ or $\partial_{k_y} h(\bsl{k})$ in $f_{i}(\bsl{k})$, based on the symmetry properties.
Then, $\partial_{i} h(\bsl{k})$ contains $\partial_{k_i} P_{n}(\bsl{k})$ which corresponds to the geometric property of the Bloch wavefunction, where $P_{n}(\bsl{k})$ is the electron projection matrix defined in \eqnref{eq:P_U}.
As mentioned at the beginning of \appref{app:geo_EPC_symmetry-rep}, the part of $f_{i}(\bsl{k})$ that involves $\partial_{k_i} P_{n}(\bsl{k})$ is just the geometric part, while the part of $f_{i}(\bsl{k})$ that involves $\partial_{k_i} E_{n}(\bsl{k})$ is the energetic part, where $E_{n}(\bsl{k})$ is the electron band  defined in \eqnref{eq:el_eigen}.

Let us be more specific in the following.
By comparing \eqnref{eq:g_beta_sym} to \eqnref{eq:t_h_hc}, \eqnref{eq:t_h_sym_g} and \eqref{eq:t_h_sym_TR}, we find that $\widetilde{f}_{\bsl{\tau}_1 \bsl{\tau}_2, \shpa}(\bsl{R}_1 + \bsl{\tau}_1 - \bsl{R}_2 - \bsl{\tau}_2 )$ in \eqnref{eq:g_beta_R} has the same symmetry properties as the electron hopping $t_{\bsl{\tau}_1 \bsl{\tau}_2}(\bsl{R}_1 + \bsl{\tau}_1 - \bsl{R}_2 - \bsl{\tau}_2 )$ in \eqnref{eq:H_el_gen}, and $ \widetilde{f}_{\shpa}(\bsl{k})$ has the same symmetry properties as the electron matrix Hamiltonian $h(\bsl{k})$ in \eqnref{eq:H_el_gen}.
In principle, we can always find the most general symmetry-allowed form of $h(\bsl{k})$, which reads 
\eq{
\label{eq:h_hhat_gen_k}
h(\bsl{k}) = \sum_{a} t_a \hat{h}_a(\bsl{k}) \text{ with }\frac{1}{N}\sum_{\bsl{k}}^{\BZ} \Tr[\hat{h}_a^\dagger(\bsl{k}) \hat{h}_{a'}(\bsl{k})] = 0 \text{ for $a\neq a'$}\ ,
}
where $t_a$ labels all the independent free parameters, and $\hat{h}_a(\bsl{k})$ is non-vanishing and has no tuning parameters in it.
Then, we know the most general symmetry-allowed form of $ \widetilde{f}_{\shpa}(\bsl{k})$ under the same approximations reads
\eq{
\label{eq:g_shpa_gen_k}
 \widetilde{f}_{\shpa}(\bsl{k}) = \sum_{a} \hat{\gamma}_{\shpa,a} \hat{h}_a(\bsl{k})\ .
}
Comparing \eqnref{eq:g_shpa_gen_k} to \eqnref{eq:h_hhat_gen_k}, it is straightforward to see that 
\eq{
\label{eq:g_shpa_h_relation}
 \widetilde{f}_{\shpa}(\bsl{k}) =  \L_{\shpa} [h(\bsl{k})] \ ,
}
where
\eq{
\label{eq:L_shpa}
\L_{\shpa} [ x ] = \sum_{a} \hat{\gamma}_{\shpa,a} \partial_{t_a} x\ .
}
Moreover, the linear operator $\L_{\shpa}$ in \eqnref{eq:L_shpa} does not depend on the choice of $\hat{h}_a(\bsl{k})$.
Specifically, if we change the choice of $\hat{h}_a(\bsl{k})$ to $\hat{h}_a'(\bsl{k})$, then $\hat{h}_a'(\bsl{k})$ must be related to $\hat{h}_a(\bsl{k})$ by
\eq{
\label{eq:hhat_hhatprime}
\hat{h}_a(\bsl{k}) = \sum_{a'}\hat{h}_{a'}'(\bsl{k})X_{a'a}\ ,
}
where $X$ is a invertible matrix since $\hat{h}_a(\bsl{k})$ and $\hat{h}_a'(\bsl{k})$ are nonvanishing in 1BZ.
As a result, according to 
\eqa{
& h(\bsl{k}) = \sum_{a'} t_a' \hat{h}_a'(\bsl{k})\\
&  \widetilde{f}_{\shpa}(\bsl{k}) = \sum_{a'} \hat{\gamma}_{\shpa,a'} \hat{h}_a'(\bsl{k})\ ,
}
we have
\eqa{
\label{eq:t_tprime}
&  t_{a'}' = \sum_{a} t_a X_{a'a} \\
&  \hat{\gamma}_{\shpa,a'}' = \sum_{a}\hat{\gamma}_{\shpa,a} X_{a'a} \ ,
}
leading to 
\eqa{
\L_{\shpa}' &= \sum_{a'} \hat{\gamma}_{a'}' \partial_{t_{a'}'}  = \sum_{a'} \sum_{a_1}\hat{\gamma}_{\shpa,a_1} X_{a'a_1}  \sum_{a_2} \frac{\partial t_{a_2}}{\partial t_{a'}' } \partial_{t_{a_2}} = \sum_{a'} \sum_{a_1}\hat{\gamma}_{\shpa,a_1} X_{a'a_1}  \sum_{a_2} \frac{\partial \sum_{a_1'}[X^{-1}]_{a_2 a_1' }t_{a_1'}'}{\partial t_{a'}' } \partial_{t_{a_2}}\\
& = \sum_{a'} \sum_{a_1}\hat{\gamma}_{\shpa,a_1} X_{a'a_1}  \sum_{a_2}  [X^{-1}]_{a_2 a' } \partial_{t_{a_2}} = \sum_{a_1} \sum_{a_2} \hat{\gamma}_{\shpa,a_1}    \partial_{t_{a_2}} \delta_{a_1 a_2} = \sum_{a} \hat{\gamma}_{\shpa,a}    \partial_{t_{a}} = \L_{\shpa}\ .
}
Therefore, we can choose any $\hat{h}_a(\bsl{k})$ to use \eqnref{eq:g_shpa_h_relation}, as long as the choice of $\hat{h}_a(\bsl{k})$ is the same for both \eqnref{eq:h_hhat_gen_k} and \eqnref{eq:g_shpa_gen_k}.

Now, we move on to $ \widetilde{f}_{\perp}(\bsl{k})$.
Due to the factor $z_{R,\perp}$ (defined in \eqnref{eq:z_factor}) of the symmetry requirement of $ \widetilde{f}_{\perp}(\bsl{k})$ in \eqnref{eq:g_beta_sym}, we should study $ \widetilde{f}_{\perp}(\bsl{k})$ in two cases: (i) all spatial symmetries have $\det(R_{2D})=1$ for \eqnref{eq:z_factor} (\eg, rotations along $z$), and (ii) there are spatial symmetries that have $\det(R_{2D})=-1$ for \eqnref{eq:z_factor}.
In case (i), $ \widetilde{f}_{\perp}(\bsl{k})$ has the same form as \eqnref{eq:h_hhat_gen_k}: 
\eq{
 \widetilde{f}_{\perp}(\bsl{k}) = \sum_{a} \hat{\gamma}_{\perp,a} \hat{h}_a(\bsl{k})\ ,
}
which means that 
\eq{
\label{eq:g_perp_h_relation_poss1}
 \widetilde{f}_{\perp}(\bsl{k}) = \L_{\perp}[h(\bsl{k})]\ ,
}
where
\eq{
\label{eq:L_perp_1}
\L_{\perp}[x]=\sum_{a} \hat{\gamma}_{\perp,a} \partial_{t_a} x\ .
}
Similar to \eqnref{eq:L_shpa}, $\L_{\perp}$ in \eqnref{eq:L_perp_1} is independent of the choice of $\hat{h}_a(\bsl{k})$.

In case (ii), we cannot use $\L_{\perp}$ in \eqnref{eq:L_perp_1}, because none of the terms in \eqnref{eq:h_hhat_gen_k} can appear in $ \widetilde{f}_{\perp}(\bsl{k})$ due to the different transformation properties for symmetries with $\det(R_{2D})=-1$.
Nevertheless, we can still try to express $\widetilde{f}_{\perp}(\bsl{k})$ in the following form:
\eq{
\label{eq:g_perp_h_relation}
 \widetilde{f}_{\perp}(\bsl{k}) = \L_{\perp}[h(\bsl{k})] + \Delta  \widetilde{f}_{\perp}(\bsl{k})\ ,
}
where $\L_{\perp}$ is different from \eqnref{eq:L_perp_1} but $\L_{\perp}$ is still a $\bsl{k}$-independent linear operator.
Here $\Delta  \widetilde{f}_{\perp}(\bsl{k})$ includes the terms beyond $\L_{\perp}[h(\bsl{k})]$, and is zero in case (i).
Unfortunately, we currently cannot give a universal expression for the $\L_{\perp}$ in case (ii), which we leave as the future work.
Let us only focus on $\L_{\perp}$ for graphene and {\mgb}, both of which belong to the case (ii).
As will be discussed in \appref{app:graphene} and \appref{app:MgB2}, both graphene and {\mgb} can be viewed as special cases of the TR-invariant systems that satisfy
\eq{
    \label{eq:Q_h_ortho_rel}
    \frac{1}{N}\sum_{\bsl{k}}^{\BZ}\Tr[(Q_l \hat{h}_a(\bsl{k}) + \hat{h}_a(\bsl{k}) Q_l^\dagger) (Q_{l'} \hat{h}_{a'}(\bsl{k}) + \hat{h}_{a'}(\bsl{k}) Q_{l'}^\dagger)] = 0 \text{ for $(l,a)\neq (l',a')$,}
}
where $Q_l$ are TR-invariant constant matrices that satisfy $U_g Q_l U_g^\dagger = Q_l z_{R,\perp}$ for any crystalline symmetry $g=\{ R | \bsl{d} \}$ of the system,
\eq{
    \label{eq:Q_l_otho_rel}
    \Tr[Q_l^\dagger Q_{l'}]=0 \text{ for }l'\neq l \ ,
}
and $\hat{h}_a(\bsl{k})$ satisfies \eqnref{eq:h_hhat_gen_k}.
We can see $(Q_l \hat{h}_a(\bsl{k}) + \hat{h}_a(\bsl{k}) Q_l^\dagger)$ should appear in $ \widetilde{f}_{\perp}(\bsl{k})$ since they have the same symmetry properties. 
Then, we have
\eq{
\label{eq:g_2D_perp_Q}
 \widetilde{f}_{\perp}(\bsl{k}) = \sum_{l,a} \hat{\gamma}_{\perp,l,a} (Q_l \hat{h}_a(\bsl{k}) + \hat{h}_a(\bsl{k}) Q_l^\dagger) + \Delta  \widetilde{f}_{\perp}(\bsl{k}) \ ,
}
where $\hat{\gamma}_{\perp,l,a} = 0 $ is required if $Q_l \hat{h}_a(\bsl{k}) + \hat{h}_a(\bsl{k}) Q_l^\dagger = 0$, and $\Delta  \widetilde{f}_{\perp}(\bsl{k})$ is the term in $ \widetilde{f}_{\perp}(\bsl{k})$ that does not involve $Q_l \hat{h}_a(\bsl{k}) + \hat{h}_a(\bsl{k}) Q_l^\dagger$.
As a result, \eqnref{eq:g_2D_perp_Q} infers
\eq{
\label{eq:L_perp_2}
\L_{\perp}[x]=\sum_{l,a} \hat{\gamma}_{\perp,l,a} (Q_l \partial_{t_a}  x + \partial_{t_a} x Q_l^\dagger)\ ,
}
and $t_a$ satisfies \eqnref{eq:h_hhat_gen_k}. 

In sum, for $\widetilde{f}_{\perp}(\bsl{k})$, we have \eqnref{eq:g_perp_h_relation} with $\L_{\perp}$ having the expression of \eqnref{eq:L_perp_1} in case (i) or \eqnref{eq:L_perp_2} for the special TR-invariant systems considered in the case (ii).

Once \eqnref{eq:g_shpa_h_relation} and \eqnref{eq:g_perp_h_relation} are achieved, we can re-write $f_{i}(\bsl{k})$ and obtain
\eqa{
\label{eq:g_i_2D_g_E_geo}
  f_{i}(\bsl{k})  &  =  \left[ \ii \partial_{k_{i}} \L_{\shpa}[h(\bsl{k})] + \ii \sum_{i'=x,y} \epsilon_{i'i}   \partial_{k_{i'}} \L_{\perp}[h(\bsl{k})]  + \ii \sum_{i'=x,y} \epsilon_{i'i}  \partial_{k_{i'}} \Delta f_\perp(\bsl{k})\right](\delta_{ix} + \delta_{iy}) +\delta_{iz} f_{z}(\bsl{k}) \\
  &  =  \left[ \ii \L_{\shpa}[\partial_{k_{i}} h(\bsl{k})] + \ii \sum_{i'=x,y} \epsilon_{i'i}    \L_{\perp}[\partial_{k_{i'}}h(\bsl{k})]  \right](\delta_{ix} + \delta_{iy}) + \Delta f_{i}(\bsl{k}) \\
   &  =  \left[ \ii \L_{\shpa}[\partial_{k_{i}} \sum_{n} E_n(\bsl{k})P_{n}(\bsl{k})] + \ii \sum_{i'=x,y} \epsilon_{i'i}    \L_{\perp}[\partial_{k_{i'}}\sum_{n} E_n(\bsl{k})P_{n}(\bsl{k})]  \right](\delta_{ix} + \delta_{iy}) + \Delta f_{i}(\bsl{k}) \\
  &  =  f_{i}^E(\bsl{k}) + f_{i}^{geo}(\bsl{k}) + \Delta f_{i}(\bsl{k})\ ,
}
where
\eq{
\label{eq:Delta_g_i_2D}
\Delta f_{i}(\bsl{k}) = \ii \sum_{i'=x,y} \epsilon_{i'i}  \partial_{k_{i'}} \Delta f_\perp(\bsl{k})(\delta_{ix} + \delta_{iy}) +\delta_{iz} f_{z}(\bsl{k}) \ ,
}
\eqa{
\label{eq:g_2D_E_geo}
f_{i}^E(\bsl{k}) & = \left[ \ii \L_{\shpa}[\sum_{n} P_{n}(\bsl{k}) \partial_{k_{i}} E_n(\bsl{k})] + \ii \sum_{i'=x,y} \epsilon_{i'i}    \L_{\perp}[\sum_{n} P_{n}(\bsl{k}) \partial_{k_{i'}}E_n(\bsl{k})]  \right](\delta_{ix} + \delta_{iy})  \\
f_{i}^{geo}(\bsl{k}) & = \left[ \ii \L_{\shpa}[ \sum_{n} E_n(\bsl{k})\partial_{k_{i}}P_{n}(\bsl{k})] + \ii \sum_{i'=x,y} \epsilon_{i'i}    \L_{\perp}[\sum_{n} E_n(\bsl{k})\partial_{k_{i'}}P_{n}(\bsl{k})]  \right](\delta_{ix} + \delta_{iy}) \ ,
}
$\L_{\shpa}$ is in \eqnref{eq:L_shpa}, $\L_{\perp}$ defined in \eqnref{eq:L_perp_1} or \eqnref{eq:L_perp_2} (or in other ways depending on the system), and we have used the fact that $[\partial_{k_i}, \L_{\shpa} ] =0 $ and $[\partial_{k_i}, \L_{\perp} ] =0 $.
We call $f_{i}^E(\bsl{k})$ the energetic part of $f_{i}(\bsl{k})$ since $f_{i}^E(\bsl{k})$ vanishes for systems with all electron bands exactly flat.
We call $f_{i}^{geo}(\bsl{k})$ the geometric part of $f_{i}(\bsl{k})$.
Eventually, we arrive at
\eq{
\label{eq:lambda_E_geo_E-geo_Deltalambda}
\lambda = \lambda_E + \lambda_{geo} + \lambda_{E-geo} + \Delta\lambda \ ,
}
where $\lambda_E$, $\lambda_{geo}$ and  $\lambda_{E-geo}$ are defined in \eqnref{eq:lambda_E}, \eqnref{eq:lambda_geo}, and \eqnref{eq:lambda_E_geo}, respectively, and $\Delta\lambda $ is the additional term given by a nonvanishing $\Delta f_{i}(\bsl{k})$ in \eqnref{eq:Delta_g_i_2D}.

\eqnref{eq:g_2D_E_geo} is just a formal way of defining the energetic and geometric parts of the EPC based on the symmetry properties of EPC and the Hamiltonian.
Based on \eqnref{eq:L_shpa} and \eqnref{eq:L_perp_1} (or \eqnref{eq:L_perp_2}), we can clearly see that \eqnref{eq:g_2D_E_geo} needs explicit information on the $t$-dependence of the Hamiltonian, not just the momentum dependence. 
In other words, without further approximations, knowing the Hamiltonian at the fixed values of the hopping parameters ($t_a$ in \eqnref{eq:h_hhat_gen_k}) is not enough for using \eqnref{eq:g_2D_E_geo}, because \eqnref{eq:g_2D_E_geo} does not involve any information of the locality.
Therefore, in practice, \eqnref{eq:g_2D_E_geo} needs to be used together with other local approximations (like short-range approximation for the hopping). 
In graphene and {\mgb}, we find that the symmetry-rep method combined with the short-range hoppings give exactly the same results as the GA, as discussed in \appref{app:graphene} and \appref{app:MgB2}.

\section{Two properties of $\lambda$}

In this section, we discuss two properties of $\lambda$.
In the first part, we will discuss how $\lambda$ bounds the mean-field superconducting $T_c$ in TR-invariant 2D or 3D systems if neglecting Coulomb. (The mean-field theory fails in 1D.)
In the second part, we show that for systems with negligible spin-orbit coupling, we can directly neglect the spin index in the evaluation of $\lambda$ in \eqnref{eq:lambda}, and obtain the right value without any missing factors such as $2$. 

\subsection{Lower Bound of the Mean-field Superconducting Critical Temperature from $\lambda$}
\label{app:lambda_bound_SC}

To show the lower bound, let us first introduce the expression of the effective electron-electron interaction induced by phonons in general.
The effective interaction is derived by performing the Schrieffer-Wolff transformation~\cite{SchriefferWolffTransformation1966} $e^S$ with $S$ having the following expression 
\eq{
S = \frac{1}{\sqrt{N}}\sum_{\bsl{k},\bsl{k}_1,l,n,m} \gamma^\dagger_{\bsl{k},n} \gamma_{\bsl{k}_1,m} \frac{ \widetilde{G}_{nml}(\bsl{k},\bsl{k}_1) }{ (E_n(\bsl{k}) - E_m(\bsl{k}_1) )^2 - \hbar^2 \omega_l^2(\bsl{k}_1-\bsl{k}) }\left[ (E_n(\bsl{k})-E_m(\bsl{k}_1))\widetilde{u}^\dagger_{\bsl{k}_1-\bsl{k},l} + \ii \hbar \widetilde{P}_{\bsl{k}_1-\bsl{k},l}^\dagger \right]\ ,
}
where $\widetilde{u}^\dagger_{\bsl{k}_1-\bsl{k},l}$ and $\widetilde{P}_{\bsl{k}_1-\bsl{k},l}^\dagger$ are defined right below \eqnref{eq:H_ph_gen_k}, $\widetilde{G}_{nml}(\bsl{k},\bsl{k}_1-\bsl{k})$ is defined in \eqnref{eq:Gt_nml}, $\gamma^\dagger_{\bsl{k},n}$ is the creation operator for the electron Bloch states, $E_n(\bsl{k})$ is the electron bands, $\omega_l(\bsl{k}_1-\bsl{k})$ is the phonon band, and $(\bsl{k},\bsl{k}_1-\bsl{k},l,n,m)$ that satisfies $(E_n(\bsl{k}) - E_m(\bsl{k}_1) )^2 -  \hbar^2 \omega_l^2(\bsl{k}_1-\bsl{k}) = 0$ is excluded from the sum since they usually occur in zero-measure regions and do not give Dirac-delta-type contributions.
Since $S^\dagger = -S$, $e^S$ is a unitary operator.
Then, imposing the unitary transformation $e^S$ on the Hamiltonian $H$ gives
\eq{
\label{eq:transformed_H}
e^S H e^{-S}  = H_{el} + H_{ph} + H_{eff-int} + ...\ ,
}
where
\eqa{
\label{eq:eff_int_ph_induced}
H_{eff-int} & = \frac{1}{N} \sum_{\bsl{k}_1,l}\sum_{nn'mm'}\sum_{\bsl{k},\bsl{k}'}\frac{G_{n'm'l}(\bsl{k}',\bsl{k}'+\bsl{k}-\bsl{k}_1) G_{nml}(\bsl{k},\bsl{k}_1) \hbar \omega_l(\bsl{k}_1-\bsl{k}) }{(E_n(\bsl{k}) - E_m(\bsl{k}_1))^2 -(\hbar \omega_l(\bsl{k}_1-\bsl{k}))^2 } \gamma^\dagger_{\bsl{k},n} \gamma_{\bsl{k}_1,m} \gamma^\dagger_{\bsl{k}',n'} \gamma_{\bsl{k}'+\bsl{k}-\bsl{k}_1,m'}\ ,
}
and ``..." does not contain the terms in the form $H_{el-ph}$ and only contains higher-order terms.
\eqnref{eq:eff_int_ph_induced} suggests that the induced interaction must be attractive when both $E_n(\bsl{k})$ and $E_m(\bsl{k}_1)$ are equal to the Fermi energy.

With the effective interaction among electrons in \eqnref{eq:eff_int_ph_induced}, we now discuss the mean-field superconducting critical temperature $T_c$.
The mean-field $T_c$ is determined by solving the linearized gap equation~\cite{BCS1957SC}.
More specifically, we solve the linearized gap equation for the temperature $T$ and the nonvanishing pairing gap function.
The highest $T$ given by the solutions is $T_c$.
Let us consider a subset of solutions whose pairing gap functions correspond to the zero-total-momentum Cooper pairings, and label the highest $T$ given by this subset of solutions as $T_c^{ZTM}$.
Then, we know 
\eq{
\label{eq:T_ZTM}
T_c \geq T_{c}^{ZTM}\ .
}
$T_c > T_{c}^{ZTM}$ happens if the system enters a pairing-density-wave phase when $T$ goes right below $T_c$.
According to \eqnref{eq:T_ZTM}, the lower bound of $T_{c}^{ZTM}$ is the lower bound of $T_c$.
In the following, we focus on the $T_{c}^{ZTM}$, which is for the zero-total-momentum Cooper pairings.

Since we only focus on the zero-total-momentum Cooper pairings, we only consider the channel of the attractive interaction in \eqnref{eq:eff_int_ph_induced} that accounts for such pairings, resulting in the following total Hamiltonian
\eq{
\label{eq:H_zerokCP_gen}
H_{ZTM} = \sum_{\bsl{k},n_1,n_2} \gamma^\dagger_{\bsl{k},n_1} \gamma_{\bsl{k},n_2} h_{n_1 n_2}(\bsl{k}) + \frac{1}{2}\sum_{\bsl{k},\bsl{k}_1} \sum_{n_1,n_2,n_3,n_4} U_{n_1 n_2 n_3 n_4}(\bsl{k},\bsl{k}_1) \gamma^\dagger_{\bsl{k}, n_1} \gamma^\dagger_{-\bsl{k}, n_2} \gamma_{-\bsl{k}_1, n_3} \gamma_{\bsl{k}_1, n_4}\ ,
}
where
\eq{
U_{n_1 n_2 n_3 n_4}(\bsl{k},\bsl{k}_1) = \frac{2}{N} \sum_{l}\frac{G_{n_2 n_3 l}(-\bsl{k},-\bsl{k}_1) G_{n_1 n_4 l}(\bsl{k},\bsl{k}_1) \hbar \omega_l(\bsl{k}_1-\bsl{k}) }{(E_{n_1}(\bsl{k}) - E_{n_4}(\bsl{k}_1))^2 -(\hbar \omega_l(\bsl{k}_1-\bsl{k}))^2 }\ ,
}
and we have neglect the Coulomb interaction.
Hermiticity gives that
\eq{
\label{eq:Hermiticity_U}
U_{n_4 n_3 n_2 n_1}^*(\bsl{k}_1,\bsl{k}) = U_{n_1 n_2 n_3 n_4}(\bsl{k},\bsl{k}_1)
}
Generally, neglecting the Coulomb interaction is only valid in the low energies after the renormalization, which means that \eqnref{eq:H_zerokCP_gen} is only valid within certain energy cutoff $\epsilon_c$.
From the interacting Hamiltonian \eqnref{eq:H_zerokCP_gen}, the mean-field Hamiltonian reads
\eq{
\label{eq:H_zerokCP_MF_gen}
H_{ZTM,MF} = \sum_{\bsl{k},n_1,n_2} \gamma^\dagger_{\bsl{k},n_1} \gamma_{\bsl{k},n_2} (h_{n_1 n_2}(\bsl{k}) -\mu \delta_{n_1n_2}) + \frac{1}{2}\sum_{\bsl{k}} \sum_{n_1,n_2} \Delta_{n_1 n_2}(\bsl{k}) \gamma^\dagger_{\bsl{k}, n_1} \gamma^\dagger_{-\bsl{k}, n_2} + \frac{1}{2}\sum_{\bsl{k}} \sum_{n_1,n_2} \Delta_{n_2 n_1}^*(\bsl{k}) \gamma_{-\bsl{k}, n_1} \gamma_{\bsl{k}, n_2} \ ,
}
where the order parameter $\Delta_{n_1 n_2}(\bsl{k})$ satisfies the following self-consistent equation
\eq{
\label{eq:SGE_gen}
\Delta_{n_1 n_2}(\bsl{k}) = \sum_{\bsl{k}_1,n_3 n_4} U_{n_1 n_2 n_3 n_4}(\bsl{k},\bsl{k}_1) \frac{\Tr\left[ e^{-\beta H_{ZTM,MF}} \gamma_{-\bsl{k}_1,n_3} \gamma_{\bsl{k}_1,n_4} \right]}{\Tr\left[ e^{-\beta H_{ZTM,MF}} \right]}\ ,
}
$\beta = 1/(k_B T)$, $k_B$ is the Boltzmann constant, and $T$ is the temperature.
As the superconductivity transition in 3D is normally a second-order phase transition near which $\Delta_{n_1 n_2}(\bsl{k})$ is infinitesimally small, we can expand the right-hand side of the self-consistent equation (\eqnref{eq:SGE_gen}) in series of $\Delta_{n_1 n_2}(\bsl{k})$ and only keep terms to first order in $\Delta_{n_1 n_2}(\bsl{k})$, resulting in the linearized gap equation~\cite{Sigrist1991SC}
\eqa{
\label{eq:LGE_gen}
& \Delta_{n_1n_2}(\bsl{k})  = \frac{1}{\beta}\sum_{\bsl{k}_1,\omega,n_3,n_4} U_{n_1n_2n_3n_4}(\bsl{k},\bsl{k}_1)   \frac{1}{\ii \omega - E_{n_4}(\bsl{k}_1) +\mu}  \frac{1}{\ii \omega + E_{n_3}(-\bsl{k}_1) -\mu}  \Delta_{n_4n_3}(\bsl{k}_1) \\
& \Leftrightarrow \Delta_{n_1n_2}(\bsl{k})  = -\sum_{\bsl{k}_1,n_3,n_4} U_{n_1n_2n_3n_4}(\bsl{k},\bsl{k}_1) \zeta(E_{n_4}(\bsl{k}_1)-\mu, -E_{n_3}(-\bsl{k}_1)+\mu)    \Delta_{n_4n_3}(\bsl{k}_1) \ ,
}
where $\omega\in (2\dsZ +1) \frac{\pi}{\beta}$ is the fermionic Matsubara frequency, and 
\eq{
\label{eq:zeta}
\zeta( a, b) = \frac{\frac{1}{e^{\beta b }+1} -\frac{1}{e^{\beta a }+1}  }{a-b}
}
which is derived from~\cite{altland2010condensed}
\eq{
\frac{1}{\beta}\sum_{\omega} \frac{1}{\ii \omega - E } =   \frac{1}{e^{\beta E }+1}\ .
}
The mean-field superconducting critical temperature $T_c^{ZTM}$ is determined by the linearized gap equation (\eqnref{eq:LGE_gen}).
Explicitly, we solve the linearized gap equation for the temperature $T$ and the nonvanishing pairing gap function, and the highest $T$ given by the solutions is $T_c^{ZTM}$.

$\zeta( a, b)$ in the linearized gap equation (\eqnref{eq:zeta}) is always positive for finite $\beta>0$ and finite $a,b$, since $\frac{1}{e^{\beta b }+1} > \frac{1}{e^{\beta a }+1}$ for $a>b$, and 
\eq{
\lim_{\epsilon\rightarrow 0} \zeta( a, a+ \epsilon) =  - \lim_{\epsilon\rightarrow 0}  \frac{\frac{1}{e^{\beta (a+\epsilon)} +1}-\frac{1}{e^{\beta a }+1}  }{\epsilon} =   \frac{\beta e^{\beta a } }{(e^{\beta a }+1)^2} >0\ .
}
Therefore, the linearized gap equation in \eqnref{eq:LGE_gen} is equivalent to 
\eqa{
\label{eq:LGE_gen_alt}
& \Delta_{n_1n_2}(\bsl{k}) \sqrt{\zeta(E_{n_1}(\bsl{k})-\mu, -E_{n_2}(-\bsl{k})+\mu)}  \\
& = -\sum_{\bsl{k}_1,n_3,n_4} \sqrt{\zeta(E_{n_1}(\bsl{k})-\mu, -E_{n_2}(-\bsl{k})+\mu)}  U_{n_1n_2n_3n_4}(\bsl{k},\bsl{k}_1) \zeta(E_{n_4}(\bsl{k}_1)-\mu, -E_{n_3}(-\bsl{k}_1)+\mu)    \Delta_{n_4n_3}(\bsl{k}_1) \ .
}
The form of the linearized gap equation in \eqnref{eq:LGE_gen_alt} is useful for our discussion.
Inspired by \eqnref{eq:LGE_gen_alt}, let us define a vector $X$ and a matrix $M$ to be 
\eqa{
\label{eq:X_M}
& X_{n_1n_2\bsl{k}} = \Delta_{n_1n_2}(\bsl{k}) \sqrt{\zeta(E_{n_1}(\bsl{k})-\mu, -E_{n_2}(-\bsl{k})+\mu)} \\
& M_{n_1n_2\bsl{k}, n_4n_3\bsl{k}_1} = -\sqrt{\zeta(E_{n_1}(\bsl{k})-\mu, -E_{n_2}(-\bsl{k})+\mu)}  U_{n_1n_2n_3n_4}(\bsl{k},\bsl{k}_1) \sqrt{\zeta(E_{n_4}(\bsl{k}_1)-\mu, -E_{n_3}(-\bsl{k}_1)+\mu)}\ .
}
With \eqnref{eq:X_M}, \eqnref{eq:LGE_gen_alt} becomes 
\eqa{
\label{eq:LGE_gen_M_X}
X = M X\ .
}
Hermiticity of $U$ in \eqnref{eq:Hermiticity_U} gives that
\eqa{
M_{n_4n_3\bsl{k}_1,n_1n_2\bsl{k}}^* & = -\sqrt{\zeta(E_{n_4}(\bsl{k}_1)-\mu, -E_{n_3}(-\bsl{k}_1)+\mu)}  U_{n_4n_3n_2n_1}^*(\bsl{k}_1,\bsl{k}) \sqrt{\zeta(E_{n_1}(\bsl{k})-\mu, -E_{n_2}(-\bsl{k})+\mu)} \\
& = -\sqrt{\zeta(E_{n_1}(\bsl{k})-\mu, -E_{n_2}(-\bsl{k})+\mu)}  U_{n_1n_2n_3n_4}(\bsl{k},\bsl{k}_1) \sqrt{\zeta(E_{n_4}(\bsl{k}_1)-\mu, -E_{n_3}(-\bsl{k}_1)+\mu)} \\
& =  M_{n_1n_2\bsl{k}, n_4n_3\bsl{k}_1}\ ,
}
meaning that $M$ is Hermitian.

With the reformulated linearized gap equation \eqnref{eq:X_M} and the fact that $M$ is Hermitian, we now have a direct equation for the critical temperature:
\eq{
\label{eq:LGE_M_T}
\det(M-1) = 0\ ,
}
which is solved for the temperature $T$.
More explicitly, $M$ has real eigenvalues, labelled as $m_0\geq m_1 \geq m_2 ...$.
Each eigenvalue of $M$ should be a continuous function of $T$, since  $M$ continuously depends on the temperature (\eqnref{eq:X_M}).
When a certain eigenvalue of $M$ intersects with $1$, the corresponding temperature at the intersection point is a solution to \eqnref{eq:LGE_M_T}.
The largest $T$ that satisfies \eqnref{eq:LGE_M_T} is $T_c^{ZTM}$.

Now we show $T_c^{ZTM}$ is the largest $T$ that satisfies $m_0 = 1$, where $m_0$ is the largest eigenvalue of $M$.
To see this, first note that $\zeta( a, b)$ in \eqnref{eq:zeta} limits to zero as $T\rightarrow \infty$, and thus $M$ limits to zero as $T\rightarrow \infty$ according to \eqnref{eq:X_M}.
So all eigenvalues of $M$ limit to zero as $T\rightarrow \infty$.
Suppose $T'$ is a generic solution to $\det(M-1) = 0$, then there exists an eigenvalue $m_l$ of $M$ such that $T'$ satisfies $m_l =1$.
Since $m_l\leq m_0$, we have $m_0\geq 1$ at $T'$.
Combined with the fact that $\lim_{T\rightarrow \infty} m_0 = 0$ and $m_0$ is a continuous function of $T'$, $m_0=1$ will happen in $[T',\infty)$, which means that $T'$ is no larger than the largest $T$ that satisfies $m_0 = 1$.
Therefore, the largest $T$ that satisfies $m_0 = 1$ is the largest $T$ that satisfies $\det(M-1) = 0$, which is equal to $T_c^{ZTM}$.

With this preparation, now we show how $T_c^{ZTM}$ is bounded from below by $\lambda$  in \eqnref{eq:lambda} for TR-invariant systems.
As a good approximation for many mean-field superconductors, we assume that the cutoff $\epsilon_c$ of the model is much larger than the temperature of interest $\epsilon_c\gg k_B T $, and the electron bands are dispersive with a large Fermi velocity.
(With this approximation, our results cannot be applied to flat-band superconductors.)
The TR symmetry acts on $\gamma_{n,\bsl{k}}$ as 
\eq{
\TR \gamma_{n,\bsl{k}} \TR^{-1} = \gamma_{n,-\bsl{k}} e^{\ii \phi_{n}(\bsl{k})} \text{ for $\bsl{k}$ not at TR-invariant points}\ ,
}
and then the EPC $G_{nml}(\bsl{k},\bsl{k}_1)$ satisfies
\eq{
\label{eq:Gnml_TR}
e^{\ii \phi_{n}(\bsl{k})} G_{nml}^*(\bsl{k},\bsl{k}_1) e^{-\ii \phi_{m}(\bsl{k}_1)} = G_{nml}(-\bsl{k},-\bsl{k}_1) \text{ for $\bsl{k}$ and $\bsl{k}_1$ not at TR-invariant points}\ ,
}
owing to the phonon TR transformation that we choose in \eqnref{eq:TR_phonon}.
Next we will use the fact that for any nonvanishing $Y$, we always have the variational principle
\eq{
\label{eq:variational_princeple}
\frac{ Y^\dagger M Y }{ Y^\dagger Y} \leq m_0\ ,
}
where recall that $m_0$ is the largest eigenvalue of $M$.
To relate to $\lambda$, let us choose a special $Y=X'$, where 
\eq{
\label{eq:X'}
X'_{n_1 n_2 \bsl{k}} =  \delta_{n_1 n_2 }e^{\ii \phi_{n_1}(\bsl{k})} \sqrt{\zeta(E_{n_1}(\bsl{k})-\mu, - E_{n_2}(-\bsl{k}) +\mu) }\ .
}
Note that we have $E_n(\bsl{k})=E_n(-\bsl{k})$ for TR-invariant system, and we have
\eq{
\zeta(a,-a) = \frac{\tanh(\beta a /2)}{2a}\ .
}
The TR-invariant points are just measure zero subset of 1BZ, and thus can be ruled out from the summation of $\bsl{k}$.
Then, we obtain
\eqa{
(X')^\dagger X' &  = \sum_{n_1 n_2 \bsl{k}} \delta_{n_1 n_2 }   \zeta(E_{n_1}(\bsl{k})-\mu, - E_{n_2}(-\bsl{k}) +\mu) =   \sum_{n \bsl{k}} \frac{\tanh(\beta (E_{n}(\bsl{k})-\mu) /2)}{2(E_{n}(\bsl{k})-\mu)}  \\
& = \int_{-\epsilon_c}^{\epsilon_c} d\xi \frac{\tanh(\beta\xi /2)}{2\xi}\sum_{n \bsl{k}} \delta( E_{n}(\bsl{k})-\mu -\xi)\approx \int_{-\epsilon_c}^{\epsilon_c} d\xi \frac{\tanh(\beta\xi /2)}{2\xi} \sum_{n \bsl{k}} \delta( E_{n}(\bsl{k})-\mu) \approx  \log\left[\frac{2 e^{\gamma_{Euler}} \epsilon_c }{\pi k_B T}\right] D(\mu)\ ,
}
where $\gamma_{Euler} = 0.5772...$ is the Euler constant, and we have used the assumption that the electron bands are dispersive with a large Fermi velocity for the first ``$\approx$", and used $\epsilon_c\gg k_B T $ for the second ``$\approx$".
Moreover, 
\eqa{
(X')^\dagger M X' &  = \sum_{n_1 n_2 \bsl{k}} \sum_{n_4 n_3 \bsl{k}_1} (X'_{n_1 n_2 \bsl{k}})^* M_{n_1 n_2 \bsl{k}, n_4 n_3 \bsl{k}_1} X'_{n_4 n_3 \bsl{k}_1}  \\
& = - \sum_{n_1 n_2 \bsl{k}} \sum_{n_4 n_3 \bsl{k}_1} \delta_{n_1 n_2 }e^{\ii \phi_{n_1}(\bsl{k})} \zeta(E_{n_1}(\bsl{k})-\mu, -E_{n_2}(-\bsl{k})+\mu)  U_{n_1n_2n_3n_4}(\bsl{k},\bsl{k}_1) \\
& \qquad \times \zeta(E_{n_4}(\bsl{k}_1)-\mu, -E_{n_3}(-\bsl{k}_1)+\mu) \delta_{n_4 n_3 }e^{-\ii \phi_{n_4}(\bsl{k}_1)} \\
& = - \sum_{n_1 \bsl{k}} \sum_{n_4 \bsl{k}_1}  e^{\ii \phi_{n_1}(\bsl{k})} \frac{\tanh(\beta (E_{n_1}(\bsl{k})-\mu) /2)}{2(E_{n_1}(\bsl{k})-\mu)}  U_{n_1n_1n_4n_4}(\bsl{k},\bsl{k}_1) \frac{\tanh(\beta (E_{n_4}(\bsl{k}_1)-\mu) /2)}{2(E_{n_4}(\bsl{k}_1)-\mu)} e^{-\ii \phi_{n_4}(\bsl{k}_1)} \\
& =  -\int_{-\epsilon_c}^{\epsilon_c} d\xi_1 \frac{\tanh(\beta\xi_1 /2)}{2\xi_1} \int_{-\epsilon_c}^{\epsilon_c} d\xi_2 \frac{\tanh(\beta\xi_1 /2)}{2\xi_2}(-) \sum_{n_1  \bsl{k}} \sum_{n_4 \bsl{k}_1} \delta(E_{n_1}(\bsl{k})-\mu-\xi_1)\delta(E_{n_4}(\bsl{k}_1)-\mu-\xi_4) \\
& \qquad  \times e^{\ii \phi_{n_1}(\bsl{k})}  U_{n_1n_1n_4n_4}(\bsl{k},\bsl{k}_1) e^{-\ii \phi_{n_4}(\bsl{k}_1)}\\
& \approx  -\left(\log\left[\frac{2 e^{\gamma_{Euler}} \epsilon_c }{\pi k_B T}\right] \right)^2 \sum_{n_1  \bsl{k}} \sum_{n_4 \bsl{k}_1} \delta(E_{n_1}(\bsl{k})-\mu)\delta(E_{n_4}(\bsl{k}_1)-\mu) \\   
& \qquad  \times \frac{2}{N} \sum_{l}\frac{G_{n_1 n_4 l}(-\bsl{k},-\bsl{k}_1) G_{n_1 n_4 l}(\bsl{k},\bsl{k}_1) \hbar \omega_l(\bsl{k}_1-\bsl{k}) }{(E_{n_1}(\bsl{k}) - E_{n_4}(\bsl{k}_1))^2 -(\hbar \omega_l(\bsl{k}_1-\bsl{k}))^2 }e^{\ii \phi_{n_1}(\bsl{k})}e^{-\ii \phi_{n_4}(\bsl{k}_1)} \ ,
}
where we again have used the approximation that the electron bands are dispersive with a large Fermi velocity for the first ``$\approx$" and used $\epsilon_c\gg k_B T $.
Combined with \eqnref{eq:Gnml_TR} and \eqnref{eq:lambda}, we arrive at
\eqa{
(X')^\dagger M X' & \approx \left(\log\left[\frac{2 e^{\gamma_{Euler}} \epsilon_c }{\pi k_B T}\right] \right)^2 \sum_{n_1  \bsl{k}} \sum_{n_4 \bsl{k}_1} \delta(E_{n_1}(\bsl{k})-\mu)\delta(E_{n_4}(\bsl{k}_1)-\mu) \frac{2}{N} \sum_{l}\frac{|G_{n_1 n_4 l}(\bsl{k},\bsl{k}_1)|^2 }{\hbar \omega_l(\bsl{k}_1-\bsl{k}) } \\
& = \left(\log\left[\frac{2 e^{\gamma_{Euler}} \epsilon_c }{\pi k_B T}\right] \right)^2 \lambda D(\mu)\ .
}
Therefore, we have
\eq{
\frac{(X')^\dagger M X'}{(X')^\dagger X'} = \log\left[\frac{2 e^{\gamma_{Euler}} \epsilon_c }{\pi k_B T}\right]   \lambda  \ .
}
Owing to $m_0 \geq \frac{(X')^\dagger M X'}{(X')^\dagger X'}$, the solution to $\frac{(X')^\dagger M X'}{(X')^\dagger X'}=1$ must be no larger than the largest $T$ that satisfies $m_0 = 1$, which is $T_{c}^{ZTM}$.
As  $\frac{(X')^\dagger M X'}{(X')^\dagger X'}=1$ gives 
\eq{
T = \frac{2 e^{\gamma_{Euler}} \epsilon_c }{\pi k_B } e^{-\frac{1}{\lambda}}\ ,
}
we arrive at 
\eq{
\label{eq:Tc_lower_bound_lambda}
T_c \geq T_{c}^{ZTM} \geq  \frac{2 e^{\gamma_{Euler}} \epsilon_c }{\pi k_B } e^{-\frac{1}{\lambda}} = 1.13\frac{ \epsilon_c }{ k_B } e^{-\frac{1}{\lambda}}\ .
}
We note that in the lower bound \eqnref{eq:Tc_lower_bound_lambda}, $T_c$ can be given by any kind of pairing form, even if it is not a uniform $s$-wave pairing (like pairing density waves).
It is because we never assume any pairing form for $T_c$, although we choose a special $Y=X'$ in the variational principle (\eqnref{eq:variational_princeple}) to derive the lower bound \eqnref{eq:Tc_lower_bound_lambda}.

\subsection{Neglecting spin for evaluating $\lambda$ in the presence of spin $\SU(2)$ symmetry}
\label{app:spin_SU2}

The discussion in \appref{app:general_EPC}, \appref{app:2center} and \appref{app:geo_EPC_symmetry-rep} holds for both spinful and spinless fermions.
However, in reality, we cannot have spinless fermions due to spin statistics~\cite{Srednicki2007QFT}.
At best, what we have is spinful fermions with spin $\SU(2)$ symmetry, \ie, the spin-orbit coupling is negligible.
In this part, we will discuss $\lambda$ (\eqnref{eq:lambda}) in the presence of spin $\SU(2)$ symmetry (\ie, the spin-orbit coupling is negligible).

When the system has spin $\SU(2)$ symmetry, we can neglect the spin when evaluating $\lambda$ in \eqnref{eq:lambda} and $\mcomega$ in \eqnref{eq:McMillan_omega_square_ave}.
Based on the general form of the EPC Hamiltonian \eqnref{eq:H_el_ph_gen}, spin $\SU(2)$ symmetry means that the spinful band index $n_{\text{spinful}}$ can be split to the spinless band index $n_{\text{spinless}}$ and the spin index $s$, \ie, $n_{\text{spinful}}=(n_{\text{spinless}},s)$.
Then, the spinful EPC matrix reduces as
\eqa{
  G_{n_{\text{spinful}} m_{\text{spinful}}l}^{\text{spinful}}(\bsl{k},\bsl{k}') = G_{n_{\text{spinless}}s m_{\text{spinless}}s'l}^{\text{spinful}}(\bsl{k},\bsl{k}') =  G_{n_{\text{spinless}}m_{\text{spinless}}l}^{\text{spinless}}(\bsl{k},\bsl{k}') \delta_{ss'}\ .
}
Combined with the spin-double degeneracy of the electron bands, we obtain 
\eqa{
& \lambda_{\text{spinful}}  = \lambda_{\text{spinless}}\\
& \mcomega_{\text{spinful}}  = \mcomega_{\text{spinless}}\ .
}
For graphene and {\mgb} discussed in \appref{app:graphene} and \appref{app:MgB2}, the spin-orbit coupling is always negligible, and we will always use spinless models.

\section{Geometric and Topological Contributions to EPC Constant in Graphene}

\label{app:graphene}

In this section, we discuss the EPC in graphene.
We neglect the spin-orbit coupling in graphene and assume spin $\SU(2)$ symmetry.
Therefore, we will use a spinless model throughout this section.
As a result, all the symmetry operations that we consider are spinless.

\subsection{Review: Electron Hamiltonian and Electron Band Topology}

In this part, we review the electron Hamiltonian of graphene and electron band topology in it following \refcite{Neto2009GrapheneRMP} and \refcite{Ahn2019TBGFragile}.

\subsubsection{Review: Electron Hamiltonian of Graphene}

Besides the charge $\U(1)$ symmetry and 2D lattice translations, the relevant symmetry group of the spinless model for graphene is generated by the six-fold rotation $C_6$, the mirror $m_y$ that flips the $y$ direction, the mirror $m_z$ with mirror plane lying in the plane of graphene, and TR symmetry $\TR$, where the spatial symmetries form the space group P6/mmm.
The model of graphene is constructed from $C_6$, $m_y$, $m_z$ and $\TR$.

Graphene has two atoms per unit cell, and at low energies, we only need to consider one $p_z$-like orbital at each atom. $p_z$-like means that the orbital behaves the same as a $p_z$ orbital under the symmetry operations of interest.
Specifically, according to the convention defined in \appref{app:EPC_real_space}, we have $\bsl{\tau}\in\{ \bsl{\tau}_A,\bsl{\tau}_B\}$ for the sublattice and the orbital index $\alpha_{\bsl{\tau}}$ can be omitted since it only takes one value---$p_z$, where 
\eqa{
& \bsl{\tau}_A = \frac{a}{\sqrt{3}}(\frac{\sqrt{3}}{2}, -\frac{1}{2} , 0)^T \\
& \bsl{\tau}_B = \frac{a}{\sqrt{3}}(\frac{\sqrt{3}}{2}, \frac{1}{2} , 0)^T \ ,
}
and $a$ is the lattice constant of graphene.
Then, the creation operators for electrons are labelled as $c^{\dagger}_{\bsl{R}+\bsl{\tau}}$, where $\bsl{R}\in \bsl{a}_1\dsZ + \bsl{a}_2\dsZ $ with
\eqa{
\label{eq:a1_a2_graphene}
& \bsl{a}_1 = a (\frac{1}{2} , \frac{\sqrt{3}}{2} , 0)^T\\
& \bsl{a}_2 = a (-\frac{1}{2} , \frac{\sqrt{3}}{2} , 0)^T\ .
}

The symmetry reps furnished by $c^{\dagger}_{\bsl{R}+\bsl{\tau}}$ are
\eqa{
&
C_6 c^{\dagger}_{\bsl{R}+\bsl{\tau}} C_6^{-1} = c^{\dagger}_{C_6(\bsl{R}+\bsl{\tau})} = c^{\dagger}_{ \bsl{R}_{C_6,\bsl{\tau}} + \bsl{\tau}_{C_6}} \\
& 
m_y c^{\dagger}_{\bsl{R}+\bsl{\tau}} m_y^{-1} = c^{\dagger}_{m_y(\bsl{R}+\bsl{\tau})} = c^{\dagger}_{m_y \bsl{R} + \bsl{\tau}_{m_y}} \\
&
m_z c^{\dagger}_{\bsl{R}+\bsl{\tau}} m_z^{-1} = -c^{\dagger}_{\bsl{R}+\bsl{\tau}} \\
&
\TR  c^{\dagger}_{\bsl{R}+\bsl{\tau}}  \TR^{-1} = c^{\dagger}_{\bsl{R}+\bsl{\tau}}\ ,
}
where 
\eqa{
& \bsl{\tau}_{C_6} = \bsl{\tau}_{m_y} = \left\{ 
\begin{array}{ll}
    \bsl{\tau}_B & \text{ if } \bsl{\tau} = \bsl{\tau}_A  \\
    \bsl{\tau}_A & \text{ if } \bsl{\tau} = \bsl{\tau}_B 
\end{array}
\right. \\ 
& \bsl{R}_{C_6,\bsl{\tau}} = \left\{ 
\begin{array}{ll}
    C_6 \bsl{R} & \text{ if } \bsl{\tau} = \bsl{\tau}_A  \\
    C_6 \bsl{R} +\bsl{a}_2 & \text{ if } \bsl{\tau} = \bsl{\tau}_B 
\end{array}
\right. \ .
}
According to the convention in \eqnref{eq:sym_rep_g_R}, we know
\eqa{
\label{eq:sym_rep_graphene}
& U_{C_6}^{\bsl{\tau}_A \bsl{\tau}_B} = U_{C_6}^{ \bsl{\tau}_B \bsl{\tau}_A} = 1 \Rightarrow U_{C_6} = \tau_x \\
& U_{m_y}^{\bsl{\tau}_A \bsl{\tau}_B} = U_{m_y}^{ \bsl{\tau}_B \bsl{\tau}_A} = 1 \Rightarrow U_{m_y} = \tau_x \\
& U_{m_z}^{\bsl{\tau} \bsl{\tau}} = -1 \Rightarrow U_{m_z} = -\tau_0 \\
& U_{\TR}^{\bsl{\tau} \bsl{\tau}} = 1 \Rightarrow U_{\TR} = \tau_0 \ ,
}
where $\tau_0$ is the identity matrix in the sublattice space, and $\tau_x$, $\tau_y$ and $\tau_z$ are the Pauli matrices in the sublattice space.
By using \eqnref{eq:FT_rule}, we have the electron basis in momentum space as 
\eq{
c^\dagger_{\bsl{k}} = ( c^\dagger_{\bsl{k},\bsl{\tau}_A}, c^\dagger_{\bsl{k},\bsl{\tau}_B} )\ ,
}
and the symmetry reps read
\eqa{
& 
C_6 c^{\dagger}_{\bsl{k}} C_6^{-1} =c^{\dagger}_{C_6\bsl{k}} \tau_x\\
& 
m_y c^{\dagger}_{\bsl{k}} m_y^{-1} = c^{\dagger}_{m_y\bsl{k}} \tau_x\\
&
m_z c^{\dagger}_{\bsl{k}} m_z^{-1} = -c^{\dagger}_{\bsl{k}} \\
&
\TR  c^{\dagger}_{\bsl{k}}  \TR^{-1} = c^{\dagger}_{-\bsl{k}}\ .
}

According to \eqnref{eq:H_el_gen}, the electron Hamiltonian of graphene in general reads
\eq{
H_{el} = \sum_{\bsl{R}_1\bsl{R}_2}\sum_{\bsl{\tau}_1\bsl{\tau}_2} t_{\bsl{\tau}_1\bsl{\tau}_2}(\bsl{R}_1+ \bsl{\tau}_1 -\bsl{R}_2 - \bsl{\tau}_2) c^\dagger_{\bsl{R}_1+ \bsl{\tau}_1}c_{\bsl{R}_2+ \bsl{\tau}_2}\ .
}
In this work, we only consider the electron hopping up to the nearest neighbors for graphene, \ie,
\eq{
\label{eq:NN_t}
t_{\bsl{\tau}_1\bsl{\tau}_2}(\bsl{R}_1+ \bsl{\tau}_1 -\bsl{R}_2 - \bsl{\tau}_2) = 0\ \forall\ |\bsl{R}_1+ \bsl{\tau}_1 -\bsl{R}_2 - \bsl{\tau}_2|>\frac{a}{\sqrt{3}}\ ,
}
where $t_{\bsl{\tau}_1\bsl{\tau}_2}(\bsl{R}_1+ \bsl{\tau}_1 -\bsl{R}_2 - \bsl{\tau}_2)$ is defined in \eqnref{eq:H_el_gen}.
Then, the terms that are allowed to be zero are $t_{\bsl{\tau}\bsl{\tau}}(0)$ (the onsite energy), $t_{\bsl{\tau}_A\bsl{\tau}_B}(\bsl{\delta}_j)$ and $t_{\bsl{\tau}_B\bsl{\tau}_A}(-\bsl{\delta}_j)$,
where 
\eq{
\label{eq:delta_j_graphene}
\bsl{\delta}_j = C_3^{j}\frac{a}{\sqrt{3}} (0,-1,0)^T \text{ with }j=0,1,2\ ,
}
and $C_3= C_6^2$.

Deriving the forms of $t_{\bsl{\tau}\bsl{\tau}}(0)$, $t_{\bsl{\tau}_A\bsl{\tau}_B}(\bsl{\delta}_j)$ and $t_{\bsl{\tau}_B\bsl{\tau}_A}(-\bsl{\delta}_j)$ based on symmetries has been well-described in the literature (\eg,  \refcite{Neto2009GrapheneRMP}).
Here we will recap it as a comparison to the later derivation for EPC.
By using \eqnref{eq:t_h_sym_g}, \eqnref{eq:t_h_sym_TR} and \eqnref{eq:sym_rep_graphene}, we obtain 
\eqa{
\label{eq:t_sym_graphene_1}
& C_6: t_{\bsl{\tau}_A\bsl{\tau}_A}(0) = t_{\bsl{\tau}_B\bsl{\tau}_B}(0) \\
& m_y: t_{\bsl{\tau}_A\bsl{\tau}_A}(0) = t_{\bsl{\tau}_B\bsl{\tau}_B}(0) \\
& m_z: \text{no constraints} \\
& \TR: t_{\bsl{\tau}_A\bsl{\tau}_A}(0)\ ,\  t_{\bsl{\tau}_B\bsl{\tau}_B}(0) \in \dsR\\
& h.c.: t_{\bsl{\tau}_A\bsl{\tau}_A}(0)\ ,\  t_{\bsl{\tau}_B\bsl{\tau}_B}(0) \in \dsR
}
and
\eqa{
\label{eq:t_sym_graphene_2}
& C_6: t_{\bsl{\tau}_A\bsl{\tau}_B}(\bsl{\delta}_0) = t_{\bsl{\tau}_B\bsl{\tau}_A}(-\bsl{\delta}_2) =  t_{\bsl{\tau}_A\bsl{\tau}_B}(\bsl{\delta}_1)=  t_{\bsl{\tau}_B\bsl{\tau}_A}(-\bsl{\delta}_0) = t_{\bsl{\tau}_A\bsl{\tau}_B}(\bsl{\delta}_2) =  t_{\bsl{\tau}_B\bsl{\tau}_A}(-\bsl{\delta}_1) \\
& m_y: t_{\bsl{\tau}_A\bsl{\tau}_B}(\bsl{\delta}_0) = t_{\bsl{\tau}_B\bsl{\tau}_A}(-\bsl{\delta}_0),\ t_{\bsl{\tau}_A\bsl{\tau}_B}(\bsl{\delta}_1)=  t_{\bsl{\tau}_B\bsl{\tau}_A}(-\bsl{\delta}_2),\ t_{\bsl{\tau}_A\bsl{\tau}_B}(\bsl{\delta}_2) =  t_{\bsl{\tau}_B\bsl{\tau}_A}(-\bsl{\delta}_1) \\
& m_z: \text{no constraints} \\
& \TR: t_{\bsl{\tau}_A\bsl{\tau}_B}(\bsl{\delta}_j) \ ,\ t_{\bsl{\tau}_B\bsl{\tau}_A}(-\bsl{\delta}_j) \in \dsR\\
& h.c.: t_{\bsl{\tau}_A\bsl{\tau}_B}(\bsl{\delta}_j)= t_{\bsl{\tau}_B\bsl{\tau}_A}^*(-\bsl{\delta}_j)\ ,
}
resulting in 
\eqa{
& t_{\bsl{\tau}_A\bsl{\tau}_A}(0) = t_{\bsl{\tau}_B\bsl{\tau}_B}(0) = \epsilon_0 \in\dsR \\
& t_{\bsl{\tau}_A\bsl{\tau}_B}(\bsl{\delta}_j)=t_{\bsl{\tau}_B\bsl{\tau}_A}(-\bsl{\delta}_j)= t \in\dsR\ .
}
Then, based on the \eqnref{eq:H_el_gen_k}, $H_{el}$ for graphene in momentum space becomes
\eq{
\label{eq:H_el_graphene}
H_{el} = \sum_{\bsl{k}} c^\dagger_{\bsl{k}} h(\bsl{k}) c_{\bsl{k}}\ ,
}
where 
\eq{
\label{eq:h_k_graphene}
h(\bsl{k}) = \epsilon_0 \tau_0 + t \sum_{j} \mat{ 0 & e^{-\ii \bsl{\delta}_j\cdot \bsl{k}} \\ e^{\ii \bsl{\delta}_j\cdot \bsl{k} } & 0}\ ,
}
and $\bsl{\delta}_j$ is defined in \eqnref{eq:delta_j_graphene}.

\subsubsection{Review: Electron Band Topology in Graphene}

Now we turn to the topological invariants in the low-energy model of graphene.

Let us first consider a generic 2D or 3D $P \TR$-invariant spinless Bloch Hamiltonian $H_{\bsl{k}}$
\eq{
\label{eq:H_k_gen}
H_{\bsl{k}} = e^{\ii \bsl{k}\cdot \hat{\bsl{r}}} H_{el}  e^{-\ii \bsl{k}\cdot \hat{\bsl{r}}}
}
with $\hat{\bsl{r}}$ the position operator, where $P$ is the inversion symmetry and $H_{el}$ is a generic single-particle electron Hamiltonian with lattice translation symmetries.
Suppose two bands (whose Bloch states are labelled as $e^{\ii \bsl{k}\cdot \hat{\bsl{r}}}\ket{u_{\bsl{k},1}}$ and $e^{\ii \bsl{k}\cdot \hat{\bsl{r}}}\ket{u_{\bsl{k},2}}$) are isolated in a region $D\in\BZ$.
Then, owing to $ (P \TR)^2 = 1$, we can always choose $\ket{u_{\bsl{k}}}= ( \ket{u_{\bsl{k},1}}, \ket{u_{\bsl{k},2}} ) R_{\bsl{k}} $ such that $\ket{u_{\bsl{k}}}$ is smooth in $D$ and $P \TR \ket{u_{\bsl{k}}} = \ket{u_{\bsl{k}}}\tau_x$.
The rep of $P \TR$ requires the projected Hamiltonian on $\ket{u_{\bsl{k}}}$ to have the following form
\eq{
\bra{u_{\bsl{k}}} H_{\bsl{k}} \ket{u_{\bsl{k}}} = d_0(\bsl{k}) \tau_0 + d_x(\bsl{k}) \tau_x  + d_y(\bsl{k}) \tau_y\ ,
}
where $d_0(\bsl{k})$, $d_x(\bsl{k})$ and $d_y(\bsl{k})$ are smooth in $D$.
Then, for any closed 1D loop $\L$ in $D$ such that $d_x(\bsl{k})^2 + d_y(\bsl{k})^2 \neq 0\ \forall\bsl{k}\in\L$, a winding number can be defined as
\eq{
\label{eq:PT_winding_number}
W_{\L} = \frac{1}{2\pi}\int_{\L} d\bsl{k}\cdot \ii e^{\ii \theta(\bsl{k})}\bsl{\nabla}_{\bsl{k}}e^{-\ii \theta(\bsl{k})}\ .
}
In general, only $|W_{\L}|$ is well-defined since $\ket{u_{\bsl{k}}}\rightarrow \ket{u_{\bsl{k}}}\tau_x$ leads to $W_{\L}\rightarrow - W_{\L}$.
Nevertheless, if we have two loops $\L_1$ and $\L_2$, the relative signs between $W_{\L_1}$ and $W_{\L_2}$ are also well-defined.
\eqnref{eq:PT_winding_number} is the $P\TR$ protected winding number proposed in \refcite{Ahn2019TBGFragile}.
Nonzero $|W_{\L}|$ indicates that $\L$ encloses at least a gapless point for 2D or a gapless line in 3D.
A numerical method to calculate \eqnref{eq:PT_winding_number} can be found in \refcite{Aris2014Wilsonloop,Yu2021EOCP}.

Going back to \eqnref{eq:h_k_graphene} explicitly for graphene,  $P\TR$ is equivalent to $C_2\TR$ symmetry for 2D spinless system (if we neglect the $z$ direction).
For graphene, $C_2\TR$ is represented as
\eq{
\label{eq:psi_C2T}
C_2 \TR c^{\dagger}_{\bsl{k}} \left(C_2 \TR\right)^{-1} =c^{\dagger}_{\bsl{k}}  \tau_x
}
with $ c^{\dagger}_{\bsl{k}} =  (c^{\dagger}_{\bsl{k},\bsl{\tau}_A},  c^{\dagger}_{\bsl{k},\bsl{\tau}_B})$.
Owing to the $C_2\TR$, $h(\bsl{k})$ in \eqnref{eq:H_el_graphene} always have the form of 
\eq{
h(\bsl{k})= d_0(\bsl{k}) \tau_0 + d_x(\bsl{k})\tau_x + d_y(\bsl{k}) \tau_y  = d_0(\bsl{k}) \tau_0 + \frac{\Delta E(\bsl{k})}{2} \mat{ & e^{-\ii \theta(\bsl{k})}  \\ e^{\ii \theta(\bsl{k})} &  }\ ,
}
$\Delta E(\bsl{k}) = 2 \sqrt{d_x^2(\bsl{k})+d_y^2(\bsl{k})}$ is the electron gap, and
\eq{
\label{eq:theta_graphene}
e^{\ii \theta(\bsl{k})} = \frac{d_x(\bsl{k})+\ii d_y(\bsl{k})}{\sqrt{d_x^2(\bsl{k})+d_y^2(\bsl{k})}}\ .
}
In addition, because of $C_2\TR$ symmetry, we also have the following useful relations 
\eqa{
\label{eq:E_P_graphene}
& \Delta E(\bsl{k}) = 2 \sqrt{d_x^2(\bsl{k})+d_y^2(\bsl{k})}\ ,\ E_n(\bsl{k}) = d_0(\bsl{k}) + (-)^n \frac{\Delta E(\bsl{k})}{2}\\
& P_{n}(\bsl{k}) = \frac{1}{2} \left[ 1 + (-)^n  \frac{d_x(\bsl{k})\tau_x + d_y(\bsl{k}) \tau_y}{\sqrt{d_x^2(\bsl{k})+d_y^2(\bsl{k})}}\right] =  \frac{1}{2} \left[ 1 + (-)^n  \mat{ & e^{-\ii \theta(\bsl{k})}  \\ e^{\ii \theta(\bsl{k})} &  }\right] \Rightarrow \tau_z P_{n}(\bsl{k}) \tau_z = 1 -P_{n}(\bsl{k})\ ,
}
where $n=1,2$.

Explicitly for graphene, we have
\eq{
\label{eq:d_graphene}
d_0(\bsl{k}) = \epsilon_0\ ,\  d_x(\bsl{k}) = t   \sum_{j} \cos\left(\ii \bsl{\delta}_j\cdot \bsl{k}\right)\ ,\  d_y (\bsl{k}) = t   \sum_{j} \sin\left(\ii \bsl{\delta}_j\cdot \bsl{k}\right)\ .
}
Clearly, we have two bands that are isolated over the entire $\BZ$, which means $D=\BZ$.
In particular, there are two gapless points at $\pm\K$, where 
\eq{
\label{eq:K_graphene}
\K = \frac{4\pi}{3 a} (1,0,0)^T\ .
}
If the loop $\L$ only encloses $\K$ or $-\K$ once, the winding in \eqnref{eq:PT_winding_number} has the value $|W_{\L}| = 1$.
We label $W_{\K} = W_{\L}$ for any loop $\L$ that encloses $\K$ once, and label $W_{-\K} = W_{\L}$ for any loop $\L$ that encloses $-\K$ once, meaning that $|W_{\K}| = |W_{-\K}| = 1$.
Specifically, since $D = \BZ$, the relative sign between $W_{\K}$ and $W_{-\K}$ is well-defined, and we have $W_{\K} = - W_{-\K}$, featuring the opposite chiralities.
In sum, we know graphene is a topological semimetal according to the definition in \refcite{Bradlyn2017TQC}.

\subsection{EPC Hamiltonian of Graphene}

The EPC Hamiltonian of graphene has been derived in \refcite{Thingstad20191211EPCSCGraphenehBN} based on symmetries.
In this part, we will rederive the EPC Hamiltonian by using the formalism in \appref{app:geo_EPC_symmetry-rep}, since the resultant form will be convenient for our further study of geometric and topological contributions.

According to \eqnref{eq:g_i_g_beta}, what we need to derive are $ \widetilde{f}_{\beta}(\bsl{k})$ and $f_z(\bsl{k})$.
First, owing to \eqnref{eq:g_k_sym} and \eqnref{eq:sym_rep_graphene}, $m_z$ requires $f_z(\bsl{k})$ to be zero, \ie,
\eq{
\label{eq:f_z_zero_graphene}
f_z(\bsl{k}) = - f_z(\bsl{k}) \Rightarrow f_z(\bsl{k}) = 0\ .
}
Therefore, \eqnref{eq:g_i_g_beta} for graphene becomes
\eq{
\label{eq:g_i_g_beta_grpahene_ini}
 f_{i}(\bsl{k})  = \left[ \ii \partial_{k_i}  \widetilde{f}_{\shpa}( \bsl{k} )  + \ii \sum_{i'=x,y} \epsilon_{i'i}\partial_{k_{i'}}    \widetilde{f}_{\perp}(\bsl{k} ) \right](\delta_{ix} + \delta_{iy})\ ,
}
meaning that we only need to consider $ \widetilde{f}_{\beta}(\bsl{k})$.

Similar to the electron hopping, we again assume that 
\eq{
\label{eq:NN_g}
f_{\bsl{\tau}_1\bsl{\tau}_2,i}(\bsl{R}_1+\bsl{\tau}_1-\bsl{R}_2 - \bsl{\tau}_2) = 0 \ \forall\ |\bsl{R}_1+\bsl{\tau}_1-\bsl{R}_2 - \bsl{\tau}_2|>\frac{a}{\sqrt{3}}\ ,
}
which means 
\eq{
f_{\bsl{\tau}_1\bsl{\tau}_2,\beta}(\bsl{R}_1+\bsl{\tau}_1-\bsl{R}_2 - \bsl{\tau}_2) = 0 \ \forall\ |\bsl{R}_1+\bsl{\tau}_1-\bsl{R}_2 - \bsl{\tau}_2|>\frac{a}{\sqrt{3}}\ .
}

According to \eqnref{eq:g_beta_sym}, $\widetilde{f}_{\bsl{\tau}_1\bsl{\tau}_2,\shpa}(\bsl{R}_1+\bsl{\tau}_1-\bsl{R}_2 - \bsl{\tau}_2)$ should have the same symmetry properties as $t_{\bsl{\tau}_1\bsl{\tau}_2}(\bsl{R}_1+\bsl{\tau}_1-\bsl{R}_2 - \bsl{\tau}_2)$ in \eqnref{eq:t_sym_graphene_1}-\eqref{eq:t_sym_graphene_2}.
Specifically, the constraints on $\widetilde{f}_{\bsl{\tau}_1\bsl{\tau}_2,\shpa}(\bsl{R}_1+\bsl{\tau}_1-\bsl{R}_2 - \bsl{\tau}_2\neq 0)$ reads
\eqa{
& C_6: \widetilde{f}_{\bsl{\tau}_A\bsl{\tau}_B,\shpa}(\bsl{\delta}_0) = \widetilde{f}_{\bsl{\tau}_B\bsl{\tau}_A,\shpa}(-\bsl{\delta}_2) =  \widetilde{f}_{\bsl{\tau}_A\bsl{\tau}_B,\shpa}(\bsl{\delta}_1)=  \widetilde{f}_{\bsl{\tau}_B\bsl{\tau}_A,\shpa}(-\bsl{\delta}_0) = \widetilde{f}_{\bsl{\tau}_A\bsl{\tau}_B,\shpa}(\bsl{\delta}_2) =  \widetilde{f}_{\bsl{\tau}_B\bsl{\tau}_A,\shpa}(-\bsl{\delta}_1) \\
& m_y: \widetilde{f}_{\bsl{\tau}_A\bsl{\tau}_B,\shpa}(\bsl{\delta}_0) = \widetilde{f}_{\bsl{\tau}_B\bsl{\tau}_A,\shpa}(-\bsl{\delta}_0),\ \widetilde{f}_{\bsl{\tau}_A\bsl{\tau}_B,\shpa}(\bsl{\delta}_1)=  \widetilde{f}_{\bsl{\tau}_B\bsl{\tau}_A,\shpa}(-\bsl{\delta}_2),\ \widetilde{f}_{\bsl{\tau}_A\bsl{\tau}_B,\shpa}(\bsl{\delta}_2) =  \widetilde{f}_{\bsl{\tau}_B\bsl{\tau}_A,\shpa}(-\bsl{\delta}_1) \\
& m_z: \text{no constraints} \\
& \TR: \widetilde{f}_{\bsl{\tau}_A\bsl{\tau}_B,\shpa}(\bsl{\delta}_j) \ ,\ \widetilde{f}_{\bsl{\tau}_B\bsl{\tau}_A,\shpa}(-\bsl{\delta}_j) \in \dsR\\
& h.c.: \widetilde{f}_{\bsl{\tau}_A\bsl{\tau}_B,\shpa}(\bsl{\delta}_j)= \widetilde{f}_{\bsl{\tau}_B\bsl{\tau}_A,\shpa}^*(-\bsl{\delta}_j)\ ,
}
resulting in 
\eqa{
\label{eq:g_shpa_R_graphene}
& \widetilde{f}_{\bsl{\tau}_A\bsl{\tau}_B,\shpa}(\bsl{\delta}_j)=\widetilde{f}_{\bsl{\tau}_B\bsl{\tau}_A,\shpa}(-\bsl{\delta}_j)= \hat{\gamma} \in\dsR\ .
}

In addition, according to \eqnref{eq:g_beta_sym}, the symmetry constraints of $ \widetilde{f}_{\bsl{\tau}_1 \bsl{\tau}_2,\perp}(\bsl{R}_1+\bsl{\tau}_1-\bsl{R}_2 - \bsl{\tau}_2)$ only differ by a minus sign for $m_y$ from those of  $t_{\bsl{\tau}_1\bsl{\tau}_2}(\bsl{R}_1+\bsl{\tau}_1-\bsl{R}_2 - \bsl{\tau}_2)$ in \eqnref{eq:t_sym_graphene_1}-\eqref{eq:t_sym_graphene_2}.
Specifically, the constraints on $\widetilde{f}_{\bsl{\tau}_1\bsl{\tau}_2,\perp}(\bsl{R}_1+\bsl{\tau}_1-\bsl{R}_2 - \bsl{\tau}_2\neq 0)$ read
\eqa{
& C_6: \widetilde{f}_{\bsl{\tau}_A\bsl{\tau}_B,\perp}(\bsl{\delta}_0) = \widetilde{f}_{\bsl{\tau}_B\bsl{\tau}_A,\perp}(-\bsl{\delta}_2) =  \widetilde{f}_{\bsl{\tau}_A\bsl{\tau}_B,\perp}(\bsl{\delta}_1)=  \widetilde{f}_{\bsl{\tau}_B\bsl{\tau}_A,\perp}(-\bsl{\delta}_0) = \widetilde{f}_{\bsl{\tau}_A\bsl{\tau}_B,\perp}(\bsl{\delta}_2) =  \widetilde{f}_{\bsl{\tau}_B\bsl{\tau}_A,\perp}(-\bsl{\delta}_1) \\
& m_y: \widetilde{f}_{\bsl{\tau}_A\bsl{\tau}_B,\perp}(\bsl{\delta}_0) = -\widetilde{f}_{\bsl{\tau}_B\bsl{\tau}_A,\perp}(-\bsl{\delta}_0),\ \widetilde{f}_{\bsl{\tau}_A\bsl{\tau}_B,\perp}(\bsl{\delta}_1)=  -\widetilde{f}_{\bsl{\tau}_B\bsl{\tau}_A,\perp}(-\bsl{\delta}_2),\ \widetilde{f}_{\bsl{\tau}_A\bsl{\tau}_B,\perp}(\bsl{\delta}_2) =  -\widetilde{f}_{\bsl{\tau}_B\bsl{\tau}_A,\perp}(-\bsl{\delta}_1) \\
& m_z: \text{no constraints} \\
& \TR: \widetilde{f}_{\bsl{\tau}_A\bsl{\tau}_B,\perp}(\bsl{\delta}_j) \ ,\ \widetilde{f}_{\bsl{\tau}_B\bsl{\tau}_A,\perp}(-\bsl{\delta}_j) \in \dsR\\
& h.c.: \widetilde{f}_{\bsl{\tau}_A\bsl{\tau}_B,\perp}(\bsl{\delta}_j)= \widetilde{f}_{\bsl{\tau}_B\bsl{\tau}_A,\perp}^*(-\bsl{\delta}_j)\ .
}
From $C_6$, we know $\widetilde{f}_{\bsl{\tau}_A\bsl{\tau}_B,\perp}(\bsl{\delta}_j)=\widetilde{f}_{\bsl{\tau}_B\bsl{\tau}_A,\perp}(-\bsl{\delta}_j)=f_{NN,\perp}$ for all $j=0,1,2$.
From $m_y$, we know $\widetilde{f}_{\bsl{\tau}_A\bsl{\tau}_B,\perp}(\bsl{\delta}_0) = -\widetilde{f}_{\bsl{\tau}_B\bsl{\tau}_A,\perp}(-\bsl{\delta}_0) $, which, combined with $C_6$, means $f_{NN,\perp}=0$.
Therefore, we have
\eqa{
\label{eq:g_perp_R_graphene}
& \widetilde{f}_{\bsl{\tau}_A\bsl{\tau}_B,\perp}(\bsl{\delta}_j)=\widetilde{f}_{\bsl{\tau}_B\bsl{\tau}_A,\perp}(-\bsl{\delta}_j)= 0\ .
}

In addition, according to the convention that we choose in \eqnref{eq:g_beta_onsite_zero}, we have $\widetilde{f}_{\bsl{\tau}_A\bsl{\tau}_A,\beta}(0) = \widetilde{f}_{\bsl{\tau}_B\bsl{\tau}_B,\beta}(0) = 0$.
By substituting \eqnref{eq:g_shpa_R_graphene} and \eqnref{eq:g_perp_R_graphene} with \eqnref{eq:g_beta_k_block}, we obtain 
\eqa{
\label{eq:g_beta_k_graphene}
 &   \widetilde{f}_{\shpa}(\bsl{k}) = \hat{\gamma} \sum_{j} \mat{ 0 & e^{-\ii \bsl{\delta}_j\cdot \bsl{k}} \\ e^{\ii \bsl{\delta}_j\cdot \bsl{k} } & 0} \\
 &   \widetilde{f}_{\perp}(\bsl{k}) = 0\ .
}
$f_i(\bsl{k})$ in \eqnref{eq:g_i_g_beta_grpahene_ini} now becomes
\eqa{
\label{eq:g_i_g_beta_grpahene}
 f_{i}(\bsl{k})  = \ii \partial_{k_i}  \widetilde{f}_{\shpa}( \bsl{k} ) (\delta_{ix} + \delta_{iy})\ .
}

By using \eqnref{eq:H_el-ph_k}, \eqnref{eq:f_k_2center} and \eqnref{eq:g_i_g_beta_grpahene}, $H_{el-ph}$ for graphene in momentum space becomes
\eq{
\label{eq:H_el-ph_k_graphene}
H_{el-ph} = \frac{1}{\sqrt{N}} \sum_{\bsl{\tau},i} \sum_{\bsl{k}_1,\bsl{k}_2}^{\text{\BZ}}  c^\dagger_{\bsl{k}_1}F_{\bsl{\tau}i}(\bsl{k}_1,\bsl{k}_2)c_{\bsl{k}_2} u_{\bsl{k}_2-\bsl{k}_1,\bsl{\tau},i}^\dagger \ ,
}
where 
\eq{
\label{eq:f_k_graphene}
F_{\bsl{\tau}i}(\bsl{k}_1,\bsl{k}_2) = \ii  \chi_{\bsl{\tau}} \partial_{k_{2,i}}  \widetilde{f}_{\shpa}( \bsl{k}_2 ) - \ii \partial_{k_{1,i}}  \widetilde{f}_{\shpa}( \bsl{k}_1 ) \chi_{\bsl{\tau}}\ ,
}
and $ \widetilde{f}_{\shpa}(\bsl{k})$ is in \eqnref{eq:g_beta_k_graphene}.

\subsection{Symmetry-Rep Method: Energetic and Geometric Parts of EPC Hamiltonian of Graphene}
\label{app:geo_EPC_graphene}

Now we identify the energetic and geometric parts of $f_{i}(\bsl{k})$ in \eqnref{eq:g_i_g_beta_grpahene}.
First, for $\L_\perp$ in \eqnref{eq:g_perp_h_relation}, we try to find the constant matrices $Q_l$ (labelled by $l$) that (i) are TR-invariant constant, (ii) satisfy $U_g Q_l U_g^\dagger = Q_l z_{R,\perp}$ for any crystalline symmetry $g=\{ R | \bsl{d} \}$ of graphene, and (iii) given rise to \eqnref{eq:Q_h_ortho_rel}.
Based on the expressions of $U_g$ in \eqnref{eq:sym_rep_graphene}, we find that the only $Q_l$ that satisfies the condition is $Q_l = 0$ with $l$ only taking one value.
Therefore, we have $\L_\perp = 0$ for \eqnref{eq:g_perp_h_relation}.
Combined with $ \widetilde{f}_{\perp}(\bsl{k}) = 0$ as shown in \eqnref{eq:g_beta_k_graphene}, we also have $\Delta  \widetilde{f}_{\perp} = 0 $ in \eqnref{eq:g_perp_h_relation}.
Since we also show $f_z(\bsl{k}) = 0 $ in \eqnref{eq:f_z_zero_graphene}, we have $\Delta f_i (\bsl{k}) = 0$ in \eqnref{eq:Delta_g_i_2D}.
Thus, we only need to care about $ \widetilde{f}_{\shpa}(\bsl{k})$.

According to \eqnref{eq:g_shpa_h_relation},  we try to re-write $ \widetilde{f}_{\shpa}(\bsl{k})$ in \eqnref{eq:g_i_g_beta_grpahene} as 
\eq{
\widetilde{f}_{\shpa}(\bsl{k})=\hat{\gamma}_0 \partial_{\epsilon_0} h(\bsl{k}) + \hat{\gamma} \partial_{t} h(\bsl{k})\ , 
}
where $h(\bsl{k})$ is the electron Hamiltonian in \eqnref{eq:h_k_graphene}.
By comparing \eqnref{eq:g_beta_k_graphene} to \eqnref{eq:h_k_graphene}, $\hat{\gamma}_0$ in $ \widetilde{f}_{\shpa}(\bsl{k})$ is zero, and then we obtain
\eq{
 \widetilde{f}_{\shpa}(\bsl{k}) = \hat{\gamma} \partial_t  h(\bsl{k})\ .
}
For all $\bsl{k}$ with $\Delta E(\bsl{k})\neq 0$ ($\Delta E(\bsl{k})$ is the difference between two electron bands of the system defined in \eqnref{eq:E_P_graphene}), \eqnref{eq:E_P_graphene} and \eqnref{eq:d_graphene} show that both $\sum_{n} \partial_{k_i} E_n(\bsl{k}) P_{n}(\bsl{k})  (\delta_{ix} + \delta_{iy})$ and $\sum_{n} E_n(\bsl{k}) \partial_{k_i}  P_{n}(\bsl{k})  (\delta_{ix} + \delta_{iy})$ are proportional to $t$.
Explicitly,
\eqa{
& \sum_{n} \partial_{k_i} E_n(\bsl{k}) P_{n}(\bsl{k})  (\delta_{ix} + \delta_{iy}) \\
& = \sum_{n} \partial_{k_i} \left[ \epsilon_0 + (-)^n \sqrt{d_x^2(\bsl{k})+d_y^2(\bsl{k})} \right]  \frac{1}{2} \left[ 1 + (-)^n  \frac{d_x(\bsl{k})\tau_x + d_y(\bsl{k}) \tau_y}{\sqrt{d_x^2(\bsl{k})+d_y^2(\bsl{k})}}\right] (\delta_{ix} + \delta_{iy}) \\
& = \sum_{n}\partial_{k_i} \left[  (-)^n \sqrt{d_x^2(\bsl{k})+d_y^2(\bsl{k})} \right] \frac{1}{2} \left[  (-)^n  \frac{d_x(\bsl{k})\tau_x + d_y(\bsl{k}) \tau_y}{\sqrt{d_x^2(\bsl{k})+d_y^2(\bsl{k})}}\right] (\delta_{ix} + \delta_{iy}) }
is proportional to $t$ since $d_x(\bsl{k})$ and $d_y(\bsl{k})$ are proportional to $t$ as shown by \eqnref{eq:d_graphene}, and 
\eqa{
& \sum_{n} E_n(\bsl{k}) \partial_{k_i} P_{n}(\bsl{k})  (\delta_{ix} + \delta_{iy}) \\
& = \sum_{n} \left[ \epsilon_0 + (-)^n \sqrt{d_x^2(\bsl{k})+d_y^2(\bsl{k})} \right]  \frac{1}{2} \partial_{k_i} \left[ 1 + (-)^n  \frac{d_x(\bsl{k})\tau_x + d_y(\bsl{k}) \tau_y}{\sqrt{d_x^2(\bsl{k})+d_y^2(\bsl{k})}}\right] (\delta_{ix} + \delta_{iy}) \\
& = \sum_{n} \left[  (-)^n \sqrt{d_x^2(\bsl{k})+d_y^2(\bsl{k})} \right] \frac{1}{2} \partial_{k_i}\left[  (-)^n  \frac{d_x(\bsl{k})\tau_x + d_y(\bsl{k}) \tau_y}{\sqrt{d_x^2(\bsl{k})+d_y^2(\bsl{k})}}\right] (\delta_{ix} + \delta_{iy})
}
is proportional to $t$ since $d_x(\bsl{k})$ and $d_y(\bsl{k})$ are proportional to $t$ as shown by \eqnref{eq:d_graphene}.
Then, we can directly replace $\partial_t$ by $1/t$ in $f^E_i$ and $f^{geo}_i$ in \eqnref{eq:g_2D_E_geo}, resulting in
\eqa{
\label{eq:g_E_graphene}
f^E_i(\bsl{k}) & = \ii \hat{\gamma} \partial_t  \sum_{n} \partial_{k_i} E_n(\bsl{k}) P_{n}(\bsl{k})  (\delta_{ix} + \delta_{iy})\\
 & = \ii \frac{\hat{\gamma}}{t}  \sum_{n} \partial_{k_i} E_n(\bsl{k}) P_{n}(\bsl{k}) (\delta_{ix} + \delta_{iy})\\
 & = \ii \gamma \sum_{n} \partial_{k_i} E_n(\bsl{k}) P_{n}(\bsl{k}) (\delta_{ix} + \delta_{iy})
}
\eqa{
\label{eq:g_geo_graphene}
f^{geo}_i(\bsl{k}) & = \ii \hat{\gamma} \partial_t  \sum_{n} E_n(\bsl{k}) \partial_{k_i}  P_{n}(\bsl{k})  (\delta_{ix} + \delta_{iy})\\
& = \ii \frac{\hat{\gamma}}{t}  \sum_{n} E_n(\bsl{k}) \partial_{k_i}  P_{n}(\bsl{k})  (\delta_{ix} + \delta_{iy})\\
 & = \ii  \gamma \sum_{n} E_n(\bsl{k}) \partial_{k_i} P_{n}(\bsl{k}) (\delta_{ix} + \delta_{iy}) \ ,
}
where we define 
\eq{
\label{eq:graphene_gamma}
\gamma = \frac{\hat{\gamma}}{t}\ .
}
As a result, we obtain 
\eq{
\label{eq:g_E_g_geo_graphene}
 f_{i}(\bsl{k})  =  f^E_i(\bsl{k}) + f^{geo}_i(\bsl{k}) \ ,
}
where 
\eqa{
\label{eq:g_E_g_geo_graphene_explicit}
& f^E_i(\bsl{k}) = \ii \gamma \sum_{n} \partial_{k_i} E_n(\bsl{k}) P_{n}(\bsl{k}) (\delta_{ix} + \delta_{iy})\\
& f^{geo}_i(\bsl{k}) = \ii  \gamma \sum_{n} E_n(\bsl{k}) \partial_{k_i} P_{n}(\bsl{k})(\delta_{ix} + \delta_{iy})\ .
}

We emphasize that we have used the fact that the hopping in graphene is nearest-neighboring (NN).
In \appref{app:further_range_hopping_graphene}, we will show that the symmetry-rep method can still separate the EPC $f_i$ into $f^E_i+f^{geo}_i$  with the hopping derivatives replaced by coefficients even if we include the next-nearest-neighboring (NNN) hopping, but will fail if we have 3rd NN hopping.

\subsection{Gaussian Approximation: Energetic and Geometric Parts of EPC Hamiltonian of Graphene}
\label{app:geo_EPC_graphene_Gaussian}

In \appref{app:geo_EPC_graphene}, we have derived the energetic and geometric parts of EPC Hamiltonian of graphene using the symmetry-rep method.
In this part, we will show that \eqnref{eq:g_E_g_geo_graphene} can also be derived from the GA introduced in \appref{app:gaussian}. 

As shown in \eqnref{eq:f_z_zero_graphene}, it is clear that the out-of-plane motions of ions cannot couple to electrons, and thus we only need to consider the in-plane motions of ions.
This means that we can safely treat $p_z$ orbitals as $s$-like orbitals for graphene, since $p_z$ behaves the same as $s$ orbital in the 2D $x-y$ plane.
Combined with the fact that we only have one kind of atoms for graphene, the hopping function $t(\bsl{r})$ has no sublattice or orbital indices.
As a result, under the tight-binding and the Frohlich two-center approximation, the EPC Hamiltonian is the same as  \eqnref{eq:H_elph_Frohlich}.
Then, with the GA $t(\bsl{r}) = t_0 \exp[\gamma \frac{x^2 + y^2}{2}]$, \eqnref{eq:g_k_Frolich_Gaussian_s} shows that $f_i(\bsl{k})$ in \eqnref{eq:g_i_g_beta_grpahene} should read
\eq{
\label{eq:f_i_GA_graphene}
 f_{i}(\bsl{k}) = \ii \gamma \partial_{k_i} h(\bsl{k}) (\delta_{ix} + \delta_{iy}) \ ,
}
where $h(\bsl{k})$ is the electron matrix Hamiltonian for graphene, and $(\delta_{ix} + \delta_{iy})$ comes from the fact that all atoms of the graphene lie in the $x-y$ plane.
As a result, \eqnref{eq:g_E_Gaussian_s} and \eqnref{eq:g_geo_Gaussian_s} suggest that the GA gives the same expression for the energetic and geometric parts of EPC as those in \eqnref{eq:g_E_g_geo_graphene}.

\subsection{General Symmetry-Allowed Hopping Form: Consistent with Gaussian Approximation}

Now We show that even if we use generic hopping function $t(\bsl{r})$ instead of GA, we would give the same expression as \eqnref{eq:f_i_GA_graphene}.
To see this, we use the idea of linear combinations of atomic orbitals proposed in \refcite{Slater1954LCAO}.
Nevertheless, instead of using $\O(3)$ symmetry in \refcite{Slater1954LCAO}, let us use $\O(2)$ and $m_z$ symmetries, since $p_z$ is not a rep of $\O(3)$ group but an irrep of $\O(2)$ and $m_z$ symmetries.
Since the electron do not couple to the ion motions along $z$ to the linear order, we only look at $(x,y)$.
As a result, owing to $\O(2)$ symmetry, we get
\eq{
t(\bsl{r}) = t(\sqrt{x^2 + y^2})\ .
}
As a result, we have
\eq{
\partial_{i} t(\bsl{r}) = \frac{r_i}{r_\shpa} \partial_{r_\shpa} t(r_\shpa) \left( \delta_{ix} + \delta_{iy} \right)\ ,
}
where $r_\shpa = \sqrt{x^2 + y^2}$.
As we know, the decay of the hopping is very strong in graphene, and thus we only keep the nearest-neighboring term, \ie, $\sqrt{x^2 + y^2} = \frac{a}{\sqrt{3}}$.
Then, we eventually have 
\eq{
\partial_{i} t(\bsl{r}) = \frac{1 }{r_\shpa} \partial_{r_\shpa} \log\left( |t(r_\shpa)| \right)  r_i t(r_\shpa) \left( \delta_{ix} + \delta_{iy} \right) \approx  \gamma r_i  t(\bsl{r})  \left( \delta_{ix} + \delta_{iy} \right)\ ,
}
where
\eq{
 \gamma = \left. \frac{1 }{r_\shpa} \partial_{r_\shpa} \log\left( |t(r_\shpa)| \right) \right|_{r_\shpa = \frac{a}{\sqrt{3}}}\ .
}
Therefore, we obtain the same form of $\partial_{i} t(\bsl{r})$ as \eqnref{eq:t_gradient_Gaussian_s} derived from Gaussian approximation, and eventually leads to the $\partial_{k_i}h(\bsl{k})$ form of $f_i(\bsl{k})$  in \eqnref{eq:f_i_GA_graphene}, as well as the energetic and geometric parts of EPC as those in \eqnref{eq:g_E_g_geo_graphene}.

\subsection{Analytical Geometric and Topological Contributions to EPC constant in Graphene}
\label{app:labmda_contributions_graphene}

With \eqnref{eq:g_E_g_geo_graphene}, we are now ready to analytically derive the geometric contribution to EPC constant $\lambda$ in graphene, as well as the topological lower bound of the geometric contribution.
For this purpose, we choose $\mu\neq \epsilon_0$, meaning that two energy bands do not touch (\ie, $\Delta E(\bsl{k})\neq 0$) on the Fermi surface, \ie, the Fermi surface does not include the Dirac points.
(Recall that $\epsilon_0$ is the onsite energy of the electron as shown in \eqnref{eq:h_k_graphene}.)

Before deriving the bounds, let us first simplify $\left\langle \Gamma \right\rangle$ in \eqnref{eq:Gamma_ave_mu_2center} for graphene.
First, \eqnref{eq:sym_rep_graphene}, we have 
\eq{
C_3 c^\dagger_{\bsl{k}} C_3^{-1} = c^\dagger_{C_3\bsl{k}}\ ,
}
which means 
\eq{
P_{n}(C_3\bsl{k}) = P_{n}(\bsl{k})\ ,\  E_{n}(C_3\bsl{k}) = E_{n}(\bsl{k})
}
for all $\bsl{k}$ such that $\Delta E(\bsl{k})\neq 0$ (\ie, $\bsl{k}$ not at $\pm \K$), according to \eqnref{eq:E_P_graphene}.
Furthermore, based on the second line of \eqnref{eq:g_k_sym}, we have
\eq{
\label{eq:C_3_f_i_graphene}
\sum_{i'=x,y}f_{i'}(\bsl{k}) \mat{\frac{-1}{2} & -\frac{\sqrt{3}}{2} \\ \frac{\sqrt{3}}{2} & \frac{-1}{2}}_{ii'} = f_{i} (C_3\bsl{k}) \ \forall i=x,y\ ,
}
where $\mat{\frac{-1}{2} & -\frac{\sqrt{3}}{2} \\ \frac{\sqrt{3}}{2} & \frac{-1}{2}}$ is the 2D part of the rotation matrix of $C_3$.
Then, combined with the fact that we choose $\mu$ not at the Dirac point (meaning that $\Delta E(\bsl{k})\neq 0$ on the Fermi surface), we obtain
\eqa{
& \sum_{\bsl{k}}^{\BZ}\sum_{n}  \delta\left(\mu - E_n(\bsl{k}) \right) \chi_{\bsl{\tau}} f_{i}(\bsl{k})  P_{n}(\bsl{k}) \\
& = \sum_{\bsl{k}\in\BZ}^{\Delta E(\bsl{k})\neq 0}\sum_{n}  \delta\left(\mu - E_n(C_3\bsl{k}) \right) \chi_{\bsl{\tau}} f_{i}(C_3\bsl{k})  P_{n}(C_3\bsl{k}) \\
& = \sum_{i'=x,y} \sum_{\bsl{k}\in\BZ}^{\Delta E(\bsl{k})\neq 0}\sum_{n}  \delta\left(\mu - E_n(\bsl{k}) \right) \chi_{\bsl{\tau}} f_{i'}(\bsl{k})   P_{n}(\bsl{k})  \mat{\frac{-1}{2} & -\frac{\sqrt{3}}{2} \\ \frac{\sqrt{3}}{2} & \frac{-1}{2}}_{ii'} 
}
for $i=x,y$, which means that
\eq{
\label{eq:Gamma_ave_sim_1}
 \sum_{\bsl{k}}^{\BZ}\sum_{n}  \delta\left(\mu - E_n(\bsl{k}) \right) \chi_{\bsl{\tau}} f_{i}(\bsl{k})  P_{n}(\bsl{k}) = 0\ .
}
$\chi_{\bsl{\tau}}$ is the projection matrix to the $\bsl{\tau}$ sublattice defined in \eqnref{eq:chi_gen}.
In addition, owing to \eqnref{eq:E_P_graphene}, we have
\eq{
\label{eq:Gamma_ave_sim_2}
\sum_{\bsl{\tau}} \chi_{\bsl{\tau}} P_{n}(\bsl{k}) \chi_{\bsl{\tau}}  = \frac{1}{2} (\chi_{\bsl{\tau}_A}+\chi_{\bsl{\tau}_B}) P_{n}(\bsl{k}) (\chi_{\bsl{\tau}_A}+\chi_{\bsl{\tau}_B}) + \frac{1}{2} (\chi_{\bsl{\tau}_A}-\chi_{\bsl{\tau}_B}) P_{n}(\bsl{k}) (\chi_{\bsl{\tau}_A}-\chi_{\bsl{\tau}_B}) =  \frac{1}{2} P_{n}(\bsl{k}) + \frac{1}{2}\tau_z P_{n}(\bsl{k}) \tau_z = \frac{1}{2}\ ,
}
for all $\bsl{k}$ such that $\Delta E(\bsl{k})\neq 0$, which comes from the fact that the electron Hamiltonian (besides the identity term) has the sublattice chiral symmetry.
By substituting \eqnref{eq:f_z_zero_graphene}, \eqnref{eq:Gamma_ave_sim_1} and \eqnref{eq:Gamma_ave_sim_2} into \eqnref{eq:Gamma_ave_mu_2center}, we arrive at
\eqa{
\label{eq:Gamma_ave_ini_graphene}
  \left\langle \Gamma \right\rangle & = - \frac{\hbar}{D^2(\mu) } \sum_{\bsl{\tau}}\sum_{i=x,y}  \frac{1}{m_{\bsl{\tau}}}  \sum_{\bsl{k}_1,\bsl{k}_2}^{\BZ}\sum_{n,m} \delta\left(\mu - E_n(\bsl{k}_1) \right) \delta\left(\mu - E_m(\bsl{k}_2) \right)  \Tr\left[ f_{i}(\bsl{k}_1) P_{n}(\bsl{k}_1)  f_{i}(\bsl{k}_1)\chi_{\bsl{\tau}} P_{m}(\bsl{k}_2)   \chi_{\bsl{\tau}}  \right] \\
  & =  - \frac{\hbar}{D^2(\mu) m_{\text{C}}} \sum_{i=x,y}   \sum_{\bsl{k}_1\in\BZ}^{\Delta E(\bsl{k}_1)\neq 0}\sum_{\bsl{k}_2\in\BZ}^{\Delta E(\bsl{k}_2)\neq 0}\sum_{n,m} \delta\left(\mu - E_n(\bsl{k}_1) \right) \delta\left(\mu - E_m(\bsl{k}_2) \right)  \\
  & \quad \times \Tr\left[ f_{i}(\bsl{k}_1) P_{n}(\bsl{k}_1)  f_{i}(\bsl{k}_1)\sum_{\bsl{\tau}}\chi_{\bsl{\tau}} P_{m}(\bsl{k}_2)   \chi_{\bsl{\tau}}  \right] \\ 
  & =  - \frac{\hbar}{2 D(\mu) m_{\text{C}}} \sum_{i=x,y}   \sum_{\bsl{k}\in\BZ}^{\Delta E(\bsl{k})\neq 0}\sum_{n} \delta\left(\mu - E_n(\bsl{k}) \right)  \Tr\left[ f_{i}(\bsl{k}) P_{n}(\bsl{k})  f_{i}(\bsl{k}) \right] \ ,
}
where $m_{\text{C}}$ is the mass of the carbon atom, and we used \eqnref{eq:D_mu} for the third equality.

Now we substitute \eqnref{eq:g_E_g_geo_graphene} into \eqnref{eq:Gamma_ave_ini_graphene} and obtain
\eqa{
\label{eq:Gamma_ave_mu_graphene_ini_2}
\left\langle \Gamma \right\rangle = \left\langle \Gamma \right\rangle^{E-E} + \left\langle \Gamma \right\rangle^{E-geo} + \left\langle \Gamma \right\rangle^{geo-geo}\ ,
}
where 
\eqa{
& \left\langle \Gamma \right\rangle^{E-E} = - \frac{\hbar}{2 D(\mu) m_{\text{C}}} \sum_{i=x,y}   \sum_{\bsl{k}\in\BZ}^{\Delta E(\bsl{k})\neq 0}\sum_{n} \delta\left(\mu - E_n(\bsl{k}) \right)  \Tr\left[ f_{i}^E(\bsl{k}) P_{n}(\bsl{k})  f_{i}^E(\bsl{k}) \right] \\
& \left\langle \Gamma \right\rangle^{E-geo} = - \frac{\hbar}{2 D(\mu) m_{\text{C}}} \sum_{i=x,y}   \sum_{\bsl{k}\in\BZ}^{\Delta E(\bsl{k})\neq 0}\sum_{n} \delta\left(\mu - E_n(\bsl{k}) \right)  \Tr\left[ f_{i}^E(\bsl{k}) P_{n}(\bsl{k})  f_{i}^{geo}(\bsl{k}) \right] + c.c.\\
& \left\langle \Gamma \right\rangle^{geo-geo} = - \frac{\hbar}{2 D(\mu) m_{\text{C}}} \sum_{i=x,y}   \sum_{\bsl{k}\in\BZ}^{\Delta E(\bsl{k})\neq 0}\sum_{n} \delta\left(\mu - E_n(\bsl{k}) \right)  \Tr\left[ f_{i}^{geo}(\bsl{k}) P_{n}(\bsl{k})  f_{i}^{geo}(\bsl{k}) \right] \ .
}
Let us first analyze $\left\langle \Gamma \right\rangle^{E-geo}$, which reads
\eqa{
&
\left\langle \Gamma \right\rangle^{E-geo} 
\\
& 
= \gamma^2 \frac{\hbar}{2 D(\mu) m_{\text{C}}} \sum_{i=x,y}   \sum_{\bsl{k}\in\BZ}^{\Delta E(\bsl{k})\neq 0}\sum_{n} \delta\left(\mu - E_n(\bsl{k}) \right)  \Tr\left[ \sum_{n_1} \partial_{k_i} E_{n_1} (\bsl{k}) P_{n_1} (\bsl{k}) P_{n}(\bsl{k})  \sum_{n_2} E_{n_2}(\bsl{k}) \partial_{k_i} P_{n_2}(\bsl{k}) \right] + c.c. 
\\
&
= \gamma^2 \frac{\hbar}{2 D(\mu) m_{\text{C}}} \sum_{i=x,y}   \sum_{\bsl{k}\in\BZ}^{\Delta E(\bsl{k})\neq 0}\sum_{n} \delta\left(\mu - E_n(\bsl{k}) \right)  \Tr\left[ \partial_{k_i} E_{n} (\bsl{k})  P_{n}(\bsl{k})  \sum_{n_2} E_{n_2}(\bsl{k}) \partial_{k_i} P_{n_2}(\bsl{k}) \right] + c.c. 
\\
&
= \gamma^2 \frac{\hbar}{2 D(\mu) m_{\text{C}}} \sum_{i=x,y}   \sum_{\bsl{k}\in\BZ}^{\Delta E(\bsl{k})\neq 0}\sum_{n} \delta\left(\mu - E_n(\bsl{k}) \right)  \partial_{k_i} E_{n} (\bsl{k}) \sum_{n_2} E_{n_2}(\bsl{k})  \Tr\left[ P_{n}(\bsl{k}) \partial_{k_i} P_{n_2}(\bsl{k}) \right] + c.c. \ .
}
For $\Tr\left[ P_{n}(\bsl{k}) \partial_{k_i} P_{n_2}(\bsl{k}) \right]$, we have 
\eqa{
& \Tr\left[ P_{n}(\bsl{k}) \partial_{k_i} P_{n_2}(\bsl{k}) \right] \\
& = \Tr\left[ P_{n}(\bsl{k}) (\partial_{k_i} U_{n_2}(\bsl{k})) U_{n_2}^\dagger(\bsl{k})  \right] + \Tr\left[ P_{n}(\bsl{k})  U_{n_2}(\bsl{k}) (\partial_{k_i}U_{n_2}^\dagger(\bsl{k}))  \right]\\
& = \delta_{n,n_2}\Tr\left[ (\partial_{k_i} U_{n_2}(\bsl{k})) U_{n_2}^\dagger(\bsl{k})  \right] + \delta_{n,n_2} \Tr\left[  U_{n_2}(\bsl{k}) (\partial_{k_i}U_{n_2}^\dagger(\bsl{k}))  \right]\\
& = \delta_{n,n_2} \partial_{k_i} 1 = 0\ ,
}
resulting in 
\eq{
\label{eq:Gamma_Egeo_ave_mu_graphene}
\left\langle \Gamma \right\rangle^{E-geo} = 0\ .
}
\eqnref{eq:Gamma_Egeo_ave_mu_graphene} means that the geometric part of EPC does not couple with the energetic part of EPC in $\lambda$. 
The underlying reasons of \eqnref{eq:Gamma_Egeo_ave_mu_graphene} are the $C_3$ symmetry (\eqnref{eq:C_3_f_i_graphene}) and the sublattice chiral symmetry (\eqnref{eq:Gamma_ave_sim_2}).
\eqnref{eq:Gamma_ave_mu_graphene_ini_2} is simplified to 
\eq{
\left\langle \Gamma \right\rangle = \left\langle \Gamma \right\rangle^{E-E} + \left\langle \Gamma \right\rangle^{geo-geo}\ .
}

Now we turn to $\left\langle \Gamma \right\rangle^{E-E}$.
\eqa{
& \left\langle \Gamma \right\rangle^{E-E} \\
& = - \frac{\hbar}{2 D(\mu) m_{\text{C}}} \sum_{i=x,y}   \sum_{\bsl{k}\in\BZ}^{\Delta E(\bsl{k})\neq 0}\sum_{n} \delta\left(\mu - E_n(\bsl{k}) \right)  \Tr\left[ f_{i}^E(\bsl{k}) P_{n}(\bsl{k})  f_{i}^E(\bsl{k}) \right] \\
& = \gamma^2 \frac{\hbar}{2 D(\mu) m_{\text{C}}} \sum_{i=x,y}   \sum_{\bsl{k}\in\BZ}^{\Delta E(\bsl{k})\neq 0}\sum_{n} \delta\left(\mu - E_n(\bsl{k}) \right)  \Tr\left[ \sum_{n_1} \partial_{k_i} E_{n_1}(\bsl{k}) P_{n_1}(\bsl{k}) P_{n}(\bsl{k}) \sum_{n_2} \partial_{k_i} E_{n_2}(\bsl{k}) P_{n_2}(\bsl{k})  \right] \\
& = \gamma^2 \frac{\hbar}{2 D(\mu) m_{\text{C}}}   \sum_{\bsl{k}\in\BZ}^{\Delta E(\bsl{k})\neq 0}\sum_{n} \delta\left(\mu - E_n(\bsl{k}) \right) |\nabla_{\bsl{k}} E_{n}(\bsl{k})|^2 \Tr\left[ P_{n}(\bsl{k})\right] \\
& = \gamma^2 \frac{\hbar}{2 D(\mu) m_{\text{C}}}   \sum_{\bsl{k}\in\BZ}^{\Delta E(\bsl{k})\neq 0}\sum_{n} \delta\left(\mu - E_n(\bsl{k}) \right) |\nabla_{\bsl{k}} E_{n}(\bsl{k})|^2 \\
& = \gamma^2 \frac{\hbar}{2 D(\mu) m_{\text{C}}}  \frac{\A}{(2\pi)^2}  \int_{FS} d\sigma_{\bsl{k}}\frac{1}{|\nabla_{\bsl{k}} E_{n_F}(\bsl{k})|}  |\nabla_{\bsl{k}} E_{n_F}(\bsl{k})|^2 \\
& = \gamma^2 \frac{\hbar}{2 D(\mu) m_{\text{C}}}  \frac{\A}{(2\pi)^2}  \int_{FS} d\sigma_{\bsl{k}} |\nabla_{\bsl{k}} E_{n_F}(\bsl{k})| \ ,
}
resulting in 
\eq{
\label{eq:Gamma_EE_ave_mu_graphene}
\left\langle \Gamma \right\rangle^{E-E} = \gamma^2 \frac{\hbar}{2 D(\mu) m_{\text{C}}}  \frac{\A}{(2\pi)^2}  \int_{FS} d\sigma_{\bsl{k}} |\nabla_{\bsl{k}} E_{n_F}(\bsl{k})|\ ,
}
where $E_{n_F}(\bsl{k})$ is the electron band that is cut by the chemical potential, $n_F=1$ for $\mu<\epsilon_0$ and $n_F=2$ for $\mu>\epsilon_0$ according to \eqnref{eq:E_P_graphene}, $d\sigma_{\bsl{k}}$ is the measure on the Fermi surface, $\A$ is the area of the sample, and we have used \eqnref{eq:sum_delta_E_n_mu}.

For $\left\langle \Gamma \right\rangle^{geo-geo}$, we first note that $f^{geo}_i(\bsl{k})$ can be simplified by \eqnref{eq:E_P_graphene}:
\eqa{
\label{eq:g_geo_sim_graphene}
f^{geo}_i(\bsl{k})  & = \ii  \gamma \sum_{n'} E_{n'}(\bsl{k}) \partial_{k_i} P_{n'}(\bsl{k})(\delta_{ix} + \delta_{iy}) \\
& = \ii  \gamma\left[ d_0(\bsl{k}) \sum_{n'}  \partial_{k_i} P_{n'}(\bsl{k}) + (-)^n \frac{\Delta E(\bsl{k})}{2}  \partial_{k_i} (P_{n}(\bsl{k})-1+P_{n}(\bsl{k}))\right](\delta_{ix} + \delta_{iy}) \\
& = \ii  \gamma  (-)^n \Delta E(\bsl{k}) \partial_{k_i} P_{n}(\bsl{k})(\delta_{ix} + \delta_{iy}) \ ,
}
for any $n$.
By substituting \eqnref{eq:g_geo_sim_graphene} in to $\left\langle \Gamma \right\rangle^{geo-geo}$ in \eqnref{eq:Gamma_ave_mu_graphene_ini_2}, we obtain
\eqa{
\left\langle \Gamma \right\rangle^{geo-geo} & = \gamma^2 \frac{\hbar}{2 D(\mu) m_{\text{C}}} \sum_{i=x,y}   \sum_{\bsl{k}\in\BZ}^{\Delta E(\bsl{k})\neq 0}\sum_{n} \delta\left(\mu - E_n(\bsl{k}) \right)  \Tr\left[   (-)^n \Delta E(\bsl{k}) \partial_{k_i} P_{n}(\bsl{k})  P_{n}(\bsl{k})    (-)^n \Delta E(\bsl{k}) \partial_{k_i} P_{n}(\bsl{k}) \right] \\
& = \gamma^2 \frac{\hbar}{2 D(\mu) m_{\text{C}}} \frac{\A}{(2\pi)^2} \sum_{i=x,y}  \int_{FS}d\sigma_{\bsl{k}} \frac{\Delta E^2(\bsl{k})}{|\nabla_{\bsl{k}} E_{n_F}(\bsl{k})|}  \Tr\left[    \partial_{k_i} P_{n_F}(\bsl{k})  P_{n_F}(\bsl{k})   \partial_{k_i} P_{n_F}(\bsl{k}) \right]  \\
& = \gamma^2 \frac{\hbar}{2 D(\mu) m_{\text{C}}} \frac{\A}{(2\pi)^2} \sum_{i=x,y}  \int_{FS}d\sigma_{\bsl{k}} \frac{\Delta E^2(\bsl{k})}{|\nabla_{\bsl{k}} E_{n_F}(\bsl{k})|}  \frac{1}{2} \Tr\left[    \partial_{k_i} P_{n_F}(\bsl{k}) \partial_{k_i} P_{n_F}(\bsl{k}) \right]  \ ,
}
resulting in 
\eq{
\label{eq:Gamma_geogeo_ave_mu_graphene}
\left\langle \Gamma \right\rangle^{geo-geo} = \gamma^2 \frac{\hbar}{2 D(\mu) m_{\text{C}}} \frac{\A}{(2\pi)^2} \int_{FS}d\sigma_{\bsl{k}} \frac{\Delta E^2(\bsl{k})}{|\nabla_{\bsl{k}} E_{n_F}(\bsl{k})|} \Tr\left[g_{n_F}(\bsl{k})\right] \ ,
}
where 
\eq{
\left[g_{n}(\bsl{k})\right]_{j_1 j_2} = \frac{1}{2}\Tr[\partial_{k_{j_1}} P_{n}(\bsl{k}) \partial_{k_{j_2}} P_{n}(\bsl{k})]  
}
is the FSM according to \eqnref{eq:FS_metric_U}.
By substituting \eqnref{eq:Gamma_EE_ave_mu_graphene} , \eqnref{eq:Gamma_Egeo_ave_mu_graphene} and \eqnref{eq:Gamma_geogeo_ave_mu_graphene} into \eqnref{eq:Gamma_ave_mu_graphene_ini_2}, we obtain
\eq{
\label{eq:Gamma_ave_mu_graphene}
\left\langle \Gamma \right\rangle = \gamma^2 \frac{\hbar}{2 D(\mu) m_{\text{C}}}  \frac{\A}{(2\pi)^2}  \int_{FS} d\sigma_{\bsl{k}} \left\{ |\nabla_{\bsl{k}} E_{n_F}(\bsl{k})| + \frac{\Delta E^2(\bsl{k})}{|\nabla_{\bsl{k}} E_{n_F}(\bsl{k})|} \Tr\left[g_{n_F}(\bsl{k})\right]\right\}\ .
}

Now we go back to $\lambda$ in \eqnref{eq:lambda_omegabar}.
Recall that according to \eqnref{eq:lambda_E_geo_E-geo_Deltalambda}, $\lambda$ in general has four contributions: the energetic contribution $\lambda_E$ in \eqnref{eq:lambda_E}, the geometric contribution $\lambda_{geo}$ in \eqnref{eq:lambda_geo}, the cross contribution $\lambda_{E-geo}$ in \eqnref{eq:lambda_E_geo}, and the other contribution $\Delta\lambda$.
Based on \eqnref{eq:Gamma_ave_mu_graphene_ini_2} and$\Delta  \widetilde{f}_{\perp} = 0 $ shown by \eqnref{eq:g_beta_k_graphene}, we have $\lambda_{E-geo}=\Delta\lambda =0$, and
we arrive at 
\eq{
\label{eq:lambda_E_geo_graphene}
\lambda = \lambda_E + \lambda_{geo}
}
with
\eqa{
\lambda_E & =  \frac{ \Omega \gamma^2 }{(2\pi)^2 m_{\text{C}} \mcomega}  \int_{FS} d\sigma_{\bsl{k}} |\nabla_{\bsl{k}} E_{n_F}(\bsl{k})| \\
\lambda_{geo}  &  =  \frac{ \Omega \gamma^2  }{(2\pi)^2 m_{\text{C}}  \mcomega} \int_{FS}d\sigma_{\bsl{k}} \frac{\Delta E^2(\bsl{k})}{|\nabla_{\bsl{k}} E_{n_F}(\bsl{k})|} \Tr\left[g_{n_F}(\bsl{k})\right] =  \frac{ \Omega \gamma^2  |2(\mu-\epsilon_0)|^2}{(2\pi)^2 m_{\text{C}}  \mcomega} \int_{FS}d\sigma_{\bsl{k}} \frac{1}{|\nabla_{\bsl{k}} E_{n_F}(\bsl{k})|} \Tr\left[g_{n_F}(\bsl{k})\right] \ ,
}
where $\Omega$ is the area of the unit cell, and $\Delta E(\bsl{k}) = |2(\mu-\epsilon_0)|$ for $\bsl{k}$ on the Fermi surface.
Based on the discussion of the GA (\appref{app:gaussian}), we can see the direct appearance of the FSM in the geometric contribution to $\lambda$ can be understood as 
\eqa{
a_{\bsl{\tau}} = \frac{1}{D(\mu) } \sum_m \sum_{\bsl{k}_2}^{\BZ}\delta\left(\mu - E_m(\bsl{k}_2) \right) \left[ P_{m}(\bsl{k}_2) \right]_{\bsl{\tau}\bsl{\tau}} = \frac{1}{2}\ ,
}
(derived from the $C_2\TR$ symmetry \eqnref{eq:E_P_graphene} and $a_{\bsl{\tau}}$ is defined in \eqnref{eq:a_tau_expression_GA}), and $\lambda_{geo,2}=0$ from the $C_3$ symmetry.

Now we show that the geometric part $\lambda_{geo}$ in \eqnref{eq:lambda_E_geo_graphene} is bounded by band topology from below.
First, \eqnref{eq:E_P_graphene} yields
\eqa{
[g_{n}(\bsl{k})]_{j_1 j_2} & = \frac{1}{2}\Tr[\partial_{k_{j_1}} P_{n}(\bsl{k}) \partial_{k_{j_2}} P_{n}(\bsl{k})]  \\
& = \frac{1}{2}\Tr[\partial_{k_{j_1}} \frac{1}{2}  (-)^n  \mat{ & e^{-\ii \theta(\bsl{k})}  \\ e^{\ii \theta(\bsl{k})} &  }   \partial_{k_{j_2}} \frac{1}{2}  (-)^n  \mat{ & e^{-\ii \theta(\bsl{k})}  \\ e^{\ii \theta(\bsl{k})} &  } ] \\
& = \frac{1}{8}\left[ \partial_{k_{j_1}} e^{-\ii \theta(\bsl{k})}  \partial_{k_{j_2}} e^{\ii \theta(\bsl{k})} +  \partial_{k_{j_1}} e^{\ii \theta(\bsl{k})} \partial_{k_{j_2}} e^{-\ii \theta(\bsl{k})} \right]\ ,
}
resulting in 
\eq{
\label{eq:g_theta}
\Tr[g_{n}(\bsl{k})] = \frac{1}{4} \sum_{i=x,y}\left|\ii e^{\ii \theta(\bsl{k})} \partial_{k_i} e^{-\ii \theta(\bsl{k})}\right|^2\ ,
}
where $\theta$ is defined in \eqnref{eq:theta_graphene}.
Then, \eqnref{eq:g_theta} leads to 
\eqa{
\label{eq:holder_use_graphene}
& \int_{FS} d\sigma_{\bsl{k}} \frac{\Delta E^2(\bsl{k})}{|\nabla_{\bsl{k}}E_{n_F}(\bsl{k})|}  \Tr[g_{n_F}(\bsl{k})] = \frac{ \int_{FS}d\sigma_{\bsl{k}} \frac{\Delta E^2(\bsl{k})}{|\nabla_{\bsl{k}} E_{n_F}(\bsl{k})|} \Tr\left[g_{n_F}(\bsl{k})\right] \int_{FS}d\sigma_{\bsl{k}_1} |\nabla_{\bsl{k}_1} E_{n_F}(\bsl{k}_1)|/\Delta^2(\bsl{k}_1) }{ \int_{FS}d\sigma_{\bsl{k}_2} |\nabla_{\bsl{k}_2} E_{n_F}(\bsl{k}_2)|/\Delta^2(\bsl{k}_2) } \\
& \geq \frac{ \left[ \int_{FS}d\sigma_{\bsl{k}}  \sqrt{\Tr\left[g_{n_F}(\bsl{k})\right]} \right]^2 }{ \int_{FS}d\sigma_{\bsl{k}_2} |\nabla_{\bsl{k}_2} E_{n_F}(\bsl{k}_2)| /\Delta^2(\bsl{k}_2)} = \frac{\left[ \sum_{\mathcal{\L}} \int_{\mathcal{\L}} d\sigma_{\bsl{k}}  \sqrt{\sum_j |\ii e^{\ii \theta(\bsl{k})} \partial_{k_j}e^{-\ii \theta(\bsl{k})} |^2}  \right]^2 }{ 4 \int_{FS}d\sigma_{\bsl{k}_2} |\nabla_{\bsl{k}_2} E_{n_F}(\bsl{k}_2)|/\Delta^2(\bsl{k}_2) }  \geq  (2\pi)^2 \frac{ \left[ \sum_{\mathcal{\L}} |W_{\mathcal{\L}}| \right]^2 }{ 4 \int_{FS} d\sigma_{\bsl{k}}  |\nabla_{\bsl{k}}E_{n_F}(\bsl{k})|/\Delta E^2(\bsl{k}) } \ ,
}
where $\L$ ranges over the connected loop that form the Fermi surface, $W_{\mathcal{\L}}$ is the winding number defined in \eqnref{eq:PT_winding_number}, and the first inequality is derived from the H\"older's inequality which reads
\eq{
\label{eq:holder}
\left(\int |f(x) g(x)| dx \right)^2 \leq \int |f(x)|^2 dx \int |g(x)|^2 dx\ .
}
By further defining 
\eq{
\label{eq:lambda_topo_graphene}
\lambda_{topo} =    \frac{ \Omega \gamma^2  }{ 4 m_{\text{C}}  \mcomega}    \frac{ \left( \sum_{\mathcal{L}} |W_{\mathcal{L}}| \right)^2 }{ \int_{FS} d\sigma_{\bsl{k}}  |\nabla_{\bsl{k}}E_n(\bsl{k})|/\Delta E^2(\bsl{k}) } =    \frac{ \Omega \gamma^2  |2(\mu-\epsilon_0)|^2}{ 4 m_{\text{C}}  \mcomega}    \frac{ \left( \sum_{\mathcal{L}} |W_{\mathcal{L}}| \right)^2 }{ \int_{FS} d\sigma_{\bsl{k}}  |\nabla_{\bsl{k}}E_n(\bsl{k})| } \ ,
}
we have
\eqa{
\lambda = \lambda_E + \lambda_{geo} \geq   \lambda_E + \lambda_{topo}
}
and 
\eqa{
\lambda \geq \lambda_{geo} \geq   \lambda_{topo}\ ,
}
where the later is the geometric and topological lower bounds of $\lambda$.

\subsection{Analytical Geometric and Topological Contributions to EPC constant in Graphene:  $(\mu-\epsilon_0) \rightarrow 0$}

\label{app:lambda_ratio_mu0}

To estimate the ratio among $\lambda$, $\lambda_E$, $\lambda_{geo}$, and $\lambda_{topo}$, let us consider $|\mu-\epsilon_0| \rightarrow 0$ limit.
In this limit, we can perform the linear approximation for the electron Hamiltonian:
\eq{
H_{el} = H_{el}^{\K} +  H_{el}^{-\K} \ ,
}
where 
\eq{
\label{eq:linear_el_graphene}
H_{el}^{\pm\K} = \sum_{\bsl{p}}^{\Lambda} \gamma^\dagger_{\pm\K+\bsl{p}} \left[ \epsilon_0 + v (\pm p_x \tau_x + p_y \tau_y ) \right] \gamma^\dagger_{\pm\K+\bsl{p}}\ ,
}
and $v=-\frac{\sqrt{3}}{2}a t$.
From the linear approximation for the electron Hamiltonian, we have the dispersion as
\eq{
\label{eq:E_linear}
E_{n}(\pm \K + \bsl{p}) = (-)^n |v| |\bsl{p}| \Rightarrow  |\nabla_{\bsl{p}} E_{n}(\pm \K + \bsl{p})  |= |v|
}
and have the Fermi surface as $FS= FS_{+} \cup FS_{-}$ with   
\eq{
\label{eq:FS_linear}
FS_{\pm} = \{ \bsl{p} \pm \K\ | \  |\bsl{p}| = |\mu-\epsilon_0|/|v| \}\ ,
}
which means that
\eq{
\label{eq:W_linear}
\left( \sum_{\mathcal{L}} |W_{\mathcal{L}}| \right)^2  = 4\ ,
}
and
\eqa{
\label{eq:int_nablaE_linear}
 \int_{FS} d\sigma_{\bsl{k}} |\nabla_{\bsl{k}} E_{n_F}(\bsl{k})| & =  \sum_{\alpha =\pm} \int_{FS_{\alpha}} d\sigma_{\bsl{k}} |\nabla_{\bsl{k}} E_{n_F}(\bsl{k})| \\
& =   \sum_{\alpha =\pm} \frac{|\mu-\epsilon_0|}{|v|} \int_{0}^{2\pi} d\phi  |\nabla_{\bsl{p}} E_{n_F}(\alpha\K + \bsl{p})| \\
& =   2  \frac{|\mu-\epsilon_0|}{|v|} 2\pi |v| \\
& = 4 \pi |\mu-\epsilon_0|  \ ,
}
where $\bsl{p} = |\bsl{p}|(\cos(\phi),\sin(\phi))$.

Based on \eqnref{eq:d_graphene}, $e^{-\ii \theta(\bsl{k})}$  in \eqnref{eq:E_P_graphene} now becomes 
\eq{
e^{-\ii \theta( \pm \K + \bsl{p})} = \frac{ \pm p_x - \ii p_y }{|\bsl{p}|}\ ,
}
which gives 
\eqa{
\label{eq:Trg_linear}
\Tr[g_{n}(\pm \K + \bsl{p})] & = \frac{1}{4} \sum_{i=x,y}\left|\ii e^{\ii \theta(\pm \K + \bsl{p})} \partial_{p_i} e^{-\ii \theta(\pm \K + \bsl{p})}\right|^2 = \frac{1}{4} \sum_{i=x,y}\left|\ii \frac{ \pm p_x - \ii p_y }{|\bsl{p}|} \partial_{p_i} \frac{ \pm p_x + \ii p_y }{|\bsl{p}|}\right|^2 \\
& = \frac{1}{4} \sum_{i=x,y}\left|\ii \frac{ \pm p_x - \ii p_y }{|\bsl{p}|} \left[ (\partial_{p_i}\frac{1}{|\bsl{p}|}) (\pm p_x + \ii p_y ) + \frac{1}{|\bsl{p}|} \partial_{p_i}(\pm p_x + \ii p_y ) \right]\right|^2 \\
& = \frac{1}{4} \sum_{i=x,y}\left| \left[ |\bsl{p}| (\partial_{p_i}\frac{1}{|\bsl{p}|})  + \frac{1}{|\bsl{p}|^2}( \pm p_x - \ii p_y ) \partial_{p_i}(\pm p_x + \ii p_y ) \right]\right|^2 \\
& = \frac{1}{4}  \left|   - \frac{p_x}{|\bsl{p}|^2}  + \frac{1}{|\bsl{p}|^2} p_x \mp \ii   \frac{1}{|\bsl{p}|^2} p_y \right|^2  + \frac{1}{4}  \left|   - \frac{p_y}{|\bsl{p}|^2}  +
\frac{1}{|\bsl{p}|^2}( \pm  \ii p_x + p_y )\right|^2 \\
& = \frac{1}{4}  \frac{1}{|\bsl{p}|^2} 
}
for $n=1,2$.

Now we look at $\lambda_E$, $\lambda_{geo}$, and $\lambda_{topo}$.
By substituting \eqnref{eq:int_nablaE_linear} and \eqnref{eq:FS_linear} into \eqnref{eq:lambda_E_geo_graphene}, we obtain
\eqa{
\label{eq:lambda_E_linear}
\lambda_E  =  \frac{ \Omega \gamma^2 }{(2\pi)^2 m_{\text{C}} \mcomega} \int_{FS } d\sigma_{\bsl{k}} |\nabla_{\bsl{k}} E_{n_F}(\bsl{k})| = \frac{ \Omega \gamma^2 |\mu-\epsilon_0| }{ \pi m_{\text{C}} \mcomega}\ ,
}
where $n_F=1$ for $\mu<\epsilon_0$ and $n_F=2$ for $\mu>\epsilon_0$ according to \eqnref{eq:E_P_graphene}.
By substituting \eqnref{eq:E_linear}, \eqnref{eq:Trg_linear} and \eqnref{eq:FS_linear} into \eqnref{eq:lambda_E_geo_graphene}, we obtain
\eqa{
\label{eq:lambda_geo_linear}
\lambda_{geo}  &  =  \frac{ \Omega \gamma^2 |2(\mu-\epsilon_0)|^2}{(2\pi)^2 m_{\text{C}}  \mcomega} \int_{FS}d\sigma_{\bsl{k}} \frac{1}{|\nabla_{\bsl{k}} E_{n_F}(\bsl{k})|} \Tr\left[g_{n_F}(\bsl{k})\right] \\
& =\frac{ \Omega \gamma^2 |2(\mu-\epsilon_0)|^2}{(2\pi)^2 m_{\text{C}}  \mcomega} \sum_{\alpha = \pm}\int_{FS_{\alpha}}d\sigma_{\bsl{k}} \frac{1}{|\nabla_{\bsl{k}} E_{n_F}(\bsl{k})|} \Tr\left[g_{n_F}(\bsl{k})\right] \\
& =\frac{ \Omega \gamma^2 |2(\mu-\epsilon_0)|^2}{(2\pi)^2 m_{\text{C}}  \mcomega} \sum_{\alpha = \pm} \frac{|\mu-\epsilon_0|}{|v|} \int_{0}^{2\pi} d\phi \frac{1}{|v|} \frac{1}{4}  \frac{|v|^2}{|\mu-\epsilon_0|^2}  \\
& =\frac{ \Omega \gamma^2 |\mu-\epsilon_0|   }{  \pi  m_{\text{C}}  \mcomega}  \ .
} 
By substituting \eqnref{eq:W_linear}, \eqnref{eq:int_nablaE_linear} and \eqnref{eq:FS_linear} into \eqnref{eq:lambda_topo_graphene}, we obtain
\eqa{
\label{eq:lambda_topo_linear}
\lambda_{topo} =    \frac{ \Omega \gamma^2 |(\mu-\epsilon_0)| }{ m_{\text{C}}  \mcomega  \pi   }     \ .
}
Therefore,
\eq{
\label{eq:lambda_ratios_linear}
\mu\rightarrow 0: \lambda_E = \lambda_{geo} = \lambda_{topo} =  \frac{1}{2}\lambda\ .
}

\subsection{Numerical Calculations}
\label{app:numerics_graphene}

\begin{figure}[t]
    \centering
    \includegraphics[width=0.5\columnwidth]{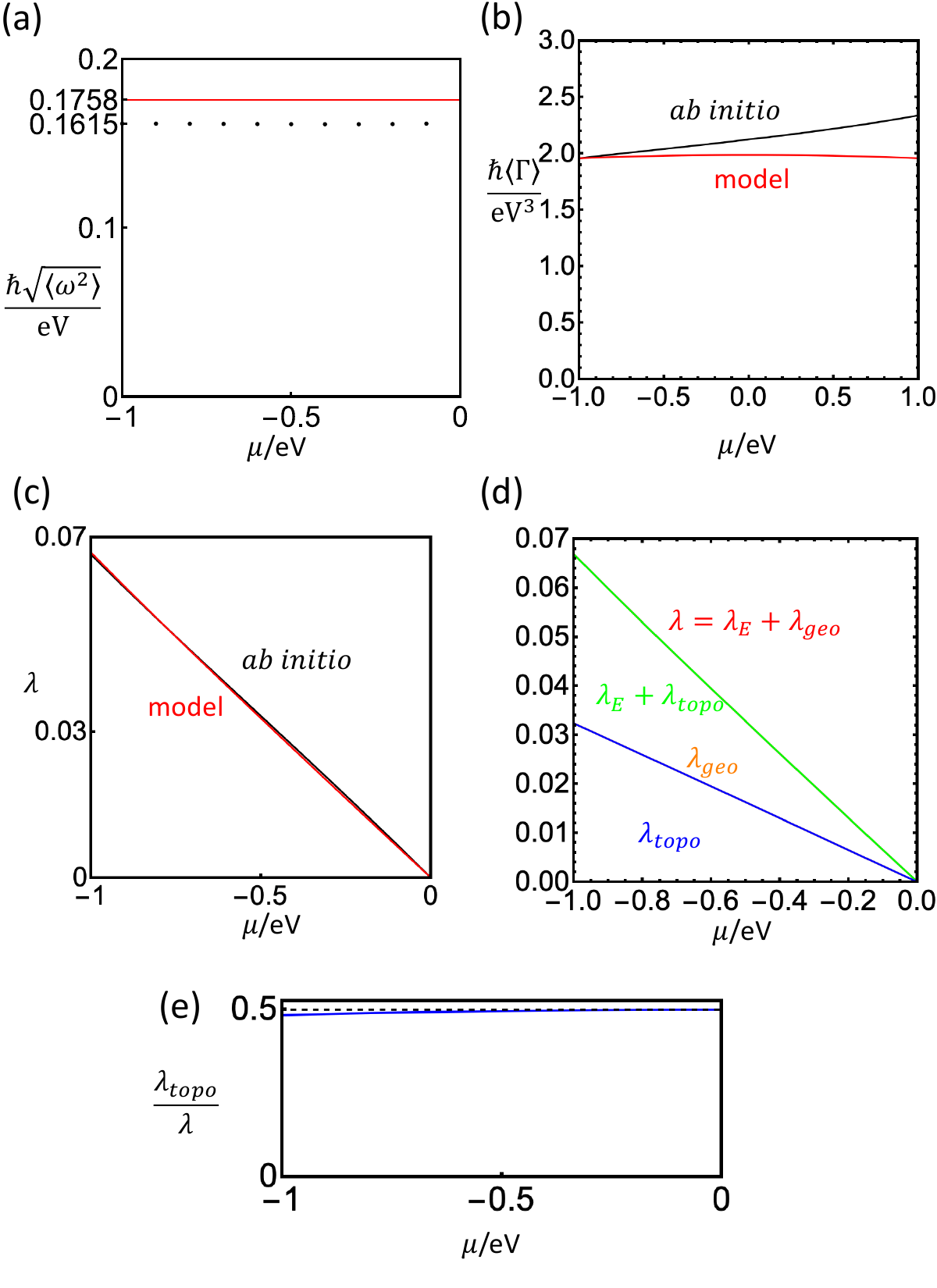}
    \caption{
    (a) The black dots are the {\abi} data of $\hbar\sqrt{\mcomega}$, and the blue line is given by the analytical expression of $\mcomega$ in \eqnref{eq:mcomega_graphene_linear} (derived for $\mu\rightarrow \epsilon_0 =0$) with \tabref{tab:para_graphene}.
    (b) The plot of $\lambda$ for the {\abi} calculation and for the \eqnref{eq:lambda_E_geo_graphene} from our model with \tabref{tab:para_graphene}.
    (c) The plot of $\left\langle \Gamma \right\rangle$ for the {\abi} calculation and for the \eqnref{eq:Gamma_ave_mu_graphene} from our model with \tabref{tab:para_graphene}.
    (d) The plot of $\lambda = \lambda_{geo}+\lambda_E$, $\lambda_{topo} + \lambda_E$, $\lambda_{geo}$, and $\lambda_{topo}$ for graphene according to \eqnref{eq:lambda_E_geo_graphene} and \eqnref{eq:lambda_topo_graphene} with \tabref{tab:para_graphene}.
    (e) The plot of $\lambda_{topo}/\lambda$ with $\lambda = \lambda_{geo}+\lambda_E$ for graphene according to \eqnref{eq:lambda_E_geo_graphene} and \eqnref{eq:lambda_topo_graphene} with \tabref{tab:para_graphene}.
    }
    \label{fig:graphene_EPC}
\end{figure}

In this part, we test our model against {\abi} calculations in order to check whether our approximations, \ie,  tight-binding approximation, two-center approximation (\asmref{asm:2center}), and nearest-neighbor approximations (\eqnref{eq:NN_t} and \eqref{eq:NN_g}), are reasonable.

In total, we have four key independent parameters: $\epsilon_0$ and $t$ in the electron Hamiltonian \eqnref{eq:H_el_graphene}, $\hat{\gamma}$ in the EPC Hamiltonian \eqnref{eq:g_i_g_beta_grpahene} (or equivalently $\gamma = \hat{\gamma}/t$ in \eqnref{eq:graphene_gamma}), and $\mcomega$ in \eqnref{eq:McMillan_omega_square_ave}.
Without loss of generality, we can always choose 
\eq{
\epsilon_0 = 0\ .
}
As well discussed in the literature (\eg, \refcite{Neto2009GrapheneRMP}), the electron bands in $[-1\eV,1\eV]$ can be well captured by the model in \eqnref{eq:H_el_graphene} for 
\eq{
\label{eq:t_value_graphene}
t= -2.751\eV \ ,
}
which is obtained by fitting the electron band structure.

For $\mcomega$ in \eqnref{eq:McMillan_omega_square_ave}, the {\abi} calculation shows a  $\hbar \sqrt{\mcomega} = 0.1615 \eV$ that is almost independent of $\mu$ as shown in \figref{fig:graphene_EPC}(a), though in general the dependence of $\mu$ is possible for  $\mcomega$ due to the dependence of $\mu$ in the Eliashberg function.

For the remaining $\hat{\gamma}$ in EPC Hamiltonian, we can directly determine its value from the EPC Hamiltonian in the real space from the {\abi} calculation  (detailed procedure in \appref{app:abi}).
Basically, $p_z$ orbitals are also Wannier functions of graphene, and we can project the EPC Hamiltonian onto those Wannier functions to get the matrix elements in the real space.
As a result, we get 
\eq{
\label{eq:gamma_hat_graphene}
\hat{\gamma} = 20.11 \text{eV}/a^2\ ,
}
which, combined with the value of $t$ in \eqnref{eq:t_value_graphene}, leads to 
\eq{
\gamma = \frac{\hat{\gamma}}{t} = -7.308 a^{-2}\ .
}
We note that the value of $\gamma$ determined in this way depends on the Wannier function obtained during the Wannierization, and thus different Wannier functions might give slightly different values of $\gamma$, but the error should be small.
All relevant parameter values of the models are summarized in \tabref{tab:para_graphene}.

To check the value of $\gamma$ or $\hat{\gamma}$, we look at $F_{\bsl{\tau}i}(\K,\K)$ and $F_{\bsl{\tau}i}(\K,-\K)$, since the approximations that we make should be the best for the low-energy electrons at $\pm K$, where  $F_{\bsl{\tau}i}(\bsl{k}_1,\bsl{k}_2)$ in \eqnref{eq:f_k_graphene} and $\K$ is defined in \eqnref{eq:K_graphene}.
The concise expression of $F_{\bsl{\tau}i}(\K,\pm\K)$ within our approximation is in \eqnref{eq:f_graphene_2center}.
Specifically, we have
\eqa{
& F_{\bsl{\tau}i}(\K,\K) = -\ii \hat{\gamma} \frac{\sqrt{3} a}{2} \left( \chi_{\bsl{\tau}}\tau_i-\tau_i\chi_{\bsl{\tau}} \right) (\delta_{ix} + \delta_{iy}) \\
& F_{\bsl{\tau}i}(\K,-\K) = -\ii  \hat{\gamma} \frac{\sqrt{3} a}{2} \left( \chi_{\bsl{\tau}}(-)^i\tau_i-\tau_i\chi_{\bsl{\tau}} \right) (\delta_{ix} + \delta_{iy})\ ,
}
where $(-)^x = - (-)^y =-1$, leading to
\eqa{
\label{eq:f_graphene_TB_2cetner}
& F_{\bsl{\tau}_A x}(\K,\K) = \hat{\gamma} \frac{\sqrt{3} a}{2} \tau_y \ ,\  F_{\bsl{\tau}_A y}(\K,\K) = \hat{\gamma} \frac{\sqrt{3} a}{2} (-\tau_x) \ ,\  F_{\bsl{\tau}_B x}(\K,\K) = \hat{\gamma} \frac{\sqrt{3} a}{2} (-\tau_y) \ ,\  F_{\bsl{\tau}_B y}(\K,\K) = \hat{\gamma} \frac{\sqrt{3} a}{2} \tau_x \\
& F_{\bsl{\tau} z}(\K,\K) = 0\\
& F_{\bsl{\tau}_A x}(\K,-\K) = \hat{\gamma} \frac{\sqrt{3} a}{2} \ii \tau_x\ ,\  F_{\bsl{\tau}_A y}(\K,-\K) = \hat{\gamma} \frac{\sqrt{3} a}{2} (-\tau_x)\\
& F_{\bsl{\tau}_B x}(\K,-\K) = \hat{\gamma} \frac{\sqrt{3} a}{2} (\ii \tau_x)\ ,\  F_{\bsl{\tau}_B y}(\K,-\K) = \hat{\gamma} \frac{\sqrt{3} a}{2} \tau_x \ ,\   F_{\bsl{\tau} z}(\K,-\K) = 0\ ,
}
where 
\eq{
\label{eq:beta_graphene}
\hat{\gamma} \frac{\sqrt{3} a}{2} = 17.41  \eV/a\ ,
}
based on \eqnref{eq:gamma_hat_graphene}.

On the other hand, by using \eqnref{eq:from_G_to_f}, the {\abi} calculation gives
\eqa{
\label{eq:f_graphene_abi}
& F^{ab\ initio}_{\bsl{\tau}_A x}(\K,\K) = \beta_1 \tau_y \ ,\  F^{ab\ initio}_{\bsl{\tau}_A y}(\K,\K) = \beta_1 (-\tau_x) \ ,\  F^{ab\ initio}_{\bsl{\tau}_B x}(\K,\K) = \beta_1 (-\tau_y) \ ,\  F^{ab\ initio}_{\bsl{\tau}_B y}(\K,\K) = \beta_1 \tau_x \\
& F^{ab\ initio}_{\bsl{\tau} z}(\K,\K) = 0\\
& F^{ab\ initio}_{\bsl{\tau}_A x}(\K,-\K) = \beta_2 \ii \tau_x + \ii \beta_3 (\chi_{\bsl{\tau}_A}-1)\ ,\  F^{ab\ initio}_{\bsl{\tau}_A y}(\K,-\K) = \beta_2 (-\tau_x) + \beta_3 (\chi_{\bsl{\tau}_A}-1) \\
& F^{ab\ initio}_{\bsl{\tau}_B x}(\K,-\K) = \beta_2 (\ii \tau_x) + \ii \beta_3 (\chi_{\bsl{\tau}_B}-1)\ ,\  F^{ab\ initio}_{\bsl{\tau}_B y}(\K,-\K) = \beta_2 \tau_x - \beta_3 (\chi_{\bsl{\tau}_B}-1)\ ,\   F^{ab\ initio}_{\bsl{\tau} z}(\K,-\K) = 0\ ,
}
where 
\eq{
\label{eq:f_graphene_abi_beta}
\beta_1 = 16.37 \eV/a\ ,\ \beta_2 = 16.54 \eV/a \ ,\ \beta_3 = 6.138 \eV/a\ ,
}
and the procedure of the calculation is explained in \appref{app:abi}.
By comparing \eqnref{eq:f_graphene_abi} to \eqnref{eq:f_graphene_TB_2cetner}, our approximations require $\beta_1 = \beta_2$ and $\beta_3 =0$.
Therefore, only the mean value of $\beta_1$ and $\beta_2$, \ie, $\frac{\beta_1 + \beta_2 }{2 }$ is within the our approximations, which gives 
\eq{
\hat{\gamma}^{\text{alternative}} \frac{\sqrt{3} a}{2} = \frac{\beta_1 + \beta_2 }{2 } = 16.46 \eV/a\ .
}
By comparing the value of $\hat{\gamma}^{\text{alternative}} \frac{\sqrt{3} a}{2} $ to $\hat{\gamma} \frac{\sqrt{3} a}{2} $ in \eqnref{eq:beta_graphene}, the relative error is smaller than $6\%$, which should come from the larger-range terms that are still within the two-center approximation.

\begin{table}[t]
    \centering
    \begin{tabular}{|c|c|c|c|c|c|c|c|}
    \hline
        $ a$        &     $\epsilon_0$   & $t $       & $\gamma$ & $m_{\text{C}}$ & $\hbar \sqrt{\mcomega} $ & $\hbar \omega_{E_{2g}}(\Gamma)$ &  $\hbar \omega_{A_1'}(\K)$ \\
         \hline
         2.46 \AA &      0    &   $-2.751\eV$       &    $-7.308 a^{-2}$          & $12$ amu  & $0.1615 \eV$  & $0.1935 \eV$ & $0.1622 \eV$\\
         \hline
    \end{tabular}
    \caption{The values of the parameters for graphene. 
    $a$ is the lattice constant.
    $\epsilon_0$ and $t$ are in \eqnref{eq:H_el_graphene}, and $ \gamma$ is in \eqnref{eq:graphene_gamma}.
    $m_{\text{C}}$ is the mass of the Carbon atom, and $\text{amu} =1.67377 \times 10^{-27}$kg.
    }
    \label{tab:para_graphene}
\end{table}

\begin{figure}
    \centering
    \includegraphics[width=\columnwidth]{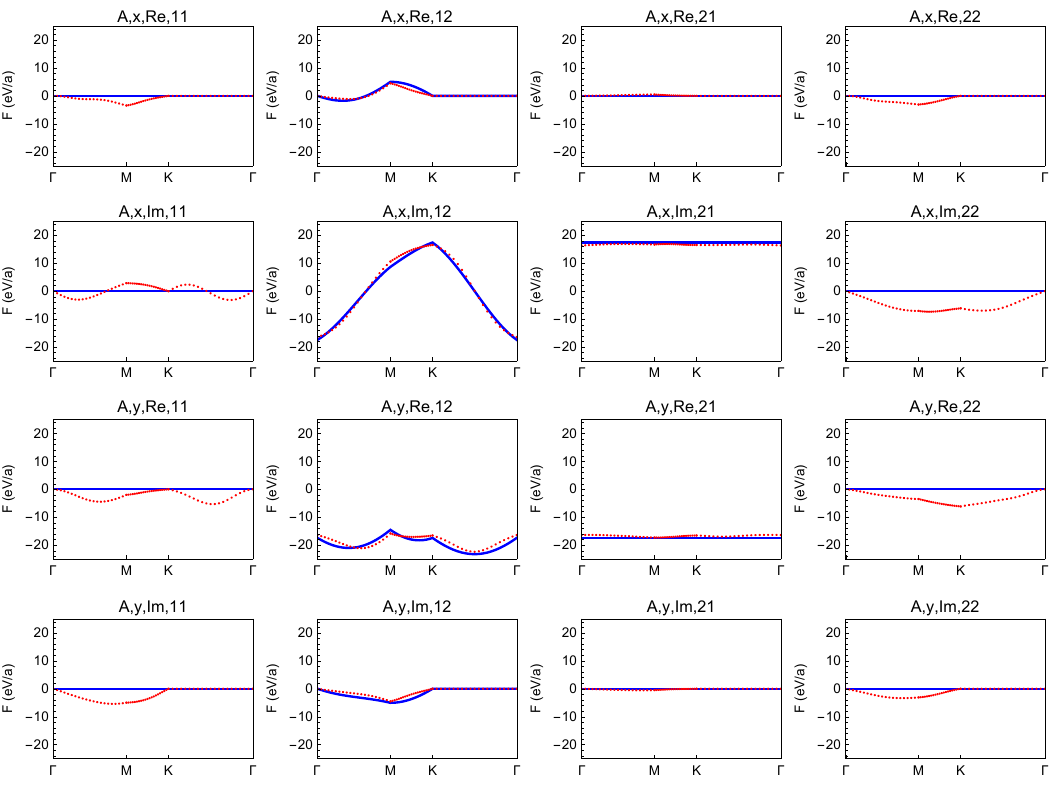}
    \caption{The values of $\left[ F_{\bsl{\tau} i }(K, K+\bsl{q}) \right]_{\bsl{\tau}_1\bsl{\tau}_2}$ (\eqnref{eq:H_el-ph_k_graphene}) for $\bsl{\tau}=\bsl{\tau}_A$ and $i=x,y$.
    In the cation of each plot, ``A" means it is for  $\bsl{\tau} =\bsl{\tau}_A$, and ``x" and ``y" means $i=x$ and $i=y$, respectively.
    ``Re" and ``Im" mean the real and imaginary parts of $\left[ F_{\bsl{\tau} i }(K, K+\bsl{q}) \right]_{\bsl{\tau}_1\bsl{\tau}_2}$, and $11$, $12$, $21$ and $22$ correspond to $\bsl{\tau}_1\bsl{\tau}_2= \bsl{\tau}_A\bsl{\tau}_A, \bsl{\tau}_A\bsl{\tau}_B, \bsl{\tau}_B\bsl{\tau}_A, \bsl{\tau}_B\bsl{\tau}_B$, respectively.
    The red dots are the {\abi} data, while the blue lines come from the model  \eqnref{eq:f_k_graphene} with the parameter values in \tabref{tab:para_graphene}.
    }
    \label{fig:F_compare_A_graphene}
\end{figure}

\begin{figure}
    \centering
    \includegraphics[width=\columnwidth]{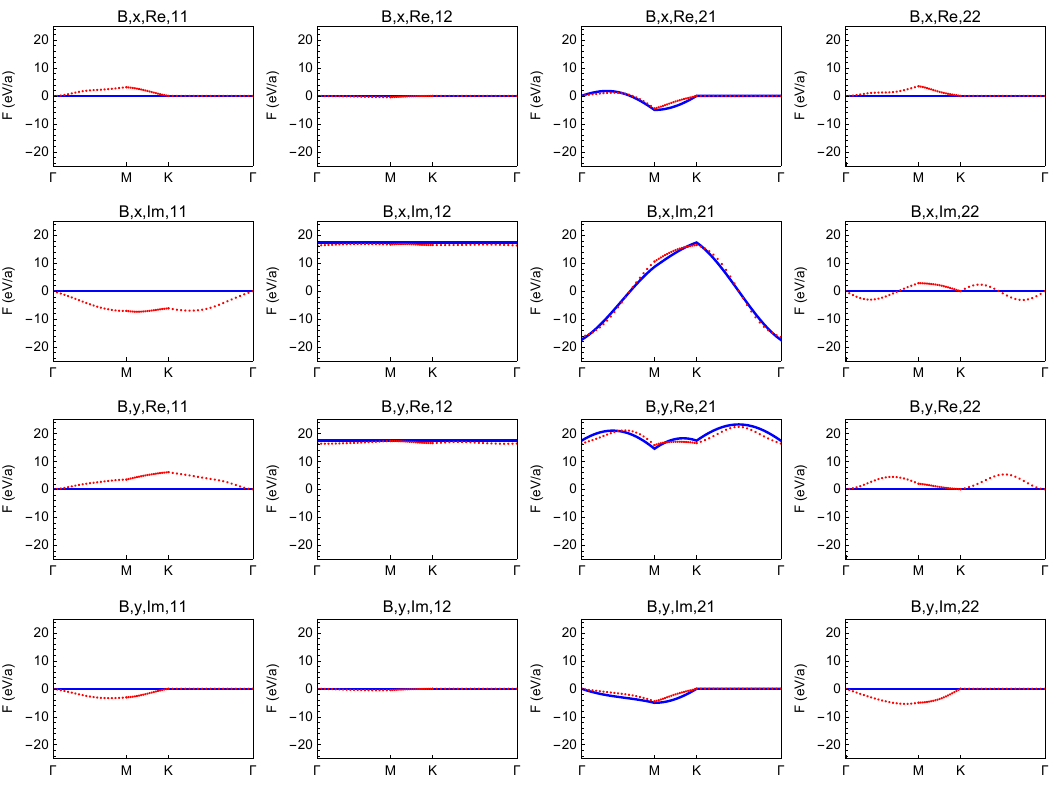}
    \caption{The values of $\left[ F_{\bsl{\tau} i }(K, K+\bsl{q}) \right]_{\bsl{\tau}_1\bsl{\tau}_2}$ (\eqnref{eq:H_el-ph_k_graphene}) for $\bsl{\tau}=\bsl{\tau}_B$ and $i=x,y$.
    In the cation of each plot, ``B" means it is for  $\bsl{\tau} =\bsl{\tau}_B$, and ``x" and ``y" means $i=x$ and $i=y$, respectively.
    ``Re" and ``Im" mean the real and imaginary parts of $\left[ F_{\bsl{\tau} i }(K, K+\bsl{q}) \right]_{\bsl{\tau}_1\bsl{\tau}_2}$, and $11$, $12$, $21$ and $22$ correspond to $\bsl{\tau}_1\bsl{\tau}_2= \bsl{\tau}_A\bsl{\tau}_A, \bsl{\tau}_A\bsl{\tau}_B, \bsl{\tau}_B\bsl{\tau}_A, \bsl{\tau}_B\bsl{\tau}_B$, respectively.
    The red dots are the {\abi} data, while the blue lines come from the model  \eqnref{eq:f_k_graphene} with the parameter values in \tabref{tab:para_graphene}.
    }
    \label{fig:F_compare_B_graphene}
\end{figure}

We note that $\frac{\beta_1 - \beta_2 }{2 } $ and $\beta_3 $ in \eqnref{eq:f_graphene_abi} are beyond our approximations, and the ratios are $|\frac{\beta_1 - \beta_2 }{2 }|/|\frac{\beta_1 + \beta_2 }{2 }|\approx 0.005$ and $|\beta_3|/|\frac{\beta_1 + \beta_2 }{2 }|\approx 1/3$.
Clearly, we can neglect $\frac{\beta_1 - \beta_2 }{2 } $, since it is small compared to $\frac{\beta_1 + \beta_2 }{2 }$.
$\beta_3$ is beyond the two-center approximation, because the symmetries would require $F_{\bsl{\tau}i}(\K,-\K) = -\ii  \beta_2 \left( \chi_{\bsl{\tau}}(-)^i\tau_i-\tau_i\chi_{\bsl{\tau}} \right) (\delta_{ix} + \delta_{iy})$ as long as we adopt the tight-binding and two-center approximations (\eqnref{eq:2center}), no matter whether we assume the hopping only appears between nearest neighbors.
In terms of the EPC strengthes, $|\beta_3| \approx 0.35 |\hat{\gamma} \frac{\sqrt{3} a}{2}|$.
In \figref{fig:F_compare_A_graphene} and \figref{fig:F_compare_B_graphene}, we compare the EPC matrix elements from the {\abi} calculation and from our model \eqnref{eq:f_k_graphene} for phonon momentum away from $\Gamma$ or $\K$ with parameter values in \tabref{tab:para_graphene}.
For the dominant matrix elements (\ie, those that have maximum absolute values about $18\eV/a$), the model matches the {\abi} calculation very well, although the mismatch occurs for the subdominant matrix elements (\ie, those that have maximum absolute values about $6\eV/a$).
Therefore, even if we go away from $\bsl{q}=\Gamma,\K$, the EPC matrix elements beyond the our model are roughly 1/3 of the EPC matrix elements within our model.

Yet, we don't see a $1/3$ mismatch in the plot of $\lambda$ in \figref{fig:graphene_EPC}. 
The reason lies in the behavior of $\left\langle \Gamma \right\rangle$ as a function of the chemical potential $\mu$.
For $\mu=0$, $\beta_3$ in \eqnref{eq:f_graphene_abi} is the only term that is beyond our model.
In this case, $\beta_3$ in \eqnref{eq:f_graphene_abi} contributes to $\left\langle \Gamma \right\rangle_{\mu\rightarrow 0}$ as $\beta_3^2 \sim |\frac{\beta_1 + \beta_2 }{2 }|^2 /9 \sim |\hat{\gamma} \frac{\sqrt{3} a}{2}|^2/10 $, since $\beta_3$ couples to different matrices than $\hat{\gamma}$; this is consistent with the numerical results in \figref{fig:graphene_EPC}(b), which shows that the difference between $\left\langle \Gamma \right\rangle_{\mu\rightarrow 0}$ from the {\abi} calculation and that from \eqnref{eq:Gamma_ave_mu_graphene} is no larger than $10\%$.
Furthermore, as shown by the numerical calculation, the difference in $\left\langle \Gamma \right\rangle$ caused by the mismatch decreases as $\mu$ decreaes from $0$, and becomes nearly zero for $\mu=-1\eV$.
As a result, the absolute error of $\lambda$, which reads 
\eq{
\label{eq:lambda_error_graphene}
\lambda -\lambda^{\abi} = \frac{2 D(\mu)}{N \mcomega} (\left\langle \Gamma \right\rangle - \left\langle \Gamma \right\rangle^{\abi})\ ,
}
should be small for both large negative $\mu$ (due to the small $|\left\langle \Gamma \right\rangle - \left\langle \Gamma \right\rangle^{\abi}|$) and small negative $\mu$ (due to small electron density of states $D(\mu)/N$).
Indeed, as shown in \figref{fig:graphene_EPC}(c), the difference between $\lambda$ from the {\abi} calculation and that from \eqnref{eq:lambda_E_geo_graphene} is quite small.
Therefore, our approximations are good for the study of $\lambda$ for $\mu\in[-1,0]$eV.

In \figref{fig:graphene_EPC}(d), we plot $\lambda = \lambda_{geo}+\lambda_E$, $\lambda_{topo} + \lambda_E$, $\lambda_{geo}$, and $\lambda_{topo}$ for graphene according to \eqnref{eq:lambda_E_geo_graphene} and \eqnref{eq:lambda_topo_graphene} with \tabref{tab:para_graphene}.
We can see the topological and geometric contributions to $\lambda$ are very close to each other.
As shown in \eqnref{fig:graphene_EPC}(e), the topological contribution to $\lambda$ is roughly half of the total value and limits to exactly half as  $(\mu-\epsilon_0) \rightarrow 0$, which is consistent with the analysis in \appref{app:lambda_ratio_mu0}.

\subsection{$ \mcomega $ Approximated by $\omega_{E_{2g}}(\Gamma)$ and $\omega_{A_1'}(\K)$}

\label{app:graphene_mcomega}

In \appref{app:numerics_graphene}, we directly use the value of $\mcomega$ from the {\abi} calculation.
In this part, we will show that  $\mcomega$ can be well approximated by the combinations of the frequencies of the $E_{2g}$ phonons at $\Gamma$ (\ie, $\omega_{E_{2g}}(\Gamma)$) and the $A_1'$ phonon at $\K$ (\ie, $\omega_{A_1'}(\K)$).

To show this, we derive an analytical expression for $\mcomega$ in the  $(\mu-\epsilon_0) \rightarrow 0$ limit.
We note that we only consider small but nonzero $|\mu-\epsilon_0|$, which means that only one electron band is cut by the chemical potential.
In the  $(\mu-\epsilon_0) \rightarrow 0$ limit, the expression of $\alpha^2 F(\omega)$ in \eqnref{eq:alpha2F} is simplified to
\eqa{
& \alpha^2 F(\omega) \\
& = \frac{1}{D(\mu) N} \sum_{\bsl{k},\bsl{k}'}^{\BZ} \sum_{nml}\frac{|G_{nml}(\bsl{k},\bsl{k}')|^2}{\hbar } \delta\left(\mu - E_n(\bsl{k}) \right) \delta\left(\mu - E_m(\bsl{k}') \right) \delta(\omega - \omega_l(\bsl{k}'-\bsl{k})) \\
& = \frac{N}{D(\mu)} \left[\frac{\Omega}{(2\pi)^2}\right]^2 \int_{FS} d\sigma_{\bsl{k}} \int_{FS} d\sigma_{\bsl{k}'} \frac{1}{|\nabla_{\bsl{k}} E_{n_F}(\bsl{k})|} \frac{1}{|\nabla_{\bsl{k}'} E_{n_F}(\bsl{k}')|} \sum_{l}\frac{|G_{n_Fn_Fl}(\bsl{k},\bsl{k}')|^2}{\hbar } \delta(\omega - \omega_l(\bsl{k}'-\bsl{k})) \\
& = \frac{N}{D(\mu)} \left[\frac{\Omega}{(2\pi)^2}\right]^2 \sum_{\alpha,\alpha'=\pm} \int_{FS_{\alpha' \K}} d\sigma_{\bsl{k}} \int_{FS_{\alpha \K}} d\sigma_{\bsl{k}'} \frac{1}{|\nabla_{\bsl{k}} E_{n_F}(\bsl{k})|} \frac{1}{|\nabla_{\bsl{k}'} E_{n_F}(\bsl{k}')|} \sum_{l}\frac{|G_{n_Fn_Fl}(\bsl{k},\bsl{k}')|^2}{\hbar } \delta(\omega - \omega_l(\bsl{k}'-\bsl{k})) \\
& = \frac{2 N}{D(\mu)} \left[\frac{\Omega}{(2\pi)^2}\right]^2 \sum_{\alpha=\pm} \int_{FS_{ \K}} d\sigma_{\bsl{k}} \int_{FS_{\alpha\K}} d\sigma_{\bsl{k}'} \frac{1}{|\nabla_{\bsl{k}} E_{n_F}(\bsl{k})|} \frac{1}{|\nabla_{\bsl{k}'} E_{n_F}(\bsl{k}')|} \sum_{l}\frac{|G_{n_Fn_Fl}(\bsl{k},\bsl{k}')|^2}{\hbar } \delta(\omega - \omega_l(\bsl{k}'-\bsl{k})) \\
& = \frac{2 N}{D(\mu) \hbar} \left[\frac{\Omega}{(2\pi)^2}\right]^2 \sum_{\alpha=\pm} \frac{|\mu|^2}{|v|^4}\int_{0}^{2\pi} d \theta \int_{0}^{2\pi} d \theta' \sum_{l} \\
& \qquad \times |G_{n_F n_F l}( \K + p_F \bsl{n}_\theta, \alpha  \K + p_F \bsl{n}_{\theta'} )|^2  \delta(\omega - \omega_l(\alpha\K -  \K +  p_F  \bsl{n}_\theta - p_F  \bsl{n}_{\theta'}) ) \\
& = \frac{2 N}{D(\mu) \hbar} \left[\frac{\Omega}{(2\pi)^2}\right]^2 \sum_{\alpha=\pm} \frac{|\mu|^2}{|v|^4}\int_{0}^{2\pi} d \theta \int_{0}^{2\pi} d \theta' \sum_{l}  |G_{n_F n_F l}( \K + p_F \bsl{n}_\theta, \alpha  \K + p_F \bsl{n}_{\theta'} )|^2  \delta(\omega - \omega_l(\alpha\K -  \K) ) \ ,
}
where $FS_{\pm \K}$ is the Fermi surface around $\pm\K$, the 4th equality comes from TR symmetry, $p_F = |\mu - \epsilon_0|/|v|$, $\bsl{n}_{\theta} = (\cos(\theta), \sin(\theta) , 0 )$, and we use $p_F\rightarrow 0$ for $\omega_l(\alpha\K -  \K +  p_F  \bsl{n}_\theta - p_F  \bsl{n}_{\theta'})$.
To proceed, let us simplify $|G_{n_F n_F l}( \K + p_F \bsl{n}_\theta, \alpha \K + p_F \bsl{n}_{\theta'} )|^2$.
According to the expression of $G_{n m l}(\bsl{k},\bsl{k}')$ in \eqnref{eq:G_nml} (as well as \eqnref{eq:ft_l} and \eqnref{eq:Gt_nml}), we have
\eqa{
\label{eq:Gnnl_graphene_linear_intermediate}
 & |G_{n_F n_F l}( \K + p_F \bsl{n}_\theta, \alpha\K + p_F \bsl{n}_{\theta'} )|^2 \\
 & = \frac{\hbar \Tr[ P_{n_F}( \K + p_F \bsl{n}_{\theta}) \widetilde{F}_l( \K + p_F \bsl{n}_{\theta} , \alpha \K + p_F \bsl{n}_{\theta'}) P_{n_F}(\alpha \K + p_F \bsl{n}_{\theta'}) \widetilde{F}_l^\dagger( \K + p_F \bsl{n}_{\theta} , \alpha \K + p_F \bsl{n}_{\theta'})]}{2 \omega_l(\alpha \K + p_F \bsl{n}_{\theta'} - \K - p_F \bsl{n}_\theta)}  \\
 & = \frac{\hbar}{2 \omega_l(\alpha \K  - \K )} \Tr[ P_{n_F, \K}( \theta ) \widetilde{F}_l( \K  , \alpha \K ) P_{n_F, \alpha\K}( \theta' ) \widetilde{F}_l^\dagger(\K , \alpha \K )] \ ,
}
where 
\eq{
\label{eq:ft_pmK_pmK_graphene}
\widetilde{F}_l(\K  , \alpha \K ) = \frac{1}{\sqrt{m_C}} \sum_{\bsl{\tau}, i} F_{\bsl{\tau}i}(\K  , \alpha \K )  \left[ v_l^*(\alpha \K - \K  ) \right]_{\bsl{\tau}i}
}
according to \eqnref{eq:ft_l}, $v_l(\bsl{q})$ is the phonon eigenvector for the $l$th phonon band,
\eq{
P_{n, \alpha \K}( \theta ) = P_{n}( \alpha \K + p_F \bsl{n}_{\theta} ) =\frac{1}{2} + \frac{(-1)^n}{2} \left[ \alpha \cos(\theta) \tau_x + \sin(\theta) \tau_y \right]
}
for the linear Hamiltonian \eqnref{eq:linear_el_graphene}, and we have used $p_F\rightarrow 0$ for the last equality.

The expression of $\widetilde{F}_l(\alpha \K  , \K )$ in \eqnref{eq:Gnnl_graphene_linear} are derived from $F_{\bsl{\tau}i}( \K  , \pm \K )$ and the phonon eigenvectors. 
From \eqnref{eq:f_k_graphene} and \eqnref{eq:g_beta_k_graphene}, we can derive the expression of $F_{\bsl{\tau}i}( \K  , \pm \K )$, which reads
\eqa{
\label{eq:f_graphene_2center}
& F_{\bsl{\tau}i}(\K,\K) = -\ii \hat{\gamma} \frac{\sqrt{3} a}{2} \left( \chi_{\bsl{\tau}}\tau_i-\tau_i\chi_{\bsl{\tau}} \right) (\delta_{ix} + \delta_{iy}) \\
& F_{\bsl{\tau}i}(\K,-\K) = -\ii  \hat{\gamma} \frac{\sqrt{3} a}{2} \left( \chi_{\bsl{\tau}}(-)^i\tau_i-\tau_i\chi_{\bsl{\tau}} \right) (\delta_{ix} + \delta_{iy})\ ,
}
where $(-)^x = - (-)^y =-1$.

For $F_{\bsl{\tau}i}( \K  , \K )$, we care about the phonon modes at $\bsl{q} = \K  - \K = 0$, which have eigenvectors 
\eqa{
\label{eq:Gamma_phonons_graphene}
v_1(\Gamma) = \frac{1}{\sqrt{2}} \mat{ 1 \\ 0 \\ 0 \\ 1 \\ 0 \\ 0}\ ,\ v_2(\Gamma) = \frac{1}{\sqrt{2}} \mat{ 0 \\ 1 \\ 0 \\ 0 \\ 1 \\ 0}\ ,\ v_3(\Gamma) = \frac{1}{\sqrt{2}} \mat{ 1 \\ 0 \\ 0 \\ -1 \\ 0 \\ 0 }\ ,\ v_4(\Gamma) = \frac{1}{\sqrt{2}} \mat{ 0 \\ 1 \\ 0 \\ 0 \\ -1 \\ 0 }\ ,
}
where the basis is 
\eq{
(\bsl{\tau}_A,x), (\bsl{\tau}_A,y), (\bsl{\tau}_A,z), (\bsl{\tau}_B,x), (\bsl{\tau}_B,y), (\bsl{\tau}_B,z)\ ,
}
and $v_1(\Gamma)$ and $v_2(\Gamma)$ are two acoustic modes with $\omega_1(\Gamma) = \omega_2(\Gamma) = 0$.
$v_3(\Gamma)$ and $v_4(\Gamma)$ are two optical modes that form the $E_{2g}$ irrep of the point group $D_{6h}$, and they have the same frequency $\omega_3(\Gamma) = \omega_4(\Gamma) = \omega_{E_{2g}}(\Gamma)$.

For $F_{\bsl{\tau}i}( \K  , -\K )$, we care about the phonon modes at $\bsl{q} = -\K-\K = -2\K $ (note that $-2\K$ and $\K$ are related by a reciprocal lattice vector), which have eigenvectors 
\eqa{
\label{eq:K_phonons_graphene}
v_1(-2\K) = \frac{1}{2} \mat{ 1 \\ \ii \\ 0 \\ 1 \\ -\ii \\ 0}\ ,\ v_2(-2\K) = \frac{1}{2} \mat{ 1 \\ -\ii \\ 0 \\ -1 \\ -\ii \\ 0}\ ,\ v_3(-2\K) = \frac{1}{2} \mat{ 1 \\ -\ii \\ 0 \\ 1 \\ \ii \\ 0 }\ ,\ v_4(-2\K) = \frac{1}{2} \mat{ -\ii \\ 1 \\ 0 \\ \ii \\ 1 \\ 0 }\ ,
}
where $v_1(-2\K)$ furnishes the $A_1'$ irrep with frequency $\omega_1(-2\K) = \omega_{A_1'}(\K)$, $v_2(-2\K)$ furnishes the $A_2'$ irrep with frequency $\omega_2(-2\K) = \omega_{A_2'}(\K)$,  and $v_3(-2\K)$ and $v_4(-2\K)$ furnish the $E'$ irrep with  frequency $\omega_3(-2\K) = \omega_4(-2\K) = \omega_{E'}(\K)$.
Here $A_1'$, $A_2'$ and $E'$ are irreps of $D_{3h}$ point group that leaves $\K$ invariant up to reciprocal lattice vectors.
Here we only consider the phonon modes with in-plane ion motions, since (i) the EPC is zero for the out-of-plane modes (\eqnref{eq:f_z_zero_graphene}), and (ii) the out-of-plane modes are decoupled from the in-plane modes in the phonon Hamiltonian owing to $m_z$ symmetry.
By substituting \eqnref{eq:f_graphene_2center}, \eqnref{eq:Gamma_phonons_graphene} and \eqnref{eq:K_phonons_graphene} into \eqnref{eq:ft_pmK_pmK_graphene}, we arrive at
\eqa{
\label{eq:ft_K_graphene_explicit}
& \widetilde{F}_1(  \K  ,  \K ) = \frac{1}{\sqrt{2 m_C}} (-\ii) \hat{\gamma} \frac{\sqrt{3} a}{2} \left( \tau_x-\tau_x \right) = 0 \\
& \widetilde{F}_2(  \K  ,  \K ) = \frac{1}{\sqrt{2 m_C}} (-\ii) \hat{\gamma} \frac{\sqrt{3} a}{2} \left( \tau_y-\tau_y \right) = 0 \\
& \widetilde{F}_3(  \K  ,  \K ) = \frac{1}{\sqrt{2 m_C}} (-\ii) \hat{\gamma} \frac{\sqrt{3} a}{2} \left( \tau_z \tau_x-\tau_x \tau_z  \right) = \frac{a \sqrt{3}}{\sqrt{ 2 m_C}}  \hat{\gamma}  \tau_y  \\
& \widetilde{F}_4(  \K  ,  \K ) = \frac{1}{\sqrt { m_C}} (-\ii) \hat{\gamma} \frac{\sqrt{3} a}{2} \left( \tau_z \tau_y-\tau_y \tau_z  \right) =  - \frac{\sqrt{3} a}{\sqrt{ 2 m_C}}  \hat{\gamma}  \tau_x  \\
& \widetilde{F}_1(  \K  ,  -\K ) = \frac{1}{2 \sqrt{m_C}} (-\ii) \hat{\gamma} \frac{\sqrt{3} a}{2} \left(  -\tau_x - \tau_x - \ii \tau_z \tau_y  + \ii \tau_y \tau_z \right) =  \ii \frac{\sqrt{3} a}{ \sqrt{m_C}}  \hat{\gamma}  \tau_x\\
& \widetilde{F}_2(  \K  ,  -\K ) = \frac{1}{2 \sqrt{m_C}} (-\ii) \hat{\gamma} \frac{\sqrt{3} a}{2} \left(  - \tau_z \tau_x - \tau_x \tau_z + \ii \tau_y  - \ii \tau_y  \right) =  0\\
& \widetilde{F}_3(  \K  , -\K ) = \frac{1}{2 \sqrt{m_C}} (-\ii) \hat{\gamma} \frac{\sqrt{3} a}{2} \left(  -\tau_x - \tau_x + \ii \tau_z \tau_y  - \ii \tau_y \tau_z \right) =  0\\
& \widetilde{F}_4(  \K  ,  -\K ) = -\ii \frac{1}{2 \sqrt{m_C}} (-\ii) \hat{\gamma} \frac{\sqrt{3} a}{2} \left(  - \ii  \tau_z \tau_x - \ii  \tau_x \tau_z +  \tau_y  - \tau_y  \right) =  0\ .
}
\eqnref{eq:ft_K_graphene_explicit} clearly shows that only the optical phonon at $\Gamma$ and the $A_1'$ phonon at $\K$ have nonzero contribution to $\widetilde{F}_l(  \K  ,  \pm \K )$.

With \eqnref{eq:ft_K_graphene_explicit}, nonzero $|G_{n_F n_F l}|^2$ in \eqnref{eq:Gnnl_graphene_linear_intermediate} becomes
\eqa{
\label{eq:Gnnl_graphene_linear}
 & |G_{n_F n_F 3}(\K + p_F \bsl{n}_\theta, \K + p_F \bsl{n}_{\theta'} )|^2 = \frac{3 \hbar a^2 \hat{\gamma}^2 }{4 m_C   \omega_{E_{2g}}(\Gamma) } \Tr[ P_{n_F, \K}( \theta ) \tau_y  P_{n_F, \K}( \theta' ) \tau_y  ] \\
 & \quad = \frac{3 \hbar a^2 \hat{\gamma}^2 }{8 m_C   \omega_{E_{2g}}(\Gamma) }  \left( 1 - \cos(\theta) \cos(\theta') + \sin(\theta) \sin(\theta') \right) \\
 & |G_{n_F n_F 4}(\K + p_F \bsl{n}_\theta, \K + p_F \bsl{n}_{\theta'} )|^2 = \frac{3 \hbar a^2 \hat{\gamma}^2}{4 m_C   \omega_{E_{2g}}(\Gamma) } \Tr[ P_{n_F, \K}( \theta ) \tau_x  P_{n_F, \K}( \theta' ) \tau_x  ]   \\
 & \quad = \frac{3 \hbar a^2 \hat{\gamma}^2}{8 m_C   \omega_{E_{2g}}(\Gamma) } \left( 1 + \cos(\theta) \cos(\theta') - \sin(\theta) \sin(\theta') \right)\\
 & |G_{n_F n_F 1}(\K + p_F \bsl{n}_\theta, -\K + p_F \bsl{n}_{\theta'} )|^2 = \frac{3 \hbar a^2 \hat{\gamma}^2}{2 m_C   \omega_{A_1'}(\K) } \Tr[ P_{n_F, \K}( \theta ) \tau_x  P_{n_F, -\K}( \theta' ) \tau_x  ]   \\
 & \quad = \frac{3 \hbar a^2 \hat{\gamma}^2}{4 m_C   \omega_{A_1'}(\K) }  \left( 1 + \cos(\theta) \cos(\theta') - \sin(\theta) \sin(\theta') \right)\ .
}
Then, we have 
\eqa{
\label{eq:alpha2F_graphene_linear}
\alpha^2 F(\omega) & = \frac{2 N}{D(\mu) \hbar} \left[\frac{\Omega}{(2\pi)^2}\right]^2 \sum_{\alpha=\pm} \frac{|\mu|^2}{|v|^4}\int_{0}^{2\pi} d \theta \int_{0}^{2\pi} d \theta' \sum_{l}  |G_{n_F n_F l}( \K + p_F \bsl{n}_\theta, \alpha  \K + p_F \bsl{n}_{\theta'} )|^2  \delta(\omega - \omega_l(\alpha\K -  \K) ) \\
& = \frac{2 N}{D(\mu) \hbar} \left[\frac{\Omega}{(2\pi)^2}\right]^2  \frac{|\mu|^2}{|v|^4}\int_{0}^{2\pi} d \theta \int_{0}^{2\pi} d \theta' \left[ \sum_{l=3,4}|G_{n_F n_F l}(\K + p_F \bsl{n}_\theta, \K + p_F \bsl{n}_{\theta'} )|^2 \delta(\omega - \omega_{E_{2g}}(\Gamma) ) \right.\\
& \qquad \left. + |G_{n_F n_F 1}(\K + p_F \bsl{n}_\theta, -\K + p_F \bsl{n}_{\theta'} )|^2  \delta(\omega - \omega_{A_1'}(\K) )\right]  \\
& = \frac{2 N}{D(\mu) \hbar} \frac{\Omega^2}{(2\pi)^2} \frac{|2\pi\mu|^2}{|v|^4} \frac{3 \hbar a^2 \hat{\gamma}^2 }{4 m_C  }\left[ \frac{ \delta(\omega - \omega_{E_{2g}}(\Gamma) ) }{ \omega_{E_{2g}}(\Gamma) }  +  \frac{ \delta(\omega - \omega_{A_1'}(\K) ) }{\omega_{A_1'}(\K) }  \right] \\
& = \frac{D(\mu)}{2 N} \frac{3  a^2 \hat{\gamma}^2 }{4 m_C  }\left[ \frac{ \delta(\omega - \omega_{E_{2g}}(\Gamma) ) }{ \omega_{E_{2g}}(\Gamma) }  +  \frac{ \delta(\omega - \omega_{A_1'}(\K) ) }{\omega_{A_1'}(\K) }  \right]
}
As a result, we arrive at the following approximated expression of $\mcomega$ for graphene in the $(\mu-\epsilon_0) \rightarrow 0$ limit:
\eq{
\label{eq:mcomega_graphene_linear}
\mcomega = \frac{ 2 }{ \frac{1}{\omega_{A_1'}^2(\K)}+  \frac{1}{\omega_{E_{2g}}^2(\Gamma)}} = \frac{2 \omega_{E_{2g}}^2(\Gamma) \omega_{A_1'}^2(\K) }{\omega_{E_{2g}}^2(\Gamma) + \omega_{A_1'}^2(\K)}\ ,
}
which is independent of the EPC.
The expression is approximated because we neglect the acoustic phonons in the denominator, which is justified by the fact that acoustic phonons only contribute about 17\% of the total EPC constant for $\mu=-0.02$eV suggested by the {\abi} numerical calculations.

The {\abi} calculation shows that  $\hbar \omega_{E_{2g}}(\Gamma) = 0.1935 \eV$ and $\hbar\omega_{A_1'}(\K) = 0.1622 \eV$.
Then, according to the expression of $\mcomega$ (\eqnref{eq:mcomega_graphene_linear}) derived from our model for $\mu \rightarrow 0$, we have an approximate value $\hbar \sqrt{\mcomega} = 0.1758 \eV$, which only has about $9\%$ error for $\mu\in [-1\eV, -0.1\eV]$, according to the {\abi} value of $\mcomega$ show in \figref{fig:graphene_EPC}(a).
The fact that the expression of $\mcomega$ in \eqnref{eq:mcomega_graphene_linear} approximately holds up to $\mu=-1$eV means that the linear approximation is good even if $\mu-\epsilon_0$ is away from zero but not too far.

\subsection{Further-Range Hopping}
\label{app:further_range_hopping_graphene}

In \appref{app:geo_EPC_graphene}, we have discussed how the symmetry-rep method can separate the EPC Hamiltonian into a energetic and geometric parts, if we only include terms up to the nearest neighbors.
In this part, we will show that the separation will work (with the hopping derivatives replaced by coefficients) if we include the NNN hoppings, but will fail if we include 3rd NN hoppings.

We first include the NNN terms in the electron Hamiltonian and the EPC Hamiltonian.
The NNN hoppings have the form as $t_{\bsl{\tau}\bsl{\tau}}(C_6^l\bsl{a}_1)$ with $\bsl{a}_1$ the primitive lattice vector in \eqnref{eq:a1_a2_graphene} and $l=0,1,...,5$.
By using the symmetry properties of the hoppings \eqnref{eq:t_h_sym_g}, \eqnref{eq:t_h_sym_TR} and the symmetry rep \eqnref{eq:sym_rep_graphene}, we obtain 
\eqa{
\label{eq:t_sym_graphene_NNN}
& C_6: t_{\bsl{\tau}_A\bsl{\tau}_A}(\bsl{a}_1) = t_{\bsl{\tau}_B\bsl{\tau}_B}(C_6\bsl{a}_1) = t_{\bsl{\tau}_A\bsl{\tau}_A}(C_6^2\bsl{a}_1) = t_{\bsl{\tau}_B\bsl{\tau}_B}(C_6^3\bsl{a}_1) = t_{\bsl{\tau}_A\bsl{\tau}_A}(C_6^4\bsl{a}_1) = t_{\bsl{\tau}_B\bsl{\tau}_B}(C_6^5\bsl{a}_1) \\
& C_6: t_{\bsl{\tau}_B\bsl{\tau}_B}(\bsl{a}_1) = t_{\bsl{\tau}_A\bsl{\tau}_A}(C_6\bsl{a}_1) = t_{\bsl{\tau}_B\bsl{\tau}_B}(C_6^2\bsl{a}_1) = t_{\bsl{\tau}_A\bsl{\tau}_A}(C_6^3\bsl{a}_1) = t_{\bsl{\tau}_B\bsl{\tau}_B}(C_6^4\bsl{a}_1) = t_{\bsl{\tau}_A\bsl{\tau}_A}(C_6^5\bsl{a}_1) \\
& m_y: t_{\bsl{\tau}_A\bsl{\tau}_A}(C_6^l\bsl{a}_1) = t_{\bsl{\tau}_B\bsl{\tau}_B}(m_y C_6^l\bsl{a}_1) \text{ for l=0,1,...,5}\\
& m_z: \text{no constraints} \\
& \TR: t_{\bsl{\tau}\bsl{\tau}}(C_6^l\bsl{a}_1) \in \dsR \text{ for $\bsl{\tau}=\bsl{\tau}_A,\bsl{\tau}_B$ and $l=0,1,...,5$}\\
& h.c.: t_{\bsl{\tau}\bsl{\tau}}(C_6^l\bsl{a}_1) = t_{\bsl{\tau}\bsl{\tau}}^*(-C_6^l\bsl{a}_1)\text{ for $\bsl{\tau}=\bsl{\tau}_A,\bsl{\tau}_B$ and $l=0,1,...,5$}\ ,
}
resulting in 
\eqa{
& t_{\bsl{\tau}\bsl{\tau}}(C_6^l\bsl{a}_1)= t_1 \in\dsR
}
for $\bsl{\tau}=\bsl{\tau}_A,\bsl{\tau}_B$ and $n=0,1,...,5$.
Then, the matrix $h(\bsl{k})$ in $H_{el}$ for graphene in momentum space (\eqnref{eq:H_el_graphene}) becomes
\eq{
\label{eq:h_k_graphene_with_NNN}
h(\bsl{k}) = \left(\epsilon_0 + t_1 \sum_{l=0}^5 e^{-\ii \bsl{k}\cdot C_6^l\bsl{a}_1} \right) \tau_0 + t \sum_{j} \mat{ 0 & e^{-\ii \bsl{\delta}_j\cdot \bsl{k}} \\ e^{\ii \bsl{\delta}_j\cdot \bsl{k} } & 0}\ ,
}
where $\bsl{\delta}_j$ is defined in \eqnref{eq:delta_j_graphene}.
The bands and projection matrices still have the form in \eqnref{eq:E_P_graphene} with 
\eq{
\label{eq:d_NNN_graphene}
d_0(\bsl{k}) = \epsilon_0+t_1 \sum_{l=0}^5 e^{-\ii \bsl{k}\cdot C_6^l\bsl{a}_1}\ ,\  d_x(\bsl{k}) = t   \sum_{j} \cos\left(\ii \bsl{\delta}_j\cdot \bsl{k}\right)\ ,\  d_y (\bsl{k}) = t   \sum_{j} \sin\left(\ii \bsl{\delta}_j\cdot \bsl{k}\right)\ ,
}
which has a different $d_0(\bsl{k})$ compared to \eqnref{eq:d_graphene}.

Now we include the NNN terms in the EPC Hamiltonian.
As shown in \eqnref{eq:g_2D_partial_k}, the EPC $f_i(\bsl{k})$ is determined by $\widetilde{f}_{\bsl{\tau}_1 \bsl{\tau}_2,\shpa}(\Delta\bsl{R} + \bsl{\tau}_1 - \bsl{\tau}_2 )$ and $\widetilde{f}_{\bsl{\tau}_1 \bsl{\tau}_2,\perp}(\Delta\bsl{R} + \bsl{\tau}_1 - \bsl{\tau}_2 )$.
The NNN terms in the EPC Hamiltonian are given by $\widetilde{f}_{\bsl{\tau} \bsl{\tau},\beta=\shpa/\perp}(C_6^l\bsl{a}_1)$ with $l=1,...,5$.
According to \eqnref{eq:g_beta_sym}, $\widetilde{f}_{\bsl{\tau}\bsl{\tau},\shpa}(C_6^l\bsl{a}_1)$ should have the same symmetry properties as $t_{\bsl{\tau}\bsl{\tau}}(C_6^l\bsl{a}_1)$ in \eqnref{eq:t_sym_graphene_NNN}, resulting in 
\eq{
\label{eq:g_shpa_R_NNN_graphene}
\widetilde{f}_{\bsl{\tau}\bsl{\tau},\shpa}(C_6^l\bsl{a}_1) = \hat{\gamma}_1\ .
}
On the other hand, \eqnref{eq:g_beta_sym} also shows that the symmetry constraints of $\widetilde{f}_{\bsl{\tau}\bsl{\tau},\perp}(C_6^l\bsl{a}_1)$ differ by a minus sign for $m_y$ from those of  $t_{\bsl{\tau}\bsl{\tau}}(C_6^l\bsl{a}_1)$ in \eqnref{eq:t_sym_graphene_NNN}, leading to
\eqa{
\label{eq:ftilde_perp_sym_graphene_NNN}
& C_6: \widetilde{f}_{\bsl{\tau}_A\bsl{\tau}_A,\perp}(\bsl{a}_1) = \widetilde{f}_{\bsl{\tau}_B\bsl{\tau}_B,\perp}(C_6\bsl{a}_1) = \widetilde{f}_{\bsl{\tau}_A\bsl{\tau}_A,\perp}(C_6^2\bsl{a}_1) = \widetilde{f}_{\bsl{\tau}_B\bsl{\tau}_B,\perp}(C_6^3\bsl{a}_1) = \widetilde{f}_{\bsl{\tau}_A\bsl{\tau}_A,\perp}(C_6^4\bsl{a}_1) = \widetilde{f}_{\bsl{\tau}_B\bsl{\tau}_B,\perp}(C_6^5\bsl{a}_1) \\
& C_6: \widetilde{f}_{\bsl{\tau}_B\bsl{\tau}_B,\perp}(\bsl{a}_1) = \widetilde{f}_{\bsl{\tau}_A\bsl{\tau}_A,\perp}(C_6\bsl{a}_1) = \widetilde{f}_{\bsl{\tau}_B\bsl{\tau}_B,\perp}(C_6^2\bsl{a}_1) = \widetilde{f}_{\bsl{\tau}_A\bsl{\tau}_A,\perp}(C_6^3\bsl{a}_1) = \widetilde{f}_{\bsl{\tau}_B\bsl{\tau}_B,\perp}(C_6^4\bsl{a}_1) = \widetilde{f}_{\bsl{\tau}_A\bsl{\tau}_A,\perp}(C_6^5\bsl{a}_1) \\
& m_y: \widetilde{f}_{\bsl{\tau}_A\bsl{\tau}_A,\perp}(C_6^l\bsl{a}_1) = -\widetilde{f}_{\bsl{\tau}_B\bsl{\tau}_B,\perp}(m_y C_6^l\bsl{a}_1) \text{ for l=0,1,...,5}\\
& m_z: \text{no constraints} \\
& \TR: \widetilde{f}_{\bsl{\tau}\bsl{\tau},\perp}(C_6^l\bsl{a}_1) \in \dsR \text{ for $\bsl{\tau}=\bsl{\tau}_A,\bsl{\tau}_B$ and $l=0,1,...,5$}\\
& h.c.: \widetilde{f}_{\bsl{\tau}\bsl{\tau},\perp}(C_6^l\bsl{a}_1) = \widetilde{f}_{\bsl{\tau}\bsl{\tau},\perp}^*(-C_6^l\bsl{a}_1)\text{ for $\bsl{\tau}=\bsl{\tau}_A,\bsl{\tau}_B$ and $l=0,1,...,5$}\ .
}
According to \eqnref{eq:ftilde_perp_sym_graphene_NNN}, Hermiticity, $\TR$ and $C_6$ together give $\widetilde{f}_{\bsl{\tau}\bsl{\tau},\perp}(C_6^l\bsl{a}_1) = \hat{\gamma}'$ for all $\bsl{\tau}$ and $l$.
Combined with $m_y$, we know  $\hat{\gamma}'=0$.
Therefore, we know 
\eq{
\label{eq:g_perp_R_NNN_graphene}
\widetilde{f}_{\bsl{\tau}\bsl{\tau},\perp}(C_6^l\bsl{a}_1) = 0\ .
}
By substituting \eqnref{eq:g_shpa_R_NNN_graphene} and \eqnref{eq:g_perp_R_NNN_graphene} into \eqnref{eq:g_beta_k_block}, we obtain 
\eqa{
\label{eq:g_beta_k_NNN_graphene}
 &   \widetilde{f}_{\shpa}(\bsl{k}) = \hat{\gamma}_1 \sum_{l=0}^5 e^{-\ii \bsl{k}\cdot C_6^l\bsl{a}_1} \tau_0 + \hat{\gamma} \sum_{j} \mat{ 0 & e^{-\ii \bsl{\delta}_j\cdot \bsl{k}} \\ e^{\ii \bsl{\delta}_j\cdot \bsl{k} } & 0} \\
 &   \widetilde{f}_{\perp}(\bsl{k}) = 0\ ,
}
and thus $ f_{i}(\bsl{k}) $ still has the form in \eqnref{eq:g_i_g_beta_grpahene} but with the expression of $\widetilde{f}_{\shpa}( \bsl{k} )$ in \eqnref{eq:g_beta_k_NNN_graphene}.

Now we identify the energetic and geometric parts of $f_{i}(\bsl{k})$ in \eqnref{eq:g_i_g_beta_grpahene}.
Same as \appref{app:geo_EPC_graphene}, $\L_{\perp} = 0$ still holds.
Since $ \widetilde{f}_{\perp}(\bsl{k}) = 0$ still holds as shown in \eqnref{eq:g_beta_k_NNN_graphene}, we still have $\Delta  \widetilde{f}_{\perp} = 0 $ in \eqnref{eq:g_perp_h_relation}.
$\Delta f_i (\bsl{k}) = 0$ still holds in \eqnref{eq:Delta_g_i_2D} owing to $f_z(\bsl{k}) = 0 $ in \eqnref{eq:f_z_zero_graphene}.
Therefore, we still only need to care about $ \widetilde{f}_{\shpa}(\bsl{k})$.

According to \eqnref{eq:g_shpa_h_relation},  we try to re-write $ \widetilde{f}_{\shpa}(\bsl{k})$ in \eqnref{eq:g_beta_k_NNN_graphene} as 
\eq{
\widetilde{f}_{\shpa}(\bsl{k})=\hat{\gamma}_0 \partial_{\epsilon_0} h(\bsl{k}) + \hat{\gamma} \partial_{t} h(\bsl{k}) + \hat{\gamma}_1 \partial_{t_1} h(\bsl{k})\ , 
}
where $h(\bsl{k})$ is the electron Hamiltonian in \eqnref{eq:h_k_graphene_with_NNN}.
By comparing \eqnref{eq:g_beta_k_NNN_graphene} to \eqnref{eq:h_k_graphene_with_NNN}, $\hat{\gamma}_0$ in $ \widetilde{f}_{\shpa}(\bsl{k})$ should still be zero, and then we obtain
\eq{
 \widetilde{f}_{\shpa}(\bsl{k}) = \hat{\gamma} \partial_t  h(\bsl{k})+ \hat{\gamma}_1 \partial_{t_1} h(\bsl{k}) \ ,
}
resulting in 
\eq{
 f_i(\bsl{k}) = f_i^{E}(\bsl{k}) + f_i^{geo}(\bsl{k})\ ,
}
where 
\eqa{
\label{eq:g_2D_E_geo_NNN_graphene}
f_{i}^E(\bsl{k}) & =  \ii (\hat{\gamma} \partial_t+\hat{\gamma}_1 \partial_{t_1}) \sum_{n} P_{n}(\bsl{k}) \partial_{k_{i}} E_n(\bsl{k})(\delta_{ix} + \delta_{iy})  \\
f_{i}^{geo}(\bsl{k}) &  =  \ii (\hat{\gamma} \partial_t+\hat{\gamma}_1 \partial_{t_1}) \sum_{n} E_n(\bsl{k}) \partial_{k_{i}}  P_{n}(\bsl{k})(\delta_{ix} + \delta_{iy})
}
according to \eqnref{eq:g_2D_E_geo}.

To address the hopping derivatives, note that 
\eqa{
& \sum_{n} \frac{(-1)^n}{2}  \partial_{k_i}\Delta E(\bsl{k}) P_{n}(\bsl{k})  (\delta_{ix} + \delta_{iy}) \\
& = \sum_{n} \partial_{k_i} \left[ (-)^n \sqrt{d_x^2(\bsl{k})+d_y^2(\bsl{k})} \right]  \frac{1}{2} \left[ 1 + (-)^n  \frac{d_x(\bsl{k})\tau_x + d_y(\bsl{k}) \tau_y}{\sqrt{d_x^2(\bsl{k})+d_y^2(\bsl{k})}}\right] (\delta_{ix} + \delta_{iy}) \\
& =  \sum_{n}\partial_{k_i} \left[  (-)^n \sqrt{d_x^2(\bsl{k})+d_y^2(\bsl{k})} \right] \frac{1}{2} \left[  (-)^n  \frac{d_x(\bsl{k})\tau_x + d_y(\bsl{k}) \tau_y}{\sqrt{d_x^2(\bsl{k})+d_y^2(\bsl{k})}}\right] (\delta_{ix} + \delta_{iy}) } 
is proportional to $t$ since $d_x(\bsl{k})$ and $d_y(\bsl{k})$ are proportional to $t$ as shown by \eqnref{eq:d_NNN_graphene}, 
\eqa{
& \sum_{n} \partial_{k_i} d_0(\bsl{k}) P_{n}(\bsl{k})  (\delta_{ix} + \delta_{iy})  = \partial_{k_i} d_0(\bsl{k}) (\delta_{ix} + \delta_{iy})
} 
is proportional to $t_1$ according to \eqnref{eq:d_NNN_graphene},
and 
\eqa{
& \sum_{n} E_n(\bsl{k}) \partial_{k_i} P_{n}(\bsl{k})  (\delta_{ix} + \delta_{iy}) \\
& = \sum_{n} \left[ d_0(\bsl{k}) + (-)^n \sqrt{d_x^2(\bsl{k})+d_y^2(\bsl{k})} \right]  \frac{1}{2} \partial_{k_i} \left[ 1 + (-)^n  \frac{d_x(\bsl{k})\tau_x + d_y(\bsl{k}) \tau_y}{\sqrt{d_x^2(\bsl{k})+d_y^2(\bsl{k})}}\right] (\delta_{ix} + \delta_{iy} )\\
& = \sum_{n} \left[  (-)^n \sqrt{d_x^2(\bsl{k})+d_y^2(\bsl{k})} \right] \frac{1}{2} \partial_{k_i}\left[  (-)^n  \frac{d_x(\bsl{k})\tau_x + d_y(\bsl{k}) \tau_y}{\sqrt{d_x^2(\bsl{k})+d_y^2(\bsl{k})}}\right] (\delta_{ix} + \delta_{iy})
}
is proportional to $t$ since $d_x(\bsl{k})$ and $d_y(\bsl{k})$ are proportional to $t$ as shown by \eqnref{eq:d_graphene}.
Then, we have 
\eqa{
\label{eq:g_E_NNN_graphene}
f^E_i(\bsl{k}) & = \ii (\hat{\gamma} \partial_t+\hat{\gamma}_1 \partial_{t_1}) \sum_{n} \partial_{k_i} E_n(\bsl{k}) P_{n}(\bsl{k})  (\delta_{ix} + \delta_{iy})\\
 & = \ii (\hat{\gamma} \partial_t+\hat{\gamma}_1 \partial_{t_1}) \left[\sum_{n} \partial_{k_i} d_0(\bsl{k}) P_{n}(\bsl{k})+\sum_{n}\frac{(-1)^n}{2} \partial_{k_i} \Delta E(\bsl{k}) P_{n}(\bsl{k}) \right] (\delta_{ix} + \delta_{iy})\\
  & = \ii \left[ \frac{\hat{\gamma}_1}{t_1} \partial_{k_i} d_0(\bsl{k})+\frac{\hat{\gamma}}{t} \sum_{n}\frac{(-1)^n}{2} \partial_{k_i} \Delta E(\bsl{k}) P_{n}(\bsl{k}) \right] (\delta_{ix} + \delta_{iy})\\
  & = \ii \left[ \gamma_1 \partial_{k_i} d_0(\bsl{k})+ \gamma \sum_{n}\frac{(-1)^n}{2} \partial_{k_i} \Delta E(\bsl{k}) P_{n}(\bsl{k}) \right] (\delta_{ix} + \delta_{iy})\\
}
\eqa{
\label{eq:g_geo_NNN_graphene}
f^{geo}_i(\bsl{k}) & = \ii \hat{\gamma} \partial_t  \sum_{n} E_n(\bsl{k}) \partial_{k_i}  P_{n}(\bsl{k})  (\delta_{ix} + \delta_{iy})\\
& = \ii \frac{\hat{\gamma}}{t}  \sum_{n} E_n(\bsl{k}) \partial_{k_i}  P_{n}(\bsl{k})  (\delta_{ix} + \delta_{iy})\\
 & = \ii  \gamma \sum_{n} E_n(\bsl{k}) \partial_{k_i} P_{n}(\bsl{k}) (\delta_{ix} + \delta_{iy}) \ ,
}
where $\gamma$ is defined in \eqnref{eq:graphene_gamma}, and 
\eq{
\gamma_1 = \frac{\hat{\gamma}_1}{t_1}\ .
}
As a result, we obtain 
\eq{
\label{eq:g_E_g_geo_NNN_graphene}
 f_{i}(\bsl{k})  =  f^E_i(\bsl{k}) + f^{geo}_i(\bsl{k}) \ ,
}
where 
\eqa{
\label{eq:g_E_g_geo_NNN_graphene_explicit}
& f^E_i(\bsl{k}) = \left[ \ii  \gamma_1 \partial_{k_i} d_0(\bsl{k})+ \ii \gamma \sum_{n}\frac{(-1)^n}{2} \partial_{k_i} \Delta E(\bsl{k}) P_{n}(\bsl{k})\right] (\delta_{ix} + \delta_{iy})\\
& f^{geo}_i(\bsl{k}) = \ii  \gamma \sum_{n} E_n(\bsl{k}) \partial_{k_i} P_{n}(\bsl{k})(\delta_{ix} + \delta_{iy})\ .
}
Thus, even with NNN terms, we can still split $ f_{i}(\bsl{k})$ into $f^E_i(\bsl{k}) + f^{geo}_i(\bsl{k}) $ with the hopping derivatives replaced by coefficients in $f^E_i(\bsl{k})$ and $f^{geo}_i(\bsl{k})$.

If we have 3rd NN hopping in addition to onsite and NN terms, then electron matrix Hamiltonian $h(\bsl{k})$ would become
\eq{
h(\bsl{k}) = \epsilon_0 \tau_0 + t \sum_{j} \mat{ 0 & e^{-\ii \bsl{\delta}_j\cdot \bsl{k}} \\ e^{\ii \bsl{\delta}_j\cdot \bsl{k} } & 0} + t' \sum_{j} \mat{ 0 & e^{\ii 2 \bsl{\delta}_j\cdot \bsl{k}} \\ e^{-\ii 2 \bsl{\delta}_j\cdot \bsl{k} } & 0}\ .
}
Then, $\sum_{n} \partial_{k_i} E_n(\bsl{k}) P_{n}(\bsl{k})  (\delta_{ix} + \delta_{iy})$ and $\sum_{n} E_n(\bsl{k}) \partial_{k_i}  P_{n}(\bsl{k})  (\delta_{ix} + \delta_{iy})$ cannot be proportional to just $t$ or $t'$.
As a result, we cannot replace $\partial_t$ by $1/t$ in \eqnref{eq:g_E_graphene} and \eqnref{eq:g_geo_graphene}, and we cannot derive \eqnref{eq:g_E_g_geo_graphene_explicit}.

\Or{
\subsection{Quantum Geometry, EPC and Mean-field Superconducting Critical Temperature in Magic-Angle Twisted Bilayer Graphene}

In this part, we will briefly discuss the case of the magic-angle twisted bilayer graphene (MATBG), and we will estimate the mean-field superconducting critical temperature of MATBG based on EPC and quantum geometry.
For convenience, we will choose the unit system (unless specified otherwise) as
\eq{
\label{eq:unit_system}
\hbar = 1\ ,\ \epsilon_0 = 1\ ,\ k_\theta = 1 \ ,\ v_0 = 1\ ,
}
where $k_\theta =  \frac{4\pi}{3 a }2 \sin(\frac{\theta}{2})$, $\epsilon_0$ is the vacuum permittivity, $a$ is the lattice constant of graphene, $v_0$ is the Fermi velocity of the monolayer graphene, and $\theta$ is the twist angle.

We will focus on the Bistritzer-MacDonald (BM) model~\cite{Bistritzer2011BMModel} in the first chiral limit~\cite{Tarnopolsky2019MagicAngleChiralLimit,BAB2021TBGIII,Wang2021ChiralMATBG}, which, in the $+$ valley, reads
\eq{
\label{eq:H0+_r}
H_{0,+}=\int d^2r\ \widetilde{\psi}^\dagger_{+,\bsl{r}}
\mat{
-\ii \bsl{\sigma}\cdot \bsl{\nabla} & T(\bsl{r}) & \\
T^\dagger(\bsl{r}) & -\ii \bsl{\sigma}\cdot \bsl{\nabla} }\otimes s_0\ 
\widetilde{\psi}_{+,\bsl{r}}\ ,
}
where $\psi^\dagger_{+,\bsl{r},l} = (\psi^\dagger_{+,\bsl{r},l,A,\uparrow},\psi^\dagger_{+,\bsl{r},l,A,\downarrow},\psi^\dagger_{+,\bsl{r},l,B,\uparrow},\psi^\dagger_{+,\bsl{r},l,B,\downarrow})$ with $l=t,b$ labelling the layer. 
Since we are considering the first chiral limit, there is only one term in the interlayer coupling $T(\bsl{r})$, which reads
$T(\bsl{r}) = \sum_{j=1,2,3} T_j e^{\ii \bsl{r} \cdot \bsl{q}_j}$,
\eq{
T_j =  w_1 \left[\cos(\frac{2\pi}{3} (j-1)) \sigma_x + \sin (\frac{2\pi}{3}(j-1) )\sigma_y \right]
}
with $\bsl{q}_1 =(0,1)^T$, $\bsl{q}_2 =(-\frac{\sqrt{3}}{2},- \frac{1}{2})^T $, $\bsl{q}_3 =(\frac{\sqrt{3}}{2}, - \frac{1}{2})^T$, and $\sigma_{x,y,z}$ and $s_{0,x,y,z}$ are Pauli matrices for sublattice and spin, respectively.
The model in the $-$ valley can be obtained by the TR symmetry.

In the first chiral limit, the model has two exactly flat bands per spin per valley, and two flat bands in one valley and one spin has well-defined but different Chern numbers $e_Y = \pm 1$.
Therefore, we can use $\gamma^\dagger_{\bsl{k}, e_Y, \eta, s}$ to label the creation operators for the flat bands, where $\eta = \pm $ and $s = \uparrow, \downarrow$ label the valley and spin, respectively. 
The expression of $\gamma^\dagger_{\bsl{k}, e_Y, \eta, s}$ is most straightforward to express in the topological heavy-fermion basis~\cite{Song20211110MATBGHF}.
As discussed in \refcite{Song20211110MATBGHF}, the topological heavy-fermion model in the first chiral limit for the $\eta$ valley reads
\eq{
h_{\eta}^{THF}(\bsl{k}) 
= 
\mat{
0_{2\times 2} & \gamma \tau_0 &  0_{2\times 2}  \\
\gamma \tau_0 &  0_{2\times 2} & v_{\star}(\eta k_x \tau_0 + \ii k_y \tau_z) \\
0_{2\times 2} & v_{\star}(\eta k_x \tau_0 - \ii k_y \tau_z) &  0_{2\times 2}
} \otimes s_0 \ ,
}
where the basis is $(f^\dagger_{\eta,\bsl{k}}, c^\dagger_{\eta,\bsl{k},\Gamma_3}, c^\dagger_{\eta,\bsl{k},\Gamma_1\Gamma_2})$ with 
\eqa{
& f^\dagger_{\eta,\bsl{k}} = (f^\dagger_{\eta,\bsl{k},1,\uparrow}, f^\dagger_{\eta,\bsl{k},1,\downarrow}, f^\dagger_{\eta,\bsl{k},2,\uparrow}, f^\dagger_{\eta,\bsl{k},2,\downarrow})\\
& c_{\eta,\bsl{k},\Gamma_3}^\dagger = (c_{\eta,\bsl{k},1, \uparrow}^\dagger, c_{\eta,\bsl{k},1,\downarrow}^\dagger, c_{\eta,\bsl{k},2, \uparrow}^\dagger, c_{\eta,\bsl{k},2, \downarrow}^\dagger) \\ 
& c_{\eta,\bsl{k},\Gamma_1\Gamma_2}^\dagger = (c_{\eta,\bsl{k},\Gamma_1, \uparrow}^\dagger, c_{\eta,\bsl{k},\Gamma_1,\downarrow}^\dagger, c_{\eta,\bsl{k},\Gamma_2, \uparrow}^\dagger, c_{\eta,\bsl{k},\Gamma_2, \downarrow}^\dagger)\ ,
}
and $\Gamma_1$, $\Gamma_2$ and $\Gamma_3$ are irreps of $D_3$ group to indicate the symmetry properties of the $c$ modes, and $s_{0,x,y,z}$ are identity and Pauli matrices for spin.
Here $\gamma$ is the coupling between $f$ and $c$, and $v_{*}$ is the velocity of the $c$ modes.
As a result, the expression of $\gamma^\dagger_{\bsl{k},e_Y, \eta, s}$ in the $\eta=+$ valley reads~\cite{CXL2023ElKPhCouplingTBG}
\eqa{
\label{eq:wavefun_flatbands_TBG}
& \gamma^\dagger_{\bsl{k},-, +, s} = \frac{1}{\sqrt{|\bsl{k}|^2 + b^2}} \left( - b \frac{k_x -\ii k_y}{|\bsl{k}|} c^\dagger_{\Gamma_1,\bsl{k},+,s} + |\bsl{k}| f^\dagger_{1,\bsl{k},+,s} \right)\\
& \gamma^\dagger_{\bsl{k}, +, +, s} = \frac{1}{\sqrt{|\bsl{k}|^2 + b^2}} \left( - b \frac{k_x + \ii k_y}{|\bsl{k}|} c^\dagger_{\Gamma_2,\bsl{k},+,s} + |\bsl{k}| f^\dagger_{2,\bsl{k},+,s} \right)\ ,
}
where $b=\gamma/v_{\star}$.
The flat-band wavefunction in \eqnref{eq:wavefun_flatbands_TBG} has FSM
\eq{
\label{eq:FSM_flat_band_MATBG}
\Tr[g_{\bsl{k}}] = \frac{ 2 b^2}{\left( |\bsl{k}|^2 +  b^2 \right)^2}\ ,
}
and the gap between the flat band and the remote bands are $\Delta E_{\bsl{k}}=\sqrt{|\bsl{k}|^2 v_\star^2 + \gamma^2}$.

We only consider the inter-valley EPC, which is experimentally shown to be strong~\cite{Chen2023ElKPhCouplingTBG}; the dominant contribution to this is the intralayer EPC of graphene for the phonons at $\K$ or $\K'$, \ie, $\widetilde{F}_{1,...,4}(K,-K)$ in \eqnref{eq:ft_K_graphene_explicit} and its TR related $\widetilde{F}_{1,...,4}(-K,K)$, which only couples to the $A_1'$ phonons at $\K$ or $\K'$.
After projecting to the flat bands, \refcite{CXL2023ElKPhCouplingTBG} shows that the EPC is diagonal in the Chern basis and has the following form
\eqa{
\label{eq:HEPC}
H_{EPC} = \frac{1}{ \sqrt{N_M}} \sum_{l,s}\sum_{\bsl{k}, \bsl{k}', e_Y} \sum_{\bsl{Q}\in \mathcal{Q}_{\bar{l},\eta} } G_{\bsl{k}, \bsl{k}', \bsl{Q}  }^{\eta e_Y l} \gamma_{\bsl{k},e_Y,\eta, s}^\dagger \gamma_{\bsl{k}',e_Y,-\eta, s} (b_{-\eta \K +\bsl{k}-\bsl{k}'-\bsl{Q},l} +b_{\eta \K -\bsl{k}+\bsl{k}'+\bsl{Q},l}^\dagger) \ ,
}
where $N_M$ is the number of moir\'e unit cells, $\bsl{k},\bsl{k}'$ are sumed over the Moir\'e BZ, $\mathcal{Q}_{l,\eta} = \{ \bsl{G}_M + \eta (-1)^l \bsl{q}_1 \}$, $(-)^t=-(-)^b = 1$, $\bsl{G}\in \bsl{b}_{M,1} \mathds{Z} + \bsl{b}_{M,2} \mathds{Z}$ are the moir\'e reciprocal lattice vectors with $\bsl{b}_{\mathrm{M},1}= \bsl{q}_3 - \bsl{q}_1$ and  $\bsl{b}_{\mathrm{M},2}= \bsl{q}_3 - \bsl{q}_2$, $\bar{l}=t,b$ for $l=b,t$, and the phonon creation operators correspond to $A_1'$ phonons at $\K/\K'$. 
The expression of $G_{\bsl{k}, \bsl{k}', \bsl{Q}  }^{\eta e_Y l}$ reads

\refcite{CXL2023ElKPhCouplingTBG} shows that the intervalley-Cooper-pairing channel of effective attractive interaction mediated by EPC reads
\eqa{
H_{ee}=-\frac{1}{N_M} \sum_{\bsl{k},\bsl{k}',s, s', e_Y, e_Y'} V^{\eta, e_Y, e_Y'}_{\bsl{k}, \bsl{k}'} \gamma_{\bsl{k}, e_Y,\eta, s}^\dagger  \gamma_{-\bsl{k}, e'_{Y}, - \eta, s'}^\dagger \gamma_{-\bsl{k}', e'_{Y},\eta, s'}  \gamma_{\bsl{k}', e_Y,-\eta, s}, 
}
where 
\eqa{
V^{\eta, e_Y, e_Y'}_{\bsl{k}, \bsl{k}'} & =\frac{1}{N_0\hbar\omega_{A_1'}} \sum_{\bsl{G}_M, l} G_{\bsl{k},\bsl{k}', - (-)^l \eta \bsl{q}_1+ \bsl{G}_M}^{\eta, e_Y, l}  
G_{-\bsl{k},-\bsl{k}', (-)^l \eta \bsl{q}_1 - \bsl{G}_M}^{-\eta,  e_{Y}',   l} 
}
with $N_0$ the number of atoms in one moir\'e unit cell and $\omega_{A_1'}$ is the frequency of the $A_1'$ phonon at $\K/\K'$.
To estimate the critical temperature, let us consider the inter-Chern pairing, for which the corresponding leading-order term of the interaction reads~\cite{CXL2023ElKPhCouplingTBG}
\eq{
V^{\eta, e_Y, -e_Y}_{\bsl{k}, \bsl{k}'} = A  \left\{ \frac{b^2 }{ |\bsl{k}|^2+b^2 } \frac{ b^2 }{ |\bsl{k}'|^2+b^2} + \frac{ |\bsl{k}|^2 }{|\bsl{k}|^2+b^2 } \frac{  |\bsl{k}'|^2 }{ |\bsl{k}'|^2+b^2 } \right\} \ ,
}
where 
\eq{
A = \frac{a^2 \hat{\gamma}^2}{ m_C \omega^2_{A_1'}(\K) N_0}
}
with $N_0$ is the number of the graphene unit cell in the moir\'e unit cell and $\hat{\gamma}$ in \eqnref{eq:gamma_hat_graphene}.
For $1.1^\circ$ twist angle, we estimate
\eq{
A \approx 0.33 \text{meV}\ .
}
(\refcite{CXL2023ElKPhCouplingTBG} has a slightly larger estimate of $A$ being $0.44$meV, since they choose a largeer $\hat{\gamma}$ than our \eqnref{eq:gamma_hat_graphene}.)
Based on the interacting Hamiltonian, the mean-field critical temperature can be determined by the following linearized gap equation~\cite{CXL2023ElKPhCouplingTBG}~\cite{Cao2018TBGSC}
\eqa{
2k_B T\Delta^{\eta,i_S}_{\bsl{k},e_{Y}, e_{Y}'}=\frac{1}{N_M} \sum_{\bsl{k}'}V^{\eta e_{Y} e_{Y}'}_{\bsl{k},\bsl{k}'} \Delta^{-\eta,i_S}_{\bsl{k}',e_{Y}, e_{Y}'}\ ,
}
where $\Delta^{\eta,i_S}_{\bsl{k};e_{Y_1} e_{Y_2}}$ is the pairing order parameter, and $i_S$ labels the spin channel (one spin-singlet components or three spin-triplet components).
%
%
Reducing to the inter-Chern channel, the linearized gap equation becomes 
\eqa{
 & 2k_B T\Delta^{\eta,i_S}_{\bsl{k},+, -}=\frac{1}{N_M} \sum_{\bsl{k}'}V^{\eta + -}_{\bsl{k},\bsl{k}'} \Delta^{-\eta,i_S}_{\bsl{k}',+, -}\\
 & \Leftrightarrow 2k_B T \mat{ \Delta^{+,i_S}_{\bsl{k},+, -} \\ \Delta^{-,i_S}_{\bsl{k},+, -}} =\frac{1}{N_M} \sum_{\bsl{k}'} \mat{ 0 & V^{+ + -}_{\bsl{k},\bsl{k}'} \\ V^{- + -}_{\bsl{k},\bsl{k}'} & 0 } \mat{ \Delta^{+,i_S}_{\bsl{k}',+, -} \\ \Delta^{-,i_S}_{\bsl{k}',+, -}} \\
 & \Leftrightarrow 
2k_BT  \mat{ \Delta_1 \\ \Delta_2 \\ \Delta_3 \\ \Delta_4} 
= A \left(\begin{array}{cccc}
0&f_{00}&0& f_{01}\\
f_{00}&0& f_{01}&0\\
0& f_{10}&0& f_{11}\\
f_{10}&0&f_{11}&0
\end{array}\right)
 \mat{ \Delta_1 \\ \Delta_2 \\ \Delta_3 \\ \Delta_4} 
 }
with
\eqa{ 
& f_{00}=\frac{1}{N_M}\sum_{\bsl{k}} \frac{ b^4}{(|\bsl{k}|^2+b^2)^2} \\
& f_{01}=f_{10}=\frac{1}{N_M}\sum_{\bsl{k}} \frac{ b^2 |\bsl{k}|^2}{(|\bsl{k}|^2+b^2)^2} \\
& f_{11}=\frac{1}{N_M}\sum_{\bsl{k}} \frac{  |\bsl{k}|^4}{(|\bsl{k}|^2+b^2)^2} \ . 
}
To derive a lower bound for $T_c$, we consider
\eqa{
\mat{ \Delta_1 & \Delta_2 & \Delta_3 & \Delta_4} =  \frac{1}{\sqrt{3}} \mat{ 1 & 1 & 1 & 0} \ ,
}
and then
\eqa{
2k_B T_c \geq 2k_B T & = A \mat{ \Delta_1 \\ \Delta_2 \\ \Delta_3 \\ \Delta_4}^\dagger
 \left(\begin{array}{cccc}
0&f_{00}&0& f_{01}\\
f_{00}&0& f_{01}&0\\
0& f_{10}&0& f_{11}\\
f_{10}&0&f_{11}&0
\end{array}\right)
 \mat{ \Delta_1 \\ \Delta_2 \\ \Delta_3 \\ \Delta_4}  \\
 & = \frac{ A}{3} (2 f_{00}+ 2 f_{01} )   = \frac{ A}{3} \frac{1}{N_M}\sum_{ \bsl{k}} \frac{2 b^2 }{(b^2 + |\bsl{k}|^2)^2} (b^2+|\bsl{k}|^2) = \frac{ A}{3} \frac{1}{N_M v_\star^2} \sum_{\bsl{k}} \Delta E_{\bsl{k}}^2 \Tr[g_{\bsl{k}}]\ ,
}
since $T_c$ corresponds to the largest eigenvalue of the matrix.
As a result, we have  
\eqa{
 k_B T_c \geq  k_B T  = \frac{ A}{6} \frac{1}{N_M v_\star^2} \sum_{\bsl{k}} \Delta E_{\bsl{k}}^2 \Tr[g_{\bsl{k}}]\ .
}
For $\gamma\approx 100$meV in the first chiral limit and $v_\star \approx v_0$, $T$ gives an estimate of 
\eq{
T \approx 0.6 \K\ ,
}
which is close to the experimental $T_c$ around $1\sim2$K~\cite{Cao2018TBGSC}.
For the estimate, we use the form of FSM in \eqnref{eq:FSM_flat_band_MATBG} and the expression of $\Delta E_{\bsl{k}}$ below it, and we perform the integration over $\left| \bsl{k} \right|\leq 1$.
}

%
%
%
%

\section{Orbital-Selective FSM}
\label{app:OFSM}

Before discussing \mgb, we first discuss the Orbital-selective FSM (OFSM), which will be useful for the discussion of $\mgb$.

Recall that the conventional FSM has the following two equal expressions:
\eqa{
\label{eq:FSM_g_two_expressions}
\left[g_{n}(\bsl{k})\right]_{ij} & = \frac{1}{2}\Tr\left[ \partial_{k_i} P_n(\bsl{k})  \partial_{k_j} P_n(\bsl{k}) \right] \\
& = \frac{1}{2}\Tr\left[ \partial_{k_i} P_n(\bsl{k}) P_n(\bsl{k}) \partial_{k_j} P_n(\bsl{k}) \right] + (i\leftrightarrow j)\ ,
}
where $P_n(\bsl{k})$ is the projection matrix on the $n$th band of an generic electron matrix Hamiltonian $h(\bsl{k})$ in the basis \eqnref{eq:FT_rule}.
If the $n$th electron band is degenerate in a region of the \BZ, $P_{n}(\bsl{k})$ would have rank more than 1 in that region.
According to the definition, $g_{n}(\bsl{k})$ must be a real symmetric positive semi-definite matrix.
Often in realistic materials, $g_{n}(\bsl{k})$ does not have zero eigenvalues, and thus is positive definite.
That's why $g_{n}(\bsl{k})$ is called a metric.
The two ways of writing the FSM in \eqnref{eq:FSM_g_two_expressions} inspire us to define the OFSM in two ways.

\subsection{Orbital-Selective FSM: Version 1}

For the first version of the OFSM, we are inspired by the first line of \eqnref{eq:FSM_g_two_expressions}, \ie, by 
\eq{
\label{eq:FSM_g_first_expression}
\left[g_{n}(\bsl{k})\right]_{ij} = \frac{1}{2}\Tr\left[ \partial_{k_i} P_n(\bsl{k})  \partial_{k_j} P_n(\bsl{k}) \right]\ .
}
OFSM is given by inserting matrix in the trace operation in \eqnref{eq:FSM_g_first_expression}.
The most general way of inserting matrix is the following
\eq{
\left[Q_{n}(\bsl{k},A,B)\right]_{ij} = \frac{1}{2}\Tr\left[ A \partial_{k_i} P_n(\bsl{k})  B \partial_{k_j} P_n(\bsl{k}) \right]\ ,
}
where $A,B$ are two matrices.
In general, $Q_{n}(\bsl{k},A,B)$ is not symmetric, \ie, $Q_{n}^T(\bsl{k},A,B)\neq Q_{n}(\bsl{k},A,B)$.
Since we care about metric in this work, let us consider the symmetric part of $Q_{n}(\bsl{k},A,B)$, which reads
\eqa{
\label{eq:g_tilde}
\left[\widetilde{g}_{n}(\bsl{k},A,B) \right]_{ij} & = \frac{1}{2}\left[Q_{n}(\bsl{k},A,B)\right]_{ij} + (i\leftrightarrow j) \\
& = \frac{1}{4}\Tr\left[ A \partial_{k_i} P_n(\bsl{k})  B \partial_{k_j} P_n(\bsl{k}) \right] + (i\leftrightarrow j) \ .
}
For generic matrices $A$ and $B$, the symmetric $\widetilde{g}_{n}(\bsl{k},A,B)$ is not guaranteed to be real and positive semidefinite, and thus is not necessarily a metric.
However, $\widetilde{g}_{n}(\bsl{k},A,B)$ can always be expressed as the linear combination of metrics, which will be the OFSM.
To see this, we note that $A=A_1 + \ii A_2$ and $B = B_1 + \ii B_2$, where $A_1$, $A_2$, $B_1$ and $B_2$ are Hermitian matrices.
Then, $\left[\widetilde{g}_{n}(\bsl{k},A,B) \right]_{ij}$ becomes
\eqa{
\left[\widetilde{g}_{n}(\bsl{k},A,B) \right]_{ij} & = \frac{1}{4}\Tr\left[ A_1 \partial_{k_i} P_n(\bsl{k})  B_1 \partial_{k_j} P_n(\bsl{k}) \right] + \ii \frac{1}{4}\Tr\left[ A_1 \partial_{k_i} P_n(\bsl{k})  B_2 \partial_{k_j} P_n(\bsl{k}) \right] \\
& \quad + \ii  \frac{1}{4}\Tr\left[ A_2 \partial_{k_i} P_n(\bsl{k})  B_1 \partial_{k_j} P_n(\bsl{k}) \right] - \frac{1}{4}\Tr\left[ A_2 \partial_{k_i} P_n(\bsl{k})  B_2 \partial_{k_j} P_n(\bsl{k}) \right] \\
& \quad + (i\leftrightarrow j) \ .
}
We now use the spectral decomposition of Hermitian matrices, \ie, 
\eq{
\label{eq:Hermtian_matrix_spetral_decomposition}
X = \sum_{\alpha_X} a_{\alpha_X} \xi_{\alpha_X}\xi_{\alpha_X}^\dagger
}
for $X=A_1,A_2,B_1,B_2$, where $\xi_{\alpha_X}$ is the eigenvector of $X$ with the eigenvalue $a_{\alpha_X}$, and $\alpha_X$ labels all the orthonormal eigenvectors of $X$.
$\xi_{\alpha_X}$ corresponds to a linear combination of electron degrees of freedom in one unit cell including orbitals and spins.
With \eqnref{eq:Hermtian_matrix_spetral_decomposition}, $\widetilde{g}_{n}(\bsl{k},A,B)$ becomes
\eqa{
\left[\widetilde{g}_{n}(\bsl{k},A,B) \right]_{ij} & =\sum_{\alpha_{A_1} \alpha_{B_1}} a_{\alpha_{A_1}} a_{\alpha_{B_1}}\frac{1}{4}\Tr\left[ \xi_{\alpha_{A_1}}\xi_{\alpha_{A_1}}^\dagger \partial_{k_i} P_n(\bsl{k})  \xi_{\alpha_{B_1}}\xi_{\alpha_{B_1}}^\dagger \partial_{k_j} P_n(\bsl{k}) \right] \\
& \quad + \ii\sum_{\alpha_{A_1} \alpha_{B_2}}   a_{\alpha_{A_1}} a_{\alpha_{B_2}} \frac{1}{4} \Tr\left[ \xi_{\alpha_{A_1}}\xi_{\alpha_{A_1}}^\dagger \partial_{k_i} P_n(\bsl{k}) \xi_{\alpha_{B_2}}\xi_{\alpha_{B_2}}^\dagger \partial_{k_j} P_n(\bsl{k}) \right] \\
& \quad + \ii  \sum_{\alpha_{A_2} \alpha_{B_1}}   a_{\alpha_{A_2}} a_{\alpha_{B_1}} \frac{1}{4}\Tr\left[ \xi_{\alpha_{A_2}}\xi_{\alpha_{A_2}}^\dagger \partial_{k_i} P_n(\bsl{k}) \xi_{\alpha_{B_1}}\xi_{\alpha_{B_1}}^\dagger \partial_{k_j} P_n(\bsl{k}) \right] \\
& \quad -  \sum_{\alpha_{A_2} \alpha_{B_2}}   a_{\alpha_{A_2}} a_{\alpha_{B_2}} \frac{1}{4}\Tr\left[  \xi_{\alpha_{A_2}}\xi_{\alpha_{A_2}}^\dagger  \partial_{k_i} P_n(\bsl{k})  \xi_{\alpha_{B_2}}\xi_{\alpha_{B_2}}^\dagger  \partial_{k_j} P_n(\bsl{k}) \right] \\
& \quad + (i\leftrightarrow j) \ .
}
Now we define 
\eq{
\label{eq:orbital_selective_FS_metric}
\left[ g_{n,\alpha\alpha'}( \bsl{k} ) \right]_{ij}= \frac{1}{4}\Tr\left[ \xi_\alpha \xi^\dagger_\alpha \partial_{k_i} P_{n}(\bsl{k})  \xi_{\alpha'} \xi^\dagger_{\alpha'}  \partial_{k_j} P_{n}(\bsl{k})       \right] + (i\leftrightarrow j)\ ,
}
where $\xi_\alpha$ is a normalized vector that represents certain linear combination of the electronic degrees of freedom in one unit cell.
Finally, we arrive at
\eqa{
\left[\widetilde{g}_{n}(\bsl{k},A,B) \right]_{ij} & =\sum_{\alpha_{A_1} \alpha_{B_1}} a_{\alpha_{A_1}} a_{\alpha_{B_1}} \left[ g_{n,\alpha_{A_1} \alpha_{B_1}}( \bsl{k} ) \right]_{ij} + \ii\sum_{\alpha_{A_1} \alpha_{B_2}}   a_{\alpha_{A_1}} a_{\alpha_{B_2}} \left[ g_{n,\alpha_{A_1} \alpha_{B_2}}( \bsl{k} ) \right]_{ij}  \\
& \quad + \ii  \sum_{\alpha_{A_2} \alpha_{B_1}}   a_{\alpha_{A_2}} a_{\alpha_{B_1}} \left[ g_{n,\alpha_{A_2} \alpha_{B_1}}( \bsl{k} ) \right]_{ij}  -  \sum_{\alpha_{A_2} \alpha_{B_2}}   a_{\alpha_{A_2}} a_{\alpha_{B_2}} \left[ g_{n,\alpha_{A_1} \alpha_{B_1}}( \bsl{k} ) \right]_{ij}  \ .
}
Therefore, we can see the general expression $\widetilde{g}_{n}(\bsl{k},A,B)$ can always be expressed as linear combinations of a set of $g_{n,\alpha\alpha'}( \bsl{k} )$.

Now we show $g_{n,\alpha\alpha'}( \bsl{k} )$ in \eqnref{eq:orbital_selective_FS_metric} with fixed values of $\alpha,\alpha'$ is real symmetric and positive semidefinite.
According to \eqnref{eq:orbital_selective_FS_metric}, it is obvious that $g_{n,\alpha\alpha'}( \bsl{k} )$ is symmetric.
$g_{n,\alpha\alpha'}( \bsl{k} )$ is a positive semidefinite matrix since
\eqa{
& \sum_{i,j} v_i^* v_j \left[ g_{n,\alpha\alpha'}( \bsl{k} ) \right]_{ij} \\
& = \frac{1}{4}\Tr\left[ \xi_\alpha \xi^\dagger_\alpha \sum_{i} v_i^* \partial_{k_i} P_{n}(\bsl{k})  \xi_{\alpha'} \xi^\dagger_{\alpha'}  \sum_{j}v_j \partial_{k_j} P_{n}(\bsl{k})       \right] +  \frac{1}{4}\Tr\left[ \xi_\alpha \xi^\dagger_\alpha  \sum_{j}v_j \partial_{k_j}P_{n}(\bsl{k})  \xi_{\alpha'} \xi^\dagger_{\alpha'}  \sum_{i} v_i^* \partial_{k_i} P_{n}(\bsl{k})       \right] \\
& = \frac{1}{4}\Tr\left[  \xi^\dagger_\alpha \sum_{i} v_i^* \partial_{k_i} P_{n}(\bsl{k})  \xi_{\alpha'} \xi^\dagger_{\alpha'}  \sum_{j}v_j \partial_{k_j} P_{n}(\bsl{k})   \xi_\alpha     \right] +  \frac{1}{4}\Tr\left[  \xi^\dagger_\alpha  \sum_{j}v_j \partial_{k_j}P_{n}(\bsl{k})  \xi_{\alpha'} \xi^\dagger_{\alpha'}  \sum_{i} v_i^* \partial_{k_i} P_{n}(\bsl{k})  \xi_\alpha     \right] \\
& = \frac{1}{4}\Tr\left[  \xi^\dagger_\alpha \sum_{i} v_i^* \partial_{k_i} P_{n}(\bsl{k})  \xi_{\alpha'} (\xi^\dagger_\alpha \sum_{j} v_j^* \partial_{k_j} P_{n}(\bsl{k})  \xi_{\alpha'})^\dagger     \right] \\
& \quad +  \frac{1}{4}\Tr\left[  \xi^\dagger_\alpha  \sum_{j}v_j \partial_{k_j}P_{n}(\bsl{k})  \xi_{\alpha'} (\xi^\dagger_\alpha  \sum_{i}v_i^* \partial_{k_i}P_{n}(\bsl{k})  \xi_{\alpha'})^\dagger     \right] \\
& \geq 0\ .
}
$g_{n,\alpha\alpha'}( \bsl{k} )$ is a real matrix since
\eqa{
\left[ g_{n,\alpha\alpha'}( \bsl{k} ) \right]_{ij}^* & = \frac{1}{4}\Tr\left[\partial_{k_j} P_{n}(\bsl{k})  \xi_{\alpha'} \xi^\dagger_{\alpha'} \partial_{k_i} P_{n}(\bsl{k})     \xi_\alpha \xi^\dagger_\alpha     \right] + (i\leftrightarrow j)\\
& = \frac{1}{4}\Tr\left[ \xi_\alpha \xi^\dagger_\alpha  \partial_{k_j} P_{n}(\bsl{k})  \xi_{\alpha'} \xi^\dagger_{\alpha'} \partial_{k_i} P_{n}(\bsl{k})  \right] + (i\leftrightarrow j)\\
& = \frac{1}{4}\Tr\left[ \xi_\alpha \xi^\dagger_\alpha  \partial_{k_i} P_{n}(\bsl{k})  \xi_{\alpha'} \xi^\dagger_{\alpha'} \partial_{k_j} P_{n}(\bsl{k})  \right] + (i\leftrightarrow j)\\
& = \left[ g_{n,\alpha\alpha'}( \bsl{k} ) \right]_{ij}\ .
}
Therefore, it is highly possible that $g_{n,\alpha\alpha'}( \bsl{k} )$ becomes real symmetric positive definite in realistic cases, and thus $g_{n,\alpha\alpha'}( \bsl{k} )$ can be called a metric.

We call $g_{n,\alpha\alpha'}( \bsl{k} )$ the OFSM.
If we sum OFSM over a complete set of all orthonormal combinations of the orbitals, we obtain the normal FSM in \eqnref{eq:FS_metric}:
\eqa{
\sum_{\alpha,\alpha'} \left[g_{n,\alpha\alpha'}( \bsl{k} )\right]_{ij} & = \sum_{\alpha,\alpha'} \frac{1}{4}\Tr\left[ \xi_\alpha \xi^\dagger_\alpha \partial_{k_i} P_{n}(\bsl{k})  \xi_{\alpha'} \xi^\dagger_{\alpha'}  \partial_{k_j} P_{n}(\bsl{k})       \right] + (i\leftrightarrow j) \\
& = \frac{1}{4}\Tr\left[ \partial_{k_i} P_{n}(\bsl{k})  \partial_{k_j} P_{n}(\bsl{k})       \right] + (i\leftrightarrow j)\\
& = \frac{1}{2}\Tr\left[ \partial_{k_i} P_{n}(\bsl{k})  \partial_{k_j} P_{n}(\bsl{k})       \right] = \left[g_{n }( \bsl{k} )\right]_{ij}\ ,
}
where $\alpha,\alpha'$ are summed over complete sets of all orthonormal combinations of the electron degrees of freedom in one unit cell.
The OFSM proposed in \refcite{Torma2018SelectiveQuantumMetric} is compartible with the version of OFSM in \eqnref{eq:orbital_selective_FS_metric}.

\subsection{Orbital-Selective FSM: Version 2}

For the second version of the OFSM, we are inspired by the second line of \eqnref{eq:FSM_g_two_expressions}, \ie, by 
\eq{
\label{eq:FSM_g_second_expression}
\left[g_{n}(\bsl{k})\right]_{ij} = \frac{1}{2}\Tr\left[ \partial_{k_i} P_n(\bsl{k})  P_n(\bsl{k}) \partial_{k_j} P_n(\bsl{k}) \right] + (i\leftrightarrow j )\ .
}
More directly, we are inspired by the term in the two-band $\lambda_{geo}$ shown in \eqnref{eq:lambda_geo_1}, which contains a matrix insertion as
\eq{
\label{eq:gwidetilde_n_second_expression}
\left[\widetilde{g}_{n}(\bsl{k},M)\right]_{ij}  = \frac{1}{2}\Tr\left[ M \partial_{k_i} P_n(\bsl{k})  P_n(\bsl{k}) \partial_{k_j} P_n(\bsl{k}) \right] + (i\leftrightarrow j )\ ,
}
where $M$ is a generic matrix.
Similar to the discussion on the version 1, we define $M=M_1 + \ii M_2$ with $M_1,M_2$ Hermitian, and 
use the spectral decomposition to get
\eqa{
\left[\widetilde{g}_{n}(\bsl{k},M)\right]_{ij} & =\sum_{\alpha_{M_1} } a_{\alpha_{M_1}} \frac{1}{2}\Tr\left[ \xi_{\alpha_{M_1}}\xi_{\alpha_{M_1}}^\dagger \partial_{k_i} P_n(\bsl{k})  P_n(\bsl{k}) \partial_{k_j} P_n(\bsl{k}) \right] \\
& \quad + \ii \sum_{\alpha_{M_2} } a_{\alpha_{M_2}} \frac{1}{2}\Tr\left[ \xi_{\alpha_{M_2}}\xi_{\alpha_{M_2}}^\dagger \partial_{k_i} P_n(\bsl{k})  P_n(\bsl{k}) \partial_{k_j} P_n(\bsl{k}) \right] + (i\leftrightarrow j) \ ,
}
where $\xi_{\alpha_X}$ is the eigenvector of $X$ with the eigenvalue $a_{\alpha_X}$ for $X=M_1,M_2$.

We can then define 
\eq{
\label{eq:orbital_selective_FS_metric_alt}
\left[ g_{n,\alpha}( \bsl{k} ) \right]_{ij}= \frac{1}{2}\Tr\left[ \xi_\alpha \xi^\dagger_\alpha \partial_{k_i} P_{n}(\bsl{k})  P_{n}(\bsl{k}) \partial_{k_j} P_{n}(\bsl{k})       \right] + (i\leftrightarrow j)\ ,
}
resulting in 
\eqa{
\widetilde{g}_{n}(\bsl{k},M) =\sum_{\alpha_{M_1} } a_{\alpha_{M_1}}  g_{n,\alpha_{M_1}}( \bsl{k} ) + \ii \sum_{\alpha_{M_2} } a_{\alpha_{M_2}} g_{n,\alpha_{M_2}}( \bsl{k} )  \ .
}
It means that \eqnref{eq:gwidetilde_n_second_expression} (contained in \eqnref{eq:lambda_geo_1} in two-band $\lambda_{geo}$) must be linear combination of $ g_{n,\alpha}( \bsl{k} ) $ defined in \eqnref{eq:orbital_selective_FS_metric_alt}.

As $ g_{n,\alpha}( \bsl{k} ) $ can be derived by replacing $\xi_{\alpha'} \xi^\dagger_{\alpha'} $ in \eqnref{eq:orbital_selective_FS_metric} by $P_{n}(\bsl{k})$, $g_{n,\alpha}(\bsl{k})$ must a real symmetric positive-semidefinite matrix for any fixed values of $n$ and $\alpha$.
Since $\xi_{\alpha}$ in \eqnref{eq:orbital_selective_FS_metric_alt} again represents a certain combination of the electron degrees of freedom in one unit cell, we also call $ g_{n,\alpha}( \bsl{k} ) $ a OFSM. 
Again, summing OFSM over a complete set of $\alpha$ restores the original FSM:
\eqa{
\sum_{\alpha} \left[g_{n,\alpha}( \bsl{k} )\right]_{ij} & = \frac{1}{2}\Tr\left[  \partial_{k_i} P_{n}(\bsl{k})  P_{n}(\bsl{k}) \partial_{k_j} P_{n}(\bsl{k})       \right] +  \frac{1}{2}\Tr\left[  P_{n}(\bsl{k}) \partial_{ki} P_{n}(\bsl{k})    \partial_{k_j} P_{n}(\bsl{k})    \right] \\
& =  \frac{1}{2}\Tr\left[  \partial_{k_i} P_{n}(\bsl{k})  \partial_{k_j} P_{n}(\bsl{k})       \right] = \left[g_{n}(\bsl{k})\right]_{ij}\ .
}
Owing to the close relation between $g_{n,\alpha}( \bsl{k} )$ in \eqnref{eq:orbital_selective_FS_metric_alt} and $\lambda_{geo}$ shown in \eqnref{eq:lambda_geo_1}, we will use \eqnref{eq:orbital_selective_FS_metric_alt} in the following.

\section{Geometric and Topological Contributions to EPC Constant in \mgb}

\label{app:MgB2}

{\mgb} becomes a superconductor under $39$K~\cite{Jun03012001MgB2SC}, and the superconductivity should mainly originate from EPC as indicated by the significant isotope effect~\cite{Budko02262001MgB2Isotope,Jorgensen05012001MgB2Isotope}.
In this section, we discuss the EPC in {\mgb}.
Again, we focus on the EPC constant defined based on the many-body electron-phonon-coupled Hamiltonian of {\mgb} (not the mean-field Hamiltonian in the superconducting phase).
Similar to graphene, it is reasonable for us to neglect the spin-orbit coupling in {\mgb} and assume spin $\SU(2)$ symmetry, in order to study the physics not too far away from the Fermi level, since both Mg and B are light atoms.
Therefore, we will use a spinless model throughout this section.
As a results, all the symmetry operations that we consider are spinless.

\begin{figure}[t]
    \centering
    \includegraphics[width=0.8\columnwidth]{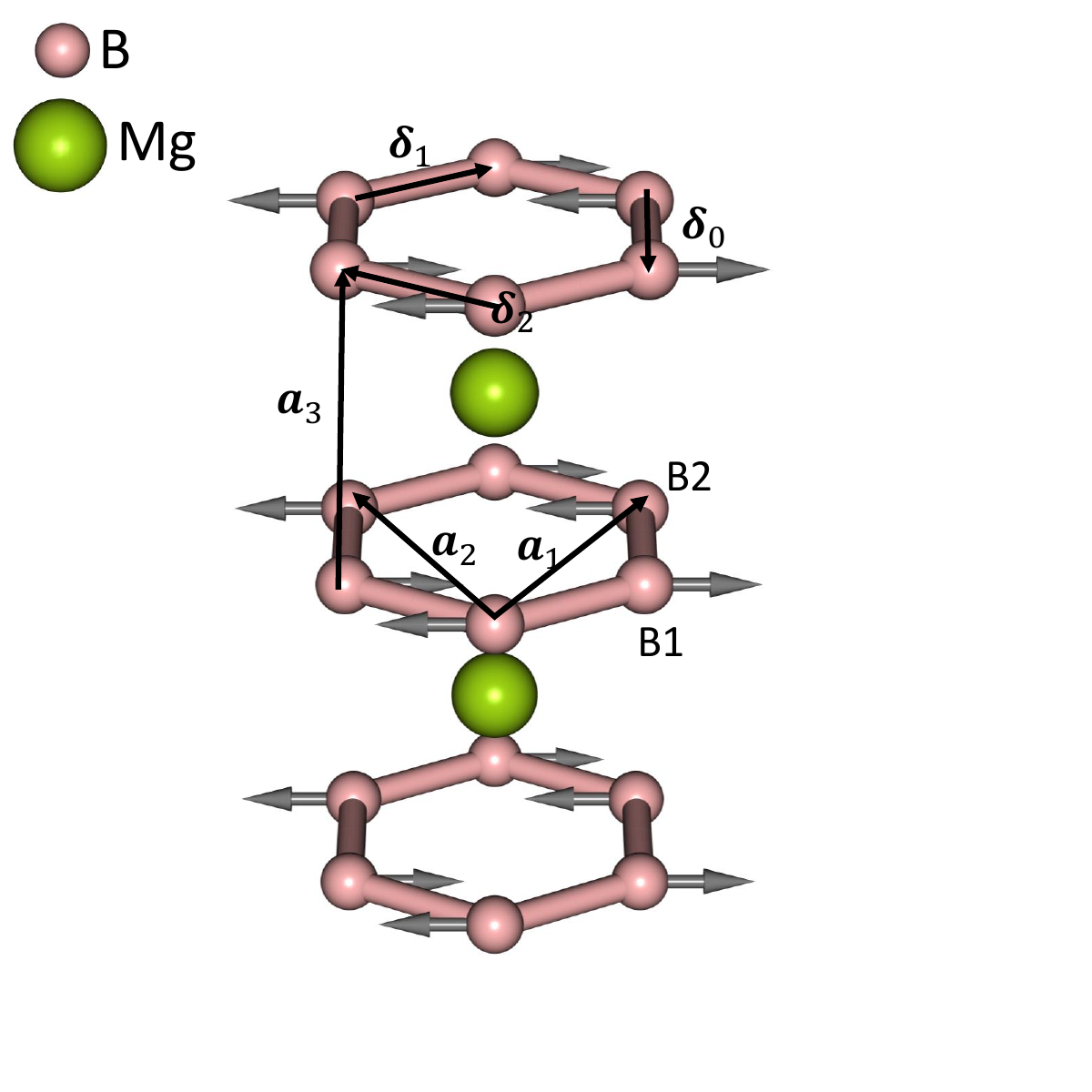}
    \caption{
    The structure of {\mgb}.
    Expressions of $\bsl{a}_1$, $\bsl{a}_2$ and $\bsl{a}_3$ are in \eqnref{eq:lattice_MgB2}.
    Expressions of $\bsl{\delta}_0$, $\bsl{\delta}_1$ and $\delta_2$ are in \eqnref{eq:delta_j_graphene}.
    B1 and B2 label two B atoms in one unit cell.
    The gray arrows show one type of ion motions of the $E_{2g}$ phonon at $\Gamma$.
    }
    \label{fig:MgB2_structure}
\end{figure}

\subsection{Electron Hamiltonian and Electron Band Topology in {\mgb}}

In this part, we will discuss the electron Hamiltonian of {\mgb} and its electron band topology.
Since we only consider the EPC constant eventually, we will not consider the Coulomb interaction among electrons, although the Coulomb interaction is crucial for the study of superconductivity.

\subsubsection{Electron Hamiltonian of {\mgb}}
\label{app:H_el_MgB2}

{\mgb} has the same symmetries as graphene---space group P6/mmm and TR symmetry, along with the charge $\U(1)$ symmetry.
The model of graphene is constructed from $C_6$, $m_y$, $m_z$ and $\TR$ in addition to the lattice translations and the charge $\U(1)$ symmetry.
   
{\mgb} has two B atoms at the 2c Wykoff position and one Mg at the 1b position per unit cell~\cite{Jun03012001MgB2SC}, as shown in \figref{fig:MgB2_structure}.
We choose the sublattice vectors to be
\eqa{
\label{eq:sublattice_MgB2}
& \bsl{\tau}_{\text{B1}} = \frac{a}{\sqrt{3}}(\frac{\sqrt{3}}{2}, -\frac{1}{2} , 0)^T \\
& \bsl{\tau}_{\text{B2}} = \frac{a}{\sqrt{3}}(\frac{\sqrt{3}}{2}, \frac{1}{2} , 0)^T \\
& \bsl{\tau}_{\text{Mg}} =  c (0,0,\frac{1}{2})^T \ ,
}
where $a$ is the lattice constant in the $x$-$y$ plane, and $c$ is the lattice constant along $z$.
According to the orbital projection in \refcite{Lordi05102018MgB2ElectronBands}, in order to study the low-energy physics, we only need to consider $s$, $p_x$, $p_y$ and $p_z$ orbitals at each B atom, and an $s$ orbital at each Mg atom.
Therefore, although each B layer in {\mgb} arranges like a sheet of graphene, we need to include more orbitals (extra $s,p_x,p_y$ orbitals) for B than for graphene.
Specifically, according to the convention defined in \appref{app:EPC_real_space}, we have $\bsl{\tau}\in\{ \bsl{\tau}_{\text{B1}},\bsl{\tau}_{\text{B2}},\bsl{\tau}_{\text{Mg}}\}$,  $\alpha_{\bsl{\tau}_{\text{B1}}},\alpha_{\bsl{\tau}_{\text{B2}}}\in\{s,p_x,p_y,p_z\}$, and  $\alpha_{\bsl{\tau}_{\text{Mg}}}$ can be omitted since it only takes one value---$s$.
Then, the creation operators for electrons are labelled as $c^{\dagger}_{\bsl{R}+\bsl{\tau}}$, where 
\eqa{
& c^{\dagger}_{\bsl{R}+\bsl{\tau}_{\text{B1}}} = (c^{\dagger}_{\bsl{R}+\bsl{\tau}_{\text{B1}},s},c^{\dagger}_{\bsl{R}+\bsl{\tau}_{\text{B1}},p_x},c^{\dagger}_{\bsl{R}+\bsl{\tau}_{\text{B1}},p_y},c^{\dagger}_{\bsl{R}+\bsl{\tau}_{\text{B1}},p_z}) \\
& c^{\dagger}_{\bsl{R}+\bsl{\tau}_{\text{B2}}} = (c^{\dagger}_{\bsl{R}+\bsl{\tau}_{\text{B2}},s},c^{\dagger}_{\bsl{R}+\bsl{\tau}_{\text{B2}},p_x},c^{\dagger}_{\bsl{R}+\bsl{\tau}_{\text{B2}},p_y},c^{\dagger}_{\bsl{R}+\bsl{\tau}_{\text{B2}},p_z}) \ ,
}
and $\bsl{R}\in \bsl{a}_1\dsZ + \bsl{a}_2\dsZ + \bsl{a}_3\dsZ $ with
\eqa{
\label{eq:lattice_MgB2}
& \bsl{a}_1 = a (\frac{1}{2} , \frac{\sqrt{3}}{2} , 0)^T\\
& \bsl{a}_2 = a (-\frac{1}{2} , \frac{\sqrt{3}}{2} , 0)^T\\
& \bsl{a}_3 = c (0 , 0 , 1)^T\ .
}

According to \eqnref{eq:sym_rep_g_R} and \eqnref{eq:sym_rep_TR_R}, the symmetry reps furnished by $c^{\dagger}_{\bsl{R}+\bsl{\tau}}$ are determined by $\bsl{\tau}_g$, $\bsl{R}_{\bsl{\tau},g}$, $U_g^{\bsl{\tau}_g\bsl{\tau}}$ and  $U_{\TR}^{\bsl{\tau}\bsl{\tau}}$, where $g=C_6,m_y,m_z$.
Specifically, we have
\eqa{
& \bsl{\tau}_{C_6} = \bsl{\tau}_{m_y} = \left\{ 
\begin{array}{ll}
    \bsl{\tau}_{\text{B2}} & \text{ if } \bsl{\tau} = \bsl{\tau}_{\text{B1}}  \\
    \bsl{\tau}_{\text{B1}} & \text{ if } \bsl{\tau} = \bsl{\tau}_{\text{B2}}  \\
    \bsl{\tau}_{\text{Mg}} & \text{ if } \bsl{\tau} = \bsl{\tau}_{\text{Mg}}  \\
\end{array}
\right. \\ 
& \bsl{\tau}_{m_z} = \bsl{\tau}\\
& \bsl{R}_{C_6,\bsl{\tau}} = \left\{ 
\begin{array}{ll}
    C_6 \bsl{R} & \text{ if } \bsl{\tau} = \bsl{\tau}_{\text{B1}}  \\
    C_6 \bsl{R} +\bsl{a}_2 & \text{ if } \bsl{\tau} = \bsl{\tau}_{\text{B2}} \\
    C_6 \bsl{R}  & \text{ if } \bsl{\tau} = \bsl{\tau}_{\text{Mg}} \\ 
\end{array}
\right. \\
& \bsl{R}_{m_y,\bsl{\tau}} = \bsl{R} \\
& \bsl{R}_{m_z,\bsl{\tau}} = \left\{ 
\begin{array}{ll}
    m_z \bsl{R} & \text{ if } \bsl{\tau} = \bsl{\tau}_{\text{B1}},\ \bsl{\tau}_{\text{B2}}  \\
    m_z \bsl{R}  + \bsl{a}_3 & \text{ if } \bsl{\tau} = \bsl{\tau}_{\text{Mg}} \\ 
\end{array}
\right. \ ,
}
and 
\eqa{
\label{eq:sym_rep_MgB2}
& U_{C_6}^{\bsl{\tau}_{\text{B1}} \bsl{\tau}_{\text{B2}}} = U_{C_6}^{ \bsl{\tau}_{\text{B2}} \bsl{\tau}_{\text{B1}}} = 
\mat{ 
1   &   &   &   \\ 
    & \frac{1}{2}  &  - \frac{\sqrt{3}}{2} &   \\
    & \frac{\sqrt{3}}{2} & \frac{1}{2}  &   \\
    &   &   &  1 
}
\ \&\ U_{C_6}^{\bsl{\tau}_{\text{Mg}} \bsl{\tau}_{\text{Mg}}} = 1  \Rightarrow 
U_{C_6} 
= 
\left(
\begin{array}{c|c|c}
U_{C_6}^{\bsl{\tau}_{\text{Mg}} \bsl{\tau}_{\text{Mg}}} & & \\ 
\hline
  &  & U_{C_6}^{\bsl{\tau}_{\text{B1}} \bsl{\tau}_{\text{B2}}} \\
\hline
  & U_{C_6}^{ \bsl{\tau}_{\text{B2}} \bsl{\tau}_{\text{B1}}} & 
\end{array}
\right) \\
& U_{m_y}^{\bsl{\tau}_{\text{B1}} \bsl{\tau}_{\text{B2}}} = U_{m_y}^{ \bsl{\tau}_{\text{B2}} \bsl{\tau}_{\text{B1}}} = 
\mat{ 
1   &   &   &   \\ 
    & 1  &   &   \\
    &    & -1  &   \\
    &   &   &  1 
}
\ \&\ U_{m_y}^{\bsl{\tau}_{\text{Mg}} \bsl{\tau}_{\text{Mg}}} = 1  \Rightarrow 
U_{m_y} 
= 
\left(
\begin{array}{c|c|c}
U_{m_y}^{\bsl{\tau}_{\text{Mg}} \bsl{\tau}_{\text{Mg}}} & & \\ 
\hline
  &  & U_{m_y}^{\bsl{\tau}_{\text{B1}} \bsl{\tau}_{\text{B2}}} \\
\hline
  & U_{m_y}^{ \bsl{\tau}_{\text{B2}} \bsl{\tau}_{\text{B1}}} & 
\end{array}
\right)  \\
& U_{m_z}^{\bsl{\tau}_{\text{B1}}\bsl{\tau}_{\text{B1}}} =  U_{m_z}^{\bsl{\tau}_{\text{B2}}\bsl{\tau}_{\text{B2}}}  = \mat{ 
1   &   &   &   \\ 
    & 1  &   &   \\
    &    & 1  &   \\
    &   &   &  -1 
} \ \&\ U_{m_z}^{\bsl{\tau}_{\text{Mg}} \bsl{\tau}_{\text{Mg}}} = 1  \Rightarrow 
U_{m_z} 
= 
\left(
\begin{array}{c|c|c}
U_{m_z}^{\bsl{\tau}_{\text{Mg}} \bsl{\tau}_{\text{Mg}}} & & \\ 
\hline
  &  U_{m_z}^{\bsl{\tau}_{\text{B1}} \bsl{\tau}_{\text{B1}}} & \\
\hline
  & & U_{m_z}^{ \bsl{\tau}_{\text{B2}} \bsl{\tau}_{\text{B2}}} 
\end{array}
\right)  \\
& U_{\TR}^{\bsl{\tau}_{\text{B1}}\bsl{\tau}_{\text{B1}}} =  U_{\TR}^{\bsl{\tau}_{\text{B2}}\bsl{\tau}_{\text{B2}}}  = \mat{ 
1   &   &   &   \\ 
    & 1  &   &   \\
    &    & 1  &   \\
    &   &   &  1 
} \ \&\ U_{\TR}^{\bsl{\tau}_{\text{Mg}} \bsl{\tau}_{\text{Mg}}} = 1  \Rightarrow 
U_{\TR} 
= 
\left(
\begin{array}{c|c|c}
U_{\TR}^{\bsl{\tau}_{\text{Mg}} \bsl{\tau}_{\text{Mg}}} & & \\ 
\hline
  &  U_{\TR}^{\bsl{\tau}_{\text{B1}} \bsl{\tau}_{\text{B1}}} & \\
\hline
  & & U_{\TR}^{ \bsl{\tau}_{\text{B2}} \bsl{\tau}_{\text{B2}}} 
\end{array}
\right)  \ ,
}
where the basis for $U_g$ and $U_{\TR}$ is
\eq{
c^\dagger_{\bsl{k}} = ( c^\dagger_{\bsl{k},\bsl{\tau}_{\text{Mg}}}, c^{\dagger}_{\bsl{k},\bsl{\tau}_{\text{B1}},s},c^{\dagger}_{\bsl{k},\bsl{\tau}_{\text{B1}},p_x},c^{\dagger}_{\bsl{k},\bsl{\tau}_{\text{B1}},p_y},c^{\dagger}_{\bsl{k},\bsl{\tau}_{\text{B1}},p_z}, c^{\dagger}_{\bsl{k},\bsl{\tau}_{\text{B2}},s},c^{\dagger}_{\bsl{k},\bsl{\tau}_{\text{B2}},p_x},c^{\dagger}_{\bsl{k},\bsl{\tau}_{\text{B2}},p_y},c^{\dagger}_{\bsl{k},\bsl{\tau}_{\text{B2}},p_z})\ .
}

In this work, we choose the following approximations for the electron hopping.
{\mgb} can be viewed as stacking B layers and Mg layers alternatively along the $z$ direction.
Between two atoms of the same kind, the hopping occurs (i) only if the two atoms are in the same layer (B layer or Mg layer) and are nearest neighbors or (ii) only if the two atoms are in different layers and are separated by $\pm \bsl{a}_3$. (See \figref{fig:MgB2_structure} for $\bsl{a}_3$.)
Between one B atom and one Mg atom, the hopping occurs only if they are the nearest neighbors.
Specifically, by using the convention in \eqnref{eq:H_el_gen}, we have the following constraints
\eqa{
\label{eq:t_constraint_MgB2}
& t_{\bsl{\tau}_{1}\bsl{\tau}_2}(\bsl{R}_1+ \bsl{\tau}_1 -\bsl{R}_2 - \bsl{\tau}_2) = 0\ \text{if}\ \bsl{\tau}_{1},\bsl{\tau}_{2}\in \{\bsl{\tau}_{\text{B1}},\bsl{\tau}_{\text{B2}}\}\ \&\ |\bsl{R}_1+ \bsl{\tau}_1 -\bsl{R}_2 - \bsl{\tau}_2|\geq \sqrt{\frac{a^2}{3}+c^2} \\
& t_{\bsl{\tau}_1\bsl{\tau}_{\text{Mg}}}(\bsl{R}_1+ \bsl{\tau}_1 -\bsl{R}_2 - \bsl{\tau}_{\text{Mg}}) = 0\  \text{if}\ \bsl{\tau}_{1}\in \{\bsl{\tau}_{\text{B1}},\bsl{\tau}_{\text{B2}}\}\ \&\ |\bsl{R}_1+ \bsl{\tau}_1 -\bsl{R}_2 -  \bsl{\tau}_{\text{Mg}}| > \sqrt{\frac{a^2}{3}+\frac{c^2}{4}} \\
& t_{\bsl{\tau}_{\text{Mg}}\bsl{\tau}_1}(\bsl{R}_1+ \bsl{\tau}_{\text{Mg}} -\bsl{R}_2 - \bsl{\tau}_2) = 0\  \text{if}\ \bsl{\tau}_{2}\in \{\bsl{\tau}_{\text{B1}},\bsl{\tau}_{\text{B2}}\}\ \&\ |\bsl{R}_1+  \bsl{\tau}_{\text{Mg}} -\bsl{R}_2 - \bsl{\tau}_2| > \sqrt{\frac{a^2}{3}+\frac{c^2}{4}} \\
& t_{\bsl{\tau}_{\text{Mg}}\bsl{\tau}_{\text{Mg}}}(\bsl{R}_1  -\bsl{R}_2) = 0\  \text{if}\  |\bsl{R}_1-\bsl{R}_2 | \geq \sqrt{a^2+c^2} \ ,
}
meaning that we only consider $t_{\bsl{\tau}\bsl{\tau}}(0)$, $t_{\bsl{\tau}\bsl{\tau}}(\pm\bsl{a}_3)$,  $t_{\bsl{\tau}_{\text{B1}}\bsl{\tau}_{\text{B2}}}(\bsl{\delta}_j)$,  $t_{\bsl{\tau}_{\text{B2}}\bsl{\tau}_{\text{B1}}}(-\bsl{\delta}_j)$, $t_{\bsl{\tau}_{\text{Mg}}\bsl{\tau}_{\text{Mg}}}(C_6^n \bsl{a}_1)$ with $n=0,1,2,...,5$, $t_{\bsl{\tau}_{\text{B1}}\bsl{\tau}_{\text{Mg}}}(-\bsl{\delta}_j\pm\bsl{a}_3/2)$, $t_{\bsl{\tau}_{\text{B2}}\bsl{\tau}_{\text{Mg}}}(\bsl{\delta}_j\pm\bsl{a}_3/2)$, $t_{\bsl{\tau}_{\text{Mg}}\bsl{\tau}_{\text{B1}}}(\bsl{\delta}_j\pm\bsl{a}_3/2)$, and $t_{\bsl{\tau}_{\text{Mg}}\bsl{\tau}_{\text{B2}}}(-\bsl{\delta}_j\pm\bsl{a}_3/2)$, where $\bsl{\delta}_j$ with $j=0,1,2$ is defined in \eqnref{eq:delta_j_graphene} and $\bsl{a}_{1,2,3}$ are defined in \eqnref{eq:lattice_MgB2}.
In the following, we derive the form of those terms.

By using \eqnref{eq:t_h_sym_g}, \eqnref{eq:t_h_sym_TR} and \eqnref{eq:sym_rep_MgB2}, we obtain the symmetry properties of the onsite terms for B atoms:
\eqa{
\label{eq:t_sym_MgB2_1}
& C_6: U_{C_6}^{\bsl{\tau}_{\text{B2}} \bsl{\tau}_{\text{B1}}} t_{\bsl{\tau}_{\text{B1}}\bsl{\tau}_{\text{B1}}}(0) \left[U_{C_6}^{\bsl{\tau}_{\text{B2}} \bsl{\tau}_{\text{B1}}}\right]^\dagger = t_{\bsl{\tau}_{\text{B2}}\bsl{\tau}_{\text{B2}}}(0) \\
& m_y: U_{m_y}^{\bsl{\tau}_{\text{B2}} \bsl{\tau}_{\text{B1}}} t_{\bsl{\tau}_{\text{B1}}\bsl{\tau}_{\text{B1}}}(0)\left[ U_{m_y}^{\bsl{\tau}_{\text{B2}} \bsl{\tau}_{\text{B1}}} \right]^\dagger = t_{\bsl{\tau}_{\text{B2}}\bsl{\tau}_{\text{B2}}}(0) \\
& m_z:  U_{m_z}^{\bsl{\tau}_{\text{B1}} \bsl{\tau}_{\text{B1}}} t_{\bsl{\tau}_{\text{B1}}\bsl{\tau}_{\text{B1}}}(0) \left[ U_{m_z}^{\bsl{\tau}_{\text{B1}} \bsl{\tau}_{\text{B1}}} \right]^\dagger= t_{\bsl{\tau}_{\text{B1}}\bsl{\tau}_{\text{B1}}}(0)\ ,\   U_{m_z}^{\bsl{\tau}_{\text{B2}} \bsl{\tau}_{\text{B2}}}  t_{\bsl{\tau}_{\text{B2}}\bsl{\tau}_{\text{B2}}}(0) \left[ U_{m_z}^{\bsl{\tau}_{\text{B2}} \bsl{\tau}_{\text{B2}}} \right]^\dagger = t_{\bsl{\tau}_{\text{B2}}\bsl{\tau}_{\text{B2}}}(0)\\
& \TR: t_{\bsl{\tau}_{\text{B1}}\bsl{\tau}_{\text{B1}}}(0)\ ,\  t_{\bsl{\tau}_{\text{B2}}\bsl{\tau}_{\text{B2}}}(0) \in \dsR^{4\times 4 }\\
& h.c.: t_{\bsl{\tau}_{\text{B1}}\bsl{\tau}_{\text{B1}}}(0)\ ,\  t_{\bsl{\tau}_{\text{B2}}\bsl{\tau}_{\text{B2}}}(0) \text{ are Hermitian}
}
for $t_{\bsl{\tau}_{\text{B1}}\bsl{\tau}_{\text{B1}}}(0)$ and $t_{\bsl{\tau}_{\text{B2}}\bsl{\tau}_{\text{B2}}}(0)$, which gives 
\eq{
\label{eq:t_MgB2_part_1}
t_{\bsl{\tau}_{\text{B1}}\bsl{\tau}_{\text{B1}}}(0) = t_{\bsl{\tau}_{\text{B2}}\bsl{\tau}_{\text{B2}}}(0) = 
\mat{
E_{\text{B},s,0} & & & \\
 & E_{\text{B},p_x p_y, 0} & & \\
 & & E_{\text{B},p_x p_y, 0} & \\
 & & & E_{p_z} 
}\ ;
}
we obtain the symmetry properties of the onsite terms and NN hopping along $z$ for Mg atoms:
\eqa{
\label{eq:t_sym_MgB2_2}
& C_6: \text{no constraints} \\
& m_y: \text{no constraints} \\
& m_z: t_{\bsl{\tau}_{\text{Mg}}\bsl{\tau}_{\text{Mg}}}(\bsl{a}_3) = t_{\bsl{\tau}_{\text{Mg}}\bsl{\tau}_{\text{Mg}}}(-\bsl{a}_3)\\
& \TR: t_{\bsl{\tau}_{\text{Mg}}\bsl{\tau}_{\text{Mg}}}(0), t_{\bsl{\tau}_{\text{Mg}}\bsl{\tau}_{\text{Mg}}}(\pm\bsl{a}_3)\in \dsR \\
& h.c.: t_{\bsl{\tau}_{\text{Mg}}\bsl{\tau}_{\text{Mg}}}(0)\in \dsR, t_{\bsl{\tau}_{\text{Mg}}\bsl{\tau}_{\text{Mg}}}(\bsl{a}_3) = t_{\bsl{\tau}_{\text{Mg}}\bsl{\tau}_{\text{Mg}}}^*(-\bsl{a}_3)
}
for $t_{\bsl{\tau}_{\text{Mg}}\bsl{\tau}_{\text{Mg}}}(0)$, which gives 
\eq{
\label{eq:t_MgB2_part_2}
t_{\bsl{\tau}_{\text{Mg}}\bsl{\tau}_{\text{Mg}}}(0) = E_{\text{Mg},s}\ ,\ t_{\bsl{\tau}_{\text{Mg}}\bsl{\tau}_{\text{Mg}}}(\pm\bsl{a}_3) = t_{\text{Mg},s,z};
}
we obtain the symmetry properties of the NN hopping terms along $z$ for B atoms:
\eqa{
\label{eq:t_sym_MgB2_3}
& C_6: U_{C_6}^{\bsl{\tau}_{\text{B2}} \bsl{\tau}_{\text{B1}}} t_{\bsl{\tau}_{\text{B1}}\bsl{\tau}_{\text{B1}}}(\pm\bsl{a}_3) \left[U_{C_6}^{\bsl{\tau}_{\text{B2}} \bsl{\tau}_{\text{B1}}}\right]^\dagger = t_{\bsl{\tau}_{\text{B2}}\bsl{\tau}_{\text{B2}}}(\pm\bsl{a}_3) \\
& \qquad  U_{C_6}^{\bsl{\tau}_{\text{B1}} \bsl{\tau}_{\text{B2}}} t_{\bsl{\tau}_{\text{B2}}\bsl{\tau}_{\text{B2}}}(\pm\bsl{a}_3) \left[U_{C_6}^{\bsl{\tau}_{\text{B1}} \bsl{\tau}_{\text{B2}}}\right]^\dagger = t_{\bsl{\tau}_{\text{B1}}\bsl{\tau}_{\text{B1}}}(\pm\bsl{a}_3) \\
& m_y: U_{m_y}^{\bsl{\tau}_{\text{B2}} \bsl{\tau}_{\text{B1}}} t_{\bsl{\tau}_{\text{B1}}\bsl{\tau}_{\text{B1}}}(\pm\bsl{a}_3)\left[ U_{m_y}^{\bsl{\tau}_{\text{B2}} \bsl{\tau}_{\text{B1}}} \right]^\dagger = t_{\bsl{\tau}_{\text{B2}}\bsl{\tau}_{\text{B2}}}(\pm\bsl{a}_3) \\
& \qquad U_{m_y}^{\bsl{\tau}_{\text{B1}} \bsl{\tau}_{\text{B2}}} t_{\bsl{\tau}_{\text{B2}}\bsl{\tau}_{\text{B2}}}(\pm\bsl{a}_3)\left[ U_{m_y}^{\bsl{\tau}_{\text{B1}} \bsl{\tau}_{\text{B2}}} \right]^\dagger = t_{\bsl{\tau}_{\text{B1}}\bsl{\tau}_{\text{B1}}}(\pm\bsl{a}_3) \\
& m_z:  U_{m_z}^{\bsl{\tau}_{\text{B1}} \bsl{\tau}_{\text{B1}}} t_{\bsl{\tau}_{\text{B1}}\bsl{\tau}_{\text{B1}}}(\pm\bsl{a}_3) \left[ U_{m_z}^{\bsl{\tau}_{\text{B1}} \bsl{\tau}_{\text{B1}}} \right]^\dagger= t_{\bsl{\tau}_{\text{B1}}\bsl{\tau}_{\text{B1}}}(\mp\bsl{a}_3)\ ,\   U_{m_z}^{\bsl{\tau}_{\text{B2}} \bsl{\tau}_{\text{B2}}}  t_{\bsl{\tau}_{\text{B2}}\bsl{\tau}_{\text{B2}}}(\pm\bsl{a}_3) \left[ U_{m_z}^{\bsl{\tau}_{\text{B2}} \bsl{\tau}_{\text{B2}}} \right]^\dagger = t_{\bsl{\tau}_{\text{B2}}\bsl{\tau}_{\text{B2}}}(\mp\bsl{a}_3)\\
& \TR: t_{\bsl{\tau}_{\text{B1}}\bsl{\tau}_{\text{B1}}}(\pm\bsl{a}_3)\ ,\  t_{\bsl{\tau}_{\text{B2}}\bsl{\tau}_{\text{B2}}}(\pm\bsl{a}_3) \in \dsR^{4\times 4 }\\
& h.c.: t_{\bsl{\tau}_{\text{B1}}\bsl{\tau}_{\text{B1}}}^\dagger(\bsl{a}_3) = t_{\bsl{\tau}_{\text{B1}}\bsl{\tau}_{\text{B1}}}(-\bsl{a}_3)\ ,\  t_{\bsl{\tau}_{\text{B2}}\bsl{\tau}_{\text{B2}}}^\dagger(\bsl{a}_3) = t_{\bsl{\tau}_{\text{B2}}\bsl{\tau}_{\text{B2}}}(-\bsl{a}_3)
}
for $t_{\bsl{\tau}_{\text{B1}}\bsl{\tau}_{\text{B1}}}(\pm\bsl{a}_3)$ and $t_{\bsl{\tau}_{\text{B2}}\bsl{\tau}_{\text{B2}}}(\pm\bsl{a}_3)$, which gives 
\eq{
\label{eq:t_MgB2_part_3}
t_{\bsl{\tau}_{\text{B1}}\bsl{\tau}_{\text{B1}}}(\bsl{a}_3) = t_{\bsl{\tau}_{\text{B2}}\bsl{\tau}_{\text{B2}}}(\bsl{a}_3) = 
t_{\bsl{\tau}_{\text{B1}}\bsl{\tau}_{\text{B1}}}^\dagger(-\bsl{a}_3) =
t_{\bsl{\tau}_{\text{B2}}\bsl{\tau}_{\text{B2}}}^\dagger(-\bsl{a}_3) =
\mat{
t_{\text{B},s,z} & & & t_{\text{B},s-p_z, z}\\
 & t_{\text{B},p_x p_y,z} & & \\
 & & t_{\text{B},p_x p_y,z} & \\
- t_{\text{B},s-p_z, z}  & & & t_{p_z,z} 
}\ ;
}
we obtain the symmetry properties of the NN hopping terms in $x-y$ plane for B atoms:
\eqa{
\label{eq:t_sym_MgB2_4}
& C_6: U_{C_6}^{\bsl{\tau}_{\text{B2}} \bsl{\tau}_{\text{B1}}} t_{\bsl{\tau}_{\text{B1}}\bsl{\tau}_{\text{B2}}}(\bsl{\delta}_0) \left[ U_{C_6}^{\bsl{\tau}_{\text{B1}} \bsl{\tau}_{\text{B2}}} \right]^\dagger = t_{\bsl{\tau}_{\text{B2}}\bsl{\tau}_{\text{B1}}}(-\bsl{\delta}_2)\ ,\  U_{C_6}^{\bsl{\tau}_{\text{B1}}\bsl{\tau}_{\text{B2}} } t_{\bsl{\tau}_{\text{B2}}\bsl{\tau}_{\text{B1}}}(-\bsl{\delta}_2) \left[ U_{C_6}^{ \bsl{\tau}_{\text{B2}} \bsl{\tau}_{\text{B1}}} \right]^\dagger =  t_{\bsl{\tau}_{\text{B1}}\bsl{\tau}_{\text{B2}}}(\bsl{\delta}_1)\\
 & \qquad   U_{C_6}^{\bsl{\tau}_{\text{B2}} \bsl{\tau}_{\text{B1}}} t_{\bsl{\tau}_{\text{B1}}\bsl{\tau}_{\text{B2}}}(\bsl{\delta}_1) \left[ U_{C_6}^{\bsl{\tau}_{\text{B1}} \bsl{\tau}_{\text{B2}}} \right]^\dagger = t_{\bsl{\tau}_{\text{B2}}\bsl{\tau}_{\text{B1}}}(-\bsl{\delta}_0)\ ,\  U_{C_6}^{\bsl{\tau}_{\text{B1}}\bsl{\tau}_{\text{B2}} } t_{\bsl{\tau}_{\text{B2}}\bsl{\tau}_{\text{B1}}}(-\bsl{\delta}_0) \left[ U_{C_6}^{ \bsl{\tau}_{\text{B2}} \bsl{\tau}_{\text{B1}}} \right]^\dagger =  t_{\bsl{\tau}_{\text{B1}}\bsl{\tau}_{\text{B2}}}(\bsl{\delta}_2)\\
 & \qquad  U_{C_6}^{\bsl{\tau}_{\text{B2}} \bsl{\tau}_{\text{B1}}} t_{\bsl{\tau}_{\text{B1}}\bsl{\tau}_{\text{B2}}}(\bsl{\delta}_2) \left[ U_{C_6}^{\bsl{\tau}_{\text{B1}} \bsl{\tau}_{\text{B2}}} \right]^\dagger = t_{\bsl{\tau}_{\text{B2}}\bsl{\tau}_{\text{B1}}}(-\bsl{\delta}_1)\ ,\  U_{C_6}^{\bsl{\tau}_{\text{B1}}\bsl{\tau}_{\text{B2}} } t_{\bsl{\tau}_{\text{B2}}\bsl{\tau}_{\text{B1}}}(-\bsl{\delta}_1) \left[ U_{C_6}^{ \bsl{\tau}_{\text{B2}} \bsl{\tau}_{\text{B1}}} \right]^\dagger =  t_{\bsl{\tau}_{\text{B1}}\bsl{\tau}_{\text{B2}}}(\bsl{\delta}_0)\\
& m_y:  U_{m_y}^{\bsl{\tau}_{\text{B2}} \bsl{\tau}_{\text{B1}}} t_{\bsl{\tau}_{\text{B2}} \bsl{\tau}_{\text{B1}}}(\bsl{\delta}_0) \left[ U_{m_y}^{\bsl{\tau}_{\text{B1}} \bsl{\tau}_{\text{B2}}} \right]^\dagger = t_{\bsl{\tau}_{\text{B2}} \bsl{\tau}_{\text{B1}}}(-\bsl{\delta}_0) \\
& \qquad U_{m_y}^{\bsl{\tau}_{\text{B2}} \bsl{\tau}_{\text{B1}}} t_{\bsl{\tau}_{\text{B2}} \bsl{\tau}_{\text{B1}}}(\bsl{\delta}_1) \left[ U_{m_y}^{\bsl{\tau}_{\text{B1}} \bsl{\tau}_{\text{B2}}} \right]^\dagger = t_{\bsl{\tau}_{\text{B2}} \bsl{\tau}_{\text{B1}}}(-\bsl{\delta}_2) \\ 
& \qquad U_{m_y}^{\bsl{\tau}_{\text{B2}} \bsl{\tau}_{\text{B1}}} t_{\bsl{\tau}_{\text{B2}} \bsl{\tau}_{\text{B1}}}(\bsl{\delta}_2) \left[ U_{m_y}^{\bsl{\tau}_{\text{B1}} \bsl{\tau}_{\text{B2}}} \right]^\dagger = t_{\bsl{\tau}_{\text{B2}} \bsl{\tau}_{\text{B1}}}(-\bsl{\delta}_1) \\
& m_z: U_{m_z}^{\bsl{\tau}_{\text{B1}} \bsl{\tau}_{\text{B1}}} t_{\bsl{\tau}_{\text{B1}} \bsl{\tau}_{\text{B2}}}(\bsl{\delta}_j) U_{m_z}^{\bsl{\tau}_{\text{B2}} \bsl{\tau}_{\text{B2}}} = t_{\bsl{\tau}_{\text{B1}} \bsl{\tau}_{\text{B2}}}(\bsl{\delta}_j) \ ,\ U_{m_z}^{\bsl{\tau}_{\text{B2}} \bsl{\tau}_{\text{B2}}} t_{\bsl{\tau}_{\text{B2}} \bsl{\tau}_{\text{B1}}}(\bsl{\delta}_j) U_{m_z}^{\bsl{\tau}_{\text{B1}} \bsl{\tau}_{\text{B1}}} = t_{\bsl{\tau}_{\text{B2}} \bsl{\tau}_{\text{B1}}}(-\bsl{\delta}_j) \\
& \TR: t_{\bsl{\tau}_{\text{B1}} \bsl{\tau}_{\text{B2}}}(\bsl{\delta}_j) \ ,\ t_{\bsl{\tau}_{\text{B2}} \bsl{\tau}_{\text{B1}}}(-\bsl{\delta}_j) \in \dsR^{4\times4}\\
& h.c.: t_{\bsl{\tau}_{\text{B1}} \bsl{\tau}_{\text{B2}}}(\bsl{\delta}_j)= t_{\bsl{\tau}_{\text{B2}} \bsl{\tau}_{\text{B1}}}^\dagger(-\bsl{\delta}_j)\ ,
}
for $t_{\bsl{\tau}_{\text{B1}}\bsl{\tau}_{\text{B2}}}(\bsl{\delta}_j)$ and  $t_{\bsl{\tau}_{\text{B2}}\bsl{\tau}_{\text{B1}}}(-\bsl{\delta}_j)$, which gives
\eqa{
\label{eq:t_MgB2_part_4}
 & t_{\bsl{\tau}_{\text{B1}}\bsl{\tau}_{\text{B2}}}(\bsl{\delta}_0) 
 = 
 \mat{ 
 t_1 & & t_4 & \\
  & t_2 + t_3 & & \\
 -t_4 & & t_2 - t_3 & \\
  & & & t_{p_z} \\
 } \ ,\ U_{C_3}^{\bsl{\tau}_{\text{B1}} \bsl{\tau}_{\text{B1}}} t_{\bsl{\tau}_{\text{B1}}\bsl{\tau}_{\text{B2}}}(\bsl{\delta}_j) U_{C_3}^{\bsl{\tau}_{\text{B2}} \bsl{\tau}_{\text{B2}}} = t_{\bsl{\tau}_{\text{B1}}\bsl{\tau}_{\text{B2}}}(\bsl{\delta}_{j+1})\\
&  t_{\bsl{\tau}_{\text{B1}} \bsl{\tau}_{\text{B2}}}(\bsl{\delta}_j)= t_{\bsl{\tau}_{\text{B2}} \bsl{\tau}_{\text{B1}}}^\dagger(-\bsl{\delta}_j) \text{ with } U_{C_3}^{\bsl{\tau}_{\text{B1}} \bsl{\tau}_{\text{B1}}} = U_{C_6}^{\bsl{\tau}_{\text{B1}} \bsl{\tau}_{\text{B2}}} U_{C_6}^{\bsl{\tau}_{\text{B2}} \bsl{\tau}_{\text{B1}}} \text{ and } U_{C_3}^{\bsl{\tau}_{\text{B1}} \bsl{\tau}_{\text{B1}}} = U_{C_3}^{\bsl{\tau}_{\text{B2}} \bsl{\tau}_{\text{B2}}}\ ;
 }
 we obtain the symmetry properties of the NN hopping terms in the $x-y$ plane for Mg atoms:
\eqa{
\label{eq:t_sym_MgB2_5}
& C_6: t_{\bsl{\tau}_{\text{Mg}} \bsl{\tau}_{\text{Mg}}}(C_6^0 \bsl{a}_1) = t_{\bsl{\tau}_{\text{Mg}} \bsl{\tau}_{\text{Mg}}}(C_6^1 \bsl{a}_1) = t_{\bsl{\tau}_{\text{Mg}} \bsl{\tau}_{\text{Mg}}}(C_6^2 \bsl{a}_1) = t_{\bsl{\tau}_{\text{Mg}} \bsl{\tau}_{\text{Mg}}}(C_6^3 \bsl{a}_1) = t_{\bsl{\tau}_{\text{Mg}} \bsl{\tau}_{\text{Mg}}}(C_6^4 \bsl{a}_1)  = t_{\bsl{\tau}_{\text{Mg}} \bsl{\tau}_{\text{Mg}}}(C_6^5 \bsl{a}_1) \\
& m_y:  t_{\bsl{\tau}_{\text{Mg}} \bsl{\tau}_{\text{Mg}}}(C_6^0 \bsl{a}_1) =  t_{\bsl{\tau}_{\text{Mg}} \bsl{\tau}_{\text{Mg}}}(C_6^4 \bsl{a}_1)\ ,\ t_{\bsl{\tau}_{\text{Mg}} \bsl{\tau}_{\text{Mg}}}(C_6^3 \bsl{a}_1) =  t_{\bsl{\tau}_{\text{Mg}} \bsl{\tau}_{\text{Mg}}}(C_6^1 \bsl{a}_1)\\
& m_z: \text{no constraints}\\
& \TR: t_{\bsl{\tau}_{\text{Mg}}\bsl{\tau}_{\text{Mg}}}(C_6^n \bsl{a}_1)\in \dsR \\
& h.c.: t_{\bsl{\tau}_{\text{Mg}}\bsl{\tau}_{\text{Mg}}}(C_6^n \bsl{a}_1)\in \dsR 
}
for $t_{\bsl{\tau}_{\text{Mg}}\bsl{\tau}_{\text{Mg}}}(C_6^n \bsl{a}_1)$, which gives 
\eq{
\label{eq:t_MgB2_part_5}
t_{\bsl{\tau}_{\text{Mg}}\bsl{\tau}_{\text{Mg}}}(C_6^n \bsl{a}_1) = t_{\text{Mg},s}\ ;
}
we obtain the symmetry properties of the NN hopping terms from a Mg atom to a B atom:
\eqa{
\label{eq:t_sym_MgB2_6}
& C_6: U_{C_6}^{\bsl{\tau}_{\text{B2}} \bsl{\tau}_{\text{B1}}} t_{\bsl{\tau}_{\text{B1}}\bsl{\tau}_{\text{Mg}}}(-\bsl{\delta}_0\pm\bsl{a}_3/2) = t_{\bsl{\tau}_{\text{B2}}\bsl{\tau}_{\text{Mg}}}(\bsl{\delta}_2\pm\bsl{a}_3/2)\ ,\  U_{C_6}^{\bsl{\tau}_{\text{B1}}\bsl{\tau}_{\text{B2}} } t_{\bsl{\tau}_{\text{B2}}\bsl{\tau}_{\text{Mg}}}(\bsl{\delta}_2\pm\bsl{a}_3/2) =  t_{\bsl{\tau}_{\text{B1}}\bsl{\tau}_{\text{Mg}}}(-\bsl{\delta}_1\pm\bsl{a}_3/2)\\
 & \qquad   U_{C_6}^{\bsl{\tau}_{\text{B2}} \bsl{\tau}_{\text{B1}}} t_{\bsl{\tau}_{\text{B1}}\bsl{\tau}_{\text{Mg}}}(-\bsl{\delta}_1\pm\bsl{a}_3/2) = t_{\bsl{\tau}_{\text{B2}}\bsl{\tau}_{\text{Mg}}}(\bsl{\delta}_0\pm\bsl{a}_3/2)\ ,\  U_{C_6}^{\bsl{\tau}_{\text{B1}}\bsl{\tau}_{\text{B2}} } t_{\bsl{\tau}_{\text{B2}}\bsl{\tau}_{\text{Mg}}}(\bsl{\delta}_0\pm\bsl{a}_3/2) =  t_{\bsl{\tau}_{\text{B1}}\bsl{\tau}_{\text{Mg}}}(-\bsl{\delta}_2\pm\bsl{a}_3/2)\\
 & \qquad  U_{C_6}^{\bsl{\tau}_{\text{B2}} \bsl{\tau}_{\text{B1}}} t_{\bsl{\tau}_{\text{B1}}\bsl{\tau}_{\text{Mg}}}(-\bsl{\delta}_2\pm\bsl{a}_3/2) = t_{\bsl{\tau}_{\text{B2}}\bsl{\tau}_{\text{Mg}}}(\bsl{\delta}_1\pm\bsl{a}_3/2)\ ,\  U_{C_6}^{\bsl{\tau}_{\text{B1}}\bsl{\tau}_{\text{B2}} } t_{\bsl{\tau}_{\text{B2}}\bsl{\tau}_{\text{Mg}}}(\bsl{\delta}_1\pm\bsl{a}_3/2) =  t_{\bsl{\tau}_{\text{B1}}\bsl{\tau}_{\text{Mg}}}(-\bsl{\delta}_0\pm\bsl{a}_3/2)\\ 
& m_y:  U_{m_y}^{\bsl{\tau}_{\text{B2}} \bsl{\tau}_{\text{B1}}} t_{\bsl{\tau}_{\text{B1}}\bsl{\tau}_{\text{Mg}}}(-\bsl{\delta}_0\pm\bsl{a}_3/2) = t_{\bsl{\tau}_{\text{B2}}\bsl{\tau}_{\text{Mg}}}(\bsl{\delta}_0\pm\bsl{a}_3/2) \\
& \qquad U_{m_y}^{\bsl{\tau}_{\text{B2}} \bsl{\tau}_{\text{B1}}} t_{\bsl{\tau}_{\text{B1}}\bsl{\tau}_{\text{Mg}}}(-\bsl{\delta}_1\pm\bsl{a}_3/2) = t_{\bsl{\tau}_{\text{B2}}\bsl{\tau}_{\text{Mg}}}(\bsl{\delta}_2\pm\bsl{a}_3/2) \\
& \qquad U_{m_y}^{\bsl{\tau}_{\text{B2}} \bsl{\tau}_{\text{B1}}} t_{\bsl{\tau}_{\text{B1}}\bsl{\tau}_{\text{Mg}}}(-\bsl{\delta}_2\pm\bsl{a}_3/2) = t_{\bsl{\tau}_{\text{B2}}\bsl{\tau}_{\text{Mg}}}(\bsl{\delta}_1\pm\bsl{a}_3/2) \\
& m_z: U_{m_z}^{\bsl{\tau}_{\text{B1}} \bsl{\tau}_{\text{B1}}} t_{\bsl{\tau}_{\text{B1}}\bsl{\tau}_{\text{Mg}}}(-\bsl{\delta}_j\pm\bsl{a}_3/2) = t_{\bsl{\tau}_{\text{B1}}\bsl{\tau}_{\text{Mg}}}(-\bsl{\delta}_j\mp\bsl{a}_3/2) \ ,\  U_{m_z}^{\bsl{\tau}_{\text{B2}} \bsl{\tau}_{\text{B2}}} t_{\bsl{\tau}_{\text{B2}}\bsl{\tau}_{\text{Mg}}}(\bsl{\delta}_j\pm\bsl{a}_3/2) = t_{\bsl{\tau}_{\text{B2}}\bsl{\tau}_{\text{Mg}}}(\bsl{\delta}_j\mp\bsl{a}_3/2) \\
& \TR:t_{\bsl{\tau}_{\text{B1}}\bsl{\tau}_{\text{Mg}}}(-\bsl{\delta}_j\pm\bsl{a}_3/2) \ ,\ t_{\bsl{\tau}_{\text{B2}}\bsl{\tau}_{\text{Mg}}}(\bsl{\delta}_j\pm\bsl{a}_3/2) \in \dsR^{4\times 1}\\
& h.c.: t_{\bsl{\tau}_{\text{B1}}\bsl{\tau}_{\text{Mg}}}(-\bsl{\delta}_j\pm\bsl{a}_3/2)= t_{\bsl{\tau}_{\text{Mg}}\bsl{\tau}_{\text{B1}}}^\dagger(\bsl{\delta}_j\mp\bsl{a}_3/2)\ \ ,\ t_{\bsl{\tau}_{\text{B2}}\bsl{\tau}_{\text{Mg}}}(\bsl{\delta}_j\pm\bsl{a}_3/2)= t_{\bsl{\tau}_{\text{Mg}}\bsl{\tau}_{\text{B2}}}^\dagger(-\bsl{\delta}_j\mp\bsl{a}_3/2)\ ,
}
for $t_{\bsl{\tau}_{\text{B1}}\bsl{\tau}_{\text{Mg}}}(-\bsl{\delta}_j\pm\bsl{a}_3/2)$, $t_{\bsl{\tau}_{\text{B2}}\bsl{\tau}_{\text{Mg}}}(\bsl{\delta}_j\pm\bsl{a}_3/2)$, $t_{\bsl{\tau}_{\text{Mg}}\bsl{\tau}_{\text{B1}}}(\bsl{\delta}_j\pm\bsl{a}_3/2)$, and $t_{\bsl{\tau}_{\text{Mg}}\bsl{\tau}_{\text{B2}}}(-\bsl{\delta}_j\pm\bsl{a}_3/2)$, which gives
\eqa{
\label{eq:t_MgB2_part_6}
& t_{\bsl{\tau}_{\text{B1}}\bsl{\tau}_{\text{Mg}}}(-\bsl{\delta}_0 + \bsl{a}_3/2) = ( t_{\text{Mg-B},1} , 0, t_{\text{Mg-B},2} ,t_{\text{Mg-B},3})^T\ ,\ U_{C_3}^{\bsl{\tau}_{\text{B1}} \bsl{\tau}_{\text{B1}}} t_{\bsl{\tau}_{\text{B1}}\bsl{\tau}_{\text{Mg}}}(-\bsl{\delta}_j + \bsl{a}_3/2) = t_{\bsl{\tau}_{\text{B1}}\bsl{\tau}_{\text{Mg}}}(-\bsl{\delta}_{j+1} + \bsl{a}_3/2)\\
& t_{\bsl{\tau}_{\text{B2}}\bsl{\tau}_{\text{Mg}}}(\bsl{\delta}_j + \bsl{a}_3/2) = U_{C_2}^{\bsl{\tau}_{\text{B2}} \bsl{\tau}_{\text{B1}}} t_{\bsl{\tau}_{\text{B1}}\bsl{\tau}_{\text{Mg}}}(-\bsl{\delta}_j + \bsl{a}_3/2) \text{ with } U_{C_2}^{\bsl{\tau}_{\text{B2}} \bsl{\tau}_{\text{B1}}} = \diag(1,-1,-1,1) \\
&  t_{\bsl{\tau}_{\text{B1}}\bsl{\tau}_{\text{Mg}}}(-\bsl{\delta}_j-\bsl{a}_3/2) = U_{m_z}^{\bsl{\tau}_{\text{B1}} \bsl{\tau}_{\text{B1}}} t_{\bsl{\tau}_{\text{B1}}\bsl{\tau}_{\text{Mg}}}(-\bsl{\delta}_j+\bsl{a}_3/2)  \ ,\  t_{\bsl{\tau}_{\text{B2}}\bsl{\tau}_{\text{Mg}}}(\bsl{\delta}_j-\bsl{a}_3/2) = U_{m_z}^{\bsl{\tau}_{\text{B2}} \bsl{\tau}_{\text{B2}}} t_{\bsl{\tau}_{\text{B2}}\bsl{\tau}_{\text{Mg}}}(\bsl{\delta}_j+\bsl{a}_3/2) \\
 & t_{\bsl{\tau}_{\text{Mg}}\bsl{\tau}_{\text{B1}}}(\bsl{\delta}_j\pm\bsl{a}_3/2)= t_{\bsl{\tau}_{\text{B1}}\bsl{\tau}_{\text{Mg}}}^\dagger(-\bsl{\delta}_j\mp\bsl{a}_3/2) \ \ ,\ t_{\bsl{\tau}_{\text{Mg}}\bsl{\tau}_{\text{B2}}}^\dagger(-\bsl{\delta}_j\pm\bsl{a}_3/2)= t_{\bsl{\tau}_{\text{B2}}\bsl{\tau}_{\text{Mg}}}(\bsl{\delta}_j\mp\bsl{a}_3/2) \ .
}
In sum, we have 17 real parameters for the electron Hamiltonian:
\eq{
E_{\text{B},s,0}, E_{\text{B},p_x p_y, 0}, E_{p_z}, t_1, t_2 , t_3 ,t_4, t_{p_z}, t_{\text{B},p_x p_y,z},  t_{p_z,z},  t_{\text{B},s-p_z, z}, E_{\text{Mg},s}, t_{\text{Mg},s}, t_{\text{B},s,z}, t_{\text{Mg-B},1} , t_{\text{Mg-B},2} ,t_{\text{Mg-B},3}\ .
}

By substituting \eqnref{eq:t_MgB2_part_1}, \eqnref{eq:t_MgB2_part_2}, \eqnref{eq:t_MgB2_part_3}, \eqnref{eq:t_MgB2_part_4}, \eqnref{eq:t_MgB2_part_5} and \eqnref{eq:t_MgB2_part_6} into \eqnref{eq:H_el_gen_k}, we arrive at $H_{el}$ for {\mgb} in momentum space:
\eq{
\label{eq:H_el_MgB2}
H_{el} = H_{el}^{\text{B},sp_xp_y} + H_{el}^{\text{B},p_z} + H_{el}^{\text{Mg}} + H_{el}^{\text{B}, sp_xp_y - p_z} + H_{el}^{\text{Mg-B}}  \ .
}
In the following, we show the forms of $H_{el}^{\text{B},sp_xp_y}$, $H_{el}^{\text{B},p_z}$, $H_{el}^{\text{Mg}}$, $H_{el}^{\text{B}, sp_xp_y - p_z}$ and $H_{el}^{\text{Mg-B}}$.

First, the form of $H_{el}^{\text{B},sp_xp_y}$ in \eqnref{eq:H_el_MgB2} reads
\eq{
\label{eq:H_el_MgB2_B_sp2}
H_{el}^{\text{B},sp_xp_y} = \sum_{\bsl{k}}^{\BZ} c^\dagger_{\bsl{k},\text{B}, sp_xp_y} h_{sp_xp_y}(\bsl{k})   c_{\bsl{k},\text{B}, sp_xp_y}\ ,
}
where
$c^\dagger_{\bsl{k},\text{B}, sp_xp_y} =  (c^{\dagger}_{\bsl{k},\bsl{\tau}_{\text{B1}},s},c^{\dagger}_{\bsl{k},\bsl{\tau}_{\text{B1}},p_x},c^{\dagger}_{\bsl{k},\bsl{\tau}_{\text{B1}},p_y},c^{\dagger}_{\bsl{k},\bsl{\tau}_{\text{B2}},s},c^{\dagger}_{\bsl{k},\bsl{\tau}_{\text{B2}},p_x},c^{\dagger}_{\bsl{k},\bsl{\tau}_{\text{B2}},p_y}) $,
\eq{
h_{sp_xp_y}(\bsl{k})  = \mat{ h_{\text{B1}-\text{B1}, sp_xp_y}(\bsl{k}) & h_{\text{B1}-\text{B2}, sp_xp_y}(\bsl{k}) \\
h_{\text{B2}-\text{B1}, sp_xp_y}(\bsl{k}) & h_{\text{B2}-\text{B2}, sp_xp_y}(\bsl{k}) }\ ,
}
\eqa{
\label{eq:hopping_el_MgB2_B_sp2}
& h_{\text{B1}-\text{B1}, sp_xp_y}(\bsl{k}) = \mat{
E_{\text{B},s,0} & &  \\
 & E_{\text{B},p_x p_y, 0} &  \\
 & & E_{\text{B},p_x p_y, 0}  
} + \mat{
t_{\text{B},s,z} 2 \cos(\bsl{a}_3\cdot \bsl{k}) & & \\
 & t_{\text{B},p_x p_y,z} 2 \cos(\bsl{a}_3\cdot \bsl{k}) & \\
 & & t_{\text{B},p_x p_y,z} 2 \cos(\bsl{a}_3\cdot \bsl{k})  
} \\
& h_{\text{B2}-\text{B2}, sp_xp_y}(\bsl{k}) = \mat{ 1 & & \\ & -1 & \\ & & -1} h_{\text{B1}-\text{B1}, sp_xp_y}(-\bsl{k}) \mat{ 1 & & \\ & -1 & \\ & & -1}\\
& h_{\text{B1}-\text{B2}, sp_xp_y}(\bsl{k}) = h_{\text{B2}-\text{B1}, sp_xp_y}^\dagger(\bsl{k}) =  \sum_{j=0,1,2} e^{-\ii \bsl{\delta}_j\cdot\bsl{k}} \mat{ 
 1 & &  \\
  & -\frac{1}{2} & -\frac{\sqrt{3}}{2} \\
  & \frac{\sqrt{3}}{2} & -\frac{1}{2}
 }  ^j\mat{ 
 t_1 & & t_4 \\
  & t_2 + t_3 & \\
 -t_4 & & t_2 - t_3 
 } \mat{ 
 1 & &  \\
  & -\frac{1}{2} & -\frac{\sqrt{3}}{2} \\
  & \frac{\sqrt{3}}{2} & -\frac{1}{2}
 }^{-j}\ ,
}
and $\bsl{\delta}_j$ is defined in \eqnref{eq:delta_j_graphene}.

Second, the form of $H_{el}^{\text{B},p_z}$ in \eqnref{eq:H_el_MgB2} reads
\eq{
\label{eq:H_el_MgB2_B_pz}
H_{el}^{\text{B},p_z} =  \sum_{\bsl{k}}^{\BZ} c^\dagger_{\bsl{k},\text{B}, p_z}  h_{p_z}(\bsl{k})  c_{\bsl{k},\text{B}, p_z}\ ,
}
where 
$c^\dagger_{\bsl{k},\text{B}, p_z} =  (c^{\dagger}_{\bsl{k},\bsl{\tau}_{\text{B1}},p_z},c^{\dagger}_{\bsl{k},\bsl{\tau}_{\text{B2}},p_z}) $,
\eq{
h_{p_z}(\bsl{k}) =\mat{ h_{\text{B1}-\text{B1}, p_z}(\bsl{k}) & h_{\text{B1}-\text{B2}, p_z}(\bsl{k}) \\
h_{\text{B2}-\text{B1}, p_z}(\bsl{k}) & h_{\text{B2}-\text{B2}, p_z}(\bsl{k}) } \ ,
}
and
\eqa{
& h_{\text{B1}-\text{B1}, p_z}(\bsl{k}) = E_{p_z} +  t_{p_z,z} 2 \cos(\bsl{k}\cdot \bsl{a}_3)  \\
& h_{\text{B2}-\text{B2}, p_z}(\bsl{k}) = h_{\text{B1}-\text{B1}, p_z}(-\bsl{k})\\
& h_{\text{B1}-\text{B2}, p_z}(\bsl{k}) = h_{\text{B2}-\text{B1}, p_z}^\dagger (\bsl{k}) = \sum_{j=0,1,2} e^{-\ii \bsl{\delta}_j\cdot\bsl{k}}   t_{p_z}  \ .
}
We note that \eqnref{eq:H_el_MgB2_B_pz} was used in \refcite{Liu10192019MgB2Dirac} to fit the $p_z$ electron bands.

Third, the form of $H_{el}^{\text{Mg}}$ in \eqnref{eq:H_el_MgB2} reads
\eq{
\label{eq:H_el_MgB2_Mg}
H_{el}^{\text{Mg}} =  \sum_{\bsl{k}}^{\BZ} c^\dagger_{\bsl{k},\text{Mg}} h_{\text{Mg}}(\bsl{k}) c_{\bsl{k},\text{Mg}}\ ,
}
where
\eq{
h_{\text{Mg}}(\bsl{k}) = E_{\text{Mg},s} + 2 t_{\text{Mg},s,z} \cos(\bsl{k}\cdot \bsl{a}_3) + t_{\text{Mg},s} \sum_{n=0,1,..,5} e^{-\ii (C_6^n \bsl{a}_1)\cdot\bsl{k}}\ ,
}
$\bsl{a}_{1,2,3}$ are defined in \eqnref{eq:lattice_MgB2}, and recall that each layer of Mg forms a triangular lattice.

Fourth, the form of $H_{el}^{\text{B}, sp_xp_y - p_z} $ in \eqnref{eq:H_el_MgB2} reads
\eq{
\label{eq:H_el_MgB2_B_sp2-pz}
H_{el}^{\text{B}, sp_xp_y - p_z} =  \sum_{\bsl{k}}^{\BZ} c^\dagger_{\bsl{k},\text{B}, sp_xp_y} \mat{ h_{\text{B1}-\text{B1},sp_xp_y - p_z}(\bsl{k}) & h_{\text{B1}-\text{B2}, sp_xp_y - p_z }(\bsl{k}) \\
h_{\text{B2}-\text{B1}, sp_xp_y - p_z }(\bsl{k}) & h_{\text{B2}-\text{B2}, sp_xp_y - p_z }(\bsl{k}) } c_{\bsl{k},\text{B}, p_z} + h.c.\ ,
}
where
\eqa{
& h_{\text{B1}-\text{B1}, sp_xp_y - p_z}(\bsl{k}) =\mat{
2 (-\ii)  t_{\text{B},s-p_z, z} \sin(\bsl{a}_3 \cdot \bsl{k})\\
0 \\
0 
} \\
& h_{\text{B1}-\text{B2}, sp_xp_y - p_z}(\bsl{k}) = 0  \\
& h_{\text{B2}-\text{B1}, sp_xp_y - p_z}(\bsl{k}) = - \mat{ 1 & & \\ & -1 & \\ & & -1}  h_{\text{B1}-\text{B2}, sp_xp_y - p_z}(-\bsl{k})\\
& h_{\text{B2}-\text{B2}, sp_xp_y - p_z}(\bsl{k}) = - \mat{ 1 & & \\ & -1 & \\ & & -1}  h_{\text{B1}-\text{B1}, sp_xp_y - p_z}(-\bsl{k}) \ ,
}
and $\bsl{a}_{1,2,3}$ are defined in \eqnref{eq:lattice_MgB2}.

Fifth, the form of $H_{el}^{\text{Mg-B}} $ in \eqnref{eq:H_el_MgB2} reads
\eq{
\label{eq:H_el_MgB2_Mg-B}
H_{el}^{\text{Mg-B}} = \sum_{\bsl{k}}^{\BZ} c^\dagger_{\bsl{k},\text{Mg}}
\mat{ 
h_{\text{Mg}-\text{B1}}(\bsl{k}) & h_{\text{Mg}-\text{B2}}(\bsl{k})
}   c_{\bsl{k},\text{B}} + h.c.\ ,
}
$c^\dagger_{\bsl{k},\text{B}} =  (c^{\dagger}_{\bsl{k},\bsl{\tau}_{\text{B1}},s},c^{\dagger}_{\bsl{k},\bsl{\tau}_{\text{B1}},p_x},c^{\dagger}_{\bsl{k},\bsl{\tau}_{\text{B1}},p_y}, c^{\dagger}_{\bsl{k},\bsl{\tau}_{\text{B1}},p_z},c^{\dagger}_{\bsl{k},\bsl{\tau}_{\text{B2}},s},c^{\dagger}_{\bsl{k},\bsl{\tau}_{\text{B2}},p_x},c^{\dagger}_{\bsl{k},\bsl{\tau}_{\text{B2}},p_y} ,c^{\dagger}_{\bsl{k},\bsl{\tau}_{\text{B2}},p_z}) $,
\eqa{
& h_{\text{Mg}-\text{B1}}(\bsl{k}) = \sum_{j=0,1,2} \sum_{\alpha=0,1} e^{-\ii \bsl{k}\cdot (\bsl{\delta}_j - (-1)^\alpha \bsl{a}_3/2 )}\mat{  t_{\text{Mg-B},1} &  0 &  t_{\text{Mg-B},2} & t_{\text{Mg-B},3} } \mat{ 
 1 & & & \\
  & -\frac{1}{2} & \frac{\sqrt{3}}{2} & \\
  &  -\frac{\sqrt{3}}{2} & -\frac{1}{2} & \\
  & & & 1
 }^j  \mat{ 
 1 & & & \\
  & 1 &  & \\
  & & 1 & \\
  & & & -1
 }^\alpha  \\
&  h_{\text{Mg}-\text{B2}}(\bsl{k}) = h_{\text{Mg}-\text{B1}}(-\bsl{k}) \mat{ 1 & & & \\ & -1 & & \\ & & -1 & \\ & & & -1}  \ ,
}  
$\bsl{\delta}_j$ is defined in \eqnref{eq:delta_j_graphene}, and $\bsl{a}_{1,2,3}$ are defined in \eqnref{eq:lattice_MgB2}.

\begin{figure}[t]
    \centering
    \includegraphics[width=0.8\columnwidth]{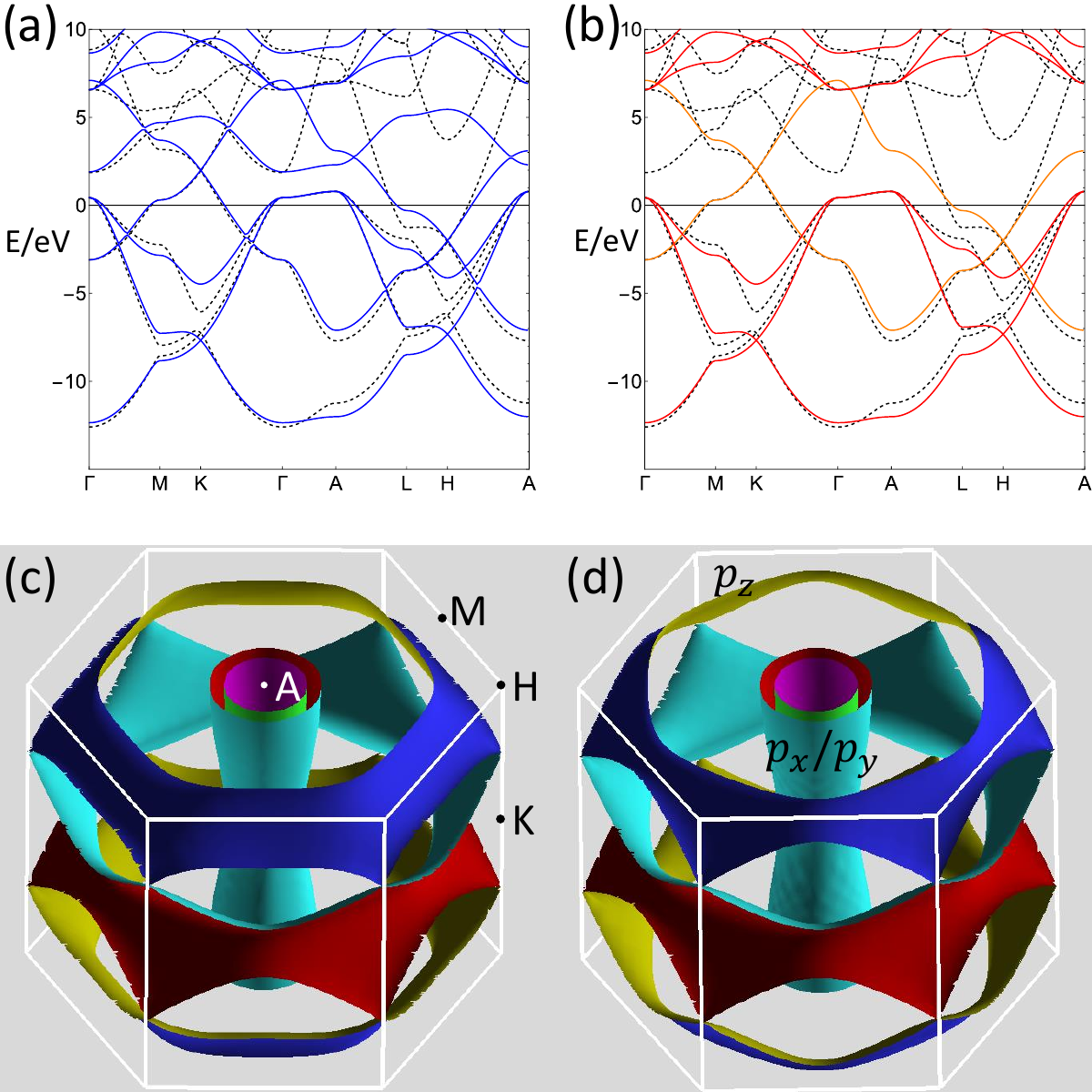}
    \caption{ 
    In both (a) and (b), the {\abi} band structure is plotted as the black dashed lines, and the Fermi level is at zero energy.
    The high-symmetry points are defined as $\Gamma=(0,0,0)$, $\K = (\frac{4 \pi }{3 a},0,0)$, $\text{M}= 2\pi (\frac{\sqrt{3}}{2 a},\frac{1}{2 a},0)$, $\text{A}=(0,0,\pi/c)$, $\text{H} = (\frac{4 \pi }{3 a},0,\pi/c)$, and $\text{L}=  (\frac{2\pi \sqrt{3}}{2 a},\frac{\pi}{a},\pi/c)$, 
    (a) We plot the band structure of the full tight-binding model (\eqnref{eq:H_el_MgB2}) with the parameter values in \eqnref{eq:MgB2_TB_values} as blue lines. 
    (b) We plot the band structure of the simplified full tight-binding model (\eqnref{eq:H_el_MgB2_sim}) with the parameter values in \eqnref{eq:MgB2_TB_values_sim} as solid lines, where the red lines are from $H_{el}^{\text{B},sp_xp_y}$ and the orange lines are from $H_{el}^{\text{B},p_z}$ (\ie, the $p_z$ orbitals of the B atoms).
    The crossings of the orange lines at K and H are two Dirac points.
    (c) We plot Fermi surface of the full tight-binding model (\eqnref{eq:H_el_MgB2}) with the parameter values in \eqnref{eq:MgB2_TB_values}. 
    (d) We plot Fermi surface of the simplified full tight-binding model (\eqnref{eq:H_el_MgB2_sim}) with the parameter values in \eqnref{eq:MgB2_TB_values_sim}. 
    The Fermi surface around K-H (and its TR partner) mainly comes from the $p_z$ orbital, while the cylindrical Fermi surfaces around $\Gamma$-A mainly originate from $p_x/p_y$ orbitals. 
    }
    \label{fig:MgB2_el}
\end{figure}

We plot the band structure of \eqnref{eq:H_el_MgB2} as the blue line in \figref{fig:MgB2_el}(a) for 
\eqa{
\label{eq:MgB2_TB_values}
& E_{\text{B},s,0} = -1.68\ ,\ E_{\text{B},p_x p_y, 0} = 3.68\ ,\ E_{p_z} = 0\ ,\ t_1 = -3.5\ ,\ t_2 = 1.022\ ,\ t_3 = -2.66\ ,\ t_4 = 3.8\ ,\ t_{p_z} = -1.7\\
& t_{\text{B},s,z} = -0.085\ ,\ t_{\text{B},p_x p_y,z} = -0.089\ ,\ t_{p_z,z} = 1\ ,\ t_{\text{B},s-p_z,z} = 0\ ,\ E_{\text{Mg},s} = 4.2\ ,\ t_{\text{Mg},s} = -0.35\ ,\ t_{\text{Mg},s,z} = -0.1 \\
& t_{\text{Mg-B},1} = 0\ ,\   t_{\text{Mg-B},2} = 0.5\ ,\  t_{\text{Mg-B},3} = 0.9\ ,
}
which matches with the {\abi} calculation quite well within $0.5$eV from the Fermi level.
\eqnref{eq:MgB2_TB_values} is in unit of eV.

However, \eqnref{eq:H_el_MgB2} is too complicated for later analytical study.
We now simplify \eqnref{eq:H_el_MgB2}.
First, in the fitting, we directly fix $t_{\text{B},s-p_z,z} = 0$, since we find that the $t_{\text{B},s-p_z,z}$ mainly affects the bands far ($\sim 2$eV) away from the Fermi energy, which is consistent with orbital projection in \refcite{Jorgensen05012001MgB2Isotope}. 
Indeed, we see that the band structure given by fixing $t_{\text{B},s-p_z,z} = 0$ is good near the Fermi level.
Thus, we can safely neglect $H_{el}^{\text{B}, sp_xp_y - p_z} $ in \eqnref{eq:H_el_MgB2} for the study of the states near the Fermi energy.

Furthermore, according to the orbital projection in \refcite{Lordi05102018MgB2ElectronBands}, the electron orbitals near the Fermi level are mainly B boron orbitals, which mean we should also be able to safely neglect $H_{el}^{\text{Mg}}$ and $H_{el}^{\text{Mg-B}}$ in \eqnref{eq:H_el_MgB2}.
To show this, we consider the following simplified Hamiltonian
\eq{
\label{eq:H_el_MgB2_sim}
H_{el} = H_{el}^{\text{B},sp_xp_y} + H_{el}^{\text{B},p_z}
}
and plot the band structure in \figref{fig:MgB2_el}(b) with 
\eqa{
\label{eq:MgB2_TB_values_sim}
& E_{\text{B},s,0} = -1.68\ ,\ E_{\text{B},p_x p_y, 0} = 3.68\ ,\ E_{p_z} = 0\ ,\ t_1 = -3.5\ ,\ t_2 = 1.022\ ,\ t_3 = -2.66\ ,\ t_4 = 3.8\ ,\ t_{p_z} = -1.7\\
& t_{\text{B},s,z} = -0.085\ ,\ t_{\text{B},p_x p_y,z} = -0.089\ ,\ t_{p_z,z} = 1\ ,
}
which are in the unit of eV.
We find that within $0.5$eV from the Fermi level, the band structure of \eqnref{eq:H_el_MgB2_sim} matches the {\abi} calculation quite well except for the $p_z$ band along L-H.
However, this change only causes a small change of the Fermi surface as shown in \figref{fig:MgB2_el}(c-d), and a small change of the density of states at the Fermi level---$0.37\eV^{-1}$ and $0.38\eV^{-1}$ per unit cell for \eqnref{eq:H_el_MgB2} and \eqnref{eq:H_el_MgB2_sim}, respectively, which are both close to the {\abi} value $0.35\eV^{-1}$ per unit cell.

We note that the values of $E_{B,s}$, $t_1$ and $t_{B,s,z}$ in \eqnref{eq:MgB2_TB_values_sim} might not be quantitatively reliable.
This is because as shown in \refcite{Lordi05102018MgB2ElectronBands}, the electron states of $H_{el}^{\text{B},sp_xp_y}$ near the Fermi level should be dominated by the $p_x$ and $p_y$ orbitals of B atoms, while the bands dominated by the $s$ orbitals of B atoms are at high energies (larger than $>5$eV away from the Fermi level)~\cite{Lordi05102018MgB2ElectronBands}. 
Therefore, we do not need to have precise values of $E_{B,s}$, $t_1$ and $t_{B,s,z}$ in \eqnref{eq:MgB2_TB_values_sim} to describe the physics near the Fermi level.
Nevertheless, we still need to include the $s$ orbitals in the model for the study of topology, since lowest three bands (that involve $s$ and $p_x p_y$) of $H_{el}^{\text{B},sp_xp_y}$ in \eqnref{eq:H_el_MgB2_sim} are connected.
Then, for the study of topology, the values of $E_{B,s}$, $t_1$ and $t_{B,s,z}$ in \eqnref{eq:MgB2_TB_values_sim} are good enough since topology is robust against small changes of the parameters that leave the gap (in \eqnref{fig:MgB2_el_mz_even}(a)) open.

We will always use the simplified model in \eqnref{eq:H_el_MgB2_sim} instead of the full model in \eqnref{eq:H_el_MgB2}.

\subsubsection{Electron Band Geometry and Topology in \mgb: $p_z$}
\label{app:topo_el_MgB2_pz}

\begin{figure}[t]
    \centering
    \includegraphics[width=\columnwidth]{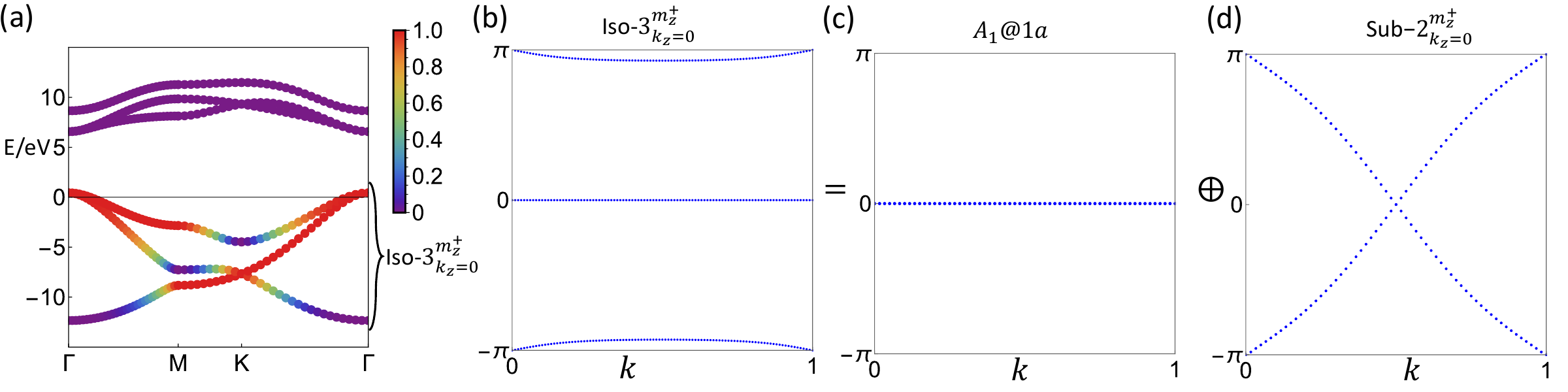}
    \caption{ 
     (a) We plot the bands of $H_{el}^{\text{B},sp_xp_y}$ in \eqnref{eq:H_el_MgB2} at $k_z=0$, which are all in the $m_z$-even subspace.
     The isolated set of lowest three bands in (a) are labelled as $\text{Iso-3}_{k_z=0}^{m_z^+}$, which contains an EBR $(A_{1g}@1a)$ and a rank-2 topological subspace $\text{Sub-2}_{k_z=0}^{m_z^+}$.
     The color indicates the probability of the corresponding eigenstates in the subspace $\text{Sub-2}_{k_z=0}^{m_z^+}$.
     (b), (c) and (d) are the Wilson loop spectrum of $\text{Iso-3}_{k_z=0}^{m_z^+}$, $A_{1g}@1a$ and $\text{Sub-2}_{k_z=0}^{m_z^+}$, respectively.
     The Wilson loop is defined as $W(k) = \bra{u_{k \bsl{b}_1}}\prod_{k_2\in [0,1]} P(k \bsl{b}_1 + k_2 \bsl{b}_2 )\ket{u_{k \bsl{b}_1 + \bsl{b}_2}}$, where $\bsl{b}_{1} = \frac{4\pi}{\sqrt{3}}(\frac{\sqrt{3}}{2},1/2,0)$, $\bsl{b}_{2} = \frac{4\pi}{\sqrt{3}}(\frac{-\sqrt{3}}{2},1/2,0)$, $\ket{u_{ \bsl{k}} }= (\ket{u_{ \bsl{k},1} }, ... )$ is the basis of the $\bsl{k}$-dependent Hilbert space of interest, and $P_{ \bsl{k}}  = \ket{u_{ \bsl{k}} } \bra{u_{ \bsl{k}} }$.
     In (b), (c) and (d), phases of the eigenvalues of $W(k)$ are plotted.
    }
    \label{fig:MgB2_el_mz_even}
\end{figure}

As discussed in \refcite{Liu10192019MgB2Dirac}, the $p_z$ part of the Hamiltonian ($H_{el}^{\text{B},p_z} $ in \eqnref{eq:H_el_MgB2_B_pz}) there are two $P\TR$-protected nodal lines along 
\eq{
\label{eq:nodal_line_pz}
line_{\pm\K} = \left\{ \pm \text{K} + (0,0,k_z)| k_z \in (-\pi/c , \pi/c] \right\}\ .
}
Since $H_{el}^{\text{B},p_z}$ is nothing but the graphene model \eqnref{eq:h_k_graphene} with an extra dispersion along $k_z$, the nodal lines are just the graphene Dirac cones dispersing along $k_z$.
Specifically, by expanding \eqnref{eq:H_el_MgB2_B_pz} around $\pm\K$, the effective Hamiltonian around the nodal lines reads
\eq{
\label{eq:MgB2_pz_nodal_line_h}
h_{p_z}(\pm \K+(p_x,p_y,k_z))= (E_{p_z,0} + t_{p_z,z} 2 \cos(k_z c)) \tau_0 + \frac{-\sqrt{3}t_{p_z}a}{2}\left(\pm p_x\tau_x + p_y \tau_y \right)\ .
}
Therefore, the nodal lines they have $P\TR$-protected winding number $\W_{\L}=1$ if the small loop $\L$ is enclosing $line_{\K}$ or $line_{-\K}$ once, where $\W_{\L}$ is defined in \eqnref{eq:PT_winding_number}.
We define $W_{\text{K}}  =  W_{\L}$ if the small loop $\L$ is enclosing $line_{\K}$ once, and define $W_{-\text{K}}  =  W_{\L}$ if the small loop $\L$ is enclosing $line_{-\K}$ once.
So we have $ |W_{\pm\text{K}}| = 1$.
The nontrivial topology of the nodal lines is detectable on the Fermi surface near $line_{\pm\K}$, because the Fermi lines for fixed $k_z$ would become disconnected close loops that enclose $line_{\K}$ or $line_{-\K}$ when $k_z$ is close to the values where the nodal lines intersect the Fermi level.
More concretely, $E_{p_z,0} + 2  t_{p_z,z} \cos(k_z c) = 0$ in \eqnref{eq:MgB2_pz_nodal_line_h} happens at $k_z c =\pm \pi/2$ based on the parameter values in \eqnref{eq:MgB2_TB_values}.
So the Fermi surface around the $K-H$ and its TR partner should be like Dirac cones at $k_z c =\pm \pi/2$, as shown in \figref{fig:MgB2_el_mz_even}(c-d).
Therefore, the Fermi lines for fixed $k_z$ would become disconnected close loops that enclose $line_{\K}$ or $line_{-\K}$ when $k_z$ is at $k_z c =\pm \pi/2$.
The geometric properties of the electron wavefunctions are constrained by the winding numbers $|W_{\pm\text{K}}|$ similar to the discussion for graphene in \appref{app:graphene}. (See details in \appref{app:lambda_pz_topo_geo}.)
Although the topology is discussed within the model $H_{el}^{\text{B},p_z}$, the nodal lines still exist even if we include other parts of the Hamiltonian, since the other parts of the Hamiltonian do not influence much the states near the nodal lines.
After including other parts of the Hamiltonian, $ W_{\pm\text{K}}$ can still be defined using Wilson lines~\cite{Aris2014Wilsonloop,Ahn2019TBGFragile,Yu2021EOCP}.

\subsubsection{Electron Band Geometry and Topology in \mgb: $sp_xp_y$}
\label{app:topo_el_MgB2_sp2}

Besides $line_{\pm\K}$ from the $p_z$ orbitals (\eqnref{eq:MgB2_pz_nodal_line_h}), we now discuss topological properties of the electron wavefunctions (different from the nodal lines) of the $sp_xp_y$ orbitals.

Let us consider the $k_z = 0 $ plane, on which momenta are invariant under $m_z$.
Then, we can split the states with $k_z = 0$ into two subspaces: $m_z$-even and $m_z$-odd; in our model, the two subspaces contain different numbers of bands.
For the tight-binding model without Mg (\eqnref{eq:H_el_MgB2_sim}), the $m_z$-even subspace at $k_z =0 $ is governed by $H_{el}^{\text{B},sp_xp_y}$, while the $m_z$-odd subspace at $k_z =0 $ is governed by $H_{el}^{\text{B},p_z}$
In the following, we focus on the $m_z$-even subspace at $k_z =0 $, and thus we will use $H_{el}^{\text{B},sp_xp_y}$.
Nevertheless, our topological discussion will be valid even if we include the Mg atoms, since (i) the $m_z$-even subspace at $k_z =0 $ is well-defined after including Mg and (ii) including Mg atoms does not affect the band gap (in \figref{fig:MgB2_el_mz_even}(a)).

As shown in \figref{fig:MgB2_el_mz_even}(a), the lowest three bands of $H_{el}^{\text{B},sp_xp_y}$ in \eqnref{eq:H_el_MgB2_sim} at $k_z=0$ are isolated.
We label the isolated set of three bands as $\text{Iso-3}_{k_z=0}^{m_z^+}$.
The symmetry reps of the isolated set of three bands are
\eq{
\Gamma: \Gamma_1^+\oplus\Gamma_5^+\ ,\ \text{M}: \text{M}_1^+\oplus \text{M}_3^- \oplus \text{M}_4^-\ ,\ \text{K}: \K_1 \oplus \K_5\ ,
}
where the symmetry reps are labelled according to the \emph{Bilbao Crystallographic Server}~\cite{Aroyo2006Bilbao}, $\Gamma_5^+$ and $\K_5$ are 2D irreps while others are 1D irreps, ``Iso" in $\text{Iso-3}_{k_z=0}^{m_z^+}$ means the states are eigenstates of an isolated set of bands, ``$3$" in $\text{Iso-3}_{k_z=0}^{m_z^+}$ means it has three bands, ``$k_z=0$" in $\text{Iso-3}_{k_z=0}^{m_z^+}$ means it is on the $k_z=0$ plane, and ``$m_z^+$" in $\text{Iso-3}_{k_z=0}^{m_z^+}$ means its states are even under $m_z$.
By comparing to the symmetry reps of the elementary band representations (EBR) of $P6/mmm$ in \emph{Bilbao Crystallographic Server}, we can see $\text{Iso-3}_{k_z=0}^{m_z^+}$ has the same symmetry irreps as $k_z = 0$ part of the EBR $A_{1g}$@3f.
We use Wannier90~\cite{wannier90} and numerically verify that there is a smooth gauge with physical reps for $\text{Iso-3}_{k_z=0}^{m_z^+}$, meaning that $\text{Iso-3}_{k_z=0}^{m_z^+}$ is an obstructed atomic set of bands (topologically equivalent to the $k_z = 0$ part of the EBR $A_{1g}$@3f).
The topologically-trivial nature of $\text{Iso-3}_{k_z=0}^{m_z^+}$ is also reflected in the gapped Wilson loop spectrum in \figref{fig:MgB2_el_mz_even}(b).
Nevertheless, the numbers of $\pi$ crossing in \figref{fig:MgB2_el_mz_even}(b) is 1.
Combined with the fact that the berry phases along any close loop are zero for  $\text{Iso-3}_{k_z=0}^{m_z^+}$, $\text{Iso-3}_{k_z=0}^{m_z^+}$ has nonzero $P\TR$-protected second Stiefel-Whitney class $w_2 =1 $.
(See general discussion on the second Stiefel-Whitney class in \refcite{Ahn2018MonopoleNLSM}.)

The nonzero $w_2 = 1$  of $\text{Iso-3}_{k_z=0}^{m_z^+}$ (which is $A_{1g}$@3f) can be understood as the EBR $(A_{1g}\oplus E_{1u})@1a$ having a band inversion at $\Gamma$ with the EBR $(B_{1u}\oplus E_{2g})@1a$.
To show this, we first decompose $\text{Iso-3}_{k_z=0}^{m_z^+}$ into two subspaces.
Specifically, according to \emph{Bilbao Crystallographic Server}, the symmetry reps of $\text{Iso-3}_{k_z=0}^{m_z^+}$ do not prevent the decomposition into well-defined subspaces (\ie, it can be a split EBR), though the bands of $\text{Iso-3}_{k_z=0}^{m_z^+}$ are connected.
Indeed, we find that the obstructed atomic $\text{Iso-3}_{k_z=0}^{m_z^+}$ can be split into a rank-1 obstructed atomic EBR and a rank-2 topological set of bands.
Specifically, we use Wannier90~\cite{wannier90} to find that the rank-1 subspace of $\text{Iso-3}_{k_z=0}^{m_z^+}$ is topologically equivalent to obstructed atomic $A_{1g}@1a$ of irreps $\Gamma_1^+$, $\text{M}_1^+$ and $\K_1$.
So the rank-1 subspace is just the $A_{1g}@1a$ EBR.
The remaining rank-2 subsapce, labelled as $\text{Sub-2}_{k_z=0}^{m_z^+}$ where ``Sub" means subspace, carries the following reps
\eq{
\Gamma:  \Gamma_5^+\ ,\ \text{M}:  M_3^- \oplus M_4^-\ ,\ \text{K}:  \K_5\ ,
}
which indicates its nontrivial topology. 
Since $\text{Iso-3}_{k_z=0}^{m_z^+}$ is an atomic limit, the nontrivial topology of $\text{Sub-2}_{k_z=0}^{m_z^+}$ is fragile~\cite{Po2018FragileTopo,Cano2018DisEBR}..
In short, we have
\eq{
\label{eq:iso-3=A1@1a+sub-2}
\text{Iso-3}_{k_z=0}^{m_z^+} = (A_{1g}@1a) \oplus \text{Sub-2}_{k_z=0}^{m_z^+}\ .
}
We note that the states in $A_{1g}@1a$ or $\text{Sub-2}_{k_z=0}^{m_z^+}$ may not be energy eigenstates, as the bands of $\text{Iso-3}_{k_z=0}^{m_z^+}$ are connected.
As an illustration, we show the the overlap between $\text{Sub-2}_{k_z=0}^{m_z^+}$ with the energy eigenstates in \figref{fig:MgB2_el_mz_even}(a).

The nontrivial topology of the rank-2 set of bands can also be seen from the Wilson loop.
In \figref{fig:MgB2_el_mz_even}(c) and (d), we plot the Wilson loop spectrum of $A_{1g}@1a$ and $\text{Sub-2}_{k_z=0}^{m_z^+}$, respectively.
 \figref{fig:MgB2_el_mz_even}(c) has no $\pi$ crossing, consistent of the trivial nature of $A_{1g}@1a$.
The number of $\pi$ crossing in \figref{fig:MgB2_el_mz_even}(d) is 1.
Combined with the fact that the berry phases along any close loop are zero for $\text{Sub-2}_{k_z=0}^{m_z^+}$, we know that $\text{Sub-2}_{k_z=0}^{m_z^+}$ has nonzero $P\TR$-protected Euler number $\N=1$.
The Euler number $\N=1$ is consistent with the fact that $\text{Sub-2}_{k_z=0}^{m_z^+}$ is fragile topological.

\begin{figure}[t]
    \centering
    \includegraphics[width=0.9\columnwidth]{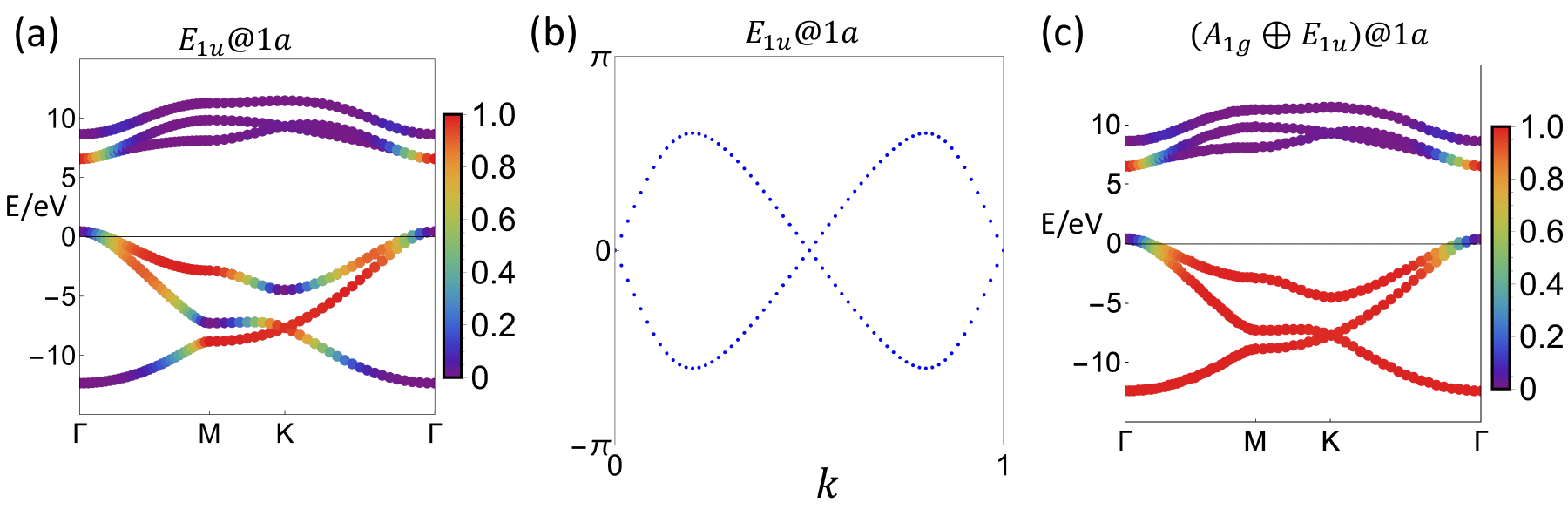}
    \caption{ 
     (a) The colors show the probability of the eigenstates in the $E_{1u}@1a$ space on the bands of $H_{el}^{\text{B},sp_xp_y}$ in \eqnref{eq:H_el_MgB2} at $k_z=0$, which are all in the $m_z$-even subspace.
     (b) is the Wilson loop spectrum of the $E_{1u}@1a$ space.
     Similar to \figref{fig:MgB2_el_mz_even}, $k$ is again along $\bsl{b}_1$.
     (c) The colors show the probability of the eigenstates in the $(A_{1g}\oplus E_{1u})@1A$ space on the bands of $H_{el}^{\text{B},sp_xp_y}$ in \eqnref{eq:H_el_MgB2} at $k_z=0$, which are all in the $m_z$-even subspace.
    }
    \label{fig:MgB2_el_E1u@1a}
\end{figure}

If we replace the $p_x p_y$ states of $\text{Sub-2}_{k_z=0}^{m_z^+}$ near $\Gamma$ by the $p_x p_y$ states around $7$eV near $\Gamma$, we will get an EBR $E_{1u}@1a$, which has zero Euler number, as shown in \figref{fig:MgB2_el_E1u@1a}(a-b).
Thus, replacing the $p_x p_y$ states of $\text{Iso-3}_{k_z=0}^{m_z^+}$ (which is $A_{1g}$@3f) near $\Gamma$ by the $p_x p_y$ states around $7$eV near $\Gamma$ leads to $(A_{1g}\oplus E_{1u})@1a$, as shown in \figref{fig:MgB2_el_E1u@1a}(c).
The complement of $(A_{1g}\oplus E_{1u})@1a$ in the entire $k_z=0$ and $m_z$-even space of $H_{el}^{\text{B},sp_xp_y}$ is $(B_{1u}\oplus E_{2g})@1a$.
Therefore, $\text{Iso-3}_{k_z=0}^{m_z^+}$ (which is $A_{1g}$@3f) can be understood as the band inversion between $(A_{1g}\oplus E_{1u})@1a$ and $(B_{1u}\oplus E_{2g})@1a$.

The band inversion can also be captured by an effective model under the basis of $(A_{1g}\oplus E_{1u})@1a$ and $(B_{1u}\oplus E_{2g})@1a$, which is not the atomic basis in \eqnref{eq:FT_rule}.
Explicitly, we use a $6\times 6$ unitary matrix $R(\bsl{k}_\shpa)$ to label the Bloch vectors $(B_{1u}\oplus E_{2g})@1a$ and $(A_{1g}\oplus E_{1u})@1a$, where the first three columns of $R(\bsl{k}_\shpa)$ correspond to $(B_{1u}\oplus E_{2g})@1a$ and the last three columns of $R(\bsl{k}_\shpa)$ correspond to  $(A_{1g}\oplus E_{1u})@1a$.
Since $(B_{1u}\oplus E_{2g})@1a$ and $(A_{1g}\oplus E_{1u})@1a$ are equivalent to EBRs, $R(\bsl{k})$ is smooth.
Then, we get a new basis $\widetilde{c}^\dagger_{\bsl{k}_\shpa,\text{B}, sp_xp_y}$ by transforming the basis $c^\dagger_{\bsl{k},\text{B}, sp_xp_y}$ at $k_z=0$ of $H_{el}^{\text{B},sp_xp_y}$ in \eqnref{eq:H_el_MgB2_B_sp2}:
\eq{
\widetilde{c}^\dagger_{\bsl{k}_\shpa} = (\widetilde{c}^\dagger_{\bsl{k}_\shpa,1},\widetilde{c}^\dagger_{\bsl{k}_\shpa,2},\widetilde{c}^\dagger_{\bsl{k}_\shpa,3},\widetilde{c}^\dagger_{\bsl{k}_\shpa,4},\widetilde{c}^\dagger_{\bsl{k}_\shpa,5},\widetilde{c}^\dagger_{\bsl{k}_\shpa,6}) =  c^\dagger_{\bsl{k}_\shpa,k_z=0,\text{B}, sp_xp_y} R(\bsl{k}_\shpa)\ ,
}
and the symmetry representation of furnished by $\widetilde{c}^\dagger_{\bsl{k}_\shpa}$ reads
\eqa{
& C_6 \widetilde{c}^\dagger_{\bsl{k}_\shpa} C_6^{-1} = \widetilde{c}^\dagger_{C_6\bsl{k}_\shpa} 
\mat{ 
 -1 & & & \\
 & -e^{-\ii \sigma_y 2 \pi / 6 } & & \\
&  & 1 & \\
& & & e^{-\ii \sigma_y 2 \pi / 6 }
} \ ,\  m_y \widetilde{c}^\dagger_{\bsl{k}_\shpa} m_y^{-1} = \widetilde{c}^\dagger_{m_y\bsl{k}_\shpa} \mat{ -1 & & & \\ 
 & -\sigma_z & & \\
 & & 1 & \\
& & &  \sigma_z
} \ ,\\
& P \widetilde{c}^\dagger_{\bsl{k}_\shpa} P^{-1} = \widetilde{c}^\dagger_{-\bsl{k}_\shpa} \mat{ 
-1 & & & \\
 & \sigma_0 & & \\
 & & 1 & \\
 & & & -\sigma_0 
} \ ,\  \TR \widetilde{c}^\dagger_{\bsl{k}_\shpa} \TR^{-1} = \widetilde{c}^\dagger_{-\bsl{k}_\shpa} \ .
}
We can then transforms the matrix Hamiltonian at $k_z = 0$ to $\widetilde{h}(\bsl{k}_\shpa) = R^\dagger(\bsl{k}_\shpa) h_{sp_xp_y}(\bsl{k}_\shpa,k_z =0) R(\bsl{k}_\shpa)$, where $h_{sp_xp_y}(\bsl{k})$ is defined in \eqnref{eq:H_el_MgB2_B_sp2}.
Since $\widetilde{h}(0) =\diag(8.77,0.51,0.51,-12.35,6.43,6.43)$eV, $\widetilde{c}^\dagger_{\bsl{k}_\shpa=0}$ create eigenstates of $H_{el}^{B,sp_xp_y}$.
Therefore, the band inversion can be seen by expanding $\widetilde{h}(\bsl{k}_\shpa)$ to the second order in $\bsl{k}_\shpa$ and project out $\widetilde{c}^\dagger_{\bsl{k}_\shpa,1}$ and $\widetilde{c}^\dagger_{\bsl{k}_\shpa,4}$ via second-order perturbation theory, leading to a $4\times 4$ effective matrix Hamiltonian that reads
\eqa{
\widetilde{h}_{4\times 4}(\bsl{k}_\shpa) &  = \left(
\begin{array}{cc}
 (\epsilon_{++}+b_{++} (\bsl{k}_\shpa a)^2 ) \sigma_0 & \ii c_{+-} (k_y a \sigma_z + k_x a \sigma_x) \\
  -\ii c_{+-} (k_y a \sigma_z + k_x a \sigma_x) & ( \epsilon_{--}+b_{--} (\bsl{k}_\shpa a)^2 ) \sigma_0
\end{array}
\right) \\
& \qquad + \left(
\begin{array}{cc}
  c_{++} \left[ ((k_xa)^2-(k_ya)^2)\sigma_z +2 (k_xa k_ya)\sigma_x \right] & \\
   & c_{--} \left[((k_xa)^2-(k_ya)^2)\sigma_z +2 (k_xa k_ya)\sigma_x \right]
\end{array}
\right) 
}
with $\epsilon_{++} = 0.52$eV, $\epsilon_{--} = 6.43$eV, $b_{++} = 6.06$eV, $b_{--} =-6.04$eV, $c_{+-}=-6.69$eV, $c_{++} = 0.04$eV and $c_{--} =-0.03$eV.
Clearly, $\epsilon_{--}-\epsilon_{++}$ has opposite signs as $b_{--}-b_{++}$ while $c_{++}$ and $c_{--}$ are negligible, indicating the band inversion, and the lower 2 bands of $\widetilde{h}_{4\times 4}(\bsl{k}_\shpa)$ have Euler number $\Delta \N = 1$, which we call the effective Euler number.

The discussion on the effective Euler number relies on the transformed basis $\widetilde{c}^\dagger_{\bsl{k}_\shpa}$ which is not the atomic basis in \eqnref{eq:FT_rule} (or the linear combination of them with $\bsl{k}$-independent coefficients).
However, the discussion on EPC uses the atomic basis, and thus it is better to connect the effective Euler number to the quantities in the atomic basis (or the linear combination of them with $\bsl{k}$-independent coefficients).
To do so, we first note that the effective Euler number must equal to the difference between the Euler numbers of $\text{Sub-2}_{k_z=0}^{m_z^+}$ and the $E_{1u}@1a$, since (i) the two states only differ from each other around $\Gamma$ point according to \figref{fig:MgB2_el_mz_even}(a) and \figref{fig:MgB2_el_E1u@1a}(a), and (ii) the vector representations of $E_{1u}@1a$ have zero Euler number in both the atomic basis $c^\dagger_{\bsl{k}_\shpa,k_z =0 ,\text{B}, sp_xp_y}$ and the non-atomic transformed basis $\widetilde{c}^\dagger_{\bsl{k}_\shpa}$.
Therefore, we have 
\eq{
\label{eq:Euler_Difference}
\Delta\N = \left|\N - \N_{E_{1u}@1a}\right| = 1\ ,
}
where $\N - \N_{E_{1u}@1a}$ is calculated in the atomic basis.

$\N - \N_{E_{1u}@1a}$ can be understood from an effective model with the basis being the linear combination of the atomic basis with $\bsl{k}$-independent coefficients.
Explicitly, we construct an effective model around $\Gamma-$A for only the $p_x/p_y$ orbitals on B atoms, since the states near Fermi level around $\Gamma$ are mainly from the $p_x/p_y$ orbitals on B atoms according to \refcite{Lordi05102018MgB2ElectronBands}. 
We choose the basis to be parity eigenstates:
\eq{
\label{eq:MgB2_eff_basis}
c^\dagger_{\bsl{k},eff} = \frac{1}{\sqrt{2}} ( c^\dagger_{\bsl{k},\text{B1},p_x}-c^\dagger_{\bsl{k},\text{B2},p_x} , c^\dagger_{\bsl{k},\text{B1},p_y}-c^\dagger_{\bsl{k},\text{B2},p_y} , c^\dagger_{\bsl{k},\text{B1},p_x}+c^\dagger_{\bsl{k},\text{B2},p_x} ,  c^\dagger_{\bsl{k},\text{B1},p_y}+c^\dagger_{\bsl{k},\text{B2},p_y} )\ .
}
The symmetry reps in the basis \eqnref{eq:MgB2_eff_basis} read
\eqa{
\label{eq:MgB2_eff_sym_rep}
& C_6 c^\dagger_{\bsl{k},eff} C_6^{-1} = c^\dagger_{C_6\bsl{k},eff} 
\mat{ 
- e^{-\ii \sigma_y 2 \pi / 6 } & \\
 & e^{-\ii \sigma_y 2 \pi / 6 } \\
} \\
& m_y c^\dagger_{\bsl{k},eff} m_y^{-1} = c^\dagger_{m_y\bsl{k},eff} \mat{ 
- \sigma_z & \\
 & \sigma_z
} \\
& P c^\dagger_{\bsl{k},eff} P^{-1} = c^\dagger_{-\bsl{k},eff} \mat{ 
\sigma_0 & \\
 & -\sigma_0 
} \\
& \TR c^\dagger_{\bsl{k},eff} \TR^{-1} = c^\dagger_{-\bsl{k},eff} \ ,
}
where we replace $m_z$ by the inversion symmetry $P= C_6^3 m_z$.
We project the $s p_x p_y$ Hamiltonian \eqnref{eq:H_el_MgB2_sim} to $c^\dagger_{\bsl{k},eff} $, and expand the resultant projected model to the first order in $\bsl{k}_\shpa$, resulting in the following effective matrix Hamiltonian:
\eq{
\label{eq:MgB2_H_eff_mat_form}
h_{eff}(\bsl{k}) = \epsilon_0(k_z) + m \tau_z\sigma_0 + d_x(\bsl{k}_\shpa) \tau_y \sigma_x + d_y(\bsl{k}_\shpa) \tau_y \sigma_z \ ,
} 
where 
\eq{
\label{eq:d_form_MgB2_eff}
d_x(\bsl{k}_\shpa) = v k_x a \ ,\ d_y(\bsl{k}_\shpa) = v k_y a \ ,
}
\eqa{
\label{eq:MgB2_H_eff_para}
& E_+ = E_{B,p_x p_y} - 3 t_2 \ ,\ E_- = E_{B,p_x p_y} + 3 t_2 \ ,\ v = - \frac{\sqrt{3}}{2}  t_3 \\
& \epsilon_0(k_z) = (E_+ + E_-)/2+ 2  t_{\text{B},p_x p_y,z} \cos( k_z c )\ ,\ m = (E_+ - E_-)/2\ .
}
The eigenvalues of \eqnref{eq:MgB2_H_eff_mat_form} read 
\eq{
\label{eq:E_eff_MgB2}
E_{eff,n}(\bsl{k}) = \epsilon_0(k_z) + (-1)^n  \sqrt{m^2 + |\bsl{d}(\bsl{k}_\shpa)|^2}\ ,
}
where $|\bsl{d}(\bsl{k}_\shpa)|^2 = d_x^2(\bsl{k}_\shpa) + d_y^2(\bsl{k}_\shpa) $, and $E_{eff,n}(\bsl{k})$ is doubly degenerate.
The projection matrices for $E_{eff,n}(\bsl{k})$ read
\eq{
\label{eq:P_eff_MgB2}
P_{eff,n}(\bsl{k}_\shpa) = \frac{1}{2} +(-1)^n \frac{ m \tau_z\sigma_0 + d_x(\bsl{k}_\shpa) \tau_y \sigma_x + d_y(\bsl{k}_\shpa) \tau_y \sigma_z}{2 \sqrt{m^2 + |\bsl{d}(\bsl{k}_\shpa)|^2}} \ .
}
With the parameter values in \eqnref{eq:MgB2_TB_values_sim}, $E_{eff,2}(\bsl{k})$ at $k_z=0$ corresponds to the second and third bands counted from the top around $\Gamma$ in \figref{fig:MgB2_el_mz_even}(a), and $E_{eff,1}(\bsl{k})$ at $k_z=0$ corresponds to the the two bands that are closest to the Fermi level around $\Gamma$ in \figref{fig:MgB2_el_mz_even}(a).
Note that the double degenerate $E_{eff,\pm}(\bsl{k})$ cannot capture the splitting of the corresponding bands in \figref{fig:MgB2_el_mz_even}(a), since the splitting happens at $\bsl{k}_{\shpa}^2$ order.
Nevertheless, $O(\bsl{k}_{\shpa})$ order is good enough for our study for $\N - \N_{E_{1u}@1a}$ as shown in the following.

The two eigenvectors for $E_{eff,1}(\bsl{k}_\shpa)$ should well capture the states that are closest to the Fermi level around $\Gamma$-A in \figref{fig:MgB2_el_mz_even}(a).
Explicitly, we can choose the eigenvectors for $E_{eff,1}(\bsl{k})$ to have the following form
\eqa{
\label{eq:MgB2_H_eff_eigenvecs}
& U_{1,1}(\bsl{k}_\shpa)=\frac{1}{\sqrt{2}}\left(\frac{d_x(\bsl{k}_\shpa)}{|\bsl{d}(\bsl{k}_\shpa)|} \sqrt{1-\frac{m}{\sqrt{|\bsl{d}(\bsl{k}_\shpa)|^2+m^2}}},-\frac{d_y(\bsl{k}_\shpa)}{|\bsl{d}(\bsl{k}_\shpa)|} \sqrt{1-\frac{m}{\sqrt{|\bsl{d}(\bsl{k}_\shpa)|^2+m^2}}},0,-\ii \sqrt{\frac{m}{\sqrt{|\bsl{d}(\bsl{k}_\shpa)|^2+m^2}}+1}\right)^T\\ 
& U_{1,2}(\bsl{k}_\shpa)=\frac{1}{\sqrt{2}}\left(\frac{d_y(\bsl{k}_\shpa)}{|\bsl{d}(\bsl{k}_\shpa)|} \sqrt{1-\frac{m}{\sqrt{|\bsl{d}(\bsl{k}_\shpa)|^2+m^2}}},\frac{d_x(\bsl{k}_\shpa)}{|\bsl{d}(\bsl{k}_\shpa)|} \sqrt{1-\frac{m}{\sqrt{|\bsl{d}(\bsl{k}_\shpa)|^2+m^2}}},-\ii \sqrt{\frac{m}{\sqrt{|\bsl{d}(\bsl{k}_\shpa)|^2+m^2}}+1},0\right)^T \ .
}
Clearly, the eigenvectors for $E_{eff,1}(\bsl{k})$ around $\Gamma$-A ($U_{1,1}(\bsl{k}_\shpa)$ and $U_{1,2}(\bsl{k}_\shpa)$) do not depend on the momentum along $k_z$ to first order in $\bsl{k}_\shpa$.

With all of these preparation, we now discuss how to understand $\left|\N - \N_{E_{1u}@1a}\right|$ in \eqnref{eq:Euler_Difference} from the effective model  $h_{eff}(\bsl{k})$ in \eqnref{eq:MgB2_H_eff_mat_form}.
According to the form of $\bsl{d}$ (\eqnref{eq:d_form_MgB2_eff}) in $h_{eff}(\bsl{k})$, $\bsl{d}(\bsl{k}_\shpa)$ must have nonzero winding number along any loop $\L$ around $\Gamma$ unless fine tuned, \ie,
\eq{
\label{eq:d_winding_MgB2_eff}
 \mathcal{W}_d = \left| \frac{1}{2\pi} \int_{\L} d\bsl{k}_\shpa \cdot  \nabla_{\bsl{k}_\shpa} \theta_{d}(\bsl{k}_\shpa) \right| = 1\ ,
}
where 
\eq{
\label{eq:d_radial_theta}
\bsl{d}(\bsl{k}_\shpa) = |\bsl{d}(\bsl{k}_\shpa)| (\cos(\theta_{d}(\bsl{k}_\shpa)), \sin(\theta_{d}(\bsl{k}_\shpa))) \ .
}

The nonzero winding number of $\bsl{d}(\bsl{k}_\shpa)$ in \eqnref{eq:d_form_MgB2_eff} must equal to the difference between the Euler numbers of $\text{Sub-2}_{k_z=0}^{m_z^+}$ and $E_{1u}@1a$, as discussed in the following.
According to \eqnref{eq:MgB2_TB_values_sim}, we have $m<0$.
We first show that if the mass $m$ in \eqnref{eq:MgB2_H_eff_mat_form} hypothetically changes sign, the Euler number of $\text{Sub-2}_{k_z=0}^{m_z^+}$ in the space of the lowest three bands of $H^{B,sp_xp_y}_{el}$ would change by $\mathcal{W}_d$ in \eqnref{eq:d_winding_MgB2_eff}, which is $1$.
According to \figref{fig:MgB2_el_mz_even}(a), the states of $\text{Sub-2}_{k_z=0}^{m_z^+}$ close to $\Gamma$ point near Fermi level should be well-captured by the eigenvectors for $E_{eff,1}(\bsl{k})$ in \eqnref{eq:MgB2_H_eff_eigenvecs}.
Based on symmetry reps \eqnref{eq:MgB2_eff_sym_rep}, the sign flipping of the mass $m$ in \eqnref{eq:MgB2_H_eff_mat_form} would change the inversion parities of two states of $\text{Sub-2}_{k_z=0}^{m_z^+}$ at $\Gamma$, which means the sign flipping must change the Euler number of $\text{Sub-2}_{k_z=0}^{m_z^+}$ by an odd number~\cite{Ahn2019TBGFragile}.
To show that this odd number is precisely $\mathcal{W}_d$, let us consider the effective model  $h_{eff}(\bsl{k})$ in \eqnref{eq:MgB2_H_eff_mat_form} at $k_z=0$ in the space of $(m,k_x,k_y)$.
We know $h_{eff}(\bsl{k}_\shpa,k_z = 0)$ is zero at $(m,k_x,k_y) = 0$.
Let us consider a cylindrical surface 
\eq{
\mathcal{M}=\{ (m,k_x,k_y) | m\in(-\infty,\infty), |\bsl{k}_\shpa| =\epsilon >0  \}
}
that encloses $(m,k_x,k_y) = 0$, and the Euler number for $U_{1,1}(\bsl{k}_\shpa)$ and $U_{1,2}(\bsl{k}_\shpa)$ on $\mathcal{M}$ is just the change of the Euler number induced by the sign flipping of $m$.
$U_{1,1}(\bsl{k}_\shpa)$ and $U_{1,2}(\bsl{k}_\shpa)$ in \eqnref{eq:MgB2_H_eff_eigenvecs} are invariant under the $P\TR$ symmetry and are smooth on $\mathcal{M}$ except for $m\rightarrow -\infty$.
Then, by defining $\bsl{k}_\shpa = |\bsl{k}_\shpa| (\cos(\theta),\sin (\theta))$, the Euler number for $U_{1,1}(\bsl{k}_\shpa)$ and $U_{1,2}(\bsl{k}_\shpa)$ on $\mathcal{M}$ reads
\eqa{
\label{eq:N_change_W_d_MgB2}
\N_{\mathcal{M}} & = \left| \frac{1}{2\pi}\int_{-\infty}^{\infty} d m \int_{0}^{2\pi}  d\theta \left[ \partial_m U_{1,1}^\dagger (\bsl{k}_\shpa) \partial_{\theta} U_{1,2} (\bsl{k}_\shpa) -\partial_{\theta} U_{1,1}^\dagger (\bsl{k}_\shpa)  \partial_mU_{-,2} (\bsl{k}_\shpa) \right] \right| \\
&  =  \left|\frac{1}{2\pi}\int_{-\infty}^{\infty} d m \int_{0}^{2\pi}  d\theta \frac{d_y(\bsl{k}_\shpa ) \partial_{\theta} d_x(\bsl{k}_\shpa )-d_x(\bsl{k}_\shpa ) \partial_{\theta}  d_y(\bsl{k}_\shpa )}{2 \left(d_x(\bsl{k}_\shpa )^2+d_y(\bsl{k}_\shpa )^2+m^2\right)^{3/2}} \right|\\
&  =   \left| \frac{1}{2\pi} \int_{0}^{2\pi}  d\theta \frac{d_y(\bsl{k}_\shpa ) \partial_{\theta} d_x(\bsl{k}_\shpa )-d_x(\bsl{k}_\shpa ) \partial_{\theta}  d_y(\bsl{k}_\shpa )}{ d_x(\bsl{k}_\shpa )^2+d_y(\bsl{k}_\shpa )^2 } \right|\\
&  = \left| \frac{1}{2\pi} \int_{|\bsl{k}_\shpa|=\epsilon}  d\bsl{k}_\shpa \cdot  \nabla_{\bsl{k}_\shpa} \theta_d(\bsl{k}_\shpa) \right| = \mathcal{W}_d \ .
}
where we used \eqnref{eq:d_radial_theta}.
Combined with the fact that the hypothetical sign flipping of $m$ would turn $\text{Sub-2}_{k_z=0}^{m_z^+}$ in the space of the lowest three bands of $H^{B,sp_xp_y}_{el}$ into $E_{1u}@1a$, we arrive at 
\eq{
\label{eq:W_d_DeltaN}
\mathcal{W}_d = \Delta\N = \left|\N - \N_{E_{1u}@1a}\right|\ .
}
We emphasize that the validity of \eqnref{eq:W_d_DeltaN} relies on the existence of band inversion.
If the band is not inverted (\ie, $m>0$), $\text{Sub-2}_{k_z=0}^{m_z^+}$ is just $E_{1u}@1a$, and we have $\Delta\N = \left|\N - \N_{E_{1u}@1a}\right|=0$, while $\mathcal{W}_d = 1$.
In this case,  we would have $\mathcal{W}_d > \Delta\N = \left|\N - \N_{E_{1u}@1a}\right|$.
Therefore, for both $m>0$ and $m<0$, we always have $\mathcal{W}_d \geq \Delta\N = \left|\N - \N_{E_{1u}@1a}\right|$.

In the study of EPC, we will use the effective model $h_{eff}(\bsl{k})$ in \eqnref{eq:MgB2_H_eff_mat_form}.
Since we care about the winding number of $\bsl{d}(\bsl{k}_\shpa)$, we will always keep the label $\bsl{d}(\bsl{k}_\shpa)$ instead of directly using its explicit form \eqnref{eq:d_form_MgB2_eff} in order to keep track of the winding number of $\bsl{d}(\bsl{k}_\shpa)$, unless specified otherwise.
Specifically,  unless specified otherwise, we keep $\bsl{d}(\bsl{k}_\shpa)$ and only use two properties of $\bsl{d}(\bsl{k}_\shpa)$: (i) $\bsl{d}(\bsl{k}_\shpa)$ is linear in $\bsl{k}_\shpa$ and (ii) $\bsl{d}(\bsl{k}_\shpa)$ has winding number being 1 as shown in \eqnref{eq:d_winding_MgB2_eff}.
We will use \eqnref{eq:N_change_W_d_MgB2} to derive the topological contribution to the EPC constant in \appref{app:lambda_sp2_topo_geo}.

From \eqnref{eq:MgB2_H_eff_para} for the effective model in \eqnref{eq:MgB2_H_eff_mat_form}, we can see the momentum dependence of the Hamiltonian near $\Gamma$-A for $p_xp_y$ only comes from two hopping terms: $t_3$ and $t_{\text{B},p_x p_y,z}$ in \eqnref{eq:H_el_MgB2_sim}.
In particular, only $t_3$ accounts for the geometric properties of the Bloch states in \eqnref{eq:MgB2_H_eff_mat_form} to first order in $|\bsl{k}_\shpa a|$.
To understand this, we recall that as shown in \eqnref{eq:hopping_el_MgB2_B_sp2}, there are only four parameters that are completely in the $p_x p_y$ subspace: namely the onsite $E_{\text{B},p_x p_y, 0}$, the $\bsl{a}_3$ hopping $t_{\text{B},p_x p_y,z}$, and two hoppings $t_2$ and $t_3$ among  $p_x p_y$ orbitals on NN hopping.
In the Bloch Hamiltonian, any hopping term is the symmetry-invariant combination of the momentum functions and the matrices; nevertheless, the momentum functions of a hopping term can form a high-dimensional irrep of the symmetry group.
The momentum functions for $E_{\text{B},p_x p_y, 0}$, $t_{\text{B},p_x p_y,z}$ and $t_{\text{B},p_x p_y,z}$ and  $t_2$ are all $C_3$-invariant, and they cannot have any $O(|\bsl{k}_\shpa a|)$ term.
Specifically, the momentum function for the on-site term $E_{\text{B},p_x p_y, 0}$ is constant, and the momentum function for $t_{\text{B},p_x p_y,z}$ only depends on $k_z$.
Moreover, since $t_2$ is equal hopping for $p_x$ and $p_y$, its matrix is invariant under $C_3$, and thus its momentum function is $C_3$-invariant.
Therefore, only $t_3$, which is the difference between the hopping for $p_x$ and $p_y$ along $y$, can appear in the off-diagonal part of \eqnref{eq:MgB2_H_eff_mat_form} at first order in $|\bsl{k}_\shpa a|$. 
This fact is very important for the discussion of the EPC in \appref{app:lambda_sp2_topo_geo}.

\subsection{EPC Hamiltonian of {\mgb}}

In this part, we discuss the form of the EPC Hamiltonian.

As mentioned in \eqnref{eq:H_el_MgB2_sim}, neglecting Mg atoms has little effect on the electron states near the Fermi level.
For the EPC Hamiltonian, we adopt the same approximation---we neglect the Mg components in the electron basis for the EPC Hamiltonian (\eqnref{eq:H_el-ph_2center_R}).
Under this approximation, the two-center approximation \eqnref{eq:2center} shows that the motions of Mg atoms do not affect the electron states, \ie, $\bsl{\tau}_1,\bsl{\tau}_2$ in $f_{\bsl{\tau}_1\bsl{\tau}_2,i}( \bsl{R}_1 + \bsl{\tau}_1 - \bsl{R}_2 - \bsl{\tau}_2 )$ only ranges over the two B atoms in one unit cell.
The approximation is supported by the experimental observation that the isotope effect of Mg atoms is much smaller than that of the B atoms~\cite{Jorgensen05012001MgB2Isotope}.
Another reason for our approximation is two-fold: (i) the key phonon modes for the EPC consist of the $E_{2}$ phonons along $\Gamma$-A (enhanced to $E_{2g}$ at $\Gamma$ and A) and the phonons near $\Gamma$-A that are similar to $E_{2}$ phonons~\cite{Kong02272001MgB2EPC}, and (ii) the $E_{2}$ phonons do not involve Mg atoms at all---$E_{2}$ solely comes from the in-plane motion of the B atoms.

Besides neglecting Mg atoms, since the dominant phonons for EPC are around $\Gamma$-A (mainly $E_{2}$ phonons along $\Gamma$-A), those dominant phonons have small in-plane momenta.
(We note that the K-phonons which are specially important in graphene are not special here anymore since the Fermi surfaces of $p_z$ are not always close to $\pm\K$.)
However, the $p_z$ Fermi surfaces (around $\K$-H and its TR-partner) have large in-plane momentum difference from the $p_xp_y$ Fermi surface (around $\Gamma$-A), as shown in \figref{fig:MgB2_el}(b).
Therefore, the cross contribution to $\lambda$ that involve both $p_xp_y$ and $p_z$ should be negligible.
In other words, we should be able to use a EPC Hamiltonian that has no coupling between $sp_xp_y$ and $p_z$.
In the following, we will construct such an EPC Hamiltonain.

As discussed in \appref{app:H_el_MgB2}, we only consider the NN terms for the in-plane (in $x-y$ plane) hopping and the terms along $\pm\bsl{a}_3$ for the out-of-plane (along $z$) hopping.
We adopt the same approximation for the EPC Hamiltonian, \ie, $f_{\bsl{\tau}_1\bsl{\tau}_2,i}( \Delta{\bsl{r}} )$ is allowed to be nonzero only if (i) $\left(\Delta{\bsl{r}} \in \{ \pm \bsl{\delta}_j | j=0,1,2 \}\ \&\ \bsl{\tau}_1\neq \bsl{\tau}_2 \right)$ or (ii) $\left( \Delta{\bsl{r}} \in \{ \pm \bsl{a}_3 \}\ \&\ \bsl{\tau}_1 = \bsl{\tau}_2 \right)$, where $\bsl{\delta}_j$ is defined in \eqnref{eq:delta_j_graphene}.

For the in-plane ion motions, we only need to  consider their effect on the in-plane hoppings, as discussed in the following.
First, the $\bsl{a}_3$-hopping among $p_x/p_y$ orbitals is very small due to the small dispersion along $\Gamma$-A near the Fermi level in \figref{fig:MgB2_el}(a), and the $\bsl{a}_3$-hopping among $p_z$ orbitals does not couple to the in-plane motions owing to $C_3$ symmetry, where $\bsl{a}_3$ is shown in \figref{fig:MgB2_structure}.
Second, the $\bsl{a}_3$-hopping that involves $s$ orbitals does not affect the electron states around the Fermi level.
Finally, although $\bsl{a}_3$-hopping between the $p_x/p_y$ orbitals and $p_z$ orbital is restricted to zero by $C_3$ symmetry, the in-plane ion motions can still change the hopping away from zero.
However, the hopping change should be small for the combinations of $p_x/p_y$ orbitals near the Fermi level, since the combinations should be strongly localized within each B layer.
Therefore, we only consider the effect of the in-plane motions of the B atoms on the in-plane hopping, which should take into account the main contribution to the EPC strength.

We further include the effect of the out-of-plane motions on the $\bsl{a}_3$-hopping, since the out-of-plane motions can change the considerable $\bsl{a}_3$-hopping among the $p_z$ orbital.
Such consideration is just for completeness, since the out-of-plane motions only have sub-leading contributions to the EPC strength.
We will not consider the effect of the out-of-plane motions on the in-plane hopping term, since we have already considered the effect of the in-plane motions on the in-plane hopping term, which is the major effect.
As a result, we have the following approximation 
\eqa{
& f_{\bsl{\tau}_1\bsl{\tau}_2,x}( \pm \bsl{a}_3 ) = f_{\bsl{\tau}_1\bsl{\tau}_2,y}( \pm \bsl{a}_3 ) = 0 \\
& f_{\bsl{\tau}\bsl{\tau},z}( \pm \bsl{\delta}_j ) = 0 \ ,
}
leaving only $f_{\bsl{\tau}_1\bsl{\tau}_2,x}( \pm \bsl{\delta}_j )$, $f_{\bsl{\tau}_1\bsl{\tau}_2,y}( \pm \bsl{\delta}_j )$ and $f_{\bsl{\tau}\bsl{\tau},z}( \pm \bsl{a}_3)$ allowed to be nonzero.

In the following, we will transform the EPC $f_i$ to the momentum space.
Let us consider $f_{\bsl{\tau}_1\bsl{\tau}_2,i=x,y}( \bsl{R}_1 + \bsl{\tau}_1 - \bsl{R}_2 - \bsl{\tau}_2 )$ first.
Since $f_{\bsl{\tau}_1\bsl{\tau}_2,i=x,y}( \bsl{R}_1 + \bsl{\tau}_1 - \bsl{R}_2 - \bsl{\tau}_2 ) = 0$ if $(\bsl{R}_1 + \bsl{\tau}_1 - \bsl{R}_2 - \bsl{\tau}_2)_z = 0$, $f_{\bsl{\tau}_1\bsl{\tau}_2,i=x,y}( \bsl{R}_1 + \bsl{\tau}_1 - \bsl{R}_2 - \bsl{\tau}_2 )$ is effectively 2D, and thus we can use the methodology in \appref{app:geo_EPC_symmetry-rep} to rewrite the form of the $f_{\bsl{\tau}_1\bsl{\tau}_2,i=x,y}( \bsl{R}_1 + \bsl{\tau}_1 - \bsl{R}_2 - \bsl{\tau}_2 )$.
Specifically, we can replace $\Delta\bsl{R}$ in \eqnref{eq:g_2D_partial_k} by $\Delta \bsl{R}_\shpa \in \bsl{a}_1 \dsZ +\bsl{a}_2 \dsZ$, and obtain $\forall i=x,y$, 
\eqa{
  & f_{\bsl{\tau}_1 \bsl{\tau}_2,i}(\bsl{k}_\shpa)  \\
  & = \sum_{\Delta\bsl{R}_\shpa }^{\Delta\bsl{R}_\shpa + \bsl{\tau}_1 - \bsl{\tau}_2  \neq 0} e^{- \ii \bsl{k}_\shpa \cdot (\Delta\bsl{R}_\shpa+ \bsl{\tau}_1 - \bsl{\tau}_2)}   f_{\bsl{\tau}_1 \bsl{\tau}_2,i}(\Delta\bsl{R}_\shpa + \bsl{\tau}_1 - \bsl{\tau}_2 )\\
  & = \sum_{\Delta\bsl{R}_\shpa }^{\Delta\bsl{R}_\shpa + \bsl{\tau}_1 - \bsl{\tau}_2  \neq 0} e^{- \ii \bsl{k}_\shpa \cdot (\Delta\bsl{R}_\shpa + \bsl{\tau}_1 - \bsl{\tau}_2)}  \sum_{i'=x,y}   f_{\bsl{\tau}_1 \bsl{\tau}_2, i'}(\Delta\bsl{R}_\shpa + \bsl{\tau}_1 - \bsl{\tau}_2 )   \delta_{ii'}  \\
  & = \sum_{\Delta\bsl{R}_\shpa }^{\Delta\bsl{R}_\shpa + \bsl{\tau}_1 - \bsl{\tau}_2  \neq 0} e^{- \ii \bsl{k}_\shpa \cdot (\Delta\bsl{R}_\shpa + \bsl{\tau}_1 - \bsl{\tau}_2)}  \sum_{i'=x,y}   f_{\bsl{\tau}_1 \bsl{\tau}_2,i'}(\Delta\bsl{R}_\shpa + \bsl{\tau}_1 - \bsl{\tau}_2 )  \\
  & \quad \times  \left[ \left(\frac{\Delta\bsl{R}_\shpa + \bsl{\tau}_1  - \bsl{\tau}_2}{|\Delta\bsl{R}_\shpa + \bsl{\tau}_1  - \bsl{\tau}_2|_{\shpa}}\right)_i \left(\frac{\Delta\bsl{R}_\shpa + \bsl{\tau}_1  - \bsl{\tau}_2}{|\Delta\bsl{R}_\shpa + \bsl{\tau}_1  - \bsl{\tau}_2|_{\shpa}}\right)_{i'}  + \left(\frac{\bsl{e}_z\times (\Delta\bsl{R}_\shpa + \bsl{\tau}_1  - \bsl{\tau}_2)}{|\Delta\bsl{R}_\shpa + \bsl{\tau}_1  - \bsl{\tau}_2|}\right)_i \left(\frac{\bsl{e}_z\times (\Delta\bsl{R}_\shpa + \bsl{\tau}_1  - \bsl{\tau}_2)}{|\Delta\bsl{R}_\shpa + \bsl{\tau}_1  - \bsl{\tau}_2|}\right)_{i'} \right] \\
& = \sum_{\Delta\bsl{R}_\shpa }^{\Delta\bsl{R}_\shpa + \bsl{\tau}_1 - \bsl{\tau}_2  \neq 0} e^{- \ii \bsl{k}_\shpa \cdot (\Delta\bsl{R}_\shpa + \bsl{\tau}_1 - \bsl{\tau}_2)}   \left(  \widetilde{f}_{\bsl{\tau}_1\bsl{\tau}_2,\shpa}(\Delta\bsl{R}_\shpa + \bsl{\tau}_1 - \bsl{\tau}_2 )   \left[ \Delta\bsl{R}_\shpa + \bsl{\tau}_1  - \bsl{\tau}_2 \right]_i \right. \\
& \quad +\left.    \widetilde{f}_{\bsl{\tau}_1 \bsl{\tau}_2,\perp}(\Delta\bsl{R}_\shpa+ \bsl{\tau}_1 - \bsl{\tau}_2 )   \left[\bsl{e}_z\times (\Delta\bsl{R}_\shpa + \bsl{\tau}_1  - \bsl{\tau}_2)\right]_i \right) \\
& = \sum_{\Delta\bsl{R}_\shpa }  e^{- \ii \bsl{k}_\shpa \cdot (\Delta\bsl{R}_\shpa + \bsl{\tau}_1 - \bsl{\tau}_2)}   \left(   \widetilde{f}_{\bsl{\tau}_1\bsl{\tau}_2,\shpa}(\Delta\bsl{R}_\shpa + \bsl{\tau}_1 - \bsl{\tau}_2 )   \left[ \Delta\bsl{R}_\shpa + \bsl{\tau}_1  - \bsl{\tau}_2 \right]_i \right. \\
& \quad +\left.    \widetilde{f}_{\bsl{\tau}_1 \bsl{\tau}_2,\perp}(\Delta\bsl{R}_\shpa + \bsl{\tau}_1 - \bsl{\tau}_2 )   \left[\bsl{e}_z\times (\Delta\bsl{R}_\shpa + \bsl{\tau}_1  - \bsl{\tau}_2)\right]_i \right)
  \ ,
}
where we have used \eqnref{eq:f_zero_onsite}, $\bsl{k}_{\shpa} = (k_x,k_y,0)$, and
\eqa{
\label{eq:g_beta_R_MgB2}
&    \widetilde{f}_{\bsl{\tau}_1\bsl{\tau}_2,\shpa}(\Delta\bsl{R}_\shpa + \bsl{\tau}_1 - \bsl{\tau}_2\neq 0 )  = 
\sum_{i'=x,y}   f_{\bsl{\tau}_1 \bsl{\tau}_2, i'}(\Delta\bsl{R}_\shpa + \bsl{\tau}_1 - \bsl{\tau}_2 )  \left(\frac{\Delta\bsl{R}_\shpa + \bsl{\tau}_1  - \bsl{\tau}_2}{|\Delta\bsl{R}_\shpa + \bsl{\tau}_1  - \bsl{\tau}_2|^2}\right)_{i'}  \\
&    \widetilde{f}_{\bsl{\tau}_1 \bsl{\tau}_2,\perp}(\Delta\bsl{R}_\shpa + \bsl{\tau}_1 - \bsl{\tau}_2 \neq 0)   = 
 \sum_{i'=x,y} f_{\bsl{\tau}_1 \bsl{\tau}_2,i'}(\Delta\bsl{R}_\shpa + \bsl{\tau}_1 - \bsl{\tau}_2 ) \left(\frac{\bsl{e}_z\times (\Delta\bsl{R}_\shpa + \bsl{\tau}_1  - \bsl{\tau}_2)}{|\Delta\bsl{R}_\shpa + \bsl{\tau}_1  - \bsl{\tau}_2|^2}\right)_{i'} \\
 & \widetilde{f}_{\bsl{\tau} \bsl{\tau},\shpa}(0)   =  \widetilde{f}_{\bsl{\tau} \bsl{\tau},\perp}(0)   = 0 \ .
}
By further defining 
\eq{
\label{eq:g_beta_k_block_Mgb2}
  \widetilde{f}_{\bsl{\tau}_1 \bsl{\tau}_2,\beta}( \bsl{k}_\shpa )  =  \sum_{\Delta\bsl{R}_\shpa }  e^{- \ii \bsl{k}_\shpa \cdot (\Delta\bsl{R}_\shpa + \bsl{\tau}_1 - \bsl{\tau}_2)}  \widetilde{f}_{\bsl{\tau}_1 \bsl{\tau}_2,\beta}(\Delta\bsl{R}_\shpa + \bsl{\tau}_1 - \bsl{\tau}_2 ) \ \forall\ \beta = \shpa,\perp\ ,
} 
we eventually have
\eq{
\label{eq:g_i_g_beta_Mgb2}
\forall i=x,y\ , f_{i}(\bsl{k}_\shpa)  =  \ii \partial_{k_i}  \widetilde{f}_{\shpa}( \bsl{k}_\shpa )  + \ii \sum_{i'=x,y} \epsilon_{i'i}\partial_{k_{i'}}    \widetilde{f}_{\perp}(\bsl{k}_\shpa ) \ .
}
where 
\eq{
 \left[  \widetilde{f}_{\beta}( \bsl{k} ) \right]_{\bsl{\tau}_1\alpha_{\bsl{\tau}_1} ,\bsl{\tau}_2\alpha_{\bsl{\tau}_2}'} = \left[   \widetilde{f}_{\bsl{\tau}_1 \bsl{\tau}_2,\beta}( \bsl{k} ) \right]_{\alpha_{\bsl{\tau}_1}\alpha_{\bsl{\tau}_2}'} \ \forall\ \beta = \shpa,\perp\ .
} 
For the convenience of later discussion, we choose the basis for the matrix $ \widetilde{f}_{\beta}( \bsl{k} )$ to be 
\eq{
(c^\dagger_{\bsl{k},\text{B}, sp_xp_y} , c^\dagger_{\bsl{k},\text{B}, p_z})
}
with $c^\dagger_{\bsl{k},\text{B}, sp_xp_y}$ right below \eqnref{eq:H_el_MgB2_B_sp2} and $c^\dagger_{\bsl{k},\text{B}, p_z}$ defined right below \eqnref{eq:H_el_MgB2_B_pz}; then  
\eq{
\label{eq:g_beta_k_Mgb2}
 \widetilde{f}_{\beta}( \bsl{k} ) = \mat{ f_{\beta,sp_xp_y}( \bsl{k} ) & f_{\beta,sp_xp_y-p_z}( \bsl{k} )\\ f_{\beta,p_z-sp_xp_y}( \bsl{k} ) & f_{\beta,p_z}( \bsl{k} ) }\ .
}

To further derive the expression of $ \widetilde{f}_{\bsl{\tau}_1 \bsl{\tau}_2,\beta}( \bsl{k} )$, we note that $P6/mmm$ of {\mgb} is compatible with 2D systems, and thus $ \widetilde{f}_{\bsl{\tau}_1 \bsl{\tau}_2,\beta}(\Delta\bsl{R}_\shpa + \bsl{\tau}_1 - \bsl{\tau}_2 )$ should have the symmetry constraints specified in \eqnref{eq:g_beta_sym}.
Then, $\widetilde{f}_{\bsl{\tau}_{B1} \bsl{\tau}_{B2},\shpa}(\bsl{\delta}_j)$ and $\widetilde{f}_{\bsl{\tau}_{B2}\bsl{\tau}_{B1},\shpa}(-\bsl{\delta}_j)$ should have the same expressions as $t_{\bsl{\tau}_{B1} \bsl{\tau}_{B2}}(\bsl{\delta}_j)$ and $t_{\bsl{\tau}_{B2}\bsl{\tau}_{B1}}(-\bsl{\delta}_j)$ in \eqnref{eq:t_MgB2_part_4}, respectively,  leading to
\eqa{
\label{eq:g_shpa_MgB2}
\widetilde{f}_{\shpa,sp_xp_y}( \bsl{k} ) &  =  \left(\begin{array}{c|c} 
& \sum_{j=0,1,2} e^{-\ii \bsl{\delta}_j\cdot\bsl{k}} \mat{ 
 1 & &  \\
  & -\frac{1}{2} & -\frac{\sqrt{3}}{2} \\
  & \frac{\sqrt{3}}{2} & -\frac{1}{2}
 }  ^j\mat{ 
 \hat{\gamma}_1 & & \hat{\gamma}_4 \\
  & \hat{\gamma}_2 + \hat{\gamma}_3 & \\
 -\hat{\gamma}_4 & & \hat{\gamma}_2 - \hat{\gamma}_3 
 } \mat{ 
 1 & &  \\
  & -\frac{1}{2} & -\frac{\sqrt{3}}{2} \\
  & \frac{\sqrt{3}}{2} & -\frac{1}{2}
 }^{-j} \\
 \hline
h.c. & 
\end{array}\right)\\
\widetilde{f}_{\shpa,p_z}( \bsl{k} ) & = \mat{ 0 & \sum_{j=0,1,2} e^{-\ii \bsl{\delta}_j\cdot\bsl{k}}   \hat{\gamma}_5 \\ h.c. & 0 } \\
\widetilde{f}_{\shpa,sp_xp_y-p_z}( \bsl{k} ) &  = 0 \\ 
\widetilde{f}_{\shpa,p_z-sp_xp_y}( \bsl{k} ) &  = 0 
}
with $\hat{\gamma}_{1}$, $\hat{\gamma}_{2}$, $\hat{\gamma}_{3}$, $\hat{\gamma}_{4}$ and $\hat{\gamma}_{5}$ being real parameters.
On the other hand, based on the $z_{R,\perp}$ defined in \eqnref{eq:z_factor}, we know $z_{C_6,\perp}=-z_{m_y,\perp}=z_{m_z,\perp}=1$, and thus $\widetilde{f}_{\bsl{\tau}_{B1} \bsl{\tau}_{B2},\perp}(\bsl{\delta}_j)$ and $\widetilde{f}_{\bsl{\tau}_{B2}\bsl{\tau}_{B1},\perp}(-\bsl{\delta}_j)$ satisfies 
\eqa{
\label{eq:g_perp_sym_MgB2}
& C_6: U_{C_6}^{\bsl{\tau}_{\text{B2}} \bsl{\tau}_{\text{B1}}} \widetilde{f}_{\bsl{\tau}_{\text{B1}}\bsl{\tau}_{\text{B2}},\perp}(\bsl{\delta}_0) \left[ U_{C_6}^{\bsl{\tau}_{\text{B1}} \bsl{\tau}_{\text{B2}}} \right]^\dagger = \widetilde{f}_{\bsl{\tau}_{\text{B2}}\bsl{\tau}_{\text{B1}},\perp}(-\bsl{\delta}_2)\ ,\  U_{C_6}^{\bsl{\tau}_{\text{B1}}\bsl{\tau}_{\text{B2}} } \widetilde{f}_{\bsl{\tau}_{\text{B2}}\bsl{\tau}_{\text{B1}},\perp}(-\bsl{\delta}_2) \left[ U_{C_6}^{ \bsl{\tau}_{\text{B2}} \bsl{\tau}_{\text{B1}}} \right]^\dagger =  \widetilde{f}_{\bsl{\tau}_{\text{B1}}\bsl{\tau}_{\text{B2}},\perp}(\bsl{\delta}_1)\\
 & \qquad   U_{C_6}^{\bsl{\tau}_{\text{B2}} \bsl{\tau}_{\text{B1}}} \widetilde{f}_{\bsl{\tau}_{\text{B1}}\bsl{\tau}_{\text{B2}},\perp}(\bsl{\delta}_1) \left[ U_{C_6}^{\bsl{\tau}_{\text{B1}} \bsl{\tau}_{\text{B2}}} \right]^\dagger = \widetilde{f}_{\bsl{\tau}_{\text{B2}}\bsl{\tau}_{\text{B1}},\perp}(-\bsl{\delta}_0)\ ,\  U_{C_6}^{\bsl{\tau}_{\text{B1}}\bsl{\tau}_{\text{B2}} } \widetilde{f}_{\bsl{\tau}_{\text{B2}}\bsl{\tau}_{\text{B1}},\perp}(-\bsl{\delta}_0) \left[ U_{C_6}^{ \bsl{\tau}_{\text{B2}} \bsl{\tau}_{\text{B1}}} \right]^\dagger =  \widetilde{f}_{\bsl{\tau}_{\text{B1}}\bsl{\tau}_{\text{B2}},\perp}(\bsl{\delta}_2)\\
 & \qquad  U_{C_6}^{\bsl{\tau}_{\text{B2}} \bsl{\tau}_{\text{B1}}} \widetilde{f}_{\bsl{\tau}_{\text{B1}}\bsl{\tau}_{\text{B2}},\perp}(\bsl{\delta}_2) \left[ U_{C_6}^{\bsl{\tau}_{\text{B1}} \bsl{\tau}_{\text{B2}}} \right]^\dagger = \widetilde{f}_{\bsl{\tau}_{\text{B2}}\bsl{\tau}_{\text{B1}},\perp}(-\bsl{\delta}_1)\ ,\  U_{C_6}^{\bsl{\tau}_{\text{B1}}\bsl{\tau}_{\text{B2}} } \widetilde{f}_{\bsl{\tau}_{\text{B2}}\bsl{\tau}_{\text{B1}},\perp}(-\bsl{\delta}_1) \left[ U_{C_6}^{ \bsl{\tau}_{\text{B2}} \bsl{\tau}_{\text{B1}}} \right]^\dagger =  \widetilde{f}_{\bsl{\tau}_{\text{B1}}\bsl{\tau}_{\text{B2}},\perp}(\bsl{\delta}_0)\\
& m_y:  U_{m_y}^{\bsl{\tau}_{\text{B2}} \bsl{\tau}_{\text{B1}}} \widetilde{f}_{\bsl{\tau}_{\text{B1}} \bsl{\tau}_{\text{B2}},\perp}(\bsl{\delta}_0) \left[ U_{m_y}^{\bsl{\tau}_{\text{B1}} \bsl{\tau}_{\text{B2}}} \right]^\dagger = -\widetilde{f}_{\bsl{\tau}_{\text{B2}} \bsl{\tau}_{\text{B1}},\perp}(-\bsl{\delta}_0) \\
& \qquad U_{m_y}^{\bsl{\tau}_{\text{B2}} \bsl{\tau}_{\text{B1}}} \widetilde{f}_{\bsl{\tau}_{\text{B1}} \bsl{\tau}_{\text{B2}},\perp}(\bsl{\delta}_1) \left[ U_{m_y}^{\bsl{\tau}_{\text{B1}} \bsl{\tau}_{\text{B2}}} \right]^\dagger = -\widetilde{f}_{\bsl{\tau}_{\text{B2}} \bsl{\tau}_{\text{B1}},\perp}(-\bsl{\delta}_2) \\ 
& \qquad U_{m_y}^{\bsl{\tau}_{\text{B2}} \bsl{\tau}_{\text{B1}}} \widetilde{f}_{\bsl{\tau}_{\text{B1}} \bsl{\tau}_{\text{B2}},\perp}(\bsl{\delta}_2) \left[ U_{m_y}^{\bsl{\tau}_{\text{B1}} \bsl{\tau}_{\text{B2}}} \right]^\dagger = -f_{\bsl{\tau}_{\text{B2}} \bsl{\tau}_{\text{B1}},\perp}(-\bsl{\delta}_1) \\
& m_z: U_{m_z}^{\bsl{\tau}_{\text{B1}} \bsl{\tau}_{\text{B1}}} \widetilde{f}_{\bsl{\tau}_{\text{B1}} \bsl{\tau}_{\text{B2}},\perp}(\bsl{\delta}_j) U_{m_z}^{\bsl{\tau}_{\text{B2}} \bsl{\tau}_{\text{B2}}} = \widetilde{f}_{\bsl{\tau}_{\text{B1}} \bsl{\tau}_{\text{B2}},\perp}(\bsl{\delta}_j) \ ,\ U_{m_z}^{\bsl{\tau}_{\text{B2}} \bsl{\tau}_{\text{B2}}} \widetilde{f}_{\bsl{\tau}_{\text{B2}} \bsl{\tau}_{\text{B1}},\perp}(-\bsl{\delta}_j) U_{m_z}^{\bsl{\tau}_{\text{B1}} \bsl{\tau}_{\text{B1}}} = \widetilde{f}_{\bsl{\tau}_{\text{B2}} \bsl{\tau}_{\text{B1}},\perp}(-\bsl{\delta}_j) \\
& \TR: \widetilde{f}_{\bsl{\tau}_{\text{B1}} \bsl{\tau}_{\text{B2}},\perp}(\bsl{\delta}_j) \ ,\ \widetilde{f}_{\bsl{\tau}_{\text{B2}} \bsl{\tau}_{\text{B1}},\perp}(-\bsl{\delta}_j) \in \dsR^{4\times4}\\
& h.c.: \widetilde{f}_{\bsl{\tau}_{\text{B1}} \bsl{\tau}_{\text{B2}},\perp}(\bsl{\delta}_j)= \widetilde{f}_{\bsl{\tau}_{\text{B2}} \bsl{\tau}_{\text{B1}}}^\dagger(-\bsl{\delta}_j)\ ,
}
leading to 
\eqa{
\label{eq:f_perp_spxpy}
& \widetilde{f}_{\bsl{\tau}_{\text{B1}}\bsl{\tau}_{\text{B2}},\perp}(\bsl{\delta}_0) = 
\mat{
 0 & \hat{\gamma}_6 &  0 &  0 \\ 
- \hat{\gamma}_6 &  0 & \hat{\gamma}_7 &  0 \\ 
 0 & \hat{\gamma}_7 &  0 &  0 \\ 
 0 & 0 & 0 & 0 \\ 
} \text{ for basis $(s,p_x,p_y,p_z)$}\ ,
}
resulting in
\eqa{
\label{eq:g_perp_MgB2}
\widetilde{f}_{\perp,sp_xp_y}( \bsl{k} ) &  =  \left(\begin{array}{c|c} 
& \sum_{j=0,1,2} e^{-\ii \bsl{\delta}_j\cdot\bsl{k}} \mat{ 
 1 & &  \\
  & -\frac{1}{2} & -\frac{\sqrt{3}}{2} \\
  & \frac{\sqrt{3}}{2} & -\frac{1}{2}
 }  ^j\mat{ 
  0 & \hat{\gamma}_6 &  0 \\ 
- \hat{\gamma}_6 &  0 & \hat{\gamma}_7 \\ 
 0 & \hat{\gamma}_7 &  0  
 } \mat{ 
 1 & &  \\
  & -\frac{1}{2} & -\frac{\sqrt{3}}{2} \\
  & \frac{\sqrt{3}}{2} & -\frac{1}{2}
 }^{-j} \\
 \hline
h.c. & 
\end{array}\right)\\
\widetilde{f}_{\shpa,p_z}( \bsl{k} ) & = 0 \\
\widetilde{f}_{\shpa,sp_xp_y-p_z}( \bsl{k} ) &  = 0 \\ 
\widetilde{f}_{\shpa,p_z-sp_xp_y}( \bsl{k} ) &  = 0 \ ,
}
with $\hat{\gamma}_{6}$ and $\hat{\gamma}_{7}$ being real parameters.

Now we move to $f_{\bsl{\tau}\bsl{\tau},z}( \pm \bsl{a}_3)$.
According to \eqnref{eq:g_k}, we have
\eqa{
  f_{\bsl{\tau} \bsl{\tau},z}(k_z)  
  & = \sum_{w=\pm} f_{\bsl{\tau}\bsl{\tau},z}( w \bsl{a}_3) e^{-\ii w c k_z} \\
  & = \sum_{w=\pm} \frac{w}{c} f_{\bsl{\tau}\bsl{\tau},z}( w \bsl{a}_3) \ii \partial_{k_z} e^{-\ii w c k_z} \\
  & = \ii \partial_{k_z} \widetilde{f}_{\bsl{\tau}\bsl{\tau}, \shpa}'( k_z)\ ,
}
where 
\eq{
\widetilde{f}_{\bsl{\tau}\bsl{\tau}, \shpa}'(k_z) = \sum_{w=\pm}  \widetilde{f}_{\bsl{\tau}\bsl{\tau}, \shpa}'(w \bsl{a}_3) e^{-\ii w c k_z}\ .
}
and 
\eq{
\widetilde{f}_{\bsl{\tau}\bsl{\tau}, \shpa}'(w \bsl{a}_3) = \frac{w}{c} f_{\bsl{\tau}\bsl{\tau},z}( w \bsl{a}_3)= \frac{(w \bsl{a}_3)\cdot\bsl{e}_c}{c^2} f_{\bsl{\tau}\bsl{\tau},z}( w \bsl{a}_3) 
}
To derive the form of $\widetilde{f}_{\bsl{\tau}\bsl{\tau}, \shpa}'( k_z) $, let us first specify the symmetry properties of $\widetilde{f}_{\bsl{\tau}\bsl{\tau}, \shpa}'(w \bsl{a}_3)$ according to \eqnref{eq:g_R_sym}.
\eqa{
& C_6: U_{C_6}^{\bsl{\tau}_{\text{B2}} \bsl{\tau}_{\text{B1}}} \widetilde{f}_{\bsl{\tau}_{\text{B1}}\bsl{\tau}_{\text{B1}},\shpa}'( w \bsl{a}_3) \left[ U_{C_6}^{\bsl{\tau}_{\text{B2}} \bsl{\tau}_{\text{B1}}} \right]^\dagger = \widetilde{f}_{\bsl{\tau}_{\text{B2}}\bsl{\tau}_{\text{B2}},\shpa}'(w \bsl{a}_3)\\
& \qquad U_{C_6}^{\bsl{\tau}_{\text{B1}} \bsl{\tau}_{\text{B2}}} \widetilde{f}_{\bsl{\tau}_{\text{B2}}\bsl{\tau}_{\text{B2}},\shpa}'( w \bsl{a}_3) \left[ U_{C_6}^{\bsl{\tau}_{\text{B1}} \bsl{\tau}_{\text{B2}}} \right]^\dagger = \widetilde{f}_{\bsl{\tau}_{\text{B1}}\bsl{\tau}_{\text{B1}},\shpa}'(w \bsl{a}_3)\\
& m_y:  U_{m_y}^{\bsl{\tau}_{\text{B2}} \bsl{\tau}_{\text{B1}}} \widetilde{f}_{\bsl{\tau}_{\text{B1}}\bsl{\tau}_{\text{B1}},\shpa}'( w \bsl{a}_3) \left[ U_{m_y}^{\bsl{\tau}_{\text{B2}} \bsl{\tau}_{\text{B1}}} \right]^\dagger = \widetilde{f}_{\bsl{\tau}_{\text{B2}}\bsl{\tau}_{\text{B2}},\shpa}'( w \bsl{a}_3) \\
& \qquad U_{m_y}^{\bsl{\tau}_{\text{B1}} \bsl{\tau}_{\text{B2}}} \widetilde{f}_{\bsl{\tau}_{\text{B2}}\bsl{\tau}_{\text{B2}},\shpa}'( w \bsl{a}_3) \left[ U_{m_y}^{\bsl{\tau}_{\text{B1}} \bsl{\tau}_{\text{B2}}} \right]^\dagger = \widetilde{f}_{\bsl{\tau}_{\text{B1}}\bsl{\tau}_{\text{B1}},\shpa}'( w \bsl{a}_3) \\
& m_z: U_{m_z}^{\bsl{\tau} \bsl{\tau}} \widetilde{f}_{\bsl{\tau}\bsl{\tau},\shpa}'( w \bsl{a}_3) \left[ U_{m_z}^{\bsl{\tau} \bsl{\tau}} \right]^\dagger = \widetilde{f}_{\bsl{\tau}\bsl{\tau},\shpa}'( -w \bsl{a}_3)\\
& \TR: \widetilde{f}_{\bsl{\tau}\bsl{\tau},\shpa}'( w \bsl{a}_3) \in \dsR^{4\times4}\\
& h.c.: \widetilde{f}_{\bsl{\tau}\bsl{\tau},\shpa}'( w \bsl{a}_3)= \left[\widetilde{f}_{\bsl{\tau}\bsl{\tau},\shpa}'( -w \bsl{a}_3)\right]^\dagger\ ,
}
which means that $\widetilde{f}_{\bsl{\tau}\bsl{\tau}, \shpa}'(w \bsl{a}_3)$ has the same symmetry constraints as $t_{\bsl{\tau}\bsl{\tau}}(w \bsl{a}_3)$ in \eqnref{eq:t_sym_MgB2_3}, resulting in 
\eq{
\widetilde{f}_{\bsl{\tau}_{\text{B1}}\bsl{\tau}_{\text{B1}}, \shpa}'(\bsl{a}_3) = \widetilde{f}_{\bsl{\tau}_{\text{B2}}\bsl{\tau}_{\text{B2}}, \shpa}'(\bsl{a}_3) = 
\left[\widetilde{f}_{\bsl{\tau}_{\text{B1}}\bsl{\tau}_{\text{B1}}, \shpa}'(-\bsl{a}_3)\right]^\dagger =
\left[\widetilde{f}_{\bsl{\tau}_{\text{B2}}\bsl{\tau}_{\text{B2}}, \shpa}'(-\bsl{a}_3)\right]^\dagger =
\mat{
\hat{\gamma}_{8} & & & \hat{\gamma}_{11}\\
 & \hat{\gamma}_{9} & & \\
 & & \hat{\gamma}_{9} & \\
- \hat{\gamma}_{11}  & & & \hat{\gamma}_{10}
}
}
for basis $(s,p_x,p_y,p_z)$, where $\hat{\gamma}_{8}$, $\hat{\gamma}_{9}$, $\hat{\gamma}_{10}$ and $\hat{\gamma}_{11}$ are real parameters.
As shown in \eqnref{eq:MgB2_TB_values}, the hopping between $s$ and $p_z$ can be neglected along $\pm\bsl{a}_3$ without changing the physics near the Fermi level much (\figref{fig:MgB2_el}(a-b)), since the states near the Fermi level have little dependence on the $s$ orbital~\cite{Lordi05102018MgB2ElectronBands}.
In the evaluation of $\lambda$ in \eqnref{eq:lambda}, we eventually will project the EPC Hamiltonian to the Fermi surface.
Therefore, we can still choose $\hat{\gamma}_{11} = 0$ to neglect the coupling between $s$ and $p_z$ for $\widetilde{f}_{\bsl{\tau}\bsl{\tau}, \shpa}'(w \bsl{a}_3)$. 
(In principle, we may neglect any terms in the EPC Hamiltonian that involves the $s$ orbital in the Hamiltonian without changing $\lambda$ much. But we do not choose to do so for our purpose here, since completely neglecting $s$ orbitals in the electron Hamiltonian may change the topology of the occupied bands as the $m_z$-even 3 occupied bands in the $k_z=0$ plane contains a rank-2 subspace with fragile topology.)
Therefore, by defining $ \left[ \widetilde{f}_{\shpa}'( k_z)\right]_{\bsl{\tau}\alpha_{\bsl{\tau}},\bsl{\tau}\alpha_{\bsl{\tau}}'} = \left[\widetilde{f}_{\bsl{\tau}\bsl{\tau}, \shpa}'( k_z) \right]_{\alpha_{\bsl{\tau}}\alpha_{\bsl{\tau}}'}$ and $ \left[ \widetilde{f}_{\shpa}'( k_z)\right]_{\bsl{\tau}_1\alpha_{\bsl{\tau}_1},\bsl{\tau}_2\alpha_{\bsl{\tau}_2}'} = 0\ \forall \bsl{\tau}_1\neq \bsl{\tau}_2$, $\widetilde{f}_{\bsl{\tau}\bsl{\tau}, \shpa}'( k_z)$ should have the following form as 
\eq{
\label{eq:g_shpa_prime_MgB2}
\widetilde{f}_{\shpa}'( k_z) = \mat{ \widetilde{f}_{\shpa,sp_xp_y}'( k_z ) & 0\\ 0 & \widetilde{f}_{\shpa,p_z}'( k_z ) }\ ,
}
where the basis is chosen to be 
\eq{
(c^\dagger_{\bsl{k},\text{B}, sp_xp_y} , c^\dagger_{\bsl{k},\text{B}, p_z})
}
with $c^\dagger_{\bsl{k},\text{B}, sp_xp_y}$ defined right below \eqnref{eq:H_el_MgB2_B_sp2} and $c^\dagger_{\bsl{k},\text{B}, p_z}$ defined right below \eqnref{eq:H_el_MgB2_B_pz}, 
\eq{
\widetilde{f}_{\shpa,sp_xp_y}'( k_z ) = \left(\begin{array}{c|c}
\mat{
\hat{\gamma}_{8} 2 \cos(k_z c) & & \\
 & \hat{\gamma}_{9} 2 \cos(k_z c) & \\
 & & \hat{\gamma}_{9} 2 \cos(k_z c)
 } & 0_{3\times 3} \\
 \hline
 0_{3\times 3} & \mat{
\hat{\gamma}_{8} 2 \cos(k_z c) & & \\
 & \hat{\gamma}_{9} 2 \cos(k_z c) & \\
 & & \hat{\gamma}_{9} 2 \cos(k_z c)
 }
\end{array}\right)
}
and
\eq{
\label{eq:g_shpa_prime_pz_MgB2}
\widetilde{f}_{\shpa,p_z}'( k_z ) = \hat{\gamma}_{10} 2 \cos(k_z c) \tau_0\ .
}

In sum, $f_i(\bsl{k})$ in \eqnref{eq:f_k_2center} has the following form
\eq{
\label{eq:g_i_MgB2}
 f_i(\bsl{k}) = \left[ \ii \partial_{k_i}  \widetilde{f}_{\shpa}( \bsl{k}_\shpa )  + \ii \sum_{i'=x,y} \epsilon_{i'i}\partial_{k_{i'}}    \widetilde{f}_{\perp}(\bsl{k}_\shpa ) \right] (\delta_{ix} + \delta_{iy} )  + \ii \partial_{k_z }  \widetilde{f}_{\shpa}'( k_z ) \delta_{iz}\ ,
}
where $ \widetilde{f}_{\shpa}( \bsl{k}_\shpa )$ is given in \eqnref{eq:g_shpa_MgB2}, $ \widetilde{f}_{\perp}( \bsl{k}_\shpa )$ is given in \eqnref{eq:g_perp_MgB2}, and $ \widetilde{f}_{\shpa}'( \bsl{k}_\shpa )$ is given in \eqnref{eq:g_shpa_prime_MgB2}.
According to \eqnref{eq:g_shpa_MgB2}, \eqnref{eq:g_perp_MgB2} and \eqnref{eq:g_shpa_prime_MgB2}, $ f_i(\bsl{k}) $ is block-diagonal: 
\eq{
f_i(\bsl{k}) = \mat{ f_{i,sp_xp_y}(\bsl{k}) & \\ & f_{i,p_z}(\bsl{k})}\ ,
}
where the basis is again 
\eq{
\label{eq:psi_basis_EPC}
(c^\dagger_{\bsl{k},\text{B}, sp_xp_y} , c^\dagger_{\bsl{k},\text{B}, p_z})
}
with $c^\dagger_{\bsl{k},\text{B}, sp_xp_y}$ defined right below \eqnref{eq:H_el_MgB2_B_sp2} and $c^\dagger_{\bsl{k},\text{B}, p_z}$ defined right below \eqnref{eq:H_el_MgB2_B_pz}, and
\eq{
\label{eq:g_X_MgB2}
f_{i, X}(\bsl{k}) = \left[ \ii \partial_{k_i} \widetilde{f}_{\shpa, X}( \bsl{k}_\shpa )  + \ii \sum_{i'=x,y} \epsilon_{i'i}\partial_{k_{i'}}   \widetilde{f}_{\perp, X}(\bsl{k}_\shpa ) \right] (\delta_{ix} + \delta_{iy} )  + \ii \partial_{k_z } \widetilde{f}_{\shpa, X}'( k_z ) \delta_{iz}\ , 
}
in the basis $c^\dagger_{\bsl{k},\text{B}, X}$ for $X\in\{ sp_xp_y , p_z\}$, $\widetilde{f}_{\shpa, X}( \bsl{k}_\shpa )$ is given in \eqnref{eq:g_shpa_MgB2}, $\widetilde{f}_{\perp, X}(\bsl{k}_\shpa )$ is given in \eqnref{eq:g_perp_MgB2}, and $\widetilde{f}_{\shpa, X}'( k_z )$ is given in \eqnref{eq:g_shpa_prime_MgB2}.
The block-diagonal $f_i(\bsl{k}) $ mainly comes from three approximations: (i) we only keep the short-range terms for the EPC Hamiltonian, (ii) we only choose in-plane/out-of-plane motions to affect only in-plane/out-of-plane hoppings, and (iii) we choose $s-p_z$ coupling to be zero along $z$. 
Then, \eqnref{eq:f_k_2center} shows that $F_{\bsl{\tau},i}(\bsl{k}_1,\bsl{k}_2)$ is also block-dagonal in the basis:
\eq{
F_{\bsl{\tau},i}(\bsl{k}_1,\bsl{k}_2) = \mat{ F_{\bsl{\tau},i,sp_xp_y}(\bsl{k}_1,\bsl{k}_2) & \\ & F_{\bsl{\tau},i,p_z}(\bsl{k}_1,\bsl{k}_2)}\ ,
}
where 
\eq{
\label{eq:f_X_MgB2}
F_{\bsl{\tau},i,X}(\bsl{k}_1,\bsl{k}_2) = \chi_{\bsl{\tau}}^X  f_{i, X}(\bsl{k}_2) - f_{i, X}(\bsl{k}_1) \chi_{\bsl{\tau}}^X  \text{ for } X= sp_xp_y , p_z\ , 
}
\eq{
\label{eq:chi_sp2_MgB2}
\chi_{\bsl{\tau}_{\text{B1}}}^{sp_xp_y} = \mat{ \mathds{1}_3 & \\ & 0_{3\times 3} }\ ,\ \chi_{\bsl{\tau}_{\text{B2}}}^{sp_xp_y} = \mat{ 0_{3\times 3} & \\ & \mathds{1}_3 }\ ,
}
and
\eq{
\chi_{\bsl{\tau}_{\text{B1}}}^{p_z} = \mat{ 1 & \\ & 0 }\ ,\ \chi_{\bsl{\tau}_{\text{B2}}}^{p_z} = \mat{ 0 & \\ & 1 }\ .
}
As the electron Hamiltonian in \eqnref{eq:H_el_MgB2_sim} is also block-diagonal in the basis \eqnref{eq:psi_basis_EPC}, we can define the projection matrix $P_{X,,n}(\bsl{k})$ for the band $E_{X,n}(\bsl{k})$ of the Hamiltonian $H_{el}^{B,X}$ in \eqnref{eq:H_el_MgB2_sim}, where $X\in\{sp_xp_y,p_z\}$.
Then, we can further define $\Gamma_{nm}^X$ according to \eqnref{eq:Gamma_nm} for the $X$ block:
\eq{
\label{eq:Gamma_X_MgB2}
\Gamma_{n m }^X(\bsl{k}_1,\bsl{k}_2) = \frac{\hbar}{2} \sum_{\bsl{\tau}\in\{\text{B1}, \text{B2}\},i}  \frac{1}{m_{\bsl{\tau}}} \Tr\left[ P_{X,n}(\bsl{k}_1)  F_{\bsl{\tau}i,X}(\bsl{k}_1,\bsl{k}_2) P_{X,m}(\bsl{k}_2)   F_{\bsl{\tau}i,X}^\dagger(\bsl{k}_1,\bsl{k}_2) \right]\ ,
}
and define 
\eqa{
\label{eq:Gamma_ave_X_MgB2}
& \left\langle\Gamma\right\rangle^{p_z} = \frac{1}{D^2_\pi(\mu)} \sum_{\bsl{k}_1,\bsl{k}_2}^{\BZ}\sum_{n,m} \delta\left(\mu - E_{p_z,n}(\bsl{k}_1) \right) \delta\left(\mu - E_{p_z,m}(\bsl{k}_2) \right) \Gamma_{n m }^{p_z}(\bsl{k}_1,\bsl{k}_2)\\
& \left\langle\Gamma\right\rangle^{sp_xp_y} = \frac{1}{D^2_\sigma(\mu)} \sum_{\bsl{k}_1,\bsl{k}_2}^{\BZ}\sum_{n,m} \delta\left(\mu - E_{sp_xp_y,n}(\bsl{k}_1) \right) \delta\left(\mu - E_{sp_xp_y,m}(\bsl{k}_2) \right) \Gamma_{n m }^{sp_xp_y}(\bsl{k}_1,\bsl{k}_2)\ ,  
}
where 
\eqa{
& D_{\pi}(\mu) = \sum_{n} \delta\left(\mu - E_{p_z,n}(\bsl{k}) \right) \\
& D_{\sigma}(\mu) = \sum_{n} \delta\left(\mu - E_{sp_xp_y,n}(\bsl{k}) \right) \\
} are the electron density of states at the Fermi level for the $p_z$ and $sp_xp_y$ blocks, respectively.
Here $\pi$ stands for the $\pi$ bonding among the $p_z$ orbital since the $\pi$ bonding mainly accounts for the $p_z$ electron states near the Fermi energy, and $\sigma$ stands for the $\sigma$ bonding among the $p_x p_y$ orbitals since the $\sigma$-bonding mainly accounts for the $p_x p_y$ electron states near the Fermi energy.
With \eqnref{eq:Gamma_ave_X_MgB2}, we split the total $\left\langle \Gamma \right\rangle $ in \eqnref{eq:lambda_omegabar} into two terms according to the two blocks: 
\eq{
\left\langle \Gamma \right\rangle = \frac{D_{\pi}^2(\mu)}{D^2(\mu)} \left\langle\Gamma\right\rangle^{p_z} + \frac{D_{\sigma}^2(\mu)}{D^2(\mu)} \left\langle\Gamma\right\rangle^{s p_x p_y}  \ ,
}
where $D(\mu)$ is the total electron density of states at the Fermi level.
Eventually, we can define 
\eqa{
\label{eq:lambda_X_MgB2}
& \lambda_{\sigma} = \frac{2}{ N} \frac{1}{\hbar \mcomega} D_{\sigma}(\mu) \frac{D_{\sigma}(\mu)}{D(\mu)} \left\langle\Gamma\right\rangle^{s p_x p_y} \\
& \lambda_{\pi} = \frac{2}{ N} \frac{1}{\hbar \mcomega} D_{\pi}(\mu) \frac{D_{\pi}(\mu)}{D(\mu)} \left\langle\Gamma\right\rangle^{p_z}\ ,
} 
and have
\eq{
\lambda = \lambda_{\sigma} + \lambda_{\pi}\ .
}
Clearly, under our approximations, $\lambda$ does not have the cross contribution that involve both $sp_xp_y$ and $p_z$, consistent with \refcite{Kong02272001MgB2EPC} which numerically shows that the cross contribution should be small.
In the following, we will study $\lambda_{\sigma}$ and $\lambda_{\pi}$ separately.

\subsection{Analytical Geometric and Topological Lower Bounds of EPC constant in {\mgb}: $\lambda_{\pi}$}
\label{app:lambda_pz_topo_geo}

In this part, we derive the analytical geometric and topological lower bounds of $\lambda_{\pi}$ in \eqnref{eq:lambda_X_MgB2}.
The derivation has a lot of similarity compared to that on graphene in \appref{app:graphene}.
According to the discussion in \appref{app:EPC_Constant}, the key quantity that we will bound is $\left\langle \Gamma \right\rangle^{p_z}$ in \eqnref{eq:lambda_X_MgB2}, where $\Gamma_{n_{p_z} m_{p_z} }^{p_z}(\bsl{k}_1,\bsl{k}_2)$ is defined in \eqnref{eq:Gamma_ave_X_MgB2} together with \eqnref{eq:f_X_MgB2} and \eqnref{eq:g_X_MgB2}.

\subsubsection{Symmetry-Rep Method: Energetic and Geometric parts of the EPC}

To derive the geometric and topological lower bounds of $\left\langle \Gamma \right\rangle^{p_z}$, we should first find the energetic and geometric parts of the EPC.
In this part, we use the symmetry-rep method, while the GA will be used in the next part.
To do so, let us first recall the electron matrix Hamiltonian of $p_z$ orbitals of B atoms, which reads 
\eq{
\label{eq:h_B_pz}
h_{p_z}(\bsl{k}) = E_{p_z,0} +  t_{p_z,z} 2 \cos(k_z c) + 
t_{p_z} \mat{ & \sum_{j=0,1,2} e^{-\ii \bsl{\delta}_j\cdot\bsl{k}}    \\ 
 \sum_{j=0,1,2} e^{\ii \bsl{\delta}_j\cdot\bsl{k}} & } 
} 
as shown in \eqnref{eq:H_el_MgB2_B_pz} with $\bsl{\delta}_j$ defined in \eqnref{eq:delta_j_graphene}.
Compared to \eqnref{eq:H_el_graphene}, \eqnref{eq:h_B_pz} is nothing but the Hamiltonian of graphene with an extra $k_z$ dispersion.
Then, similar to graphene, $h_{p_z}(\bsl{k})$ has the form 
\eq{
h_{p_z}(\bsl{k})= (E_{p_z,0} + \epsilon_{p_z,0}(k_z)) \tau_0 + d_{p_z,x}(\bsl{k}_\shpa)\tau_x + d_{p_z,y}(\bsl{k}_\shpa) \tau_y \ ,
}
where $\epsilon_{p_z,0}(k_z) =  t_{p_z,z} 2 \cos(k_z c)$, and 
\eq{
\label{eq:d_pz}
d_{p_z,x}(\bsl{k}_\shpa) - \ii  d_{p_z,y}(\bsl{k}_\shpa) = t_{p_z} \sum_{j=0,1,2} e^{-\ii \bsl{\delta}_j\cdot\bsl{k}}\ .
}
As a result, the bands $E_{p_z,n}(\bsl{k})$ of $h_{p_z}(\bsl{k})$ and the corresponding projection matrices $P_{p_z,n}(\bsl{k})$ have the forms
\eqa{
\label{eq:E_P_MgB2_pz}
& \Delta E_{p_z}(\bsl{k}_\shpa) = 2 \sqrt{d_{p_z,x}^2(\bsl{k}_\shpa)+d_{p_z,y}^2(\bsl{k}_\shpa)}\ ,\ E_{p_z,n}(\bsl{k}) = \epsilon_{p_z,0}(k_z) + E_{p_z,n}^{2D}(\bsl{k}_\shpa) \ ,\ E_{p_z,n}^{2D}(\bsl{k}_\shpa) = E_{p_z,0} + (-)^n \frac{\Delta E_{p_z}(\bsl{k}_\shpa)}{2}\\
& P_{p_z,n}(\bsl{k}_\shpa) = \frac{1}{2} \left[ 1 + (-)^n  \frac{d_{p_z,x}(\bsl{k}_\shpa)\tau_x + d_{p_z,y}(\bsl{k}_\shpa) \tau_y}{\sqrt{d_{p_z,x}^2(\bsl{k}_\shpa)+d_{p_z,y}^2(\bsl{k}_\shpa)}}\right] \Rightarrow \tau_z P_{p_z,n}(\bsl{k}_\shpa) \tau_z = 1 -P_{p_z,n}(\bsl{k}_\shpa)\ ,
}
where $\tau_z P_{p_z,n}(\bsl{k}_\shpa) \tau_z = 1 -P_{p_z,n}(\bsl{k}_\shpa)$ can be chosen to hold even if $\Delta E_{p_z}(\bsl{k}_\shpa) = 0$. 
We note that $E_{p_z,n}^{2D}(\bsl{k}_\shpa)$, $\Delta E_{p_z}(\bsl{k}_\shpa)$ and $P_{p_z,n}(\bsl{k}_\shpa)$ have the same forms as $E_{n}(\bsl{k})$, $\Delta E(\bsl{k}_\shpa)$ and $P_{n}(\bsl{k}_\shpa)$ in \eqnref{eq:E_P_graphene} for graphene, respectively.

Now we are ready to specify the energetic and geometric parts of the EPC.
Following the same spirit of \appref{app:geo_EPC_symmetry-rep}, we have used the momentum derivative to carry the $i$ index in $f_i(\bsl{k})$ in terms of the momentum derivatives as shown in \eqnref{eq:g_X_MgB2}, and we just need to reexpress $\widetilde{f}_{\shpa,p_z}(\bsl{k}_\shpa)$, $\widetilde{f}_{\perp,p_z}(\bsl{k}_\shpa)$ and $\widetilde{f}_{\shpa,p_z}'(k_z)$.
By comparing \eqnref{eq:h_B_pz} to \eqnref{eq:g_shpa_MgB2}, \eqnref{eq:g_perp_MgB2} and \eqnref{eq:g_shpa_prime_pz_MgB2}, we have
\eqa{
\label{eq:g_E_geo_MgB2_pz_intermediate}
 & \widetilde{f}_{\shpa,p_z}(\bsl{k}_\shpa) = \hat{\gamma}_5 \partial_{t_{p_z}}  h_{p_z}(\bsl{k}) \\
 & \widetilde{f}_{\perp,p_z}(\bsl{k}_\shpa) = 0 \\
 & \widetilde{f}_{\shpa,p_z}'(k_z) = \hat{\gamma}_{10} \partial_{t_{p_z,z}}  h_{p_z}(\bsl{k}) \ ,
}
resulting in 
\eqa{
f_{i, p_z}(\bsl{k}) & = \left[ \ii \partial_{k_i} (\frac{\hat{\gamma}_5}{2t_{p_z}} \left[ h_{p_z}(\bsl{k}) - \tau_z h_{p_z}(\bsl{k}) \tau_z \right])  \right] (\delta_{ix} + \delta_{iy} )  + \ii \partial_{k_z } (\frac{\hat{\gamma}_{10}}{2t_{p_z,z}} \left[ h_{p_z}(\bsl{k}) + \tau_z h_{p_z}(\bsl{k}) \tau_z  - 2 E_{p_z} \right]) \delta_{iz}\\
&  =  \frac{\hat{\gamma}_5}{2t_{p_z}}  \ii \partial_{k_i} \sum_{n=1,2} E_{p_z,n}(\bsl{k}) (P_{p_z,n}(\bsl{k}_\shpa) - \tau_z h_{p_z}(\bsl{k}) \tau_z )  (\delta_{ix} + \delta_{iy} )  + \ii \partial_{k_z } \frac{\hat{\gamma}_{10}}{2 t_{p_z,z}} \sum_{n=1,2} E_{p_z,n}(\bsl{k}) (P_{p_z,n}(\bsl{k}_\shpa) + \tau_z h_{p_z}(\bsl{k}) \tau_z ) \delta_{iz} \\
&  =   \hat{\gamma}_5 \partial_{t_{p_z}} \ii \partial_{k_i} \sum_{n=1,2} E_{p_z,n}(\bsl{k})   P_{p_z,n}(\bsl{k}_\shpa)   (\delta_{ix} + \delta_{iy} )  + \ii \hat{\gamma}_{10} \partial_{t_{p_z,z}} \partial_{k_z} \sum_{n=1,2} E_{p_z,n}(\bsl{k}) P_{p_z,n}(\bsl{k}_\shpa)\delta_{iz}\ ,
}
where we used \eqnref{eq:E_P_MgB2_pz}.
As discussed in \appref{app:symmetry-rep_detials}, the derivatives with respect to hopping parameters in \eqnref{eq:g_E_geo_MgB2_pz_intermediate} is in general very hard to deal with.
However, owing to the short-range nature of the  electron Hamiltonian (\eqnref{eq:h_B_pz}), the momentum dependence of the electron Hamiltonian only comes from two hopping parameters, $t_{p_z,z}$ and $t_{p_z}$, and they have different matrix forms.
As a result, the derivatives with respect to hopping parameters in \eqnref{eq:g_E_geo_MgB2_pz_intermediate} can eventually be converted to $1/t_{p_z}$ or $1/t_{p_z,z}$.
Explicitly, after incorporating the form of the electron Hamiltonian in \eqnref{eq:h_B_pz}, the energetic and geometric parts of $f_{i, p_z}(\bsl{k})$ read
\eq{
\label{eq:g_E_geo_MgB2_pz}
f_{i, p_z}(\bsl{k}) = f_{i, p_z}^{E}(\bsl{k}) + f_{i, p_z}^{geo}(\bsl{k})\ ,
}
where 
\eqa{
f_{i, p_z}^{E}(\bsl{k}) & = \hat{\gamma}_5 \partial_{t_{p_z}} \ii  \sum_{n=1,2} \partial_{k_i} E_{p_z,n}(\bsl{k})   P_{p_z,n}(\bsl{k}_\shpa)   (\delta_{ix} + \delta_{iy} )  + \ii \hat{\gamma}_{10} \partial_{t_{p_z,z}} \sum_{n=1,2} \partial_{k_z}  E_{p_z,n}(\bsl{k}) P_{p_z,n}(\bsl{k}_\shpa)\delta_{iz} \\
& = \frac{\hat{\gamma}_5}{t_{p_z}} \ii \partial_{k_i} \frac{\Delta E_{p_z}(\bsl{k}_\shpa)}{2}
 \sum_{n=1,2} (-1)^n P_{p_z,n}(\bsl{k}_\shpa)   (\delta_{ix} + \delta_{iy} )  + \ii \frac{\hat{\gamma}_{10}}{t_{p_z,z}} \sum_{n=1,2} \partial_{k_z} \epsilon_{p_z,0}(k_z)\delta_{iz} \\
& = \gamma_{\pi,\shpa}  \ii \partial_{k_i} \frac{\Delta E_{p_z}(\bsl{k}_\shpa)}{2}
 \sum_{n=1,2} (-1)^n P_{p_z,n}(\bsl{k}_\shpa)   (\delta_{ix} + \delta_{iy} )  + \ii \gamma_{\pi,z} \sum_{n=1,2} \partial_{k_z} \epsilon_{p_z,0}(k_z)\delta_{iz} \\
}
\eqa{
f_{i, p_z}^{geo}(\bsl{k}) & =  \hat{\gamma}_5 \partial_{t_{p_z}} \ii  \sum_{n=1,2} E_{p_z,n}(\bsl{k})   \partial_{k_i} P_{p_z,n}(\bsl{k}_\shpa)   (\delta_{ix} + \delta_{iy} ) \\
 & =  \frac{\hat{\gamma}_5}{t_{p_z}}  \frac{\Delta E_{p_z}(\bsl{k}_\shpa)}{2} \ii  \sum_{n=1,2} (-1)^n \partial_{k_i} P_{p_z,n}(\bsl{k}_\shpa)   (\delta_{ix} + \delta_{iy} )\\
& =  \gamma_{\pi,\shpa}  \frac{\Delta E_{p_z}(\bsl{k}_\shpa)}{2} \ii  \sum_{n=1,2} (-1)^n \partial_{k_i} P_{p_z,n}(\bsl{k}_\shpa)   (\delta_{ix} + \delta_{iy} ) \ ,
}
\eq{
\label{eq:gamma_pz_MgB2}
\gamma_{\pi,\shpa} = \frac{\hat{\gamma}_5}{t_{p_z}}\ , \ \gamma_{\pi,z} = \frac{\hat{\gamma}_{10}}{ t_{p_z,z}}\ ,
}
$\Delta E_{p_z}(\bsl{k}_\shpa)$ and  $ \epsilon_{p_z,0}(k_z)$ are defined in \eqnref{eq:E_P_MgB2_pz}, and we heavily used \eqnref{eq:E_P_MgB2_pz} and \eqnref{eq:d_pz}.
We note that 
\eq{
f_{i, p_z}^{E}(\bsl{k}) = f_{i, p_z}^{E}(\bsl{k}_\shpa)\ ,\ f_{i, p_z}^{geo}(\bsl{k}) = f_{i, p_z}^{geo}(\bsl{k}_\shpa)\ ,\ f_{i, p_z}(\bsl{k}) = f_{i, p_z}(\bsl{k}_\shpa) \text{ for }i=x,y\ ,
}
and $f_{i=x,y, p_z}^{E}(\bsl{k}_\shpa)$ and $f_{i=x,y, p_z}^{geo}(\bsl{k}_\shpa)$ have the same forms as $f_{i=x,y}^{E}(\bsl{k}_\shpa)$ and $f_{i=x,y}^{geo}(\bsl{k}_\shpa)$ in \eqnref{eq:g_E_g_geo_graphene} for graphene, respectively, after replacing $\gamma_{\pi,\shpa} $ by $\gamma$.

\subsubsection{Gaussian Approximation: Energetic and Geometric parts of the EPC}

\label{app:geo_EPC_MgB2_pi_Gaussian}

In this part, we will show that the energetic and geometric parts of the EPC in \eqnref{eq:g_E_geo_MgB2_pz} have exactly the same forms as those derived from the GA with extra constraints in \eqnref{eq:E_P_MgB2_pz}.

Since we are now considering $p_z$ orbitals in a 3D system, they differ from $s$-like orbitals. 
We need to first study the form of the hopping function $t(\bsl{r})$ between any two $p_z$ orbitals separated by $\bsl{r}$ in 3D.
To do so, we use the idea of linear combinations of atomic orbitals proposed in \refcite{Slater1954LCAO}.
Nevertheless, instead of using $\O(3)$ symmetry in \refcite{Slater1954LCAO}, let us use $\O(2)$ and $m_z$ symmetries, since $p_z$ is not a rep of $\O(3)$ group but an irrep of $\O(2)$ and $m_z$ symmetries.
Explicitly, $O(2)$ and $m_z$ symmetries give that 
\eq{
t(\bsl{r}) = t(C_\theta \bsl{r}) = t(m_y \bsl{r}) = t(m_z \bsl{r})\ ,
}
where $C_\theta$ labels the continuous rotation symmetry along the $z$ axis.
As a result, 
\eq{
t(\bsl{r}) = t(\sqrt{x^2 + y^2}, 0, |z|)\ ,
}
where $\bsl{r} = (x,y,z) $.
Then, since the dominant phonon modes are in-plane ion motions, we choose the GA for the in-plane directions:
\eq{
\label{eq:hopping_function_Gaussian_Mgb2_pi}
t(\bsl{r}) = t(\sqrt{x^2 + y^2}, 0, |z|) = t_0 F(z^2)\exp\left[ \gamma_{\pi,\shpa} (x^2 + y^2)  \right]\ , 
}
which leads to
\eq{
\label{eq:gradient_hopping_function_Gaussian_Mgb2_pi_intermediate}
\nabla_{\bsl{r}} t(\bsl{r}) = \left[ \gamma_{\pi,\shpa} (x \bsl{e}_x + y \bsl{e}_y) + \widetilde{\gamma}(z^2) z \bsl{e}_z \right] t(\bsl{r}) \ ,
}
where 
\eq{
\widetilde{\gamma}(z^2) = \frac{1}{z} \partial_z \log(|F(z^2)|) = 2 \partial_{z^2} \log(|F(z^2)|) \ .
}
Nevertheless, as shown in \eqnref{eq:h_B_pz}, we only consider the $\pm \bsl{a}_3$ hopping for  B  atoms in different layers.
Thus, as a good approximation, we can replace $\widetilde{\gamma}(z^2)$ in \eqnref{eq:gradient_hopping_function_Gaussian_Mgb2_pi_intermediate} by
\eq{
\gamma_{\pi,z} = \widetilde{\gamma}(c^2)\ ,
}
leading to 
\eq{
\label{eq:gradient_hopping_function_Gaussian_Mgb2_pi}
\nabla_{\bsl{r}} t(\bsl{r}) = \left[ \gamma_{\pi,\shpa} (x \bsl{e}_x + y \bsl{e}_y) + \gamma_{\pi,z} z \bsl{e}_z \right] t(\bsl{r}) \ .
}
Here $c$ is the lattice constant along $z$.

By following the derivation in \appref{app:gaussian}, we know the electron Hamiltonian reads 
\eq{
H_{el}^{\text{B},p_z} =  \sum_{\bsl{k}}^{\BZ} c^\dagger_{\bsl{k},\text{B}, p_z}  h_{p_z}(\bsl{k})  c_{\bsl{k},\text{B}, p_z}\ ,
}
where $c^\dagger_{\bsl{k},\text{B}, p_z} =  (c^{\dagger}_{\bsl{k},\bsl{\tau}_{\text{B1}},p_z},c^{\dagger}_{\bsl{k},\bsl{\tau}_{\text{B2}},p_z}) $ and 
\eq{
[h_{p_z}(\bsl{k})]_{\bsl{\tau}\bsl{\tau}'} = \sum_{\bsl{R}} e^{-\ii (\bsl{R}+\bsl{\tau}-\bsl{\tau}')\cdot \bsl{k} }   t(\bsl{R}+\bsl{\tau}-\bsl{\tau}')\ .
}
With \eqnref{eq:gradient_hopping_function_Gaussian_Mgb2_pi}, the EPC Hamiltonian reads
\eq{
 H_{el-ph} = \frac{1}{\sqrt{N}}\sum_{\bsl{k}_1}^{\BZ} \sum_{\bsl{k}_2}^{\BZ}  \sum_{ \bsl{\tau}\in\{\text{B1},\text{B2}\} ,  i}     c^\dagger_{\bsl{k}_1 ,\text{B}, p_z} \left[ \chi_{\bsl{\tau}} f_{i}(\bsl{k}_2) - f_{i}(\bsl{k}_1)\chi_{\bsl{\tau}} \right] c_{\bsl{k}_2,\text{B}, p_z}    u^\dagger_{\bsl{k}_2 - \bsl{k}_1,\bsl{\tau},i}\ ,
}
where 
\eqa{
 &  \left[f_{i}(\bsl{k})\right]_{\bsl{\tau}_1 \bsl{\tau}_2 } \\
 & = \sum_{ \bsl{R} } e^{- \ii \bsl{k} \cdot ( \bsl{R}+ \bsl{\tau}_1 - \bsl{\tau}_2)}   \left. \partial_{r_i} t(\bsl{r})\right|_{\bsl{r} = \bsl{R}+\bsl{\tau}_1-\bsl{\tau}_2} \\
& = \sum_{ \bsl{R} } e^{- \ii \bsl{k} \cdot ( \bsl{R}+ \bsl{\tau}_1 - \bsl{\tau}_2)}   \left[\gamma_{\pi,\shpa} (\bsl{R}+\bsl{\tau}_1-\bsl{\tau}_2)_x \delta_{ix} + \gamma_{\pi,\shpa}  (\bsl{R}+\bsl{\tau}_1-\bsl{\tau}_2)_y \delta_{iy} + \gamma_{\pi,z} (\bsl{R}+\bsl{\tau}_1-\bsl{\tau}_2)_z \delta_{iz} \right] t(\bsl{R}+\bsl{\tau}_1-\bsl{\tau}_2) \\
& =  \ii \left[\delta_{ix} \gamma_{\pi,\shpa} \partial_{k_x}  + \delta_{iy} \gamma_{\pi,\shpa}  \partial_{k_y}  + \delta_{iz} \gamma_{\pi,z}\partial_{k_z}   \right] [h_{p_z}(\bsl{k})]_{\bsl{\tau}_1\bsl{\tau}_2} \\
& \Leftrightarrow  f_{i}(\bsl{k}) =\ii  \left[\delta_{ix} \gamma_{\pi,\shpa} \partial_{k_x}  + \delta_{iy} \gamma_{\pi,\shpa}  \partial_{k_y}  + \delta_{iz} \gamma_{\pi,z}\partial_{k_z}   \right] h_{p_z}(\bsl{k})
\ ,
}
and $i=x,y,z$ labels the spatial direction.
As a result, we have
\eq{
f_{i}(\bsl{k}) = f_{i}^E(\bsl{k}) + f_{i}^{geo}(\bsl{k}) \ ,
}
where 
\eqa{
\label{eq:g_E_g_geo_Gaussian_Mgb2_pi}
& f_{i}^E(\bsl{k}) =\sum_{n} \ii P_n(\bsl{k})  \left[\delta_{ix} \gamma_{\pi,\shpa} \partial_{k_x}  + \delta_{iy} \gamma_{\pi,\shpa}  \partial_{k_y}  + \delta_{iz} \gamma_{\pi,z}\partial_{k_z}   \right] E_n(\bsl{k}) \\
& f_{i}^{geo}(\bsl{k}) =\sum_{n} \ii E_n(\bsl{k}) \left[\delta_{ix} \gamma_{\pi,\shpa} \partial_{k_x}  + \delta_{iy} \gamma_{\pi,\shpa}  \partial_{k_y}  + \delta_{iz} \gamma_{\pi,z}\partial_{k_z}   \right] P_n(\bsl{k}) \ .
}

\eqnref{eq:g_E_g_geo_Gaussian_Mgb2_pi} are the forms of the energetic and geometric parts of EPC derived from the GA \eqnref{eq:hopping_function_Gaussian_Mgb2_pi}.
Now we include the extra constraints of the electron bands and electron projection  matrices in \eqnref{eq:E_P_MgB2_pz}.
As a result, we have
\eqa{
f_{i}^E(\bsl{k}) &  =\sum_{n} \ii P_n(\bsl{k})  \left[\delta_{ix} \gamma_{\pi,\shpa} \partial_{k_x}  + \delta_{iy} \gamma_{\pi,\shpa}  \partial_{k_y}  + \delta_{iz} \gamma_{\pi,z}\partial_{k_z}   \right] E_n(\bsl{k}) \\
&  =\sum_{n} \ii P_n(\bsl{k}_\shpa)  \left[ \frac{(-1)^n}{2}\delta_{ix} \gamma_{\pi,\shpa} \partial_{k_x} \Delta E_{p_z} (\bsl{k}_\shpa)  + \frac{(-1)^n}{2} \delta_{iy} \gamma_{\pi,\shpa}  \partial_{k_y} \Delta E_{p_z} (\bsl{k}_\shpa) + \delta_{iz} \gamma_{\pi,z}\partial_{k_z} \epsilon_{p_z,0}(k_z)\right] \\
&  =\ii \gamma_{\pi,\shpa}\sum_{n}  (-1)^n P_n(\bsl{k}_\shpa) \partial_{k_i} \frac{\Delta E_{p_z} (\bsl{k}_\shpa)}{2} \left(  \delta_{ix}  + \delta_{iy} \right) +\ii   \delta_{iz} \gamma_{\pi,z}\partial_{k_z} \epsilon_{p_z,0}(k_z) 
}
and 
\eqa{
f_{i}^{geo}(\bsl{k}) &  = \sum_{n} \ii E_n(\bsl{k}) \left[\delta_{ix} \gamma_{\pi,\shpa} \partial_{k_x}  + \delta_{iy} \gamma_{\pi,\shpa}  \partial_{k_y} \right] P_n(\bsl{k}_\shpa) \\
&  = \ii \gamma_{\pi,\shpa} \frac{ \Delta E_{p_z}(\bsl{k}_\shpa)}{2} \sum_{n} (-1)^n \partial_{k_i}P_n(\bsl{k}_\shpa) \left( \delta_{ix} + \delta_{iy}   \right) \ ,
}
which are the same as the energetic and geometric parts of the EPC in \eqnref{eq:g_E_geo_MgB2_pz}.

\subsubsection{General Symmetry-Allowed Hopping Form: Consistent with Gaussian Approximation}

We now show that similar to the discussion on graphene in \appref{app:geo_EPC_graphene_Gaussian}, even if we use generic symmetry-allowed hopping function $t(\bsl{r})$ instead of GA, we would give the same expression as the energetic and geometric parts of the EPC in \eqnref{eq:g_E_geo_MgB2_pz}.
Without GA, we should use the generic symmetry-allowed hopping form (instead of the last expression in \eqnref{eq:hopping_function_Gaussian_Mgb2_pi}), which reads
\eq{
\label{eq:hopping_function_Gaussian_Mgb2_pi_NoGA}
t(\bsl{r}) = t(r_\shpa, 0, |z|) \ , 
}
where $r_\shpa = \sqrt{x^2 + y^2}$.
Then, $\nabla_{\bsl{r}} t(\bsl{r})$ becomes
\eq{
\label{eq:gradient_hopping_function_Gaussian_Mgb2_pi_intermediate_NoGA}
\nabla_{\bsl{r}} t(\bsl{r}) = \left[ \widetilde{\gamma}_{\pi,\shpa}(r_\shpa, |z|) (x \bsl{e}_x + y \bsl{e}_y) + \widetilde{\gamma}_{\pi,z}(r_\shpa, |z|) z \bsl{e}_z \right] t(\bsl{r}) \ ,
}
where 
\eqa{
&\widetilde{\gamma}_{\pi,\shpa}(r_\shpa, |z|)  = \frac{1}{r_\shpa} \partial_{r_\shpa} \log(\left| t(r_\shpa, 0, |z|)\right|) \\
&\widetilde{\gamma}_{\pi,z}(r_\shpa, |z|) = \frac{1}{z} \partial_z \log(\left| t(r_\shpa, 0, |z|)\right|) = 2 \partial_{z^2} \log(\left| t(r_\shpa, 0, |z|)\right|) \ .
}
As shown in \eqnref{eq:h_B_pz}, we only consider the NN hopping for B atoms in the same layer and the $\pm \bsl{a}_3$ hopping for  B  atoms in different layers.
Thus, as a good approximation, we can approximate \eqnref{eq:gradient_hopping_function_Gaussian_Mgb2_pi_intermediate_NoGA} as 
\eq{
\nabla_{\bsl{r}} t(\bsl{r}) = \left[ \widetilde{\gamma}_{\pi,\shpa}(\frac{a}{\sqrt{3}},0) (x \bsl{e}_x + y \bsl{e}_y) + \widetilde{\gamma}_{\pi,z}(0,c) z \bsl{e}_z \right] t(\bsl{r}) = \left[ \gamma_{\pi,\shpa}(x \bsl{e}_x + y \bsl{e}_y) + \gamma_{\pi,z} z \bsl{e}_z \right] t(\bsl{r}) \ ,
}
where
\eqa{
&\gamma_{\pi,\shpa} = \widetilde{\gamma}_{\pi,\shpa}\left(\frac{a}{\sqrt{3}},0\right)\\
&\gamma_{\pi,z} = \widetilde{\gamma}_{\pi,z}\left(0,c\right)\ .
}
Therefore, we have obtained exactly the same form of $\nabla_{\bsl{r}} t(\bsl{r})$ in \eqnref{eq:gradient_hopping_function_Gaussian_Mgb2_pi} which is for GA, and by following the rest of the derivation for the GA, we would obtain the same expression as the energetic and geometric parts of the EPC in \eqnref{eq:g_E_geo_MgB2_pz}.

\subsubsection{Geometric and Topological Lower Bounds of $\lambda_{\pi}$}

Now we are ready to move to $\left\langle \Gamma \right\rangle^{p_z}$.
By substituting \eqnref{eq:Gamma_X_MgB2} and \eqnref{eq:f_X_MgB2} into \eqnref{eq:Gamma_ave_mu_2center}, we obtain the following expression for $\left\langle \Gamma \right\rangle^{p_z}$:
\eqa{
\label{eq:Gamma_ave_pz_MgB2}
\left\langle \Gamma \right\rangle^{p_z}  & = \frac{1}{D^2_{p_z}(\mu)} \sum_{\bsl{k}_1,\bsl{k}_2}^{\BZ}\sum_{n,m} \delta\left(\mu - E_{p_z,n}(\bsl{k}_1) \right) \delta\left(\mu - E_{p_z,m}(\bsl{k}_2) \right) \Gamma_{n m  }^{p_z}(\bsl{k}_1,\bsl{k}_2) \\ 
& = - 2 \frac{1}{D^2_{p_z}(\mu)} \sum_{\bsl{k}_1,\bsl{k}_2}^{\BZ}\sum_{n,m} \delta\left(\mu - E_{p_z,n}(\bsl{k}_1) \right) \delta\left(\mu - E_{p_z,m}(\bsl{k}_2) \right) \frac{\hbar}{2 m_{\text{B}}}  \\
& \qquad \times   \sum_i \Tr\left[     f_{i, p_z}(\bsl{k}_1)  P_{p_z,n}(\bsl{k}_1)     f_{i, p_z}(\bsl{k}_1) \sum_{\bsl{\tau}\in\{\text{B1}, \text{B2}\}} \chi_{\bsl{\tau}}^{p_z}  P_{p_z,n}(\bsl{k}_2) \chi_{\bsl{\tau}}^{p_z} \right] \\ 
& \quad +  \frac{1}{D^2_{p_z}(\mu)} \sum_{\bsl{k}_1,\bsl{k}_2}^{\BZ}\sum_{n,m} \delta\left(\mu - E_{p_z,n}(\bsl{k}_1) \right) \delta\left(\mu - E_{p_z,m}(\bsl{k}_2) \right) \frac{\hbar}{2 m_{\text{B}}}   \\
& \qquad \times \sum_{\bsl{\tau}\in\{\text{B1}, \text{B2}\},i} \left( \Tr\left[  \chi_{\bsl{\tau}}^{p_z} f_{i, p_z}(\bsl{k}_1) P_{p_z,n}(\bsl{k}_1)   \chi_{\bsl{\tau}}^{p_z}  f_{i, p_z}(\bsl{k}_2)  P_{p_z,m}(\bsl{k}_2)  \right] + c.c.\right)\ ,
}
where $m_{\text{B}}$ is the mass of the B atom.

To further simplify \eqnref{eq:Gamma_ave_pz_MgB2}, we note that owing to the symmetry representation in \eqnref{eq:sym_rep_MgB2}, we have 
\eqa{
& C_3 c^\dagger_{\bsl{k},\text{B}, p_z} C_3^{-1} = c^\dagger_{C_3\bsl{k},\text{B}, p_z}\\
& m_z c^\dagger_{\bsl{k},\text{B}, p_z} m_z^{-1} = c^\dagger_{m_z\bsl{k},\text{B}, p_z} (-)\ ,
}
where $c^\dagger_{\bsl{k},\text{B}, p_z}$ is defined in \eqnref{eq:H_el_MgB2_B_pz}.
Then, combined with \eqnref{eq:g_k_sym}, we have 
\eqa{
& f_{i, p_z}(C_3 \bsl{k}) = \sum_{i'} f_{i', p_z}( \bsl{k}) (C_{3})_{i'i}\ ,\ P_{p_z,n}(\bsl{k}) = P_{p_z,n}(C_3 \bsl{k})\ ,\ E_{p_z,n}(\bsl{k}) = E_{p_z,n}( C_3 \bsl{k})\\
& f_{i, p_z}(m_z \bsl{k}) = \sum_{i'} f_{i', p_z}( \bsl{k}) (m_z)_{i'i}\ ,\ P_{p_z,n}(\bsl{k}) = P_{p_z,n}(m_z \bsl{k})\ ,\ E_{p_z,n}(\bsl{k}) = E_{p_z,n}( m_z \bsl{k})\ ,
}
leading to 
\eqa{ 
& \sum_{\bsl{k}_1}^{\BZ} \sum_{n} \delta\left(\mu - E_{p_z,n}(\bsl{k}_1) \right) \chi_{\bsl{\tau}}^{p_z} f_{i, p_z}(\bsl{k}_1) P_{p_z,n}(\bsl{k}_1)  
= 
\sum_{i'} \sum_{\bsl{k}_1}^{\BZ} \sum_{n} \delta\left(\mu - E_{p_z,n}(\bsl{k}_1) \right) \chi_{\bsl{\tau}}^{p_z} f_{i', p_z}(\bsl{k}_1) P_{p_z,n}(\bsl{k}_1) (C_{3})_{i'i} \\
& \sum_{\bsl{k}_1}^{\BZ} \sum_{n} \delta\left(\mu - E_{p_z,n}(\bsl{k}_1) \right) \chi_{\bsl{\tau}}^{p_z} f_{i, p_z}(\bsl{k}_1) P_{p_z,n}(\bsl{k}_1)  
= 
\sum_{i'} \sum_{\bsl{k}_1}^{\BZ} \sum_{n} \delta\left(\mu - E_{p_z,n}(\bsl{k}_1) \right) \chi_{\bsl{\tau}}^{p_z} f_{i', p_z}(\bsl{k}_1) P_{p_z,n}(\bsl{k}_1) (m_z)_{i'i}
}
which means that 
\eq{
\label{eq:Gamma_ave_sim_1_pz}
\sum_{\bsl{k}_1}^{\BZ} \sum_{n} \delta\left(\mu - E_{p_z,n}(\bsl{k}_1) \right) \chi_{\bsl{\tau}}^{p_z} f_{i, p_z}(\bsl{k}_1) P_{p_z,n}(\bsl{k}_1)    = 0\ .
}
\eqnref{eq:Gamma_ave_sim_1_pz} for $i=x,y$ is the same as \eqnref{eq:Gamma_ave_sim_1} for graphene.
In addition, owing to the $\P \TR$ symmetry, the matrix Hamiltonian for $p_z$ in \eqnref{eq:h_B_pz} only has two Pauli matrices besides the identity part; as a result, we have another simplification similar to \eqnref{eq:Gamma_ave_sim_2}, which reads
\eq{
\label{eq:Gamma_ave_sim_2_pz}
\sum_{\bsl{\tau}\in\{\text{B1}, \text{B2}\}} \chi_{\bsl{\tau}}^{p_z} P_{p_z,n}(\bsl{k}_\shpa) \chi_{\bsl{\tau}}^{p_z}  =  \frac{1}{2} P_{p_z,n}(\bsl{k}_\shpa) + \frac{1}{2}\tau_z P_{p_z,n}(\bsl{k}_\shpa) \tau_z = \frac{1}{2}\ .
}
Substitute \eqnref{eq:Gamma_ave_sim_1_pz} and \eqnref{eq:Gamma_ave_sim_2_pz} into \eqnref{eq:Gamma_ave_pz_MgB2}, we arrive at
\eqa{
\label{eq:Gamma_ave_pz_sim_MgB2}
\left\langle \Gamma \right\rangle^{p_z} & = -   \frac{\hbar}{2 m_{\text{B}}} \frac{1}{D_{\pi}(\mu)} \sum_{\bsl{k}}^{\BZ}\sum_{n} \delta\left(\mu - E_{p_z,n}(\bsl{k}) \right)\sum_i \Tr\left[     f_{i, p_z}(\bsl{k})  P_{p_z,n}(\bsl{k}_\shpa)     f_{i, p_z}(\bsl{k})  \right] \ .
}

Based on \eqnref{eq:g_E_geo_MgB2_pz}, we can split $\left\langle \Gamma \right\rangle^{p_z}$ into three terms
\eqa{
&   \left\langle \Gamma \right\rangle^{p_z} = \left\langle \Gamma \right\rangle^{p_z,E-E} +  \left\langle \Gamma \right\rangle^{p_z,E-geo} +  \left\langle \Gamma \right\rangle^{p_z,geo-geo} \ ,
}
where
\eqa{
 & \left\langle \Gamma \right\rangle^{p_z,E-E} =  (-)  \frac{\hbar}{2 m_{\text{B}}} \frac{1}{D_{\pi}(\mu)} \sum_{\bsl{k}}^{\BZ}\sum_{n} \delta\left(\mu - E_{p_z,n}(\bsl{k}) \right) \sum_{i} \Tr\left[     f_{i, p_z}^E(\bsl{k})  P_{p_z,n}(\bsl{k}_\shpa)     f_{i, p_z}^E(\bsl{k})  \right]  \\
 & \left\langle \Gamma \right\rangle^{p_z,geo-geo} =  (-)  \frac{\hbar}{2 m_{\text{B}}} \frac{1}{D_{\pi}(\mu)} \sum_{\bsl{k}}^{\BZ}\sum_{n} \delta\left(\mu - E_{p_z,n}(\bsl{k}) \right)\sum_{i=x,y} \Tr\left[     f_{i, p_z}^{geo}(\bsl{k}_\shpa)  P_{p_z,n}(\bsl{k}_\shpa)     f_{i, p_z}^{geo}(\bsl{k}_\shpa)  \right]  \\
 & \left\langle \Gamma \right\rangle^{p_z,E-geo} =  (-)  \frac{\hbar}{2 m_{\text{B}}} \frac{1}{D_{\pi}(\mu)} \sum_{\bsl{k}}^{\BZ}\sum_{n} \delta\left(\mu - E_{p_z,n}(\bsl{k}) \right) \sum_{i=x,y} \Tr\left[     f_{i, p_z}^E(\bsl{k}_\shpa)  P_{p_z,n}(\bsl{k}_\shpa)     f_{i, p_z}^{geo}(\bsl{k}_\shpa)  \right] + c.c. \ .
}
Similar to \eqnref{eq:Gamma_Egeo_ave_mu_graphene} for graphene, we have 
\eq{
\label{eq:gamma_ave_Egeo_pz}
\left\langle \Gamma \right\rangle^{p_z,E-geo} = 0
}
since  $\Tr\left[     f_{i, p_z}^E(\bsl{k}_\shpa)  P_{p_z,n}(\bsl{k}_\shpa)     f_{i, p_z}^{geo}(\bsl{k}_\shpa)  \right]$ has the same form as $\Tr\left[     f_{i,}^E(\bsl{k})  P_{n}(\bsl{k})     f_{i}^{geo}(\bsl{k})  \right]$ in \eqnref{eq:Gamma_ave_mu_graphene_ini_2} for graphene and for $i=x,y$.

For $E-E$ part, we have
\eq{
\sum_{i=x,y} \Tr[ f_{i, p_z}^E(\bsl{k}_\shpa)  P_{p_z,n}(\bsl{k}_\shpa)     f_{i, p_z}^E(\bsl{k}_\shpa) ]  = - \gamma_{\pi,\shpa}^2 |\nabla_{\bsl{k}_\shpa}E_{p_z,n}(\bsl{k})|^2
}
and 
\eq{
\Tr[ f_{z, p_z}^E(\bsl{k}_\shpa)  P_{p_z,n}(\bsl{k}_\shpa)     f_{z, p_z}^E(\bsl{k}_\shpa) ] = -\gamma_{\pi,z}^2 |\partial_{k_z}\epsilon_{k_z,0}(k_z)|^2
}
resulting in
\eqa{
\label{eq:gamma_ave_EE_pz}
\left\langle \Gamma \right\rangle^{p_z,E-E} & = \frac{\hbar}{2 m_{\text{B}}} \frac{1}{D_{\pi}(\mu)} \sum_{\bsl{k}}^{\BZ}\sum_{n} \delta\left(\mu - E_{p_z,n}(\bsl{k}) \right)  \left[     \gamma_{\pi,\shpa}^2 |\nabla_{\bsl{k}_\shpa}E_{p_z,n}(\bsl{k})|^2 + \gamma_{\pi,z}^2 |\partial_{k_z}\epsilon_{k_z,0}(k_z)|^2  \right] \\
& = \frac{\hbar}{2 m_{\text{B}}} \frac{\V}{(2\pi)^3 D_{\pi}(\mu)} \sum_{n} \int_{FS_{p_z,n}} d\sigma_{\bsl{k}}\frac{1}{|\nabla_{\bsl{k}}  E_{p_z,n}(\bsl{k})|}   \left[     \gamma_{\pi,\shpa}^2 |\nabla_{\bsl{k}_\shpa}E_{p_z,n}(\bsl{k})|^2 + \gamma_{\pi,z}^2 |\partial_{k_z}\nabla_{\bsl{k}_\shpa}E_{p_z,n}(\bsl{k})|^2  \right] \ ,
}
where $\V$ is the volume of the system, and $FS_{p_z,n}$ is the FS surface given by the $n$th band.
Compared to the expression of $\left\langle \Gamma \right\rangle^{E-E}$ in graphene (\appref{app:labmda_contributions_graphene}), the major difference in $\left\langle \Gamma \right\rangle^{p_z,E-E}$ is the appearance of the $k_z$-derivative of the energy dispersion.

For $geo-geo$ part, we have
\eqa{
\label{eq:gamma_ave_geogeo_pz}
\left\langle \Gamma \right\rangle^{p_z,geo-geo} & =  (-)  \frac{\hbar}{2 m_{\text{B}}} \frac{1}{D_{\pi}(\mu)} \sum_{\bsl{k}}^{\BZ}\sum_{n} \delta\left(\mu - E_{p_z,n}(\bsl{k}) \right)\sum_{i=x,y} \Tr\left[     f_{i, p_z}^{geo}(\bsl{k}_\shpa)  P_{p_z,n}(\bsl{k}_\shpa)     f_{i, p_z}^{geo}(\bsl{k}_\shpa)  \right]  \\
& =  \gamma_{\pi,\shpa}^2\frac{\hbar}{2 m_{\text{B}}} \frac{1}{D_{\pi}(\mu)} \sum_{\bsl{k}}^{\BZ}\sum_{n} \delta\left(\mu - E_{p_z,n}(\bsl{k}) \right) \Delta E_{p_z}^2(\bsl{k}_\shpa) \sum_{i=x,y} \Tr\left[    P_{p_z,n}(\bsl{k}_\shpa)     \partial_{k_i} P_{p_z,n}(\bsl{k}_\shpa)  \partial_{k_i} P_{p_z,n}(\bsl{k}_\shpa)    \right]  \\
& = \gamma_{\pi,\shpa}^2\frac{\hbar}{2 m_{\text{B}}} \frac{\V}{(2\pi)^3 D_{\pi}(\mu)}  \sum_{n} \int_{FS_{p_z,n}} d\sigma_{\bsl{k}}\frac{\Delta E_{p_z}^2(\bsl{k}_\shpa)}{|\nabla_{\bsl{k}}  E_{p_z,n}(\bsl{k})|}  \Tr[g_{p_z,n}(\bsl{k}_\shpa)] \ ,
}
where $\Delta E_{p_z}^2(\bsl{k}_\shpa)$ is the absolute difference between two $p_z$ bands defined in \eqnref{eq:E_P_MgB2_pz}, and
\eq{
[g_{p_z,n}(\bsl{k}_\shpa)]_{ij}= \frac{1}{2}\Tr\left[   \partial_{k_i} P_{p_z,n}(\bsl{k}_\shpa)  \partial_{k_j} P_{p_z,n}(\bsl{k}_\shpa)    \right]\ .
}
Compared to the expression of $\left\langle \Gamma \right\rangle^{geo-geo}$ in graphene (\appref{app:labmda_contributions_graphene}), the major difference in $\left\langle \Gamma \right\rangle^{p_z,geo-geo}$ is again the appearance of the $k_z$-derivative of the energy dispersion in $|\nabla_{\bsl{k}}  E_{p_z,n}(\bsl{k})|$.

Eventually, by substituting \eqnref{eq:gamma_ave_Egeo_pz}, \eqnref{eq:gamma_ave_EE_pz}, \eqnref{eq:gamma_ave_geogeo_pz} into \eqnref{eq:lambda_X_MgB2}, we arrive at 
\eq{
\label{eq:lambda_pz_final}
\lambda_{\pi} = \lambda_{\pi,E} + \lambda_{\pi,geo}\ ,
}
where
\eqa{
\label{eq:lambda_E_pz}
\lambda_{\pi,E} & = \frac{2}{ N} \frac{1}{\hbar \mcomega} D_{\pi}(\mu) \frac{D_{\pi}(\mu)}{D(\mu)} \left\langle \Gamma \right\rangle^{p_z,E-E} \\
& = \frac{D_{\pi}(\mu)}{D(\mu)}\frac{1}{ m_{\text{B}} \mcomega} \frac{\Omega}{(2\pi)^3 }   \sum_{n} \int_{FS_{p_z,n}} d\sigma_{\bsl{k}}\frac{1}{|\nabla_{\bsl{k}}  E_{p_z,n}(\bsl{k})|}   \left[     \gamma_{\pi,\shpa}^2 |\nabla_{\bsl{k}_\shpa}E_{p_z,n}(\bsl{k})|^2 + \gamma_{\pi,z}^2 |\partial_{k_z}E_{p_z,n}(\bsl{k})|^2  \right]
}
is the energetic contribution, and
\eqa{
\label{eq:lambda_geo_pz}
\lambda_{\pi,geo} & = \frac{2}{ N} \frac{1}{\hbar \mcomega} D_{\pi}(\mu) \frac{D_{\pi}(\mu)}{D(\mu)} \left\langle \Gamma \right\rangle^{p_z,geo-geo} \\
& = \frac{1}{ m_{\text{B}} \mcomega} \frac{\Omega}{(2\pi)^3 } \frac{D_{\pi}(\mu)}{D(\mu)}  \gamma_{\pi,\shpa}^2 \sum_{n} \int_{FS_{p_z,n}} d\sigma_{\bsl{k}}\frac{\Delta E_{p_z}^2(\bsl{k}_\shpa)}{|\nabla_{\bsl{k}}  E_{p_z,n}(\bsl{k})|}  \Tr[g_{p_z,n}(\bsl{k}_\shpa)]
}
is the geometric contribution.
Compared to the expression of $\lambda_E$ and $\lambda_{geo}$ in graphene (\eqnref{eq:lambda_E_geo_graphene}), the major change in $\lambda_{\pi,E}$ is the appearance of the $\gamma_{\pi,z}$ term that involves the $k_z$-derivative of the energy dispersion, while the form of $\lambda_{\pi,geo}$ does not change much.

Similar to \eqnref{eq:g_theta}, we have
\eq{
\Tr[g_{p_z,n}(\bsl{k}_\shpa)] = \frac{1}{4} \sum_{i=x,y}\left|\ii e^{\ii \theta_{p_z}(\bsl{k}_\shpa)} \partial_{k_i} e^{-\ii \theta_{p_z}(\bsl{k}_\shpa)}\right|^2\ ,
}
where 
\eq{
e^{-\ii \theta_{p_z}(\bsl{k}_\shpa)} = \frac{d_{p_z,x}(\bsl{k}_\shpa) - \ii  d_{p_z,y}(\bsl{k}_\shpa) }{|d_{p_z,x}(\bsl{k}_\shpa) - \ii  d_{p_z,y}(\bsl{k}_\shpa) |}
}
with $d_{p_z,x}(\bsl{k}_\shpa) - \ii  d_{p_z,y}(\bsl{k}_\shpa) $ defined in \eqnref{eq:d_pz}.
Then, 
\eqa{
& \int_{FS_{p_z,n}} d\sigma_{\bsl{k}}\frac{\Delta E_{p_z}^2(\bsl{k}_\shpa)}{|\nabla_{\bsl{k}}  E_{p_z,n}(\bsl{k})|}  \Tr[g_{p_z,n}(\bsl{k}_\shpa)] \geq \frac{\left(\int_{FS_{p_z,n}} d\sigma_{\bsl{k}} \left|\ii e^{\ii \theta_{p_z}(\bsl{k}_\shpa)} \nabla_{\bsl{k}_{\shpa}} e^{-\ii \theta_{p_z}(\bsl{k}_\shpa)}\right|\right)^2}{ 4 \int_{FS_{p_z,n}} d\sigma_{\bsl{k}}\frac{|\nabla_{\bsl{k}}  E_{p_z,n}(\bsl{k})|}{\Delta E_{p_z}^2(\bsl{k}_\shpa)}} \\
& = \frac{ \int_{- \pi/c}^{ \pi/c} d k_z\left(\int_{FS_{p_z,n}^{2D,k_z}} d \sigma_{\bsl{k}_\shpa}  |\ii e^{\ii \theta_{p_z}(\bsl{k}_\shpa)} \nabla_{\bsl{k}_{\shpa}} e^{-\ii \theta_{p_z}(\bsl{k}_\shpa)}|\right)^2}{ 4 \int_{FS_{p_z,n}} d\sigma_{\bsl{k}}\frac{|\nabla_{\bsl{k}}  E_{p_z,n}(\bsl{k})|}{\Delta E_{p_z}^2(\bsl{k}_\shpa)}}\ ,
}
where $FS_{p_z,n}^{2D,k_z}$ is the intersection between $FS_{p_z,n}$ and the fixed-$k_z$ plane, and the first inequality is derived in a similar way to \eqnref{eq:holder_use_graphene} by using the H\"older's inequality in \eqnref{eq:holder}.
According to \eqnref{eq:PT_winding_number} and \eqnref{eq:h_B_pz}, if $FS_{p_z,n}^{2D,k_z}$ consists of two disconnected TR-related loops with each loop enclosing one nodal line in \eqnref{eq:nodal_line_pz},  
\eq{
\int_{FS_{p_z,n}^{2D,k_z}} d \sigma_{\bsl{k}_\shpa}  |\ii e^{\ii \theta_{p_z}(\bsl{k}_\shpa)} \nabla_{\bsl{k}_{\shpa}} e^{-\ii \theta_{p_z}(\bsl{k}_\shpa)}|\geq 2\pi (|W_{\K}| + |W_{-\K}| ) = 4\pi\ , 
}
where $W_{\pm\K}$ are the winding numbers of the nodal lines \eqnref{eq:nodal_line_pz}.
Suppose $FS_{p_z,n}^{2D,k_z}$ consists of two disconnected TR-related loops with each loop enclosing one nodal line in \eqnref{eq:nodal_line_pz} if and only if $k_z \in S_n$.
We can then label the length of  $S_n$ by $|S_n|$, and have  
\eqa{
& \int_{FS_{p_z,n}} d\sigma_{\bsl{k}}\frac{\Delta E_{p_z}^2(\bsl{k}_\shpa)}{|\nabla_{\bsl{k}}  E_{p_z,n}(\bsl{k})|}  \Tr[g_{p_z,n}(\bsl{k}_\shpa)] \geq \frac{|S_n| (\pi)^2 (|W_{\K}| + |W_{-\K}|)^2}{ \int_{FS_{p_z,n}} d\sigma_{\bsl{k}}\frac{|\nabla_{\bsl{k}}  E_{p_z,n}(\bsl{k})|}{\Delta E_{p_z}^2(\bsl{k}_\shpa)}}\ .
}
Eventually, we can can define 
\eqa{
\label{eq:lambda_topo_pz}
\lambda_{\pi,topo} = \frac{1}{ m_{\text{B}} \mcomega} \frac{\Omega}{(2\pi)^3 } \frac{D_{\pi}(\mu)}{D(\mu)}  \gamma_{\pi,\shpa}^2 \sum_{n} \frac{|S_n| (\pi)^2 }{ \int_{FS_{p_z,n}} d\sigma_{\bsl{k}}\frac{|\nabla_{\bsl{k}}  E_{p_z,n}(\bsl{k})|}{\Delta E_{p_z}^2(\bsl{k}_\shpa)}} (|W_{\K}| + |W_{-\K}|)^2\ ,
}
which satisfies 
\eq{
\lambda_{\pi,geo} \geq  \lambda_{\pi,topo} \ .
}
Numerical calculations of $\lambda_{\pi,E}$, $\lambda_{\pi,geo}$ and $\lambda_{\pi,topo}$ are in \appref{app:numerics_MgB2}.

\subsection{$p_x p_y$-Gaussian Approximation: Analytical Geometric and Topological Contributions to $\lambda_{\sigma}$}

\label{app:4-band_Gaussian_lambda_spxpy}

In this part, we derive the energetic and geometric contributions to $\lambda_{\sigma}$ from the GA.
We will only consider $p_x$ and $p_y$ orbitals, since we know the electron states near the Fermi level around $\Gamma$-A mainly originates from the $p_x$ and $p_y$ orbitals.
Therefore, we consider a $2\times 2$ hopping matrix function $t(\bsl{r})$, whose basis are $(p_x,p_y)$ of B atoms.
Specifically, $t_{\alpha_1\alpha_2}(\bsl{r})$ labels the hopping from the $\alpha_1$ orbital at $\bsl{r}_1$ to the $\alpha_2$ orbital at $\bsl{r}_1+\bsl{r}$, where $\alpha_1$, $\alpha_2$ takes values of $p_x$ and $p_y$ orbitals.
From the hopping function, the electron matrix Hamiltonian reads
\eqa{
\label{eq:h_pxpy_MgB2}
\left[h_{p_xp_y}(\bsl{k})\right]_{\bsl{\tau}_1\alpha_1, \bsl{\tau}_2\alpha_2} &  = \sum_{\bsl{R}} e^{-\ii (\bsl{R}+\bsl{\tau}_1-\bsl{\tau}_2)\cdot\bsl{k}}  t_{\alpha_1\alpha_2}(\bsl{R}+\bsl{\tau}_1-\bsl{\tau}_2)\ ,
}
where the basis is 
\eq{
\label{eq:MgB2_pxpy_basis}
c^\dagger_{p_xp_y,\bsl{k}} = ( c^\dagger_{\bsl{k},\text{B1},p_x} , c^\dagger_{\bsl{k},\text{B1},p_y}, c^\dagger_{\bsl{k},\text{B2},p_x} , c^\dagger_{\bsl{k},\text{B2},p_y} )\ .
}
With only $p_x$ and $p_y$ orbitals, the EPC  $F_{\bsl{\tau}i,p_xp_y}$ (general definition is in \eqnref{eq:H_el-ph_k}) reads
\eq{
\label{eq:f_p_xp_y_MgB2}
F_{\bsl{\tau}i,p_xp_y}(\bsl{k}_1,\bsl{k}_2) = \chi_{\bsl{\tau}}^{p_xp_y}  f_{i, p_xp_y}(\bsl{k}_2) - f_{i, p_xp_y}(\bsl{k}_1) \chi_{\bsl{\tau}}^{p_xp_y} \ , 
}
where
\eq{
\label{eq:chi_p_xp_y_MgB2}
\chi_{\bsl{\tau}_{\text{B1}}}^{p_xp_y} = \mat{ \mathds{1}_2 & \\ & 0_{2\times 2} }\ ,\ \chi_{\bsl{\tau}_{\text{B2}}}^{p_xp_y} = \mat{ 0_{2\times 2} & \\ & \mathds{1}_2 }\ ,
}
and
\eqa{
\left[f_{i,p_xp_y}(\bsl{k})\right]_{\bsl{\tau}_1\alpha_1, \bsl{\tau}_2\alpha_2} &  = \sum_{\bsl{R}} e^{-\ii (\bsl{R}+\bsl{\tau}_1-\bsl{\tau}_2)\cdot\bsl{k}} \left[\left.\partial_{r_i} t(\bsl{r})\right|_{\bsl{r}=\bsl{R}+\bsl{\tau}_1-\bsl{\tau}_2}\right]_{\alpha_1\alpha_2} \ .
}

Now we introduce the GA.
Similar to \appref{app:geo_EPC_MgB2_pi_Gaussian}, we use the idea of the linear combinations of atomic orbitals to derive the form of the hopping function $t(\bsl{r})$.
Since $p_x$ and $p_y$ orbitals do not form a rep of $\O(3)$, we again consider $\O(2)$ and $m_z$ symmetries for $t(\bsl{r})$.
Then, $\O(2)$ and $m_z$ symmetries gives
\eqa{
& \SO(2):  
\mat{ 
\cos(\theta) & -\sin(\theta) \\ 
\sin(\theta) & \cos(\theta) 
} t(\bsl{r})
\mat{ 
\cos(\theta) & \sin(\theta) \\ 
-\sin(\theta) & \cos(\theta) 
} = t(C_\theta \bsl{r}) \\
& m_y : 
\mat{ 
 1 & 0 \\ 
 0 & -1 
} t(\bsl{r})
\mat{ 
 1 & 0 \\ 
 0 & -1 
} = t(m_y \bsl{r}) \\
& m_x : 
\mat{ 
 -1 & 0 \\ 
 0 & 1 
} t(\bsl{r})
\mat{ 
 -1 & 0 \\ 
 0 & 1 
} = t(m_x \bsl{r}) \\
& m_z: t(\bsl{r}) = t(m_z \bsl{r}) \ , 
}
where 
\eq{
C_\theta= \mat{ 
\cos(\theta) & -\sin(\theta) & 0 \\ 
\sin(\theta) & \cos(\theta) & 0 \\
 0 & 0 & 1
}\ .
}
Combined with the Hermitian condition that $t^\dagger(\bsl{r}) = t(-\bsl{r})$, we arrive at
\eq{
\label{eq:hopping_function_Gaussian_MgB2_pxpy}
t(\bsl{r})=\mat{ 
\frac{-y}{\sqrt{x^2+y^2}} & \frac{-x}{\sqrt{x^2+y^2}} \\ 
\frac{x}{\sqrt{x^2+y^2}} & \frac{-y}{\sqrt{x^2+y^2}}
} t(0,-\sqrt{x^2+y^2},|z|) \mat{ 
\frac{-y}{\sqrt{x^2+y^2}} & \frac{x}{\sqrt{x^2+y^2}}\\ 
\frac{-x}{\sqrt{x^2+y^2}} & \frac{-y}{\sqrt{x^2+y^2}}
}\ ,
}
where 
\eq{
t(0,-\sqrt{x^2+y^2},|z|) = \mat{ 
pp_1(\sqrt{x^2+y^2},|z|) + pp_2(\sqrt{x^2+y^2},|z|) & 0 \\
0 & pp_1(\sqrt{x^2+y^2},|z|) - pp_2(\sqrt{x^2+y^2},|z|)
}\ ,
}
where $pp_1$ and $pp_2$ have the meaning of $(pp\pi+pp\sigma)/2$ and $(pp\pi-pp\sigma)/2$ bonding, respectively, $t(0,-\sqrt{x^2+y^2},|z|)$ is along $y$ since two B atoms in our chosen unit cell are on the $y$ axis, $pp\pi$ and $pp\sigma$ are respectively the $\pi$ bonding and $\sigma$ bonding among $p$ orbitals defined in \refcite{Slater1954LCAO}.
More explicitly, 
\eqa{
\label{eq:hopping_function_Gaussian_MgB2_pxpy_explicit}
t(\bsl{r}) & = pp_1(\sqrt{x^2+y^2},|z|) \mat{ 
 1 & 0 \\
 0 & 1
} + pp_2(\sqrt{x^2+y^2},|z|) \mat{ 
 \frac{y^2-x^2}{x^2+y^2} & \frac{-2 x y}{x^2+y^2}\\
 \frac{-2xy}{x^2+y^2} & \frac{x^2-y^2}{x^2+y^2}
}\ .
}
Based on \eqnref{eq:hopping_function_Gaussian_MgB2_pxpy_explicit}, we use the Gaussian function for the hopping decay in the $x-y$ plane owing to the fact that the dominant phonons originate from the in-plane ion motions, and introduce the following expression for the hopping function:
\eqa{
\label{eq:hopping_function_Gaussian_Form_MgB2_pxpy}
& pp_1(\sqrt{x^2+y^2},|z|) = t_{0} F(z^2) \exp\left[ \gamma_{\shpa} \frac{x^2+y^2}{2} \right]  \\
& pp_2(\sqrt{x^2+y^2},|z|) =t_{0}' F(z^2)  \frac{x^2+y^2}{a^2} \exp\left[ \gamma_{\shpa} \frac{x^2+y^2}{2} \right]\ ,
}
where the prefactor $x^2+y^2$ in $pp_2(\sqrt{x^2+y^2},|z|)$ are included to makes sure $t(\bsl{r})$ is smooth, and $F(z^2)$ is the decay function along $z$.
We note that we cannot choose $F(z^2)$ to be Gaussian, because $t_2 = pp_1(a/\sqrt{3},0)$ and $t_{B,p_xp_y,z}= pp_1(0,c)$ have different sign according to \eqnref{eq:MgB2_TB_values_sim}.

Next we derive $\nabla_{\bsl{r}} t(\bsl{r})$.
To do so, first we note that
\eqa{
& \partial_{r_i} \mat{ 
\frac{-y}{\sqrt{x^2+y^2}} & \frac{-x}{\sqrt{x^2+y^2}}\\ 
\frac{x}{\sqrt{x^2+y^2}} & \frac{-y}{\sqrt{x^2+y^2}}
} \\
& =  \mat{ 
\frac{-\delta_{iy}}{\sqrt{x^2+y^2}} & \frac{-\delta_{ix}}{\sqrt{x^2+y^2}}\\ 
\frac{\delta_{ix}}{\sqrt{x^2+y^2}} & \frac{-\delta_{iy}}{\sqrt{x^2+y^2}}
} - \frac{r_i}{x^2+y^2}(\delta_{ix}+\delta_{iy}) \mat{ 
\frac{-y}{\sqrt{x^2+y^2}} & \frac{-x}{\sqrt{x^2+y^2}}\\ 
\frac{x}{\sqrt{x^2+y^2}} & \frac{-y}{\sqrt{x^2+y^2}}
}\\
& = \delta_{ix}\frac{1}{x^2+y^2} \mat{ 
\frac{x y}{\sqrt{x^2+y^2}} & \frac{-y^2}{\sqrt{x^2+y^2}}\\ 
\frac{y^2}{\sqrt{x^2+y^2}} & \frac{ x y}{\sqrt{x^2+y^2}}
} + \delta_{iy} \frac{1}{x^2+y^2} \mat{ 
\frac{-x^2}{\sqrt{x^2+y^2}} & \frac{x y}{\sqrt{x^2+y^2}}\\
\frac{-x y}{\sqrt{x^2+y^2}} & \frac{-x^2}{\sqrt{x^2+y^2}}
}\\
& = \delta_{ix}\frac{-y}{x^2+y^2} \mat{ 
\frac{-x }{\sqrt{x^2+y^2}} & \frac{y}{\sqrt{x^2+y^2}}\\ 
\frac{-y}{\sqrt{x^2+y^2}} & \frac{ -x }{\sqrt{x^2+y^2}}
} + \delta_{iy} \frac{x}{x^2+y^2} \mat{ 
\frac{-x}{\sqrt{x^2+y^2}} & \frac{ y}{\sqrt{x^2+y^2}}\\ 
\frac{- y}{\sqrt{x^2+y^2}} & \frac{-x}{\sqrt{x^2+y^2}}
}\\
& = (\delta_{ix}+\delta_{iy})\frac{(\bsl{e}_z\times \bsl{r})_i}{x^2+y^2} \mat{  
 0 & -1\\ 
 1 & 0
} \mat{ 
 \frac{-y }{\sqrt{x^2+y^2}} & \frac{-x}{\sqrt{x^2+y^2}}\\ 
\frac{x}{\sqrt{x^2+y^2}} & \frac{ -y }{\sqrt{x^2+y^2}}
}\\
& = (\delta_{ix}+\delta_{iy})\frac{(\bsl{e}_z\times \bsl{r})_i}{x^2+y^2} (-\ii) \sigma_y \mat{ 
 \frac{-y }{\sqrt{x^2+y^2}} & \frac{-x}{\sqrt{x^2+y^2}}\\ 
\frac{x}{\sqrt{x^2+y^2}} & \frac{ -y }{\sqrt{x^2+y^2}}
}\ .
}
Furthermore, 
\eqa{
\partial_{r_i}t(0,-\sqrt{x^2+y^2},|z|) & =
(\delta_{i x}\gamma_{\shpa}  + \delta_{i y} \gamma_{\shpa}+ \delta_{i z} \widetilde{\gamma}(z^2)) r_i 
\mat{   
  pp_1(\sqrt{x^2+y^2},|z|) & 0 \\
  0 & pp_1(\sqrt{x^2+y^2},|z|)
}\\ 
& \quad +
(\delta_{i x}\gamma_{\shpa}  + \delta_{i y} \gamma_{\shpa}+ \delta_{i z} \widetilde{\gamma}(z^2)) r_i 
\mat{   
pp_2(\sqrt{x^2+y^2},|z|) & 0 \\
0 & -pp_2(\sqrt{x^2+y^2},|z|)
} \\
& \quad +
2 (\delta_{i x} + \delta_{i y})\frac{ r_i}{x^2+y^2} 
\mat{  
pp_2(\sqrt{x^2+y^2},|z|) & 0 \\
 0 & -pp_2(\sqrt{x^2+y^2},|z|)
} \\
 & =
(\delta_{i x}\gamma_{\shpa}  + \delta_{i y} \gamma_{\shpa}+ \delta_{i z} \widetilde{\gamma}(z^2))\ r_i\  t(0,-\sqrt{x^2+y^2},|z|) \\
& \quad +
(\delta_{i x} + \delta_{i y})\frac{ r_i}{x^2+y^2} \left[t(0,-\sqrt{x^2+y^2},|z|) - \sigma_y  t(0,-\sqrt{x^2+y^2},|z|) \sigma_y \right]\ ,
}
where
\eq{
\label{eq:gamma_t(z)}
\widetilde{\gamma}(z^2) = 2 \partial_{z^2} \log(|F(z^2)|)\ .
}
Combined with the fact that 
\eq{
\left[\sigma_y, \mat{ 
\frac{-y}{\sqrt{x^2+y^2}} & \frac{-x}{\sqrt{x^2+y^2}}\\ 
\frac{x}{\sqrt{x^2+y^2}} & \frac{-y}{\sqrt{x^2+y^2}}
}\right] = 0\ ,
} 
we arrive at  
\eqa{
\label{eq:gradient_hopping_function_Gaussian_MgB2_pxpy}
\partial_{r_i} t(\bsl{r}) & = \left[ \partial_{r_i} \mat{ 
 \frac{-y}{\sqrt{x^2+y^2}} & \frac{-x}{\sqrt{x^2+y^2}}\\
 \frac{x}{\sqrt{x^2+y^2}} & \frac{-y}{\sqrt{x^2+y^2}}
} \right] t(0,-\sqrt{x^2+y^2},|z|) \mat{ 
 \frac{-y}{\sqrt{x^2+y^2}} & \frac{x}{\sqrt{x^2+y^2}}\\ 
 \frac{-x}{\sqrt{x^2+y^2}} & \frac{-y}{\sqrt{x^2+y^2}}
}\\
& +\mat{ 
\frac{-y}{\sqrt{x^2+y^2}} & \frac{-x}{\sqrt{x^2+y^2}}\\
\frac{x}{\sqrt{x^2+y^2}} & \frac{-y}{\sqrt{x^2+y^2}}
}  t(0,-\sqrt{x^2+y^2},|z|)  \left[ \partial_{r_i} \mat{ 
\frac{-y}{\sqrt{x^2+y^2}} & \frac{x}{\sqrt{x^2+y^2}}\\ 
\frac{-x}{\sqrt{x^2+y^2}} & \frac{-y}{\sqrt{x^2+y^2}}
}\right]\\
& + \mat{ 
\frac{-y}{\sqrt{x^2+y^2}} & \frac{-x}{\sqrt{x^2+y^2}}\\ 
\frac{x}{\sqrt{x^2+y^2}} & \frac{-y}{\sqrt{x^2+y^2}}
}  \left[ \partial_{r_i} t(0,-\sqrt{x^2+y^2},|z|)\right]   \mat{ 
\frac{-y}{\sqrt{x^2+y^2}} & \frac{x}{\sqrt{x^2+y^2}}\\
\frac{-x}{\sqrt{x^2+y^2}} & \frac{-y}{\sqrt{x^2+y^2}}
}\\
& =  (\delta_{i x}\gamma_{\shpa}  + \delta_{i y} \gamma_{\shpa}+ \delta_{i z} \widetilde{\gamma}(z^2))\ r_i\  t(\bsl{r}) + (\delta_{ix}+\delta_{iy})\frac{(\bsl{e}_z\times \bsl{r})_i}{x^2+y^2} (-\ii)  [\sigma_y ,t(\bsl{r})] \\
& \qquad +
(\delta_{i x} + \delta_{i y})\frac{ r_i}{x^2+y^2} \left[t(\bsl{r}) - \sigma_y  t(\bsl{r}) \sigma_y \right]\ .
}

From \eqnref{eq:gradient_hopping_function_Gaussian_MgB2_pxpy}, now we can derive $f_{i,p_xp_y}(\bsl{k})$, which reads
\eqa{
\label{eq:g_Gaussian_MgB2_sp2_intermediate}
\left[f_{i,p_xp_y}(\bsl{k})\right]_{\bsl{\tau}_1\alpha_1, \bsl{\tau}_2\alpha_2} &  = \sum_{\bsl{R}} e^{-\ii (\bsl{R}+\bsl{\tau}_1-\bsl{\tau}_2)\cdot\bsl{k}} \left[\left.\partial_{r_i} t(\bsl{r})\right|_{\bsl{r}=\bsl{R}+\bsl{\tau}_1-\bsl{\tau}_2}\right]_{\alpha_1\alpha_2} \\
&  = \ii   (\delta_{ix} +\delta_{iy})\gamma_{i} \partial_{k_i}
   \left[h_{p_xp_y}(\bsl{k})\right]_{\bsl{\tau}_1\alpha_1,\bsl{\tau}_2\alpha_2} + f_{z,p_xp_y}(\bsl{k}) + \left[\Delta f_{i,p_xp_y}(\bsl{k})\right]_{\bsl{\tau}_1\alpha_1, \bsl{\tau}_2\alpha_2} \ ,
}
where 
\eq{
\label{eq:gamma_xy_pxpy_MgB2}
\gamma_x=\gamma_y = \gamma_{\sigma,\shpa} \ , 
}
\eqa{
\label{eq:gz_pxpy_MgB2}
 & f_{z,p_xp_y}(\bsl{k}) \\
 & \quad = \sum_{\bsl{R}} e^{-\ii (\bsl{R}+\bsl{\tau}_1-\bsl{\tau}_2)\cdot\bsl{k}} \left[\left.\delta_{i z} \widetilde{\gamma}(z^2) z\  t(\bsl{r}) \right|_{\bsl{r}=\bsl{R}+\bsl{\tau}_1-\bsl{\tau}_2}\right]_{\alpha_1\alpha_2}
}
and 
\eqa{
\label{eq:Delta_g_nonGaussian_pxpy_MgB2}
\left[\Delta f_{i,p_xp_y}(\bsl{k})\right]_{\bsl{\tau}_1\alpha_1, \bsl{\tau}_2\alpha_2} &  =   \sum_{\bsl{R}} e^{-\ii (\bsl{R}+\bsl{\tau}_1-\bsl{\tau}_2)\cdot\bsl{k}} \left[  \frac{ \left(\bsl{e}_z\times \bsl{r}\right)_i }{x^2+y^2} \left( -\ii  [\sigma_y ,t(\bsl{r})] \right)_{\alpha_1\alpha_2} +  \frac{(x, y, 0)_i}{x^2+y^2} \left( t(\bsl{r})  - \sigma_y t(\bsl{r})   \sigma_y \right)_{\alpha_1\alpha_2}\right]_{\bsl{r}=\bsl{R}+\bsl{\tau}_1-\bsl{\tau}_2}.
}
In \eqnref{eq:g_Gaussian_MgB2_sp2_intermediate}, the $\partial_{k_i} h_{p_xp_y}(\bsl{k})$ term comes from the Gaussian factor of the hopping in \eqnref{eq:hopping_function_Gaussian_MgB2_pxpy_explicit}, while $f_{z,p_xp_y}(\bsl{k})$ and $\Delta f_{i,p_xp_y}(\bsl{k})$ do not.
As the final step of the in-plane GA, we need to neglect $\Delta f_{i,p_xp_y}(\bsl{k})$.
Next we will show that (i) under the short-ranged hopping approximation, $f_{z,p_xp_y}(\bsl{k})$ will have the $\partial_{k_z} h_{p_xp_y}(\bsl{k})$ form, and (ii) if we further adopt small $|\bsl{k}_\shpa|$ approximation, $\Delta f_{i,p_xp_y}(\bsl{k})$ is negligible.

\subsubsection{Short-range Hopping and Small $|\bsl{k}_\shpa|$ Approximation}

As discussed in \appref{app:H_el_MgB2}, the short-ranged hopping approximation works well for electron states near the Fermi energy in \mgb.
Explicitly, we only need to consider hopping among B atoms along in-plane $\pm \bsl{\delta}_j$ in \eqnref{eq:delta_j_graphene} and out-of-plane $\pm\bsl{a}_3$ (\figref{fig:MgB2_structure}).
Within the GA, it means that $\gamma_{\sigma,\shpa}$ and $\gamma_{\sigma,z}$ in \eqnref{eq:hopping_function_Gaussian_MgB2_pxpy_explicit} are so large that longer-range hopping are very close to zero and are negligible.

Let us discuss $f_{z,p_xp_y}(\bsl{k})$ first.
From the expression of $f_{z,p_xp_y}(\bsl{k})$ in \eqnref{eq:gz_pxpy_MgB2}, $f_{z,p_xp_y}(\bsl{k})$ cannot have contribution from $\pm \bsl{\delta}_j$, since $\pm \bsl{\delta}_j$ have zero $z$ components.
So $f_{z,p_xp_y}(\bsl{k})$ only has contribution from $\pm\bsl{a}_3 = \pm (0,0,c)$.
As a result, we can safely replace $\widetilde{\gamma}(z^2)$ in \eqnref{eq:gz_pxpy_MgB2} by $\widetilde{\gamma}(c^2)$, and obtain
\eqa{
\label{eq:gz_pxpy_MgB2_short_ranged}
f_{z,p_xp_y}(\bsl{k}) & = \sum_{\bsl{R}} e^{-\ii (\bsl{R}+\bsl{\tau}_1-\bsl{\tau}_2)\cdot\bsl{k}} \left[\left.\delta_{i z} \widetilde{\gamma}(z^2) z\  t(\bsl{r}) \right|_{\bsl{r}=\bsl{R}+\bsl{\tau}_1-\bsl{\tau}_2}\right]_{\alpha_1\alpha_2} \\
& = \sum_{\bsl{R}} e^{-\ii (\bsl{R}+\bsl{\tau}_1-\bsl{\tau}_2)\cdot\bsl{k}} \left[\left.\delta_{i z} \widetilde{\gamma}(c^2) z\  t(\bsl{r}) \right|_{\bsl{r}=\bsl{R}+\bsl{\tau}_1-\bsl{\tau}_2}\right]_{\alpha_1\alpha_2} \\
& = \delta_{i z} \gamma_{\sigma,z}\ii \partial_{k_z}\left[h_{p_xp_y}(\bsl{k})\right]_{\bsl{\tau}_1\alpha_1,\bsl{\tau}_2\alpha_2} \ ,
}
where
\eq{
\label{eq:gamma_z_pxpy_MgB2}
\gamma_{\sigma,z} = \widetilde{\gamma}(c^2) \ .
}

Now we discuss $\Delta f_{i,p_xp_y}(\bsl{k})$ first.
According to the form of the hopping function in \eqnref{eq:hopping_function_Gaussian_MgB2_pxpy} and \eqnref{eq:hopping_function_Gaussian_MgB2_pxpy_explicit}, $[t(\bsl{r}),\sigma_y]$ and $ t(\bsl{r}) -\sigma_y t(\bsl{r})\sigma_y$ approach zero as $x^2+y^2$, when $x^2+y^2$ limits to zero.
Then, we have 
\eq{
\lim_{x^2+y^2\rightarrow 0 }\left[\frac{ \left(\bsl{e}_z\times \bsl{r}\right)_i }{x^2+y^2}    [\sigma_y ,t(\bsl{r})]  +  \frac{(x, y, 0)_i}{x^2+y^2}  \left( t(\bsl{r})  - \sigma_y t(\bsl{r})   \sigma_y \right) \right]= 0\ ,
}
meaning that $\Delta f_{i,p_xp_y}(\bsl{k})$ cannot have contribution from $\pm\bsl{a}_3$.
So the only contribution to \eqnref{eq:Delta_g_nonGaussian_pxpy_MgB2} comes from the in-plane $\pm \bsl{\delta}_j$ hopping terms.
Since $\pm \bsl{\delta}_j$ all give $x^2 + y^2 = a^2 /3$, we can safely replace the $\frac{1}{x^2 + y^2}$ factor in \eqnref{eq:Delta_g_nonGaussian_pxpy_MgB2} by $3/a^2$, resulting in 
\eqa{
& \left[\Delta f_{i,p_xp_y}(\bsl{k})\right]_{\bsl{\tau}_1\alpha_1, \bsl{\tau}_2\alpha_2} \\
& =   \sum_{\bsl{R}} e^{-\ii (\bsl{R}+\bsl{\tau}_1-\bsl{\tau}_2)\cdot\bsl{k}} \left[  \frac{ \left(\bsl{e}_z\times \bsl{r}\right)_i }{x^2+y^2} \left( -\ii [\sigma_y ,t(\bsl{r})] \right)_{\alpha_1\alpha_2} +  \frac{(x, y, 0)_i}{x^2+y^2} \left( t(\bsl{r})  - \sigma_y t(\bsl{r})   \sigma_y \right)_{\alpha_1\alpha_2}\right]_{\bsl{r}=\bsl{R}+\bsl{\tau}_1-\bsl{\tau}_2}\\
& =   \sum_{\bsl{R}} e^{-\ii (\bsl{R}+\bsl{\tau}_1-\bsl{\tau}_2)\cdot\bsl{k}} \left[  \frac{ \left(\bsl{e}_z\times \bsl{r}\right)_i }{a^2/3} \left( -\ii  [\sigma_y ,t(\bsl{r})] \right)_{\alpha_1\alpha_2} +  \frac{(x, y, 0)_i}{a^2/3} \left( t(\bsl{r})  - \sigma_y t(\bsl{r})   \sigma_y \right)_{\alpha_1\alpha_2}\right]_{\bsl{r}=\bsl{R}+\bsl{\tau}_1-\bsl{\tau}_2} \\
& = \frac{3 \ii}{a^3} \left[ \left(\bsl{e}_z\times \nabla_{\bsl{k}}\right)_i  \left( -\ii [\sigma_y ,\left[h_{p_xp_y}(\bsl{k})\right]_{\bsl{\tau}_1\bsl{\tau}_2}] \right)_{\alpha_1\alpha_2} + (\partial_{k_x}, \partial_{k_y}, 0)_i \left( \left[h_{p_xp_y}(\bsl{k})\right]_{\bsl{\tau}_1\bsl{\tau}_2}  - \sigma_y \left[h_{p_xp_y}(\bsl{k})\right]_{\bsl{\tau}_1\bsl{\tau}_2}  \sigma_y \right)_{\alpha_1\alpha_2}\right]\ ,
}
where $h_{p_xp_y}(\bsl{k})$ is defined in \eqnref{eq:h_pxpy_MgB2}, and $\left[h_{p_xp_y}(\bsl{k})\right]_{\bsl{\tau}_1\bsl{\tau}_2}$ is the $\bsl{\tau}_1\bsl{\tau}_2$ block ($2 \times 2$) of $h_{p_xp_y}(\bsl{k})$.
In the matrix form, we have
\eq{
\label{eq:Delta_g_nonGaussian_pxpy_MgB2_short_range}
 \Delta f_{i,p_xp_y}(\bsl{k}) = \frac{3 \ii}{a^3}  \left[ -\ii\left(\bsl{e}_z\times \nabla_{\bsl{k}}\right)_i   [\tau_0\sigma_y , h_{p_xp_y}(\bsl{k}) ] + (\partial_{k_x}, \partial_{k_y}, 0)_i \left(  h_{p_xp_y}(\bsl{k})  - \tau_0 \sigma_y  h_{p_xp_y}(\bsl{k})   \tau_0\sigma_y \right) \right] \ ,
}
where $\tau_0$ is the identity matrix in the sublattice subspace.

Now we adopt the small $|\bsl{k}_\shpa|$ approximation.
As discussed in \appref{app:H_el_MgB2}, the Fermi surfaces of the $\sigma$-binding electron states are around $\Gamma$-A, which have small $|\bsl{k}_\shpa|$.
Then, we consider $h_{p_xp_y}(\bsl{k})$ to first order in $\bsl{k}_\shpa$, which is related to the effective Hamiltonian $h_{eff}(\bsl{k})$ in \eqnref{eq:MgB2_H_eff_mat_form} by a basis transformation $R_{eff}$.
(The reason for us to use different basis for $h_{p_xp_y}(\bsl{k})$ and $h_{eff}(\bsl{k})$ is that the atomic basis (\eqnref{eq:MgB2_pxpy_basis}) for $h_{p_xp_y}(\bsl{k})$ is convenient for the study of EPC since EPC cares about atomic motions, while the parity-eigenbasis (\eqnref{eq:MgB2_eff_basis}) for $h_{eff}(\bsl{k})$ is convenient for band topology and band geometry since the parity-eigenbasis are energy eigenstates for $\bsl{k}_\shpa = 0$.)
Explicitly, under the first-order $|\bsl{k}_\shpa|$ perturbation, we have
\eq{
\label{eq:h_pxpy_short_range_small_kshpa}
h_{p_xp_y}(\bsl{k}) = R_{eff} h_{eff}(\bsl{k}) R_{eff}^\dagger\ ,
}
where $ h_{eff}(\bsl{k})$ is in \eqnref{eq:MgB2_H_eff_mat_form}, 
\eq{
\label{eq:R_eff}
R_{eff} = \left(
\begin{array}{cccc}
 \frac{1}{\sqrt{2}} & 0 & \frac{1}{\sqrt{2}} & 0 \\
 0 & \frac{1}{\sqrt{2}} & 0 & \frac{1}{\sqrt{2}} \\
 -\frac{1}{\sqrt{2}} & 0 & \frac{1}{\sqrt{2}} & 0 \\
 0 & -\frac{1}{\sqrt{2}} & 0 & \frac{1}{\sqrt{2}} \\
\end{array}
\right) = \frac{1}{\sqrt{2}} \left[ \tau_0 \sigma_0 + \ii \tau_y \sigma_0 \right]\ ,
}
and $c^\dagger_{p_xp_y,\bsl{k}} R_{eff} = c^\dagger_{eff,\bsl{k}}$ with $c^\dagger_{p_xp_y,\bsl{k}}$ in 
\eqnref{eq:MgB2_pxpy_basis} and $c^\dagger_{eff,\bsl{k}}$ in \eqnref{eq:MgB2_eff_basis}.
Then, $\Delta f_{i,p_xp_y}(\bsl{k})$ in \eqnref{eq:Delta_g_nonGaussian_pxpy_MgB2_short_range} becomes 
\eqa{
\label{eq:Delta_g_nonGaussian_pxpy_MgB2_short_range_small_kshpa}
R_{eff}^\dagger \Delta f_{i,p_xp_y}(\bsl{k})  R_{eff} &  = \frac{3 \ii}{a^3}   \left[ -\ii \left(\bsl{e}_z\times \nabla_{\bsl{k}}\right)_i   [\tau_0\sigma_y , h_{eff}(\bsl{k}) ] + (\partial_{k_x}, \partial_{k_y}, 0)_i \left(  h_{eff}(\bsl{k})  - \tau_0 \sigma_y  h_{eff}(\bsl{k})   \tau_0\sigma_y \right)  \right] \\
&  = \frac{3 \ii}{a^3} (\delta_{ix}+\delta_{iy}) \left[-\ii \sum_{i'=x,y}\epsilon^{i' i} \partial_{k_{i'}}   [\tau_0\sigma_y , v a (k_x \tau_y \sigma_x+k_y \tau_y \sigma_z) ] + 2   v a  \partial_{k_i} (k_x \tau_y \sigma_x+k_y \tau_y \sigma_z)  \right]\\
&  = \frac{3 \ii}{a^3} (\delta_{ix}+\delta_{iy}) \left[ -\ii \sum_{i'=x,y}\epsilon^{i' i} v a  [\tau_0\sigma_y ,   ( \tau_y \sigma_x , \tau_y \sigma_z)_{i'}  ] + 2   v a  ( \tau_y \sigma_x , \tau_y \sigma_z)_i  \right]\\
&  = \frac{ \ii 6 v a}{a^3} (\delta_{ix}+\delta_{iy}) \left[ \sum_{i'=x,y}\epsilon^{i' i}   ( - \tau_y \sigma_z , \tau_y \sigma_x)_{i'}  + ( \tau_y \sigma_x , \tau_y \sigma_z)_i  \right]\\
&  = \frac{ \ii 6 v a}{a^3} (\delta_{ix}+\delta_{iy}) \left[ -( \tau_y \sigma_x , \tau_y \sigma_z)_i  + ( \tau_y \sigma_x , \tau_y \sigma_z)_i  \right]\\
& = 0\ .
}
Therefore, under the short-range hopping and small $|\bsl{k}_\shpa|$ approximation, $\Delta f_{i,p_xp_y}(\bsl{k})$ in \eqnref{eq:g_Gaussian_MgB2_sp2_intermediate} can be neglected owing to the form of the GA that we choose (\eqnref{eq:hopping_function_Gaussian_Form_MgB2_pxpy}).
Combined with \eqnref{eq:gz_pxpy_MgB2_short_ranged}, we arrive at
\eq{
\label{eq:g_Gaussian_MgB2_eff}
f_{i,eff}(\bsl{k}) = R_{eff}^\dagger f_{i,p_xp_y}(\bsl{k}) R_{eff}   = \ii  \gamma_{i} \partial_{k_i}   h_{eff}(\bsl{k}) \ ,
}
where $\gamma_{i=x,y}$ is in \eqnref{eq:gamma_xy_pxpy_MgB2}, $\gamma_{\sigma,z}$ is in \eqnref{eq:gamma_z_pxpy_MgB2}, $R_{eff}$ is in \eqnref{eq:R_eff}, and $h_{eff}(\bsl{k})$ is in \eqnref{eq:MgB2_H_eff_mat_form}.

We now show that the GA in \eqnref{eq:hopping_function_Gaussian_Form_MgB2_pxpy} reproduces exactly the electron and EPC Hamiltonian under the short-ranged hopping and small-$|\bsl{k}_\shpa|$ approximations.
Specifically, the parameters in the electron and EPC Hamiltonian should be related to the parameters in the GA \eqnref{eq:hopping_function_Gaussian_Form_MgB2_pxpy}, and the question is whether \eqnref{eq:hopping_function_Gaussian_Form_MgB2_pxpy} imposes any extra constraints on the hopping after adopting the short-ranged-hopping and small-$|\bsl{k}_\shpa|$ approximations.
In \eqnref{eq:MgB2_H_eff_mat_form}, there are three independent parameters besides the global energy shift: $t_{B,p_x p_y, z}$, $m$, $v$.
They are related to the parameters in \eqnref{eq:hopping_function_Gaussian_Form_MgB2_pxpy} by 
\eqa{
& t_{B,p_x p_y, z} = t_0 F(c^2) \ ,\ m = -3 t_2 = t_0 F(0) \exp[\frac{\gamma_{\sigma,\shpa}}{6} a^2]\ ,\ v= -\frac{\sqrt{3}}{2} t_3 = -\frac{1}{2\sqrt{3}} t_0' F(0) \exp[\frac{\gamma_{\sigma,\shpa}}{6} a^2]\ .
}
On the other hand, for EPC, the GA gives 
\eq{
\label{eq:g_i_MgB2_eff_explicit}
f_{i,eff}(\bsl{k}) = \ii   2 c \gamma_{\sigma,z} t_{B,p_xp_y,z}  \sin(k_z c) \delta_{iz} + \ii \gamma_{\sigma,\shpa} v a \tau_y \sigma_x \delta_{ix} + \ii \gamma_{\sigma,\shpa} v a \tau_y \sigma_z \delta_{iy} \ .
}
By projecting the EPC $f_{i,s p_x p_y}(\bsl{k})$ in \eqnref{eq:g_X_MgB2} to the basis \eqnref{eq:MgB2_eff_basis} and by only considering the zeroth order of $\bsl{k}_\shpa$, we obtain 
\eq{
f_{i,eff}(\bsl{k}) = -\ii 2 c \hat{\gamma}_9  \sin(k_z c) \delta_{iz} + (-\ii) \frac{\sqrt{3}}{2} a (\hat{\gamma}_3 -\hat{\gamma}_7 )\tau_y \sigma_x \delta_{ix} + (-\ii) \frac{\sqrt{3}}{2} a (\hat{\gamma}_3 -\hat{\gamma}_7 ) \tau_y \sigma_z \delta_{iy} \ ,
}
meaning that 
\eq{
\label{eq:gamma_gamma_hat_Gaussian_pxpy_MgB2}
\hat{\gamma}_9 = -\gamma_{\sigma,z}  t_0 F(c^2)\ ,\    \hat{\gamma}_3 -\hat{\gamma}_7  =    \frac{1}{3} \gamma_{\sigma,\shpa} t_0' F(0) \exp[\frac{\gamma_{\sigma,\shpa}}{6} a^2] \ . 
}
Since $ \gamma_{\sigma,\shpa} , \gamma_{\sigma,z} , F(c^2) , F(0) ,  t_0'$ are independent parameters, the GA  \eqnref{eq:hopping_function_Gaussian_Form_MgB2_pxpy} does not impose any extra constraints on the Hamiltonian and EPC after adopting the short-ranged-hopping and small-$|\bsl{k}_\shpa|$ approximations.
Therefore, the GA \eqnref{eq:hopping_function_Gaussian_Form_MgB2_pxpy} is exact after adopting the short-ranged-hopping and small-$|\bsl{k}_\shpa|$ approximations, meaning that the electron Hamiltonian and EPC can always be precisely reproduced by GA after adopting the short-ranged-hopping and small-$|\bsl{k}_\shpa|$ approximations. 
From \eqnref{eq:gamma_gamma_hat_Gaussian_pxpy_MgB2}, we have the following relations:
\eq{
\label{eq:gamma_z_sp2_MgB2}
\gamma_{\sigma,z} =\frac{\hat{\gamma}_9}{t_{B,p_x p_y,z}} 
}
and
\eq{
\label{eq:gamma_spxpy_geo}
\gamma_{\sigma,\shpa} =  \frac{\hat{\gamma}_3-\hat{\gamma}_7}{t_3}\ .
}

Based on the form of $f_{i,eff}(\bsl{k})$ in \eqnref{eq:g_Gaussian_MgB2_eff}, we can define the energetic and geometric parts of $f_{i,eff}(\bsl{k})$, which read
\eqa{
\label{eq:g_i_MgB2_eff_E_geo}
& f_{i,eff}^{E}(\bsl{k}) = \ii  \gamma_{i} \sum_{n=1,2} [\partial_{k_i}   E_{eff,n}(\bsl{k})] P_{eff,n}(\bsl{k}_\shpa) \\
& f_{i,eff}^{geo}(\bsl{k}) = \ii  \gamma_{i} \sum_{n=1,2}    E_{eff,n}(\bsl{k}) [\partial_{k_i} P_{eff,n}(\bsl{k}_\shpa) ] \ ,
}
where $E_{eff,n}(\bsl{k})$ are the doubly degenerate bands in \eqnref{eq:E_eff_MgB2}, and $P_{eff,n}(\bsl{k})$ are the projection matrices in \eqnref{eq:P_eff_MgB2}.
To define the energetic and geometric contributions to $\lambda$, let us first look at the expression of $\left\langle \Gamma \right\rangle^{sp_xp_y}$ in \eqnref{eq:Gamma_ave_X_MgB2} under the Gaussian, short-ranged-hopping and small-$|\bsl{k}_\shpa|$ approximation with basis of $h^{eff}(\bsl{k})$ in \eqnref{eq:MgB2_eff_basis}.
Specifically, under the Gaussian, short-ranged-hopping and small-$|\bsl{k}_\shpa|$ approximation, we have
\eqa{
\label{eq:Gamma_eff_pz_MgB2}
\left\langle \Gamma \right\rangle^{sp_xp_y}  & = \frac{1}{D^2_{\sigma}(\mu)} \sum_{\bsl{k}_1,\bsl{k}_2}^{\BZ}\delta\left(\mu - E_{eff,1}(\bsl{k}_1) \right) \delta\left(\mu - E_{eff,1}(\bsl{k}_2) \right) \Gamma_{11  }^{eff}(\bsl{k}_1,\bsl{k}_2) \ ,
}
where
\eq{
D_{\sigma}(\mu) = 2 \sum_{\bsl{k}}^{\BZ}\delta\left(\mu - E_{eff,1}(\bsl{k}) \right)
}
with the factor $2$ comes from the double degeneracy,
\eq{
\Gamma_{11}^{eff}(\bsl{k}_1,\bsl{k}_2) = \frac{\hbar}{2 m_{B}} \sum_{\delta=\pm,i}   \Tr\left[ P_{eff,1}(\bsl{k}_{1,\shpa})  F_{\delta i,eff}(\bsl{k}_1,\bsl{k}_2) P_{eff,1}(\bsl{k}_{2,\shpa})   F_{\delta i,eff}^\dagger(\bsl{k}_1,\bsl{k}_2) \right]\ ,
}
and
\eq{
F_{\delta i,eff}(\bsl{k}_1,\bsl{k}_2) = \chi^{eff}_{\delta} f_{i,eff}(\bsl{k}_2) - f_{i,eff}(\bsl{k}_1) \chi^{eff}_{\delta}\ .
}
Moreover, $\chi^{eff}_{\pm} = R_{eff}^\dagger \frac{1}{\sqrt{2}} (\chi^{p_x p_y}_{\text{B1}} \pm \chi^{p_x p_y}_{\text{B2}})  R_{eff}$, meaning that
\eq{
\label{eq:chi_pm_eff}
\chi^{eff}_{+} = \frac{1}{\sqrt{2}} \ ,\ \chi^{eff}_{-} = \frac{1}{\sqrt{2}} \tau_x\sigma_0\ .
}
Then, we have
\eqa{
\label{eq:Gamma_eff_parts_MgB2}
& \Gamma_{11}^{eff,E-E}(\bsl{k}_1,\bsl{k}_2) \\
& \quad = \frac{\hbar}{2 m_{B}} \sum_{\delta=\pm,i}   \Tr\left[ P_{eff,1}(\bsl{k}_{1,\shpa})  (\chi^{eff}_{\delta} f_{i,eff}^{E}(\bsl{k}_2) - f_{i,eff}^{E}(\bsl{k}_1) \chi^{eff}_{\delta}) P_{eff,1}(\bsl{k}_{2,\shpa})    (\chi^{eff}_{\delta} f_{i,eff}^{E}(\bsl{k}_1) -f_{i,eff}^{E}(\bsl{k}_2) \chi^{eff}_{\delta}  ) \right]\\
& \Gamma_{11}^{eff,E-geo}(\bsl{k}_1,\bsl{k}_2) \\
& \quad = \frac{\hbar}{2 m_{B}} \sum_{\delta=\pm,i}   \Tr\left[ P_{eff,1}(\bsl{k}_{1,\shpa})  (\chi^{eff}_{\delta} f_{i,eff}^{E}(\bsl{k}_2) - f_{i,eff}^{E}(\bsl{k}_1) \chi^{eff}_{\delta}) P_{eff,1}(\bsl{k}_{2,\shpa})    (\chi^{eff}_{\delta} f_{i,eff}^{geo}(\bsl{k}_1) - f_{i,eff}^{geo}(\bsl{k}_2) \chi^{eff}_{\delta}) \right]+c.c.\\
& \Gamma_{11}^{eff,geo-geo}(\bsl{k}_1,\bsl{k}_2) \\
& \quad = \frac{\hbar}{2 m_{B}} \sum_{\delta=\pm,i}   \Tr\left[ P_{eff,1}(\bsl{k}_{1,\shpa})  (\chi^{eff}_{\delta} f_{i,eff}^{geo}(\bsl{k}_2) - f_{i,eff}^{geo}(\bsl{k}_1) \chi^{eff}_{\delta}) P_{eff,1}(\bsl{k}_{2,\shpa})    (\chi^{eff}_{\delta} f_{i,eff}^{geo}(\bsl{k}_1) -f_{i,eff}^{geo}(\bsl{k}_2)\chi^{eff}_{\delta}  ) \right]\ ,
}
resulting in
\eq{
\label{eq:Gamma_ave_sp2_parts}
\left\langle \Gamma \right\rangle^{sp_xp_y, label}  = \frac{1}{D^2_{\sigma}(\mu)} \sum_{\bsl{k}_1,\bsl{k}_2}^{\BZ}\delta\left(\mu - E_{eff,1}(\bsl{k}_1) \right) \delta\left(\mu - E_{eff,1}(\bsl{k}_2) \right) \Gamma_{11  }^{eff, label}(\bsl{k}_1,\bsl{k}_2)\ ,
}
where $label=E-E, E-geo, geo-geo$.

Finally, we arrive at
\eq{
\label{eq:lambda_sp2_E_geo}
\lambda_{\sigma} = \lambda_{\sigma,E} + \lambda_{\sigma,geo} + \lambda_{\sigma,E-geo}\ , 
}
where 
\eqa{
\label{eq:lambda_sp2_parts}
& \lambda_{\sigma,E} = \frac{2}{ N} \frac{1}{\hbar \mcomega} D_{\sigma}(\mu) \frac{D_{\sigma}(\mu)}{D(\mu)} \left\langle \Gamma \right\rangle^{sp_xp_y,E-E} \\
& \lambda_{\sigma,E-geo} = \frac{2}{ N} \frac{1}{\hbar \mcomega} D_{\sigma}(\mu) \frac{D_{\sigma}(\mu)}{D(\mu)} \left\langle \Gamma \right\rangle^{sp_xp_y,E-geo}\\
& \lambda_{\sigma,geo} = \frac{2}{ N} \frac{1}{\hbar \mcomega} D_{\sigma}(\mu) \frac{D_{\sigma}(\mu)}{D(\mu)} \left\langle \Gamma \right\rangle^{sp_xp_y,geo-geo}\ ,
}
and $\frac{D_{\sigma}(\mu)}{D(\mu)}$ represents the portion of the electron DOS at the Fermi level for $sp_xp_y$ orbitals.

As shown in \eqnref{eq:g_i_MgB2_eff_explicit}, the EPC coupling $f_{i,eff}(\bsl{k})$ does not depend on $\bsl{k}_\shpa$.
On the other hand, $f_{i,eff}^{E}(\bsl{k})$ and $f_{i,eff}^{geo}(\bsl{k})$ defined in \eqnref{eq:g_i_MgB2_eff_E_geo} have $\bsl{k}_\shpa$ dependence, since the expression of the energy bands (\eqnref{eq:E_eff_MgB2}) and the projection matrix (\eqnref{eq:P_eff_MgB2}) involves infinite order of $\bsl{k}_\shpa$.
Nevertheless, the $|\bsl{k}_\shpa|$-independent $f_{i,eff}(\bsl{k})$ means that $f_{i,eff}^{E/geo}(\bsl{k})$ as well as  $\Gamma_{11}^{eff,E-E/E-geo/geo-geo}(\bsl{k}_1,\bsl{k}_2)$ in \eqnref{eq:Gamma_eff_parts_MgB2} are only reliable to zeroth order in $\bsl{k}_\shpa$, since the high-order-$\bsl{k}_\shpa$ correction of $f_{i,eff}(\bsl{k})$ would change the higher-order-$\bsl{k}_\shpa$  terms in $f_{i,eff}^{E/geo}(\bsl{k})$ and $\Gamma_{11}^{eff,E-E/E-geo/geo-geo}(\bsl{k}_1,\bsl{k}_2)$. 

In the following, we may use the full expression of $f_{i,eff}^{E}(\bsl{k})$ and $f_{i,eff}^{geo}(\bsl{k})$ in \eqnref{eq:g_i_MgB2_eff_E_geo} to derive $\Gamma_{11}^{eff,E-E/E-geo/geo-geo}(\bsl{k}_1,\bsl{k}_2)$ as intermediate steps.
But eventually, we will only keep the energetic and geometric contributions to $\lambda_{\sigma}$ from the zeroth-order-$|\bsl{k}_\shpa|$ part of $\Gamma_{11}^{eff,E-E/E-geo/geo-geo}(\bsl{k}_1,\bsl{k}_2)$.

\subsubsection{$\lambda_{\sigma,E}$}
To address $\Gamma_{11}^{eff,E-E}$, we first note that
\eq{
\label{eq:chi_-_Peff_MgB2}
\chi_-^{eff} P_{eff,1}(\bsl{k}_\shpa) = P_{eff,2}(\bsl{k}) \chi_-^{eff}\ ,
}
according to the expression of $\chi_-^{eff}$ in \eqnref{eq:chi_pm_eff} and $P_{eff,n}(\bsl{k}_\shpa)$ in \eqnref{eq:P_eff_MgB2}.
The simple reason for \eqnref{eq:chi_-_Peff_MgB2} is that $\chi_-^{eff}$ anti-commutes with the matrix part of $h_{eff}(\bsl{k})$ in \eqnref{eq:MgB2_H_eff_mat_form}.
Then, we have
\eqa{
& \Gamma_{11}^{eff,E-E}(\bsl{k}_1,\bsl{k}_2) \\
& \quad = \frac{\hbar}{2 m_{B}} \sum_{\delta=\pm,i}   \Tr\left[ P_{eff,1}(\bsl{k}_{1,\shpa})  (\chi^{eff}_{\delta} f_{i,eff}^{E}(\bsl{k}_2) - f_{i,eff}^{E}(\bsl{k}_1) \chi^{eff}_{\delta}) P_{eff,1}(\bsl{k}_{2,\shpa})    (\chi^{eff}_{\delta} f_{i,eff}^{E}(\bsl{k}_1) - f_{i,eff}^{E}(\bsl{k}_2) \chi^{eff}_{\delta} ) \right] \\
& \quad = \left\{ -\frac{\hbar}{2 m_{B}} \sum_{\delta=\pm,i}   \Tr\left[ P_{eff,1}(\bsl{k}_{1,\shpa})  f_{i,eff}^{E}(\bsl{k}_1) \chi^{eff}_{\delta} P_{eff,1}(\bsl{k}_{2,\shpa})     \chi^{eff}_{\delta} f_{i,eff}^{E}(\bsl{k}_1)  \right] + (\bsl{k}_1\leftrightarrow \bsl{k}_2)\right\} \\
&  \qquad + \left\{ \frac{\hbar}{2 m_{B}} \sum_{\delta=\pm,i}   \Tr\left[ P_{eff,1}(\bsl{k}_{1,\shpa})  f_{i,eff}^{E}(\bsl{k}_1) \chi^{eff}_{\delta} P_{eff,1}(\bsl{k}_{2,\shpa})    f_{i,eff}^{E}(\bsl{k}_2)  \chi^{eff}_{\delta} \right] + c.c.\right\} \ .
}
Combined with 
\eq{
\sum_{\delta=\pm } \chi_{\delta}^{eff} P_{eff,n}(\bsl{k}_\shpa)   \chi_{\delta}^{eff}= \frac{1}{2}
}
from \eqnref{eq:chi_-_Peff_MgB2} and 
\eq{
P_{eff,n}(\bsl{k})  f_{i,eff}^{E}(\bsl{k}) =  f_{i,eff}^{E}(\bsl{k}) P_{eff,n}(\bsl{k}) = \ii  \gamma_{i} [\partial_{k_i}   E_{eff,n}(\bsl{k})] P_{eff,n}(\bsl{k})\ ,
}
we further obtain
\eqa{
\label{eq:Gamma_eff_E-E_MgB2_kshpa0}
& \Gamma_{11}^{eff,E-E}(\bsl{k}_1,\bsl{k}_2) \\
& \quad =\frac{\hbar}{2 m_{B}}  \sum_{i}  \gamma_{i}^2 [\partial_{k_{1,i}}   E_{eff,1}(\bsl{k}_1)]^2 + \frac{\hbar}{2 m_{B}}  \sum_{i}  \gamma_{i}^2 [\partial_{k_{2,i}}   E_{eff,1}(\bsl{k}_2)]^2  \\
&  \qquad - \gamma_{i}^2 \left\{ \frac{\hbar}{2 m_{B}} \sum_{\delta=\pm,i}  [\partial_{k_{1,i}}   E_{eff,1}(\bsl{k}_1)] [\partial_{k_{2,i}}   E_{eff,1}(\bsl{k}_2)] \Tr\left[ P_{eff,1}(\bsl{k}_{1,\shpa})   \chi^{eff}_{\delta} P_{eff,1}(\bsl{k}_{2,\shpa})     \chi^{eff}_{\delta} \right] + c.c.\right\}\\
& \quad =\frac{\hbar}{2 m_{B}}   \sum_{i}  \gamma_{i}^2 [\partial_{k_{1,i}}   E_{eff,1}(\bsl{k}_1)]^2  + \frac{\hbar}{2 m_{B}}   \sum_{i}  \gamma_{i}^2 [\partial_{k_{2,i}}   E_{eff,1}(\bsl{k}_2)]^2  \\
& \qquad - \sum_{i}  \gamma_{i}^2  \frac{2 \hbar}{2 m_{B}}  [\partial_{k_{1,i}}   E_{eff,1}(\bsl{k}_1)] [\partial_{k_{2,i}}   E_{eff,1}(\bsl{k}_2)]\\
& = \frac{\hbar}{2 m_{B}}  \gamma_{\sigma,z}^2 [\partial_{k_{1,z}}   E_{eff,1}(\bsl{k}_1)]^2  + \frac{\hbar}{2 m_{B}}   \gamma_{\sigma,z}^2 [\partial_{k_{2,z}}   E_{eff,1}(\bsl{k}_2)]^2  \\
& \qquad -  \gamma_{\sigma,z}^2  \frac{2 \hbar}{2 m_{B}}  [\partial_{k_{1,z}}   E_{eff,1}(\bsl{k}_1)] [\partial_{k_{2,z}}   E_{eff,1}(\bsl{k}_2)] + O(|\bsl{k}_\shpa|^2)\ ,
}
where $\partial_{k_{z}}   E_{eff,1}(\bsl{k}) = \partial_{k_z} \epsilon_0(k_z)$ as shown in \eqnref{eq:E_eff_MgB2} is independent of $\bsl{k}_\shpa$.
As a result, by only keeping the zeroth-order-$|\bsl{k}_\shpa|$ term of $\Gamma_{11}^{eff,E-E}(\bsl{k}_1,\bsl{k}_2)$, we obtain
\eqa{
\left\langle \Gamma \right\rangle^{sp_xp_y,E-E}  & = \frac{1}{D^2_{\sigma}(\mu)} \sum_{\bsl{k}_1,\bsl{k}_2}^{\BZ}\delta\left(\mu - E_{eff,1}(\bsl{k}_1) \right) \delta\left(\mu - E_{eff,1}(\bsl{k}_2) \right) \Gamma_{11  }^{eff,E-E}(\bsl{k}_1,\bsl{k}_2) \\
& = \frac{1}{D^2_{\sigma}(\mu)} \sum_{\bsl{k}_1,\bsl{k}_2}^{\BZ}\delta\left(\mu - E_{eff,1}(\bsl{k}_1) \right) \delta\left(\mu - E_{eff,1}(\bsl{k}_2) \right)  \frac{\hbar}{ m_{B}}   \gamma_{\sigma,z}^2 [\partial_{k_{1,z}}   E_{eff,1}(\bsl{k}_1)]^2 \\
& \qquad - \frac{1}{D^2_{\sigma}(\mu)} \gamma_{\sigma,z}^2  \frac{2 \hbar}{2 m_{B}} \left[ \sum_{\bsl{k}_1}^{\BZ}\delta\left(\mu - E_{eff,1}(\bsl{k}_1) \right) \partial_{k_{1,z}}   E_{eff,1}(\bsl{k}_1)  \right]^2\\
& = \frac{\hbar}{2  m_{B}} \frac{1}{D_{\sigma}(\mu)} \sum_{\bsl{k}}^{\BZ}\delta\left(\mu - E_{eff,1}(\bsl{k}) \right)    \gamma_{\sigma,z}^2 [\partial_{k_{z}}   E_{eff,1}(\bsl{k})]^2 \ ,
}
where we used 
\eq{
\sum_{\bsl{k}_1}^{\BZ}\delta\left(\mu - E_{eff,1}(\bsl{k}_1) \right) \partial_{k_{1,i}}   E_{eff,1}(\bsl{k}_1) = 0
}
for the second equality, which is derived from both $C_3$ and $m_z$ symmetries.
Eventually, we have
\eq{
 \left\langle \Gamma \right\rangle^{sp_xp_y,E-E} = \frac{\hbar}{2  m_{B}} \frac{1}{D_{\sigma}(\mu)} \frac{\V}{(2\pi)^3} \int_{FS_{eff,1}} d\sigma_{\bsl{k}} \frac{ \gamma_{\sigma,z}^2 [\partial_{k_{z}}   E_{eff,1}(\bsl{k})]^2}{\left|\nabla_{\bsl{k}}E_{eff,1}(\bsl{k})\right|}\ ,    
}
and 
\eqa{
\label{eq:lambda_E_eff_MgB2}
\lambda_{\sigma,E} & = \frac{2}{ N} \frac{1}{\hbar \mcomega} D_{\sigma}(\mu) \frac{D_{\sigma}(\mu)}{D(\mu)} \frac{\hbar}{2  m_{B}} \frac{1}{D_{\sigma}(\mu)} \frac{\V}{(2\pi)^3} \int_{FS_{eff,1}} d\sigma_{\bsl{k}} \frac{ \gamma_{\sigma,z}^2 [\partial_{k_{z}}   E_{eff,1}(\bsl{k})]^2}{\left|\nabla_{\bsl{k}}E_{eff,1}(\bsl{k})\right|} \\
& = \frac{D_{\sigma}(\mu)}{D(\mu)} \frac{ \gamma_{\sigma,z}^2}{   m_{B} \mcomega}   \frac{\Omega}{(2\pi)^3} \int_{FS_{eff,1}} d\sigma_{\bsl{k}} \frac{  [\partial_{k_{z}}   E_{eff,1}(\bsl{k})]^2}{\left|\nabla_{\bsl{k}}E_{eff,1}(\bsl{k})\right|} \\
}

\subsubsection{$\lambda_{\sigma,E-geo}$}

For $\Gamma_{11}^{eff,E-geo}$, we have
\eqa{
& \Gamma_{11}^{eff,E-geo}(\bsl{k}_1,\bsl{k}_2) \\
&  = \frac{\hbar}{2 m_{B}} \sum_{\delta=\pm,i}   \Tr\left[ P_{eff,1}(\bsl{k}_{1,\shpa})  (\chi^{eff}_{\delta} f_{i,eff}^{E}(\bsl{k}_2) - f_{i,eff}^{E}(\bsl{k}_1) \chi^{eff}_{\delta}) P_{eff,1}(\bsl{k}_{2,\shpa})    (\chi^{eff}_{\delta} f_{i,eff}^{geo}(\bsl{k}_1) - f_{i,eff}^{geo}(\bsl{k}_2) \chi^{eff}_{\delta} ) \right] + c.c. \\
&  = \frac{\hbar}{2 m_{B}} \sum_{\delta=\pm,i}   \ii  \gamma_{i}  [\partial_{k_{2,i}}   E_{eff,1}(\bsl{k}_2)-\partial_{k_{1,i}}   E_{eff,1}(\bsl{k}_1)]  \\
& \qquad \times \Tr\left[  P_{eff,1}(\bsl{k}_{1,\shpa}) \chi^{eff}_{\delta} P_{eff,1}(\bsl{k}_{2,\shpa})    (\chi^{eff}_{\delta} f_{i,eff}^{geo}(\bsl{k}_1) - f_{i,eff}^{geo}(\bsl{k}_2) \chi^{eff}_{\delta} ) \right] + c.c. \\
&  = \frac{\hbar}{2 m_{B}} \sum_{\delta=\pm,i}   \ii  \gamma_{i}  [\partial_{k_{2,i}}   E_{eff,1}(\bsl{k}_2)-\partial_{k_{1,i}}   E_{eff,1}(\bsl{k}_1)]  \\
& \qquad \times \Tr\left[  P_{eff,1}(\bsl{k}_{1,\shpa}) \chi^{eff}_{\delta} P_{eff,1}(\bsl{k}_{2,\shpa})    (\chi^{eff}_{\delta} f_{i,eff}^{geo}(\bsl{k}_1) - f_{i,eff}^{geo}(\bsl{k}_2) \chi^{eff}_{\delta} ) \right] + c.c. \\
& = \frac{\hbar}{4 m_{B}} \sum_{i}   \ii  \gamma_{i}  [\partial_{k_{2,i}}   E_{eff,1}(\bsl{k}_2)-\partial_{k_{1,i}}   E_{eff,1}(\bsl{k}_1)]    \Tr\left[  P_{eff,1}(\bsl{k}_{1,\shpa}) f_{i,eff}^{geo}(\bsl{k}_1) -  P_{eff,1}(\bsl{k}_{2,\shpa}) f_{i,eff}^{geo}(\bsl{k}_2) \right] + c.c. \ .
}
Combined with
\eq{
\label{eq:Gamma_eff_E-geo_MgB2}
\Tr[P_{eff,1}(\bsl{k}) f_{i,eff}^{geo}(\bsl{k})] =  \ii  \gamma_{i} \sum_{n=1,2}    E_{eff,n}(\bsl{k}) \Tr[P_{eff,1}(\bsl{k}) \partial_{k_i} P_{eff,n}(\bsl{k}_\shpa) ] = 0\ ,
}
we arrive at 
\eq{
\Gamma_{11}^{eff,E-geo}(\bsl{k}_1,\bsl{k}_2) = 0, 
}
resulting in 
\eqa{
\label{eq:lambda_E-geo_eff_MgB2}
\lambda_{\sigma,E-geo} = 0\ .
}
The results stay the same if we only keep the zeroth-order-$|\bsl{k}_\shpa|$ term.

\subsubsection{$\lambda_{\sigma,geo}$}
\label{eq:lambda_sp_xp_y_geo_eff}

Now we discuss $\lambda_{\sigma,geo}$.
Since $f_{i,eff}^{geo}$ in \eqnref{eq:g_i_MgB2_eff_E_geo} is nonzero only for $i=x,y$, the contribution to $\lambda_{\sigma,geo}$ comes from the in-plane motions of B atoms.
In fact, to the zeroth-order of $\bsl{k}_\shpa$, the in-plane motions of B atoms only contribute to $\Gamma_{11}^{eff,geo-geo}$ in \eqnref{eq:Gamma_eff_parts_MgB2}, since the zeroth-order-$\bsl{k}_\shpa$ part of $\Gamma_{11}^{eff,E-E}$ only comes from the motion along $z$ as shown in \eqnref{eq:Gamma_eff_E-E_MgB2_kshpa0} and $\Gamma_{11}^{eff,E-geo}$ is zero as shown in \eqnref{eq:Gamma_eff_E-geo_MgB2}.

In $\Gamma_{11}^{eff,geo-geo}$ in \eqnref{eq:Gamma_eff_parts_MgB2}, $\chi_{+}^{eff}$ and $\chi_{-}^{eff}$ correspond to the same and opposite motions of two NN B atoms, respectively, according to \eqnref{eq:chi_pm_eff} and the meaning of $\chi_{\bsl{\tau}}$ shown in \eqnref{eq:H_el-ph_2center_k}.
Then, let us split $\Gamma_{11}^{eff,geo-geo}$ into two parts: 
\eq{
 \Gamma_{11}^{eff,geo-geo}(\bsl{k}_1,\bsl{k}_2) =  \Gamma_{11,+}^{eff,geo-geo}(\bsl{k}_1,\bsl{k}_2) +  \Gamma_{11,-}^{eff,geo-geo}(\bsl{k}_1,\bsl{k}_2)\ ,
}
where
\eqa{
\label{eq:Gamma_--_+_eff_geo-geo}
& \Gamma_{11,+}^{eff,geo-geo}(\bsl{k}_1,\bsl{k}_2) \\
& \quad =  \frac{\hbar}{2 m_{B}} \sum_{\delta=\pm,i}   \Tr\left[ P_{eff,1}(\bsl{k}_{1,\shpa})  (\chi^{eff}_{+} f_{i,eff}^{geo}(\bsl{k}_2) - f_{i,eff}^{geo}(\bsl{k}_1) \chi^{eff}_{+}) P_{eff,1}(\bsl{k}_{2,\shpa})    (\chi^{eff}_{+} f_{i,eff}^{geo}(\bsl{k}_1) - f_{i,eff}^{geo}(\bsl{k}_2) \chi^{eff}_{+} ) \right] + c.c.
}
corresponds to the contribution from the same motions of two NN B atoms, and \eqa{
\label{eq:Gamma_--_-_eff_geo-geo}
& \Gamma_{11,-}^{eff,geo-geo}(\bsl{k}_1,\bsl{k}_2) \\
&  =  \frac{\hbar}{2 m_{B}} \sum_{\delta=\pm,i}   \Tr\left[ P_{eff,1}(\bsl{k}_{1,\shpa})  (\chi^{eff}_{-} f_{i,eff}^{geo}(\bsl{k}_2) - f_{i,eff}^{geo}(\bsl{k}_1) \chi^{eff}_{-}) P_{eff,1}(\bsl{k}_{2,\shpa})    (\chi^{eff}_{-} f_{i,eff}^{geo}(\bsl{k}_1) - f_{i,eff}^{geo}(\bsl{k}_2) \chi^{eff}_{-} ) \right] + c.c.
}
corresponds to the contribution from the opposite motions of two NN B atoms.
Recall that the dominant phonons to the EPC constant $\lambda$ are the $E_{2}$ phonons along $\Gamma$-A (enhanced to $E_{2g}$ at $\Gamma$ and A), which originate from the opposite in-plane motions of NN B atoms.
Therefore, for small $|\bsl{k}_\shpa|$, we should expect that  $\Gamma_{11,+}^{eff,geo-geo}(\bsl{k}_1,\bsl{k}_2)$ must be small while $\Gamma_{11,-}^{eff,geo-geo}(\bsl{k}_1,\bsl{k}_2)$ is large.
Indeed, we will show that to zeroth order in $|\bsl{k}_\shpa|$, $\Gamma_{11,+}^{eff,geo-geo}(\bsl{k}_1,\bsl{k}_2)$ is zero, while $\Gamma_{11,-}^{eff,geo-geo}(\bsl{k}_1,\bsl{k}_2)$ saturates to its nonzero upper bound.

Start from $\Gamma_{11,+}^{eff,geo-geo}(\bsl{k}_1,\bsl{k}_2)$.
Since $\chi_+ = \mathds{1}_{4}/\sqrt{2}$, we have
\eqa{
& \Gamma_{11,+}^{eff,geo-geo}(\bsl{k}_1,\bsl{k}_2)  \\
& \quad =  \frac{\hbar}{4 m_{B}} \sum_{\delta=\pm,i}   \Tr\left[ P_{eff,1}(\bsl{k}_{1,\shpa})  (  f_{i,eff}^{geo}(\bsl{k}_2) - f_{i,eff}^{geo}(\bsl{k}_1) ) P_{eff,1}(\bsl{k}_{2,\shpa})    (  f_{i,eff}^{geo}(\bsl{k}_1) - f_{i,eff}^{geo}(\bsl{k}_2)  ) \right] + c.c.\\
& \quad = O(|\bsl{k}_{1,\shpa}|,|\bsl{k}_{2,\shpa}|)\ .
}
So we can neglect $\Gamma_{11,+}^{eff,geo-geo}(\bsl{k}_1,\bsl{k}_2)$.

Now we discuss $\Gamma_{11,-}^{eff,geo-geo}$.
From \eqnref{eq:Gamma_--_-_eff_geo-geo}, we have
\eqa{
\label{eq:Gamma_--_-_eff_geo-geo_Z_Y}
\Gamma_{11, - }^{eff,geo-geo}(\bsl{k}_1,\bsl{k}_2) & = -\frac{\hbar}{2} \frac{1}{m_{\text{B}}}  \sum_{i} \Tr\left[ P_{eff,1}(\bsl{k}_{1,\shpa})  \chi_{-}^{eff}  f_{i, eff}^{geo}(\bsl{k}_{2})   P_{eff,1}(\bsl{k}_{2,\shpa})    f_{i, eff}^{geo}(\bsl{k}_{2}) \chi_{-}^{eff}   \right]\\
 & \qquad  -  \frac{\hbar}{2} \frac{1}{m_{\text{B}}}  \sum_{i} \Tr\left[ P_{eff,1}(\bsl{k}_{1,\shpa})   f_{i, eff}^{geo}(\bsl{k}_{1}) \chi_{-}^{eff}  P_{eff,1}(\bsl{k}_{2,\shpa})    \chi_{-}^{eff} f_{i, eff}^{geo}(\bsl{k}_{1})    \right]\\
  & \qquad + \frac{\hbar}{2} \frac{1}{m_{\text{B}}}  \sum_{i} \Tr\left[   \chi_{-}^{eff}  f_{i, eff}^{geo}(\bsl{k}_{2})   P_{eff,1}(\bsl{k}_{2,\shpa})  \chi_{-}^{eff}  f_{i, eff}^{geo}(\bsl{k}_{1})   P_{eff,1}(\bsl{k}_{1,\shpa})  \right]\\
  & \qquad + \frac{\hbar}{2} \frac{1}{m_{\text{B}}}  \sum_{i} \Tr\left[   \chi_{-}^{eff}     P_{eff,1}(\bsl{k}_{2,\shpa}) f_{i, eff}^{geo}(\bsl{k}_{2}) \chi_{-}^{eff}    P_{eff,1}(\bsl{k}_{1,\shpa})  f_{i, eff}^{geo}(\bsl{k}_{1})  \right]\\
  & = Z(\bsl{k}_1,\bsl{k}_2) + Y(\bsl{k}_1,\bsl{k}_2) \ ,
}
where
\eqa{
\label{eq:ZandY_eff_geo}
& Z(\bsl{k}_1,\bsl{k}_2) = \left\{ -  \frac{\hbar}{2} \frac{1}{m_{\text{B}}}  \sum_{i} \Tr\left[ \chi_{-}^{eff} f_{i, eff}^{geo}(\bsl{k}_{1}) P_{eff,1}(\bsl{k}_{1,\shpa})   f_{i, eff}^{geo}(\bsl{k}_{1}) \chi_{-}^{eff}  P_{eff,1}(\bsl{k}_{2,\shpa}) \right]+ (\bsl{k}_1\leftrightarrow \bsl{k}_2) \right\} \\
& Y(\bsl{k}_1,\bsl{k}_2) = \left\{  \frac{\hbar}{2} \frac{1}{m_{\text{B}}}  \sum_{i} \Tr\left[   \chi_{-}^{eff}     P_{eff,1}(\bsl{k}_{2,\shpa}) f_{i, eff}^{geo}(\bsl{k}_{2}) \chi_{-}^{eff}    P_{eff,1}(\bsl{k}_{1,\shpa})  f_{i, eff}^{geo}(\bsl{k}_{1})  \right] + c.c. \right\}\ .
}
We separate $\Gamma_{11, - }^{eff,geo-geo}(\bsl{k}_1,\bsl{k}_2)$ into $Z(\bsl{k}_1,\bsl{k}_2)$ and $Y(\bsl{k}_1,\bsl{k}_2)$ because $Z(\bsl{k}_1,\bsl{k}_2)$ is always non-negative and provides an upper bound of $Y(\bsl{k}_1,\bsl{k}_2)$.
Specifically, non-negative $Z(\bsl{k}_1,\bsl{k}_2)$ is given by
\eqa{
& Z(\bsl{k}_1,\bsl{k}_2) \\
& = \left\{ \frac{\hbar}{2} \frac{1}{m_{\text{B}}}  \sum_{i,m n} U_{1,m}^\dagger(\bsl{k}_2)\chi_{-}^{eff} f_{i, eff}^{geo}(\bsl{k}_{1}) U_{1,n}(\bsl{k}_1)  U_{1,n}(\bsl{k}_1)^\dagger  (-  f_{i, eff}^{geo}(\bsl{k}_{1})) \chi_{-}^{eff}  U_{1,m}(\bsl{k}_2) + (\bsl{k}_1\leftrightarrow \bsl{k}_2) \right\} \\
& = \left\{ \frac{\hbar}{2} \frac{1}{m_{\text{B}}}  \sum_{i,mn} \left|U_{1,m}^\dagger(\bsl{k}_2)\chi_{-}^{eff} f_{i, eff}^{geo}(\bsl{k}_{1}) U_{1,n}(\bsl{k}_1) \right|^2 + (\bsl{k}_1\leftrightarrow \bsl{k}_2) \right\} \geq 0\ ,
}
where $P_{eff,1}(\bsl{k}) = \sum_n U_{1,n}(\bsl{k}) U_{1,n}^\dagger(\bsl{k})$ with $U_{1,n}(\bsl{k})$ in \eqnref{eq:MgB2_H_eff_eigenvecs}, and we have used 
\eq{
\label{eq:g_geo_eff_Hermiticity}
[f_{i, eff}^{geo}(\bsl{k})]^\dagger = -  f_{i, eff}^{geo}(\bsl{k})
}
given by \eqnref{eq:g_i_MgB2_eff_E_geo}.
Then, we have
\eqa{
\label{eq:ZminusY_eff_geo}
& Z(\bsl{k}_1,\bsl{k}_2) - Y(\bsl{k}_1,\bsl{k}_2) \\
& = -\frac{\hbar}{2} \frac{1}{m_{\text{B}}}  \sum_{i} \Tr\left[ P_{eff,1}(\bsl{k}_{1,\shpa})  \chi_{-}^{eff}  f_{i, eff}^{geo}(\bsl{k}_{2})   P_{eff,1}(\bsl{k}_{2,\shpa})    f_{i, eff}^{geo}(\bsl{k}_{2}) \chi_{-}^{eff}   \right]\\
 & \qquad  -  \frac{\hbar}{2} \frac{1}{m_{\text{B}}}  \sum_{i} \Tr\left[ P_{eff,1}(\bsl{k}_{1,\shpa})   f_{i, eff}^{geo}(\bsl{k}_{1}) \chi_{-}^{eff}  P_{eff,1}(\bsl{k}_{2,\shpa})    \chi_{-}^{eff} f_{i, eff}^{geo}(\bsl{k}_{1})    \right]\\
  & \qquad - \frac{\hbar}{2} \frac{1}{m_{\text{B}}}  \sum_{i} \Tr\left[   \chi_{-}^{eff}  f_{i, eff}^{geo}(\bsl{k}_{2})   P_{eff,1}(\bsl{k}_{2,\shpa})  \chi_{-}^{eff}  f_{i, eff}^{geo}(\bsl{k}_{1})   P_{eff,1}(\bsl{k}_{1,\shpa})  \right]\\
  & \qquad - \frac{\hbar}{2} \frac{1}{m_{\text{B}}}  \sum_{i} \Tr\left[   \chi_{-}^{eff}     P_{eff,1}(\bsl{k}_{2,\shpa}) f_{i, eff}^{geo}(\bsl{k}_{2}) \chi_{-}^{eff}    P_{eff,1}(\bsl{k}_{1,\shpa})  f_{i, eff}^{geo}(\bsl{k}_{1})  \right]\\
  & = - \frac{\hbar}{2} \frac{1}{m_{\text{B}}}  \sum_{i} \Tr\left[ P_{eff,1}(\bsl{k}_{1,\shpa})  (\chi_{\delta}^{eff}  f_{i, eff}^{geo}(\bsl{k}_{2}) + f_{i, eff}^{geo}(\bsl{k}_{1}) \chi_{\delta}^{eff}) \right. \\
&\quad  \left. P_{eff,1}(\bsl{k}_{2,\shpa})   (\chi_{\delta}^{eff} f_{i, eff}^{geo}(\bsl{k}_{1}) + f_{i, eff}^{geo}(\bsl{k}_{2}) \chi_{\delta}^{eff} )  \right]\\
  & = \frac{\hbar}{2} \frac{1}{m_{\text{B}}}  \sum_{i,mn}  \left| U_{1,n}^\dagger(\bsl{k}_1)  (\chi_{-}^{eff}  f_{i, eff}^{geo}(\bsl{k}_{2}) + f_{i, eff}^{geo}(\bsl{k}_{1}) \chi_{-}^{eff}) U_{1,m}(\bsl{k}_2)   \right|^2 \geq 0\ ,
}
where we have used \eqnref{eq:g_geo_eff_Hermiticity} for the last equality.
Therefore, we arrive at
\eq{
\label{eq:Gamma_geo-geo_-_upper_bound}
\Gamma_{11, - }^{eff,geo-geo}(\bsl{k}_1,\bsl{k}_2) \leq 2 Z(\bsl{k}_1,\bsl{k}_2)\ ,
}
where $Z(\bsl{k}_1,\bsl{k}_2)$ is defined in \eqnref{eq:ZandY_sp2_geo}.

Now we show that to zeroth order in $|\bsl{k}_\shpa|$, the upper bound in \eqnref{eq:Gamma_geo-geo_-_upper_bound} would saturate.
First, note that
\eq{
\label{eq:g_i_eff_geo_explicit}
f_{i,eff}^{geo}(\bsl{k}) = \ii  \gamma_i (-) \Delta E_{eff}(\bsl{k}_\shpa) \partial_{k_i} P_{eff,1}(\bsl{k})\ ,
}
where
\eq{
\label{eq:DeltaE_eff_explicit}
\Delta E_{eff}(\bsl{k}_\shpa) = E_{eff,2}(\bsl{k}) - E_{eff,1}(\bsl{k}) = 2 \sqrt{m^2 + |\bsl{d}(\bsl{k}_\shpa)|^2}\ .
}
Then, combined with \eqnref{eq:chi_-_Peff_MgB2}, we know $\chi_-^{eff}$ would anti-commute with $f_{i,eff}^{geo}(\bsl{k})$:  
\eqa{
\chi_-^{eff} f_{i,eff}^{geo}(\bsl{k}) & = \ii  \gamma_i (-) \Delta E_{eff}(\bsl{k}_\shpa) \chi_-^{eff}  \partial_{k_i} P_{eff,1}(\bsl{k}) =  \ii  \gamma_i (-) \Delta E_{eff}(\bsl{k}_\shpa)  \partial_{k_i} P_{eff,2}(\bsl{k}) \chi_-^{eff} \\
& =  \ii  \gamma_i (-) \Delta E_{eff}(\bsl{k}_\shpa)  (-) \partial_{k_i} P_{eff,1}(\bsl{k}) \chi_-^{eff} = - f_{i,eff}^{geo}(\bsl{k}) \chi_-^{eff} \ .
}
As a result, we know $Z(\bsl{k}_1,\bsl{k}_2) - Y(\bsl{k}_1,\bsl{k}_2) $ in \eqnref{eq:ZminusY_eff_geo} is zero to zeroth order in $|\bsl{k}_\shpa|$:
\eqa{
\label{eq:Z_Y_Ok1_sigma_MgB2}
Z(\bsl{k}_1,\bsl{k}_2) - Y(\bsl{k}_1,\bsl{k}_2) & = \frac{\hbar}{2} \frac{1}{m_{\text{B}}}  \sum_{i}  \left| U_{1,n}^\dagger(\bsl{k}_1)  (\chi_{-}^{eff}  f_{i, eff}^{geo}(\bsl{k}_{2}) + f_{i, eff}^{geo}(\bsl{k}_{1}) \chi_{-}^{eff}) U_{1,m}(\bsl{k}_2)   \right|^2 \\
& =  \frac{\hbar}{2} \frac{1}{m_{\text{B}}}  \sum_{i}  \left| U_{1,n}^\dagger(\bsl{k}_1)  \chi_{-}^{eff}  ( f_{i, eff}^{geo}(\bsl{k}_{2}) - f_{i, eff}^{geo}(\bsl{k}_{1})) U_{1,m}(\bsl{k}_2)   \right|^2  = O(|\bsl{k}_{1,\shpa}|, |\bsl{k}_{2,\shpa}|)\ .
}
Combined with \eqnref{eq:Gamma_--_-_eff_geo-geo_Z_Y}, we know that the upper bound in \eqnref{eq:Gamma_geo-geo_-_upper_bound} is saturated to zeroth order in $|\bsl{k}_\shpa|$, leading to 
\eq{
\label{eq:Gamma_geo-geo_-_2Z}
\Gamma_{11, - }^{eff,geo-geo}(\bsl{k}_1,\bsl{k}_2) = \left. 2 Z(\bsl{k}_1,\bsl{k}_2) \right|_{|\bsl{k}_{1,\shpa}|=|\bsl{k}_{2,\shpa}|=0} + O(|\bsl{k}_{1,\shpa}|, |\bsl{k}_{2,\shpa}|)\ .
}

Now we derive $\left. 2 Z(\bsl{k}_1,\bsl{k}_2) \right|_{|\bsl{k}_{1,\shpa}|=|\bsl{k}_{2,\shpa}|=0}$.
From \eqnref{eq:g_i_eff_geo_explicit} and \eqnref{eq:ZandY_eff_geo}, we have
\eqa{
&  2 Z(\bsl{k}_1,\bsl{k}_2)  \\
&  =   \hbar \frac{1}{m_{\text{B}}} \Delta E_{eff}^2(\bsl{k}_{1,\shpa}) \sum_{i} \gamma_i^2 \Tr\left[ \chi_{-}^{eff} \partial_{k_{1,i}}P_{eff,1}(\bsl{k}_{1,\shpa}) P_{eff,1}(\bsl{k}_{1,\shpa})   \partial_{k_{1,i}}P_{eff,1}(\bsl{k}_{1,\shpa})  \chi_{-}^{eff}  P_{eff,1}(\bsl{k}_{2,\shpa}) \right]+ (\bsl{k}_1\leftrightarrow \bsl{k}_2)  \\
&  =   \frac{\hbar}{2} \frac{1}{m_{\text{B}}} \Delta E_{eff}^2(\bsl{k}_{1,\shpa}) \sum_{i} \gamma_i^2 \Tr\left[ \partial_{k_{1,i}}P_{eff,1}(\bsl{k}_{1,\shpa}) P_{eff,1}(\bsl{k}_{1,\shpa})   \partial_{k_{1,i}}P_{eff,1}(\bsl{k}_{1,\shpa})    P_{eff,2}(\bsl{k}_{2,\shpa}) \right]+ (\bsl{k}_1\leftrightarrow \bsl{k}_2)  \ ,
}
meaning that 
\eqa{
&  \left. 2 Z(\bsl{k}_1,\bsl{k}_2) \right|_{|\bsl{k}_{1,\shpa}|=|\bsl{k}_{2,\shpa}|=0} \\
&  =   \left. \frac{\hbar}{2} \frac{1}{m_{\text{B}}} \Delta E_{eff}^2(\bsl{k}_{1,\shpa})  \gamma_{\sigma,\shpa}^2 \sum_{i=x,y} \Tr\left[ \partial_{k_{1,i}}P_{eff,1}(\bsl{k}_{1,\shpa}) P_{eff,1}(\bsl{k}_{1,\shpa})   \partial_{k_{1,i}}P_{eff,1}(\bsl{k}_{1,\shpa})    P_{eff,2}(0) \right] \right|_{|\bsl{k}_{1,\shpa}|=0} \\
& \quad + \left. \frac{\hbar}{2} \frac{1}{m_{\text{B}}} \Delta E_{eff}^2(\bsl{k}_{2,\shpa}) \gamma_{\sigma,\shpa}^2 \sum_{i} \Tr\left[ \partial_{k_{2,i}}P_{eff,1}(\bsl{k}_{2,\shpa}) P_{eff,1}(\bsl{k}_{2,\shpa})   \partial_{k_{2,i}}P_{eff,1}(\bsl{k}_{2,\shpa})    P_{eff,2}(0) \right] \right|_{|\bsl{k}_{2,\shpa}|=0}\ ,
}
Since 
\eq{
P_{eff,2}(0) = \diag(0,0,1,1)
}
owing to $m<0$ according to \eqnref{eq:MgB2_H_eff_para} and \eqnref{eq:MgB2_TB_values_sim}, $P_{eff,2}(0)$ is the projection matrix to the parity-odd combination of $p_x/p_y$ orbitals according to the basis in \eqnref{eq:MgB2_eff_basis}.
Then, we have
\eqa{
\label{eq:Z_OFSM_eff_MgB2}
&  \left. 2 Z(\bsl{k}_1,\bsl{k}_2) \right|_{|\bsl{k}_{1,\shpa}|=|\bsl{k}_{2,\shpa}|=0} \\
& = \frac{\hbar}{2} \frac{1}{m_{\text{B}}}  \gamma_{\sigma,\shpa}^2 \sum_{i=x,y} \sum_{\alpha\in \{ (p_x,-), (p_y,-)\}} \left\{ \left. \Delta E_{eff}^2(\bsl{k}_{1,\shpa})  \left[ g_{eff,1,\alpha}(\bsl{k}_{1,\shpa}) \right]_{ii} \right|_{|\bsl{k}_{1,\shpa}|=0} + \left. \Delta E_{eff}^2(\bsl{k}_{2,\shpa})   \left[g_{eff,1,\alpha}(\bsl{k}_{2,\shpa}) \right]_{ii} \right|_{|\bsl{k}_{2,\shpa}|=0}\right\}\ ,
}
where
\eq{
\label{eq:g_Obital_Selective_FS_eff_MgB2}
\left[g_{eff,1,\alpha}(\bsl{k}_{\shpa}) \right]_{ii} =  \Tr\left[ \partial_{k_{i}}P_{eff,1}(\bsl{k}_{\shpa}) P_{eff,1}(\bsl{k}_{\shpa})   \partial_{k_{i}}P_{eff,1}(\bsl{k}_{\shpa})    \xi_{eff,\alpha} \xi_{eff,\alpha}^\dagger \right]
}
is the OFSM according to the definition in \eqnref{eq:orbital_selective_FS_metric_alt}, $\alpha\in \{ (p_x,-), (p_y,-)\}$, and 
\eq{
\xi_{eff,p_x,-} = \mat{ 0 \\ 0 \\ 1 \\ 0 \\}\ ,\ \xi_{eff,p_y,-} = \mat{ 0 \\ 0 \\ 0 \\ 1 \\}
}
are parity-odd combinations of $p_x/p_y$ orbitals.
Eventually, if we neglect $O(|\bsl{k}_{1,\shpa}|, |\bsl{k}_{2,\shpa}|)$, we arrive at 
\eq{
\label{eq:Gamma_geo-geo_-_2Z_k0}
\Gamma_{11, - }^{eff,geo-geo}(\bsl{k}_1,\bsl{k}_2) = \frac{\hbar}{ m_{\text{B}}} \gamma_{\sigma,\shpa}^2 \sum_{i=x,y} \sum_{\alpha\in \{ (p_x,-), (p_y,-)\}}   \Delta E_{eff}^2(0)  \left[ g_{eff,1,\alpha}(0) \right]_{ii} \ ,
}
\eq{
\label{eq:Gamma_ave_geo-geo_-_2Z_k0}
 \left\langle \Gamma \right\rangle^{sp_xp_y,geo-geo} = \frac{\hbar}{2  m_{B}} \frac{1}{D_{\sigma}(\mu)} \frac{\V}{(2\pi)^3} \int_{FS_{eff,1}} d\sigma_{\bsl{k}} \frac{\gamma_{\sigma,\shpa}^2 \sum_{i=x,y} \sum_{\alpha\in \{ (p_x,-), (p_y,-)\}}   \Delta E_{eff}^2(0)  \left[ g_{eff,1,\alpha}(0) \right]_{ii}}{\left|\nabla_{\bsl{k}}E_{eff,1}(\bsl{k})\right|}\ ,    
}
and 
\eqa{
\label{eq:lambda_geo_eff_MgB2}
\lambda_{\sigma,geo} & = \frac{D_{\sigma}(\mu)}{D(\mu)} \frac{ \gamma_{\sigma,\shpa}^2}{  m_{B} \mcomega}   \frac{\Omega}{(2\pi)^3} \int_{FS_{eff,1}} d\sigma_{\bsl{k}} \frac{ \sum_{i=x,y} \sum_{\alpha\in \{ (p_x,-), (p_y,-)\}}   \Delta E_{eff}^2(0)  \left[ g_{eff,1,\alpha}(0) \right]_{ii}}{\left|\nabla_{\bsl{k}}E_{eff,1}(\bsl{k})\right|} \ ,
}
where $\gamma_{\sigma,\shpa}$ is defined in \eqnref{eq:gamma_spxpy_geo}.
As shown in \appref{app:numerics_MgB2}, \eqnref{eq:lambda_geo_eff_MgB2} is nonzero, meaning that the upper bound $\left. 2 Z(\bsl{k}_1,\bsl{k}_2) \right|_{|\bsl{k}_{1,\shpa}|=|\bsl{k}_{2,\shpa}|=0}$ is non-vanishing.

In the derivation of \eqnref{eq:lambda_geo_eff_MgB2}, we have used $Z(\bsl{k}_1,\bsl{k}_2)$ to approximate $Y(\bsl{k}_1,\bsl{k}_2)$ since they are the same zero momenta (\eqnref{eq:Z_Y_Ok1_sigma_MgB2}), where $Z(\bsl{k}_1,\bsl{k}_2)$ and $Y(\bsl{k}_1,\bsl{k}_2)$ are defined in \eqnref{eq:ZandY_eff_geo}.
Eventually, the OFSM comes from the $Z(\bsl{k}_1,\bsl{k}_2)$ as shown in \eqnref{eq:Z_OFSM_eff_MgB2}.
If we do not use $Z(\bsl{k}_1,\bsl{k}_2)$ to approximate $Y(\bsl{k}_1,\bsl{k}_2)$, then we would have a different geometric quantity that comes from $Y(\bsl{k}_1,\bsl{k}_2)$.
Explicitly, $Y(\bsl{k}_1,\bsl{k}_2)$ can be simplified as
\eqa{
& Y(\bsl{k}_1,\bsl{k}_2) \\
& =   \frac{\hbar}{2} \frac{1}{m_{\text{B}}}  \sum_{i} \Tr\left[   \chi_{-}^{eff}     P_{eff,1}(\bsl{k}_{2,\shpa}) f_{i, eff}^{geo}(\bsl{k}_{2}) \chi_{-}^{eff}    P_{eff,1}(\bsl{k}_{1,\shpa})  f_{i, eff}^{geo}(\bsl{k}_{1})  \right] + c.c.  \\
& = \frac{\hbar}{4} \frac{1}{m_{\text{B}}} \sum_{\alpha, \alpha'\in \{ 1,2,3,4\}}  \sum_{i} \Tr\left[   \zeta_{\alpha}\zeta_{\alpha}^\dagger     P_{eff,1}(\bsl{k}_{2,\shpa}) f_{i, eff}^{geo}(\bsl{k}_{2}) \zeta_{\alpha'}\zeta_{\alpha'}^\dagger    P_{eff,1}(\bsl{k}_{1,\shpa})  f_{i, eff}^{geo}(\bsl{k}_{1})  \right] + c.c. \\
& = -\frac{\hbar}{4} \frac{1}{m_{\text{B}}} \Delta E_{eff}(\bsl{k}_{1,\shpa}) \Delta E_{eff}(\bsl{k}_{2,\shpa}) \\
& \quad \times \sum_{\alpha, \alpha'\in \{ 1,2,3,4 \}}  \sum_{i} \gamma_i^2 \Tr\left[   \zeta_{\alpha}\zeta_{\alpha}^\dagger     P_{eff,1}(\bsl{k}_{2,\shpa}) \partial_{k_{2,i}} P_{eff,1}(\bsl{k}_{2,\shpa}) \zeta_{\alpha'}\zeta_{\alpha'}^\dagger    P_{eff,1}(\bsl{k}_{1,\shpa}) \partial_{k_{1,i}} P_{eff,1}(\bsl{k}_{1,\shpa})   \right] + c.c. \\
& = -\frac{\hbar}{4} \frac{1}{m_{\text{B}}} \Delta E_{eff}(\bsl{k}_{1,\shpa}) \Delta E_{eff}(\bsl{k}_{2,\shpa})  \sum_{\alpha, \alpha'\in \{ 1,2,3,4 \}}  \sum_{i} \gamma_i^2 \mathcal{A}_{eff, 1, \alpha \alpha', i} (\bsl{k}_{2,\shpa}) \mathcal{A}_{eff, 1, \alpha' \alpha, i} (\bsl{k}_{1,\shpa})   + c.c. \\
& = -\frac{\hbar}{4} \frac{1}{m_{\text{B}}} \Delta E_{eff}(\bsl{k}_{1,\shpa}) \Delta E_{eff}(\bsl{k}_{2,\shpa}) \gamma_{\sigma,\shpa}^2 \sum_{\alpha, \alpha'\in \{ 1,2,3,4\}}  \sum_{i=x,y}  \left[\mathcal{A}_{eff, 1, \alpha \alpha', i} (\bsl{k}_{2,\shpa}) \mathcal{A}_{eff, 1, \alpha' \alpha, i} (\bsl{k}_{1,\shpa})   + c.c. \right]\ ,
}
where 
\eq{
\label{eq:A_eff_1_alphaalpha'}
\mathcal{A}_{eff, 1, \alpha \alpha', i}(\bsl{k}_\shpa) =  \zeta_{\alpha}^\dagger     P_{eff,1}(\bsl{k}_{\shpa}) \partial_{k_{i}} P_{eff,1}(\bsl{k}_\shpa) \zeta_{\alpha'}  = \Tr\left[ \zeta_{\alpha'}\zeta_{\alpha}^\dagger     P_{eff,1}(\bsl{k}_{\shpa}) \partial_{k_{i}} P_{eff,1}(\bsl{k}_\shpa) \right]\ ,
}
and
\eq{
\zeta_1 = \frac{1}{\sqrt{2}}\mat{1 \\ 0 \\ 1 \\ 0}\ ,\ \zeta_2 = \frac{1}{\sqrt{2}}\mat{0 \\ 1 \\ 0 \\ 1}\ ,\ \zeta_3 = \frac{1}{\sqrt{2}}\mat{1 \\ 0 \\ -1 \\ 0}\ ,\ \zeta_4 = \frac{1}{\sqrt{2}}\mat{0 \\ 1 \\ 0 \\ -1}\ .
}
Thus, 
\eqa{
\left.  \Gamma_{11, - }^{eff,geo-geo}(\bsl{k}_1,\bsl{k}_2)  \right|_{|\bsl{k}_{1,\shpa}|=|\bsl{k}_{2,\shpa}|=0} & =   \left. Z(\bsl{k}_1,\bsl{k}_2)\right|_{|\bsl{k}_{1,\shpa}|=|\bsl{k}_{2,\shpa}|=0} +  \left. Y(\bsl{k}_1,\bsl{k}_2)\right|_{|\bsl{k}_{1,\shpa}|=|\bsl{k}_{2,\shpa}|=0} \\
& = \frac{\hbar}{2} \frac{ \gamma_{\sigma,\shpa}^2}{m_{\text{B}}} \Delta E_{eff}^2(0)   \sum_{i=x,y} \sum_{\alpha\in \{ (p_x,-), (p_y,-)\}}  \left[ g_{eff,1,\alpha}(0) \right]_{ii}    \\
& \quad - \frac{\hbar}{4} \frac{\gamma_{\sigma,\shpa}^2}{m_{\text{B}}} \Delta E_{eff}^2(0)   \sum_{i=x,y} \sum_{\alpha, \alpha'\in \{1,2,3,4\}}  \left[\mathcal{A}_{eff, 1, \alpha \alpha', i} (0) \mathcal{A}_{eff, 1, \alpha' \alpha, i} (0)   + c.c. \right] \ ,
}
resulting in 
\eqa{
\left\langle \Gamma \right\rangle^{sp_xp_y,geo-geo}&  = \frac{\hbar}{4  m_{B}} \frac{1}{D_{\sigma}(\mu)} \frac{\V}{(2\pi)^3} \int_{FS_{eff,1}} d\sigma_{\bsl{k}} \frac{\gamma_{\sigma,\shpa}^2 \Delta E_{eff}^2(0) }{\left|\nabla_{\bsl{k}}E_{eff,1}(\bsl{k})\right|}  \sum_{i=x,y}     \left[ \sum_{\alpha\in \{ (p_x,-), (p_y,-)\}} \left[ g_{eff,1,\alpha}(0) \right]_{ii}  \right.\\
 & \quad \left. -  \frac{1}{2}\sum_{\alpha, \alpha'\in \{1,2,3,4\}} \mathcal{A}_{eff, 1, \alpha \alpha', i} (0) \mathcal{A}_{eff, 1, \alpha' \alpha, i} (0)  - \frac{1}{2}\sum_{\alpha, \alpha'\in \{1,2,3,4\}} \mathcal{A}_{eff, 1, \alpha \alpha', i}^* (0) \mathcal{A}_{eff, 1, \alpha' \alpha, i}^* (0) \right] \ ,    
}
and 
\eqa{
\label{eq:lambda_geo_eff_MgB2_alternative}
\lambda_{\sigma,geo} & = \lambda_{\sigma,geo,-,1} + \lambda_{\sigma,geo,-,2}  \ ,  
}
where 
\eq{
\lambda_{\sigma,geo,-,1} = \frac{D_{\sigma}(\mu)}{D(\mu)} \frac{ \gamma_{\sigma,\shpa}^2}{  2 m_{B} \mcomega}   \frac{\Omega}{(2\pi)^3} \int_{FS_{eff,1}} d\sigma_{\bsl{k}} \frac{    \Delta E_{eff}^2(0)  }{\left|\nabla_{\bsl{k}}E_{eff,1}(\bsl{k})\right|}\sum_{i=x,y} \sum_{\alpha \in \{ (p_x,-), (p_y,-)\}}      \left[ g_{eff,1,\alpha}(0) \right]_{ii} \ ,  
}
and 
\eqa{
\lambda_{\sigma,geo,-,2} & = -\frac{1}{2}\frac{D_{\sigma}(\mu)}{D(\mu)} \frac{ \gamma_{\sigma,\shpa}^2}{  4 m_{B} \mcomega}   \frac{\Omega}{(2\pi)^3} \int_{FS_{eff,1}} d\sigma_{\bsl{k}} \frac{    \Delta E_{eff}^2(0)  }{\left|\nabla_{\bsl{k}}E_{eff,1}(\bsl{k})\right|} \\
& \qquad \times \sum_{i=x,y} \sum_{\alpha \alpha'\in \{1,2,3,4\}}    \left[   \mathcal{A}_{eff, 1, \alpha \alpha', i} (0) \mathcal{A}_{eff, 1, \alpha' \alpha, i} (0)   + c.c.\right] \ .
}
$\lambda_{\sigma,geo,-,1}$ and $\lambda_{\sigma,geo,-,2}$ have the same meaning as $\lambda_{geo,1}$ and $\lambda_{geo,2}$ in \appref{app:gaussian} except that we now go beyond two bands and  $\lambda_{\sigma,geo,-,1}$ and $\lambda_{\sigma,geo,-,2}$ are only in the $\chi_{eff,-}$ channel.
\eqnref{eq:lambda_geo_eff_MgB2_alternative} would be the expression of $\lambda_{\sigma,geo}$ if we do not use $Z(\bsl{k}_1,\bsl{k}_2)$ to approximate $Y(\bsl{k}_1,\bsl{k}_2)$.
Nevertheless, \eqnref{eq:Z_Y_Ok1_sigma_MgB2} shows that 
\eq{
\sum_{i=x,y} \sum_{\alpha\in \{ (p_x,-), (p_y,-)\}} \left[ g_{eff,1,\alpha}(0) \right]_{ii}  =  -\sum_{i=x,y} \sum_{\alpha \alpha'\in \{ (p_x,-), (p_y,-)\}}  \mathcal{A}_{eff, 1, \alpha \alpha', i} (0) \mathcal{A}_{eff, 1, \alpha' \alpha, i} (0)   + c.c.\ ,
}
which allows us to merge them in \eqnref{eq:lambda_geo_eff_MgB2_alternative} and restore \eqnref{eq:lambda_geo_eff_MgB2}.
In other words, $\lambda_{\sigma,geo,-,1}$ and $\lambda_{\sigma,geo,-,2}$ can be merged to $2 \lambda_{\sigma,geo,-,1}$, and thus we do not explicit use the complex vector field in \eqnref{eq:A_eff_1_alphaalpha'}.

Before moving onto the topological contribution, we note that by replacing $\chi^{eff}_-$ to $\chi^{eff}_+$ in the derivation of \eqnref{eq:Gamma_geo-geo_-_upper_bound}, we can also define the upper bound for $\Gamma_{11,+}^{eff,geo-geo}(\bsl{k}_1,\bsl{k}_2)$, which reads
\eq{
\label{eq:Gamma_geo-geo_+_upper_bound}
\Gamma_{11,+}^{eff,geo-geo}(\bsl{k}_1,\bsl{k}_2) \leq 2 Z'(\bsl{k}_1,\bsl{k}_2)\ ,
}
where
\eq{
 Z'(\bsl{k}_1,\bsl{k}_2) =  -  \frac{\hbar}{2} \frac{1}{m_{\text{B}}}  \sum_{i} \Tr\left[ \chi^{eff}_+ f_{i, eff}^{geo}(\bsl{k}_{1}) P_{eff,1}(\bsl{k}_{1,\shpa})   f_{i, eff}^{geo}(\bsl{k}_{1}) \chi^{eff}_+  P_{eff,1}(\bsl{k}_{2,\shpa}) \right]+ (\bsl{k}_1\leftrightarrow \bsl{k}_2)  \ .
}
Nevertheless, explicit evaluation gives that $\left. Z'(\bsl{k}_1,\bsl{k}_2) \right|_{|\bsl{k}_{1,\shpa}|=|\bsl{k}_{2,\shpa}|=0} = 0$, which is consistent to the fact that $\Gamma_{11,+}^{eff,geo-geo}(\bsl{k}_1,\bsl{k}_2)$ is zero to zeroth order in $|\bsl{k}_\shpa|$.
Therefore, $\Gamma_{11,+}^{eff,geo-geo}(\bsl{k}_1,\bsl{k}_2)$ also saturates to its upper bound in \eqnref{eq:Gamma_geo-geo_+_upper_bound}, but its upper bound vanishes.

\subsubsection{$\lambda_{\sigma,topo}$}

At last, we consider the topological term.
First, we evaluate $ \sum_{i=x,y} \sum_{\alpha\in \{ (p_x,-), (p_y,-)\}}   \Delta E_{eff}^2(0)  \left[ g_{eff,1,\alpha}(0) \right]_{ii}$ from the expression of $g_{eff,1,\alpha}(\bsl{k}_\shpa)$ in \eqnref{eq:g_Obital_Selective_FS_eff_MgB2} and the expression of $\Delta E_{eff}^2(\bsl{k}_\shpa)$ in \eqnref{eq:DeltaE_eff_explicit}.
Specifically, 
\eq{
\sum_{i=x,y} \sum_{\alpha\in \{ (p_x,-), (p_y,-)\}}   \Delta E_{eff}^2(0)  \left[ g_{eff,1,\alpha}(0) \right]_{ii} = 4 m^2 \frac{\sum_{i=x,y} |\partial_{k_i} \bsl{d}(\bsl{k}_\shpa)|^2}{2 m^2} = 2 \sum_{i=x,y} |\partial_{k_i} \bsl{d}(\bsl{k}_\shpa)|^2\ .
}
Note that
\eq{
|\partial_{k_i} \bsl{d}(\bsl{k}_\shpa)|^2 = |\partial_{k_i} |\bsl{d}(\bsl{k}_\shpa)| e^{\ii \theta(\bsl{k}_\shpa)}|^2 = (\partial_{k_i} |\bsl{d}(\bsl{k}_\shpa)|)^2 + |\bsl{d}(\bsl{k}_\shpa)|^2 |\partial_{k_i}\theta(\bsl{k}_\shpa)|^2 \geq |\bsl{d}(\bsl{k}_\shpa)|^2 |\partial_{k_i}\theta(\bsl{k}_\shpa)|^2\ ,
}
where $\bsl{d}(\bsl{k}_\shpa) = |\bsl{d}(\bsl{k}_\shpa)| (\cos(\theta_{\bsl{k}_\shpa}), \sin(\theta_{\bsl{k}_\shpa}))$.
Then, combined with the expression of $\left\langle \Gamma \right\rangle^{sp_xp_y,geo-geo}$ in \eqnref{eq:Gamma_ave_geo-geo_-_2Z_k0}, we arrive at
\eqa{
\left\langle \Gamma \right\rangle^{sp_xp_y,geo-geo} & = \frac{\hbar}{  m_{B}} \frac{1}{D_{\sigma}(\mu)} \frac{\V}{(2\pi)^3} \int_{FS_{eff,1}} d\sigma_{\bsl{k}} \frac{\gamma_{\sigma,\shpa}^2   \sum_{i=x,y} |\partial_{k_i} \bsl{d}(\bsl{k}_\shpa)|^2}{\left|\nabla_{\bsl{k}}E_{eff,1}(\bsl{k})\right|} \\
& \geq \frac{\hbar}{  m_{B}} \frac{1}{D_{\sigma}(\mu)} \frac{\V}{(2\pi)^3} \int_{FS_{eff,1}} d\sigma_{\bsl{k}} \frac{\gamma_{\sigma,\shpa}^2   \sum_{i=x,y}|\bsl{d}(\bsl{k}_\shpa)|^2 |\partial_{k_i}\theta(\bsl{k}_\shpa)|^2}{\left|\nabla_{\bsl{k}}E_{eff,1}(\bsl{k})\right|} \\
& = \frac{\hbar \gamma_{\sigma,\shpa}^2 }{  m_{B}} \frac{1}{D_{\sigma}(\mu)} \frac{\V}{(2\pi)^3} \int_{FS_{eff,1}} d\sigma_{\bsl{k}} \frac{  |\bsl{d}(\bsl{k}_\shpa)|^2 }{\left|\nabla_{\bsl{k}}E_{eff,1}(\bsl{k})\right|} |\partial_{\bsl{k}_\shpa}\theta(\bsl{k}_\shpa)|^2 \\
& = \frac{\hbar \gamma_{\sigma,\shpa}^2 }{  m_{B}} \frac{1}{D_{\sigma}(\mu)} \frac{\V}{(2\pi)^3} \int_{FS_{eff,1}} d\sigma_{\bsl{k}} \frac{  |\bsl{d}(\bsl{k}_\shpa)|^2 }{\left|\nabla_{\bsl{k}}E_{eff,1}(\bsl{k})\right|} |\partial_{\bsl{k}_\shpa}\theta(\bsl{k}_\shpa)|^2 \\
& \geq \frac{\hbar \gamma_{\sigma,\shpa}^2 }{  m_{B}} \frac{1}{D_{\sigma}(\mu)} \frac{\V}{(2\pi)^3} \frac{\left[\int_{FS_{eff,1}} d\sigma_{\bsl{k}}   |\partial_{\bsl{k}_\shpa}\theta(\bsl{k}_\shpa)|\right]^2}{ \int_{FS_{eff,1}} d\sigma_{\bsl{k}} \frac{\left|\nabla_{\bsl{k}}E_{eff,1}(\bsl{k})\right|} {  |\bsl{d}(\bsl{k}_\shpa)|^2 } } \ ,
}
where we have used the H\"older inequality (\eqnref{eq:holder}) for the last inequality.
The Fermi surface of $E_{eff,1}(\bsl{k})$ can be expressed as $FS_{eff,1} = \cup_{k_z c \in (-\pi,\pi]} FS_{eff,1, k_z}$ with $FS_{eff,1, k_z}$ is the intersection between $FS_{eff,1}$ and the fixed $k_z$ plane.
Then,
\eqa{
& \int_{FS_{eff,1}} d\sigma_{\bsl{k}} |\partial_{\bsl{k}_\shpa}\theta(\bsl{k}_\shpa)| = \int_{-\pi/c}^{\pi/c} d k_z \int_{FS_{eff,1,k_z}} d\sigma_{\bsl{k}_\shpa} |\partial_{\bsl{k}_\shpa}\theta(\bsl{k}_\shpa)| = \int_{-\pi/c}^{\pi/c} d k_z \int_{FS_{eff,1,k_z=0}} d\sigma_{\bsl{k}_\shpa} |\partial_{\bsl{k}_\shpa}\theta(\bsl{k}_\shpa)| \\
& \geq \int_{-\pi/c}^{\pi/c} d k_z \left|\int_{FS_{eff,1,k_z=0}} d \bsl{k}_\shpa \cdot  \partial_{\bsl{k}_\shpa}\theta(\bsl{k}_\shpa)\right| = \int_{-\pi/c}^{\pi/c} d k_z 2\pi \W_d = \int_{-\pi/c}^{\pi/c} d k_z 2\pi \Delta\N = \frac{(2\pi)^2}{c} \Delta\N\ ,
}
where we have used the fact that $FS_{eff,1,k_z=0}$ is a closed loop of $\bsl{k}_\shpa$ that encloses $\Gamma$ once, $\W_d$ is the winding number in \eqnref{eq:d_winding_MgB2_eff}, and the relation between $\W_d$ and the effective Euler number $\Delta \N$ is in \eqnref{eq:W_d_DeltaN}.
As a result, we have
\eq{
\left\langle \Gamma \right\rangle^{sp_xp_y,geo-geo} \geq \frac{\hbar \gamma_{\sigma,\shpa}^2 }{  m_{B}} \frac{1}{D_{\sigma}(\mu)}  \V \frac{2\pi}{ c^2} \frac{\left[\Delta\N\right]^2}{ \int_{FS_{eff,1}} d\sigma_{\bsl{k}} \frac{\left|\nabla_{\bsl{k}}E_{eff,1}(\bsl{k})\right|} {  |\bsl{d}(\bsl{k}_\shpa)|^2 } }\ ,
}
and
\eq{
\lambda_{\sigma ,geo} \geq \frac{D_{\sigma}(\mu)}{D(\mu)} \frac{ 2 \gamma_{\sigma,\shpa}^2}{ m_{B} \mcomega}   \frac{2 \pi\Omega}{c^2} \frac{\left[\Delta\N\right]^2}{ \int_{FS_{eff,1}} d\sigma_{\bsl{k}} \frac{\left|\nabla_{\bsl{k}}E_{eff,1}(\bsl{k})\right|} {  |\bsl{d}(\bsl{k}_\shpa)|^2 } } 
}
derived from \eqnref{eq:lambda_sp2_parts}.
By defining
\eq{
\label{eq:lambda_topo_eff_MgB2}
\lambda_{\sigma,topo} = \frac{D_{\sigma}(\mu)}{D(\mu)} \frac{ 2 \gamma_{\sigma,\shpa}^2}{ m_{B} \mcomega}   \frac{2 \pi\Omega}{c^2} \frac{\left[\Delta\N\right]^2}{ \int_{FS_{eff,1}} d\sigma_{\bsl{k}} \frac{\left|\nabla_{\bsl{k}}E_{eff,1}(\bsl{k})\right|} {  |\bsl{d}(\bsl{k}_\shpa)|^2 } } \ ,
}
we eventually have
\eq{
\label{eq:lambda_geo_>=_topo_eff_MgB2}
\lambda_{\sigma ,geo}  \geq \lambda_{\sigma,topo} \ .
}

We note that $ \sum_{i=x,y} \sum_{\alpha\in \{ (p_x,-), (p_y,-)\}} g_{eff,1,\alpha}(0)$ itself in this case is not bounded from below, because 
\eq{
\sum_{i=x,y} \sum_{\alpha\in \{ (p_x,-), (p_y,-)\}} g_{eff,1,\alpha}(0) = \frac{\sum_{i=x,y} |\partial_{k_i} \bsl{d}(\bsl{k}_\shpa)|^2}{2 m^2} = \frac{v^2 a^2}{m^2} 
}
relies on the ratio between $v^2$ and $m^2$ owing to the nonzero gap at $\Gamma$ (which is $|2m|$) for the $\sigma$-bonding states, where we have used the explicit expression of $\bsl{d}(\bsl{k}_\shpa)$ in \eqnref{eq:d_form_MgB2_eff}.
For example, if the gap limits to infinity, we should not have any geometric effect of $U_{eff,1}(\bsl{k})$.
Nevertheless, the following quantity is bounded from below:
\eqa{
\int_{FS_{eff,1}} d\sigma_{\bsl{k}} \sqrt{ \sum_{i=x,y} \sum_{\alpha\in \{ (p_x,-), (p_y,-)\}}   \frac{ \Delta E_{eff}^2(0)  \left[ g_{eff,1,\alpha}(0) \right]_{ii} } {2 |\bsl{d}(\bsl{k}_\shpa)|^2} } \geq  \int_{FS_{eff,1}} d\sigma_{\bsl{k}} |\partial_{k_i}\theta(\bsl{k}_\shpa)|  = \frac{4\pi^2}{c} \W_d\ ,
}
where the expression of $\W_d$ in \eqnref{eq:d_winding_MgB2_eff} is used.

\subsection{$p_x p_y$ General Symmetry-Allowed Hopping Form: Consistent with Gaussian Approximation}

\label{app:4-band_general_hopping_form}

In this part, we show that even if we use the general symmetry-allowed hopping form for $p_x p_y$ orbitals, we would get the same energetic and geometric contributions to $\lambda_{\sigma}$ from the GA, under the same NN and small $\bsl{k}_\shpa$ approximation.

We still consider a $2\times 2$ hopping matrix function $t(\bsl{r})$, whose basis are $(p_x,p_y)$ of B atoms, and will still use $\O(2)$ and $m_z$ symmetries for $t(\bsl{r})$.
The part of the derivations in \appref{app:4-band_Gaussian_lambda_spxpy} are valid here, and we will not repeat them.
We will focus on the difference brought by the general symmetry-allowed $t(\bsl{r})$.
The first difference brought by the general symmetry-allowed $t(\bsl{r})$ is that \eqnref{eq:hopping_function_Gaussian_Form_MgB2_pxpy} cannot be used, since \eqnref{eq:hopping_function_Gaussian_Form_MgB2_pxpy} introduces the GA.
Instead, we should use
\eqa{
& pp_1(\sqrt{x^2+y^2},|z|) = F_1(\sqrt{x^2+y^2},|z|)\\
& pp_2(\sqrt{x^2+y^2},|z|) = \frac{x^2+y^2}{a^2} F_2(\sqrt{x^2+y^2},|z|)\ ,
}
with $F_2(\sqrt{x^2+y^2},|z|)$ a smooth function.
Then $\partial_{r_i}t(0,-r_\shpa,|z|)$ (with $r_\shpa=\sqrt{x^2+y^2}$) in \eqnref{eq:gradient_hopping_function_Gaussian_MgB2_pxpy} becomes
\eqa{
\partial_{r_i}t(0,-r_\shpa,|z|) & =
\left[ \delta_{i x}\widetilde{\gamma}_{\shpa,1}(r_\shpa,|z|)  + \delta_{i y} \widetilde{\gamma}_{\shpa,1} (r_\shpa,|z|) + \delta_{i z} \widetilde{\gamma}_{z,1}(r_\shpa,|z|) \right] r_i 
\mat{   
  pp_1(r_\shpa,|z|) & 0 \\
  0 & pp_1(r_\shpa,|z|)
}\\ 
& \quad +
\left[  \delta_{i x}\widetilde{\gamma}_{\shpa,2}(r_\shpa,|z|)  + \delta_{i y} \widetilde{\gamma}_{\shpa,2} (r_\shpa,|z|) + \delta_{i z} \widetilde{\gamma}_{z,2}(r_\shpa,|z|) \right] r_i 
\mat{   
pp_2(r_\shpa,|z|) & 0 \\
0 & -pp_2(r_\shpa,|z|)
} \\
& \quad +
2 (\delta_{i x} + \delta_{i y})\frac{ r_i}{x^2+y^2} 
\mat{  
pp_2(r_\shpa,|z|) & 0 \\
 0 & -pp_2(r_\shpa,|z|)
} \\
 & =
 \frac{1}{2}\left[ \delta_{i x}\widetilde{\gamma}_{\shpa,1}(r_\shpa,|z|)  + \delta_{i y} \widetilde{\gamma}_{\shpa,1} (r_\shpa,|z|) + \delta_{i z} \widetilde{\gamma}_{z,1}(r_\shpa,|z|) \right] r_i 
\left[t(0,-r_\shpa,|z|) + \sigma_y  t(0,-r_\shpa,|z|) \sigma_y \right]\\ 
& \quad +
 \frac{1}{2} \left[  \delta_{i x}\widetilde{\gamma}_{\shpa,2}(r_\shpa,|z|)  + \delta_{i y} \widetilde{\gamma}_{\shpa,2} (r_\shpa,|z|) + \delta_{i z} \widetilde{\gamma}_{z,2}(r_\shpa,|z|) \right] r_i 
\left[t(0,-r_\shpa,|z|) - \sigma_y  t(0,-r_\shpa,|z|) \sigma_y \right] \\
& \quad +
 (\delta_{i x} + \delta_{i y})\frac{ r_i}{r_\shpa^2} 
\left[t(0,-r_\shpa,|z|) - \sigma_y  t(0,-r_\shpa,|z|) \sigma_y \right]
}
where
\eqa{
\label{eq:gamma_t_shpa_z}
& \widetilde{\gamma}_{\shpa,l}(r_\shpa,|z|) =  \frac{1}{r_\shpa} \partial_{r_\shpa} \log\left[\left| F_l(r_\shpa,|z|) \right| \right] \\
& \widetilde{\gamma}_{z,l}(r_\shpa,|z|) =  2  \partial_{z^2} \log\left[\left| F_l(r_\shpa,|z|) \right| \right] \ ,
}
and $l=1,2$.
Then, $\partial_{r_i} t(\bsl{r})$ becomes  
\eqa{
\partial_{r_i} t(\bsl{r}) & = 
 \frac{1}{2}\left[ \delta_{i x}\widetilde{\gamma}_{\shpa,1}(r_\shpa,|z|)  + \delta_{i y} \widetilde{\gamma}_{\shpa,1} (r_\shpa,|z|) + \delta_{i z} \widetilde{\gamma}_{z,1}(r_\shpa,|z|) \right] r_i 
\left[t(\bsl{r}) + \sigma_y  t(\bsl{r}) \sigma_y \right]\\ 
& \quad +
 \frac{1}{2} \left[  \delta_{i x}\widetilde{\gamma}_{\shpa,2}(r_\shpa,|z|)  + \delta_{i y} \widetilde{\gamma}_{\shpa,2} (r_\shpa,|z|) + \delta_{i z} \widetilde{\gamma}_{z,2}(r_\shpa,|z|) \right] r_i 
\left[t(\bsl{r}) - \sigma_y  t(\bsl{r}) \sigma_y \right] \\
& \quad + (\delta_{ix}+\delta_{iy})\frac{(\bsl{e}_z\times \bsl{r})_i}{x^2+y^2} (-\ii)  [\sigma_y ,t(\bsl{r})]  +
(\delta_{i x} + \delta_{i y})\frac{ r_i}{x^2+y^2} \left[t(\bsl{r}) - \sigma_y  t(\bsl{r}) \sigma_y \right]\ .
}
Since we only care about the NN hopping among the B atoms in one plane and the $\pm\bsl{a}_3$ hopping among B atoms in different layers, we can approximate 
\eqa{
& \widetilde{\gamma}_{\shpa,l}(r_\shpa,|z|) \approx \widetilde{\gamma}_{\shpa,l}(\frac{a}{\sqrt{3}},0) = \gamma_{\shpa,l} \\
& \widetilde{\gamma}_{z,l}(r_\shpa,|z|) \approx \widetilde{\gamma}_{z,l}(0,c) = \gamma_{z,l} \ ,
}
resulting in the following form of $f_{i,p_xp_y}(\bsl{k})$
\eqa{
\label{eq:g_NoGaussian_MgB2_sp2_intermediate}
\left[f_{i,p_xp_y}(\bsl{k})\right]_{\bsl{\tau}_1\alpha_1, \bsl{\tau}_2\alpha_2} & = 
 \frac{\ii}{2}\left[ \delta_{i x} \gamma_{\shpa,1}  + \delta_{i y} \gamma_{\shpa,1} + \delta_{i z} \gamma_{z,1} \right] 
   \left[\partial_{k_i} h_{p_xp_y}(\bsl{k}) + \tau_0 \sigma_y  \partial_{k_i} h_{p_xp_y}(\bsl{k}) \tau_0 \sigma_y \right]_{\bsl{\tau}_1\alpha_1,\bsl{\tau}_2\alpha_2}  \\ 
& \quad +
 \frac{\ii}{2} \left[   \delta_{i x} \gamma_{\shpa,2}  + \delta_{i y} \gamma_{\shpa,2} + \delta_{i z} \gamma_{z,2}  \right] \left[\partial_{k_i} h_{p_xp_y}(\bsl{k}) - \tau_0 \sigma_y  \partial_{k_i} h_{p_xp_y}(\bsl{k}) \tau_0 \sigma_y \right]_{\bsl{\tau}_1\alpha_1,\bsl{\tau}_2\alpha_2}  \\
& +\left[\Delta f_{i,p_xp_y}(\bsl{k})\right]_{\bsl{\tau}_1\alpha_1, \bsl{\tau}_2\alpha_2}\ ,
}
where $\Delta f_{i,p_xp_y}(\bsl{k})$ is in \eqnref{eq:Delta_g_nonGaussian_pxpy_MgB2}.

Now we adopt the linear $\bsl{k}_\shpa$ approximation for the electron Hamiltonian.
Combined with the short-range hopping, \eqnref{eq:Delta_g_nonGaussian_pxpy_MgB2_short_range_small_kshpa} show that $\Delta f_{i,p_xp_y}(\bsl{k})$ becomes zero.
Moreover, we note that 
\eq{
\partial_{k_i} h_{p_x p_y}(\bsl{k}) =  - 2 c t_{B,p_xp_y,z}  \sin(k_z c) \delta_{iz} + v a R_{eff} \tau_y \sigma_x R_{eff}^\dagger \delta_{ix} + v a R_{eff} \tau_y \sigma_z R_{eff}^\dagger \delta_{iy}\ ,
}
where the expression of the $4\times 4$ matrix $R_{eff}$ is in \eqnref{eq:R_eff}.
Combined with the fact that $R_{eff}$ commutes with $\tau_0 \sigma_y$, we have 
\eqa{
& \partial_{k_i} h_{p_xp_y}(\bsl{k}) + \tau_0 \sigma_y  \partial_{k_i} h_{p_xp_y}(\bsl{k}) \tau_0 \sigma_y = 0 \text{ for $i=x,y$}\\
& \partial_{k_i} h_{p_xp_y}(\bsl{k}) - \tau_0 \sigma_y  \partial_{k_i} h_{p_xp_y}(\bsl{k}) \tau_0 \sigma_y = 0 \text{ for $i=z$}\ .
}
It means that we can safely tune $\gamma_{\shpa,1}$ and $\gamma_{z,2}$ in \eqnref{eq:g_NoGaussian_MgB2_sp2_intermediate} without changing $f_{i,p_xp_y}(\bsl{k})$ under our approximation.
Moreover, such tuning does not change $f_{i,p_xp_y}^E(\bsl{k})$ and $f_{i,p_xp_y}^{geo}(\bsl{k})$ to the leading order either, as $f_{i,p_xp_y}^E(\bsl{k})$ and $f_{i,p_xp_y}^{geo}(\bsl{k})$ come from $i=z$ and $i=x,y$, respectively, to the leading order, because the dispersion is quadratic along $x-y$ and the eigenstates of $h_{p_xp_y}(\bsl{k})$ is independent of $k_z$.
Then, we choose $\gamma_{\shpa,1}=\gamma_{\shpa,2}=\gamma_{\shpa}$ and $\gamma_{z,1}=\gamma_{z,2}=\gamma_{z}$ in \eqnref{eq:g_NoGaussian_MgB2_sp2_intermediate}, and get the same form of  $f_{i,p_xp_y}(\bsl{k})$ as \eqnref{eq:g_Gaussian_MgB2_eff} derived from GA.
As a result, the all the later energetic and geometric contributions to $\lambda$ are the same as those derived from GA.

\subsection{Symmetry-Rep Method: Analytical Geometric and Topological Contributions to $\lambda_{\sigma}$}
\label{app:lambda_sp2_topo_geo}

 In \appref{app:4-band_Gaussian_lambda_spxpy}, we derive the analytical geometric and topological contributions to $\lambda_{\sigma}$ based on the GA with a 4-band $p_xp_y$ model.
In this part, we derive the analytical geometric and topological lower contributions to $\lambda_{\sigma}$ in \eqnref{eq:lambda_X_MgB2} from the 6-band $sp_xp_y$ model with the symmetry-rep method, in order to check our results.
We will show that the two methods give the same results.

\subsubsection{Symmetry-Rep Method: Energetic and Geometric parts of the EPC}

The key first step is again to specify the energetic and geometric parts of the EPC.
Although {\mgb} is not strictly a 2D material, we can follow the same spirit of \appref{app:geo_EPC_symmetry-rep} to identify the energetic and geometric parts of the EPC.
It is mainly because (i) the symmetry group of {\mgb} is effectively 2D, (ii) $\widetilde{f}_{\shpa,sp_xp_y}(\bsl{k}_\shpa)$ and $\widetilde{f}_{\perp,sp_xp_y}(\bsl{k}_\shpa)$ only relies on the momentum components in the 2D plane (\ie, $x y$ plane), and (iii) \eqnref{eq:g_shpa_MgB2} is generalizable to $\widetilde{f}_{\shpa,sp_xp_y}'(k_z)$.
Now we move on to the details.

Since we have used the momentum derivative to the carry $i$ index in $f_{i,sp_xp_y}(\bsl{k})$ in terms of the momentum derivatives as shown in \eqnref{eq:g_X_MgB2}, and we just need to reexpress $\widetilde{f}_{\shpa,sp_xp_y}(\bsl{k}_\shpa)$, $\widetilde{f}_{\perp,sp_xp_y}(\bsl{k}_\shpa)$ and $\widetilde{f}_{\shpa,sp_xp_y}'(k_z)$ by using the symmetry-rep method.
To do so, we need to first express $h_{sp_xp_y}(\bsl{k})$ in terms of $\hat{h}_{a}(\bsl{k})$ according to \eqnref{eq:h_hhat_gen_k}:
\eq{
h_{sp_xp_y}(\bsl{k}) = \sum_{a=1,...,8} \hat{h}_{a}(\bsl{k}) t_a \ ,
}
where 
\eq{
\label{eq:hhat_sp2}
\hat{h}_{a}(\bsl{k}) = \partial_{t_a} h_{sp_xp_y}(\bsl{k})\ ,
}
$h_{sp_xp_y}(\bsl{k})$ and $t_{1,2,3,4}$ are in \eqnref{eq:H_el_MgB2_B_sp2},
and 
\eq{
t_5 = E_{B,s,0}\ ,\  t_6= E_{B,p_x p_y,0}\ ,\ t_7 = t_{B,s,z}\ ,\ t_8 = t_{B,p_x p_y,z}\ .
}
We can explicitly verify that 
\eq{
\frac{1}{\N}\sum_{\bsl{k}\in\BZ}\Tr[\hat{h}_{a}(\bsl{k}) \hat{h}_{a'}(\bsl{k})] = 0\text{ for $a\neq a'$}\ .
}
Then, by comparing $h_{sp_xp_y}(\bsl{k})$ in \eqnref{eq:H_el_MgB2_B_sp2} to \eqnref{eq:g_shpa_MgB2}, we obtain
\eqa{
\label{eq:g_h_relation_sp2_MgB2_part_1}
 & \widetilde{f}_{\shpa,sp_xp_y}(\bsl{k}_\shpa) = \sum_{a=1,2,3,4} \hat{\gamma}_a \partial_{t_a} h_{sp_xp_y}(\bsl{k}) \\
 & \widetilde{f}_{\shpa,sp_xp_y}'(k_z) = (\hat{\gamma}_8 \partial_{t_{B,s,z}} + \hat{\gamma}_9 \partial_{t_{B,p_x p_y,z}}) h_{sp_xp_y}(\bsl{k}) \ ,
}
where we have generalized \eqnref{eq:g_shpa_MgB2} to $\widetilde{f}_{\shpa,sp_xp_y}'(k_z)$.

On the other hand, $\widetilde{f}_{\perp,sp_xp_y}(\bsl{k}_\shpa)$ is nonvanishing for {\mgb}, there are symmetries such as $m_y$ that has $z_{m_y,\perp} = -1$ according to \eqnref{eq:z_factor}, and {\mgb} is TR invariant.
Then, we should try to find $Q_l$ that satisfy \eqnref{eq:Q_h_ortho_rel}.
According to \eqnref{eq:sym_rep_MgB2}, the symmetry reps for the basis $c^\dagger_{\bsl{k},\text{B}, sp_xp_y}$ right below \eqnref{eq:H_el_MgB2_B_sp2} are
\eqa{
\label{eq:sym_rep_sp2}
& U_{C_6,sp_xp_y} = \left(
\begin{array}{cccccc}
 0 & 0 & 0 & 1 & 0 & 0 \\
 0 & 0 & 0 & 0 & \frac{1}{2} & -\frac{\sqrt{3}}{2} \\
 0 & 0 & 0 & 0 & \frac{\sqrt{3}}{2} & \frac{1}{2} \\
 1 & 0 & 0 & 0 & 0 & 0 \\
 0 & \frac{1}{2} & -\frac{\sqrt{3}}{2} & 0 & 0 & 0 \\
 0 & \frac{\sqrt{3}}{2} & \frac{1}{2} & 0 & 0 & 0 \\
\end{array}
\right)\\
& U_{m_y,sp_xp_y} = \left(
\begin{array}{cccccc}
 0 & 0 & 0 & 1 & 0 & 0 \\
 0 & 0 & 0 & 0 & 1 & 0 \\
 0 & 0 & 0 & 0 & 0 & -1 \\
 1 & 0 & 0 & 0 & 0 & 0 \\
 0 & 1 & 0 & 0 & 0 & 0 \\
 0 & 0 & -1 & 0 & 0 & 0 \\
\end{array}
\right)\\
& U_{m_z,sp_xp_y} =\mathds{1}_{6\times 6}\\
& U_{\TR,sp_xp_y} =\mathds{1}_{6\times 6} \ .
}
Then, by choosing $Q_l$ to be
\eq{
\label{eq:Q_sp2_MgB2}
Q_1 = \ii L_y\ ,\ Q_2 = \ii  \left(
\begin{array}{cccccc}
 0 & 0 & 0 & 0 & 0 & 0 \\
 0 & 0 & 0 & 0 & 0 & -\ii \\
 0 & 0 & 0 & 0 & \ii & 0 \\
 0 & 0 & 0 & 0 & 0 & 0 \\
 0 & 0 & -\ii & 0 & 0 & 0 \\
 0 & \ii & 0 & 0 & 0 & 0 \\
\end{array}
\right)
}
with
\eqa{
\label{eq:Ly}
L_y = \left(
\begin{array}{cccccc}
 0 & 0 & 0 & 0 & 0 & 0 \\
 0 & 0 & -\ii & 0 & 0 & 0 \\
 0 & \ii & 0 & 0 & 0 & 0 \\
 0 & 0 & 0 & 0 & 0 & 0 \\
 0 & 0 & 0 & 0 & 0 & -\ii \\
 0 & 0 & 0 & 0 & \ii & 0 \\
\end{array}
\right)\ ,
}
\eqnref{eq:Q_h_ortho_rel} is satisfied.
Then, we have
\eq{
\widetilde{f}_{\perp,sp_xp_y}(\bsl{k}_\shpa) = \hat{\gamma}_6 \left(Q_1 \hat{h}_{4}(\bsl{k})+ \hat{h}_{4}(\bsl{k}) Q_1^\dagger \right) - \frac{1}{2} \hat{\gamma}_7 \left(Q_1 \hat{h}_{3}(\bsl{k})+ \hat{h}_{3}(\bsl{k}) Q_1^\dagger \right)\ ,
}
which leads to
\eqa{
\label{eq:g_h_relation_sp2_MgB2_part_2}
 & \widetilde{f}_{\perp,sp_xp_y}(\bsl{k}_\shpa) =  \ii [L_y, (\hat{\gamma}_6 \partial_{t_4} - \frac{1}{2} \hat{\gamma}_7 \partial_{t_3}) h_{sp_xp_y}(\bsl{k})]
}
based on \eqnref{eq:Q_sp2_MgB2} and \eqnref{eq:hhat_sp2}.
It means that $\Delta  \widetilde{f}_{\shpa}(\bsl{k}) = 0$ in \eqnref{eq:g_perp_h_relation}.
The appearance of $L_y$ is consistent with the appearance of $\tau_0\sigma_y$ in \eqnref{eq:Delta_g_nonGaussian_pxpy_MgB2_short_range} derived from the GA.

By combining \eqnref{eq:g_h_relation_sp2_MgB2_part_1} and \eqnref{eq:g_h_relation_sp2_MgB2_part_2}, we arrive at
\eqa{
\label{eq:g_h_relation_sp2_MgB2}
 & \widetilde{f}_{\shpa,sp_xp_y}(\bsl{k}_\shpa) = \sum_{a=1,2,3,4} \hat{\gamma}_a \partial_{t_a} h_{sp_xp_y}(\bsl{k}) \\
 & \widetilde{f}_{\perp,sp_xp_y}(\bsl{k}_\shpa) =  \ii [L_y, (\hat{\gamma}_6 \partial_{t_4} - \frac{1}{2} \hat{\gamma}_7 \partial_{t_3}) h_{sp_xp_y}(\bsl{k})]\\
 & \widetilde{f}_{\shpa,sp_xp_y}'(k_z) = (\hat{\gamma}_8 \partial_{t_{B,s,z}} + \hat{\gamma}_9 \partial_{t_{B,p_x p_y,z}}) h_{sp_xp_y}(\bsl{k}) \ ,
}
where the basis is $c^\dagger_{\bsl{k},\text{B}, sp_xp_y}$ right below \eqnref{eq:H_el_MgB2_B_sp2},
\eq{
\label{eq:chi_sp2_pm}
\chi^{sp_xp_y}_\pm = \chi^{sp_xp_y}_{\bsl{\tau}_{\text{B1}}} \pm \chi^{sp_xp_y}_{\bsl{\tau}_{\text{B2}}}
}
with $\chi^{sp_xp_y}_{\bsl{\tau}}$ defined in \eqnref{eq:chi_sp2_MgB2}.
As a result, we have 
\eqa{
\label{eq:g_E_geo_sp2_MgB2_ini} 
f_{i, sp_xp_y}(\bsl{k})  = f_{i, sp_xp_y}^E(\bsl{k})  + f_{i, sp_xp_y}^{geo}(\bsl{k}) \ ,
}
where
\eqa{
\label{eq:g_E_sp2_MgB2_ini} 
f_{i, sp_xp_y}^E(\bsl{k}) & = (\delta_{ix} + \delta_{iy} ) \ii \sum_{a=1,2,3,4} \hat{\gamma}_a \partial_{t_a} \sum_{n} \partial_{k_i} E_{sp_xp_y,n}(\bsl{k}) P_{sp_xp_y,n}(\bsl{k}) \\
& \quad + (\delta_{ix} + \delta_{iy} )  \ii \sum_{i'=x,y} \epsilon_{i'i} \ii [L_y, (\hat{\gamma}_6 \partial_{t_4} - \frac{1}{2} \hat{\gamma}_7 \partial_{t_3}) \sum_{n} \partial_{k_{i'}} E_{sp_xp_y,n}(\bsl{k}) P_{sp_xp_y,n}(\bsl{k})] \\
& \quad + \ii \delta_{iz}  (\hat{\gamma}_8 \partial_{t_{B,s,z}} + \hat{\gamma}_9 \partial_{t_{B,p_x p_y,z}}) \sum_{n} \partial_{k_z }E_{sp_xp_y,n}(\bsl{k}) P_{sp_xp_y,n}(\bsl{k}) \ ,
}
\eqa{
\label{eq:g_geo_sp2_MgB2_ini} 
f_{i, sp_xp_y}^{geo}(\bsl{k})
& = (\delta_{ix} + \delta_{iy} ) \ii \sum_{a=1,2,3,4} \hat{\gamma}_a \partial_{t_a} \sum_{n} E_{sp_xp_y,n}(\bsl{k}) \partial_{k_i}  P_{sp_xp_y,n}(\bsl{k}) \\
& \quad + (\delta_{ix} + \delta_{iy} )  \ii \sum_{i'=x,y} \epsilon_{i'i} \ii [L_y, (\hat{\gamma}_6 \partial_{t_4} - \frac{1}{2} \hat{\gamma}_7 \partial_{t_3}) \sum_{n}  E_{sp_xp_y,n}(\bsl{k}) \partial_{k_{i'}} P_{sp_xp_y,n}(\bsl{k})] \\
& \quad + \ii \delta_{iz}  (\hat{\gamma}_8 \partial_{t_{B,s,z}} + \hat{\gamma}_9 \partial_{t_{B,p_x p_y,z}}) \sum_{n} E_{sp_xp_y,n}(\bsl{k}) \partial_{k_z } P_{sp_xp_y,n}(\bsl{k}) \ ,
}
and $E_{sp_xp_y,n}(\bsl{k})$ and $P_{sp_xp_y,n}(\bsl{k})$ are energies and projection matrices of $h_{sp_xp_y}(\bsl{k}) $ satisfying $h_{sp_xp_y}(\bsl{k}) P_{sp_xp_y,n}(\bsl{k}) = E_{sp_xp_y,n}(\bsl{k}) P_{sp_xp_y,n}(\bsl{k}) $.
We note that we do not simply include $f_{sp_xp_y,z}(\bsl{k})$ (\eqnref{eq:g_X_MgB2}) in $\Delta f_i(\bsl{k})$ in \eqnref{eq:Delta_g_i_2D}, since we have the momentum derivative along $z$ now and thus can obtain the energetic and geometric parts of $f_{sp_xp_y,z}(\bsl{k})$ as shown in \eqnref{eq:g_E_geo_sp2_MgB2_ini}.

\subsubsection{Contributions to $\lambda_{\sigma}$}
\label{app:parts_of_lambda_sp2_MgB2}

In this part, we will show different contributions to $\left\langle \Gamma \right\rangle^{sp_xp_y}$ in \eqnref{eq:Gamma_ave_X_MgB2} and $\lambda_{\sigma}$ in \eqnref{eq:lambda_X_MgB2}.
Let us first discuss $\Gamma_{n m }^{sp_xp_y}(\bsl{k}_1,\bsl{k}_2)$ defined in \eqnref{eq:Gamma_X_MgB2}.
First, $\Gamma_{n m }^{sp_xp_y}(\bsl{k}_1,\bsl{k}_2)$ has the following expression according to \eqnref{eq:Gamma_X_MgB2}, \eqnref{eq:f_X_MgB2} and \eqnref{eq:g_E_geo_sp2_MgB2_ini}:
\eqa{
\label{eq:Gamma_sp2_ini}
& \ \Gamma_{n m }^{sp_xp_y}(\bsl{k}_1,\bsl{k}_2)  \\
 & = \frac{\hbar}{2 m_{\text{B}}} \sum_{\bsl{\tau}\in\{\text{B1}, \text{B2}\},i}    \Tr\left[ P_{sp_xp_y ,n}(\bsl{k}_1)  F_{\bsl{\tau}i,sp_xp_y}(\bsl{k}_1,\bsl{k}_2) P_{sp_xp_y ,m}(\bsl{k}_2)   F_{\bsl{\tau}i,sp_xp_y}^\dagger(\bsl{k}_1,\bsl{k}_2) \right] \\
 & = \frac{\hbar}{2 m_{\text{B}}}\sum_{\bsl{\tau}\in\{\text{B1}, \text{B2}\},i}   \Tr\left[ P_{sp_xp_y ,n}(\bsl{k}_1)  (\chi_{\bsl{\tau}}^{sp_xp_y}  f_{i, sp_xp_y}(\bsl{k}_2) - f_{i, sp_xp_y}(\bsl{k}_1) \right.\\
 & \qquad \left. \chi_{\bsl{\tau}}^{sp_xp_y} ) P_{sp_xp_y ,m}(\bsl{k}_2)   (\chi_{\bsl{\tau}}^{sp_xp_y} f_{i, sp_xp_y}(\bsl{k}_1) -  f_{i, sp_xp_y}(\bsl{k}_2) \chi_{\bsl{\tau}}^{sp_xp_y}) \right] \\
 & =  \Gamma_{n m }^{sp_xp_y ,E-E}(\bsl{k}_1,\bsl{k}_2) + \Gamma_{n m }^{sp_xp_y, geo-geo}(\bsl{k}_1,\bsl{k}_2) + \Gamma_{n m }^{sp_xp_y ,E-geo}(\bsl{k}_1,\bsl{k}_2) \ ,
}
where
\eqa{
\label{eq:Gamma_E-E_sp2_MgB2}
&  \Gamma_{n m }^{sp_xp_y ,E-E}(\bsl{k}_1,\bsl{k}_2) \\
& \quad = \frac{\hbar}{2 m_{\text{B}}} \sum_{\bsl{\tau}\in\{\text{B1}, \text{B2}\}}\sum_i    \Tr\left[ P_{sp_xp_y,n}(\bsl{k}_1)  (\chi_{\bsl{\tau}}^{sp_xp_y} f_{i, sp_xp_y}^E(\bsl{k}_2) - f_{i, sp_xp_y}^E(\bsl{k}_1) \chi_{\bsl{\tau}}^{sp_xp_y} ) \right. \\
& \qquad \qquad\left.  P_{sp_xp_y,m}(\bsl{k}_2)   (\chi_{\bsl{\tau}}^{sp_xp_y} f_{i, sp_xp_y}^E(\bsl{k}_1) -  f_{i, sp_xp_y}^E(\bsl{k}_2) \chi_{\bsl{\tau}}^{sp_xp_y}) \right]  \ ,
}
\eqa{
\label{eq:Gamma_E-geo_sp2_MgB2}
&  \Gamma_{n m }^{sp_xp_y ,E-geo}(\bsl{k}_1,\bsl{k}_2)  \\
& \quad = \frac{\hbar}{2 m_{\text{B}}} \sum_{\bsl{\tau}\in\{\text{B1}, \text{B2}\}} \sum_i   \Tr\left[ P_{sp_xp_y,n}(\bsl{k}_1)  (\chi_{\bsl{\tau}}^{sp_xp_y} f_{i, sp_xp_y}^E(\bsl{k}_2) - f_{i, sp_xp_y}^E(\bsl{k}_1) \chi_{\bsl{\tau}}^{sp_xp_y} ) \right. \\
& \qquad \qquad \left.  P_{sp_xp_y,m}(\bsl{k}_2)   (\chi_{\bsl{\tau}}^{sp_xp_y} f_{i, sp_xp_y}^{geo}(\bsl{k}_1) -  f_{i, sp_xp_y}^{geo}(\bsl{k}_2) \chi_{\bsl{\tau}}^{sp_xp_y}) \right] + h.c. \ ,
}
and
\eqa{
\label{eq:Gamma_geo-geo_sp2_MgB2_intermedia}
&  \Gamma_{n m }^{sp_xp_y ,geo-geo}(\bsl{k}_1,\bsl{k}_2)  \\
& \quad = \frac{\hbar}{2 m_{\text{B}}} \sum_{\bsl{\tau}\in\{\text{B1}, \text{B2}\}} \sum_{i }   \Tr\left[ P_{sp_xp_y,n}(\bsl{k}_1)  (\chi_{\bsl{\tau}}^{sp_xp_y} f_{i, sp_xp_y}^{geo}(\bsl{k}_2) - f_{i, sp_xp_y}^{geo}(\bsl{k}_1) \chi_{\bsl{\tau}}^{sp_xp_y} ) \right. \\
& \qquad \qquad \left. P_{sp_xp_y,m}(\bsl{k}_2)   (\chi_{\bsl{\tau}}^{sp_xp_y} f_{i, sp_xp_y}^{geo}(\bsl{k}_1) -  f_{i, sp_xp_y}^{geo}(\bsl{k}_2) \chi_{\bsl{\tau}}^{sp_xp_y}) \right]  \ .
}
The different contributions to $\lambda_{\sigma}$ are defined in \eqnref{eq:lambda_sp2_E_geo}.

Before moving on to deriving the expressions of $\lambda_{\sigma,E}$, $\lambda_{\sigma,E-geo}$ and $\lambda_{\sigma,geo}$, we rewrite $\left\langle \Gamma \right\rangle^{sp_xp_y,geo-geo}$ for the convenience of later discussion.
Similar to the discussion in \appref{eq:lambda_sp_xp_y_geo_eff}, we can split $\Gamma_{n m }^{sp_xp_y,geo-geo}(\bsl{k}_1,\bsl{k}_2)$ into two parts:
 \eqa{
\label{eq:Gamma_as_sum_Gamma_pm_geo-geo_MgB2_sp2}
  & \Gamma_{n m }^{sp_xp_y,geo-geo}(\bsl{k}_1,\bsl{k}_2) = \sum_{\delta = \pm} \Gamma_{n m, \delta }^{sp_xp_y,geo-geo}(\bsl{k}_1,\bsl{k}_2)\ ,
 }
 where  
 \eqa{
 \label{eq:Gamma_pm_geo-geo_MgB2_sp2}
 \Gamma_{n m, \delta }^{sp_xp_y,geo-geo}(\bsl{k}_1,\bsl{k}_2) &  = \frac{\hbar}{4} \frac{1}{m_{\text{B}}}  \sum_{i} \Tr\left[ P_{sp_xp_y,n}(\bsl{k}_1)  (\chi_{\delta}^{sp_xp_y}  f_{i, sp_xp_y}^{geo}(\bsl{k}_{2}) - f_{i, sp_xp_y}^{geo}(\bsl{k}_{1}) \chi_{\delta}^{sp_xp_y}) \right. \\
 &\quad  \left. P_{sp_xp_y,m}(\bsl{k}_2)   (\chi_{\delta}^{sp_xp_y} f_{i, sp_xp_y}^{geo}(\bsl{k}_{1}) - f_{i, sp_xp_y}^{geo}(\bsl{k}_{2}) \chi_{\delta}^{sp_xp_y} )  \right]\ ,
 }
 and $\chi^{sp_xp_y}_\pm$ are defined in \eqnref{eq:chi_sp2_pm}.
Then, 
\eq{
\label{eq:Gamma_ave_geo-geo_sp2_MgB2}
\left\langle \Gamma \right\rangle^{sp_xp_y,geo-geo} = \left\langle \Gamma \right\rangle_+^{sp_xp_y,geo-geo} + \left\langle \Gamma \right\rangle_-^{sp_xp_y,geo-geo}\ ,
}
where 
\eq{
\label{eq:Gamma_ave_pm_geo-geo_sp2_MgB2}
\left\langle \Gamma \right\rangle_\pm^{sp_xp_y,geo-geo} = \frac{1}{D^2_{\sigma}(\mu)} \sum_{\bsl{k}_1,\bsl{k}_2}^{\BZ}\sum_{n ,m } \delta\left(\mu - E_{sp_xp_y,n}(\bsl{k}_1) \right) \delta\left(\mu - E_{sp_xp_y,m}(\bsl{k}_2) \right) \Gamma_{n m ,\pm}^{sp_xp_y,geo-geo}(\bsl{k}_1,\bsl{k}_2)\ ,
}
and $\chi^{sp_xp_y}_\pm$ are defined in \eqnref{eq:chi_sp2_pm}
Again,  $\Gamma_{n m, -}^{sp_xp_y,geo-geo}(\bsl{k}_1,\bsl{k}_2)$ has two parts:
\eqa{
\Gamma_{n m, - }^{sp_xp_y,geo-geo}(\bsl{k}_1,\bsl{k}_2) = Z_{nm,-}(\bsl{k}_1,\bsl{k}_2) + Y_{nm,-}(\bsl{k}_1,\bsl{k}_2) \ ,
}
where
\eqa{
\label{eq:ZandY_sp2_geo}
& Z_{nm,-}(\bsl{k}_1,\bsl{k}_2) = \left\{ -  \frac{\hbar}{4} \frac{1}{m_{\text{B}}}  \sum_{i} \Tr\left[ \chi_{-}^{sp_xp_y} f_{i, sp_xp_y}^{geo}(\bsl{k}_{1}) P_{sp_xp_y,n}(\bsl{k}_1)   f_{i, sp_xp_y}^{geo}(\bsl{k}_{1}) \chi_{-}^{sp_xp_y}  P_{sp_xp_y,m}(\bsl{k}_2) \right]+ (\bsl{k}_1\leftrightarrow \bsl{k}_2) \right\} \\
& Y_{nm,-}(\bsl{k}_1,\bsl{k}_2) = \left\{  \frac{\hbar}{4} \frac{1}{m_{\text{B}}}  \sum_{i} \Tr\left[   \chi_{-}^{sp_xp_y}     P_{sp_xp_y,m}(\bsl{k}_2) f_{i, sp_xp_y}^{geo}(\bsl{k}_{2}) \chi_{-}^{sp_xp_y}    P_{sp_xp_y,n}(\bsl{k}_1)  f_{i, sp_xp_y}^{geo}(\bsl{k}_{1})  \right] + c.c. \right\}\ .
}
Again, $Z_{nm,-}(\bsl{k}_1,\bsl{k}_2)$ is always non-negative and provides an upper bound of $Y_{nm,-}(\bsl{k}_1,\bsl{k}_2)$, \ie,
\eq{
\label{eq:Gamma_geo-geo_-}
\Gamma_{n m, - }^{sp_xp_y,geo-geo}(\bsl{k}_1,\bsl{k}_2) \leq 2 Z_{nm,-}(\bsl{k}_1,\bsl{k}_2)\ ,
}
where $Z_{nm,-}(\bsl{k}_1,\bsl{k}_2)$ is defined in \eqnref{eq:ZandY_sp2_geo}.
Similar upper bound has been derived in \eqnref{eq:Gamma_geo-geo_-_upper_bound} for the four-band effective Hamiltonian \eqnref{eq:MgB2_H_eff_mat_form}, but \eqnref{eq:Gamma_geo-geo_-} holds to all orders.

We will discuss in \appref{app:small_k_shpa_MgB2_sp2} that under the small $\bsl{k}_\shpa$ approximation, the upper bound in \eqnref{eq:Gamma_geo-geo_-} is nonzero and saturated and $\Gamma_{n m,+ }^{sp_xp_y,geo-geo}(\bsl{k}_1,\bsl{k}_2)$ is negligible, and thus we can use $2 Z_{nm,-}(\bsl{k}_1,\bsl{k}_2)$ as a good approximation for $\Gamma_{n m}^{sp_xp_y,geo-geo}(\bsl{k}_1,\bsl{k}_2)$.

\subsubsection{Symmetry-Rep Method: Zeroth-Order $|\bsl{k}_\shpa a|$ Approximation For $f_{i, sp_xp_y}(\bsl{k})$}
\label{app:small_k_shpa_MgB2_sp2}

To derive analytic expressions for $\lambda_{\sigma,E}$, $\lambda_{\sigma,E-geo}$ and $\lambda_{\sigma,geo}$ , we should derive analytic expressions for $f_{i, sp_xp_y}^E(\bsl{k})$ in \eqnref{eq:g_E_sp2_MgB2_ini} and for $f_{i, sp_xp_y}^{geo}(\bsl{k})$ in \eqnref{eq:g_geo_sp2_MgB2_ini}.
Since the Fermi surface of $H_{el}^{\text{B},sp_xp_y}$ in \eqnref{eq:H_el_MgB2_B_sp2} is around $\Gamma$-A, $\bsl{k}_\shpa a $ should be small on the Fermi surface.
Therefore, we will only explicitly derive the expressions for $f_{i, sp_xp_y}^E(\bsl{k})$ and $f_{i, sp_xp_y}^{geo}(\bsl{k})$ to zeroth order in $|\bsl{k}_\shpa a|$.
To do so, we only need the explicit forms of the energy bands $E_{sp_xp_y,n}(\bsl{k})$ and the projection matrices $P_{sp_xp_y,n}(\bsl{k})$ to first order in $|\bsl{k}_\shpa a|$ in $f_{i, sp_xp_y}^E(\bsl{k})$ (\eqnref{eq:g_E_sp2_MgB2_ini}) and in $f_{i, sp_xp_y}^{geo}(\bsl{k})$ (\eqnref{eq:g_geo_sp2_MgB2_ini}).
We emphasize that we will \emph{only} use the first-order-$|\bsl{k}_\shpa a|$ approximation for the energy bands and projection matrices $\Gamma_{n m }^{sp_xp_y ,E-E}(\bsl{k}_1,\bsl{k}_2)$, $\Gamma_{n m }^{sp_xp_y, geo-geo}(\bsl{k}_1,\bsl{k}_2)$ and $\Gamma_{n m }^{sp_xp_y ,E-geo}(\bsl{k}_1,\bsl{k}_2)$ in \eqnref{eq:Gamma_sp2_ini}, which contain $f_{i, sp_xp_y}^E(\bsl{k})$ (\eqnref{eq:g_E_sp2_MgB2_ini}), $f_{i, sp_xp_y}^{geo}(\bsl{k})$ (\eqnref{eq:g_geo_sp2_MgB2_ini}).
We do \emph{not} need to (and will \emph{not})  use the first-order-$|\bsl{k}_\shpa a|$ approximation for the energy bands and projection matrices in the other parts of $\lambda_{\sigma}$ (\eqnref{eq:lambda_sp2_E_geo}), unless specified otherwise.

To derive the expressions of $E_{sp_xp_y,n}(\bsl{k})$ and $P_{sp_xp_y,n}(\bsl{k})$ to first order in $|\bsl{k}_\shpa a|$, we can expand $h_{sp_xp_y}(\bsl{k})$ in \eqnref{eq:H_el_MgB2_B_sp2} to first order in $|\bsl{k}_\shpa|$, resulting in
\eq{
\label{eq:MgB2_H_sp2_eff}
h_{sp_xp_y}(\bsl{k}) = U_{sp_xp_y} h_{sp_xp_y,eff}(\bsl{k}) U_{sp_xp_y}^\dagger + O(|\bsl{k}_\shpa a|^2)\ ,
}
where $O(|\bsl{k}_\shpa a|^2)$ include second-order and higher-order terms,
\eq{
U_{sp_xp_y} = \left(
\begin{array}{cccccc}
 \frac{1}{\sqrt{2}} & 0 & 0 & \frac{1}{\sqrt{2}} & 0 & 0 \\
 0 & \frac{1}{\sqrt{2}} & 0 & 0 & \frac{1}{\sqrt{2}} & 0 \\
 0 & 0 & \frac{1}{\sqrt{2}} & 0 & 0 & \frac{1}{\sqrt{2}} \\
 \frac{1}{\sqrt{2}} & 0 & 0 & -\frac{1}{\sqrt{2}} & 0 & 0 \\
 0 & -\frac{1}{\sqrt{2}} & 0 & 0 & \frac{1}{\sqrt{2}} & 0 \\
 0 & 0 & -\frac{1}{\sqrt{2}} & 0 & 0 & \frac{1}{\sqrt{2}} \\
\end{array}
\right)\ ,
}
the columns of $U_{sp_xp_y}$ are eigenvectors of $h_{sp_xp_y}(\bsl{k}=0)$, and
\eqa{
\label{eq:MgB2_H_sp2_eff_explicit}
& h_{sp_xp_y,eff}(\bsl{k}) \\
&= \mat{ 
E_{+,s}  + 2  t_{\text{B},s,z} \cos( k_z c ) & (0 , 0) & 0 & (\ii v_2 k_x a , \ii v_2 k_y a)\\
 h.c. & [E_+ + 2  t_{\text{B},p_x p_y,z} \cos( k_z c )] \sigma_0  & \mat{ -\ii v_1 k_x a \\  -\ii v_1 k_y a} &    (-\ii) (d_x(\bsl{k}_\shpa)\sigma_x + d_y(\bsl{k}_\shpa)\sigma_z)  \\
  h.c.  & h.c.  &  E_{-,s}   + 2  t_{\text{B},s,z} \cos( k_z c ) & (0 , 0)  \\
h.c.  & h.c. & h.c.  & [E_-  + 2  t_{\text{B},p_x p_y,z} \cos( k_z c ) ] \sigma_0 
}\ .
}
The meaning of $\bsl{d}(\bsl{k}_\shpa)$ in \eqnref{eq:MgB2_H_sp2_eff_explicit} is the same as that in \eqnref{eq:MgB2_H_eff_mat_form}; the explicit expression of $\bsl{d}(\bsl{k}_\shpa)$ is in \eqnref{eq:d_form_MgB2_eff}.
As discussed in \appref{app:topo_el_MgB2_sp2}, we will keep $\bsl{d}(\bsl{k}_\shpa)$ instead of directly writing out its explicit form in order to keep track of its winding number.
The relation between the parameters in  \eqnref{eq:MgB2_TB_values_sim} and those in the tight-binding model (\eqnref{eq:MgB2_H_sp2_eff}) is
\eqa{
\label{eq:MgB2_H_sp2_eff_para_values}
& E_+ = E_{B,p_x p_y} - 3 t_2 = 0.614 \eV \ ,\ E_- = E_{B,p_x p_y} + 3 t_2 = 6.746 \eV\ ,\ v = - \frac{\sqrt{3}}{2}  t_3 = 2.30  \eV\\
&  v_1 = - \frac{\sqrt{3}}{2} t_4 = -2.86 \eV\ ,\ v_2 = \frac{\sqrt{3}}{2} t_4 = 2.86 \eV\ ,\ E_{+,s} = 
 E_{B,s} + 3 t_1 = -9.78\eV\ ,\ E_{-,s} = 
 E_{B,s} - 3 t_1 = 6.42 \eV\\
 & t_{B,s,z} = -0.085 \eV\ .
}

Based on \eqnref{eq:MgB2_H_sp2_eff}, the energy dispersion  reads 
\eqa{
\label{eq:E_sp2_k1}
 & E_{sp_xp_y,1}(\bsl{k}) = E_{s,+}(k_z) + O( |\bsl{k}_\shpa a|^2 )\\
 & E_{sp_xp_y,2}(\bsl{k}) = E_{p_x p_y,+}(k_z) + O( |\bsl{k}_\shpa a|^2 )\\
 & E_{sp_xp_y,3}(\bsl{k}) = E_{p_x p_y,+}(k_z) + O( |\bsl{k}_\shpa a|^2 )\\
 & E_{sp_xp_y,4}(\bsl{k}) = E_{s,-}(k_z) + O( |\bsl{k}_\shpa a|^2 )\\
 & E_{sp_xp_y,5}(\bsl{k}) = E_{p_x p_y,-}(k_z) + O( |\bsl{k}_\shpa a|^2 )\\
 & E_{sp_xp_y,6}(\bsl{k}) = E_{p_x p_y,-}(k_z)  + O( |\bsl{k}_\shpa a|^2 )
}
with
\eqa{
\label{eq:E_sp2_k1_explicit}
& E_{s,\pm}(k_z) = E_{\text{B},s,0} \pm 3 t_1 +  2  t_{\text{B},s,z} \cos( k_z c )  \\
& E_{p_x p_y,\pm}(k_z) = E_{\text{B},p_x p_y,0} \mp 3 t_2  +  2  t_{\text{B},p_x p_y,z} \cos( k_z c )  \ .
}
As a result of \eqnref{eq:E_sp2_k1}, we have 
\eq{
\label{eq:dE_sp2}
\nabla_{\bsl{k}_\shpa} E_{sp_xp_y,n}(\bsl{k}) = 0 + O( |\bsl{k}_\shpa a|^1 )\ ,
}
which means that 
\eq{
\label{eq:g_xy_sp2_E_k1}
f_{x, sp_xp_y}^E(\bsl{k}) = f_{y, sp_xp_y}^E(\bsl{k}) = 0 + O( |\bsl{k}_\shpa a|^1 ) 
}
for \eqnref{eq:g_E_geo_sp2_MgB2_ini}.
We note that we always use $E_{sp_xp_y,n}(\bsl{k})$ to label the energy bands of the tight-binding model (\eqnref{eq:H_el_MgB2_B_sp2}), and this is why we include $O( |\bsl{k}_\shpa a|^2 )$ in \eqnref{eq:E_sp2_k1}.
Therefore, $E_{sp_xp_y,n}(\bsl{k})$ is in general non-degenerate away from $\Gamma$-A.
Clearly, the difference between $E_{sp_xp_y,2}(\bsl{k})$ and $E_{sp_xp_y,3}(\bsl{k})$ happens at $O( |\bsl{k}_\shpa a|^2 )$; so do $E_{sp_xp_y,4}(\bsl{k})$ and $E_{sp_xp_y,5}(\bsl{k})$.
With the parameter values in \eqnref{eq:MgB2_H_sp2_eff_para_values}, only $E_{sp_xp_y,2}(\bsl{k})$ and $E_{sp_xp_y,3}(\bsl{k})$ are cut by the Fermi surface. 

For projection matrices, from \eqnref{eq:MgB2_H_sp2_eff} we obtain
\eqa{
\label{eq:U_sp2}
& U_{sp_xp_y,1}(\bsl{k}) = U_{s,+}(\bsl{k})  + O( |\bsl{k}_\shpa a|^2 )\\
& U_{sp_xp_y,4}(\bsl{k}) = U_{s,-}(\bsl{k})  + O( |\bsl{k}_\shpa a|^2 )\\
& \mat{ U_{sp_xp_y,2}(\bsl{k}) & U_{sp_xp_y,3}(\bsl{k}) } = U_{p_x p_y,+}(\bsl{k}) \left(
\begin{array}{cc}
 -\frac{k_y}{\sqrt{k_x^2+k_y^2}} & \frac{k_x}{\sqrt{k_x^2+k_y^2}} \\
 \frac{k_x}{\sqrt{k_x^2+k_y^2}} & \frac{k_y}{\sqrt{k_x^2+k_y^2}} \\
\end{array}
\right)  + O( |\bsl{k}_\shpa a|^2 )\\
& \mat{ U_{sp_xp_y,5}(\bsl{k}) & U_{sp_xp_y,6}(\bsl{k}) } = U_{p_x p_y,-}(\bsl{k}) \left(
\begin{array}{cc}
 -\frac{k_y}{\sqrt{k_x^2+k_y^2}} & \frac{k_x}{\sqrt{k_x^2+k_y^2}} \\
 \frac{k_x}{\sqrt{k_x^2+k_y^2}} & \frac{k_y}{\sqrt{k_x^2+k_y^2}} \\
\end{array}
\right)  + O( |\bsl{k}_\shpa a|^2 )\ ,
}
where
\eqa{
\label{eq:U_sp2_non_eigen}
 & U_{s,+}(\bsl{k}) = \left(
 \begin{array}{c}
 \frac{1}{\sqrt{2}} \\
 \frac{\ii \sqrt{\frac{3}{2}} k_x a t_4}{2 (E_{p_xp_y,-}(k_z)-E_{s,+}(k_z))} \\
 \frac{\ii \sqrt{\frac{3}{2}} k_ya t_4}{2 (E_{p_xp_y,-}(k_z)-E_{s,+}(k_z))} \\
 \frac{1}{\sqrt{2}} \\
 \frac{\ii \sqrt{\frac{3}{2}} k_x a t_4}{2 (E_{p_xp_y,-}(k_z)-E_{s,+}(k_z))} \\
 \frac{\ii \sqrt{\frac{3}{2}} k_ya t_4}{2 (E_{p_xp_y,-}(k_z)-E_{s,+}(k_z))} \\
\end{array}
\right) \ ,\ 
 U_{s,-}(\bsl{k}) = \left(
\begin{array}{c}
 \frac{1}{\sqrt{2}} \\
 -\frac{\ii \sqrt{\frac{3}{2}} k_x a t_4}{2 (E_{p_xp_y,+}(k_z)-E_{s,-}(k_z))} \\
 -\frac{\ii \sqrt{\frac{3}{2}} k_ya t_4}{2 (E_{p_xp_y,+}(k_z)-E_{s,-}(k_z))} \\
 -\frac{1}{\sqrt{2}} \\
 \frac{\ii \sqrt{\frac{3}{2}} k_x a t_4}{2 (E_{p_xp_y,+}(k_z)-E_{s,-}(k_z))} \\
 \frac{\ii \sqrt{\frac{3}{2}} k_ya t_4}{2 (E_{p_xp_y,+}(k_z)-E_{s,-}(k_z))} \\
\end{array}
\right)\\
& 
U_{p_x p_y,+}(\bsl{k}) =  \left(
\begin{array}{cc}
   0 & 0  \\
   \frac{1}{\sqrt{2}} & 0   \\
  0 & \frac{1}{\sqrt{2}}    \\
 0 & 0   \\
   -\frac{1}{\sqrt{2}} & 0  \\
  0 & -\frac{1}{\sqrt{2}}   \\
\end{array}
\right) + \left(
\begin{array}{cc}
 -\frac{\ii \sqrt{\frac{3}{2}} k_x a t_4}{2 (E_{p_xp_y,+}(k_z)-E_{s,-}(k_z))} & -\frac{\ii \sqrt{\frac{3}{2}} k_ya t_4}{2 (E_{p_xp_y,+}(k_z)-E_{s,-}(k_z))} \\
 \frac{-\ii d_y(\bsl{k}_\shpa)}{6 \sqrt{2} t_2} & -\frac{\ii d_x(\bsl{k}_\shpa)}{6 \sqrt{2} t_2} \\
 -\frac{\ii d_x(\bsl{k}_\shpa)}{6 \sqrt{2} t_2} & \frac{\ii d_y(\bsl{k}_\shpa)}{6 \sqrt{2} t_2} \\
 \frac{\ii \sqrt{\frac{3}{2}} k_x a t_4}{2 (E_{p_xp_y,+}(k_z)-E_{s,-}(k_z))} & \frac{\ii \sqrt{\frac{3}{2}} k_ya t_4}{2 (E_{p_xp_y,+}(k_z)-E_{s,-}(k_z))} \\
 -\frac{\ii d_y(\bsl{k}_\shpa)}{6 \sqrt{2} t_2} & -\frac{\ii d_x(\bsl{k}_\shpa)}{6 \sqrt{2} t_2} \\
 -\frac{\ii d_x(\bsl{k}_\shpa)}{6 \sqrt{2} t_2} & \frac{\ii d_y(\bsl{k}_\shpa))}{6 \sqrt{2} t_2} \\
\end{array}
\right) 
\\
& U_{p_x p_y,-}(\bsl{k}) =   \left(
\begin{array}{cccccc}
   0 & 0 \\
   \frac{1}{\sqrt{2}} & 0 \\
   0 & \frac{1}{\sqrt{2}} \\
 0 & 0 \\
   \frac{1}{\sqrt{2}} & 0 \\
  0 & \frac{1}{\sqrt{2}} \\
\end{array}
\right) + \left(
\begin{array}{cc}
 \frac{\ii \sqrt{\frac{3}{2}} k_x a t_4}{2 (E_{p_xp_y,-}(k_z)-E_{s,+}(k_z))} & \frac{\ii \sqrt{\frac{3}{2}} k_ya t_4}{2 (E_{p_xp_y,-}(k_z)-E_{s,+}(k_z))} \\
 \frac{ -\ii d_y(\bsl{k}_\shpa)}{6 \sqrt{2} t_2} & -\frac{\ii d_x(\bsl{k}_\shpa)}{6 \sqrt{2} t_2} \\
 -\frac{\ii d_x(\bsl{k}_\shpa)}{6 \sqrt{2} t_2} & \frac{ \ii d_y(\bsl{k}_\shpa)}{6 \sqrt{2} t_2} \\
 \frac{\ii \sqrt{\frac{3}{2}} k_x a t_4}{2 (E_{p_xp_y,-}(k_z)-E_{s,+}(k_z))} & \frac{\ii \sqrt{\frac{3}{2}} k_ya t_4}{2 (E_{p_xp_y,-}(k_z)-E_{s,+}(k_z))} \\
 \frac{ \ii d_y(\bsl{k}_\shpa)}{6 \sqrt{2} t_2} & \frac{\ii d_x(\bsl{k}_\shpa)}{6 \sqrt{2} t_2} \\
 \frac{\ii d_x(\bsl{k}_\shpa)}{6 \sqrt{2} t_2} & \frac{ -\ii d_y(\bsl{k}_\shpa)}{6 \sqrt{2} t_2} \\
\end{array}
\right)\ .
}
The expressions of the projection matrices can be obtained from $P_{sp_xp_y,n}(\bsl{k})=U_{sp_xp_y,n}(\bsl{k})U_{sp_xp_y,n}^\dagger(\bsl{k})$.
Again, we note that we always use $U_{sp_xp_y,n}(\bsl{k})$ and $P_{sp_xp_y,n}(\bsl{k})$ to respectively label the eigenvectors and projection matrices of the tight-binding model (\eqnref{eq:H_el_MgB2_B_sp2}), and this is why we include $O( |\bsl{k}_\shpa a|^2 )$ in \eqnref{eq:U_sp2}.
From the \eqnref{eq:U_sp2}, we obtain the following relations for the projection matrices, which read
\eqa{
\label{eq:P_sp2_U}
& P_{sp_xp_y,1}(\bsl{k}) = P_{s,+}(\bsl{k})  + O( |\bsl{k}_\shpa a|^2 )\\
& P_{sp_xp_y,4}(\bsl{k}) = P_{s,-}(\bsl{k})  + O( |\bsl{k}_\shpa a|^2 )\\
& P_{sp_xp_y,2}(\bsl{k}) + P_{sp_xp_y,3}(\bsl{k}) = P_{p_x p_y,+}(\bsl{k})  + O( |\bsl{k}_\shpa a|^2 )\\
& P_{sp_xp_y,5}(\bsl{k}) + P_{sp_xp_y,6}(\bsl{k}) = P_{p_x p_y,-}(\bsl{k})  + O( |\bsl{k}_\shpa a|^2 )\ ,
}
where 
\eq{
 P_{s,+}(\bsl{k})   = U_{s,+}(\bsl{k}) U_{s,+}^\dagger(\bsl{k}) =\frac{1}{2} \left(
\begin{array}{cccccc}
 1 & 0 & 0 & 1 & 0 & 0 \\
 0 & 0 & 0 & 0 & 0 & 0 \\
 0 & 0 & 0 & 0 & 0 & 0 \\
 1 & 0 & 0 & 1 & 0 & 0 \\
 0 & 0 & 0 & 0 & 0 & 0 \\
 0 & 0 & 0 & 0 & 0 & 0 \\
\end{array}
\right)-\frac{\ii \sqrt{3} t_4 \widetilde{P}_1(\bsl{k}_\shpa) }{4 (E_{p_x p_y,-}(k_z) - E_{s,+}(k_z))} \ ,
}
\eq{
 P_{s,-}(\bsl{k})   = U_{s,-}(\bsl{k}) U_{s,-}^\dagger(\bsl{k}) = \frac{1}{2} \left(
\begin{array}{cccccc}
 1 & 0 & 0 & -1 & 0 & 0 \\
 0 & 0 & 0 & 0 & 0 & 0 \\
 0 & 0 & 0 & 0 & 0 & 0 \\
 -1 & 0 & 0 & 1 & 0 & 0 \\
 0 & 0 & 0 & 0 & 0 & 0 \\
 0 & 0 & 0 & 0 & 0 & 0 \\
\end{array}
\right) + \frac{\ii \sqrt{3} t_4 \chi^{sp_xp_y}_-\widetilde{P}_1(\bsl{k}_\shpa)\chi^{sp_xp_y}_- }{4 (E_{p_x p_y,+}(k_z) - E_{s,-}(k_z))} \ ,
}
\eq{
\label{eq:P_p_xp_y+_k}
P_{p_x p_y,+}(\bsl{k}) = U_{p_x p_y,+}(\bsl{k}) U_{p_x p_y,+}^\dagger(\bsl{k}) = \left(
\begin{array}{cccccc}
 0 & 0 & 0 & 0 & 0 & 0 \\
 0 & \frac{1}{2} & 0 & 0 & -\frac{1}{2} & 0 \\
 0 & 0 & \frac{1}{2} & 0 & 0 & -\frac{1}{2} \\
 0 & 0 & 0 & 0 & 0 & 0 \\
 0 & -\frac{1}{2} & 0 & 0 & \frac{1}{2} & 0 \\
 0 & 0 & -\frac{1}{2} & 0 & 0 & \frac{1}{2} \\
\end{array}
\right) -\frac{\ii \sqrt{3} t_4 \chi^{sp_xp_y}_-\widetilde{P}_1(\bsl{k}_\shpa)\chi^{sp_xp_y}_- }{4 (E_{p_x p_y,+}(k_z) - E_{s,-}(k_z))} + \frac{ \ii   }{6 t_2} \widetilde{P}_2(\bsl{k}_\shpa)\ ,
}
\eq{
\label{eq:P_p_xp_y-_k}
P_{p_x p_y,-}(\bsl{k}) = U_{p_x p_y,-}(\bsl{k}) U_{p_x p_y,-}^\dagger(\bsl{k}) = \left(
\begin{array}{cccccc}
 0 & 0 & 0 & 0 & 0 & 0 \\
 0 & \frac{1}{2} & 0 & 0 & \frac{1}{2} & 0 \\
 0 & 0 & \frac{1}{2} & 0 & 0 & \frac{1}{2} \\
 0 & 0 & 0 & 0 & 0 & 0 \\
 0 & \frac{1}{2} & 0 & 0 & \frac{1}{2} & 0 \\
 0 & 0 & \frac{1}{2} & 0 & 0 & \frac{1}{2} \\
\end{array}
\right) + \frac{\ii \sqrt{3} t_4  \widetilde{P}_1(\bsl{k}_\shpa) }{4 (E_{p_x p_y,-}(k_z) - E_{s,+}(k_z))} - \frac{ \ii   }{6 t_2} \widetilde{P}_2(\bsl{k}_\shpa)\ ,
}
\eq{
\label{eq:Pt_1}
\widetilde{P}_1(\bsl{k}_\shpa) = \left(
\begin{array}{cccccc}
 0 & k_{x}a & k_{y}a & 0 & k_{x}a & k_{y}a \\
 -k_{x}a & 0 & 0 & -k_{x}a & 0 & 0 \\
 -k_{y}a & 0 & 0 & -k_{y}a & 0 & 0 \\
 0 & k_{x}a & k_{y}a & 0 & k_{x}a & k_{y}a \\
 -k_{x}a & 0 & 0 & -k_{x}a & 0 & 0 \\
 -k_{y}a & 0 & 0 & -k_{y}a & 0 & 0 \\
\end{array}
\right)\ ,
}
\eq{
\label{eq:Pt_2}
\widetilde{P}_2(\bsl{k}_\shpa) = \left(
\begin{array}{cccccc}
 0 & 0 & 0 & 0 & 0 & 0 \\
 0 & 0 & 0 & 0 & d_y(\bsl{k}_\shpa) & d_x(\bsl{k}_\shpa) \\
 0 & 0 & 0 & 0 & d_x(\bsl{k}_\shpa) & -d_y(\bsl{k}_\shpa) \\
 0 & 0 & 0 & 0 & 0 & 0 \\
 0 & -d_y(\bsl{k}_\shpa) & -d_x(\bsl{k}_\shpa) & 0 & 0 & 0 \\
 0 & -d_x(\bsl{k}_\shpa) & d_y(\bsl{k}_\shpa) & 0 & 0 & 0 
\end{array}
\right)\ ,
}
and $d_x(\bsl{k}_\shpa)$ and $d_y(\bsl{k}_\shpa)$ are defined in \eqnref{eq:d_form_MgB2_eff}.
\eqnref{eq:P_sp2_U} shows that 
\eq{
\label{eq:dP_sp2}
\sum_n E_{sp_xp_y,n}(\bsl{k}) \partial_{k_z } P_{sp_xp_y,n}(\bsl{k}) =  \sum_{\delta=\pm} \left[E_{s,\delta}(k_z)  \partial_{k_z } P_{s,\delta}(\bsl{k}) + E_{p_x p_y,\delta}(k_z) \partial_{k_z } P_{p_x p_y,\delta}(\bsl{k}) \right]+ O( |\bsl{k}_\shpa a |^2) = 0 + O( |\bsl{k}_\shpa a|^1 )\ ,
}
meaning that
\eq{
\label{eq:f_z_sp2_geo_k1}
f_{z, sp_xp_y}^{geo}(\bsl{k}) = 0 + O( |\bsl{k}_\shpa a|^1 )
}
for \eqnref{eq:g_E_geo_sp2_MgB2_ini}.
With \eqnref{eq:g_xy_sp2_E_k1}, \eqnref{eq:f_z_sp2_geo_k1} and \eqnref{eq:P_sp2_U}, we can write $f_{i, sp_xp_y}^E(\bsl{k})$ and $f_{i, sp_xp_y}^{geo}(\bsl{k})$ in \eqnref{eq:g_E_geo_sp2_MgB2_ini} as the following:
\eqa{
\label{eq:g_E_sp2_k0}
& f_{i, sp_xp_y}^E(\bsl{k}) \\
& = \ii \delta_{iz}  (\hat{\gamma}_8 \partial_{t_{B,s,z}} + \hat{\gamma}_9 \partial_{t_{B,p_x p_y,z}}) \sum_{n} \partial_{k_z }E_{sp_xp_y,n}(\bsl{k}) P_{sp_xp_y,n}(\bsl{k}) \\
& = \ii \delta_{iz}   (\hat{\gamma}_8 \partial_{t_{B,s,z}} + \hat{\gamma}_9 \partial_{t_{B,p_x p_y,z}})\left[  \sum_{\delta=\pm}  \partial_{k_z } E_{s,\delta}(k_z)  \diag(1,0,0,1,0,0) +  \sum_{\delta=\pm}  \partial_{k_z } E_{p_x p_y,\delta}(k_z)  \diag(0,1,1,0,1,1) \right]  + O( |\bsl{k}_\shpa a|^1 ) \\
& = \ii \delta_{iz}   \left[ \frac{\hat{\gamma}_8}{t_{B,s,z}}  \sum_{\delta=\pm}  \partial_{k_z } E_{s,\delta}(k_z)  \diag(1,0,0,1,0,0) + \frac{\hat{\gamma}_9}{t_{B,p_x p_y,z}}  \sum_{\delta=\pm}  \partial_{k_z } E_{p_x p_y,\delta}(k_z)  \diag(0,1,1,0,1,1) \right]  + O( |\bsl{k}_\shpa a|^1 ) \\
& = \ii \delta_{iz}   \left[ \frac{\hat{\gamma}_8}{t_{B,s,z}}  \sum_{\delta=\pm}  \partial_{k_z } E_{s,\delta}(k_z)  \diag(1,0,0,1,0,0) + \gamma_{\sigma,z}  \sum_{\delta=\pm}  \partial_{k_z } E_{p_x p_y,\delta}(k_z)  \diag(0,1,1,0,1,1) \right]  + O( |\bsl{k}_\shpa a|^1 ) 
}
\eqa{
\label{eq:g_geo_sp2_k0}
& f_{i, sp_xp_y}^{geo}(\bsl{k}) \\
& = (\delta_{ix} + \delta_{iy} ) \ii \sum_{a=1,2,3,4} \hat{\gamma}_a \partial_{t_a} \sum_{\delta=\pm}   \left[ E_{s,\delta}(k_z) \partial_{k_i}  P_{s,\delta}(\bsl{k}) + E_{p_x p_y,\delta}(k_z) \partial_{k_i} P_{p_x p_y,\delta}(\bsl{k}) \right] \\
& \quad + (\delta_{ix} + \delta_{iy} )  \ii \sum_{i'=x,y} \epsilon_{i'i} \ii \left[L_y, (\hat{\gamma}_6 \partial_{t_4} - \frac{1}{2} \hat{\gamma}_7 \partial_{t_3}) \sum_{\delta=\pm}   \left[ E_{s,\delta}(k_z) \partial_{k_{i'}}  P_{s,\delta}(\bsl{k}) + E_{p_x p_y,\delta}(k_z) \partial_{k_{i'}} P_{p_x p_y,\delta}(\bsl{k}) \right] \right] + O( |\bsl{k}_\shpa a|^1 )\ ,
}
where the expression of $\gamma_{\sigma,z}$ is in \eqnref{eq:gamma_z_sp2_MgB2}.
For $f_{i, sp_xp_y}^E(\bsl{k})$ in \eqnref{eq:g_E_sp2_k0}, we have converted $\partial_{t_{B,s,z}}$ to $\frac{1}{t_{B,s,z}}$ and converted $\partial_{t_{B,p_x p_y,z}}$ to $\frac{1}{t_{B,p_x p_y,z}}$, which is enabled by the short-ranged nature of the electron Hamiltonian \eqnref{eq:H_el_MgB2_B_sp2} and the small-$\bsl{k}_\shpa$ approximation.
Specifically, owing to the two approximations, the $k_z$ dependence of $E_{s,\delta}(k_z)$ and $E_{p_x p_y,\delta}(k_z)$ only relies on $t_{B,s,z}$ and $t_{B,p_x p_y,z}$, respectively, and $E_{s,\delta}(k_z)$ and $E_{p_x p_y,\delta}(k_z)$ couple to different matrices, which enables the conversion.

On the other hand, the derivatives with respect to hopping parameters still appear in $f_{i, sp_xp_y}^{geo}(\bsl{k})$ (\eqnref{eq:g_geo_sp2_k0}).
To address this issue, recall that the states of \eqnref{eq:H_el_MgB2_B_sp2} near the Fermi level around $\Gamma$-A mainly originate from the $p_x p_y$ orbitals of B atoms~\cite{Kong02272001MgB2EPC}.
Moreover, we eventually project $f_{i, sp_xp_y}^{geo}(\bsl{k})$ to the Fermi surface according to the expression of the $\left\langle \Gamma \right\rangle^{geo-geo}$ in \eqnref{eq:Gamma_ave_sp2_parts}.
Therefore, the projection of $f_{i, sp_xp_y}^{geo}(\bsl{k})$ to the $p_xp_y$ subspace is reasonable.
We will show that the projection will make $f_{i, sp_xp_y}^{geo}(\bsl{k})$ rely on just one hopping parameter and allow the conversion of hopping derivatives to the normal coefficients in $f_{i, sp_xp_y}^{geo}(\bsl{k})$.
To derive the projected $f_{i, sp_xp_y}^{geo}(\bsl{k})$, we define 
\eq{
\label{eq:U_pxpy}
U_{p_xp_y} = \mat{ U_{p_x p_y, + }(\bsl{k}= 0) & U_{p_x p_y, - }(\bsl{k} = 0)} =  \left(
\begin{array}{cccccc}
   0 & 0 & 0 & 0 \\
   \frac{1}{\sqrt{2}} & 0 & \frac{1}{\sqrt{2}} & 0 \\
  0 & \frac{1}{\sqrt{2}}  & 0 & \frac{1}{\sqrt{2}} \\
 0 & 0  & 0 & 0 \\
   -\frac{1}{\sqrt{2}} & 0  & \frac{1}{\sqrt{2}} & 0 \\
  0 & -\frac{1}{\sqrt{2}}  & 0 & \frac{1}{\sqrt{2}} \\
\end{array}
\right)\ ,
}
where $U_{p_x p_y, \pm }(\bsl{k})$ are in \eqnref{eq:U_sp2_non_eigen}.
Then, we consider $U_{p_xp_y}^\dagger f_{i, sp_xp_y}^{geo}(\bsl{k}) U_{p_xp_y}$, which reads
\eqa{
\label{eq:projected_fispxpy_intermediate}
& U_{p_xp_y}^\dagger  f_{i, sp_xp_y}^{geo}(\bsl{k}) U_{p_xp_y} \\
& = (\delta_{ix} + \delta_{iy} ) \ii \sum_{a=1,2,3,4} \hat{\gamma}_a \partial_{t_a} \sum_{\delta=\pm}   \left[ E_{s,\delta}(k_z) U_{p_xp_y}^\dagger \partial_{k_i}  P_{s,\delta}(\bsl{k})U_{p_xp_y}  + E_{p_x p_y,\delta}(k_z) U_{p_xp_y}^\dagger \partial_{k_i} P_{p_x p_y,\delta}(\bsl{k}) U_{p_xp_y} \right] \\
& \quad + (\delta_{ix} + \delta_{iy} )  \ii \sum_{i'=x,y} \epsilon_{i'i} \ii \left[U_{p_xp_y}^\dagger L_y U_{p_xp_y}, (\hat{\gamma}_6 \partial_{t_4} - \frac{1}{2} \hat{\gamma}_7 \partial_{t_3}) \sum_{\delta=\pm}   \left[ E_{s,\delta}(k_z)  U_{p_xp_y}^\dagger  \partial_{k_{i'}}P_{s,\delta}(\bsl{k}) U_{p_xp_y} \right] \right] \\
& \quad + (\delta_{ix} + \delta_{iy} )  \ii \sum_{i'=x,y} \epsilon_{i'i} \ii \left[U_{p_xp_y}^\dagger L_y U_{p_xp_y}, (\hat{\gamma}_6 \partial_{t_4} - \frac{1}{2} \hat{\gamma}_7 \partial_{t_3}) \sum_{\delta=\pm}   \left[E_{p_x p_y,\delta}(k_z) U_{p_xp_y}^\dagger \partial_{k_{i'}} P_{p_x p_y,\delta}(\bsl{k}) U_{p_xp_y}\right] \right] + O( |\bsl{k}_\shpa a|^1 )\\
& = (\delta_{ix} + \delta_{iy} ) \ii \sum_{a=1,2,3,4} \hat{\gamma}_a \partial_{t_a}    \left[ (E_{p_x p_y,+}(k_z)-E_{p_x p_y,-}(k_z)) U_{p_xp_y}^\dagger \partial_{k_i} P_{p_x p_y,+}(\bsl{k}) U_{p_xp_y} \right] \\
& \quad + (\delta_{ix} + \delta_{iy} )  \ii \sum_{i'=x,y} \epsilon_{i'i} \ii \left[U_{p_xp_y}^\dagger L_y U_{p_xp_y}, (\hat{\gamma}_6 \partial_{t_4} - \frac{1}{2} \hat{\gamma}_7 \partial_{t_3})    \left[ (E_{p_x p_y,+}(k_z)-E_{p_x p_y,-}(k_z)) U_{p_xp_y}^\dagger \partial_{k_i'} P_{p_x p_y,+}(\bsl{k}) U_{p_xp_y} \right] \right] \\
& \quad + O( |\bsl{k}_\shpa a|^1 )\\
& = \frac{\hat{\gamma}_3}{t_3}  (\delta_{ix} + \delta_{iy} ) \ii    (E_{p_x p_y,+}(k_z)-E_{p_x p_y,-}(k_z)) U_{p_xp_y}^\dagger \partial_{k_i} P_{p_x p_y,+}(\bsl{k}) U_{p_xp_y}  \\
& \quad  -\frac{\hat{\gamma}_7}{2t_3} (\delta_{ix} + \delta_{iy} )  \ii \sum_{i'=x,y} \epsilon_{i'i} \ii \left[U_{p_xp_y}^\dagger L_y U_{p_xp_y}, \left[ (E_{p_x p_y,+}(k_z)-E_{p_x p_y,-}(k_z)) U_{p_xp_y}^\dagger \partial_{k_i'} P_{p_x p_y,+}(\bsl{k}) U_{p_xp_y} \right] \right] \\
& \quad + O( |\bsl{k}_\shpa a|^1 )\ ,
}
where we have used $L_y U_{p_xp_y} = U_{p_xp_y} U_{p_xp_y}^\dagger L_y U_{p_xp_y} $ and $U_{p_xp_y}^\dagger L_y = U_{p_xp_y}^\dagger L_y U_{p_xp_y}  U_{p_xp_y}^\dagger$ for the first equality, the second equality comes from $U_{p_xp_y}^\dagger  P_{s,\delta}(\bsl{k}) U_{p_xp_y} = 0$ and
\eq{
U_{p_xp_y}^\dagger  \partial_{k_i} P_{p_x p_y,-}(\bsl{k}) U_{p_xp_y} = - U_{p_xp_y}^\dagger  \partial_{k_i} P_{p_x p_y,+}(\bsl{k}) U_{p_xp_y} \ \forall i=x,y\ .
}
The third equality in \eqnref{eq:projected_fispxpy_intermediate} is derived from
\eqa{
\label{eq:DeltaE_P_d}
& (E_{p_x p_y,+}(k_z)-E_{p_x p_y,-}(k_z)) U_{p_xp_y}^\dagger \partial_{k_i} P_{p_x p_y,+}(\bsl{k}) U_{p_xp_y} \\
& = \left(
\begin{array}{cccc}
 0 & 0 & \ii \partial_{k_i} d_y(\bsl{k}_\shpa) & \ii \partial_{k_i} d_x(\bsl{k}_\shpa) \\
 0 & 0 & \ii \partial_{k_i} d_x(\bsl{k}_\shpa) & -\ii \partial_{k_i} d_y(\bsl{k}_\shpa) \\
 -\ii \partial_{k_i} d_y(\bsl{k}_\shpa) & -\ii \partial_{k_i} d_x(\bsl{k}_\shpa) & 0 & 0 \\
 -\ii \partial_{k_i} d_x(\bsl{k}_\shpa) & \ii \partial_{k_i} d_y(\bsl{k}_\shpa) & 0 & 0 \\
\end{array}
\right)\ \forall i=x,y
}
and $\bsl{d}(\bsl{k}_\shpa)$ only depends on $t_3$ according to \eqnref{eq:d_form_MgB2_eff}.
So by projecting $f_{i, sp_xp_y}^{geo}(\bsl{k})$ to the $p_xp_y$ subspace, we have converted the derivatives with respect to hopping parameters to a scalar factor.
The conversion relies on the fact that within the $p_x p_y$ subspace, only one hopping parameter $t_3$ is responsible for the geometric properties of the wavefunction to first order in $|\bsl{k}_\shpa a|$, as discussed in \appref{app:topo_el_MgB2_sp2}.

To further simplify \eqnref{eq:projected_fispxpy_intermediate}, we note that owing to the explicit expression of $\bsl{d}(\bsl{k}_\shpa)$ in \eqnref{eq:d_form_MgB2_eff}, \eqnref{eq:DeltaE_P_d} leads to 
\eqa{
\label{eq:Two_parts_f_geo_equal_sigma_MgB2}
& \frac{1}{2} \sum_{i'=x,y} \epsilon_{i'i} \ii \left[U_{p_xp_y}^\dagger L_y U_{p_xp_y}, \left[ (E_{p_x p_y,+}(k_z)-E_{p_x p_y,-}(k_z)) U_{p_xp_y}^\dagger \partial_{k_i'} P_{p_x p_y,+}(\bsl{k}) U_{p_xp_y} \right] \right] \\
& = (E_{p_x p_y,+}(k_z)-E_{p_x p_y,-}(k_z)) U_{p_xp_y}^\dagger \partial_{k_i} P_{p_x p_y,+}(\bsl{k}) U_{p_xp_y}\ ,
}
which holds because of the $O(|\bsl{k}_\shpa|)$ approximation that we made for the electron Hamiltonian.
Then, we simplify \eqnref{eq:projected_fispxpy_intermediate} into
\eqa{
& U_{p_xp_y}^\dagger  f_{i, sp_xp_y}^{geo}(\bsl{k}) U_{p_xp_y}
= \frac{\hat{\gamma}_3-\hat{\gamma}_7}{t_3}  (\delta_{ix} + \delta_{iy} ) \ii    (E_{p_x p_y,+}(k_z)-E_{p_x p_y,-}(k_z)) U_{p_xp_y}^\dagger \partial_{k_i} P_{p_x p_y,+}(\bsl{k}) U_{p_xp_y}  + O( |\bsl{k}_\shpa a|^1 )\ .
}
With the expression of $\gamma_{\sigma,\shpa}$ in \eqnref{eq:gamma_spxpy_geo}, we arrive at 
\eq{
\label{eq:g_geo_sp2_k0_projected}
U_{p_xp_y}^\dagger  f_{i, sp_xp_y}^{geo}(\bsl{k}) U_{p_xp_y} = \ii  \gamma_{\sigma,\shpa} (\delta_{ix} + \delta_{iy} ) (E_{p_x p_y,+}(k_z)-E_{p_x p_y,-}(k_z)) U_{p_xp_y}^\dagger \partial_{k_i} P_{p_x p_y,+}(\bsl{k}) U_{p_xp_y}  + O( |\bsl{k}_\shpa a|^1 )\ .
}
In principle, one can use the expression in the last line of \eqnref{eq:projected_fispxpy_intermediate} for $U_{p_xp_y}^\dagger  f_{i, sp_xp_y}^{geo}(\bsl{k}) U_{p_xp_y}$ instead of \eqnref{eq:g_geo_sp2_k0_projected}, which might change the eventual expression of $\lambda_{\sigma,geo}$; but it cannot change the value of $\lambda_{\sigma,geo}$ due to \eqnref{eq:Two_parts_f_geo_equal_sigma_MgB2} under our approximations.
Therefore, we will use \eqnref{eq:g_geo_sp2_k0_projected}.

%
In the following, we will discuss $\lambda_{\sigma, E-geo}$, $\lambda_{\sigma,E}$ and $\lambda_{\sigma,geo}$ in \eqnref{eq:lambda_sp2_E_geo}, one by one.

\subsubsection{$\lambda_{\sigma, E-geo}$}
As shown in \eqnref{eq:g_E_sp2_k0} and \eqnref{eq:g_geo_sp2_k0}, $f_{i, sp_xp_y}^E(\bsl{k})$ only has $i=z$ component and $f_{i, sp_xp_y}^{geo}(\bsl{k})$ only has $i=x,y$ components to zeroth order in $|\bsl{k}_\shpa a|$.
As a result, $\Gamma_{n m }^{sp_xp_y ,E-geo}(\bsl{k}_1,\bsl{k}_2) $ in \eqnref{eq:Gamma_E-geo_sp2_MgB2} would be of order $O(|\bsl{k}_{1,\shpa} a|, |\bsl{k}_{2,\shpa} a|)$, meaning that 
\eq{
\label{eq:lambda_sigma_E-geo_MgB2}
\left\langle \Gamma \right\rangle^{sp_xp_y ,E-geo} = O(k_{F,\shpa}) \Rightarrow \lambda_{\sigma, E-geo}= \frac{2}{ N} \frac{1}{\hbar \mcomega} D_{\sigma}(\mu) \frac{D_{\sigma}(\mu)}{D(\mu)} \left[ O(k_{F,\shpa}) \right]\ ,
}
where $k_{F,\shpa}$ is the largest value of $|\bsl{k}_{\shpa}|$ on the Fermi surfaces of the $sp_xp_y$ orbitals.

\eqnref{eq:lambda_sigma_E-geo_MgB2} matches \eqnref{eq:lambda_E-geo_eff_MgB2} derived from the GA.

\subsubsection{$\lambda_{\sigma,E}$}

Now we discuss $\lambda_{\sigma,E}$.
Recall that $E_{sp_xp_y,n=2,3}(\bsl{k})$ are cut by the Fermi energy $\mu =0$.
Then, for $\Gamma_{nm}^{sp_xp_y, E-E}(\bsl{k}_1,\bsl{k}_2)$ in \eqnref{eq:Gamma_E-E_sp2_MgB2}, we only need to consider $n,m\in \{2,3\}$.
Since we only care about $\Gamma_{nm}^{sp_xp_y, E-E}(\bsl{k}_1,\bsl{k}_2)$ to zeroth order in $\bsl{k}_{1,\shpa}$ and $\bsl{k}_{2,\shpa}$, we will need the following expressions
\eqa{
\label{eq:P_sp2_U_EF}
& P_{sp_xp_y,2}(\bsl{k}) =   U_{p_x p_y, + }(0) P_{+}(\bsl{k}_\shpa) U_{p_x p_y, + }^\dagger(0) + O(|\bsl{k}_{\shpa}|^1) \\
& P_{sp_xp_y,3}(\bsl{k}) =   U_{p_x p_y, + }(0) P_{-}(\bsl{k}_\shpa) U_{p_x p_y, + }^\dagger(0) + O(|\bsl{k}_{\shpa}|^1)\ ,
}
where 
\eq{
\label{eq:U_pxpy_+(0)}
U_{p_x p_y, + }(0) = \left(
\begin{array}{cc}
 0 & 0 \\
 \frac{1}{\sqrt{2}} & 0 \\
 0 & \frac{1}{\sqrt{2}} \\
 0 & 0 \\
 -\frac{1}{\sqrt{2}} & 0 \\
 0 & -\frac{1}{\sqrt{2}} \\
\end{array}
\right)\ .
}
according to \eqnref{eq:U_sp2_non_eigen}, and
\eq{
\label{eq:P_pm_sp2_MgB2}
P_{\pm}(\bsl{k}_\shpa) = \frac{1}{2} \pm \frac{(k_x^2 - k_y^2) \sigma_z + 2 k_x  k_y  \sigma_x }{2 |\bsl{k}_\shpa|^2} \ .
}

Now we try to simplify \eqnref{eq:Gamma_E-E_sp2_MgB2}.
To do so, first note that
\eqa{
& U_{p_x p_y, + }^\dagger(0) \chi_{\bsl{\tau}}^{sp_xp_y} f_{z, sp_xp_y}^E(\bsl{k}) U_{p_x p_y, + }(0)  = \ii \gamma_{\sigma,z}  \sum_{\delta=\pm}  \partial_{k_z } E_{p_x p_y,\delta}(k_z) U_{p_x p_y, + }^\dagger(0) \chi_{\bsl{\tau}}^{sp_xp_y}U_{p_x p_y, + }(0)  + O(|\bsl{k}_{\shpa}|^1)\\
& U_{p_x p_y, + }^\dagger(0)  f_{z, sp_xp_y}^E(\bsl{k}) \chi_{\bsl{\tau}}^{sp_xp_y} U_{p_x p_y, + }(0)  = \ii \gamma_{\sigma,z}  \sum_{\delta=\pm}  \partial_{k_z } E_{p_x p_y,\delta}(k_z) U_{p_x p_y, + }^\dagger(0) \chi_{\bsl{\tau}}^{sp_xp_y}U_{p_x p_y, + }(0)  + O(|\bsl{k}_{\shpa}|^1)
}
according to \eqnref{eq:g_E_sp2_k0} and \eqnref{eq:U_sp2_non_eigen}.
Combined with \eqnref{eq:P_sp2_U_EF}, we are ready to deal with $\left\langle \Gamma \right\rangle^{sp_xp_y,E-E}$ in \eqnref{eq:Gamma_ave_sp2_parts} based on \eqnref{eq:Gamma_E-E_sp2_MgB2}:
\eqa{
& \left\langle \Gamma \right\rangle^{sp_xp_y,E-E} \\
& = \frac{1}{D^2_{\sigma}(\mu)} \sum_{\bsl{k}_1,\bsl{k}_2}^{\BZ}\sum_{n ,m \in \{2,3\}} \delta\left(\mu - E_{sp_xp_y,n}(\bsl{k}_1) \right) \delta\left(\mu - E_{sp_xp_y,m}(\bsl{k}_2) \right) \Gamma_{nm}^{sp_xp_y,E-E}(\bsl{k}_1,\bsl{k}_2) \\
& = \frac{1}{D^2_{\sigma}(\mu)} \sum_{\bsl{k}_1,\bsl{k}_2}^{\BZ}\sum_{n ,m \in \{2,3\}} \delta\left(\mu - E_{sp_xp_y,n}(\bsl{k}_1) \right) \delta\left(\mu - E_{sp_xp_y,m}(\bsl{k}_2) \right)  \\
& \qquad \times \left[\left. \Gamma_{nm}^{sp_xp_y,E-E}(\bsl{k}_1,\bsl{k}_2) \right|_{|\bsl{k}_{1,\shpa}|=|\bsl{k}_{2,\shpa}|=0} + O(|\bsl{k}_{1,\shpa} a|,|\bsl{k}_{2,\shpa} a|)\right] \\
 &  = \frac{1}{D^2_{\sigma}(\mu)} \frac{\hbar}{2 m_{\text{B}}}  \sum_{\bsl{k}_1,\bsl{k}_2}^{\BZ}\sum_{n ,m \in \{2,3\}} \delta\left(\mu - E_{sp_xp_y,n}(\bsl{k}_1) \right) \delta\left(\mu - E_{sp_xp_y,m}(\bsl{k}_2) \right) \\
 & \quad \times\sum_{\bsl{\tau}\in\{\text{B1}, \text{B2}\}}   \Tr\left[ P_{(-)^n}(\bsl{k}_{\shpa,1})  U_{p_x p_y, + }^\dagger(0)(\chi_{\bsl{\tau}}^{sp_xp_y} f_{z, sp_xp_y}^E(\bsl{k}_2) - f_{z, sp_xp_y}^E(\bsl{k}_1) \chi_{\bsl{\tau}}^{sp_xp_y} ) \right. \\
 & \qquad  \left. U_{p_x p_y, + }(0) P_{(-)^m}(\bsl{k}_{\shpa,2})  U_{p_x p_y, + }^\dagger(0)(\chi_{\bsl{\tau}}^{sp_xp_y} f_{z, sp_xp_y}^E(\bsl{k}_1) -  f_{z, sp_xp_y}^E(\bsl{k}_2) \chi_{\bsl{\tau}}^{sp_xp_y})U_{p_x p_y, + }(0) \right] + O(|k_{F,\shpa}|^1) \ .
 }
 We note that we only expand $\Gamma_{nm}^{sp_xp_y,E-E}(\bsl{k}_1,\bsl{k}_2)$ in series of $|\bsl{k}_{\shpa}|$, while leaving $E_{sp_xp_y,n}(\bsl{k})$ in $\delta\left(\mu - E_{sp_xp_y,n}(\bsl{k}) \right)$ untouched. 
 Therefore, $E_{sp_xp_y,n}(\bsl{k})$ in $\delta\left(\mu - E_{sp_xp_y,n}(\bsl{k}) \right)$ is not generally degenerate with other bands away from $\Gamma$-A.

For the integration on the Fermi surfaces, we have
\eqa{
 \label{eq:P_eff_pxpy_+_ave}
 & \sum_{\bsl{k}}^{\BZ} \delta\left(\mu - E_{sp_xp_y,n}(\bsl{k}) \right) F(\bsl{k}) P_{ \pm}(\bsl{k}_{\shpa})  \\
 & =  \frac{\V}{(2\pi)^3} \int_{FS_{sp_xp_y,n}} d\sigma_{\bsl{k}} \frac{1 }{|\nabla_{\bsl{k}} E_{sp_xp_y,n}(\bsl{k})|} F(\bsl{k}) \left[\frac{1}{2} \pm    \frac{1}{2} \frac{k_x^2 - k_y^2}{|\bsl{k}_\shpa|^2} \sigma_z +  \frac{1}{2} \frac{2 k_x k_y}{|\bsl{k}_\shpa|^2} \sigma_x\right] \\
 & = \frac{\V}{(2\pi)^3} \int_{FS_{sp_xp_y,n}} d\sigma_{\bsl{k}} \frac{1 }{|\nabla_{\bsl{k}} E_{sp_xp_y,n}(\bsl{k})|} F(\bsl{k}) \frac{1}{2} \\
 & = \sum_{\bsl{k}}^{\BZ} \delta\left(\mu - E_{sp_xp_y,n}(\bsl{k}) \right) F(\bsl{k}) \frac{1}{2}
 }
 for any $F$ such that $F(C_3\bsl{k}) = F(\bsl{k})$ and $F(\bsl{k}+\bsl{G}) = F(\bsl{k})$ with reciprocal lattice vector $\bsl{G}$, where $FS_{sp_xp_y,n}$ is the $C_3$-invariant Fermi surface given by $E_{sp_xp_y,n}(\bsl{k}) = \mu$.
We need $F(\bsl{k}+\bsl{G}) = F(\bsl{k})$ because we need to extend the domain of $\bsl{k}$ to $\dsR^3$ and then reduce back to $\BZ$ for the first equality in \eqnref{eq:P_eff_pxpy_+_ave}, since the Fermi surface in $\BZ$ might be disconnected. 
 Then, we obtained
  \eqa{
  & \left\langle \Gamma \right\rangle^{sp_xp_y,E-E} \\
  &  = \frac{1}{D^2_{\sigma}(\mu)} \frac{\hbar}{2 m_{\text{B}}}  \sum_{\bsl{k}_1,\bsl{k}_2}^{\BZ}\sum_{n ,m \in \{2,3\}} \delta\left(\mu - E_{sp_xp_y,n}(\bsl{k}_1) \right) \delta\left(\mu - E_{sp_xp_y,m}(\bsl{k}_2) \right) \\
  &. \quad \times \left[  \gamma_{\sigma,z}  \sum_{\delta=\pm}  (\partial_{k_{2,z} } E_{p_x p_y,\delta}(k_{2,z}) - \partial_{k_{1,z} } E_{p_x p_y,\delta}(k_{1,z}) ) \right]^2 \\
 & \quad \times\sum_{\bsl{\tau}\in\{\text{B1}, \text{B2}\}}   \Tr\left[ P_{(-)^n}(\bsl{k}_{\shpa,1})    U_{p_x p_y, + }^\dagger(0) \chi_{\bsl{\tau}}^{sp_xp_y}U_{p_x p_y, + }(0) P_{(-)^m}(\bsl{k}_{\shpa,2})    U_{p_x p_y, + }^\dagger(0) \chi_{\bsl{\tau}}^{sp_xp_y}U_{p_x p_y, + }(0) \right] + O(|k_{F,\shpa}|^1)\ .
 }
 Combined with the fact that 
 \eq{
 \sum_{\bsl{\tau}\in\{\text{B1}, \text{B2}\}}   \Tr\left[ U_{p_x p_y, + }^\dagger(0) \chi_{\bsl{\tau}}^{sp_xp_y}U_{p_x p_y, + }(0) U_{p_x p_y, + }^\dagger(0) \chi_{\bsl{\tau}}^{sp_xp_y}U_{p_x p_y, + }(0) \right] = 1 \ ,
 }
 we can further simplify \eqnref{eq:Gamma_E-E_sp2_MgB2}:
  \eqa{
& \left\langle \Gamma \right\rangle^{sp_xp_y,E-E} \\
  &  = \frac{1}{D^2_{\sigma}(\mu)} \frac{\hbar}{2 m_{\text{B}}}  \sum_{\bsl{k}_1,\bsl{k}_2}^{\BZ}\sum_{n ,m \in \{2,3\}} \delta\left(\mu - E_{sp_xp_y,n}(\bsl{k}_1) \right) \delta\left(\mu - E_{sp_xp_y,m}(\bsl{k}_2) \right) \\
 & \quad \times \left[  \gamma_{\sigma,z}  \sum_{\delta=\pm}  (\partial_{k_{2,z} } E_{p_x p_y,\delta}(k_{2,z}) - \partial_{k_{1,z} } E_{p_x p_y,\delta}(k_{1,z}) ) \right]^2 \\
 & \quad \times\frac{1}{4}\sum_{\bsl{\tau}\in\{\text{B1}, \text{B2}\}}   \Tr\left[ U_{p_x p_y, + }^\dagger(0) \chi_{\bsl{\tau}}^{sp_xp_y}U_{p_x p_y, + }(0) U_{p_x p_y, + }^\dagger(0) \chi_{\bsl{\tau}}^{sp_xp_y}U_{p_x p_y, + }(0) \right] + O(|k_{F,\shpa}|^1)\\
  &  = \frac{1}{4 D^2_{\sigma}(\mu)} \frac{\hbar}{2 m_{\text{B}}}  \sum_{\bsl{k}_1,\bsl{k}_2}^{\BZ}\sum_{n ,m \in \{2,3\}} \delta\left(\mu - E_{sp_xp_y,n}(\bsl{k}_1) \right) \delta\left(\mu - E_{sp_xp_y,m}(\bsl{k}_2) \right) \\
 & \quad \times \left[  \gamma_{\sigma,z}    (\partial_{k_{2,z} } E_{p_x p_y,+}(k_{2,z}) - \partial_{k_{1,z} } E_{p_x p_y,+}(k_{2,z})) ) \right]^2  + O(|k_{F,\shpa}|^1)\\
  &  = \frac{1}{4 D^2_{\sigma}(\mu)} \frac{\hbar}{2 m_{\text{B}}}  \sum_{\bsl{k}_1,\bsl{k}_2}^{\BZ}\sum_{n ,m \in \{2,3\}} \delta\left(\mu - E_{sp_xp_y,n}(\bsl{k}_1) \right) \delta\left(\mu - E_{sp_xp_y,m}(\bsl{k}_2) \right)  (\gamma_{\sigma,z})^2\\
  & \quad \times  \left[     (\partial_{k_{2,z} } E_{p_x p_y,+}(k_{2,z}))^2 + (\partial_{k_{1,z} } E_{p_x p_y,+}(k_{2,z}))^2- 2 \partial_{k_{2,z} } E_{p_x p_y,+}(k_{2,z})\partial_{k_{1,z} } E_{p_x p_y,+}(k_{2,z})  \right]+ O(|k_{F,\shpa}|^1)\\
  &  = \frac{1}{4D^2_{\sigma}(\mu)} \frac{\hbar}{ m_{\text{B}}}  \sum_{\bsl{k}_1,\bsl{k}_2}^{\BZ}\sum_{n ,m \in \{2,3\}} \delta\left(\mu - E_{sp_xp_y,n}(\bsl{k}_1) \right) \delta\left(\mu - E_{sp_xp_y,m}(\bsl{k}_2) \right)  (\gamma_{\sigma,z})^2 (\partial_{k_{2,z} } E_{p_x p_y,-}(k_{2,z}))^2  \\
  & \quad + O(|k_{F,\shpa}|^1)\\
  &  = \frac{1}{4 D_{\sigma}(\mu)} \frac{\hbar}{ m_{\text{B}}}  \sum_{\bsl{k}}^{\BZ}\sum_{n \in \{2,3\}} \delta\left(\mu - E_{sp_xp_y,n}(\bsl{k}) \right)  (\gamma_{\sigma,z})^2 (\partial_{k_{z} } E_{p_x p_y,-}(k_{z}))^2  + O(|k_{F,\shpa}|^1)\\
  &  = \frac{1}{4 D_{\sigma}(\mu)} \frac{\hbar (\gamma_{\sigma,z})^2 }{ m_{\text{B}}}  \sum_{n \in \{2,3\}} \frac{\V}{(2\pi)^3} \int_{FS_{sp_xp_y,n}} d\sigma_{\bsl{k}} \frac{1 }{|\nabla_{\bsl{k}} E_{sp_xp_y,n}(\bsl{k})|}  (\partial_{k_{z} } E_{p_x p_y,-}(k_{z}))^2  + O(|k_{F,\shpa}|^1)\ .
}
Eventually, combined with \eqnref{eq:E_sp2_k1} and \eqnref{eq:lambda_sp2_E_geo}, we arrive at 
\eq{
\left\langle \Gamma \right\rangle^{sp_xp_y,E-E} = \frac{1}{4 D_{\sigma}(\mu)} \frac{\hbar (\gamma_{\sigma,z})^2 }{ m_{\text{B}}}  \frac{\V}{(2\pi)^3} \sum_{n \in \{2,3\}} \int_{FS_{sp_xp_y,n}} d\sigma_{\bsl{k}} \frac{(\partial_{k_{z} } E_{sp_xp_y,n}(\bsl{k}_\shpa = 0 ,k_z))^2}{|\nabla_{\bsl{k}} E_{sp_xp_y,n}(\bsl{k})|}  + O(|k_{F,\shpa}|^1)
}
and
\eq{
\label{eq:lambda_sp2_E}
\lambda_{\sigma,E } = \frac{  (\gamma_{\sigma,z})^2  }{ 2 m_{\text{B} }\mcomega}  \frac{D_{\sigma}(\mu)}{D(\mu)}  \frac{\Omega}{(2\pi)^3} \sum_{n \in \{2,3\}} \int_{FS_{sp_xp_y,n}} d\sigma_{\bsl{k}} \frac{(\partial_{k_{z} } E_{sp_xp_y,n}(\bsl{k} ))^2}{|\nabla_{\bsl{k}} E_{sp_xp_y,n}(\bsl{k})|}  + O(|k_{F,\shpa}|^1) \ ,
}
where the expression of $\gamma_{\sigma,z}$ is in \eqnref{eq:gamma_z_sp2_MgB2}.

\eqnref{eq:lambda_sp2_E} matches \eqnref{eq:lambda_E_eff_MgB2} derived from the GA, if we neglect the $O(|k_{F,\shpa}|^1)$ part and take $E_{sp_xp_y,2}(\bsl{k}) = E_{sp_xp_y,3}(\bsl{k}) = E_{eff,1}(\bsl{k})$.

 \subsubsection{$\lambda_{\sigma,geo}$}
 \label{app:lambda_geo_sp2_MgB2}

Now we move onto $\left\langle \Gamma \right\rangle^{sp_xp_y,geo-geo}$ in \eqnref{eq:Gamma_ave_geo-geo_sp2_MgB2} and $\lambda_{\sigma ,geo}$ in \eqnref{eq:lambda_sp2_E_geo}.

As shown in \eqnref{eq:Gamma_as_sum_Gamma_pm_geo-geo_MgB2_sp2}, $\Gamma_{n m }^{sp_xp_y , geo-geo}(\bsl{k}_1,\bsl{k}_2)$ is split into two parts---$\Gamma_{n m,\pm}^{sp_xp_y , geo-geo}(\bsl{k}_1,\bsl{k}_2)$ in \eqnref{eq:Gamma_pm_geo-geo_MgB2_sp2}.
In the following, we will show that $\Gamma_{n m }^{sp_xp_y , geo-geo}(\bsl{k}_1,\bsl{k}_2)$ can be approximated by the upper bound $2Z_{nm,-}(\bsl{k}_1,\bsl{k}_2)$ of $\Gamma_{n m,-}^{sp_xp_y , geo-geo}(\bsl{k}_1,\bsl{k}_2)$, where $Z_{nm,-}(\bsl{k}_1,\bsl{k}_2)$ in \eqnref{eq:ZandY_sp2_geo} and the upper bound is shown in \eqnref{eq:Gamma_geo-geo_-}.
To show this, we define 
\eqa{
\label{eq:Delta_Gamma}
& \Delta \Gamma_{n m }^{sp_xp_y,geo-geo}(\bsl{k}_1,\bsl{k}_2) \\
& = \Gamma_{n m}^{sp_xp_y , geo-geo}(\bsl{k}_1,\bsl{k}_2) -2 Z_{nm,-}(\bsl{k}_1,\bsl{k}_2) \\
& = \frac{\hbar}{4} \frac{1}{m_{\text{B}}}   \sum_{i} \Tr\left[ P_{sp_xp_y,n}(\bsl{k}_1)  ( f_{i, sp_xp_y}^{geo}(\bsl{k}_{2}) - f_{i, sp_xp_y}^{geo}(\bsl{k}_{1}) ) P_{sp_xp_y,m}(\bsl{k}_2)   ( f_{i, sp_xp_y}^{geo}(\bsl{k}_{1}) - f_{i, sp_xp_y}^{geo}(\bsl{k}_{2}) )  \right] \\
 & \quad +  \frac{\hbar}{4} \frac{1}{m_{\text{B}}}  \sum_{i} \Tr\left[ P_{sp_xp_y,n}(\bsl{k}_1)  (\chi_{-}^{sp_xp_y}  f_{i, sp_xp_y}^{geo}(\bsl{k}_{2})+f_{i, sp_xp_y}^{geo}(\bsl{k}_{1}) \chi_{-}^{sp_xp_y}) \right.\\
 & \qquad \qquad \left. P_{sp_xp_y,m}(\bsl{k}_2)   (\chi_{-}^{sp_xp_y} f_{i, sp_xp_y}^{geo}(\bsl{k}_{1}) + f_{i, sp_xp_y}^{geo}(\bsl{k}_{2}) \chi_{-}^{sp_xp_y} )  \right]\ .
}
By substituting \eqnref{eq:g_geo_sp2_k0} into \eqnref{eq:Delta_Gamma}, we obtain
\eqa{
& \Delta \Gamma_{n m }^{sp_xp_y, geo-geo}(\bsl{k}_1,\bsl{k}_2) \\
 & = \frac{\hbar}{4} \frac{1}{m_{\text{B}}}   \sum_{i=x,y} \Tr\left[ P_{sp_xp_y,n}(\bsl{k}_1)  ( f_{i, sp_xp_y}^{geo}(\bsl{k}_{2}) - f_{i, sp_xp_y}^{geo}(\bsl{k}_{1}) ) P_{sp_xp_y,m}(\bsl{k}_2)   ( f_{i, sp_xp_y}^{geo}(\bsl{k}_{1}) - f_{i, sp_xp_y}^{geo}(\bsl{k}_{2}) )  \right] \\
 & \quad +  \frac{\hbar}{4} \frac{1}{m_{\text{B}}}  \sum_{i=x,y} \Tr\left[ P_{sp_xp_y,n}(\bsl{k}_1)  \chi_{-}^{sp_xp_y}(  f_{i, sp_xp_y}^{geo}(\bsl{k}_{2})-f_{i, sp_xp_y}^{geo}(\bsl{k}_{1})) P_{sp_xp_y,m}(\bsl{k}_2)   \chi_{-}^{sp_xp_y} ( f_{i, sp_xp_y}^{geo}(\bsl{k}_{1}) - f_{i, sp_xp_y}^{geo}(\bsl{k}_{2}) )  \right] \\
 & \quad +  O(|\bsl{k}_{1,\shpa} a|, |\bsl{k}_{2,\shpa} a|)\\
 & = O(|\bsl{k}_{1,\shpa} a|, |\bsl{k}_{2,\shpa} a|)\ ,
}
where we have used the fact that $f_{i=x/y, sp_xp_y}(\bsl{k})$ is off-diagonal in the sub-lattice basis as shown in \eqnref{eq:g_X_MgB2} with \eqnref{eq:g_shpa_MgB2} and \eqnref{eq:g_perp_MgB2}.
Then, we obtain
\eqa{
& \Gamma_{n m }^{sp_xp_y, geo-geo}(\bsl{k}_1,\bsl{k}_2) \\
& \quad = -  \frac{\hbar}{2} \frac{1}{m_{\text{B}}}  \sum_{i} \Tr\left[ \chi_{-}^{sp_xp_y} f_{i, sp_xp_y}^{geo}(\bsl{k}_{1}) P_{sp_xp_y,n}(\bsl{k}_1)   f_{i, sp_xp_y}^{geo}(\bsl{k}_{1}) \chi_{-}^{sp_xp_y}  P_{sp_xp_y,m}(\bsl{k}_2) \right]+ (\bsl{k}_1\leftrightarrow \bsl{k}_2)  + O(|\bsl{k}_{1,\shpa} a|, |\bsl{k}_{2,\shpa} a|)
}
As discussed in \appref{app:small_k_shpa_MgB2_sp2}, only $n,m\in\{ 2,3\}$ are cut by the Fermi level.
Since the deviation of $P_{sp_xp_y,2}(\bsl{k})$ and $P_{sp_xp_y,3}(\bsl{k})$ from the $p_x p_y$ subspace is at the order of $|\bsl{k}_\shpa a|$ according to \eqnref{eq:P_sp2_U_EF}, we have
\eqa{
& P_{sp_xp_y,2} =U_{p_xp_y}U_{p_xp_y}^\dagger P_{sp_xp_y,2}(\bsl{k}) U_{p_xp_y}U_{p_xp_y}^\dagger + O( |\bsl{k}_\shpa a | )\\
& P_{sp_xp_y,3} =U_{p_xp_y}U_{p_xp_y}^\dagger  P_{sp_xp_y,3}(\bsl{k}) U_{p_xp_y}U_{p_xp_y}^\dagger + O( |\bsl{k}_\shpa a | )\ ,
}
where $U_{p_xp_y}$ is in \eqnref{eq:U_pxpy}.
Then, combined with 
\eqa{
& \chi_{-}^{sp_xp_y} U_{p_xp_y} = U_{p_xp_y} U_{p_xp_y}^\dagger \chi_{-}^{sp_xp_y}  U_{p_xp_y} \\
&  U_{p_xp_y}^\dagger \chi_{-}^{sp_xp_y}  = U_{p_xp_y}^\dagger \chi_{-}^{sp_xp_y}  U_{p_xp_y} U_{p_xp_y}^\dagger
}
which are derived from  \eqnref{eq:chi_sp2_pm} and \eqnref{eq:U_pxpy}, we have for $n,m\in\{ 2,3\}$
\eqa{
\label{eq:Gamma_geo_geo_sp2_nm_FS}
& \Gamma_{n m }^{sp_xp_y, geo-geo}(\bsl{k}_1,\bsl{k}_2) \\
& \quad = -  \frac{\hbar}{2} \frac{1}{m_{\text{B}}}  \sum_{i} \Tr\left[ \chi_{-}^{sp_xp_y} f_{i, sp_xp_y}^{geo}(\bsl{k}_{1})U_{p_xp_y}U_{p_xp_y}^\dagger P_{sp_xp_y,n}(\bsl{k}_1) U_{p_xp_y}U_{p_xp_y}^\dagger  \right.\\
& \qquad \qquad \left.f_{i, sp_xp_y}^{geo}(\bsl{k}_{1}) \chi_{-}^{sp_xp_y}  U_{p_xp_y}U_{p_xp_y}^\dagger P_{sp_xp_y,m}(\bsl{k}_2) U_{p_xp_y}U_{p_xp_y}^\dagger \right]+ (\bsl{k}_1\leftrightarrow \bsl{k}_2)  + O(|\bsl{k}_{1,\shpa} a|, |\bsl{k}_{2,\shpa} a|)\\
& \quad =  \frac{\hbar}{2} \frac{1}{m_{\text{B}}}  \sum_{i} \Tr\left[ \chi_{-}^{sp_xp_y} U_{p_xp_y}U_{p_xp_y}^\dagger f_{i, sp_xp_y}^{geo}(\bsl{k}_{1})U_{p_xp_y}U_{p_xp_y}^\dagger P_{sp_xp_y,n}(\bsl{k}_1) U_{p_xp_y}  \right.\\
& \qquad \qquad \left. U_{p_xp_y}^\dagger f_{i, sp_xp_y}^{geo}(\bsl{k}_{1}) U_{p_xp_y} U_{p_xp_y}^\dagger\chi_{-}^{sp_xp_y}  U_{p_xp_y}U_{p_xp_y}^\dagger P_{sp_xp_y,m}(\bsl{k}_2) U_{p_xp_y}U_{p_xp_y}^\dagger \right]+ (\bsl{k}_1\leftrightarrow \bsl{k}_2)  + O(|\bsl{k}_{1,\shpa} a|, |\bsl{k}_{2,\shpa} a|) \\
& \quad = \frac{\hbar}{2} \frac{ \gamma_{\sigma,\shpa}^2}{m_{\text{B}}}  \sum_{i=x,y}  (E_{p_x p_y,+}(k_{1,z})-E_{p_x p_y,-}(k_{1,z}))^2 \Tr\left[ \chi_{-}^{sp_xp_y} U_{p_xp_y}U_{p_xp_y}^\dagger \partial_{k_i} P_{p_x p_y,+}(\bsl{k}_1) U_{p_xp_y}U_{p_xp_y}^\dagger P_{sp_xp_y,n}(\bsl{k}_1) U_{p_xp_y}  \right.\\
& \qquad \qquad \left. U_{p_xp_y}^\dagger\partial_{k_i} P_{p_x p_y,+}(\bsl{k}_1) U_{p_xp_y} U_{p_xp_y}^\dagger\chi_{-}^{sp_xp_y}  U_{p_xp_y}U_{p_xp_y}^\dagger P_{sp_xp_y,m}(\bsl{k}_2) U_{p_xp_y}U_{p_xp_y}^\dagger \right]+ (\bsl{k}_1\leftrightarrow \bsl{k}_2)  + O(|\bsl{k}_{1,\shpa} a|, |\bsl{k}_{2,\shpa} a|) \\
& \quad = \frac{\hbar}{2} \frac{ \gamma_{\sigma,\shpa}^2}{m_{\text{B}}}  \sum_{i=x,y}  (E_{p_x p_y,+}(k_{1,z})-E_{p_x p_y,-}(k_{1,z}))^2 \Tr\left[ \chi_{-}^{sp_xp_y}  \partial_{k_i} P_{p_x p_y,+}(\bsl{k}_1)   P_{sp_xp_y,n}(\bsl{k}_1) \right.\\
& \qquad \qquad \left.  \partial_{k_i} P_{p_x p_y,+}(\bsl{k}_1) \chi_{-}^{sp_xp_y}    P_{sp_xp_y,m}(\bsl{k}_2)  \right]+ (\bsl{k}_1\leftrightarrow \bsl{k}_2)  + O(|\bsl{k}_{1,\shpa} a|, |\bsl{k}_{2,\shpa} a|) \ ,
}
where we used \eqnref{eq:g_geo_sp2_k0_projected} for $U_{p_xp_y}^\dagger f_{i, sp_xp_y}^{geo}(\bsl{k}_{1}) U_{p_xp_y}$.

 With \eqnref{eq:Gamma_geo_geo_sp2_nm_FS}, we now discuss $\left\langle \Gamma \right\rangle^{sp_xp_y,geo-geo}$ in \eqnref{eq:Gamma_ave_sp2_parts}, which reads
\eqa{
\label{eq:Gamma_geogeo_ave_intermidiate}
& \left\langle \Gamma \right\rangle^{sp_xp_y,geo-geo} =  \frac{1}{D^2_{\sigma}(\mu)}  \sum_{\bsl{k}_1,\bsl{k}_2}^{\BZ}\sum_{n ,m } \delta\left(\mu - E_{sp_xp_y,n}(\bsl{k}_1) \right) \delta\left(\mu - E_{sp_xp_y,m}(\bsl{k}_2) \right)  \Gamma_{n m }^{sp_xp_y, geo-geo}(\bsl{k}_1,\bsl{k}_2) \\
& =  2 \frac{\hbar}{2} \frac{ \gamma_{\sigma,\shpa}^2}{m_{\text{B}}} \frac{1}{D^2_{\sigma}(\mu)}   \sum_{\bsl{k}_1,\bsl{k}_2}^{\BZ}\sum_{n ,m \in \{ 2,3\}} \delta\left(\mu - E_{sp_xp_y,n}(\bsl{k}_1) \right) \delta\left(\mu - E_{sp_xp_y,m}(\bsl{k}_2) \right) (E_{p_x p_y,+}(k_{1,z})-E_{p_x p_y,-}(k_{1,z}))^2  \\
& \quad \times    \sum_{i=x,y}  \Tr\left[ \chi_{-}^{sp_xp_y}  \partial_{k_i} P_{p_x p_y,+}(\bsl{k}_1)   P_{sp_xp_y,n}(\bsl{k}_1) \partial_{k_i} P_{p_x p_y,+}(\bsl{k}_1) \chi_{-}^{sp_xp_y}    P_{sp_xp_y,m}(\bsl{k}_2)  \right]+ O(k_{F,\shpa})  \\
& =  2 \frac{\hbar}{2} \frac{ \gamma_{\sigma,\shpa}^2}{m_{\text{B}}} \frac{1}{D^2_{\sigma}(\mu)}   \sum_{\bsl{k}_1,\bsl{k}_2}^{\BZ}\sum_{n ,m \in \{ 2,3\}} \delta\left(\mu - E_{sp_xp_y,n}(\bsl{k}_1) \right) \delta\left(\mu - E_{sp_xp_y,m}(\bsl{k}_2) \right) (E_{p_x p_y,+}(k_{1,z})-E_{p_x p_y,-}(k_{1,z}))^2  \\
& \quad \times    \sum_{i=x,y}  \Tr\left[ \chi_{-}^{sp_xp_y}  \partial_{k_i} P_{p_x p_y,+}(\bsl{k}_1)   P_{sp_xp_y,n}(\bsl{k}_1) \partial_{k_i} P_{p_x p_y,+}(\bsl{k}_1) \chi_{-}^{sp_xp_y}    U_{p_xp_y,+}(0) P_{(-)^m}(\bsl{k}_{2,\shpa}) U_{p_xp_y,+}^\dagger(0)    \right]+ O(k_{F,\shpa})  \\
& =  \hbar \frac{ \gamma_{\sigma,\shpa}^2}{m_{\text{B}}} \frac{1}{D_{\sigma}(\mu)}   \sum_{\bsl{k}}^{\BZ}\sum_{n  \in \{ 2,3\}} \delta\left(\mu - E_{sp_xp_y,n}(\bsl{k}) \right)  (E_{p_x p_y,+}(k_{1,z})-E_{p_x p_y,-}(k_{1,z}))^2  \\
& \quad \times    \sum_{i=x,y}  \Tr\left[ \chi_{-}^{sp_xp_y}  \partial_{k_i} P_{p_x p_y,+}(\bsl{k})   P_{sp_xp_y,n}(\bsl{k}) \partial_{k_i} P_{p_x p_y,+}(\bsl{k}) \chi_{-}^{sp_xp_y}    U_{p_xp_y,+}(0) U_{p_xp_y,+}^\dagger(0)  \right]+ O(k_{F,\shpa})  \ ,
}
where $U_{p_xp_y,+}(0)$ is in \eqnref{eq:U_pxpy_+(0)}, we have used \eqnref{eq:P_sp2_U_EF} for the third equality, and used \eqnref{eq:P_eff_pxpy_+_ave} for the fourth equality.
Note that again, we only expand $\Gamma_{n m }^{sp_xp_y, geo-geo}(\bsl{k}_1,\bsl{k}_2)$ in series of $|\bsl{k}_{\shpa}|$, but leave $E_{sp_xp_y,n}(\bsl{k})$ in $\delta\left(\mu - E_{sp_xp_y,n}(\bsl{k}) \right)$ untouched.

Combined with \eqnref{eq:lambda_sp2_parts}, we have
\eqa{
\label{eq:lambda_geo_MgB2_sim}
& \lambda_{\sigma ,geo} \\
& = \frac{2}{ N} \frac{1}{\hbar \mcomega} D_{\sigma}(\mu) \frac{D_{\sigma}(\mu)}{D(\mu)} \hbar  \frac{ \gamma_{\sigma,\shpa}^2}{m_{\text{B}}} \frac{1}{D_{\sigma}(\mu)}   \sum_{\bsl{k}}^{\BZ}\sum_{n  \in \{ 2,3\}} \delta\left(\mu - E_{sp_xp_y,n}(\bsl{k}) \right)  \Delta E_{p_x p_y,+}^2  \\
& \quad \times    \sum_{i=x,y}  \Tr\left[ \chi_{-}^{sp_xp_y}  \partial_{k_i} P_{p_x p_y,+}(\bsl{k})   P_{sp_xp_y,n}(\bsl{k}) \partial_{k_i} P_{p_x p_y,+}(\bsl{k}) \chi_{-}^{sp_xp_y}    U_{p_xp_y,+}(0) U_{p_xp_y,+}^\dagger(0)  \right]+ O(k_{F,\shpa}) \\
& =   \frac{D_{\sigma}(\mu)}{D(\mu)}   \frac{ \gamma_{\sigma,\shpa}^2}{ m_{\text{B}}  \mcomega}     \frac{\Omega}{(2\pi)^3}\sum_{n  \in \{ 2,3\}} \int_{FS_{sp_xp_y , n}} d\sigma_{\bsl{k}}\frac{1}{|\nabla_{\bsl{k}} E_{sp_xp_y,n}(\bsl{k})|}  \Delta E_{p_x p_y,+}^2  \\
& \quad \times    \sum_{i=x,y}  \Tr\left[ \partial_{k_i} P_{p_x p_y,+}(\bsl{k})   P_{sp_xp_y,n}(\bsl{k}) \partial_{k_i} P_{p_x p_y,+}(\bsl{k})     P_{p_xp_y,-}(0)   \right]+ O(k_{F,\shpa}) \ ,
}
where $\gamma_{\sigma,\shpa}$ is in \eqnref{eq:gamma_spxpy_geo},
\eq{
\label{eq:DeltaE_pxpy_+}
\Delta E_{p_x p_y,+} = E_{p_x p_y,+}(k_{1,z})-E_{p_x p_y,-}(k_{1,z}) =  - 6 t_2 
}
based on \eqnref{eq:E_sp2_k1_explicit}, and we, for the last equality, have used 
\eq{
\chi_{-}^{sp_xp_y}    U_{p_xp_y,+}(0) U_{p_xp_y,+}^\dagger(0)   \chi_{-}^{sp_xp_y}  = U_{p_xp_y,-}(0) U_{p_xp_y,-}^\dagger(0) = P_{p_xp_y,-}(0)
}
derived from on \eqnref{eq:chi_sp2_pm}, \eqnref{eq:P_p_xp_y-_k} and  and \eqnref{eq:U_pxpy_+(0)}.

\eqnref{eq:lambda_geo_MgB2_sim} matches \eqnref{eq:lambda_geo_eff_MgB2} derived from the GA if we neglect the $ O(k_{F,\shpa})$ term and take $E_{sp_xp_y,2}(\bsl{k}) = E_{sp_xp_y,3}(\bsl{k}) = E_{eff,1}(\bsl{k})$, since $P_{p_x p_y,+/-}(\bsl{k})$ corresponds to $P_{eff,1/2}(\bsl{k}_\shpa)$ if projecting $P_{p_x p_y,\pm}(\bsl{k})$ to the basis in \eqnref{eq:MgB2_eff_basis} according to \eqnref{eq:P_p_xp_y-_k}, \eqnref{eq:P_p_xp_y+_k} and \eqnref{eq:P_eff_MgB2}.

\subsubsection{$\lambda_{\sigma,topo}$}

At last, we consider the topological term.
From \eqnref{eq:P_sp2_U_EF}, we have for $n\in \{ 2, 3\}$ and $i=x,y$,
\eqa{
& \Tr\left[  \chi_{-}^{sp_xp_y} \partial_{k_i} P_{p_x p_y,+}(\bsl{k}) P_{sp_xp_y,n}(\bsl{k})     \partial_{k_i} P_{p_x p_y,+}(\bsl{k})  \chi_{-}^{sp_xp_y} U_{p_xp_y,+}(0) U_{p_xp_y,+}^\dagger(0)       \right]  \\
& = \Tr\left[ U_{p_xp_y,+}^\dagger(0)  \chi_{-}^{sp_xp_y} \partial_{k_i} P_{p_x p_y,+}(\bsl{k})  U_{p_xp_y,+}(0) P_{(-)^n}(\bsl{k}_\shpa)  U_{p_xp_y,+}^\dagger(0)  \chi_{-}^{sp_xp_y} \partial_{k_i} P_{p_x p_y,+}(\bsl{k})  U_{p_xp_y,+}(0)       \right] + O( |\bsl{k}_\shpa a|^1 )\\
& =   \frac{1}{2}  \Tr\left[ \left(U_{p_xp_y,+}^\dagger(0)  \chi_{-}^{sp_xp_y} \partial_{k_i} P_{p_x p_y,+}(\bsl{k})  U_{p_xp_y,+}(0)  \right)^2\right]\\
& \quad +   (-)^n \Tr\left[ \left(U_{p_xp_y,+}^\dagger(0)  \chi_{-}^{sp_xp_y} \partial_{k_i} P_{p_x p_y,+}(\bsl{k})  U_{p_xp_y,+}(0)  \right)^2  \frac{(k_x^2 - k_y^2) \sigma_z + 2 k_x  k_y  \sigma_x }{2 |\bsl{k}_\shpa|^2}   \right] + O( |\bsl{k}_\shpa a|^1 )\ ,
}
where we used the form of the projection matrix in \eqnref{eq:P_sp2_U_EF} and the expression of $P_{\pm}$ in \eqnref{eq:P_pm_sp2_MgB2}.
According to \eqnref{eq:P_p_xp_y+_k} with \eqnref{eq:Pt_1} and \eqnref{eq:Pt_2}, we have
\eq{
U_{p_xp_y,+}^\dagger(0)  \chi_{-}^{sp_xp_y} \partial_{k_i} P_{p_x p_y,+}(\bsl{k})  U_{p_xp_y,+}(0)  = - \frac{\ii}{ 6 t_2} \partial_{k_i} \left[ d_x(\bsl{k}_\shpa) \sigma_x + d_y(\bsl{k}_\shpa) \sigma_z  \right]\ .
}
Then, for $n\in \{ 2, 3\}$ and $i=x,y$, we obtain
\eqa{
& (\Delta E_{p_x p_y,+})^2   \Tr\left[  \chi_{-}^{sp_xp_y} \partial_{k_i} P_{p_x p_y,+}(\bsl{k}) P_{sp_xp_y,n}(\bsl{k})     \partial_{k_i} P_{p_x p_y,+}(\bsl{k})  \chi_{-}^{sp_xp_y} U_{p_xp_y,+}(0) U_{p_xp_y,+}^\dagger(0)       \right] \\
& = -  |\partial_{k_i} \bsl{d}(\bsl{k}_\shpa)|^2  -   (-)^n \Tr\left[ \left(\partial_{k_i} \left[ d_x(\bsl{k}_\shpa) \sigma_x + d_y(\bsl{k}_\shpa) \sigma_z  \right] \right)^2  \frac{(k_x^2 - k_y^2) \sigma_z + 2 k_x  k_y  \sigma_x }{2 |\bsl{k}_\shpa|^2}   \right] + O( |\bsl{k}_\shpa a|^1 ) ,
}
where we used \eqnref{eq:DeltaE_pxpy_+}, and $\bsl{d}(\bsl{k}_\shpa) = (d_x (\bsl{k}_\shpa) , d_y (\bsl{k}_\shpa)) $ is defined in \eqnref{eq:d_form_MgB2_eff}.
As 
\eq{
|\partial_{k_i} \bsl{d}(\bsl{k}_\shpa)|^2 = |\partial_{k_i} |\bsl{d}(\bsl{k}_\shpa)| e^{\ii \theta(\bsl{k}_\shpa)}|^2 = (\partial_{k_i} |\bsl{d}(\bsl{k}_\shpa)|)^2 + |\bsl{d}(\bsl{k}_\shpa)|^2 |\partial_{k_i}\theta(\bsl{k}_\shpa)|^2\ ,
}
where $\bsl{d}(\bsl{k}_\shpa) = |\bsl{d}(\bsl{k}_\shpa)| (\cos(\theta_{\bsl{k}_\shpa}), \sin(\theta_{\bsl{k}_\shpa}))$.
Then, for $n\in \{ 2, 3\}$ and $i=x,y$, we obtain 
\eqa{
&(\Delta E_{p_x p_y,+})^2   \Tr\left[  \chi_{-}^{sp_xp_y} \partial_{k_i} P_{p_x p_y,+}(\bsl{k}) P_{sp_xp_y,n}(\bsl{k})     \partial_{k_i} P_{p_x p_y,+}(\bsl{k})  \chi_{-}^{sp_xp_y} U_{p_xp_y,+}(0) U_{p_xp_y,+}^\dagger(0)       \right] \\
& \leq  -  |\bsl{d}(\bsl{k}_\shpa)|^2 |\partial_{k_i}\theta(\bsl{k}_\shpa)|^2 -   (-)^n \Tr\left[ \left(\partial_{k_i} \left[ d_x(\bsl{k}_\shpa) \sigma_x + d_y(\bsl{k}_\shpa) \sigma_z  \right] \right)^2  \frac{(k_x^2 - k_y^2) \sigma_z + 2 k_x  k_y  \sigma_x }{2 |\bsl{k}_\shpa|^2}   \right] + O( |\bsl{k}_\shpa a|^1 )\ .
}
Combined with \eqnref{eq:Gamma_geogeo_ave_intermidiate} and \eqnref{eq:d_form_MgB2_eff}, we have
\eqa{
\label{eq:Gamma_topo_ave_intermidiate}
& \left\langle \Gamma \right\rangle^{sp_xp_y,geo-geo} \\
& \geq   \frac{\gamma_{\sigma,\shpa}^2}{D_{\sigma}(\mu)} \frac{\hbar}{ 2 m_{\text{B}}}  \sum_{\bsl{k}}^{\BZ}\sum_{n \in \{ 2,3\}}  \delta\left(\mu - E_{sp_xp_y,n}(\bsl{k}) \right) \sum_{i=x,y}      |\bsl{d}(\bsl{k}_\shpa)|^2 |\partial_{k_i}\theta(\bsl{k}_\shpa)|^2  \\
& \quad + \frac{\gamma_{\sigma,\shpa}^2}{D_{\sigma}(\mu)} \frac{\hbar}{ 2 m_{\text{B}}}  \sum_{\bsl{k}}^{\BZ}\sum_{n \in \{ 2,3\}}  \delta\left(\mu - E_{sp_xp_y,n}(\bsl{k}) \right) \\
& \qquad\times \sum_{i=x,y}     (-)^n \Tr\left[ \left(\partial_{k_i} \left[ d_x(\bsl{k}_\shpa) \sigma_x + d_y(\bsl{k}_\shpa) \sigma_z  \right] \right)^2  \frac{(k_x^2 - k_y^2) \sigma_z + 2 k_x  k_y  \sigma_x }{2 |\bsl{k}_\shpa|^2}   \right]  + O( k_{F,\shpa} )\\
& = \frac{ \gamma_{\sigma,\shpa}^2 }{2 D_{\sigma}(\mu)}   \frac{\hbar}{  m_{\text{B}}}   \frac{\V}{(2\pi)^3}  \sum_{n \in \{ 2,3\}} \int_{FS_{sp_xp_y , n}} d\sigma_{\bsl{k}}\  \frac{|\bsl{d}(\bsl{k}_\shpa)|^2}{|\nabla_{\bsl{k}} E_{sp_xp_y,n}(\bsl{k})|} |\partial_{\bsl{k}_\shpa}\theta(\bsl{k}_\shpa)|^2 + O( k_{F,\shpa} )\\
& \geq  \frac{  \gamma_{\sigma,\shpa}^2 }{2 D_{\sigma}(\mu)} \frac{\hbar}{  m_{\text{B}}}   \frac{\V}{(2\pi)^3}    \sum_{n \in \{ 2,3\}} \frac{\left[\int_{FS_{sp_xp_y , n}} d\sigma_{\bsl{k}}\  |\partial_{\bsl{k}_\shpa}\theta(\bsl{k}_\shpa)|\right]^2 }{\int_{FS_{sp_xp_y , n}} d\sigma_{\bsl{k}}\  \frac{|\nabla_{\bsl{k}} E_{sp_xp_y,n}(\bsl{k})|}{|\bsl{d}(\bsl{k}_\shpa)|^2} }+ O( k_{F,\shpa} )\ ,
}
where we have used $\partial_{k_i} \left[ d_x(\bsl{k}_\shpa) \sigma_x + d_y(\bsl{k}_\shpa) \sigma_z  \right] = (\delta_{ix} \sigma_x +\delta_{iy}\sigma_z) v a $ from \eqnref{eq:d_form_MgB2_eff} and 
\eq{
\Tr\left[ \left(\partial_{k_i} \left[ d_x(\bsl{k}_\shpa) \sigma_x + d_y(\bsl{k}_\shpa) \sigma_z  \right] \right)^2  \frac{(k_x^2 - k_y^2) \sigma_z + 2 k_x  k_y  \sigma_x }{2 |\bsl{k}_\shpa|^2}   \right] = 2 (v a)^2 (\delta_{ix} +\delta_{iy}) \Tr\left[  \frac{(k_x^2 - k_y^2) \sigma_z + 2 k_x  k_y  \sigma_x }{2 |\bsl{k}_\shpa|^2}   \right] = 0
}
for first equality, we use the H\"odal inequallity in \eqnref{eq:holder} for the last inequality, and $\gamma_{\sigma,\shpa}^2$ is defined in \eqnref{eq:gamma_spxpy_geo}.
The $sp_xp_y$ Fermi surface $FS_{sp_xp_y , n}$ can be expressed as $FS_{sp_xp_y , n} = \cup_{k_z c \in (-\pi,\pi]} FS_{sp_xp_y , n, 2D, k_z}$ with $FS_{sp_xp_y , n, 2D, k_z}$ is the intersection between $FS_{sp_xp_y , n}$ and the fixed $k_z$ plane.
Then,
\eqa{
& \int_{FS_{sp_xp_y , n}} d\sigma_{\bsl{k}} |\partial_{\bsl{k}_\shpa}\theta(\bsl{k}_\shpa)| = \int_{-\pi/c}^{\pi/c} d k_z \int_{FS_{sp_xp_y , n, 2D, k_z}} d\sigma_{\bsl{k}_\shpa} |\partial_{\bsl{k}_\shpa}\theta(\bsl{k}_\shpa)| = \int_{-\pi/c}^{\pi/c} d k_z \int_{FS_{sp_xp_y , n, 2D, k_z=0}} d\sigma_{\bsl{k}_\shpa} |\partial_{\bsl{k}_\shpa}\theta(\bsl{k}_\shpa)| \\
& \geq \int_{-\pi/c}^{\pi/c} d k_z \left|\int_{FS_{sp_xp_y , n, 2D, k_z=0}} d \bsl{k}_\shpa \cdot  \partial_{\bsl{k}_\shpa}\theta(\bsl{k}_\shpa)\right| = \int_{-\pi/c}^{\pi/c} d k_z 2\pi \Delta\N = \frac{(2\pi)^2}{c} \Delta\N\ ,
}
where we have used the fact that $FS_{sp_xp_y , n, 2D, k_z=0}$ is a closed loop of $\bsl{k}_\shpa$ that encloses $\Gamma$ once as shown by the Fermi surface near $\Gamma$-A in \eqnref{fig:MgB2_el}(d), and $\Delta \N$ is the Euler number difference defined in \eqnref{eq:W_d_DeltaN}.
As a result, we have the following lower bound of $\left\langle \Gamma \right\rangle^{sp_xp_y,geo-geo}$:
\eqa{
\label{eq:Gamma_topo_ave}
& \left\langle \Gamma \right\rangle^{sp_xp_y,geo-geo}   \geq   \frac{\gamma_{\sigma,\shpa}^2 }{2 D_{\sigma}(\mu)} \frac{\hbar}{  m_{\text{B}}}   \frac{2\pi  \V}{c^2}  \sum_{n \in \{ 2,3\}} \frac{ 1 }{\int_{FS_{sp_xp_y , n}} d\sigma_{\bsl{k}}\  |\nabla_{\bsl{k}} E_{sp_xp_y,n}(\bsl{k})|/| \bsl{d}(\bsl{k}_\shpa)|^2 } \Delta\N^2+ O( k_{F,\shpa} )\ .
}
By defining
\eq{
\label{eq:lambda_topo_MgB2}
\lambda_{\sigma,topo} = \frac{D_{\sigma}(\mu)}{D(\mu)} \frac{  \gamma_{\sigma,\shpa}^2 }{  m_{\text{B}} \mcomega}   \frac{2\pi\Omega}{ c^2}   \sum_{n \in \{ 2,3\}} \frac{ 1 }{\int_{FS_{sp_xp_y , n}} d\sigma_{\bsl{k}}\  |\nabla_{\bsl{k}} E_{sp_xp_y,n}(\bsl{k})|/| \bsl{d}(\bsl{k}_\shpa)|^2 } \Delta\N^2 \ .
}
This leads to
\eq{
\lambda_{\sigma ,geo}  \geq \lambda_{\sigma,topo}  + O( k_{F,\shpa} )
}
according to \eqnref{eq:Gamma_topo_ave} and \eqnref{eq:lambda_sp2_E_geo}.
\eqnref{eq:lambda_topo_MgB2} matches \eqnref{eq:lambda_topo_eff_MgB2} derived from the GA if we take $E_{sp_xp_y,2}(\bsl{k}) = E_{sp_xp_y,3}(\bsl{k}) = E_{eff,1}(\bsl{k})$.

\subsection{Numerical Calculations of Contributions to $\lambda$}
\label{app:numerics_MgB2}

In this part, we perform numerical calculations with our model.
To do so, we need to determine the values of $\hat{\gamma}$ parameters in the EPC \eqnref{eq:g_i_MgB2}.
However, according to the expression of the $\gamma$ parameters in \eqnref{eq:gamma_pz_MgB2}, \eqnref{eq:gamma_z_sp2_MgB2} and \eqnref{eq:gamma_spxpy_geo}, we do not need to know the values of all $\hat{\gamma}$.
Explicitly, we only need to know the following $\hat{\gamma}$ parameters:
\eq{
\label{eq:gamma_hat_needed_MgB2}
\hat{\gamma}_5\ ,\ \hat{\gamma}_{10}\ ,\ \hat{\gamma}_9\ ,\ \hat{\gamma}_3-\hat{\gamma}_7\ ,
}
where $\hat{\gamma}_5$ and $\hat{\gamma}_{10}$ are for the $\lambda_{\pi}$, and $\hat{\gamma}_9$ and $\hat{\gamma}_3-\hat{\gamma}_7$ are for the $\lambda_{\sigma}$.

Compared to the calculation for graphene in \appref{app:graphene}, the {\abi} calculation for {\mgb} is more complicated, and thus we cannot directly give the form of the gauge-dependent EPC $F_{\bsl{\tau}i}(\bsl{k}_1,\bsl{k}_2)$ for {\mgb} due to the random gauges in the numerical calculations.
Instead, we will use the gauge-invariant $\Gamma_{nm}^{p_z}(\bsl{k}_1,\bsl{k}_2)$ and $\Gamma_{nm}^{sp_xp_y}(\bsl{k}_1,\bsl{k}_2)$ in \eqnref{eq:Gamma_X_MgB2}.
Specifically, for the $p_z$ part, we sum over all $p_z$ bands for $\Gamma_{nm}^{p_z}(\bsl{k}_1,\bsl{k}_2)$, and obtain
\eqa{
\label{eq:Gamma_pz_for_abi_MgB2}
& \sum_{n,m}\Gamma_{n m  }^{p_z}(\K , \K+(0,0,k_z) ) = \frac{3  \hbar a^2  \hat{\gamma}_5^2}{m_{\text{B}}} + \frac{4 \hbar  \hat{\gamma}_{10}^2 c^2}{m_{\text{B}}} (\sin(k_z c))^2\ ,
}
where we used the form of $\Gamma_{nm}^{p_z}(\bsl{k}_1,\bsl{k}_2)$ in \eqnref{eq:Gamma_X_MgB2}.
For the $sp_xp_y$ part, we only consider $\bsl{k}_1=\Gamma$ and $\bsl{k}_2$ along $\Gamma$-A, and $n,m$ of $\Gamma_{nm}^{sp_xp_y}(\Gamma,\Gamma+(0,0,k_z))$ are summed over the two degenerate bands (\ie, $n,m\in\{ 2, 3\}$ as shown in \figref{fig:MgB2_el_mz_even}(a)) along $\Gamma$-A near the Fermi level, resulting in 
\eqa{
\label{eq:Gamma_spxpy_for_abi_MgB2}
& \sum_{n,m\in\{ 2,3\}}\Gamma_{n m  }^{sp_xp_y}(\Gamma,\Gamma+(0,0,k_z)) = \frac{3  \hbar a^2}{m_{\text{B}}}  (\hat{\gamma}_3 - \hat{\gamma}_7)^2 + 2 \hbar \hat{\gamma}_9^2 \frac{c^2}{m_{\text{B}}} (\sin(k_z c))^2\ .
}

\begin{figure}
    \centering
    \includegraphics[width=0.8\columnwidth]{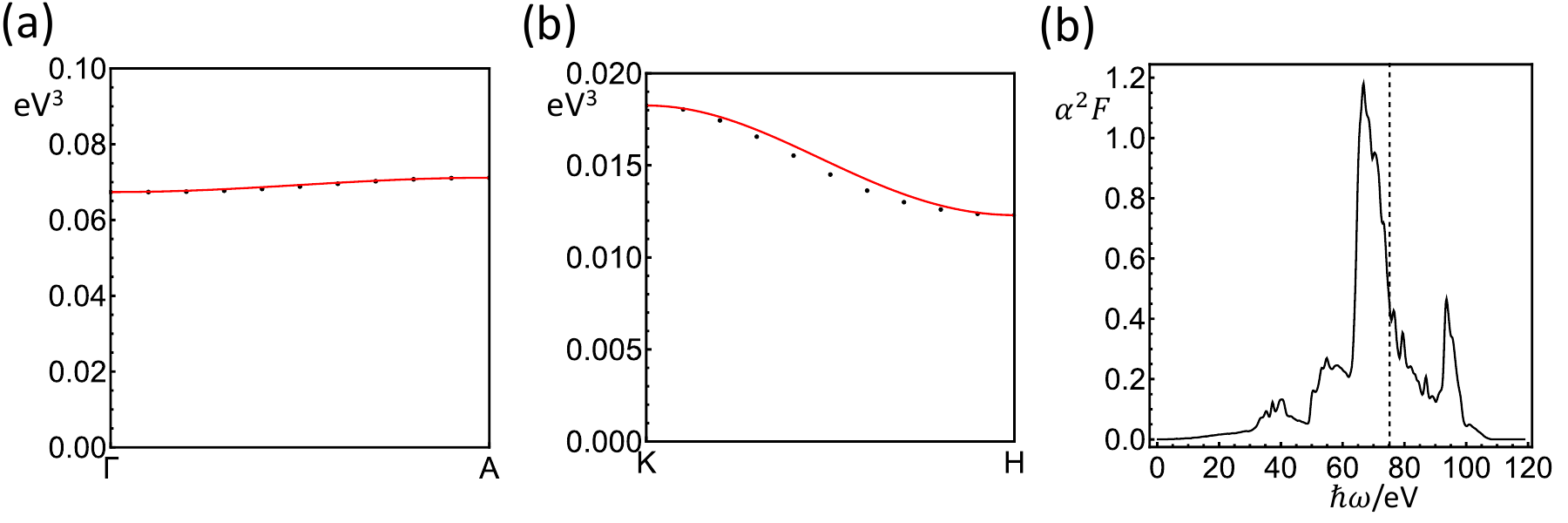}
    \caption{
    (a) The black dots are the sum of $\hbar\Gamma_{nm}^\abi(\Gamma,\Gamma+(0,0,k_z))$ over the two degenerate bands along $\Gamma$-A near the Fermi level, which is obtained from the {\abi} calculation.
    The red line is $\sum_{n,m\in\{ 2,3\}}\hbar\Gamma_{n m  }^{sp_xp_y}(\Gamma,\Gamma+(0,0,k_z))$ in \eqnref{eq:Gamma_spxpy_for_abi_MgB2_more_gen} with parameter values in \eqnref{eq:beta_values_MgB2}.
    The horizontal axis is $\Gamma+(0,0,k_z)$, which is the phonon momentum.
     (b) The black dots are the sum of $\Gamma_{nm}^\abi(\K , \K+(0,0,k_z) )$ over the two degenerate bands for the nodal line along $\K$-H.
    The red line is $\sum_{n,m}\hbar\Gamma_{n m  }^{p_z}(\K , \K+(0,0,k_z) )$ in \eqnref{eq:Gamma_pz_for_abi_MgB2_more_gen} with parameter values in \eqnref{eq:beta_values_MgB2}.
    The horizontal axis is $\K+(0,0,k_z)$, which is $\K$ plus the phonon momentum.
    (c) The Eliashberg function $\alpha^2 F(\omega)$ in \eqnref{eq:McMillan_omega_square_ave} given by the {\abi} calculation.
    The dashed line is at frequency of the $E_{2g}$ phonon at $\Gamma$, \ie, $\hbar \omega_{E_{2g}}(\Gamma) = 75.2 $meV.
    }
    \label{fig:EPC_abi_MgB2}
\end{figure}

In the {\abi} calculations (procedure described in \appref{app:abi}), we can directly evaluate $\Gamma_{nm}(\bsl{k}_1,\bsl{k}_2)$, which we label as $\Gamma_{nm}^{\abi}(\bsl{k}_1,\bsl{k}_2)$.
Then, $\sum_{n,m}\Gamma_{n m  }^{p_z}(\K , \K+(0,0,k_z) )$ in \eqnref{eq:Gamma_pz_for_abi_MgB2} corresponds to the sum of $\Gamma_{nm}^\abi(\K , \K+(0,0,k_z) )$ over the two degenerate bands for the nodal line along $\K$-H, since they mainly originate from $p_z$ orbitals.
$\sum_{n,m\in\{ 2,3\}}\Gamma_{n m  }^{s p_x p_y}(\Gamma,\Gamma+(0,0,k_z))$ in \eqnref{eq:Gamma_spxpy_for_abi_MgB2} corresponds to the sum of $\Gamma_{nm}^\abi(\Gamma,\Gamma+(0,0,k_z))$ over the two degenerate bands along $\Gamma$-A near the Fermi level, since they mainly originate from $p_xp_y$ orbitals.
We plot the {\abi} data in \figref{fig:EPC_abi_MgB2} as black dots.
To test our approximation against the {\abi} data, we derive more general forms of $\sum_{n,m}\Gamma_{n m  }^{p_z}(\K , \K+(0,0,k_z) )$ in \eqnref{eq:Gamma_pz_for_abi_MgB2} and $\sum_{n,m\in\{ 2,3\}}\Gamma_{n m  }^{sp_xp_y}(\Gamma,\Gamma+(0,0,k_z))$ in \eqnref{eq:Gamma_spxpy_for_abi_MgB2}.
Specifically, we allow longer-range hopping terms among B atoms, and then the form of the EPC of interest becomes
\eqa{
& F_{p_z, \bsl{\tau}, i=x,y}(\K+(0,0,k_z), \K+(0,0,k_z') ) = \left[ \beta_1 + \beta_{z,1} (\cos(k_z c) + \cos(k_z' c))\right] (-)^{\bsl{\tau}} (\tau_y , -\tau_x)_i \\
& F_{p_z, \bsl{\tau}, z}(K+(0,0,k_z), K+(0,0,k_z') ) =  \ii \beta_{z,2} (\sin(k_z) - \sin(k_z')) (\tau_0 + (-)^{\bsl{\tau}} \tau_z)\ ,
}
and
\eqa{
\label{eq:f_more_gen_pxpy_for_abi}
& U_{p_xp_y,+}^\dagger(0) F_{sp_xp_y, \bsl{\tau}, i=x,y}(\Gamma+(0,0,k_z), \Gamma+(0,0,k_z') ) U_{p_xp_y,+}(0) = \left[ \beta_3 + \beta_{z,3} (\cos(k_z c) + cos(k_z' c))\right] (-)^{\bsl{\tau}} (\tau_x , \tau_z)_i \\
& U_{p_xp_y,+}^\dagger(0) F_{sp_xp_y, \bsl{\tau}, z}(\Gamma+(0,0,k_z), \Gamma+(0,0,k_z') ) U_{p_xp_y,+}(0) =   \ii \beta_{z,4} (\sin(k_z c) - \sin(k_z' c))  \tau_0  \ ,
}
where $U_{p_xp_y,+}(0)$ is the basis for the parity-even combination of $p_xp_y$ orbitals at $\Gamma$ as shown in \eqnref{eq:U_pxpy_+(0)}.
As a result, the more general forms of $\sum_{n,m}\Gamma_{n m  }^{p_z}(\K , \K+(0,0,k_z) )$  and $\sum_{n,m\in\{ 2,3\}}\Gamma_{n m  }^{sp_xp_y}(\Gamma,\Gamma+(0,0,k_z))$ read
\eqa{
\label{eq:Gamma_pz_for_abi_MgB2_more_gen}
& \sum_{n,m}\Gamma_{n m  }^{p_z}(\K , \K+(0,0,k_z) )  \\
& = \frac{\hbar}{2} \sum_{\bsl{\tau}\in\{\text{B1}, \text{B2}\},i}  \frac{1}{m_{\bsl{\tau}}} \Tr\left[F_{p_z, \bsl{\tau}, i}(\K, \K+(0,0,k_z) )  F_{p_z, \bsl{\tau}, i}^\dagger(\K , \K+(0,0,k_z) ) \right]\\ 
& = \frac{4  \hbar }{m_{\text{B}}} \left[ \beta_1 + \beta_{z,1} (1+\cos(k_z c)\right]^2 + \frac{  \hbar  \beta_{z,2}^2 }{m_{\text{B}}} ( \sin(k_z c))^2
}
and
\eqa{
\label{eq:Gamma_spxpy_for_abi_MgB2_more_gen}
 & \sum_{n,m\in\{2,3\} }\Gamma_{n m }^{sp_xp_y}( \Gamma,  \Gamma+ (0,0,k_z) ) \\
 & = \frac{\hbar}{2} \sum_{\bsl{\tau}\in\{\text{B1}, \text{B2}\},i}  \frac{1}{m_{\bsl{\tau}}} \Tr\left[ U_{p_xp_y,+}(0) U_{p_xp_y,+}^\dagger(0) F_{sp_xp_y , \bsl{\tau}, i}(\Gamma, (0,0,k_z) ) \right. \\
 & \qquad \left. U_{p_xp_y,+}(0) U_{p_x p_y,+}^\dagger(0) F_{sp_xp_y , \bsl{\tau}, i}^\dagger( \Gamma, (0,0,k_z) ) \right]\\ 
& = \frac{ 4  \hbar }{m_{\text{B}}}  \left[ \beta_3 + \beta_{z,3} (1 + \cos(k_z  c))\right]^2 +    \frac{2 \hbar  \beta_{z,4}^2}{m_{\text{B}}} (\sin(k_z c))^2\ .
}
Compared to \eqnref{eq:Gamma_pz_for_abi_MgB2} and \eqnref{eq:Gamma_spxpy_for_abi_MgB2}, we know
\eqa{
& \hat{\gamma}_5^2 = \frac{4  }{3  a^2}  \beta_1^2  \ ,\  \hat{\gamma}_{10}^2  = \frac{    \beta_{z,2}^2 }{ 4 c^2}\ ,\ (\hat{\gamma}_3 - \hat{\gamma}_7)^2 =  \frac{4  \beta_3^2 }{3    a^2} \ ,\    \hat{\gamma}_9^2  = \frac{\beta_{z,4}^2}{ c^2 }      \ ,
}
while $\beta_{z,1}$ in \eqnref{eq:Gamma_pz_for_abi_MgB2_more_gen} and $\beta_{z,3}$ in \eqnref{eq:Gamma_spxpy_for_abi_MgB2_more_gen} are beyond our approximation.
By fitting \eqnref{eq:Gamma_pz_for_abi_MgB2_more_gen} and \eqnref{eq:Gamma_spxpy_for_abi_MgB2_more_gen} to the {\abi} data {\eqnref{fig:EPC_abi_MgB2}(a-b)), we obtain
\eqa{
\label{eq:beta_values_MgB2}
 & |\beta_1| = 0.055 \frac{\sqrt{ \eV^{3} m_{\text{B}} } }{\hbar} \ ,\ |\beta_{1,z}| = 0.006 \frac{\sqrt{ \eV^{3} m_{\text{B}} } }{\hbar} \ ,\ |\beta_{2,z}| = 0 \frac{\sqrt{ \eV^{3} m_{\text{B}} } }{\hbar} \\
 & |\beta_3| = 0.13 \frac{\sqrt{ \eV^{3} m_{\text{B}} } }{\hbar} \ ,\ |\beta_{3,z}| = 0.002 \frac{\sqrt{ \eV^{3} m_{\text{B}} } }{\hbar} \ ,\ |\beta_{4,z}| = 0 \frac{\sqrt{ \eV^{3} m_{\text{B}} } }{\hbar} \ .
}
Clearly, we can see that $\beta_{1,z}$ and $\beta_{3,z}$ that are beyond our approximation are very small, and can be neglected.
As a result, we obtain the values of $\hat{\gamma}$ of interests, which read
\eqa{
\label{eq:gamma_hat_values_MgB2}
& \hat{\gamma}_5^2 = 0.004 \frac{\eV^{3} m_{\text{B}} }{ a^2\hbar^2}    \ ,\  \hat{\gamma}_{10}^2  = 0\ ,\ (\hat{\gamma}_3 - \hat{\gamma}_7)^2 =  0.024 \frac{\eV^{3} m_{\text{B}} }{ a^2\hbar^2} \ ,\    \hat{\gamma}_9^2  =  0      \ .
}
Furthermore, the {\abi} value of $\mcomega$ reads
\eq{
\label{eq:mcomega_value_MgB2}
\hbar \sqrt{\mcomega} = 68 \text{meV}\ .
}
Combined with
\eq{
\frac{D_{\pi}(\mu)}{D(\mu)} = 0.577\ ,\ \frac{D_{\sigma}(\mu)}{D(\mu)} = 0.423
}
given by the numerical calculation in \refcite{Kong02272001MgB2EPC}, we can evaluate the values of $\lambda_{\pi,E}$  (\eqnref{eq:lambda_E_pz}), $\lambda_{\pi,geo}$  (\eqnref{eq:lambda_geo_pz}), $\lambda_{\pi,topo}$  (\eqnref{eq:lambda_topo_pz}), $\lambda_{\sigma,E}$  (\eqnref{eq:lambda_E_eff_MgB2}), $\lambda_{\sigma,geo}$  (\eqnref{eq:lambda_geo_eff_MgB2}), and $\lambda_{\sigma,topo}$  (\eqnref{eq:lambda_topo_eff_MgB2}), which are listed in \tabref{tab:lambda_MgB2}.

From \tabref{tab:lambda_MgB2}, $\lambda = 0.78$ from our model is close to the {\abi} value $\lambda^{\abi} = 0.67$ verifies the validity of our approximations.
Moreover, $\lambda_{\sigma}$ is much larger than $\lambda_{\pi}$, which is consistent with the understanding that the $\sigma$ bonding under $E_{2g}$ phonon modes accounts for the main contribution to the EPC constant $\lambda$ in the literature~\cite{Kong02272001MgB2EPC}.
In particular, we find that the geometric contribution is 91.7\% of the total $\lambda$, and most geometric contribution comes from the $\sigma$ bond.
On the other hand, the energetic contribution from the $\sigma$ bond ($\lambda_{\sigma,E}$) is negligible.
Therefore, it is the geometric property of the Bloch states that supports the large EPC constant from the $\sigma$ bond.
The geometric contribution is further bounded from below by the topological contribution in the bands, indicated by $\lambda_{topo}\approx 0.44 \lambda_{geo}$.

\begin{table}[t]
    \centering
    \begin{tabular}{|c|c|c|c|}
    \hline
     $\lambda$ ($\lambda^{\abi}$)   & $\lambda_{E}$ & $\lambda_{geo}$ & $\lambda_{topo}$ \\
    \hline
     0.78 (0.67) & 0.07 & 0.71 & 0.32\\
    \hline
    $\lambda_{\pi}$  & $\lambda_{\pi,E}$ & $\lambda_{\pi,geo}$ & $\lambda_{\pi,topo}$ \\
    \hline
     0.16 & 0.07 & 0.09 & 0.01   \\ 
    \hline
      $\lambda_{\sigma}$ & $\lambda_{\sigma,E}$ & $\lambda_{\sigma,geo}$ & $\lambda_{\sigma,topo}$ \\
    \hline
          0.62       & 0.00 & 0.62 & 0.31 \\
    \hline
    \end{tabular}
    \caption{The numerical values of the $\lambda$ and its various contributions for {\mgb}.
   $\lambda^{\abi} = 0.67$ in the bracket is the {\abi} value  for $\lambda$.
    }
    \label{tab:lambda_MgB2}
\end{table}

\subsubsection{Numerical Evidence for the Irrelevance of On-Site EPC Terms for the $\sigma$-bond-stretching ion Motions}

According to \refcite{Pickett2011CovalentBondsDriven}, we know the EPC mainly comes from the bond-stretching ion motions---the ion motions that stretch the bond between neighboring B atoms, \eg, the $E_2$ modes along $\Gamma$-A.
In principle, there are two possible origins of the EPC from the bond-stretching ion motions: (i) the change the hopping between two neighboring ions from the ion motions and (ii) the change of the on-site potential or coupling between the orbitals from the ion motions.
In this work, we use the two-center approximation to describe the EPC, meaning that we only consider the change of the hopping induced by the bond-stretching ion motions, while neglecting the on-site terms.
We, in this part, provide numerical evidence for our approximation.

For the bond-stretching motions, the hopping change term (or the term within two-center approximation) reads
\eqa{
\label{eq:two-center_EPC}
H_{EPC, two-center} & = \sum_{\bsl{R},i} \hat{T}_{\bsl{R}, i} (u_{\bsl{R}+\bsl{\tau}_{B1},i} - u_{\bsl{R}+\bsl{\tau}_{B2},i}) + \sum_{\bsl{R},i,j} C_3 \hat{T}_{C_3^{-1}\bsl{R}, i} C_3^{-1}  (u_{\bsl{R}+\bsl{a}_2+\bsl{\tau}_{B1},j} - u_{\bsl{R}-\bsl{a}_1+\bsl{a}_2+\bsl{\tau}_{B2},j}) \left( C_3 \right)_{ji}\\
& \quad + \sum_{\bsl{R},i,j} C_3^2 \hat{T}_{C_3^{-2}\bsl{R}, i} C_3^{-2} (u_{\bsl{R}-\bsl{a}_1+\bsl{a}_2+\bsl{\tau}_{B1},j} - u_{\bsl{R}-\bsl{a}_1+\bsl{\tau}_{B2},j}) \left( C_3^2 \right)_{ji} + h.c.\ ,
    }
where
\eqa{
 & \hat{T}_{\bsl{R}, i} = \mat{ c^\dagger_{\bsl{R}+\bsl{\tau}_{B1},p_x} & c^\dagger_{\bsl{R}+\bsl{\tau}_{B1},p_y}
} ( \widetilde{\beta}_3 \sigma_x, \widetilde{\beta}_1 \sigma_0 + \widetilde{\beta}_2 \sigma_z)_i \mat{ c_{\bsl{R}+\bsl{\tau}_{B2},p_x} \\ c_{\bsl{R}+\bsl{\tau}_{B2},p_y}}\\
 & C_3 \hat{T}_{C_3^{-1}\bsl{R}, i} C_3^{-1} = \mat{ c^\dagger_{\bsl{R}+\bsl{a}_2+\bsl{\tau}_{B1},p_x} & c^\dagger_{\bsl{R}+\bsl{a}_2+\bsl{\tau}_{B1},p_y}
} (  \widetilde{\beta}_3  e^{-\ii \sigma_y \frac{2\pi}{3}} \sigma_x e^{\ii \sigma_y \frac{2\pi}{3}} , \widetilde{\beta}_1 \sigma_0 + \widetilde{\beta}_2 e^{-\ii \sigma_y \frac{2\pi}{3}} \sigma_z  e^{\ii \sigma_y \frac{2\pi}{3}} )_i \mat{ c_{\bsl{R}+\bsl{a}_2-\bsl{a}_1+\bsl{\tau}_{B2},p_x} \\ c_{\bsl{R}+\bsl{a}_2-\bsl{a}_1+\bsl{\tau}_{B2},p_y}} \\
 & C_3^2 \hat{T}_{C_3^{-2}\bsl{R}, i} C_3^{-2} = \mat{ c^\dagger_{\bsl{R}-\bsl{a}_1+\bsl{a}_2+\bsl{\tau}_{B1},p_x} & c^\dagger_{\bsl{R}-\bsl{a}_1+\bsl{a}_2+\bsl{\tau}_{B1},p_y}
} ( \widetilde{\beta}_3  e^{-\ii \sigma_y \frac{4\pi}{3}} \sigma_x  e^{\ii \sigma_y \frac{4\pi}{3}}, \widetilde{\beta}_1 \sigma_0 + \widetilde{\beta}_2 e^{-\ii \sigma_y \frac{4\pi}{3}} \sigma_z  e^{\ii \sigma_y \frac{4\pi}{3}} )_i \mat{ c_{\bsl{R}-\bsl{a}_1+\bsl{\tau}_{B2},p_x} \\ c_{\bsl{R}-\bsl{a}_1+\bsl{\tau}_{B2},p_y}}\ ,
}
and $\bsl{\tau}_{B1}$ and $\bsl{\tau}_{B2}$ are defined in \eqnref{eq:sublattice_MgB2}.
The onsite term (which is beyond the two-center approximation) reads
\eqa{
H_{EPC, on-site} & = \sum_{\bsl{R},i} \hat{S}_{\bsl{R}, i} (u_{\bsl{R}+\bsl{\tau}_{B1},i} - u_{\bsl{R}+\bsl{\tau}_{B2},i}) + \sum_{\bsl{R},i,j} C_3 \hat{S}_{C_3^{-1}\bsl{R}, i} C_3^{-1}  (u_{\bsl{R}+\bsl{a}_2+\bsl{\tau}_{B1},j} - u_{\bsl{R}-\bsl{a}_1+\bsl{a}_2+\bsl{\tau}_{B2},j}) \left( C_3 \right)_{ji}\\
& \quad + \sum_{\bsl{R},i,j} C_3^2 \hat{S}_{C_3^{-2}\bsl{R}, i} C_3^{-2} (u_{\bsl{R}-\bsl{a}_1+\bsl{a}_2+\bsl{\tau}_{B1},j} - u_{\bsl{R}-\bsl{a}_1+\bsl{\tau}_{B2},j}) \left( C_3^2 \right)_{ji}\ ,
    }
where
\eqa{
 & \hat{S}_{\bsl{R}, i} = \sum_{\bsl{\tau}}\mat{ c^\dagger_{\bsl{R}+\bsl{\tau},p_x} & c^\dagger_{\bsl{R}+\bsl{\tau},p_y}
} (  \widetilde{\beta}'_3 \sigma_x , \widetilde{\beta}'_1 \sigma_0 + \widetilde{\beta}'_2 \sigma_z)_i \mat{ c_{\bsl{R}+\bsl{\tau},p_x} \\ c_{\bsl{R}+\bsl{\tau},p_y}}
}
and $C_3 \hat{S}_{C_3^{-1}\bsl{R}, i} C_3^{-1}$ and $C_3^2 \hat{S}_{C_3^{-2}\bsl{R}, i} C_3^{-2} $ can be obtained from symmetries.

To numerically compare the amplitudes of  $H_{EPC, two-center} $ and $H_{EPC, on-site}$, we adopt the method in \refcite{Pickett2011CovalentBondsDriven} and distort the lattice with the $\Gamma$-$E_{2g}$ bond-stretching motion (\ie, a frozen $E_{2g}$ phonon at $\Gamma$): 
\eqa{
& \bsl{u}_{\bsl{R}+\bsl{\tau}_{B2}} = -(\cos(\phi),\sin(\phi),0) \epsilon a \\
& \bsl{u}_{\bsl{R}+\bsl{\tau}_{B1}} = (\cos(\phi),\sin(\phi),0) \epsilon a\ ,
}
where $a$ is the in plane lattice constant and $\epsilon$ measures the distortion.
We note that the distortion does not break the lattice translations.
With this distortion (or the frozen phonon), the two-center EPC Hamiltonian becomes
\eqa{
H_{frozen, two-center} & = \sum_{\bsl{R},i} \hat{T}_{\bsl{R}, i} (\cos(\phi),\sin(\phi),0)_i 2 \epsilon a  + \sum_{\bsl{R},i,j} C_3 \hat{T}_{C_3^{-1}\bsl{R}, i} C_3^{-1} (\cos(\phi),\sin(\phi),0)_j 2 \epsilon a \left( C_3 \right)_{ji}\\
& \quad + \sum_{\bsl{R},i,j} C_3^2 \hat{T}_{C_3^{-2}\bsl{R}, i} C_3^{-2} (\cos(\phi),\sin(\phi),0)_j 2 \epsilon a \left( C_3^2 \right)_{ji} + h.c. \\
& = \sum_{\bsl{k}} \mat{ c^\dagger_{\bsl{k} , \bsl{\tau}_{B1}, p_x } & c^\dagger_{\bsl{k}, \bsl{\tau}_{B1}, p_y } & c^\dagger_{\bsl{k} , \bsl{\tau}_{B2}, p_x } &  c^\dagger_{\bsl{k}, \bsl{\tau}_{B2}, p_y } } h_{two-center}(\bsl{k}) \mat{ c^\dagger_{\bsl{k} , \bsl{\tau}_{B1}, p_x } \\ c^\dagger_{\bsl{k}, \bsl{\tau}_{B1}, p_y } \\ c^\dagger_{\bsl{k} , \bsl{\tau}_{B2}, p_x } \\  c^\dagger_{\bsl{k}, \bsl{\tau}_{B2}, p_y } }\ ,
}
where
\eqa{
h_{two-center}(\bsl{k}) = \mat{ 
0_{2\times 2} & 2 \epsilon a \sum_{ij}\sum_{n=0}^2  e^{-\ii \bsl{\delta}_n\cdot \bsl{k} } (\widetilde{\beta}_3  e^{-\ii \sigma_y \frac{2\pi n }{3}} \sigma_x e^{\ii \sigma_y \frac{2\pi n }{3}} , \widetilde{\beta}_1 \sigma_0 + \widetilde{\beta}_2 e^{-\ii \sigma_y \frac{2\pi n }{3}} \sigma_z  e^{\ii \sigma_y \frac{2\pi n}{3}} )_i \left( C_3^n \right)_{ji}  (\cos(\phi),\sin(\phi),0)_j \\
h.c. & 0_{2\times 2} \ .
}
}
On the other hand, the two-center EPC Hamiltonian becomes
\eqa{
H_{frozen, on-site} & = \sum_{\bsl{R},i} \hat{S}_{\bsl{R}, i} (\cos(\phi),\sin(\phi),0)_i 2 \epsilon a  + \sum_{\bsl{R},i,j} C_3 \hat{S}_{C_3^{-1}\bsl{R}, i} C_3^{-1}  (\cos(\phi),\sin(\phi),0)_j 2 \epsilon a \left( C_3 \right)_{ji}\\
& \quad + \sum_{\bsl{R},i,j} C_3^2 \hat{S}_{C_3^{-2}\bsl{R}, i} C_3^{-2} (\cos(\phi),\sin(\phi),0)_j 2 \epsilon a \left( C_3^2 \right)_{ji} \\
& = \sum_{\bsl{k}} \mat{ c^\dagger_{\bsl{k} , \bsl{\tau}_{B1}, p_x } & c^\dagger_{\bsl{k}, \bsl{\tau}_{B1}, p_y } & c^\dagger_{\bsl{k} , \bsl{\tau}_{B2}, p_x } &  c^\dagger_{\bsl{k}, \bsl{\tau}_{B2}, p_y } } h_{on-site}(\bsl{k}) \mat{ c^\dagger_{\bsl{k} , \bsl{\tau}_{B1}, p_x } \\ c^\dagger_{\bsl{k}, \bsl{\tau}_{B1}, p_y } \\ c^\dagger_{\bsl{k} , \bsl{\tau}_{B2}, p_x } \\  c^\dagger_{\bsl{k}, \bsl{\tau}_{B2}, p_y } }\ ,
}
where
\eqa{
h_{on-site}(\bsl{k}) = \tau_0\otimes \sum_{ij,n}(  \widetilde{\beta}'_3 e^{-\ii \sigma_y \frac{2n\pi}{3}} \sigma_x e^{\ii \sigma_y \frac{2n\pi}{3}}   , \widetilde{\beta}'_1 \sigma_0 + \widetilde{\beta}'_2 e^{-\ii \sigma_y \frac{2n\pi}{3}} \sigma_z e^{\ii \sigma_y \frac{2n\pi}{3}})_i  \left( C_3^n \right)_{ji}  (\cos(\phi),\sin(\phi),0)_j \ .
}
To compare the effect of $H_{frozen, two-center}$ and $H_{frozen, on-site}$, let us focus on the $\bsl{k}=\Gamma$ point: 
\eqa{
\label{eq:h_two_center_and_on-site}
& h_{two-center}(0)= \frac{3(\widetilde{\beta}_2+\widetilde{\beta}_3)}{2} 2\epsilon a\tau_x(\cos(\phi) \sigma_x + \sin(\phi) \sigma_z)\\
& h_{on-site}(0)= \frac{3(\widetilde{\beta}'_2+\widetilde{\beta}'_3)}{2} 2\epsilon a \tau_0(\cos(\phi) \sigma_x + \sin(\phi) \sigma_z) \ .
}
As we can see, only the $(\widetilde{\beta}_2+\widetilde{\beta}_3)$ matters for the two-center EPC at $\bsl{k}=\Gamma$, and only the $(\widetilde{\beta}_2'+\widetilde{\beta}_3')$ matters for the on-site EPC at $\bsl{k}=\Gamma$, under the $\Gamma$-$E_{2g}$ frozen phonon.

To specifically compare $(\widetilde{\beta}_2+\widetilde{\beta}_3)$ to $(\widetilde{\beta}_2'+\widetilde{\beta}_3')$, we choose
\eqa{
\label{eq:E2g_frozen}
& \bsl{u}_{\bsl{R}+\bsl{\tau}_{B2}} = (0,1,0) \epsilon a \\
& \bsl{u}_{\bsl{R}+\bsl{\tau}_{B1}} = (0,-1,0) \epsilon a\ ,
}
which means $\phi=-\pi/2$.
Without the distortion, the electron Hamiltonian for $p_x$ and $p_y$ orbitals at $\Gamma$ reads
\eq{
h_{p_x p_y}(0) =E_{B,p_x p_y,0} + 2 t_{B,p_x p_y,z}  +  3 t_2 \tau_x \sigma_0\ ,
}
according to \eqnref{eq:H_el_MgB2_B_sp2}.
Then, if only the two-center term matters, we have
\eq{
h_{p_x p_y}(0) + h_{two-center}(0) = E_{B,p_x p_y,0} + 2 t_{B,p_x p_y,z}  + \tau_x (3 t_2 \sigma_0 - 3(\widetilde{\beta}_2+\widetilde{\beta}_3) \epsilon a   \sigma_z)\ ,
}
which has eigenvalues and eigenvectors as
\eqa{
\label{eq:two_center_eigenvectors_eigenvalues}
& E_{B,p_x p_y,0} + 2 t_{B,p_x p_y,z}  \pm (3 t_2 -  3(\widetilde{\beta}_2+\widetilde{\beta}_3) \epsilon a): (1,0,\pm 1,0)/\sqrt{2}\\
& E_{B,p_x p_y,0} + 2 t_{B,p_x p_y,z}   \pm ( 3 t_2 +  3(\widetilde{\beta}_2+\widetilde{\beta}_3)  \epsilon a): (0,1,0,\pm 1)/\sqrt{2}\ ,
}
where the values before ``$:$" are eigenvalues and the vectors after ``$:$" are eigenvectors.
On the other hand, if only the on-site term matters, we have
\eq{
h_{p_x p_y}(0) +h_{on-site}(0) = E_{B,p_x p_y,0} + 2 t_{B,p_x p_y,z}  +  3 t_2  \tau_x \sigma_0 -  3 (\widetilde{\beta}'_2+\widetilde{\beta}'_3)\epsilon a \tau_0\sigma_z\ ,
}
where
\eqa{
\label{eq:on-site_eigenvectors_eigenvalues}
& E_{B,p_x p_y,0} + 2 t_{B,p_x p_y,z}  \pm 3 t_2 -  3(\widetilde{\beta}'_2+\widetilde{\beta}'_4)\epsilon a: (1,0,\pm 1,0)/\sqrt{2}\\
& E_{B,p_x p_y,0} + 2 t_{B,p_x p_y,z}  \pm 3 t_2 + 3(\widetilde{\beta}'_2+\widetilde{\beta}'_4)\epsilon a: (0,1,0,\pm 1)/\sqrt{2}\ .
}
 
From \eqnref{eq:two_center_eigenvectors_eigenvalues} and \eqnref{eq:on-site_eigenvectors_eigenvalues}, both the two-center and the on-site terms can open the gaps at the previously doubly degenerate energy level $E_{B,p_x p_y,0} + 2 t_{B,p_x p_y,z}  + 3 t_2$ and at the previously doubly degenerate energy level $E_{B,p_x p_y,0} + 2 t_{B,p_x p_y,z}  - 3 t_2$ .
Nevertheless, they open the gaps in different ways.
Note that $(1,0,1,0)/\sqrt{2}$ and $(1,0,-1,0)/\sqrt{2}$ are respectively the bonding and anti-bonding state of $p_x$ orbitals, and $(0,1,0,1)/\sqrt{2}$ and $(0,1,0,-1)/\sqrt{2}$ are respectively the anti-bonding state and bonding state of $p_y$ orbitals.
Then, \eqnref{eq:two_center_eigenvectors_eigenvalues} suggests that the bonding and anti-bonding states of the same orbital would have opposite energy shifts for the two-center term, while \eqnref{eq:on-site_eigenvectors_eigenvalues} suggests that the bonding and anti-bonding states of the same orbital would have same energy shifts for the on-site term.
According to \figref{fig:MgB2_bands_frozen_E2g_phonons}, we can see the the bonding and anti-bonding states of the same orbital indeed have opposite energy shifts, and barely have the equal shifts.
In addition, there are two signatures of in the band structure that come from these opposite energy shifts in \figref{fig:MgB2_bands_frozen_E2g_phonons} with the frozen $E_{2g}$ phonon in \eqnref{eq:E2g_frozen} for $\epsilon<0$: (i) the top two $p_x/p_y$ bands cross with each other along $\Gamma-M$, while the bottom two $p_x/p_y$ bands along $\Gamma-M$ have no crossing, and (ii) the bottom two $p_x/p_y$ bands cross with each other along $\Gamma-K$, while the top two $p_x/p_y$ bands along $\Gamma-K$ have no crossing
The $\Gamma-M$ signature is shown in \refcite{Fan03052002MgB2}, but the $\Gamma-K$ signature is missing in \refcite{Fan03052002MgB2}.
Moreover, the wavefunctions are not presented in \refcite{Fan03052002MgB2}.

\begin{figure}
    \centering
    \includegraphics[width=\columnwidth]{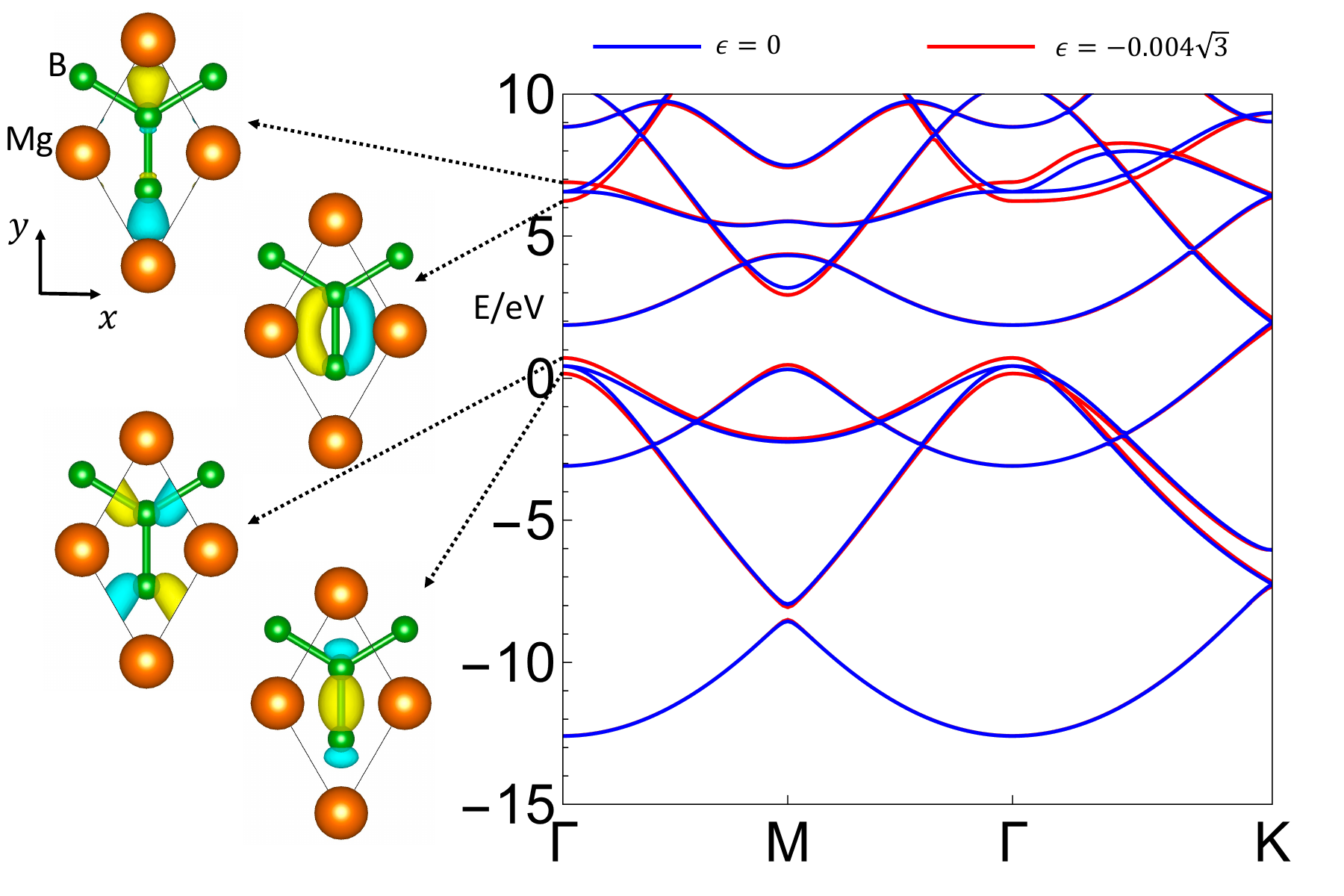}
    \caption{On the right, the plot shows the band structures for {\mgb} with no frozen phonons (blue) and with frozen $E_{2g}$ phonon in \eqnref{eq:E2g_frozen} with $\epsilon = -0.004 \sqrt{3}$.
    Both $\Gamma-M$ and $\Gamma-K$ paths have mirror symmetries, since the former is along $y$ and the latter is along $x$.
    On the left, the 4 plots are the 4 $p_x/p_y$ wave-functions at $\Gamma$ for {\mgb} with frozen $E_{2g}$ phonon in \eqnref{eq:E2g_frozen} with $\epsilon = -0.004 \sqrt{3}$.
    The signs of the blue and yellow parts of the wavefunction are opposite. 
    This figure is also shown in \refcite{yu2023reply}.
    }
    \label{fig:MgB2_bands_frozen_E2g_phonons}
\end{figure}

To more precisely compare the amplitude, we include the both distortions into the Hamiltonian at $\Gamma$ and get 
\eqa{
h_{p_x p_y}(0) + h_{two-center}(0) + h_{on-site}(0) & = E_{B,p_x p_y,0} + 2 t_{B,p_x p_y,z}  + \tau_x (3 t_2 \sigma_0 - 3(\widetilde{\beta}_2+\widetilde{\beta}_3) \epsilon a   \sigma_z)-  3 (\widetilde{\beta}'_2+\widetilde{\beta}'_3)\epsilon a \tau_0\sigma_z\ ,
}
which has eigenvalues and eigenvectors as
\eqa{
\label{eq:two_center+onsite_eigenvectors_eigenvalues}
& E_1 = E_{B,p_x p_y,0} + 2 t_{B,p_x p_y,z} + 3 t_2 -  3(\widetilde{\beta}_2+\widetilde{\beta}_3) \epsilon a - 3(\widetilde{\beta}'_2+\widetilde{\beta}'_3)\epsilon a: (1,0,1,0)/\sqrt{2}\\
& E_2 = E_{B,p_x p_y,0} + 2 t_{B,p_x p_y,z} -3t_2 + 3(\widetilde{\beta}_2+\widetilde{\beta}_3)\epsilon a -  3(\widetilde{\beta}'_2+\widetilde{\beta}'_3)\epsilon a: (1,0,-1,0)/\sqrt{2}\\
& E_3 =E_{B,p_x p_y,0} + 2 t_{B,p_x p_y,z} +3 t_2 +  3(\widetilde{\beta}_2+\widetilde{\beta}_3)\epsilon a +  3(\widetilde{\beta}'_2+\widetilde{\beta}'_3)\epsilon a: (0,1,0,1)/\sqrt{2}\\
& E_4 =E_{B,p_x p_y,0} + 2 t_{B,p_x p_y,z} -3 t_2 - 3(\widetilde{\beta}_2+\widetilde{\beta}_3)\epsilon a +  3(\widetilde{\beta}'_2+\widetilde{\beta}'_3)\epsilon a: (0,1,0,-1)/\sqrt{2}\ .
}
From DFT calcualtion for $\epsilon = -0.004 \sqrt{3}$, we get 
\eqa{
& E_1 = 6.231 \eV \\
& E_2 = 0.7202 \eV\\
& E_3 = 6.899 \eV \\
& E_4 = 0.1675\eV \ ,
}
which gives
\eqa{
& \widetilde{\beta}_2 +  \widetilde{\beta}_3 = -\frac{E_1 - E_2 + E_4 - E_3 }{12 \epsilon a } = -14.68 \eV/a \\
& \widetilde{\beta}'_2+\widetilde{\beta}'_3 = \frac{E_1 + E_2 - E_4 - E_3 }{12 \epsilon a } = -1.387 \eV/a\ .
}
Therefore, the effect of the on-site EPC term under the bond-stretching motion in \eqnref{eq:E2g_frozen} contribution is negligible compared to the two-center EPC term, serving as a numerical evidence for the irrelevance of the on-site EPC term.

By comparing \eqnref{eq:two-center_EPC} to \eqnref{eq:g_shpa_MgB2} and \eqnref{eq:f_perp_spxpy}, we find that $\hat{\gamma}_3 = -\widetilde{\beta}_2 \frac{\sqrt{3}}{a}$ and $\hat{\gamma}_7= \widetilde{\beta}_3 \frac{\sqrt{3}}{a}$, and thus
\eq{
\hat{\gamma}_3 -\hat{\gamma}_7 = -(\widetilde{\beta}_2 +\widetilde{\beta}_3 )\frac{\sqrt{3}}{a} = 25.43 \eV/a^2 = 0.1651 \frac{\eV^{3/2}}{a^2} \sqrt{\frac{m_{\text{B}} a^2  }{\hbar^2}}\ ,
}
which is close to the value of $|\hat{\gamma}_3 -\hat{\gamma}_7|= 0.1549 \frac{\eV^{3/2}}{a^2} \sqrt{\frac{m_{\text{B}} a^2  }{\hbar^2}}$ given by \eqnref{eq:gamma_hat_values_MgB2}, showing the consistency in the numerical results.

\subsection{$\sqrt{\mcomega}$ Approximated by $\omega_{E_{2g}}(\Gamma)$}

In \appref{app:numerics_MgB2}, we directly use the {\abi} value of $\mcomega$ for the calculation.
In this part, we will show that $\sqrt{\mcomega}$ can be well approximated by the frequency of the $E_{2g}$ phonons at $\Gamma$, labelled by $\omega_{E_{2g}}(\Gamma)$.

As shown in \appref{app:numerics_MgB2}, the main contribution to $\lambda$ comes from the $p_xp_y$ $\sigma$-bonding states near $\Gamma$-A.
Then, let us focus on the EPC for the $p_xp_y$ $\sigma$-bonding states exactly at $\Gamma$-A.
Moreover, as shown in \figref{fig:EPC_abi_MgB2}(a), the EPC barely changes along with the phonon momentum, and thus we will only look at EPC among the two $p_xp_y$ states at $\Gamma$ point near the Fermi level.
As shown in \eqnref{eq:f_more_gen_pxpy_for_abi}, the projection of the EPC $F_{sp_xp_y, \bsl{\tau}, i}(\Gamma, \Gamma )$ to the two $p_xp_y$ states at $\Gamma$ point near the Fermi level read
\eq{
\label{eq:f_pxpy_Gamma_MgB2}
U_{p_xp_y,+}^\dagger(0) F_{sp_xp_y, \bsl{\tau}, i }(\Gamma, \Gamma ) U_{p_xp_y,+}(0) =\hat{\beta}  (-)^{\bsl{\tau}} (\tau_x , \tau_z)_i (\delta_{ix}+ \delta_{iy}) (\delta_{\bsl{\tau}, \bsl{\tau}_{\text{B1}}}  + \delta_{\bsl{\tau}, \bsl{\tau}_{\text{B2}}})\ ,
}
where $U_{p_xp_y,+}(0)$ is in \eqnref{eq:U_pxpy_+(0)} and is formed by the two eigenvectors for the two parity-even $p_xp_y$ states at $\Gamma$ point near the Fermi level, and we include $ (\delta_{\bsl{\tau}, \bsl{\tau}_{\text{B1}}}  + \delta_{\bsl{\tau}, \bsl{\tau}_{\text{B2}}})$ because we have neglected the Mg atoms in the EPC Hamiltonian.
\eqnref{eq:f_pxpy_Gamma_MgB2} is in the atomic basis for the ion motions, which are labelled by $\bsl{\tau}, i $ according to \eqnref{eq:H_el-ph_k}.
Now we switch to the phonon eigenstates at $\Gamma$.
According to \eqnref{eq:ft_l}, in the phonon eigenbasis at $\Gamma$, we have 
\eqa{
\label{eq:ft_pxpy_Gamma_MgB2}
& U_{p_xp_y,+}^\dagger(0) \widetilde{F}_{sp_xp_y, l }(\Gamma, \Gamma ) U_{p_xp_y,+}(0) = \sum_{\bsl{\tau} i} U_{p_xp_y,+}^\dagger(0) F_{sp_xp_y, \bsl{\tau}, i }(\Gamma, \Gamma ) U_{p_xp_y,+}(0)  \frac{1}{\sqrt{m_{\bsl{\tau}}}} \left[ v_l^*(\Gamma) \right]_{\bsl{\tau} i} \\
& =   \frac{\hat{\beta}}{\sqrt{m_{B}}} \sum_{\bsl{\tau}\in \{\bsl{\tau}_{\text{B1}},\bsl{\tau}_{\text{B2}}\}} \sum_{i\in\{x,y\}}   (-)^{\bsl{\tau}} (\tau_x , \tau_z)_i   \left[ v_l^*(\Gamma) \right]_{\bsl{\tau}i}\ ,
}
where $v_l(\Gamma)$ is the eigenvector for $\Gamma$ phonons labelled by $l$.
\eqnref{eq:ft_pxpy_Gamma_MgB2} clearly shows that to have a nonzero $U_{p_xp_y,+}^\dagger(0) \widetilde{F}_{sp_xp_y, l }(\Gamma, \Gamma ) U_{p_xp_y,+}(0)$, $v_l(\Gamma)$ must have nonzero $x,y$ components that are opposite for $\text{B1}$ and $\text{B2}$ atoms, which are nothing but the $E_{2g}$ phonons at $\Gamma$.
Therefore, we should expect the Eliashberg function $\alpha^2 F(\omega)$ in the definition of $\mcomega$ (\eqnref{eq:McMillan_omega_square_ave}) to peak at the frequency of $E_{2g}$ $\Gamma$ phonons, and we should expect $\sqrt{\mcomega} \approx \omega_{E_{2g}}(\Gamma)$.

Indeed, the {\abi} calculation shows $\hbar\omega_{E_{2g}}(\Gamma)= 75.3$meV, which deviates form the {\abi} value of $\hbar \sqrt{\mcomega}$ in \eqnref{eq:mcomega_value_MgB2} by $11\%$, \ie
\eq{
\frac{ \sqrt{\mcomega} -  \omega_{E_{2g}}(\Gamma)}{  \sqrt{\mcomega}  } = 11\%\ .
}
Consistently, the Eliashberg function $\alpha^2 F(\omega)$ peaks near the frequency of $E_{2g}$ $\Gamma$ phonons as shown in \figref{fig:EPC_abi_MgB2}(c).
Therefore, with 11\% error, we can approximate $\sqrt{\mcomega}$ with $\omega_{E_{2g}}(\Gamma)$.

\section{Details on {\abi} Calculation}
\label{app:abi}

In this work, \emph{ab initio} calculations on {\mgb} were carried out with two methods.
The two methods are distinguished by whether the Wannier interpolation is used and by various other aspects like the exchange-correlation functional.
%
The calculation on graphene was done with only the Wannier-interpolation method.
In the following, we will first present the general discussion that is valid for both methods and then distinguish and compare the two methods.
We will show that the two methods give reasonably similar results.

\subsection{Electron model}

For both methods, the electronic states are modeled using a plane-wave basis set on a grid of $\bsl{k}$-points. 
Within the density functional theory (DFT) formalism, the many-body Schr\"{o}dinger equation can be approximated as a series of one-particle Kohn-Sham Hamiltonians, which relate to the single-particle Bloch state $\ket{\psi_{n \bsl{k}} } $ and energies $E_{n \bsl{k}}$ in a Schr\"{o}dinger-like manner:
\begin{equation}
    H^{\mathrm{KS}} \ket{\psi_{n \bsl{k}} }  = E_{n\bsl{k}} \ket{\psi_{n \bsl{k}} } .
    \label{eq:KS-Hamiltonian}
\end{equation}
The Kohn-Sham Hamiltonian has the general form:
\begin{equation}
    H^\mathrm{KS} = \frac{\hbar^2}{2m_e} \left( -i \nabla \right)^2 + v(\bsl{r}),
\end{equation}
where $m_e$ is the mass of an electron. The first term is the kinetic energy operator, while the second represents the potential, which generally has three contributions:
\begin{equation}
    v(\bsl{r}) = v^\mathrm{e-i} (\bsl{r}) + v^\mathrm{H} (\bsl{r}) + v^\mathrm{xc} (\bsl{r}).
\label{eq:DFT-potential} 
\end{equation}
On the right-hand side of \eqnref{eq:DFT-potential}, the first term denotes the interaction between the electron and the ions in the system. 
This interaction is commonly described using pseudopotentials~\cite{ONCV,USPP}. 

The potential $v(\bsl{r})$ in \eqnref{eq:DFT-potential} is the electrostatic Hartree potential, which accounts for the mean-field interaction felt by the electron in the presence of the other electrons in the system:
\begin{equation}
    v^{\mathrm{H}} (\bsl{r}) = e^2 \int d \bsl{r}^\prime \frac{\rho(\bsl{r}^\prime)} {|\bsl{r} - \bsl{r}^\prime | }. 
\end{equation}
Here $\rho(\bsl{r})$ is the electron density and $e$ is the electric charge of an electron.
The final term of \eqnref{eq:DFT-potential} is the exchange-correlation potential. In general it is challenging to formalize this quantum mechanical electron-electron interaction, and there have been significant research efforts devoted to finding expressions for this interaction. 
The specific approximations that we make for the exchange-correlation potential will be specified in \appref{app:method_I_wannier_interpolation} and \appref{app:method_II_NO_wannier_interpolation}.

Given the Kohn-Sham equation for a single electron in \eqnref{eq:KS-Hamiltonian}, and the ingredients for the electronic potential generally defined in \eqnref{eq:DFT-potential}, one can solve for the single particle states $\ket{\psi_{n \bsl{k}} } $ in a self-consistent manner.

\subsection{Phonon model} 

The motion of atoms in solids is often, to a good approximation, describable within the Harmonic approximation. In this approximation, phonons can emerge as normal modes of vibration for the system. 
Analogous to electrons, these phonons can be labelled by their mode index $l$ and wavevector $\bsl{q}$. Eigenvectors for phonons modes are given by diagonalizing  the dynamical matrix $\widetilde{D}(\bsl{q})$, which is defined in \eqnref{eq:dynemical_matrix} for the analytical study.
For DFT calculation, it is more conventional to define the dynamical matrix $\widetilde{D}(\bsl{q})$ as:
\begin{equation}
    \widetilde{D}(\bsl{q})_{\bsl{\tau}\bsl{\tau}',ii^\prime}(\bsl{q})  = \frac{1}{\sqrt{m_{\bsl{\tau}} m_{\bsl{\tau}}^\prime}}
     \sum_{\bsl{R}} 
     \Phi_{i,i^\prime}^{\bsl{\tau}, \bsl{\tau}^\prime} (\bsl{R}) e^{-i \bsl{q} \cdot \bsl{R}},
     \label{eq:dynamical-from-IFCs}
\end{equation}
where $\Phi$ is the interatomic force constant matrix that reads
\begin{equation}
    \Phi_{i,i^\prime}^{\bsl{\tau}, \bsl{\tau}^\prime} (\bsl{R})  = 
    \frac{\partial^2 E}{\partial u_{\bsl{\tau},i} \partial u_{\bsl{R}+\bsl{\tau}',i'}}
    = - \frac{\partial F_{i^\prime,\bsl{R}+\bsl{\tau}^\prime}}{\partial u_{\bsl{\tau},i}} \ ,
    \label{eq:interatomic-forces}
\end{equation}
and recall that $u_{\bsl{R}+\bsl{\tau},i}$ is the motion of the atom with equilibrium position $\bsl{R}+\bsl{\tau}$ along the direction $i$.
The variable $F_{i,\bsl{R}+\bsl{\tau}}$ represents the force felt by the atom with equilibrium position $\bsl{R}+\bsl{\tau}$ along the direction $i$.
In general the interatomic force constant matrix can depend on two unit cells, $\bsl{R}$ and $\bsl{R}^\prime$, but in practice only the difference $\bsl{R} - \bsl{R}^\prime$ matters for our purposes, and we therefore take one of them zero. 

In examining \eqnref{eq:interatomic-forces}, the main ingredients involve either the total energy or the forces on each atom. DFT has been widely used to theoretically study phonons in materials, in part because both of these quantities are readily available. There have been two popular approaches for describing phonons in materials, both of which we employ in our work. 

The conceptually simpler approach is known as the ``frozen-phonon'' method, which relies on finite displacements to evaluate the elements of the interatomic force constant matrix given by \eqnref{eq:interatomic-forces}. 
Specifically, we can write the force derivative as a discrete derivative:
\begin{equation}
    \frac{\partial F_{i^\prime,\bsl{R}+\bsl{\tau}^\prime}}{\partial u_{\bsl{\tau},i}}
    \approx  \frac{\Delta F_{i^\prime,\bsl{R}+\bsl{\tau}^\prime}}{\Delta u_{\bsl{\tau},i}}
\label{eq:frozen-phonon}
\end{equation}
In this approach, one can generate an interatomic force constant matrix $\Phi$ by perturbing each of the atoms of the unit cell in different directions and measuring the changes in the forces on the atoms of the system. Using \eqnref{eq:dynamical-from-IFCs}, one can also generate the dynamical matrix which can then be used to capture the phonon eigenvector and phonon frequencies for the system. Note that this approach requires a summation over unit cells $\bsl{R}$ in \eqnref{eq:dynamical-from-IFCs}. Thus it is often important to include many unit cells such that all relevant interatomic interactions are captured and also so that the dynamical matrix can be described at different $\bsl{q}$-points. 
The ``frozen-phonon'' method is used by method I with Wannier interpolation for graphene as discussed in \appref{app:method_I_wannier_interpolation}.

An alternative approach to describing the phonons of the system is known as density functional perturbation theory (DFPT). A comprehensive review of this method can be found in \refcite{baroni2001}.
The DFPT method is used by both methods for {\mgb} as discussed in \appref{app:method_I_wannier_interpolation} and \appref{app:method_II_NO_wannier_interpolation}.

\subsection{EPC} 

As outlined above, the phonon eigenvectors and frequencies are computed on a grid of $\bsl{q}$-points, along with the EPC matrix elements $g$, which read
\begin{align}
    G_{mnl}(\bsl{k}+\bsl{q},\bsl{k}) & = \sqrt{\frac{\hbar}{2 M_0 \omega_{l}(\bsl{q})}} g_{mnl}(\bsl{k},\bsl{q}) 
    \label{eq:g-scaled}
    \\ 
    g_{mnl}(\bsl{k},\bsl{q})
    & = \langle \psi_{m,\bsl{k} + \bsl{q}} | \partial_{l \bsl{q}} V  | 
    \psi_{n,\bsl{k}} \rangle ,
    \label{eq:g-KS}
\end{align}
where $G_{mnl}(\bsl{k},\bsl{k}')$ is in \eqnref{eq:G_nml}.
Here $g$ depends on the band indices $n$ and $m$ of the electronic states, the phonon mode index $l$, the wavevector of the initial electron state $\bsl{k}$ and the wavevector of the phonon emitted $\bsl{q}$. The variable $M_0$ is a reference mass.  The electronic states in our cases are obtained from Kohn-Sham DFT (\eqnref{eq:KS-Hamiltonian}). The operator $\partial_{l \bsl{q}} V (\bsl{r})$ is defined as the follows~\cite{Giustino:2007,Giustino2017}:
\begin{equation}
    \partial_{\bsl{q}l} V(\bsl{r})
    = \sum_{ \bsl{R} \bsl{\tau}} e^{\ii \bsl{q} \cdot \bsl{R} } \sum_i \left[ \widetilde{v}_{l}( \bsl{q} )\right]_{\bsl{\tau} i}  \partial_{u_{\bsl{R}+\bsl{\tau},i}} V(\bsl{r})\ ,
    \label{eq:bloch-to-wfs-perturbation}
\end{equation}
where $\left[ \widetilde{v}_{l}( \bsl{q} )\right]_{\bsl{\tau} i} = \sqrt{M_0/m_{\bsl{\tau}}} \left[ v_{l}( \bsl{q} )\right]_{\bsl{\tau} i} $. (Note the Fourier transformation is in a different convention than \eqnref{eq:FT_rule}.)
Essentially, $\partial_{l \bsl{q}} V (\bsl{r})$ describes how the self-consistent potential $V$ changes as a result of collective ionic motion arising from a phonon indexed by $l \bsl{q}$.
In the end, we need to compute the Fermi-surface averaged EPC constant $\lambda$ in \eqnref{eq:lambda}.

The EPC can be evaluated either with Wannier interpolation (method I in \appref{app:method_I_wannier_interpolation}) or directly (method II in \appref{app:method_II_NO_wannier_interpolation}).
In the following, we will describe the two methods.

\subsection{Method I: With Wannier Interpolation} 

\label{app:method_I_wannier_interpolation}

In this subsection, we describe the first method, which involves Wannier interpolation.
For the electron calculation, we use open-source DFT codes~\cite{JDFTx,qe1,qe2}.
For EPC, Wannier functions (WFs) are used to represent the \emph{ab initio} data in real space~\cite{Marzari1997,SouzaEntangled,Giustino:2007}. Below we outline the general idea of this approach. 
The real space electronic wavefunctions can be expressed using the reciprocal space wavefunctions from DFT as:
\begin{equation}
    |W_{\alpha \bsl{R}} \rangle = 
    \sum_{n\bsl{k}} e^{-i \bsl{k} \cdot \bsl{R}} U_{n\alpha,\bsl{k}}^* \ket{\psi_{n \bsl{k}} } .
    \label{eq:bloch-to-wfs}
\end{equation}
Here, $|W_{\alpha \bsl{R}} \rangle$ is the real-space Wannier function (WFs) labelled by $\alpha$ and the lattice vector $\bsl{R}$. 
The main ingredients in these expressions are the Bloch states $\ket{\psi_{n \bsl{k}} } $ and the unitary matrix $U_{\bsl{k}}$. The well-known limitation of this transformation is the fact that Bloch functions have a gauge freedom. This is not an issue in the reciprocal space Bloch representation, however this freedom causes significant changes to the nature of the real-space Wannier functions. To mitigate this issue, maximally-localized Wannier functions are used. The unitary matrix $U_{\bsl{k}}$ is defined such that the real-space spread of the Wannier functions is minimized and thus the WFs are maximally-localized~\cite{Marzari1997,SouzaEntangled,marzari2011}. Within this basis, the WFs are guaranteed to be exponentially localized~\cite{He2001}. 
This essentially gives us an \emph{ab initio} tight-binding description of our system.

With the electron Wannier function, the interpolation in \refcite{Giustino:2007,Giustino2017} works as follows.
First, we compute $g_{m n l}(\bsl{k}, \bsl{q})$ with DFT on a grid of $k$ points, labelled as $g_{m n l}^{\text{DFT}}(\bsl{k}, \bsl{q})$.
Then, we use the following expression to Fourier transform $g_{m n l}^{\text{DFT}}(\bsl{k}, \bsl{q})$ into real space:
\begin{equation}
 g_{\alpha\alpha', \bsl{\tau}i} (\bsl{R}, \bsl{R}') = 
 \frac{1}{N} 
 \sum_{n m l }
 \sum_{\bsl{k},\bsl{q}}^{DFT} 
 e^{-i (\bsl{k} \cdot \bsl{R}' + 
 \bsl{q} \cdot \bsl{R})}
 U_{m \alpha,\bsl{k} + \bsl{q}}
 g_{m n l}^{\text{DFT}}(\bsl{k}, \bsl{q}) U_{n \alpha',\bsl{k}}^* \left[ \widetilde{v}_{l}( \bsl{q} )\right]_{\bsl{\tau} i}^* \frac{m_{\bsl{\tau}}}{M_0}\ ,
\end{equation}
where $\bsl{k},\bsl{q}$ are summed over the DFT grid for $\bsl{k}$ and $\bsl{q}$.
$g_{\alpha\alpha', \bsl{\tau}i} (\bsl{R}, \bsl{R}')$ is local, \ie, $g_{\alpha\alpha', \bsl{\tau}i} (\bsl{R}, \bsl{R}')$ decays to zero for large $|\bsl{R} - \bsl{R}'|$, because if $\bsl{k},\bsl{q}$ are summed over an infinitely-fine grid, we have $g_{\alpha\alpha', \bsl{\tau}i} (\bsl{R}, \bsl{R}') = \bra{W_{\alpha 0} }\partial_{u_{\bsl{R}+\bsl{\tau},i}} V(\bsl{r}) |W_{\alpha' \bsl{R}'} \rangle $.
Therefore, we can put a cutoff on $|\bsl{R} - \bsl{R}'|$ for $g_{\alpha\alpha', \bsl{\tau}i} (\bsl{R}, \bsl{R}')$ and transform $g_{\alpha\alpha', \bsl{\tau}i} (\bsl{R}, \bsl{R}')$ back to momentum space, resulting in the interpolated $g_{m n l}(\bsl{k}, \bsl{q})$.

\subsubsection{Details for graphene}

We use the JDFTx software package~\cite{JDFTx} to calculate the electron-phonon interactions in graphene. For the DFT calculation, the electronic structure is computed using a $24\times24\times1$ k-point mesh. The 2D nature of the system is captured using a Coulomb truncation scheme~\cite{TruncatedEXX}. In short, this approach removes long-range Coulomb interactions in the aperiodic out-of-plane direction to ensure there is no artificial interactions between unit cells. 
To do this, the Coulomb potential $V(\bsl{r})$, which is normally long-ranged, is augmented by a Heaviside function $\Theta(\bsl{r}_z - \bsl{r})$ such that for positions in the out-of-plane direction beyond $\bsl{r}_z$ the Coulomb potential goes to zero. 
We use an in-plane lattice constant of $a = b = 2.46$~\AA. The electron-ion interaction is described using ultrasoft pseudopotentials~\cite{USPP}, 
and the exchange-correlation interaction is approximated using the PBEsol functional~\cite{PBEsol}. The kinetic energy cutoff for the plane-wave basis set is 40 Hartree. We use a Fermi-Dirac smearing with width of $\sigma = 0.01$~Hartree. 
As the name suggests, this smearing function is used to describe the occupation $F_{n\bsl{k}}$ of the Kohn-Sham electronic states $|\psi_{n\bsl{k}}\rangle$ using a Fermi-Dirac distribution function centered around the Fermi energy $\varepsilon_\mathrm{F}$:
\begin{equation}
    F_{n\bsl{k}} = \frac{1}{e^{(E_{n\bsl{k}} - \varepsilon_\mathrm{F})/\sigma + 1 }}.
\end{equation} 
To describe the phonons and EPC, we use the frozen phonon approach (\eqnref{eq:frozen-phonon}) on a $6\times6\times1$ supercell of the graphene unit cell. 
Then, all electronic states and EPC matrix elements are described in the basis of Maximally-localized Wannier functions, as described above. We use seven trial Wannier centers composed of two carbon $p_z$ orbitals, three Gaussian $s$-like orbitals located at the midpoint of the C-C bonds, and with two Gaussian $s$-like orbitals located above and below the center of the hexagon, respectively.

\subsubsection{Details for MgB$_2$}

In order to describe the electronic and phononic properties of MgB$_2$, we use the Quantum ESPRESSO~\cite{qe1,qe2} code. Calculations are carried out using a $12\times12\times12$ k-point grid and the plane-wave cutoff for the kinetic energy is 40 Hartree. We use a Methfessel-Paxton smearing~\cite{mpSmearing} with a width of 0.01 Hartree. 
We use norm-conserving pseudopotentials and the exchange-correlation interaction is described by the LDA. In these conditions we find the optimized unit cell parameters to be $a = b = 3.03$~\AA\ and $c = 3.47$~\AA. The phonons are modeled using DFPT~\cite{baroni2001} on a $6\times6\times6$ $\bsl{q}$-point grid. To interpolate the electron and electron-phonon interactions to much denser $\bsl{k}$ and $\bsl{q}$ grids, we use the EPW~\cite{EPW} and Wannier90~\cite{wannier90} codes. We describe five energy bands near the Fermi level starting with trial Wannier centers of two Boron $p_z$ orbitals and three $s$-like orbitals centered at positions (0,0.5,0.5), (0.5,0,0.5) and (0.5,0.5,0.5) in lattice coordinates of the unit cell. 
The Eliashberg spectral function (\eqnref{eq:alpha2F}) and EPC strength $\lambda$ (\eqnref{eq:lambda}) was then computed in EPW using a dense $60 \times 60 \times 60$ grid of both $\bsl{k}$-points and $\bsl{q}$-points. 

\subsection{Method II: No Wannier Interpolation}
\label{app:method_II_NO_wannier_interpolation}
 
 In this subsection, we describe the second method without Wannier interpolation.
 We only use this method for {\mgb} as a check.
 For method II, the DFT calculations and linear-response calculations were done using the QUANTUM-ESPRESSO package~\cite{qe1,qe2}. We used optimized norm-conserving pseudpotentials (ONCPSP)~\cite{ONCV,VANSETTEN201839}, with the Perdew-Burke-Ernzerhof (PBE) parametrization of the generalized gradient approximations (GGA) for the exchange-correlation functional~\cite{PBE}. A $84$ Ry cutoff was used for the plane-wave basis set. 
 The phonon calculation was performed on a $6 \times 6 \times 4$ mesh of the BZ for the phonon momentum, using the DFPT theory.
 The electron BZ integrals involved in DFPT were carried out in a $16 \times 16 \times 12$ mesh of BZ for the electron momentum.
 On the other hand, in order to compute the Fermi-surface average of the deformation potential for the EPC, we discretize the electron momentum on a finer mesh ($60 \times 60 \times 60$) of the BZ.

\subsection{Comparison of two Methods for {\mgb}}

The two methods give very close values of $\lambda$: $\lambda^{\abi,I} = 0.67$ from the method I and $\lambda^{\abi,II} = 0.645$ from the method II, where the relative error is $|(\lambda^{\abi,I}-\lambda^{\abi,II})/\lambda^{\abi,I}|=3.7\%$.
To further compare the results from the two methods, we plot the sum of $\hbar\Gamma_{nm}^\abi(\Gamma,\Gamma+(0,0,k_z))$ over the two degenerate bands along $\Gamma$-A near the Fermi level obtained from the two methods in \figref{fig:EPC_abi_MgB2_two_methods}, where $\Gamma_{nm}$ is defined in \eqnref{eq:Gamma_nm}.
As shown in \figref{fig:EPC_abi_MgB2_two_methods}, the two values are close---the averages of the plotted values from the two methods differ by $9.7\%$.
The consistency between the two methods indicates the strong reliability of our numerical calculations. 
 
 \begin{figure}
    \centering
    \includegraphics[width=0.5\columnwidth]{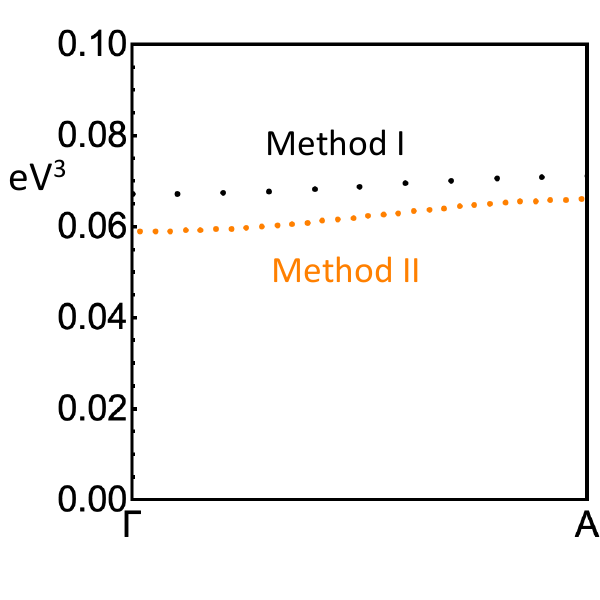} 
    \caption{
    The black and orange dots are the sum of $\hbar\Gamma_{nm}^\abi(\Gamma,\Gamma+(0,0,k_z))$ over the two degenerate bands along $\Gamma$-A near the Fermi level, obtained from the method I (\appref{app:method_I_wannier_interpolation}) and method II (\appref{app:method_II_NO_wannier_interpolation}) respectively. 
    }
    \label{fig:EPC_abi_MgB2_two_methods}
\end{figure}

 \section{Experimental Discussions}
 \label{app:experimental_discussions}
 
 In this section, we present discussions on the experimental aspects.
 
 \subsection{Graphene}
 
We first show that the energetic contribution $\lambda_E$ in \eqnref{eq:lambda_E_geo_graphene} can be directly measured from the width of the in-plane optical phonon at $\Gamma$ (\ie, the $E_{2g}$ phonons at $\Gamma$).
 
In general, for systems with spin $\SU(2)$ symmetry, the phonon linewidth $\gamma_{\bsl{q}l}$ of the $l$th phonon at $\bsl{q}$ caused by the EPC to the leading order has the following form
\eq{
\label{eq:phonon_linewidth_EPC}
\gamma_{\bsl{q}l}^{\text{EPC}} = \frac{4\pi}{N}\sum_{\bsl{k}}^{\BZ} \sum_{nm} |G_{nml}(\bsl{k},\bsl{k}+\bsl{q})|^2 \left[  n_F(E_n(\bsl{k})) - n_F(E_m(\bsl{k}+\bsl{q})) \right] \delta\left(E_{m}(\bsl{k}+\bsl{q})-E_{n}(\bsl{k})-\hbar \omega_{l}(\bsl{q}) \right)\ ,
}
where we choose the convention in \refcite{Lazzeri08292005PhononLinewidthGraphene,Bonini04072007PhononLinewidthGraphene} which have been heavily used for the phonon linewidth in graphene, $G_{nml}(\bsl{k}_1,\bsl{k}_2)$ is in \eqnref{eq:H_el_ph_gen}, $n_F$ is the Fermi-Dirac distribution, $n,m$ are spinless band indices, and the spin-double degeneracy of the electron bands has been included.
\eqnref{eq:phonon_linewidth_EPC} can be straightforwardly derived from Fermi golden rule.
For the phonons at $\Gamma$ that are Raman active, their linewidths of the phonons at $\Gamma$ can be measured in the Raman spectroscopy as the full width at half maximum (FWHM) of the corresponding peaks.

The $E_{2g}$ phonons at $\Gamma$ of Graphene are Raman active, and thus we focus on them.
Two $E_{2g}$ phonons are doubly degenerate, and thus let us consider the average linewidth of each $E_{2g}$ phonon, which reads
\eqa{
\label{eq:phonon_linewidth_EPC_graphene_Gamma_E2g_intermediate}
\gamma_{\Gamma,E_{2g}}^{EPC} & = \frac{1}{2}\sum_{l\in E_{2g}} \gamma_{\bsl{0}l}^{\text{EPC}} \\
& = \frac{2\pi}{N}\sum_{\bsl{k}}^{\BZ} \sum_{nm} \sum_{l\in E_{2g}} |G_{nml}(\bsl{k},\bsl{k})|^2 \left[ n_F(E_n(\bsl{k}))  -  n_F(E_m(\bsl{k})) \right] \delta\left(E_{m}(\bsl{k})-E_{n}(\bsl{k})-\hbar \omega_{E_{2g}}(\Gamma) \right) \\
& = \frac{2\pi}{N}\sum_{\bsl{k}}^{\BZ} \sum_{nm} \frac{\hbar}{2 \omega_{E_{2g}}(\Gamma)}\sum_{l\in E_{2g}} |\widetilde{G}_{nml}(\bsl{k},\bsl{k})|^2 \left[ n_F(E_n(\bsl{k}))  -  n_F(E_m(\bsl{k})) \right] \delta\left(E_{m}(\bsl{k})-E_{n}(\bsl{k})-\hbar \omega_{E_{2g}}(\Gamma) \right) \ ,
}
where $\widetilde{G}_{nml}(\bsl{k}_1,\bsl{k}_2)$ is defined in \eqnref{eq:Gt_nml}.
Since $\widetilde{G}_{nml}(\bsl{k},\bsl{k})=0$ for the acoustic phonons and the ion motions along $z$, we can sum $l$ in \eqnref{eq:phonon_linewidth_EPC_graphene_Gamma_E2g_intermediate} over all phonons without changing $\gamma_{\Gamma,E_{2g}}^{EPC}$, leading to
\eqa{
\gamma_{\Gamma,E_{2g}}^{EPC} & = \frac{2\pi}{N}\sum_{\bsl{k}}^{\BZ} \sum_{nm} \frac{\hbar}{2 \omega_{E_{2g}}(\Gamma)}\sum_{l} |\widetilde{G}_{nml}(\bsl{k},\bsl{k})|^2 \left[ n_F(E_n(\bsl{k}))  -  n_F(E_m(\bsl{k})) \right] \delta\left(E_{m}(\bsl{k})-E_{n}(\bsl{k})-\hbar \omega_{l}(\bsl{q}) \right) \\
& = \frac{2\pi}{N}\sum_{\bsl{k}}^{\BZ} \sum_{nm} \frac{1}{ \omega_{E_{2g}}(\Gamma)} \Gamma_{nm}(\bsl{k},\bsl{k}) \left[ n_F(E_n(\bsl{k}))  -  n_F(E_m(\bsl{k})) \right] \delta\left(E_{m}(\bsl{k})-E_{n}(\bsl{k})-\hbar \omega_{E_{2g}}(\Gamma) \right)\ ,
}
where expression of $\Gamma_{nm}(\bsl{k}_1,\bsl{k}_2)$ is in \eqnref{eq:Gamma_nm}.
Since we choose a NN hopping electron model and we choose the energy of Dirac point to be zero, the electron part has chiral symmetry.
Then, according to the convention in \appref{app:graphene}, $E_2(\bsl{k})=-E_1(\bsl{k})\geq 0$, and thus $\delta\left(E_{m}(\bsl{k})-E_{n}(\bsl{k})-\hbar \omega_{E_{2g}}(\Gamma) \right)$ is nonzero only for $m=2$ and $n=1$, leading to 
\eqa{
\label{eq:gamma_E_2g_EPC_genericT_Graphene}
\gamma_{\Gamma,E_{2g}}^{EPC} & = \frac{2\pi}{N}\sum_{\bsl{k}}^{\BZ} \frac{1}{ \omega_{E_{2g}}(\Gamma)} \Gamma_{12}(\bsl{k},\bsl{k}) \left[ n_F(E_1(\bsl{k}))  -  n_F(E_2(\bsl{k})) \right] \delta\left(E_{2}(\bsl{k})-E_{1}(\bsl{k})-\hbar \omega_{E_{2g}}(\Gamma) \right) \\ 
& = \frac{2\pi}{N}\sum_{\bsl{k}}^{\BZ} \frac{1}{ \omega_{E_{2g}}(\Gamma)} \Gamma_{12}(\bsl{k},\bsl{k}) \left[ n_F(-\frac{\hbar \omega_{E_{2g}}(\Gamma)}{2})  -  n_F(\frac{\hbar \omega_{E_{2g}}(\Gamma)}{2}) \right] \delta\left(2 E_{1}(\bsl{k})+\hbar \omega_{E_{2g}}(\Gamma) \right) \ .
}

Let us consider the case where $\mu\in ( -\frac{\hbar \omega_{E_{2g}}(\Gamma)}{2} , \frac{\hbar \omega_{E_{2g}}(\Gamma)}{2}  )$, and focus on zero temperature.
Then, the linewidth becomes 
\eqa{
\left. \gamma_{\Gamma,E_{2g}}^{EPC,T=0} \right|_{\mu\in \left( -\frac{\hbar \omega_{E_{2g}}(\Gamma)}{2} , \frac{\hbar \omega_{E_{2g}}(\Gamma)}{2}  \right)} = \frac{\pi}{ \omega_{E_{2g}}(\Gamma)} \frac{1}{N}\sum_{\bsl{k}}^{\BZ} \delta\left( E_{1}(\bsl{k})+\frac{\hbar \omega_{E_{2g}}(\Gamma)}{2} \right) \Gamma_{12}(\bsl{k},\bsl{k})  \ .
}
The relation between $\gamma_{\Gamma,E_{2g}}^{EPC}$ and $\lambda_E$ is hidden in  $\Gamma_{12}(\bsl{k},\bsl{k})$:
\eqa{
\Gamma_{12}(\bsl{k},\bsl{k}) & = \frac{\hbar}{2 m_C} \sum_{\bsl{\tau}i}\Tr\left[P_1(\bsl{k})f_{\bsl{\tau} i} (\bsl{k},\bsl{k}) P_2(\bsl{k})f_{\bsl{\tau} i}^\dagger (\bsl{k},\bsl{k}) \right] \\
& = \frac{\hbar \gamma^2 }{2 m_C}\sum_{\bsl{\tau}i} \Tr\left[P_1(\bsl{k}) (\chi_{\bsl{\tau}}\partial_{k_i}h(\bsl{k})- \partial_{k_i}h(\bsl{k}) \chi_{\bsl{\tau}}) P_2(\bsl{k})(\partial_{k_i}h(\bsl{k})\chi_{\bsl{\tau}}- \chi_{\bsl{\tau}} \partial_{k_i}h(\bsl{k}) ) \right] \\
& = \frac{\hbar \gamma^2 }{4 m_C}\sum_{i=x,y} \Tr\left[P_1(\bsl{k}) (\tau_z\partial_{k_i}h(\bsl{k})- \partial_{k_i}h(\bsl{k})\tau_z) P_2(\bsl{k})(\partial_{k_i}h(\bsl{k})\tau_z-\tau_z \partial_{k_i}h(\bsl{k}) ) \right] \\
& = \frac{\hbar \gamma^2 }{4 m_C}\sum_{i=x,y} \Tr\left[P_1(\bsl{k}) (\tau_z\partial_{k_i}h(\bsl{k})\tau_z- \partial_{k_i}h(\bsl{k})) P_1(\bsl{k})(\tau_z\partial_{k_i}h(\bsl{k})\tau_z- \partial_{k_i}h(\bsl{k}) ) \right] \ .
}
As $\partial_{k_i}h(\bsl{k})$ only involves $\tau_x$ and $\tau_y$ as shown in \eqnref{eq:h_k_graphene}, we have $\tau_z\partial_{k_i}h(\bsl{k})\tau_z = - \partial_{k_i}h(\bsl{k})$, and $\Gamma_{12}(\bsl{k},\bsl{k})$ can be further simplified to 
\eqa{
\Gamma_{12}(\bsl{k},\bsl{k})  & =  \frac{\hbar \gamma^2}{m_C} \sum_{i=x,y} \Tr\left[P_1(\bsl{k})  \partial_{k_i}h(\bsl{k}) P_1(\bsl{k})  \partial_{k_i}h(\bsl{k})  \right] \ .
}
Owing to the Hellmann–Feynman theorem, we know 
\eq{
P_1(\bsl{k})  \partial_{k_i}h(\bsl{k}) P_1(\bsl{k})   = \partial_{k_i} E_1(\bsl{k}) P_1(\bsl{k})\ ,
}
and thus
\eqa{
\Gamma_{12}(\bsl{k},\bsl{k})  & =  \frac{\hbar \gamma^2}{m_C} \sum_{i=x,y} \left( \partial_{k_i} E_1(\bsl{k}) \right)^2  = \frac{\hbar \gamma^2}{m_C} \left| \nabla_{\bsl{k}} E_1(\bsl{k}) \right|^2\ .
}
As a result, the average linewidth of each $E_{2g}$ phonons finally becomes
\eqa{
\left. \gamma_{\Gamma,E_{2g}}^{EPC,T=0} \right|_{\mu\in \left( -\frac{\hbar \omega_{E_{2g}}(\Gamma)}{2} , \frac{\hbar \omega_{E_{2g}}(\Gamma)}{2}  \right)} & = \frac{\hbar \gamma^2 \pi }{\omega_{E_{2g}}(\Gamma)  m_C} \frac{\Omega}{(2\pi)^2}\int_{E_{1}(\bsl{k}) = -\frac{\hbar \omega_{E_{2g}}(\Gamma)}{2} } d\sigma_{\bsl{k}} \frac{1}{\left| \nabla_{\bsl{k}} E_1(\bsl{k}) \right|} \left| \nabla_{\bsl{k}} E_1(\bsl{k}) \right|^2  \\
& = \frac{\hbar \gamma^2 \pi}{ \omega_{E_{2g}}(\Gamma)  m_C} \frac{\Omega}{(2\pi)^2} \int_{E_{1}(\bsl{k}) = -\frac{\hbar \omega_{E_{2g}}(\Gamma)}{2} } d\sigma_{\bsl{k}} \left| \nabla_{\bsl{k}} E_1(\bsl{k}) \right|\ ,
}
where the integral is along the line for $E_{1}(\bsl{k}) = -\frac{\hbar \omega_{E_{2g}}(\Gamma)}{2} $.
By comparing to $\lambda_E$ in \eqnref{eq:lambda_E_geo_graphene}, we arrive at
\eqa{
\left. \gamma_{\Gamma,E_{2g}}^{EPC,T=0} \right|_{\mu\in \left( -\frac{\hbar \omega_{E_{2g}}(\Gamma)}{2} , \frac{\hbar \omega_{E_{2g}}(\Gamma)}{2}  \right)}  & = \frac{\pi \hbar}{\omega_{E_{2g}}(\Gamma) } \frac{ \gamma^2}{  m_C} \frac{\Omega}{(2\pi)^2} \int_{E_{1}(\bsl{k})  = -\frac{\hbar \omega_{E_{2g}}(\Gamma)}{2} } d\sigma_{\bsl{k}} \left| \nabla_{\bsl{k}} E_1(\bsl{k}) \right|\\
& = \frac{ \pi \hbar \mcomega}{ \omega_{E_{2g}}(\Gamma) } \left. \lambda_E\right|_{\mu =-\frac{\hbar \omega_{E_{2g}}(\Gamma)}{2}} \ ,
}
which, combined with the approximated analytical expression of $\mcomega$ in \eqnref{eq:mcomega_graphene_linear}, results in
\eqa{
\label{eq:lambda_E_gamma_0_E2g_graphene}
& \left. \gamma_{\Gamma,E_{2g}}^{EPC,T=0}\right|_{\mu\in \left( -\frac{\hbar \omega_{E_{2g}}(\Gamma)}{2} , \frac{\hbar \omega_{E_{2g}}(\Gamma)}{2}  \right)} = \frac{ \hbar 2 \pi  \omega_{E_{2g}}(\Gamma) \omega_{A_1'}^2(K) }{\left( \omega_{A_1'}^2(K) + \omega_{E_{2g}}^2(\Gamma) \right) } \left. \lambda_E\right|_{\mu =-\frac{\hbar \omega_{E_{2g}}(\Gamma)}{2}} \\
& \Rightarrow \left. \lambda_E\right|_{\mu =-\frac{\hbar \omega_{E_{2g}}(\Gamma)}{2}}  = \frac{1}{2 \pi \hbar \omega_{E_{2g}}(\Gamma)} \left( 1 + \frac{\omega_{E_{2g}}^2(\Gamma)}{\omega_{A_1'}^2(K)} \right) \left. \gamma_{\Gamma,E_{2g}}^{EPC,T=0}\right|_{\mu\in \left( -\frac{\hbar \omega_{E_{2g}}(\Gamma)}{2} , \frac{\hbar \omega_{E_{2g}}(\Gamma)}{2}  \right)}\ .
}
Therefore, if we can measure the low-temperature average phonon linewidth of the $E_{2g}$ phonons at $\Gamma$ that is \emph{caused by EPC} for $\mu\in \left( -\frac{\hbar \omega_{E_{2g}}(\Gamma)}{2} , \frac{\hbar \omega_{E_{2g}}(\Gamma)}{2}  \right)$, it, combined with the measured phonon frequencies, gives a direct measurement of the energetic contribution $\lambda_E$ at $\mu =-\frac{\hbar \omega_{E_{2g}}(\Gamma)}{2}$.

The low-temperature average phonon linewidth of the $E_{2g}$ phonons at $\Gamma$ that is \emph{caused by EPC} is measurable in graphene.
In graphene, the low-temperature average phonon linewidth of the $E_{2g}$ phonons at $\Gamma$ does not solely comes from EPC---the EPC contribution is about 85\% according to \refcite{Han07012021GrapheneFWHMDFT}---the remaining contribution is from anharmonicities.
In short, we have
\eq{
\gamma_{\Gamma,E_{2g}} = \gamma_{\Gamma,E_{2g}}^{EPC} + \gamma_{\Gamma,E_{2g}}^{an}\ ,
}
where $\gamma_{\Gamma,E_{2g}}$ is the total low-temperature average phonon linewidth of one of the $E_{2g}$ phonons at $\Gamma$, and $\gamma_{\Gamma,E_{2g}}^{an}$ is the part that comes from anharmonicities.
Now the question becomes how to separate $\gamma_{\Gamma,E_{2g}}^{EPC}$ from the total linewidth $\gamma_{\Gamma,E_{2g}}$.
As shown in \eqnref{eq:gamma_E_2g_EPC_genericT_Graphene}, the zero-$T$ $\gamma_{\Gamma,E_{2g}}^{EPC}$ has strong chemical potential dependence, \ie, $\gamma_{\Gamma,E_{2g}}^{EPC,T=0} = 0$ for $|\mu|>\frac{\hbar\omega_{E_{2g}}(\Gamma)}{2}\approx 0.1 eV$ and $\gamma_{\Gamma,E_{2g}}^{EPC,T=0}$ has a discontinuous jump as $|\nu|$ is tuned continuously into the range $|\mu|<\frac{\hbar\omega_{E_{2g}}(\Gamma)}{2}\approx 0.1 eV$.
On the other hand, $\gamma_{\Gamma,E_{2g}}^{an}$ only has weak dependence on the chemical potential~\cite{Yan2007EPCGrapheneEXPRamen}.
Therefore, one can get the value of $ \gamma_{\Gamma,E_{2g}}^{EPC,T=0} $ for small chemical potential from the difference of low-temperature total linewidth $ \gamma_{\Gamma,E_{2g}}$ at $\mu=0$ and large $\mu$:
\eq{
\left. \gamma_{\Gamma,E_{2g}}^{EPC,T=0}\right|_{\mu\in \left( -\frac{\hbar \omega_{E_{2g}}(\Gamma)}{2} , \frac{\hbar \omega_{E_{2g}}(\Gamma)}{2}  \right)} \approx  \left. \gamma_{\Gamma,E_{2g}}^{\text{low T}} \right|_{\mu=0} - \left. \gamma_{\Gamma,E_{2g}}^{\text{low T}} \right|_{|\mu|>0.1\text{eV}}\ .
}
With this method, the Ramen measurement in \refcite{Yan2007EPCGrapheneEXPRamen} suggests 
\eq{
\frac{1}{h c}\left. \gamma_{\Gamma,E_{2g}}^{EPC,T=0}\right|_{\mu\in \left( -\frac{\hbar \omega_{E_{2g}}(\Gamma)}{2} , \frac{\hbar \omega_{E_{2g}}(\Gamma)}{2}  \right)} = 7\sim 13 \text{cm}^{-1}\ ,
}
where $h$ is the Planck constant, $c$ is the speed of light, and the error is considerable---about $\pm 3 \text{cm}^{-1}$.
Experimentally, $\hbar \omega_{E_{2g}}(\Gamma)  = 0.196$eV (\ie, $1582 \text{cm}^{-1}$) according to the measurement in graphene~\cite{Han07012021GrapheneFWHMDFT} and $\hbar \omega_{A_1}(K)  = 0.156$eV (\ie, $1260 \text{cm}^{-1}$) according to the measurement in graphite~\cite{Maultzsch11092003Graphite}.
With these experimental values, \eqnref{eq:lambda_E_gamma_0_E2g_graphene} suggests
\eq{
\left. \lambda_E \right|_{\mu\approx -0.1\text{eV}} = 0.0018\sim 0.0034\ ,
}
whereas the model calculation suggests
\eq{
\left. \lambda_E \right|_{\mu = -0.1\text{eV}} = 0.0032\ ,
}
which is within the experimental errors.
More precise measurement can be done in the future.

The total $\lambda$ of graphene may be measured from the Helium scattering~\cite{Benedek09082021EPCConstantGrapheneHe}.
Combined with the experimental observation of $\lambda_E$, it can serve as a probe as the geometric contribution $\lambda_{geo}$.
Furthermore, the FSM in graphene may be measured from the current noise spectrum~\cite{Neupert06102013CurrentNoiseFSM} or more generally the first-order optical response~\cite{Sipe06291999OpticalResponseFSM}, owing to the two-band nature of graphene.
Therefore, the following expression may be experimentally testable:
\eq{
\label{eq_lambda_geo_lambda_E_optical_absorption}
\frac{\lambda_{geo}}{\lambda_E} = \frac{\int_{FS} d\sigma_{\bsl{k}} \frac{\Delta E^2(\bsl{k})}{|\nabla_{\bsl{k}} E_{n_F}(\bsl{k})|} \Tr[g_{n_F}(\bsl{k})]}{\int_{FS} d\sigma_{\bsl{k}} |\nabla_{\bsl{k}} E_{n_F}(\bsl{k})|} = \frac{|\mu|}{\pi} \int_{FS} d\sigma_{\bsl{k}} \frac{\Tr[g_{n_F}(\bsl{k})]}{|\nabla_{\bsl{k}} E_{n_F}(\bsl{k})|} = \frac{h c }{2 \pi^2 e^2 } A(\omega = 2|\mu|/\hbar)\ , 
}
where 
\eq{
A(\omega) = \frac{4 \hbar |\mu|}{c h} \hbar \omega \int_{FS} d\sigma_{\bsl{k}} \frac{\Tr[g_{n_F}(\bsl{k})]}{|\nabla_{\bsl{k}} E_{n_F}(\bsl{k})|}
}
is the optical absorption coefficient for photons with frequency $\omega$ in the unit system where $1/(4\pi\epsilon_0)= 1$~\cite{Sipe06291999OpticalResponseFSM,Geim2008OpticalGraphene,Mak2008OpticalGraphene,Mak2012optical}.
Here we have used the expressions of $\lambda_{geo}$ and $\lambda_E$ below \eqnref{eq:lambda_E_geo_graphene}.
Known experiments observe $A(\omega) \approx 2.3\%$ for $\hbar \omega > 0.5 \eV$, which gives $ \frac{h c }{2 \pi^2 e^2 } A(\omega = 2|\mu|/\hbar) \approx 1.0$, which is consistent with \eqnref{eq:lambda_ratios_linear}.
We note that such test is for $|\mu|>\hbar \omega /2 >0.25$eV, while our proposal of measuring $\lambda_E$ is for $\mu \approx -0.1$eV.
Ff one can measure $A(\omega)$ down to $\hbar\omega = 0.1 \eV$ while measureing $\lambda_E$ and $\lambda_{geo}$ based on our proposals, these measurements would serve as a good test our theory, and relate the $\frac{\lambda_{geo}}{\lambda_E}$ in scattering experiments to the absorption coefficient $A(\omega)$ in the optical response.

\subsection{Other Systems}

Besides graphene, it is possible to verify the relation between quantum geometry and the EPC strength in other systems.
On the surface of the topological insulator Bi$_2$Se$_3$, the geometric properties of the Bloch states should vary along the Fermi surface, due to the hexagonal distortion~\cite{Chen07102009Bi2Te3}.
Furthermore, the coupling between the surface states and the $\Gamma$ phonons in principle should be measurable in the momentum-resolved way in time- and angle-resolved photoemission spectroscopy measurements~\cite{Hofmann04262011EPClambdaARPES,Pan09162011EPClambdaARPES,Sobota2014Bi2Se3EPCArpes,Baldini2023Ta2NiSe5EPCArpes}.
Then, we can compare the EPC strength and the geometric quantities (like FSM or OFSM) of the surface states, and check whether they have similar behaviors as varying momentum on the Fermi surface.

\end{document}